\documentclass[11pt,a4paper]{article}

\usepackage{graphicx}
\usepackage{float}
\usepackage{afterpage}
\usepackage{epsfig,cite}
\usepackage{amssymb}
\usepackage{amsmath}
\usepackage{dsfont}
\usepackage{multirow}
\usepackage{url,hyperref}

\usepackage{url}
\usepackage{hyperref}

\newcommand{\be}{\begin{equation}}
\newcommand{\ee}{\end{equation}}
\newcommand{\bea}{\begin{eqnarray}}
\newcommand{\eea}{\end{eqnarray}}
\newcommand{\bi}{\begin{itemize}}
\newcommand{\ei}{\end{itemize}}
\newcommand{\ben}{\begin{enumerate}}
\newcommand{\een}{\end{enumerate}}
\newcommand{\la}{\left\langle}
\newcommand{\ra}{\right\rangle}
\newcommand{\lc}{\left[}
\newcommand{\rc}{\right]}
\newcommand{\lp}{\left(}
\newcommand{\rp}{\right)}

\def\frac#1#2{{{#1}\over {#2}}}
\def\gsim{\mathrel{\rlap{\lower4pt\hbox{\hskip1pt$\sim$}}
    \raise1pt\hbox{$>$}}}         
\def\lsim{\mathrel{\rlap{\lower4pt\hbox{\hskip1pt$\sim$}}
    \raise1pt\hbox{$<$}}}         

\newcommand{\rep}{\mathrm{rep}}

\newcommand{\draft}[1]{}

\def\beq{\begin{equation}}  
\def\eeq{\end{equation}}  

\def \n0{N_j^{(0)}}

\def\lapprox{\lower .7ex\hbox{$\;\stackrel{\textstyle <}{\sim}\;$}}
\def\gapprox{\lower .7ex\hbox{$\;\stackrel{\textstyle >}{\sim}\;$}}

\def\GeV{{\rm GeV}}

\begin{document}
\begin{figure}[h]
\epsfig{width=0.32\textwidth,figure=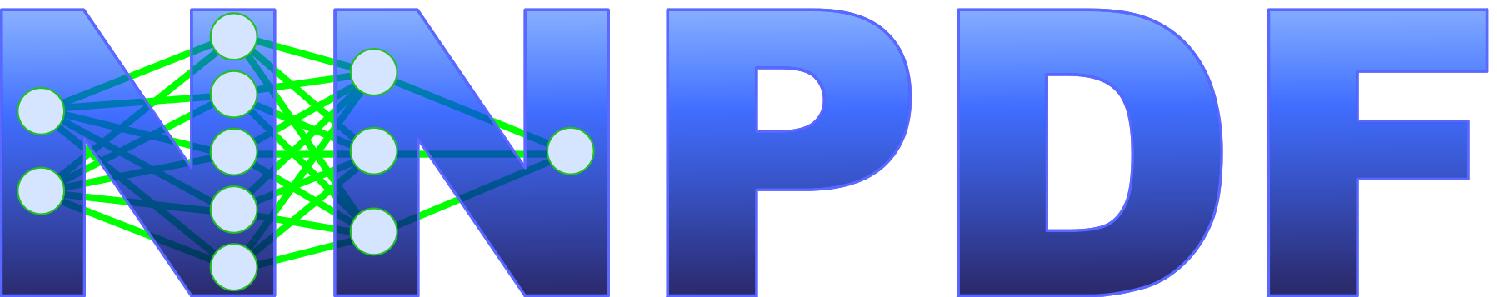}
\end{figure}
\vspace{-2.0cm}
\begin{flushright}
Edinburgh 2014/15\\
IFUM-1034-FT\\
CERN-PH-TH/2013-253\\
OUTP-14-11p \\
CAVENDISH-HEP-14-11\\
\end{flushright}

\begin{center}
{\Large \bf Parton distributions for the LHC Run II}
\vspace{.7cm}

{\bf  The NNPDF Collaboration:}\\

Richard~D.~Ball$^{1,2}$, Valerio~Bertone$^{2}$,
Stefano~Carrazza$^{4,2}$,\\ Christopher~S.~Deans$^1$,
Luigi~Del~Debbio$^1$, Stefano~Forte$^4$, Alberto~Guffanti$^{5}$,\\ 
Nathan~P.~Hartland$^1$, Jos\'e~I.~Latorre$^3$, Juan~Rojo$^{6}$ 
and Maria~Ubiali$^{7}$

\vspace{.3cm}
{\it ~$^1$ The Higgs Centre for Theoretical Physics, University of Edinburgh,\\
JCMB, KB, Mayfield Rd, Edinburgh EH9 3JZ, Scotland\\
~$^2$  PH Department, TH Unit, CERN,\\ CH-1211 Geneva 23, Switzerland\\
~$^3$ Departament d'Estructura i Constituents de la Mat\`eria, 
Universitat de Barcelona,\\ Diagonal 647, E-08028 Barcelona, Spain\\
~$^4$ Dipartimento di Fisica, Universit\`a di Milano and
INFN, Sezione di Milano,\\ Via Celoria 16, I-20133 Milano, Italy\\
~$^5$ Niels Bohr International Academy and Discovery Center, \\
Niels Bohr Institute,  University of Copenhagen, \\
Blegdamsvej 17, DK-2100 Copenhagen, Denmark\\
~$^6$ Rudolf Peierls Centre for Theoretical Physics, 1 Keble Road,\\ University of Oxford, OX1 3NP Oxford, UK\\
~$^7$ The Cavendish Laboratory, University of Cambridge,\\
J.J. Thomson Avenue, CB3 0HE, UK}
\end{center}

\vspace{0.1cm}

\begin{center}
{\bf \large Abstract}

\end{center}

We present NNPDF3.0, the first  set of parton distribution
functions (PDFs) determined
with a
methodology validated by a closure test.
NNPDF3.0 uses a global dataset including HERA-II deep-inelastic
inclusive 
cross-sections, the combined HERA charm data, jet production 
from ATLAS and CMS, vector boson rapidity and transverse momentum
distributions 
from ATLAS, CMS and LHCb, $W$+$c$ data from CMS and top quark pair
production
total cross sections from ATLAS and CMS.
Results are based on LO, NLO and NNLO QCD theory and also include
electroweak corrections. To validate our methodology, we show  
that PDFs determined from pseudo-data
generated  from a known underlying law  correctly reproduce the
statistical distributions expected on the basis of the
assumed experimental uncertainties. This closure test ensures that our
methodological uncertainties are negligible in comparison to the
generic
theoretical and experimental uncertainties of PDF determination.
This enables us to
determine with confidence  PDFs at different perturbative orders and 
using a variety of
experimental datasets ranging from HERA-only  up to a global
set including the latest LHC results, all using precisely 
the same validated methodology.
We explore some of the 
phenomenological implications of our results for the 
upcoming 13 TeV Run of the LHC,
in particular for Higgs production cross-sections.

\clearpage

\tableofcontents

\clearpage

\section{Introduction}
\label{sec:introduction}

Parton distribution functions (PDFs) are currently one of the major
sources of uncertainty in processes at hadron colliders, specifically at
the LHC. Ever since they have been quantified for the first time~\cite{dflm},
it has been recognized that PDF uncertainties stem from three
sources: the underlying data (which are affected by statistical and
systematic errors), the theory which is used to describe them
(which is typically based on the truncation of a perturbative
expansion) and the procedure which is used to extract the PDFs from
the data. This last source of uncertainty is the most elusive, and because
of it, it was thought until very recently~\cite{demortier} that PDF
uncertainties are not yet understood from a statistical point of view.

In a series of
papers~\cite{Forte:2002fg,DelDebbio:2004qj,DelDebbio:2007ee,Ball:2008by,Rojo:2008ke,Ball:2009mk,Ball:2009qv,Ball:2010de,Ball:2011mu,Ball:2011uy,Ball:2012cx}
we have introduced a methodology aimed at reducing as much as possible
this procedural uncertainty, up to the point that
it can be neglected in comparison to the remaining data and theory
uncertainty. However, as data become more abundant and precise, and
theoretical calculations more accurate, a full characterization of the
procedural uncertainties becomes necessary.

In the present paper we construct for the first time a set of PDFs
with 
an explicit characterization of procedural uncertainties. We do
so by using a methodology which has been repeatedly
suggested~\cite{demortier}, and recently used to study specific
sources of PDF uncertainty~\cite{Watt:2012tq} though never (to the best
of our knowledge) for full characterization: the so called ``closure
test''. The idea is to assume that the underlying PDFs are known, and use
this assumed set of PDFs to generate artificial data. One may then set
theoretical uncertainties to zero (by using the same theory to
generate data and to analyze them), and  fix the data
uncertainty to any desired value. By determining
PDFs from these artificial data, one can then tune the methodology
until procedural uncertainties have been removed (or at least made
small enough to be undetectable), and the ensuing PDFs faithfully
reproduce the data uncertainty. Thus
one may explicitly check that the output PDFs provide consistent and
unbiased ~\cite{cowan} estimators of the underlying truth, that
confidence levels reproduce the correct coverage (e.g., that the true
value is indeed within the 68\% interval in 68\% of cases), and so
on. Of course, a full closure test requires verifying that the
procedure is robust: so, for example, that the conclusion that there
are no procedural uncertainties is independent of the assumed
underlying set of PDFs.

This program will be fully realized in this paper, whose core is a
discussion of the closure test itself. The main motivation for doing
so at this time is, as mentioned, that data are more and more abundant
and accurate and that more and more higher-order calculations have
become available. 
In this work
 we will use a very wide dataset, which in
particular includes, besides the data used in previous global PDF fits, 
all the most recent HERA deep-inelastic scattering
data for inclusive and charm production
cross-sections, 
essentially all the most recent single-inclusive jet data and $W$
and $Z$ rapidity distributions from
the LHC, and also, for the first time in a
global fit, $W$+$c$ production data, double-differential and high-mass Drell-Yan
distributions, $W$ transverse momentum distributions, and top total
cross section data from ATLAS and CMS. 
For many of these data, higher-order calculations
and/or fast computational interfaces
have become available only recently.  The dataset on which
our PDF determination is based is summarized in Sect.~\ref{sec:expdata}, where
in particular all new data included on top
of those used for the previous NNPDF2.3 set~\cite{Ball:2012cx} are
discussed. The theoretical calculations and tools used for their
description and inclusion  in the PDF fit are summarized in
Sect.~\ref{sec:theorytools}.

The  inclusion of these new data and processes in our
PDF fitting code has required very substantial computational upgrades,
including a migration of the code to C{}\verb!++!.
Also, whereas the basic principles of our methodology are the same as in our
previous PDF determinations, namely, a Monte Carlo approach based on
using neural networks as underlying interpolating functions, several
improvements have become necessary in order to ensure the full
independence of procedural uncertainties which is required to pass the
closure test. This includes a more general way to define the basis
PDFs (needed in order to test basis-independence in the
closure test); an improved way of making sure that the
so-called preprocessing exponents which are introduced as part of the
PDF parametrization do not bias the results of the fit; an improved
version of the genetic algorithm used for PDF minimization; and an
improved form of the cross-validation procedure which is used in order
to determine the optimal fit (which in the presence of a redundant
parametrization is not the absolute minimum of the figure of merit,
since this would require fitting noise). All these methodological
aspects are summarized in Sect.~\ref{sec:methodology}, where we also summarize
the general structure and features of the new NNPDF 
 C{}\verb!++! code.

The closure test procedure is presented in
Sect.~\ref{sec:closure}. We actually perform three different kinds of
closure test, which we call respectively Level-0, Level-1 and
Level-2. In the Level-0 closure test, perfect data are generated from
the assumed underlying law, with  no uncertainty. The test is
successful if a perfect fit to the data can be produced, i.e., a fit
with vanishing $\chi^2$. By showing that our PDFs pass this test we
prove that our methodology is not limited by choice of basis
functions, functional
form, number of parameter, or minimization algorithm. Also, by looking
at the remaining uncertainty on the final PDF we can determine what we
call the ``extrapolation uncertainty'', i.e., the uncertainty due to
the fact that data points, even when infinitely precise, are not
covering the whole of phase space. In a Level-1 closure test, data
are generated with an uncertainty: this gives a full pseudodata set,
which  thus simulates a realistic
situation (but with knowledge of the underlying physical law). 
However, instead of using the NNPDF Monte Carlo methodology, whereby
pseudodata replicas are generated about the experimental data, we
simply fit PDF replicas all to the same pseudodata. Despite the fact
that the data are not fluctuating (i.e. the replicas are all fitted to
the same pseudodata set), we find that results are affected by an
uncertainty which we call  ``functional uncertainty'', and which is
due to the fact that we are reconstructing a set of functions from a
finite amount of discrete information. In a Level-2 closure test, once
pseudodata are constructed, they are treated as real data according to
the NNPDF methodology. We can then check that the closure test is
successful with a wide variety of indicators, which include
testing that the PDF replicas behave according to Bayes'
theorem upon the inclusion of new (pseudo)data. We finally check that
our results are stable upon a wide variety of changes, such as different
assumed underlying PDFs, different choices of basis, different size of
the neural network used for parametrization, and so on. 
We also show how ``self-closure'' of the NNPDF3.0 set is succesful,
by using it as the input of a Level-2 closure test.

Finally, we present the NNPDF3.0 set, which, as usual, we deliver
at LO, NLO and NNLO in QCD theory, for a variety of values of
$\alpha_s$, and  for a variety of values of the maximum value of
active flavor numbers. Also, we provide PDFs determined using   various subsets
of our full dataset. This includes sets based on HERA data only (which
may thus be compared to HERAPDF PDFs~\cite{Radescu:2010zz,CooperSarkar:2011aa},
 and may be used
for applications where one wishes to use a PDF set based on a
restrictive but consistent dataset); sets based on a ``conservative''
data subset, which is found by studying a specific statistical
indicator introduced previously by us~\cite{Ball:2010gb} which
measures the consistency of each dataset with the remaining data,
thereby allowing the removal data which are less compatible with the
bulk of the dataset; sets based on the same data as our previous
NNPDF2.3 dataset, thereby enabling us to separate the impact of new
data from that of improved methodology when comparing NNPDF2.3 to
NNPDF3.0; a set without LHC data, which is useful in order to
gauge the impact of the latter; and sets
where the HERA data are supplemented by
all available data from either ATLAS or CMS,
useful to compare with related studies presented by the LHC
collaborations.
 All these datasets and their features
are presented and compared in
Sect.~\ref{sec:results}, where we also discuss in light of our
results  specific issues of particular recent relevance in PDF
determination: jet data and their impact on the gluon distribution,
and the strange PDF which follows from both deep-inelastic and hadronic
scattering data.

Finally, in  Sect.~\ref{sec:phenomenology} we briefly discuss the 
implications of the NNPDF3.0 set for LHC
phenomenology, with special regard to Higgs production in gluon fusion, and 
a number of other representative processes, for
a center-of-mass energy of  13 TeV. 
Information 
on delivery, usage, and future developments is collected in
Sect.~\ref{sec:delivery}, while some technical results are relegated to 
appendices: in Appendix~\ref{app-ew}
we provide more details for the computation of QCD and electroweak corrections
to LHC Drell-Yan data and in Appendix~\ref{app:distances} we present
the definitions of the distance estimators we use 
to compare between different sets
of PDFs.

\section{Experimental data}
\label{sec:expdata}

Here we present the new experimental data that have been added
in the NNPDF3.0 analysis, and their treatment.
After reviewing  the NNPDF3.0 dataset, which includes most of the data on
which the previous NNPDF2.3 set was based,
we discuss each of the new datasets in turn.
 We then review our
theoretical treatment, and specifically discuss the
 computational tools used to implement
perturbative corrections, and particular issues related to 
 NNLO QCD corrections to jet production,  electroweak corrections, and
 the treatment of heavy quark mass effects.
We finally review the way the total dataset is constructed, and  in
particular which data and which cuts are included at each
perturbative order, and the treatment of experimental uncertainties,
specifically the systematics.

\subsection{Overview}
\label{sec:dataover}

The NNPDF2.3 fit~\cite{Ball:2012cx} included fixed-target deep inelastic scattering (DIS) 
data from NMC~\cite{Arneodo:1996kd,Arneodo:1996qe}, BCDMS~\cite{bcdms1,bcdms2}
and SLAC~\cite{Whitlow:1991uw}; the combined HERA-I DIS dataset~\cite{Aaron:2009aa}, HERA
$F_L$~\cite{h1fl} and the separated ZEUS and H1 $F_2^c$ structure function
data~\cite{Breitweg:1999ad,Chekanov:2003rb,Chekanov:2008yd,Chekanov:2009kj,Adloff:2001zj,
  Collaboration:2009jy,H1F2c10:2009ut}, some ZEUS HERA-II DIS
cross-sections~\cite{Chekanov:2009gm,Chekanov:2008aa};
CHORUS inclusive neutrino DIS~\cite{Onengut:2005kv}, and
NuTeV dimuon production data~\cite{Goncharov:2001qe,MasonPhD};
fixed-target E605~\cite{Moreno:1990sf} and
E866~\cite{Webb:2003ps,Webb:2003bj,Towell:2001nh} Drell-Yan production
data; CDF W asymmetry~\cite{Aaltonen:2009ta} and
CDF~\cite{Aaltonen:2010zza} and D0~\cite{Abazov:2007jy} $Z$ rapidity
distributions; CDF~\cite{Aaltonen:2008eq} and D0~\cite{D0:2008hua}
Run-II one-jet inclusive cross-sections; ATLAS~\cite{Aad:2011dm}, 
CMS~\cite{Chatrchyan:2012xt} 
and LHCb~\cite{Aaij:2012vn} data on vector boson production; 
and ATLAS one-jet inclusive cross-sections~\cite{Aad:2011fc} 
from the 2010 run.

In NNPDF3.0, several new datasets have been included.
First, we have included all relevant inclusive
cross section measurements from HERA Run 
II~\cite{Aaron:2012qi,Collaboration:2010ry,ZEUS:2012bx,Collaboration:2010xc}.
This includes the complete set of inclusive measurements from
H1 and ZEUS, as well as new H1 data at low $Q^2$ and high-$y$.
In addition, we have replaced the separate H1 and ZEUS $F_2^c$ data with the 
combined measurements of the charm production
cross section $\sigma_{cc}^{\rm red}$~\cite{Abramowicz:1900rp}.
The separate HERA-II H1 and ZEUS data, as well
as the inclusive HERA-I data, will be replaced with
the combined HERA dataset as soon as it becomes available. 
However,
we expect this replacement
to have a small impact
on the PDFs, since the neural network fit effectively performs
a dataset combination.
This was explicitly demonstrated for 
 the combined HERA-I dataset in the NNPDF2.0 analysis~\cite{Ball:2010de}.

We have  included new vector boson production data from the LHC at a center-of-mass
energy of 7 TeV from the 2010 and 2011 runs: 
the 5 fb$^{-1}$ high-mass Drell-Yan production~\cite{Aad:2013iua} and the
31 pb$^{-1}$ transverse momentum
distribution of $W$ bosons data~\cite{Aad:2011fp} from ATLAS;
the 5 fb$^{-1}$ $W$ muon asymmetry~\cite{Chatrchyan:2013mza},
the double differential Drell-Yan cross
sections~\cite{CMSDY}, and
the differential distributions for $W$ production in association with
charm quarks from CMS~\cite{Chatrchyan:2013uja};
and finally the 940 pb$^{-1}$ forward $Z\to ee$ production
data from LHCb from the 2011 run~\cite{Aaij:2012mda}. 
In comparison to the NNPDF2.3 dataset, we have removed the CDF $W$ asymmetry
data~\cite{Aaltonen:2009ta}, since more precise and cleaner data from the LHC 
(based on leptons rather than on the reconstructed $W$) covers the same
region in $x$ and extends their kinematical coverage.

%
Concerning LHC jet measurements,
we have complemented  the previously available ATLAS 7 TeV inclusive
jet production data from the 2010 run~\cite{Aad:2011fc} 
with the 7 TeV CMS 5 fb$^{-1}$ inclusive jet cross-sections
from the 2011 run~\cite{Chatrchyan:2012bja},
and the ATLAS 2.76 TeV jet cross section data~\cite{Aad:2013lpa}, 
supplemented by their
correlations with the 7 TeV cross sections.
In comparison  to NNPDF2.3, we have removed the Tevatron D0 Run II inclusive jet
cross section measurements~\cite{D0:2008hua}, which  were
obtained with the midpoint algorithm, which
is infrared unsafe~\cite{Salam:2007xv} and therefore cannot be used in conjunction with  
NNLO calculations.
On the other hand, the CDF Run II inclusive jet measurements, based on the $k_t$ algorithm, are retained in NNPDF3.0.

Finally, we have included total cross section measurements for
top quark pair production data
from ATLAS and CMS
at 7 and 8 TeV~\cite{ATLAS:2012aa,ATLAS:2011xha,TheATLAScollaboration:2013dja,
Chatrchyan:2013faa,Chatrchyan:2012bra,Chatrchyan:2012ria}, for a total of six data points. 

Besides extending the traditional dataset used in our previous global
fit, NNPDF2.3, 
based on deep-inelastic scattering, Drell-Yan, and jet data, to include
new information from HERA and the LHC (whose kinematic coverage and
resolution is greatly expanded, specifically through the inclusive
jet production and the high-mass and
double-differential gauge boson production), the NNPDF3.0 dataset 
uses for the first time in a global fit many datasets whose
importance has been repeatedly emphasized for  constraining particular PDF
combinations~\cite{Forte:2010dt,Forte:2013wc}:
 correlated pairs of
measurements at different center-of-mass energies (the ATLAS 2.76 and 7 TeV jet
measurements) which provide a handle on PDFs due to the reduced
experimental and theoretical systematics of cross-section
ratios~\cite{Mangano:2012mh}; $W$+c production which provides  direct
information on strangeness~\cite{Stirling:2012vh,Alekhin:2014sya}; top quark
pair production which provides a handle on the gluon at 
large-$x$~\cite{Czakon:2013tha,Beneke:2012wb,
Gauld:2013aja}; and the $W$ $p_T$ distribution 
which also helps to constrain the gluon and sea quarks 
in a wide range of $x$~\cite{Malik:2013kba}. 
The inclusion of these data is especially
powerful thanks to recent theoretical progress, specifically the
completion of the computation of the NNLO corrections to the
total cross section for top quark pair
production~\cite{Czakon:2012pz,Czakon:2013goa} and the computation of
NNLO corrections to jet production in the gluon
channel~\cite{Ridder:2013mf}.

%

For ease of reference, in Tabs.~\ref{tab:completedataset} (for DIS data) 
and~\ref{tab:completedataset2} (for hadronic data) we
summarize the complete list of data included in the NNPDF3.0 
analysis. 
For each dataset we provide the corresponding published reference, the
availability and treatment of systematics (to be
discussed in Sect.~\ref{sec:chi2definition} below), the
number of data points before and after cuts at NLO and NNLO (to be
discussed in Sect.~\ref{sec:exclusion} below).
  For the new
data which were not in NNPDF2.3 we also give the kinematic coverage
(which also flags the new datasets included for the first time in NNPDF3.0). 
Finally, in
Tab.~\ref{tab:completedataset3} we also summarize the data which were in
NNPDF2.3, but are no longer used in NNPDF3.0, for
the reasons described above.

\begin{figure}[t]
\centering
\epsfig{width=0.99\textwidth,figure=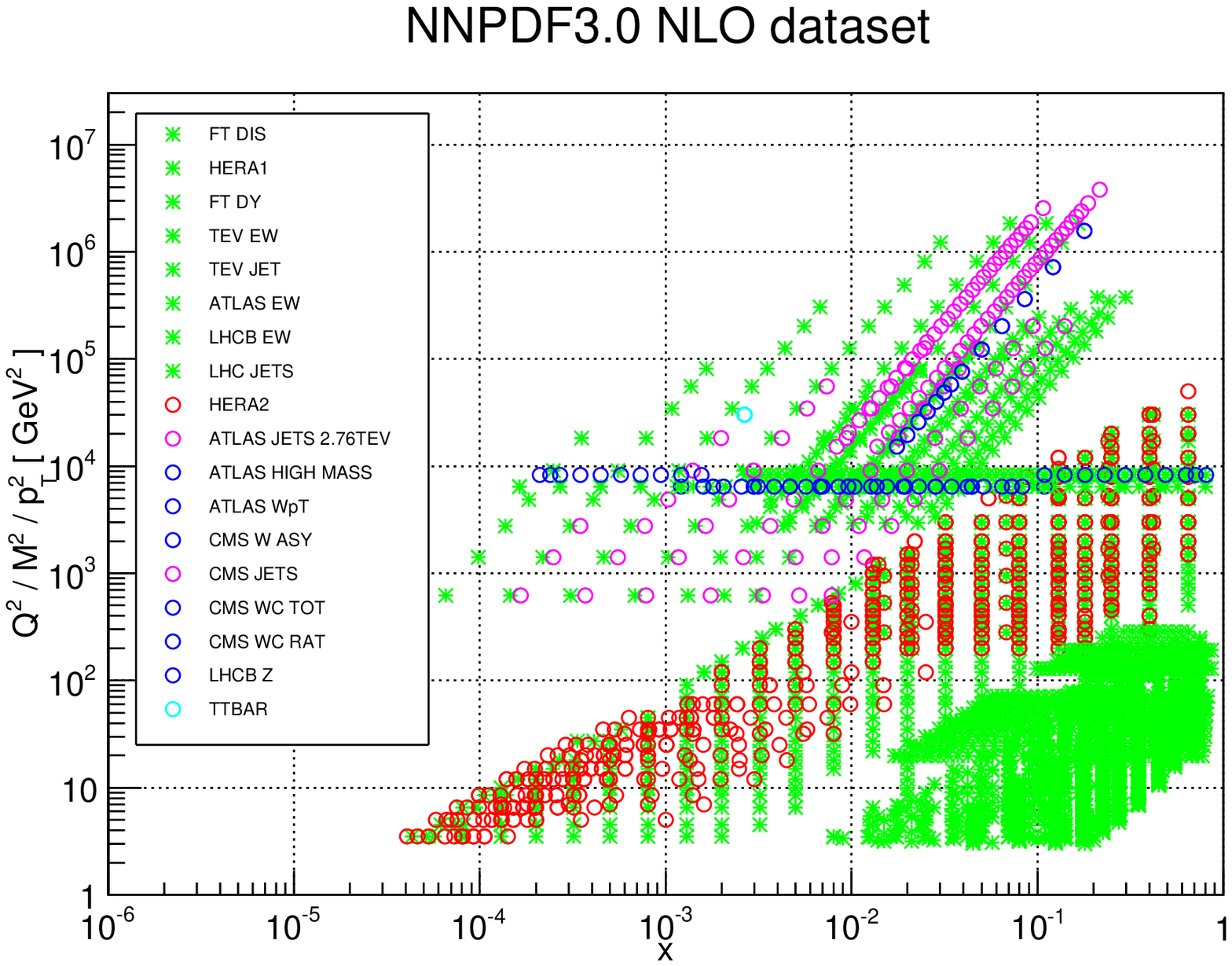}
\caption{\small 
The kinematical coverage in the $\lp x,Q^2\rp$ plane of the NNPDF3.0 dataset.
For hadronic data, leading-order kinematics have been assumed for
illustrative purposes.
The green stars mark the data already included in NNPDF2.3,
while the circles correspond to experiments that
are novel in NNPDF3.0.
 \label{fig:newkin}} cross section  
\end{figure}

The kinematical coverage in the $\lp x,Q^2\rp$ 
plane of the NNPDF3.0 dataset
is shown in the scatter plot Fig.~\ref{fig:newkin} (note that
for hadronic data  leading-order kinematics has been assumed for
illustrative purposes, as discussed in~\cite{Ball:2010de}).


\begin{table}
\footnotesize
\begin{centering}
\begin{tabular}{|c|c|c|c|c|c|c|c|}
\hline
{Experiment} & {Dataset} & {Ref.} & \multicolumn{3}{|c|}{Sys. Unc.} & $ N_{\rm \bf dat}$ no cuts & Kinematics \tabularnewline 
&  &  &   \multicolumn{3}{|c|}{}   &   (NLO/NNLO cuts)  &  \tabularnewline 
\hline
\hline
  NMC & \multicolumn{7}{c|}{}  \tabularnewline
 &   NMC $d/p$ & \cite{Arneodo:1996kd}& A & full & &  211 (132/132) & \tabularnewline
 &   NMC  $\sigma^{\rm NC,p}$ & \cite{Arneodo:1996qe}& A & full &  & 289 (224/224) & \tabularnewline
\hline
  SLAC & \multicolumn{7}{c|}{}  \tabularnewline
 &   SLAC $p$ & \cite{Whitlow:1991uw} & A & none & a & 191 (37/37) & \tabularnewline
 &   SLAC $d$ & ~\cite{Whitlow:1991uw}& A & none & a & 191 (37/37) & \tabularnewline
\hline
  BCDMS & \multicolumn{7}{c|}{}  \tabularnewline
 &   BCDMS $p$ & \cite{bcdms1}& A & full & b& 351 (333/333) & \tabularnewline
 &   BCDMS $d$ & \cite{bcdms2}& A & full & b & 254 (248/248) & \tabularnewline
\hline
  CHORUS & \multicolumn{7}{c|}{}  \tabularnewline
 &   CHORUS $\nu$ & \cite{Onengut:2005kv} & A & full & c & 572 (431/431) & \tabularnewline
 &   CHORUS $\bar{\nu}$ & \cite{Onengut:2005kv} & A & full & c & 572 (431/431)  & \tabularnewline
\hline
  NuTeV &  \multicolumn{7}{c|}{}  \tabularnewline
 &   NuTeV $\nu$ & \cite{Goncharov:2001qe,MasonPhD} & A & none & & 45 (41/41) & \tabularnewline
 &   NuTeV  $\bar{\nu}$ & \cite{Goncharov:2001qe,MasonPhD} & A & none & & 44 (38/38) &  \tabularnewline
\hline
  HERA-I & \multicolumn{7}{c|}{}  \tabularnewline
 &    HERA-I NC $e^+$ & \cite{Aaron:2009aa} & M & full & d  & 434 (379/379) & \tabularnewline
 &  HERA-I NC $e^-$  & \cite{Aaron:2009aa} & M & full & d & 145 (145/145) &\tabularnewline
 &  HERA-I CC $e^+$ & \cite{Aaron:2009aa} & M & full & d & 34 (34/34) &\tabularnewline
 &   HERA-I CC $e^-$ & \cite{Aaron:2009aa} & M & full & d & 34 (34/34) &\tabularnewline
\hline
 ZEUS HERA-II & \multicolumn{7}{c|}{}  \tabularnewline
 &   ZEUS-II NC $e^-$  & \cite{Chekanov:2009gm} & M & full & e & 90 (90/90) &\tabularnewline
 &   ZEUS-II CC $e^-$ & \cite{Chekanov:2008aa} & M & full & e& 37 (37/37)  &\tabularnewline
 &   ZEUS-II NC $e^+$ & \cite{ZEUS:2012bx} & M & full & f & 90 (90/90) &$5\,10^{-3} \le x \le 0.40$ \tabularnewline
  & & & & & & & 
$200 \le Q^2 \le 3\,10^4$ GeV$^2$
 \tabularnewline
 &   ZEUS-II CC $e^+$  & \cite{Collaboration:2010xc} & M & full & f & 35  (35/35) &
$7.8\,10^{-3} \le x \le 0.42$
 \tabularnewline
 & & & & & & & 
$280 \le Q^2 \le 3\,10^4$ GeV$^2$
 \tabularnewline
\hline
 H1 HERA-II & \multicolumn{7}{c|}{}  \tabularnewline
 &   H1-II NC $e^-$ & \cite{Aaron:2012qi} & M & full &  g& 139 (139/139) &
$2\,10^{-3} \le x \le 0.65$
\tabularnewline
 & & & & & & & 
$120 \le Q^2 \le 4\,10^4$ GeV$^2$
 \tabularnewline
 &   H1-II NC $e^+$ & \cite{Aaron:2012qi }& M & full & g& 138 (138/138) &
$2\,10^{-3} \le x \le 0.65$
\tabularnewline
 & & & & & & & 
$120 \le Q^2 \le 4\,10^4$ GeV$^2$
 \tabularnewline
 &   H1-II CC $e^-$ & \cite{Aaron:2012qi} & M & full & g& 29 (29/29) &
$8\,10^{-3} \le x \le 0.40$
\tabularnewline
 & & & & & & & 
$300 \le Q^2 \le 3\,10^4$ GeV$^2$
 \tabularnewline
 &   H1-II CC $e^+$ & \cite{Aaron:2012qi} & M & full & g & 29 (29/29) &
$8\,10^{-3} \le x \le 0.40$
\tabularnewline
 & & & & & & & 
$300 \le Q^2 \le 3\,10^4$ GeV$^2$
 \tabularnewline
 &   H1-II low $Q^2$ & \cite{Collaboration:2010ry} & M & full & & 136 (124/124) &
$2.8\,10^{-5} \le x \le 0.015$
\tabularnewline
 & & & & & & & 
$1.5 \le Q^2 \le 90$ GeV$^2$
 \tabularnewline
 &   H1-II high $y$ & \cite{Collaboration:2010ry} & M & full & & 55 (52/52) &
$2.9\,10^{-5} \le x \le 5\,10^{-3}$
\tabularnewline
 & & & & & & & 
$2.5 \le Q^2 \le 90$ GeV$^2$
 \tabularnewline
\hline
 HERA $\sigma_{\rm NC}^{c}$ & & \cite{Abramowicz:1900rp} & M & full &  & 52 (47/47) & $3\,10^{-5} \le x \le 0.05$
 \tabularnewline
  & & & & & & & 
$2.5 \le Q^2 \le 2\,10^3$ GeV$^2$
 \tabularnewline
\hline
\end{tabular}
\par\end{centering}
\caption{\small List of all the deep-inelastic scattering
data included in the NNPDF3.0 
analysis. For each dataset we provide the corresponding
reference in the third column.
The next three columns provide information on the treatment of
systematic uncertainties for each dataset.
The fourth column
 specifies the treatment of systematic uncertainties, where $M$ stands for for multiplicative
and $A$ for additive.
The fifth column 
 states how the available experimental information on
correlated uncertainties other than normalization uncertainties
is provided by the experiment: ``none'' corresponds to 
datasets for which only the sum in quadrature of
systematic uncertainties is provided, ``cov'' in case the covariance matrix is provided
and ``full" is the full breakup of systematics is provided.
The sixth column gives information on whether datasets share systematics 
uncertainties among them:
datasets that are marked with the same letter have common correlated systematics; see
text for more details.
Then, 
for each dataset, we provide the total number of data points $N_{\rm dat}$ available, as well
as the number of data points left after
kinematical cuts both in the NLO and in the NNLO fits (in parenthesis).
For the experiments which are new in NNPDF3.0, we also include the information
on their kinematical coverage in the last column.
\label{tab:completedataset}
}
\end{table}

\begin{table}[ht]
\footnotesize
\begin{centering}
\begin{tabular}{|c|c|c|c|c|c|c|c|}
\hline
{Experiment} & {Dataset} & {Ref.} & \multicolumn{3}{|c|}{Sys. Unc.} & $ N_{\rm \bf dat}$ no cuts & Kinematics \tabularnewline 
&  &  &   \multicolumn{3}{|c|}{}   &   (NLO/NNLO cuts)  &  \tabularnewline 
\hline
\hline
  DY E866 &  \multicolumn{7}{c|}{}  \tabularnewline
 &  DY E866  $d/p$ & \cite{Towell:2001nh} & M & none & & 15 (15/15) & \tabularnewline
 &  DY E866 $p$ & \cite{Webb:2003ps,Webb:2003bj} & M & none & & 184 (184/184) & \tabularnewline
\hline
  DY E605 & &  \cite{Moreno:1990sf} & M & none & & 119 (119/119) & \tabularnewline
\hline
  CDF & \multicolumn{7}{c|}{}  \tabularnewline
 &   CDF $Z$ rap & \cite{Aaltonen:2010zza}& M & full & h & 29 (29/29)  & \tabularnewline
 &   CDF Run-II $k_t$ jets & \cite{Abulencia:2007ez}& M & full & h & 76 (76/52) &  \tabularnewline
\hline
D0 & \multicolumn{7}{c|}{}  \tabularnewline
&   D0 $Z$ rap &  \cite{Abazov:2007jy} & M & full & & 28 (28/28) & \tabularnewline
\hline
  ATLAS & \multicolumn{7}{c|}{}  \tabularnewline
 &   ATLAS $W,Z$ 2010 & \cite{Aad:2011dm} & M & full & i & 30 (30/30) & \tabularnewline
 &   ATLAS 7 TeV jets 2010 & \cite{Aad:2011fc} & M & full & i,j & 90 (90/9) & \tabularnewline
 &   ATLAS 2.76 TeV jets  & \cite{Aad:2013lpa} & M & full & j & 59 (59/3) & $20 \le p_T^{\rm jet} \le 200$ GeV
\tabularnewline
& & & & & & & 
$0 \le |\eta^{\rm jet}| \le 4.4$ 
 \tabularnewline
 &   ATLAS high-mass DY  & \cite{Aad:2013iua} & M & full & & 11 (5/5) & $ 116 \le M_{ll} \le 1500$ GeV \tabularnewline
 &   ATLAS $W$ $p_T$  & \cite{Aad:2011fp} & M & full & & 11 (9/-) & 
$0 \le p_{T}^W \le 300 $ GeV
\tabularnewline
\hline
  CMS & \multicolumn{7}{c|}{}  \tabularnewline
 &   CMS $W$ electron asy & \cite{Chatrchyan:2012xt} & M & cov &  & 11 (11/11) & \tabularnewline
 &   CMS $W$ muon asy  & \cite{Chatrchyan:2013mza}& M & cov & & 11 (11/11) & 
 $0 \le |\eta_l| \le 2.4  $
\tabularnewline
 &   CMS jets 2011     & \cite{Chatrchyan:2012bja}& M & full & & 133 (133/83) &  $114 \le p_T^{\rm jet} \le 2116$ GeV \tabularnewline
& & & & & & & 
$0 \le |\eta^{\rm jet}| \le 2.5$  \tabularnewline
 &   CMS $W+c$ total  & \cite{Chatrchyan:2013uja}& M & cov & & 5 (5/5) &  $0 \le |\eta_l| \le 2.1$ \tabularnewline
 &   CMS $W+c$ ratio  & \cite{Chatrchyan:2013uja}& M & cov & & 5 (5/5) & $0 \le |\eta_l| \le 2.1$ \tabularnewline
 &   CMS 2D DY 2011      & \cite{CMSDY} & M & cov &  & 124 (88/110) & $20 \le M_{ll} \le 1200$ GeV \tabularnewline
& & & & & & & 
$0 \le |\eta_{ll}| \le 2.4$  \tabularnewline
\hline
  LHCb & \multicolumn{7}{c|}{}  \tabularnewline
 &   LHCb $W$ rapidity  & \cite{Aaij:2012vn} & M & cov &  & 10 (10/10) & \tabularnewline
 &   LHCb $Z$ rapidity & \cite{Aaij:2012mda}& M & cov & &  9 (9/9) & $2.0 \le \eta_l \le 4.5$   \tabularnewline
\hline
  $\sigma(t\bar{t})$ & \multicolumn{7}{c|}{}  \tabularnewline
 &   ATLAS $\sigma(t\bar{t})$  & \cite{ATLAS:2012aa,ATLAS:2011xha,TheATLAScollaboration:2013dja}& M & none & 
& 3 (3/3) & - \tabularnewline
 &   CMS $\sigma(t\bar{t})$   & \cite{Chatrchyan:2013faa,Chatrchyan:2012bra,Chatrchyan:2012ria}& M & none & &
 3 (3/3) & - \tabularnewline
\hline
\hline
Total & & & & & & 5179 (4276/4078) & 
 \tabularnewline
\hline
\end{tabular}
\par\end{centering}
\caption{\small Same as Table~\ref{tab:completedataset}
for fixed-target Drell-Yan production, electroweak vector
boson production data from the Tevatron and LHC data.
Again  we explicitly provide the kinematics only for the new experiments.
The $t\bar{t}$ cross-sections are new in NNPDF3.0, but being total
cross-sections no information on kinematics needs to be provided.
In the bottom row of the table we give the number of data points in 
the global fit dataset, including also the DIS numbers 
from Table~\ref{tab:completedataset}.
\label{tab:completedataset2}
}
\end{table}


\begin{table}[ht]
\footnotesize
\begin{centering}
\begin{tabular}{|c|c|c|c|c|c|}
\hline
{Experiment} & {Dataset} & {Ref.} & Sys. Unc. & $ N_{\rm \bf dat}$ no cuts & Details \tabularnewline 
&  &  &  &     &  \tabularnewline 
\hline
\hline
  H1 $F_2^c$ &  \multicolumn{5}{c|}{}  \tabularnewline
 &  H1 $F_2^c$ 01  & \cite{Adloff:2001zj} & full & 12 & Superseded by combination\tabularnewline
 &  H1 $F_2^c$ 09  & \cite{Collaboration:2009jy} & full & 6 & "\tabularnewline
 &  H1 $F_2^c$ 10  & \cite{H1F2c10:2009ut} & full & 26 & "\tabularnewline
\hline
 ZEUS $F_2^c$ &  \multicolumn{5}{c|}{}  \tabularnewline
&  ZEUS $F_2^c$ 99 &   \cite{Breitweg:1999ad} & full & 21 & Superseded by combination\tabularnewline
&  ZEUS $F_2^c$ 03 &   \cite{Chekanov:2003rb} & full & 31 & "\tabularnewline
&  ZEUS $F_2^c$ 08 &   \cite{Chekanov:2008yd} & full & 9 & " \tabularnewline
&  ZEUS $F_2^c$ 09 &   \cite{Chekanov:2009kj} & full & 8 & " \tabularnewline
\hline
  CDF & \multicolumn{5}{c|}{}  \tabularnewline
 &   CDF $W$ asymmetry & \cite{Aaltonen:2009ta}& full & 13  & Lepton-level data from LHC \tabularnewline
\hline
D0 & \multicolumn{5}{c|}{}  \tabularnewline
&   D0 Run II cone jets &  \cite{D0:2008hua} & full & 110 & Infrared unsafe at NNLO\tabularnewline
\hline
\end{tabular}
\par\end{centering}
\caption{\small Same as Table~\ref{tab:completedataset}, this time for those experiments
that were present in NNPDF2.3 but that have been excluded from NNPDF3.0.
In the last column we provide information on why these experiments are
not included anymore in NNPDF3.0.
\label{tab:completedataset3}
}
\end{table}


\subsection{New experimental data in NNPDF3.0}

We now examine in more detail the features of the
new experimental data that has been included in NNPDF3.0.
We begin with the new deep-inelastic scattering data.
In this work we have supplemented the combined
HERA-I dataset with the inclusion of all the relevant HERA-II inclusive
cross sections measurements from H1 and 
ZEUS~\cite{Aaron:2012qi,Collaboration:2010ry,ZEUS:2012bx,Collaboration:2010xc}.
These data improve the statistical and systematic precision of
medium- and high-$Q^2$ data from the HERA-I run, and thus provide
valuable information for quarks at medium and large $x$.
In addition, we have included low-$Q^2$ data from H1 that
provides additional information on the small-$x$ gluon and that
could also be sensitive to potential deviations from fixed-order DGLAP
at small-$x$.

From the H1 experiment, we have included
the new high-$Q^2$ data 
from the HERA-II run~\cite{Aaron:2012qi}, which covers the large $Q^2$ region,
$60 \le Q^2 \le 5\, 10^4$ GeV$^2$, with improved statistical
and systematic precision in comparison to Run-I.
These data, taken at the default proton beam energy of $E_p=920$ GeV used in most
of the HERA-II run, have been supplemented with inclusive cross-section measurements 
performed
at lower center-of-mass energies~\cite{Collaboration:2010ry},
 obtained with
proton beam energies  of $E_p=575$ GeV and
$E_p=460$ GeV.
These lower-energy measurements 
are the same ones used to determine the longitudinal structure
function $F_L$.
Therefore,  we exclude the previously used  $F_L$~\cite{h1fl} data 
from the present fit to avoid any double counting, and for the same
reason we do not
include any other of the updated $F_L$ extractions at 
HERA~\cite{Andreev:2013vha,Chekanov:2009na}.
For completeness, in NNPDF3.0 we also include the the high-inelasticity
data that H1 extracted from their Run II measurements ~\cite{Collaboration:2010ry}.

From the ZEUS experiment, in NNPDF2.3 we already included
some HERA-II data~\cite{Chekanov:2009gm,Chekanov:2008aa},
for  neutral- and charged-current DIS 
with an electron beam.
We now also include   
neutral- and charged-current cross sections
with a positron beam~\cite{Collaboration:2010xc,ZEUS:2012bx},
which have been published since.
As in the case of H1, ZEUS Run II inclusive cross-sections exhibit reduced statistical
and systematic uncertainties when compared to Run I data in the medium and large $Q^2$
region.
For both H1 and ZEUS, we use the data averaged over lepton beam polarizations.

Turning to semi-inclusive measurements,
in NNPDF3.0 we have replaced the separate charm structure function
data $F_2^c$ from the H1 and ZEUS 
experiments~\cite{Breitweg:1999ad,Chekanov:2003rb,Chekanov:2008yd,
Chekanov:2009kj,Adloff:2001zj,
  Collaboration:2009jy,H1F2c10:2009ut} with the combined HERA charm production
dataset~\cite{Abramowicz:1900rp}, which provides data for the reduced
cross-section (rather than structure function) and is based on a more
extensive dataset; furthermore, 
the cross-calibration between common systematics implies that
the combined data is more accurate than the separate inclusion
of the individual measurements.
The combined HERA charm production cross-sections offer a handle
on the small-$x$ gluon~\cite{Bertone:2012wi}, provide a unique testing ground
for different treatments of heavy quark mass effects, and allow
one to extract the running charm quark mass $m_c(m_c)$ with
competitive uncertainties~\cite{Alekhin:2012vu,Gao:2013wwa}.

Concerning vector boson production at the LHC,
NNPDF2.3 already included some data taken at 7 TeV,
in particular the CMS $W$ electron asymmetry data from
an integrated luminosity of 
880 pb$^{-1}$~\cite{Chatrchyan:2012xt}, the ATLAS $W^{\pm}$ and
$Z$ rapidity distributions from the 2010 run~\cite{Aad:2011dm} 
and the LHCb forward $W$ production data also from
2010~\cite{Aaij:2012vn}.
These datasets have now
been supplemented with 
essentially all vector  boson production data which have become
available since.

From the ATLAS experiment,
we include the  high-mass Drell-Yan production data
from the 2011 run~\cite{Aad:2013iua}, based on an integrated luminosity of 4.9 fb$^{-1}$.
These data, presented in terms of the invariant mass of the electron pairs produced at 
an invariant mass larger than the $Z$ peak, which extends to $M_{ll}=1.5$~TeV, 
can be used to constrain the large-$x$ antiquarks.
In addition, it was shown in Ref.~\cite{Ball:2013hta} that high-mass Drell-Yan
at the LHC can be used to constrain the photon
PDFs of the proton, and it was indeed used there to construct
the NNPDF2.3QED PDF set.
We also include now the ATLAS measurement of the $W$ boson transverse momentum distribution 
from the 2010 run of the LHC at $\sqrt{s}=$7 TeV~\cite{Aad:2011fp},
corresponding to an integrated luminosity of 31 pb$^{-1}$.
This data, unlike the 7 TeV ATLAS measurement of the $Z$ boson transverse momentum 
distribution~\cite{Aad:2011gj},
is provided with all information on correlated uncertainties and 
has the potential to constrain
the gluon and the light quark distributions in the medium-$x$ 
region~\cite{Malik:2013kba}.

From the CMS experiment,
we include the  $W$ muon asymmetry data
based on the full statistics (5 fb$^{-1}$) of the 7 TeV 
run~\cite{Chatrchyan:2013mza}, which have substantially reduced
statistical and systematic uncertainties. We also
include the double-differential distributions for Drell-Yan production
from low dilepton masses (from $M_{ll}=20$ GeV) to high dilepton
masses (up to $M_{ll}=1.5$ TeV) in bins of the dilepton invariant mass
and rapidity, from the full 2011 dataset~\cite{CMSDY}.
The  CMS data for
the production of charm quarks associated to $W$ 
bosons~\cite{Chatrchyan:2013uja}, which we also include,
provide the absolute cross sections,
differential on the lepton rapidity from the $W$ decay $\eta_l$, as
well as cross-section ratios $W^+/W^-$ also binned in $\eta_l$.
They are both included, as the former constrain the
shape and overall normalization of the
total strangeness $s+\bar{s}$ at $Q \sim M_W$ and the  
latter offer some handle on the strangeness asymmetry
in the proton, $s-\bar{s}$.
Data from this same process are available from
the ATLAS Collaboration~\cite{Aad:2014xca}, but are given at the
hadron level and thus 
cannot be directly included in our fit (though they could be
included by for example estimating a hadron-to-parton correction 
factor using {\sc\small MadGraph5\_aMC@NLO}).  

Finally, we include the LHCb $Z\to ee$ rapidity distributions from the 2011
dataset~\cite{Aaij:2012mda}, which are more precise than the previous
data from the 2010 run.
The forward kinematics of this data provide constraints on PDFs 
at smaller and larger
values of $x$ than the vector boson production data from ATLAS and CMS.
Further LHCb data from the 2011 run for $Z$ boson
rapidity distributions in the $\mu\mu$
channel~\cite{LHCb-CONF-2013-007}
and 
for low mass Drell-Yan 
production~\cite{LHCb-CONF-2012-013} are still preliminary.

Concerning inclusive jet production from ATLAS and CMS,
we include the CMS inclusive jet production measurement at 7 TeV 
from the full 5 fb$^{-1}$ dataset~\cite{Chatrchyan:2012bja}, which has been provided
with the full experimental covariance matrix, and which supersedes
previous inclusive jet measurements from CMS~\cite{CMS:2011ab}.
This data has a large kinematical coverage: 
for example, in the central rapidity region, the CMS data
reaches up to jet transverse momenta of more than 2 TeV,
thus constraining the
large-$x$ quark and gluon PDFs~\cite{Watt:2013oha,CMS-PAS-SMP-12-028}.
From ATLAS, we include the new inclusive cross-section measurement at
$\sqrt{s}=2.76$ TeV~\cite{Aad:2013lpa}, which is provided with the full correlation
matrix with the corresponding $\sqrt{s}=7$ TeV measurement.
Measuring the ratio of jet cross-sections at two different center of mass energies enhances the PDF sensitivity thanks to the partial cancellation of
theoretical (missing higher order corrections) and experimental
(jet energy scale) systematic uncertainties~\cite{Mangano:2012mh}.
On the other hand no LHC dijet data are included~\cite{Aad:2013tea}, 
since it is notoriously
difficult to achieve a good description of these
measurements~\cite{Watt:2013oha}.

Finally, we include
 six independent measurements of the total top quark pair production
cross-section from ATLAS and CMS, both at 7 TeV and at 8 TeV.
These data provides information on the large-$x$ gluon PDF, complementary
to that provided by inclusive jet production.
At 7 TeV we include 
the measurements in the dilepton channel, based on 0.70 fb$^{-1}$ integrated luminosity
from ATLAS~\cite{ATLAS:2012aa} and on 2.3 fb$^{-1}$ from CMS~\cite{Chatrchyan:2013faa}
and the measurements performed using lepton+jets events
from ATLAS~\cite{ATLAS:2011xha} and CMS~\cite{Chatrchyan:2012bra}.
At 8 TeV
we have included the  dilepton channel measurement corresponding to
an integrated luminosity of 2.4 fb$^{-1}$ by CMS~\cite{CMS:2012lba}
and the ATLAS analysis of the lepton+jets final state in a dataset corresponding to 
an integrated luminosity of 5.8 fb$^{-1}$~\cite{TheATLAScollaboration:2013dja}.

\subsection{Theoretical treatment}
\label{sec:theorytools}

NNPDF3.0 PDFs are provided at LO, NLO and NNLO in perturbative QCD.
While for most of the observables included in the fit NNLO QCD corrections 
are known, some observables are known only up to NLO, while for others 
only partial contributions to the full
NNLO corrections have been calculated.
Specifically, NNLO corrections are not available for two datasets
included in our fit: the vector boson transverse momentum distribution
and the $W+c$ rapidity distribution (since there are no NNLO calculations 
for $V+$jets and $V+$heavy quarks).  
For the jet inclusive cross section, only the $gg$-channel has been recently
computed at NNLO~\cite{Currie:2013dwa,Ridder:2013mf}, while for the
full cross section 
only an approximate NNLO prediction based on threshold resummation is
available~\cite{deFlorian:2013qia}. 
For all other observables included in the fit the cross sections are known up to NNLO.

The theoretical predictions for DIS observables have been implemented in the {\sc\small FastKernel}
framework and thoroughly benchmarked~\cite{Ball:2011uy,Bertone:2013vaa}. 
Drell-Yan cross sections, both for fixed target and
for collider experiments, are computed at 
NNLO by using the local $C$-factors computed according to the procedure
described in Ref.~\cite{Ball:2011uy}: 
we define $C$-factors as the ratio of NNLO to NLO calculations for fixed NNLO PDFs, that is
\be
\label{eq:cfact}
C^{\rm nnlo} \equiv \frac{\hat{\sigma}^{\rm nnlo} \otimes \mathcal{L}^{\rm nnlo}}{
\hat{\sigma}^{\rm nlo} \otimes \mathcal{L}^{\rm nnlo}} \ , 
\ee
where $\hat{\sigma}$ is the partonic cross section computed at
either NNLO or NLO accuracy, and $\mathcal{L}^{\rm nnlo}$ is the corresponding
parton luminosity computed with a reference set of NNLO parton
distributions. 
The tools used to compute the $C$-factors are described in 
Sect.~\ref{sec:ctools}.
For a detailed description of the procedure that we adopt to include a subset of the inclusive
jet data in our analysis, see Sect.~\ref{sec:thjetdata}.

Given that electroweak corrections can be relevant 
in the large invariant mass region covered
by some of the experimental data included in our fit (see Ref.~\cite{Campbell:2013qaa} 
for an overview), 
we provide EW corrections for all LHC vector boson production data. 
To include these corrections in our NLO and NNLO calculation, we compute
$C-$factors, $C^{\rm ew}$, defined analogously to Eq.~\eqref{eq:cfact}, with the
NNLO computation substituted by the NLO+EW one, and using NLO parton luminosities
on both numerator and denominator. All details on their computation and implementation are provided in 
Sect.~\ref{sec:ew}.
The QCD NNLO and EW $C$--factors for LHC gauge boson production are 
listed in Appendix~\ref{app-ew}.

In this work,
we do not include all-order perturbative resummation of QCD
corrections: these will be the object of a future separate study,
possibly leading to the construction of dedicated resummed sets.
Also, we do not include nuclear corrections, which are relevant
for fixed-target deuterium DIS data,  neutrino DIS data, and
fixed-target Drell-Yan data,
given the substantial uncertainties 
in their modeling and their moderate impact~\cite{Ball:2013gsa}. We
will briefly assess the  impact of this omission in Sect.~\ref{sec:model}.

The way our dataset is constructed, and in particular which data are
included at various perturbative orders, is discussed in
Sect.~\ref{sec:exclusion} below.

\subsubsection{Computational tools}
\label{sec:ctools}

The inclusion of perturbative corrections to
hadronic processes in PDF fits requires the fast computation of the
relevant cross-sections.
 Several
fast interfaces have been developed to this purpose, 
including {\sc\small APPLgrid}~\cite{Carli:2010rw}, which provides an 
interface to {\sc\small MCFM}~\cite{Campbell:2002tg,MCFMurl} 
and {\sc\small NLOjet++}\cite{Nagy:2003tz}, 
and {\sc\small FastNLO}~\cite{Kluge:2006xs,Wobisch:2011ij}, 
also interfaced to {\sc\small NLOjet++}. The {\sc\small MCgrid}~\cite{DelDebbio:2013kxa} package
interfaces the {\sc\small Rivet}~\cite{Buckley:2010ar} analysis package to {\sc\small APPLgrid}, making use of the BlackHat/Sherpa~\cite{Bern:2013zja}
prescription for NLO reweighting.

Recently, a new fast interface  has become available, namely
  {\sc\small aMCfast}~\cite{aMCfast}, a code which provides
the complete automation of fast NLO QCD calculations for PDF
fits, interfaced to 
{\sc\small MadGraph5\_aMC@NLO}~\cite{Alwall:2014hca},
which in turn achieves the complete automation of the computations of 
tree-level and next-to-leading order cross sections and of their matching to parton shower simulations.

Such tools have been used extensively in the present analysis. 
For the 7 TeV CMS jet data, we have used the {\sc\small FastNLO} calculation
with central scales $\mu_F=\mu_R=p_{T}^{\rm jet}$.
For the 2.76 ATLAS jet data, we have used instead the  {\sc\small APPLgrid}
calculation, which uses the same scale choices.
For consistency, we use exactly the same settings of the calculation, 
including the central scales, as were used in the corresponding ATLAS 7 TeV
inclusive jet analysis.
The CDF Run II $k_t$ jets have also been computed using the
{\sc\small FastNLO} calculation with the same central scales as in
the ATLAS and CMS case.

For  electroweak vector boson production data, we have used
the  {\sc\small APPLgrid} code interfaced to {\sc\small MCFM6.6} in all cases,
with a consistent choice of electroweak parameters.
We use the $G_{\mu}$ scheme, with $M_Z=91.1876$ GeV, $M_W=80.398$ GeV 
and $G_F=1.16637 \cdot 10^{-5}$ GeV$^{-2}$ set as input parameters
and $\alpha_e$, $\sin\theta_W$ derived from those. We turn off the 
Narrow-Width approximation. 
For all rapidity distributions, we set $\mu_F=\mu_R=M_V$,
with $V=W,Z$. For the $W$ $p_T$ distribution we set $\mu_F=\mu_R=M_W$, while
in the case of CMS double-differential distribution we set the scales to the central
value of the invariant mass bin.
The {\sc\small MCFM6.6} calculations have been cross-checked
with independent calculations of 
{\sc\small DYNNLO}~\cite{DYNNLOurl,Catani:2010en,dynnlo1,dynnlo2} and {\sc\small FEWZ3.1}~\cite{Gavin:2012sy,Li:2012wna} 
at NLO, with the same settings, finding perfect agreement in
all cases. 

In the NNLO fits, the NNLO $C-$factors defined in Eq.~\eqref{eq:cfact} 
have been computed with {\sc\small FEWZ3.1} and cross-checked against 
{\sc\small DYNNLO1.3}.
Very high statistics runs of all these codes have been necessary to achieve
negligible integration error in all the data bins. In order to obtain smooth $C$-factors,
the NNLO curves are smoothed with a high-degree polynomial interpolation. Notice that
the difference between smoothed and original NNLO predictions is always within
the Monte Carlo uncertainty of the code used to compute it.   
It turns out that these NNLO QCD corrections are sizable, especially
for small invariant masses of the produced lepton pairs
as it is shown on the right plot in Fig.~\ref{fig:cfactEWandNNLO}:
the NNLO $C$-factor for the CMS double differential Drell-Yan data
at low $M_{ll}$ is around 10\%, independent of the dilepton rapidity.
NNLO corrections are also important for the ATLAS high-mass Drell-Yan data,
reaching almost 10\% around $M_{ll}\sim$ 1 TeV, see left plot of 
Fig.~\ref{fig:cfactEWandNNLO}.
Details of the computation of the EW corrections 
and their size are given below in Sect.~\ref{sec:ew}.

For the computation of top quark pair production data at NLO,
we again used {\sc\small APPLgrid} interfaced to {\sc\small MCFM6.6}.
The NNLO $C-$factors have been computed by using 
the NNLO calculation of Ref.~\cite{Czakon:2013goa}, as implemented in
the {\small\sc top++} code~\cite{Czakon:2011xx}.
Finally, we have used {\sc\small aMCfast} interfaced to
{\sc\small MadGraph5\_aMC@NLO} to compute the Higgs rapidity
distributions in gluon fusion at NLO with an unphysical boson
of mass $m_h=\sqrt{5}$~GeV.
As explained below in Sect.~\ref{sec:pdfparam}, this unphysical calculation
has been used to enforce the positivity of cross-sections that
depend on the small-$x$ gluon.
More details about Higgs production at NLO in the 
{\sc\small MadGraph5\_aMC@NLO} framework can be found in
in Refs.~\cite{Frederix:2014hta,Artoisenet:2013puc}.

\subsubsection{Approximate NNLO treatment of jets}
\label{sec:thjetdata}

NNLO corrections to jet production in the the $gg$-channel have become
available recently~\cite{Currie:2013dwa,Ridder:2013mf}.
An approximate  NNLO prediction based on threshold resummation (but
including all partonic subchannels) was presented
in~\cite{deFlorian:2013qia}. Its accuracy can now be gauged against
the existing part of the full calculation.

This was done recently in a systematic
study~\cite{Carrazza:2014hra}, to which we refer for a more detailed
treatment. 
An important consequence of the analysis of
Ref.~\cite{Carrazza:2014hra} is that some of the comparisons between
the exact computations and the threshold approximation performed
hitherto~\cite{deFlorian:2013qia}  were
marred by the fact that the renormalization scale was set equal to the jet
transverse momentum $p_T$ in the threshold calculation, while the
transverse momentum of the hardest jet $p_T^{\rm lead}$ was used instead for
the exact calculation. 
It turns out that the perturbative expansion is
rather better behaved, with smaller perturbative corrections,
when the scale $p_T$ is used, and also, that the
regions of agreement are rather different if a consistent scale choice
is made: specifically, when using $p_T$ as a scale in the large
rapidity region the accuracy of the threshold approximation
appears to be very poor (while there is a better, accidental agreement when
an inconsistent scale choice is made).

\begin{figure}[t]
  \begin{centering}
    \includegraphics[scale=0.35]{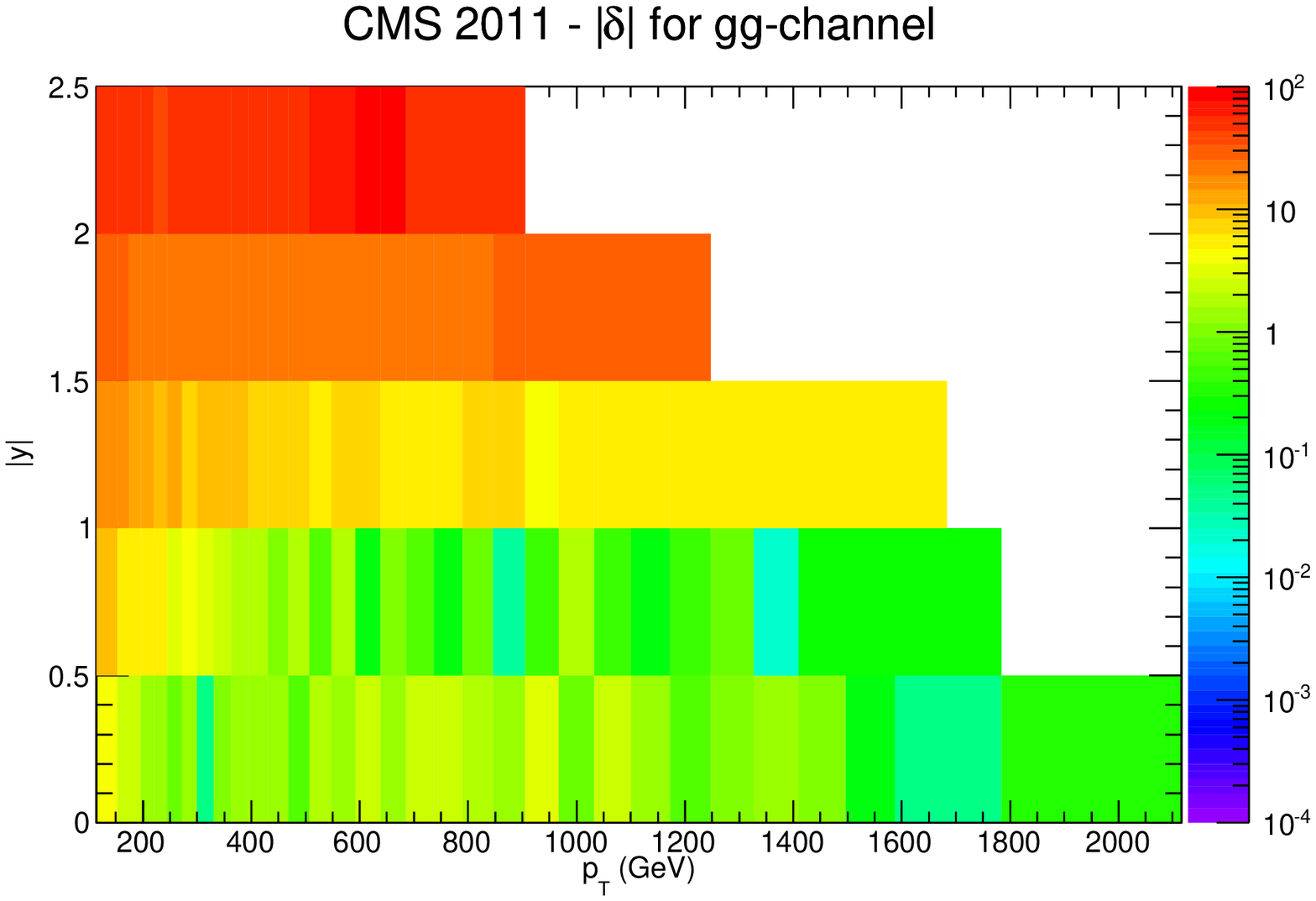}
\includegraphics[scale=0.35]{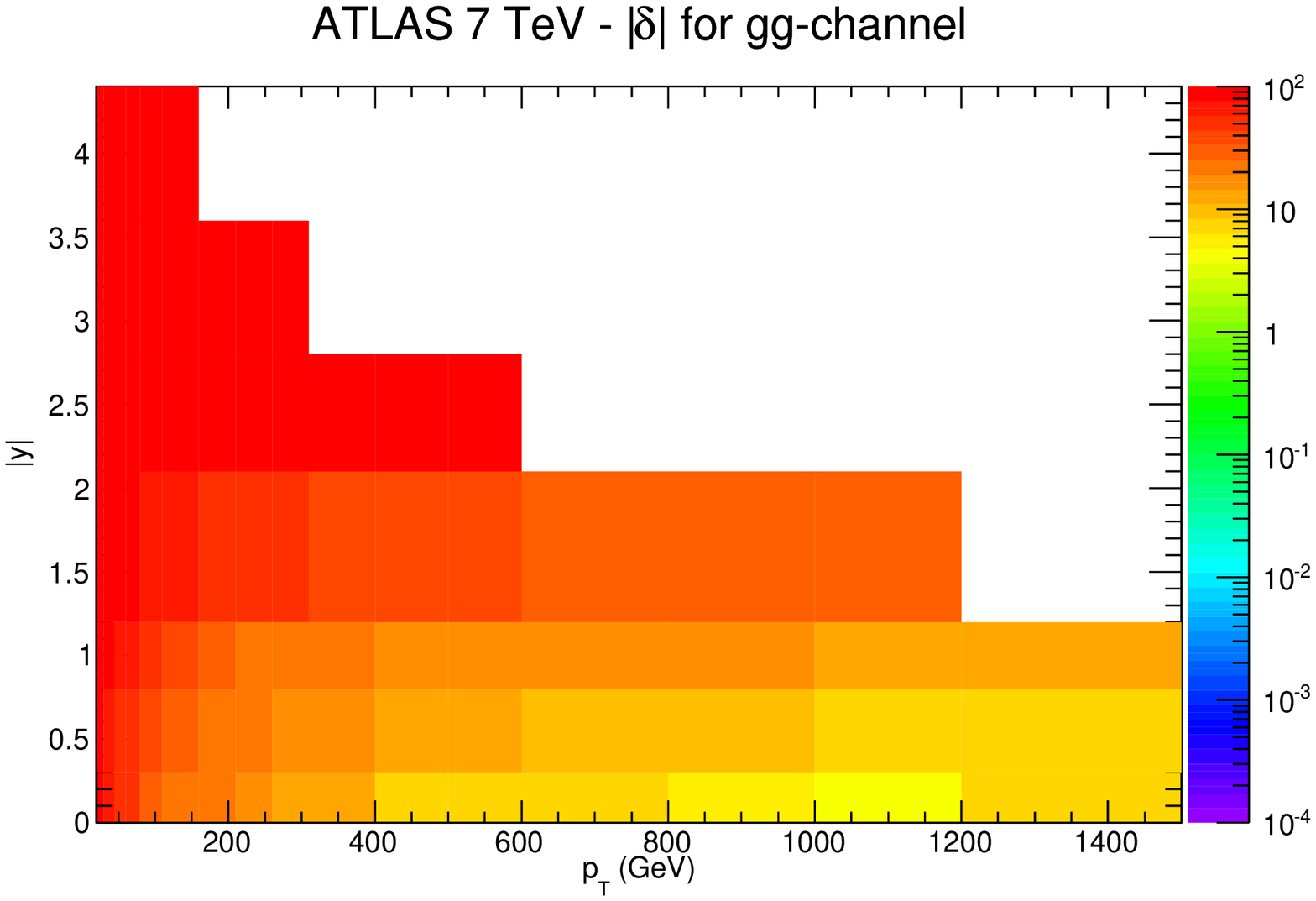}
    \par\end{centering}
  \begin{centering}
    \includegraphics[scale=0.35]{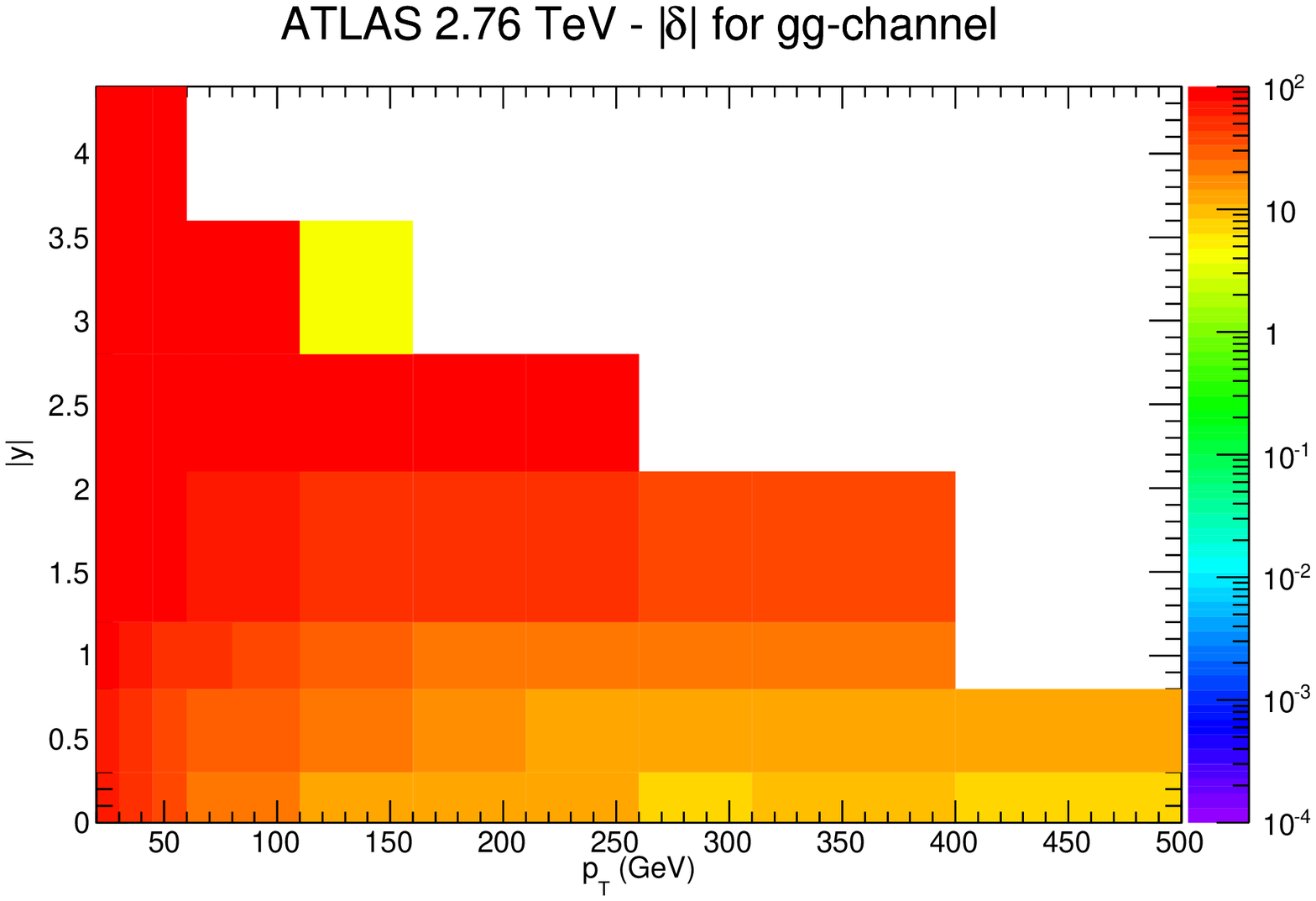}
\includegraphics[scale=0.35]{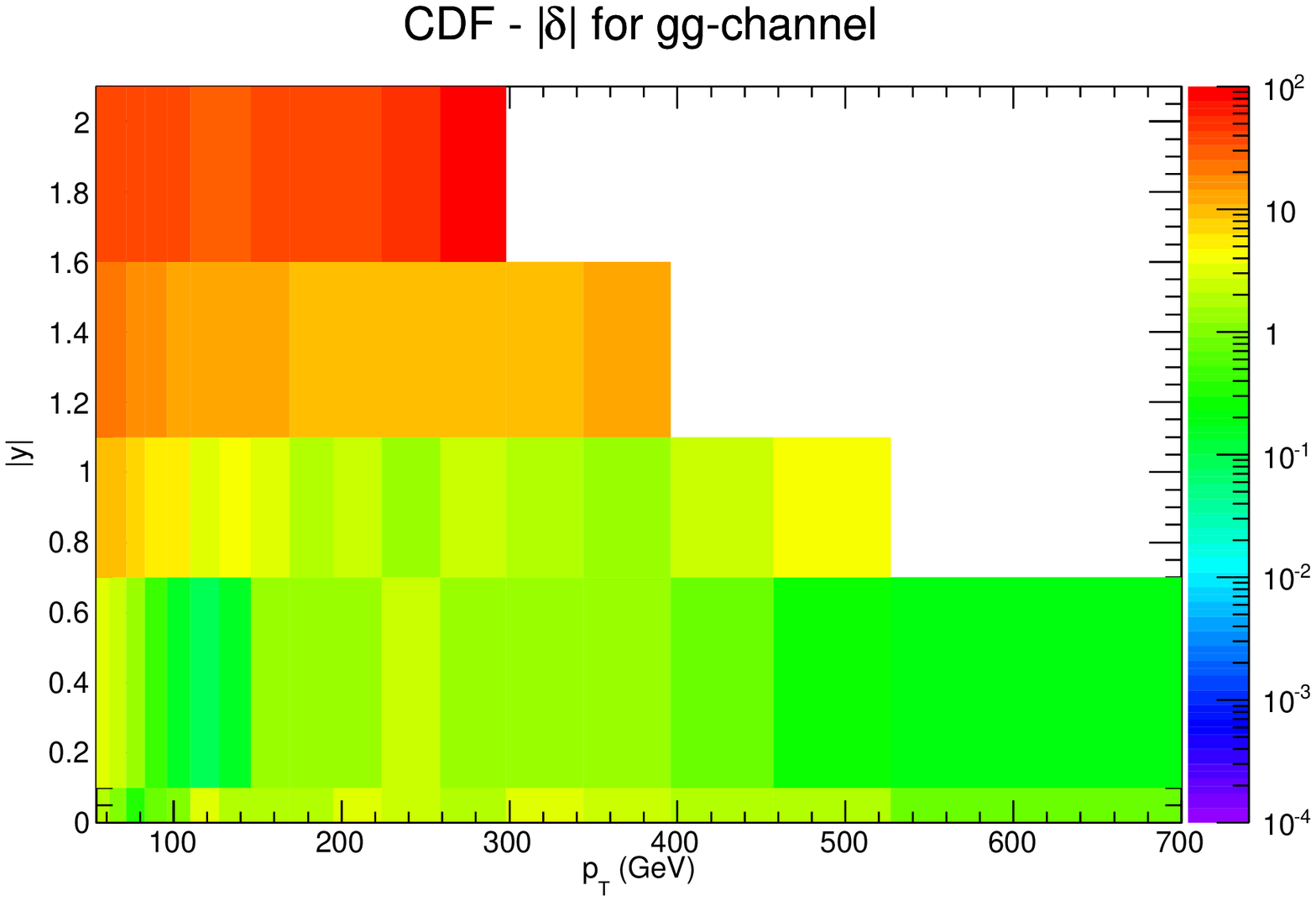}
    \par\end{centering}
  \caption{\small
    \label{fig:jets} Percentage difference
    $|\delta|$ between the exact and approximate $gg$-channel NNLO
    $C$-factors as a function of $p_{T}$ and $|y|$ for the CMS, ATLAS
    7~TeV and 2.76~TeV and CDF data that
is included in NNPDF3.0.
Each entry in the contour plot corresponds to one of the experimental bins of
the corresponding measurement.
 Differences larger than
    $|\delta|=100\%$ are shown in red.}
\end{figure} 
\begin{figure}[t]
  \begin{centering}
    \includegraphics[scale=0.35]{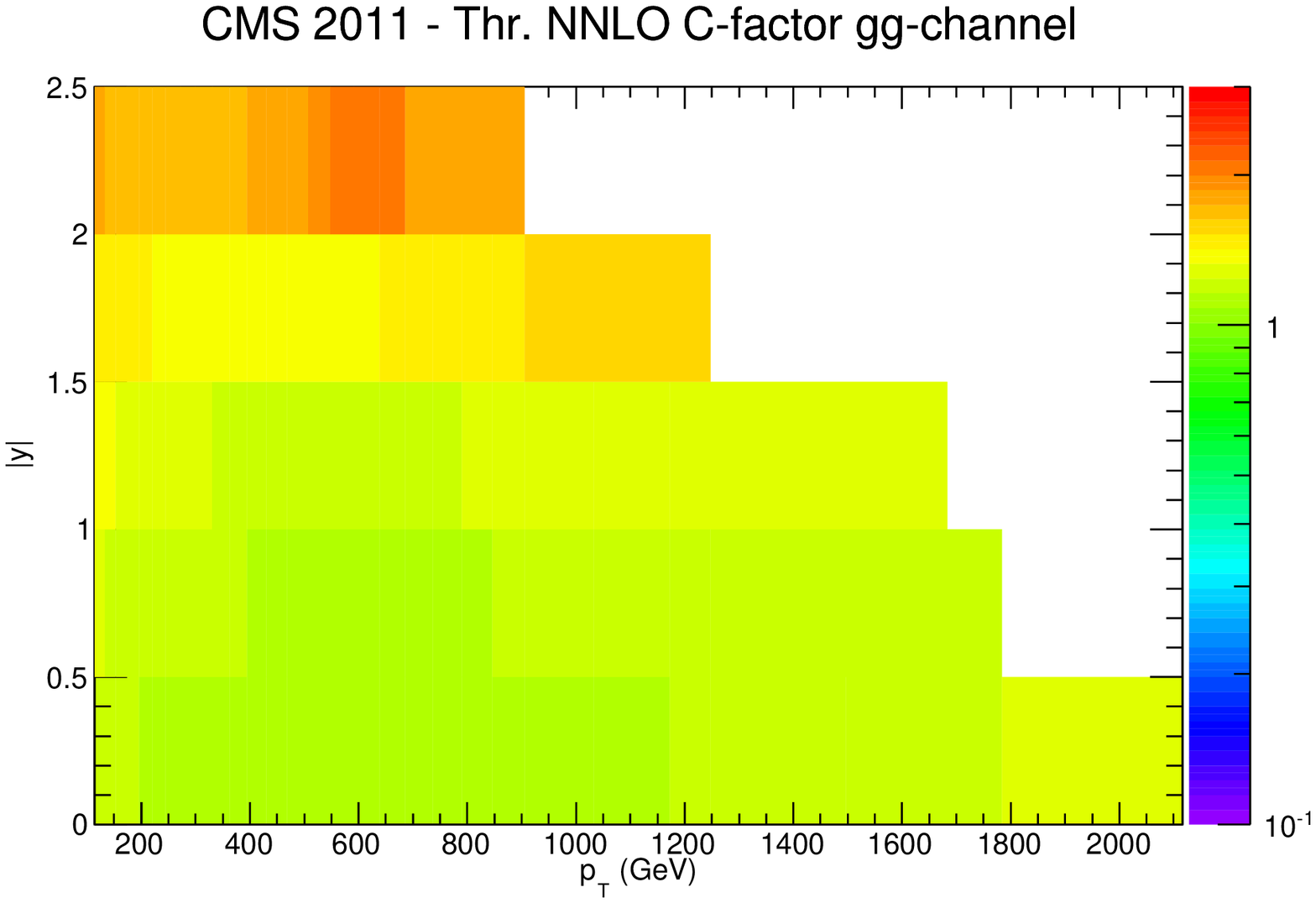}
\includegraphics[scale=0.35]{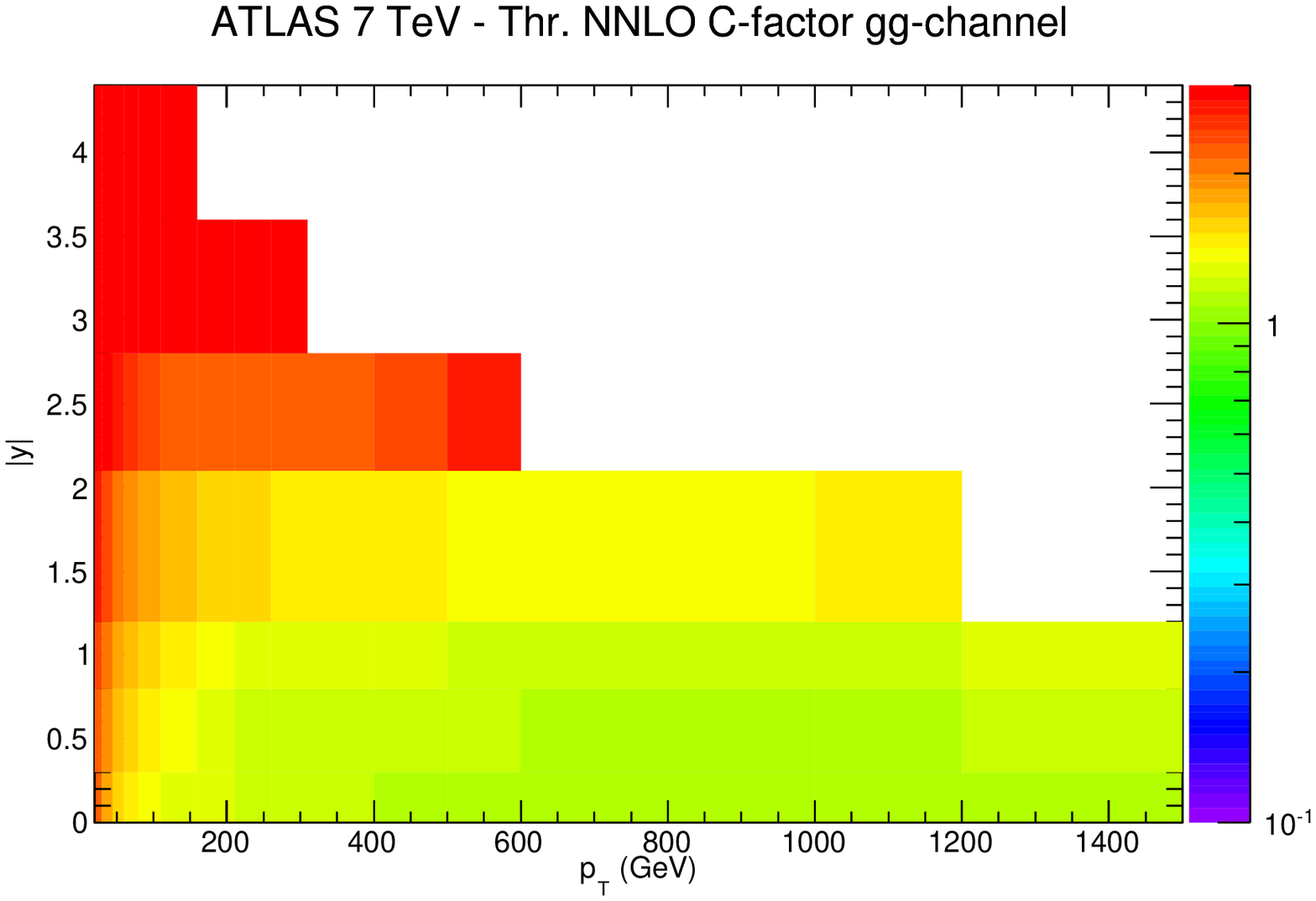}
    \par\end{centering}
  \begin{centering}
    \includegraphics[scale=0.35]{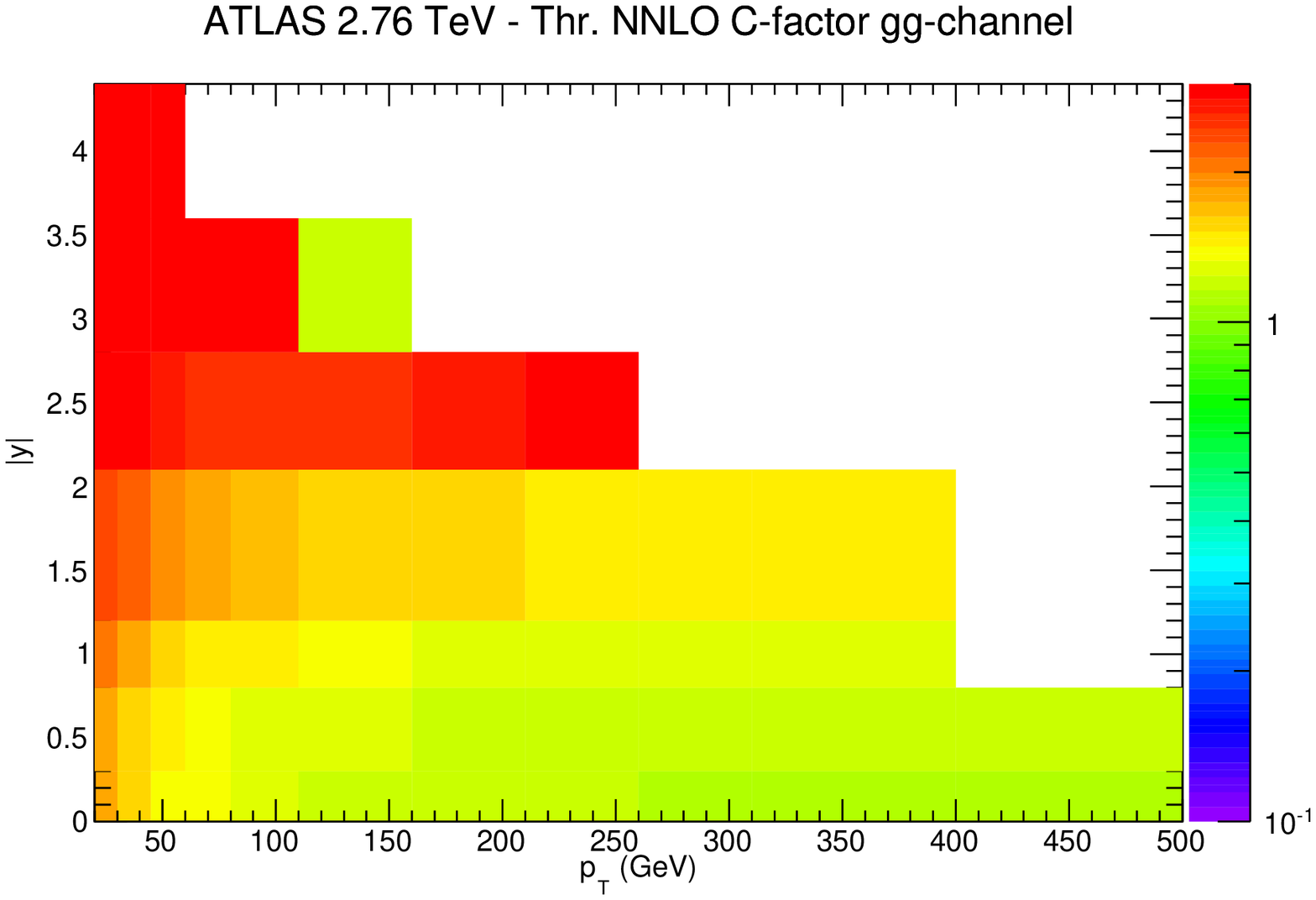}
\includegraphics[scale=0.35]{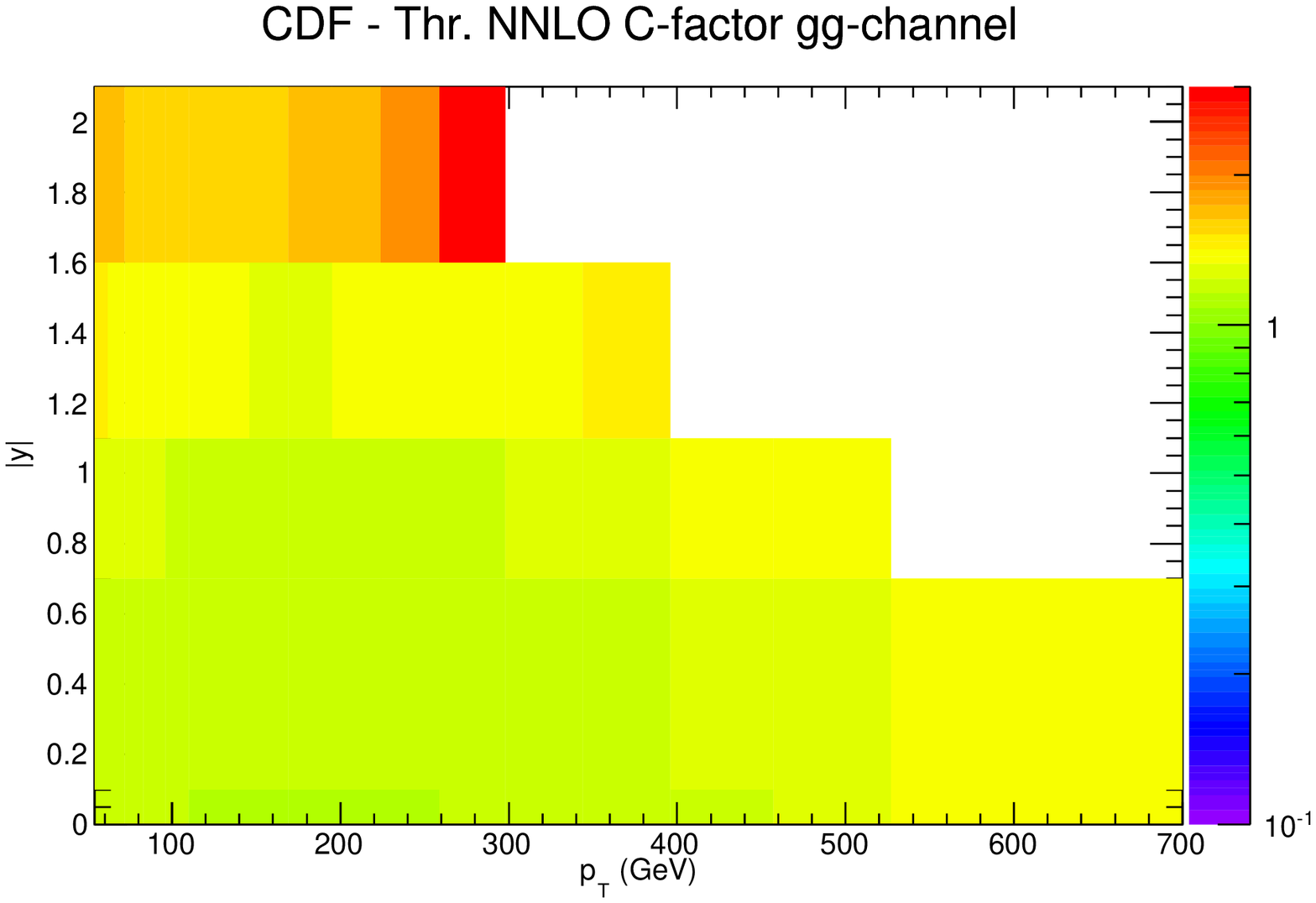}
    \par\end{centering}
  \caption{\small
    \label{fig:jetsk} Same as Fig.~\ref{fig:jets}, but now showing
the size of the approximate 
 $gg$-channel NNLO
    $C$-factors.}
\end{figure} 

The percentage deviation between the exact and approximate calculation
(with the renormalization scales set equal to $p_T$ in the two cases) is shown in
Fig.~\ref{fig:jets} in the 
($p_{T},y$) plane for the
CMS 2011~\cite{Chatrchyan:2012bja}, ATLAS 7 and 2.76
TeV~\cite{Aad:2011fc,Aad:2013lpa} and CDF Run II $k_t$~\cite{Abulencia:2007ez} 
inclusive jet data included in our fit. 
A common NNLO reference PDF set, NNPDF2.3, is used in the two calculations.
In Fig.~\ref{fig:jets}, each of the entries of the contour plot correspond to
one of the experimental data bins.
It is apparent
that the threshold approximation and the exact results are in reasonable
agreement in the high-$p_{T}$ and central-$y$ regions. 
Note also that the disagreement between the two calculations can become very large
for small transverse momenta and/or large rapidities.

In the kinematical region covered by the Tevatron and LHC 
jet cross-sections, see Fig.~\ref{fig:jets},
the gluon channel is not necessarily dominant, however
both theoretical arguments and the known NLO behavior suggest that
the pattern of agreement or disagreement of the threshold and exact
computations is similar in all partonic
channels~\cite{Carrazza:2014hra}. 
This result then  suggests that the
threshold approximation can be used provided all channels are
included, and a sufficiently conservative cut is applied in order to keep only
the regions in which  the size
of the deviation between the  exact and threshold
computations in the $gg$ channel does not exceed a suitable low threshold. 

Reassuringly, 
these are also the regions in which NNLO corrections are not too
large, as it is apparent from Fig.~\ref{fig:jetsk}, where 
we show the size of the
approximate $C$-factors. It is clear that the regions in which the
quality of the 
approximation is better are the same as those where the size of the
correction is relatively smaller, and conversely.
In particular, in the region in which the exact and
threshold NNLO calculations in the gluon channel differ by less than
10\%, the $C$-factor Eq.~(\ref{eq:cfact}) (summing over all parton
channels) is typically of order 15\%, so the expected accuracy of the
threshold approximation at NNLO is at the level of a few percent, smaller
than the experimental uncertainties.

In NNPDF3.0, we thus follow the strategy of Ref.~\cite{Carrazza:2014hra} and
compute approximate NNLO $C$-factors, Eq.~(\ref{eq:cfact}), using the threshold
calculation, while restricting the fitted dataset to the region where, thanks
to the comparison with the exact $gg$ calculation, we know the former to be reliable. This leads to the set of cuts outlined below in Sect.~\ref{sec:dataset}.

\subsubsection{Electroweak corrections}
\label{sec:ew}

Electroweak corrections, though generally small, may become large 
at high scales $Q^2 \gg M_V^2$.
 While this will certainly be an issue
for future LHC data at higher center of mass energy, already
for some high-mass data included in our analysis
the high accuracy of the experimental measurements 
may require theoretical predictions at the percent level of
precision, and the size of the EW corrections
needs to be carefully assessed.
 
\begin{figure}
\begin{center}
        \includegraphics[width=0.48\textwidth]{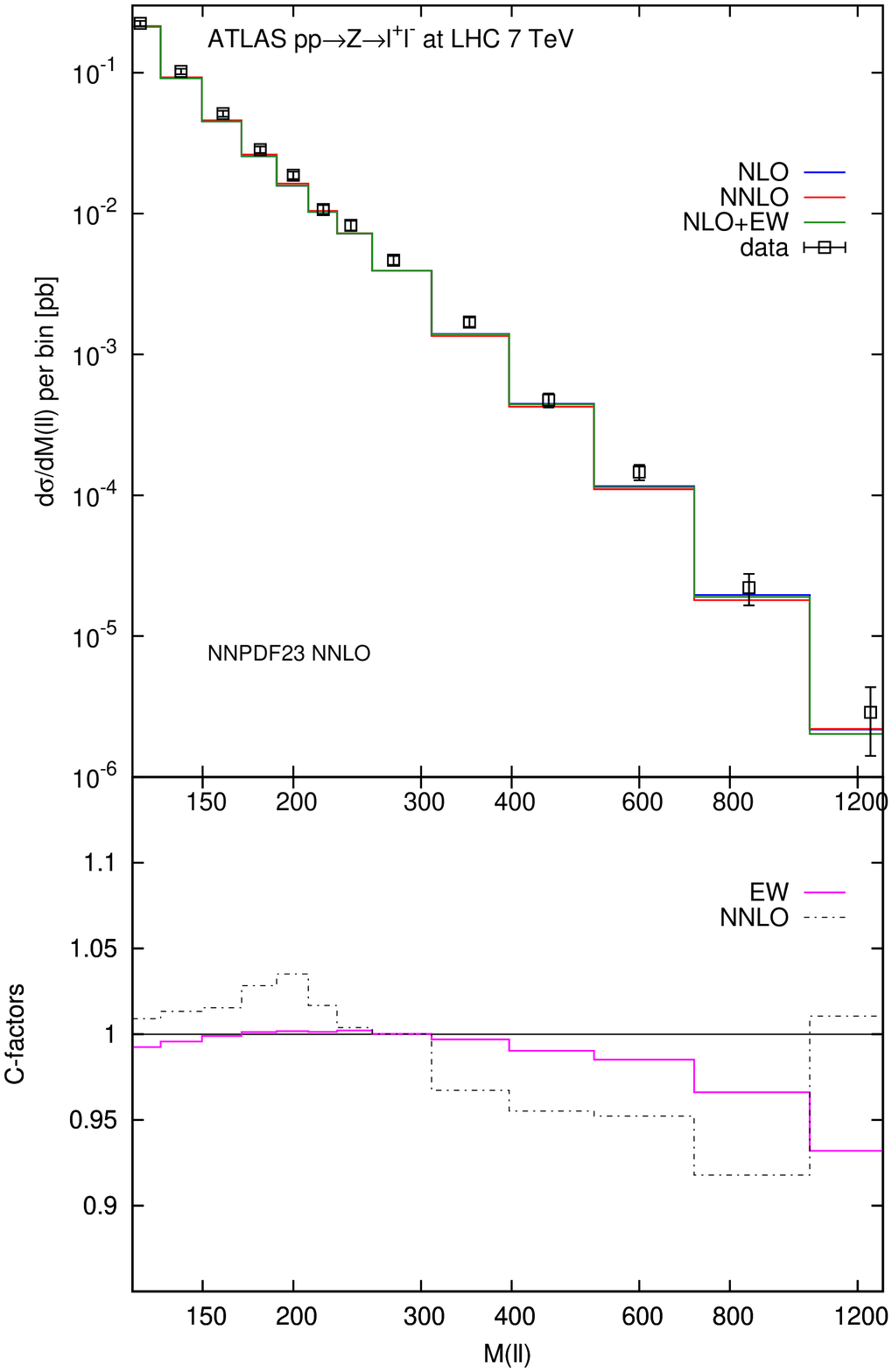}
        \includegraphics[width=0.48\textwidth]{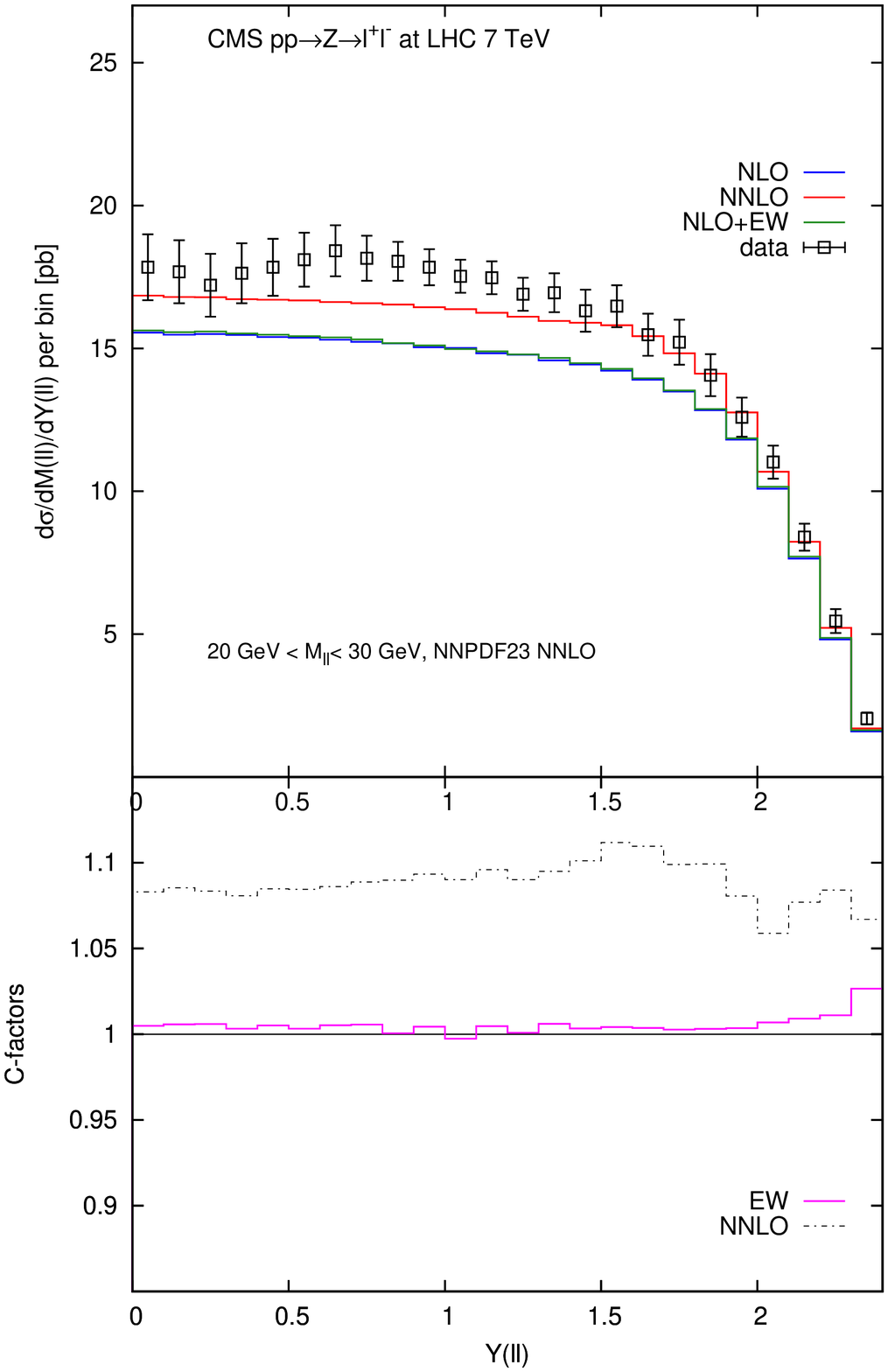}
\end{center}
        \caption{\small Left plot: the NLO, NNLO and NLO+EW predictions compared to the
          ATLAS
high mass Drell-Yan distribution data as a function of the
invariant mass of the dilepton system $M_{ll}$.
As explained in the text, only virtual pure weak corrections have been
included in the calculation (in particular no photon-initiated corrections
are included).
The three curves displayed have been computed with {\sc\small FEWZ3.1} with the same 
input PDF set, namely NNPDF2.3 with $n_f=5$ and $\alpha_s(M_Z))=0.118$.
Right plot: same comparison for
the CMS 
double differential
Drell-Yan distribution as a function of the rapidity of the lepton pair
in the lowest invariant mass bin, with $20 \le M_{ll} \le 30$ GeV.
The bottom panels show the corresponding NNLO and EW $C$-factors that
are applied to the NLO calculations.
}        
\label{fig:cfactEWandNNLO}
\end{figure}

The NLO EW one-loop corrections are
 known~\cite{CarloniCalame:2007cd,Berends:1984qa,Berends:1984xv,Dittmaier:
2009cr,Baur:1997wa,Baur:2001ze} 
and have been implemented in several
public codes such as {\sc\small HORACE}~\cite{CarloniCalame:2007cd} and 
{\sc\small ZGRAD2}~\cite{Baur:1997wa,Baur:2001ze}. 
In {\sc\small FEWZ3.1}~\cite{Gavin:2012sy,Li:2012wna} the NLO EW corrections are 
combined to the NNLO QCD corrections,
 using the complex mass scheme. 
This code allows the user to separate a gauge--invariant QED
subset of the corrections from the full EW result.
The QED subset includes initial--state QED radiation, 
final--state QED radiation (FSR) and the initial--final interference terms.
The former cannot be consistently included without taking
into account the photon PDF $\gamma(x,Q^2)$. 
Within the current uncertainties that affect
the photon PDF, as determined in Ref.~\cite{Ball:2013hta}, initial--state QED corrections
are compatible with zero for most of the data included in the analysis.
In case they may be larger, we exclude the data from the fit, as 
explained in Sect.~\ref{sec:exclusion}.
As far as final state radiation (FSR) is concerned, when available, we
use data from which 
it has already subtracted. 
These include all  
ATLAS and CMS electro-weak vector boson 
production data, which have been corrected for  FSR (though the
correction has not been applied to  LHCb data). 

We may thus consistently isolate and compute the weak component 
of the EW corrections
with {\sc\small FEWZ3.1} and include it in our calculation via the computation of
$C-$factors for all the electroweak gauge boson production.
The size of the corrections is displayed in
Fig.~\ref{fig:cfactEWandNNLO}.
 We find that the effect
of the pure EW corrections is negligible for most of the data in the Z peak
region, of the order 1\% or below. For the CMS double differential
distributions, in the smallest invariant mass bin, we find that the pure EW
corrections are small, much smaller than the NNLO QCD corrections. At large
invariant masses the pure EW corrections are rather large and negative, as
expected from the results of Ref.~\cite{Boughezal:2013cwa}. 
This can be clearly seen
in  Fig.~\ref{fig:cfactEWandNNLO} for the ATLAS high-mass
Drell-Yan data, where EW
corrections reach $\mathcal{O}(7\%)$
in the last bin of the distribution corresponding to a dilepton
invariant mass of $M_{ll}\in\,[1000,1500]$.
Note that for the 
last bin of invariant mass measured by CMS, $M_{ll}\in\,[200,1500]$ GeV, EW effects
are still moderate, since due to the steep fall-off of the 
cross-section, most of the events are from the region around $M_{ll}\sim 200$ GeV. 

In summary, although the ATLAS high mass distribution is the only measurement for which EW corrections
are required, for consistency we include them for all the $Z/\gamma^*$ production data.

\subsubsection{Treatment of heavy quarks}
\label{sec:hq}

As in previous NNPDF analyses, in NNPDF3.0, 
heavy quark structure functions have been computed using the FONLL
general-mass variable-flavor-number (GM-VFN)
scheme~\cite{Forte:2010ta}.
In this scheme, the massive fixed-order calculation (in which the
heavy quark is
only counted in the final state) and resummed calculation (in which
the heavy quark is treated as a massless parton) are consistently
matched order by order. There is some latitude in deciding at which
order the fixed-order massive result ought to be included when parton
evolution is treated at a given perturbative order, and the FONLL
method allows consistent matching in any case. Specifically, in a NLO
computation one may decide to include massive contributions to
structure functions up to  $\mathcal{O}\lp \alpha_s\rp$, (on the grounds
that this is the order at which the massless structure functions are
also computed) or  up to  $\mathcal{O}\lp \alpha_s^2\rp$, (on the grounds
that the massive structure function starts at  $\mathcal{O}\lp
\alpha_s\rp$). These are called respectively FONLL-A and FONLL-B
schemes~\cite{Forte:2010ta}.
At NNLO,
while a similar ambiguity would exist in principle, in practice the
massive coefficient function can only be included up to
$\mathcal{O}\lp \alpha_s^2\rp$ (FONLL-C) because the  $\mathcal{O}\lp
\alpha_s^3\rp$ massive result is not known (progress in this direction
is reported e.g. in~\cite{Ablinger:2014nga})

While the FONLL-A scheme was used for the NNPDF2.1 and NNPDF2.3 NLO PDF
sets, we now adopt the FONLL-B scheme for NNPDF3.0 NLO, with FONLL-C
used at NNLO. While this scheme is less systematic, in that when going
to NLO to NNLO the massless computation goes up one order but the
massive one does not, it has the advantage that massive terms at NLO
are more accurate, thereby allowing for the inclusion of a somewhat
wider set of data already at NLO, as
small-$x$ and $Q^2$ charm production data are affected 
by large $\mathcal{O}\lp \alpha_s^2\rp$
corrections, and cannot be described in the FONLL-A scheme.
For the same reason, using FONLL-B  allows for a more accurate
description of the HERA inclusive cross section data at small-$x$.

Heavy quark structure functions are computed using  the expression
which corresponds to the
pole mass definition. Note that the distinction between pole and
$\overline{\rm MS}$ mass  is only relevant at NNLO and beyond~\cite{Alekhin:2010sv}.
In this paper, we adopt
the following values for the heavy quark pole masses,
\be\label{eq:hqmv}
m_c\,=\, 1.275~{\rm GeV} \, , \qquad m_b\,=\, 4.18~{\rm GeV} \,  , \qquad m_t\,=\, 173.07~{\rm GeV} \, ,
\ee
which correspond to the current PDG values for the $\overline{\rm MS}$ masses.
Note that these values are different from the ones used in NNPDF2.3, namely
$m_c=\sqrt{2}$ GeV, $m_b=4.75$ GeV and $m_t=175$ GeV.  A full
discussion of heavy quark mass dependence will be the subject of
future work; for the time being we have checked that the dependence of
our results on the value of the heavy quark masses is moderate, in
agreement with previous findings in the framework of
NNPDF2.1~\cite{Ball:2011mu}, as we shall briefly discuss in
Sect.~\ref{sec:model} below.

In NNPDF3.0 we take as a default the $n_f=5$ scheme, in which the
number of active flavors never exceeds  $n_f=5$ (i.e.
in the fit the top quark is always treated as massive, never
as a parton), though results will also be provided in the  $n_f=3$,
$n_f=4$ and  $n_f=6$ schemes.
 Note that in NNPDF2.1 and NNPDF2.3 the
default was $n_f=6$ (with the other cases also provided). While in
previous PDF determinations
the distinction between $n_f=5$ and $n_f=6$ was relevant only for
delivery, as no data above threshold were available, now several jet
data (especially for 2011 CMS inclusive jets, which has the highest reach in $p_T$) are
above top threshold, as well as some CMS and ATLAS high-mass Drell-Yan
data.
 Close to top threshold use of an $n_f=5$ scheme might be
advantageous because the top mass is treated exactly, while the loss
of accuracy due to the fact that the $n_f=5$ running of $\alpha_s$
differs from the exact   $n_f=6$ running~\cite{Demartin:2010er} 
is a comparatively smaller effect (only visible for processes which
start at high order in $\alpha_s$, see e.g.~\cite{Cascioli:2013era}).
 Furthermore, most of the codes  which we used for NNLO computations
 (specifically 
 {\sc\small NLOjet++} and {\sc\small FEWZ})  use an $n_f=5$ scheme, and the same
 is true for many of the codes and interfaces used in the computation
 of LHC processes. With an ever increasing set of LHC data, the use of
 an $n_f=5$ both in fitting, and as a default for PDF delivery appears
 thus to be more advantageous.

\subsection{Construction of the dataset}
\label{sec:dataset}

We strive to include as much of the available experimental data as
possible in our dataset. However, cuts have to be applied in order to
ensure that only data for which the available perturbative treatment is adequate
are included in the fit. These cuts are discussed in
Sect.~\ref{sec:exclusion}. Also, we include all available information
on correlated uncertainties; however, there is some latitude in the
treatment of correlated systematics, which we discuss in
Sect.~\ref{sec:chi2definition}.

\subsubsection{Kinematic cuts}
\label{sec:exclusion}

As in previous NNPDF fits, we apply a cut in $Q^2$ and $W^2$ to
fixed-target DIS data, in order to make sure that higher-twist
corrections are not needed: 
\be
Q^2 \ge Q^2_{\rm min}=3.5~{\rm GeV}^2,\,\qquad  W^2 \ge W^2_{\rm min} =
12.5~{\rm GeV}^2 \, .
\ee
The stability of the fit with respect to these choices (and in
particular the explicit check that they eliminate the need for higher
twists) has been studied in detail in
Ref.~\cite{Ball:2013gsa}. 
Low-scale DIS data carry less weight in our current fit than they did
previously, because of the inclusion of a large amount of new HERA and
LHC data. Therefore, the impact of these choices is even less important 
than it was previously.
Note that all NNPDF fits include target-mass corrections,
followind the method of Ref.~\cite{Ball:2008by}.

As discussed in Sect.~\ref{sec:theorytools} above, 
NNLO corrections are not available for the $W$ $p_T$ distribution and
for $W+c$ production. As a consequence,
ATLAS $W$ $p_T$ distribution data are included only in the NLO
fit. However, we include the
CMS $W+c$ distributions also in the
NNLO fit, but with matrix elements computed up to NLO only (but
$\alpha_s$ running at NNLO) because these data almost only affect the
strange distribution, and
the uncertainty on strangeness is rather larger
than the typical size of  NNLO corrections, so we would rather keep the
corresponding experimental information. 

\begin{table}
\small
  \begin{centering}
    \begin{tabular}{l|c|c|c}
      \hline 
      Experiment & $N_{{\rm dat}}$ & \multicolumn{2}{c}{Exclusion regions in the ($y,p_{T}$) plane}\tabularnewline
      \hline 
      \hline 
\multirow{2}{*}{CDF Run-II $k_t$ jets~\cite{Abulencia:2007ez}} & \multirow{2}{*}{52} & $1.1<|y|<1.6$ & $p_{T}<224$ GeV, $p_{T}>298$ GeV\tabularnewline
      &  & $|y|>1.6$ & all $p_{T}$ bins\tabularnewline
          \hline 
    
      \multirow{2}{*}{ATLAS 2.76 TeV jets~\cite{Aad:2013lpa}} & \multirow{2}{*}{3} & $0.0<|y|<0.3$ & $p_{T}<260$ GeV\tabularnewline
      &  & $|y|>0.3$ & all $p_{T}$ bins\tabularnewline
      \hline 
      \multirow{3}{*}{ATLAS 7 TeV jets 2010~ \cite{Aad:2011fc}} & \multirow{3}{*}{9} & $0.0<|y|<0.3$ & $p_{T}<400$ GeV\tabularnewline
      &  &  $0.3<|y|<0.8$ & $p_{T}<800$ GeV\tabularnewline
      &  & $|y|>0.8$ & all $p_{T}$ bins\tabularnewline
      \hline 
        \multirow{2}{*}{CMS jets 2011~\cite{Chatrchyan:2012bja}} & \multirow{2}{*}{83} & $1.0<|y|<1.5$ & $p_{T}<272$ GeV\tabularnewline
      &  & $|y|>1.5$ & all $p_{T}$ bins\tabularnewline
      \hline 
        \end{tabular}
    \par\end{centering}
  \caption{\small 
    \label{tab:jet-table} Summary of the exclusion regions in the jet transverse
momentum $p_T$ and rapidity $|y|$  used in the NNPDF3.0 NNLO fits
for the inclusive jet production measurements.
As explained in the text, these exclusion regions are determined
from a cut-off in the relative difference between the exact and approximate threshold
    $C$-factors in the 
 gluon-gluon channel~\cite{Carrazza:2014hra}.
$N_{{\rm dat}}$  in the second column is the number of experimental
    data points for these jet datasets that pass the selection
cuts in the NNLO fits.}
\end{table}
%

For inclusive jet production, we include all available 
data in the NLO fit, while in the NNLO fit we include only
the data points such that the
relative difference between the exact and the approximate $gg$-channel
NNLO $C$-factors differ by less than 10\%.
As already mentioned in Sect.~\ref{sec:thjetdata} above, a cut on the
size of the $C$-factors, only accepting jet data for which the NNLO
$C$-factor does not exceed 15\%, results in approximately the same
dataset.
In Tab.~\ref{tab:jet-table} we summarize the ensuing exclusion regions
in the ($p_{T},y$) plane and the number of data points 
points $N_{\rm dat}$ which survive the cut  for each experiment. 
As discussed in
Ref.~\cite{Carrazza:2014hra} and Sect.~\ref{sec:thjetdata} above, this
choice of cut ensures that NNLO corrections are included with an
expected accuracy of a few percent for all
the bins in $(p_T,y)$ of the inclusive jet cross section data.

For the ATLAS measurement of the $W$ transverse momentum
distribution, we include only those data points with $p_T^W>25$ GeV.
This cut excludes the first two bins in $p_T$,
 and is motivated by the observation that at small
$p_T$ the perturbative series is not well-behaved and all-order resummation is 
needed (either analytically or by matching the fixed order calculation
to a parton shower).

For the neutral-current Drell-Yan measurements from ATLAS and CMS,
we include only data for which the dilepton
invariant mass satisfies $M_{ll}< 200$ GeV.
This excludes the last six bins of the ATLAS DY invariant mass distribution
and the 12 points 
in the rapidity distribution corresponding to the last bin of invariant mass,
$M_{ll}\in[200,1500]$ GeV, for the CMS measurement.
The reason behind this cut is that the pure weak corrections
that we include in our calculation, using {\sc\small FEWZ},
 do not 
include the QED subset of the total electroweak correction.
 In particular, including
the photon--initiated contributions to the dilepton cross-section
would require an initial photon PDF $\gamma(x,Q^2)$, 
which is not fitted in this analysis. 
Using the NNPDF2.3QED set, we have
found that, while for most data points this contribution is compatible
with zero within the photon PDF uncertainty, for values
of $M_{ll}$ above $M_Z$
 such correction can become sizable, more than a few
percent, and up to 20\% for the highest invariant mass values. 
On this ground, we 
exclude all neutral-current Drell-Yan data with $M_{ll}>200$ GeV from the fit,
both at NLO and at NNLO.

The final cut is imposed, for the NLO fit only, to the
lowest invariant-mass bin of the CMS Drell-Yan double differential
distributions.
As can be seen from Fig.~\ref{fig:cfactEWandNNLO} (right plot),
for the bin with invariant mass $20 \le M_{ll}\le 30$ GeV, the NNLO
$C$--factors are large, around 10\%, while experimental uncertainties
are a few percent.
Therefore, it is not possible to obtain a reasonable fit
to these data points in a NLO fit, and therefore the 24 points 
of the $20 \le M_{ll}\le 30$ GeV bin of the CMS data 
are excluded from the NLO analysis.

The number of data points before and after cuts, both in the NLO and
NNLO fit, are summarized in Tab.~\ref{tab:completedataset3}.
At LO we use the same cuts as in the NLO fit: this is a sensible choice,
given that theory uncertainties at LO are larger than experimental uncertainties,
it does not make much sense to attempt to devise a set of optimized
kinematical cuts specifically for the LO fits.

\subsubsection{Treatment of correlated systematic uncertainties}
\label{sec:chi2definition}

In NNPDF3.0
we include all available information on correlated systematic
uncertainties. This information is summarized in
Tabs.~\ref{tab:completedataset} and~\ref{tab:completedataset2}.
All
experiments provide uncorrelated statistical uncertainties; and most of them
also a correlated normalization uncertainty.
The exception of the latter case are ratio experiments such 
as NMC~$d/p$, E866~$d/p$ and the CMS $W$ electron
and muon asymmetries, and measurements which
are normalized to the fiducial cross section like the D0
$Z$ rapidity distribution, the ATLAS $W$ $p_T$
distribution and the CMS 2D Drell-Yan data.
On top of these two types of uncertainties,
most experiments also
provide information on correlated systematics, either as an overall
covariance matrix, or as a set of nuisance parameters with the
corresponding correlation. 
The last case is the most complete in that
each individual source of systematic uncertainty 
can be in principle treated independently.

In Tabs.~\ref{tab:completedataset} and~\ref{tab:completedataset2} we
indicate in the 5th column whether the information on correlated
systematics, on top of the normalization, is either not provided, provided as
a covariance matrix, or fully provided with a complete breakdown
of all the sources of systematic uncertainties. 
Furthermore, in the 6th column,
datasets which are affected by common correlated systematics are
flagged with the same letter. 
In Tab.~\ref{tab:completedataset}, all the DIS datasets that
are labelled with the same letter in this column share a common
normalization uncertainty.
In addition, the BCDMS, CHORUS, HERA-I, ZEUS HERA-II and H1 HERA-II datasets
share other correlated systematic uncertainties.
In Tab.~\ref{tab:completedataset2}, the CDF datasets share a
normalization uncertainty, the ATLAS 7 TeV jets and $W,Z$ rapidity distributions
also have a correlated normalization uncertainty, and the ATLAS inclusive jet measurements
at 2.76 TeV and 7 TeV share a number of systematic uncertainties, most
importantly the jet energy scale.
On the other hand, none of the CMS datasets has common correlated uncertainties
among them.

Normalization uncertainties are special because the
uncertainty is proportional to the measured value, i.e. they are ``multiplicative''. This poses peculiar
problems in including them in a fit, because their naive inclusion in
the covariance matrix would lead to a systematically biased
result~\cite{D'Agostini:1993uj}. A full solution to this problem based
on an iterative procedure (the $t_0$ method)
has been presented in Ref.~\cite{Ball:2009qv} and
adopted by NNPDF in all PDF determination from NNPDF2.0
onwards (previous approximate solutions such as adopted in the
MSTW08~\cite{Martin:2009iq} PDF determination lead in practice to very
similar results in most realistic situations, 
see Refs.~\cite{Ball:2012wy,Butterworth:2014efa}). 

In hadron collider experiments, not only normalization uncertainties
but most, or perhaps all other correlated sources of uncertainty are
also multiplicative. After checking with the respective experimental
collaborations, we have thus concluded that the most accurate
treatment of correlated systematics is obtained by departing from the
standard approach of fixed-target deep-inelastic experiments, in which
all systematics but normalizations are additive~\cite{dagos}, and
treating all
systematics as multiplicative for hadronic collisions.
For deep-inelastic experiments, we treat all the correlated systematics
of the HERA data as multiplicative,
while for fixed-target experiments the systematics are treated as
additive 
 and only the
normalization uncertainties as multiplicative.
This information is also
summarized in
Tabs.~\ref{tab:completedataset} and~\ref{tab:completedataset2}, where in the
4th column experiments for which correlated systematic uncertainties
are treated multiplicatively are denoted by M and others are denoted
by A (additive uncertainties); normalization uncertainties are treated
multiplicatively for all experiments.
In Sect.~\ref{sec:results} we will study the effect of a
multiplicative vs. additive treatment of systematics,
and find that it is small
though perhaps not completely negligible.
This will be especially true for the gluon PDF at large $x$,
which is determined from collider jet data, for which an accurate
treatment of experimental systematics is crucial.

As usual in the NNPDF methodology,
we use the information on correlated systematics both in order to calculate the 
$\chi^2$ used in fitting PDF to the
Monte Carlo replicas, and also  for 
generating the Monte Carlo replicas themselves.
As explained in~\cite{Ball:2008by} 
(see also Eq.~(\ref{eq:L2shift}) in Sect.~\ref{sec:closure}),
the generation of the Monte Carlo replicas requires the knowledge
of the individual sources of independent correlated systematic uncertainties.
As can be seen from Tab.~\ref{tab:completedataset2}, for some
LHC experiments this breakup is not provided and only
the experimental covariance matrix is available.
In those cases,  we first create a set of artificial systematics 
consistent with the covariance matrix.
These are obtained by diagonalizing the covariance matrix and then rescaling 
each eigenvector by the square root of its eigenvalue. 
This provides a set of $\textrm{N}_{\textrm{dat}}$ artificial systematics which 
can be used to generate replica datasets in the same way as if we had
the full breakup of
systematics uncertainties, 
and which recombine correctly to give the original covariance matrix.
This procedure effectively allows us to generate replica data according to the 
multivariate Gaussian distribution implied by the experimental covariance matrix.

\section{Improved methodology}
\label{sec:methodology}

\noindent The inclusion of a substantial number of LHC 
datasets in a global PDF determination presents a variety of technical
challenges, both in terms of the complexity of adding new hadronic observables
into the fitting machinery and of
the corresponding increase in the computational cost of 
running fits. 
The implementation of the enlarged LHC dataset is particularly demanding
for NNPDF, which, being based upon a genetic algorithm minimization in a large parameter 
space, is already more computationally intensive than competing approaches. 
To provide an optimized solution to these challenges, for NNPDF3.0
we have developed a brand new fitting code based on
C{}\verb!++! and {\tt Python}, object-oriented languages that allow us 
to streamline
the inclusion of new datasets and to achieve a highly efficient implementation
of the minimization algorithms.

In what follows we first describe the modifications of the fitting code which make it more robust and 
efficient. 
Then we discuss the new PDF parametrization and the
enforced positivity constraints. 
Finally, we discuss
the improvements in the minimization strategy.
 
\subsection{Code}

The majority of the computer time spent in performing a PDF fit arises from the calculation of 
the physical observables for the comparison with experimental data.
Indeed, any PDF determination involves an iterative procedure where all the data points
included in the fit need to be recomputed a very large number of times for
different functional forms of the input PDFs.
The computation of physical observables in the NNPDF
framework is based upon the {\sc\small FastKernel} method introduced in 
Refs.~\cite{Ball:2010de,Ball:2012cx}. We refer the reader to the original publications 
for a detailed explanation of the method; here we simply recall 
a few basic facts that are necessary to explain the structure of the new code.

\subsubsection{FastKernel methodology}
\label{sec:fk}

For a suitable choice of interpolating functions, the parton distributions at a 
given scale $Q^2$ are represented on a grid of points in $x$: 
\begin{equation}
  \label{eq:fxgrid}
  \left\{f_i(x_\alpha, Q^2); \alpha=1,\ldots, N_x\right\}\, ,
\end{equation}
where the index $i$ identifies the parton flavor, 
and the index $\alpha$ enumerates the points on the grid.
 
Deep-inelastic observables, 
which are linear in the PDFs, can be computed using a precomputed
kernel $\hat{\sigma}^{I,J}_{\alpha j}$: 
\begin{equation}
  \label{eq:disFK}
  F_I(x_J,Q_J^2) = \sum_{j=1}^{N_\mathrm{pdf}}
  \sum_{\alpha=1}^{N_x} \hat{\sigma}^{I,J}_{\alpha j}
  f_j(x_\alpha,Q_0^2)\, .
\end{equation}
The index $I$ here labels the physical observable, $x_J$ and $Q_J$ are
the corresponding kinematical variables for each particular
experimental data point,  $\alpha$ runs over the points
in the interpolating grid and $j$ is a parton flavor index.
The kernel $\hat{\sigma}$ is
referred to as an {\sc\small FK} table. 
Similarly, hadronic observables, which are written as a convolution
of two PDFs,
 can be computed in terms of an (hadronic) {\sc\small FK}
table $\hat{W}^{I,J}_{kl\gamma\delta}$; the final expression is: 
\begin{equation}
  \label{eq:colFK}
  F_I(x_J,Q_J^2) = \sum_{k,l=1}^{N_{\rm pdf}} \sum_{\gamma,\delta=1}^{N_x}
  \hat{W}^{I,J}_{kl\gamma\delta} f_k(x_\gamma,Q_0^2)
  f_l(x_\delta,Q_0^2)\, .
\end{equation}
Once again we used the index $I$ to identify the physical observable
and $J$ to label the specific experimental data point for this observable; the
indices $k,l$ run over the parton flavors, and the indices $\gamma,\delta$
count the points on the interpolating grids. 
In the fitting code, for each experimental dataset $I$ we will have
a separate {\sc\small FK} table that encodes all the theory information.

The quantities $\hat{\sigma}$
and $\hat{W}$ in Eqs.~(\ref{eq:disFK}, \ref{eq:colFK}) encode all
the information about the theoretical description of the observables
 {\it e.g.}\ the perturbative order, the value of the strong
coupling, the choice of scales, the QCD and electroweak 
perturbative corrections, or the
prescription for the evolution. 
A modification in the theoretical
formulation is reflected in a new {\sc\small FK} table. The convolutions of
the {\sc\small FK} tables with the PDFs at the initial scale are generic,
and do not require any knowledge about the theoretical framework. 
It is thus clear from their structure that
 the {\sc\small FK} tables now implement a clean separation of the
theory assumptions from the actual fitting procedure. During a fit, the
PDFs at the initial scale are varied in order to minimize the $\chi^2$
while the {\sc\small FK} tables are always kept fixed and treated
as an external input.

It should be emphasized that the main difference between fast NLO calculators such as 
{\sc\small FastNLO}, {\sc\small APPLgrid}  and {\sc\small aMCfast},
and our {\sc\small FastKernel} method is that in the latter
also the PDF evolution is included into the precomputed tables, while
the other approaches require as input the PDFs evolved at the scales
where experimental data is provided.
This optimization step is essential to reduce the computational cost of running
NNPDF fits.
Note also that the generic structure Eqs.~(\ref{eq:disFK},
\ref{eq:colFK}) holds for any fast NLO calculator as well as for
any PDF evolution code. In NNPDF3.0 we use our own internal
Mellin-space {\sc\small Fkgenerator} code for PDF evolution and
DIS observables, but using $x$-space codes such as
{\sc\small HOPPET}~\cite{Salam:2008qg} or {\sc\small APFEL}~\cite{Bertone:2013vaa} 
should also be possible.
 
The fact that the expressions Eqs.~(\ref{eq:disFK},
\ref{eq:colFK}) of observables in terms of {\sc\small FK} tables
include the appropriate PDF evolution  pre-computed in the {\sc\small FK} tables
 $\hat{\sigma}$ (for DIS data) and $\hat{W}$ (for hadronic observables)
is particularly important for processes 
for which the precomputed fast NLO matrix elements, obtained
for example from {\sc\small APPLgrid}
or {\sc\small FastNLO}, require the PDFs at a large number of $Q^2$ values.
This is specially true for inclusive jet production, where for each
experimental bin 
the range of $Q^2$ values for which PDFs need to be provided is different
(since this range depends on the jet $p_T$ and rapidity).
In these cases, the associated acceleration due to the pre-computation of 
the PDF evolution makes all the difference between a 
process being practical or impossible to include in an NNPDF fit.
The improvement due to this acceleration is quantified in 
Tab. \ref{tab:FKtimings} below.


\subsubsection{NNPDF++}

For the NNPDF3.0 determination, a new fitting toolchain has been developed 
from scratch with the {\sc\small FK} procedure at its core in order to ensure 
that the most expensive part of any fit, the calculation of the physical observables, 
is performed in the most efficient manner possible. 
As discussed above the new fitting code has been designed 
with an explicit separation between experiment and theory, with all theory 
information contained within the {\sc\small FK} tables, 
and all experimental information held within a new data file format, 
internally dubbed {\tt CommonData}. 
These two data formats have an associated class structure built in 
modular {\tt C{}\verb!++!} (which replaces the previous {\tt FORTRAN}
implementation), 
each retaining the following areas of jurisdiction:

\begin{itemize}
\item Theoretical treatment: {\sc\small FK} tables
  \begin{itemize}
  \item PDF evolution.
  \item NLO QCD hard-scattering matrix elements.
  \item Perturbative order, value of $\alpha_S$/$\alpha_{\rm EW}$ and values
of the heavy quark masses.
  \item Heavy flavour renormalization scheme.
  \item NNLO QCD C-factors.
   \item NLO electroweak C-factors.
  \end{itemize}
\item Experimental information: {\tt CommonData}
  \begin{itemize}
  \item Process type information.
  \item Experimental kinematics for each data point.
  \item Experimental central values.
  \item Full breakdown of experimental systematic uncertainties.
  \item Flexible choice of additive/multiplicative for the treatment systematic uncertainties.
  \end{itemize}
\end{itemize}

The inclusion of a new experimental dataset into an NNPDF fit is therefore 
now a matter of converting the experimental results into the {\tt CommonData} file format, 
and generating a set of {\sc\small FK} tables each with the desired choices of theory parameters. 
Once generated, the {\tt CommonData} and {\sc\small FK} files are stored, and 
accessed thereon only via a set of classes used universally throughout the NNPDF++ fitting code. 

These new code structures have enabled a number of methodological improvements  
by providing a more efficient and flexible treatment of experimental data. 
For example, performing kinematic cuts upon an experimental dataset 
can now be performed simply and algorithmically by selecting the points in 
the {\tt CommonData} format which pass the required cuts according to their 
bundled kinematic information, and matching with the equivalent points in the 
{\sc\small FK table}. This is a considerable improvement over the earlier structure 
in {\tt FORTRAN} whereby a variation of kinematical cuts necessitated a 
complete regeneration of the pre-computed theory tables due to the 
monolithic treatment of experimental data. 

While the construction of this new fitting toolchain in C{}\verb!++! has made 
it possible to streamline
any variation in the theory settings and experimental cuts, the new fitting 
code itself also achieves substantial improvements over the previous {\tt FORTRAN} 
implementation.
 A modular treatment of most of the fitting procedure has been implemented, including 
PDF parametrization (both in terms of the flavor basis and of
the specific functional form), minimization methods and 
stopping criteria.
This allows for the rapid and safe 
replacement or improvement of separate modules of the NNPDF fitting
framework
 without the need 
for a large amount of programming work, enabling methodological studies 
of a greater depth than those that we have been able to 
perform previously, as we shall report in later sections.

To ensure a fast and efficient minimization procedure, 
the calculation of observables in the {\sc\small FK} 
convolution (the main bottleneck here) has been carefully optimized. 
In particular, the  {\sc\small FK} 
class that holds the hadronic table $\hat W$ from Eq.~(\ref{eq:colFK}) has been designed 
such that the $\hat W$ table is stored with the optimal alignment in machine memory 
for use with {\tt SIMD} (Single Instruction Multiple Data) instructions, which allow for
an acceleration of the observable calculation by performing multiple numerical
operations simultaneously. 
 The large size of a typical {\sc\small FK} product makes the 
careful memory alignment of the {\sc\small FK} table and PDFs extremely beneficial. 
A number of {\tt SIMD} instruction sets are available depending on the individual
processor. 
By default we use a 16-byte memory alignment for suitability with Streaming
{\tt SIMD} Extensions (SSE) instructions, although 
this can be modified by a parameter to 32-bytes for use with 
processors enabled with Advanced
Vector Extensions (AVX).
The product itself is performed both with {\tt SIMD} instructions and, where available,  
{\tt OpenMP} is used to provide acceleration using multiple CPU cores. 
An implementation of the {\sc\small FK} 
product for GPUs, while presenting no technical
objections, has so far not been developed due to scalability concerns on 
available computing clusters. 

Thanks to the implementation of
these various conceptual and technical 
optimizations in the calculation of physical observables, the NNPDF3.0 fitting
code benefits from a
substantial speed-up with respect to alternative  calculations,
 using for example
{\sc\small} {\small\sc APPLgrid}, where the PDF evolution needs to
be performed separately from the calculation of hard-scattering matrix elements.
This performance improvement is also clearly visible when comparing with the
calculation of the hadronic convolution Eq.~(\ref{eq:colFK}) using the
optimized settings with that 
using non-optimized settings.
To illustrate this point, in 
Tab.~\ref{tab:FKtimings} we compare the timings,
for a couple of representative LHC observables,
 for the convolution performed using
 {\small\sc APPLgrid}, a naive 
{\small\sc FK} implementation,
 and the optimized {\small\sc FK} implementation using the
SSE-accelerated calculation, for two representative observables. 
From Tab.~\ref{tab:FKtimings} it is clear that a massive 
 improvement in speed is achieved by precomputing the
PDF evolution in the {\small\sc FK} table,
with further improvements
obtained by the careful
optimization of the {\small\sc FK} product, and even further gains possible when
combined with {\tt OpenMP} on a multiprocessor platform.
 
\begin{table}[t]
\begin{center}
\begin{tabular}{|c|c|c|c|c|}
\hline Observable &{\small\sc APPLgrid} & {\small\sc FK} & optimized  {\small\sc FK} \\
\hline $W^+$ production &1.03 ms & 0.41 ms (2.5x) & 0.32 ms (3.2x) \\
\hline Inclusive jet production &2.45 ms & 20.1 $\mu$s (120x) & 6.57 $\mu$s (370x) \\ 
\hline
\end{tabular}
\caption[Comparison of {\sc\small APPLgrid} and {\sc\small FK} convolution timings.]{
\small
Illustrative timings for the calculation of LHC
observables using different methods to perform the
convolution between the matrix elements and the PDFs. 
We provide results for two different
 observables:
the total cross-section for $W^+$ production and for 
inclusive jet production
for typical cuts of $p_T$ and rapidity.
 In parenthesis we show the relative speed-up compared to the the reference
convolution based on
{\small\sc APPLgrid}.
The main difference between the second and third columns is that
in the latter case the PDF evolution is included
in the precomputed {\sc\small FK} table, while the differences
between the third and fourth column arise from 
the use of 
explicit SSE acceleration in the product in the
 optimized {\sc\small  FK} convolution. \label{tab:FKtimings}}
\end{center}
\end{table}

\subsection{PDF parametrization}

\label{sec:inputbasis}

As compared to the NNPDF2.3 analysis, there have been a number of substantial 
modifications in the way PDFs are parametrized and constrained. 
These include the
choice of input scale, the parametrization basis, preprocessing, and
the implementation of PDF positivity, which we
now discuss in turn.

\subsubsection{Parametrization basis}
\label{sec:pdfparam}
In previous NNPDF fits, PDFs were parametrized at a reference scale  
$Q_0^2$ = 2~GeV$^2$.
In NNPDF3.0 instead we choose $Q_0^2$ = 1~GeV$^2$.
Of course, this choice has no effect whatsoever on the results of the
fit: indeed, changing the input scale amounts to a change in the
input PDF parametrization, of which our fits are fully
independent, as we have checked at length previously, and will verify
again here (see in particular Sects.~\ref{sec:basis}, \ref{sec:huge}, \ref{sec:bastability}).
The motivation for lowering $Q_0^2$ is to be able to span a wider range of
possible values of the charm quark mass, without having to cross
the initial evolution scale ({\it i.e.} while always having $Q_0 \le
m_c$), which will simplify future studies of heavy quark mass dependence.
At this reference scale, in previous NNPDF fits, including NNPDF2.3, 
PDFs were expressed in terms of the following set of basis functions for quark and antiquark PDFs:
\bea
\Sigma(x,Q_0^2) &=&  \lp u+ \bar{u} + d +\bar{d}+ s + \bar{s} \rp(x,Q_0^2) 
\nonumber \\ 
T_3(x,Q_0^2) &=& \lp u+ \bar{u} - d -\bar{d} \rp(x,Q_0^2) \nonumber \\ 
V(x,Q_0^2) &=& \lp u- \bar{u} + d -\bar{d} + s - \bar{s} 
\rp(x,Q_0^2)  \nonumber\\ 
\Delta_S(x,Q_0^2) &=& \lp  \bar{d} - \bar{u}  \rp(x,Q_0^2) \label{eq:nnpdf23basis} \\ 
s^+(x,Q_0^2) &=& \lp s+ \bar{s} \rp(x,Q_0^2)  \nonumber\\ 
s^-(x,Q_0^2) &=& \lp s- \bar{s} \rp(x,Q_0^2)\,,  \nonumber
\eea
and then the gluon PDF $g(x,Q_0^2)$.
This basis was chosen because it directly relates  physical
observables to PDFs, by making the leading order expression of some
physical observables in terms of the basis functions particularly simple:
for example, $T_3$ is directly related to the difference in proton and
deuteron deep-inelastic structure functions $F_2^p-F_2^d$, and  $\Delta_S$ is
simply expressed in terms of Drell-Yan production in proton-proton and
proton-deuteron collisions, for which there is data for example from the
E866 experiment.

With the widening of the experimental dataset
in NNPDF3.0, there is little reason to
favor any particular PDF combination based on data, and thus we prefer
to choose the  basis that diagonalizes the
DGLAP evolution equations. We emphasize that the only purpose of such
choices is to speed up the minimization while leaving results
unaffected: independence of our results of this basis change will be checked
explicitly in Sects.~\ref{sec:basis} and \ref{sec:results} below.
The default basis in the NNPDF3.0 fits is thus
\bea
\Sigma(x,Q_0^2) &=&  \lp u+ \bar{u} + d +\bar{d}+ s + \bar{s} \rp(x,Q_0^2) \nonumber \\ 
T_3(x,Q_0^2) &=& \lp u+ \bar{u} - d -\bar{d} \rp(x,Q_0^2) \nonumber \\ 
T_8(x,Q_0^2) &=& \lp u+ \bar{u} + d +\bar{d} -2s - 2\bar{s} \rp(x,Q_0^2) \label{eq:nnpdf30basis} \\ 
V(x,Q_0^2) &=& \lp u- \bar{u} + d -\bar{d} + s - \bar{s} 
\rp(x,Q_0^2) \nonumber \\ 
V_3(x,Q_0^2) &=& \lp u-  \bar{u} - d + \bar{d} \rp(x,Q_0^2) \nonumber \\ 
V_8(x,Q_0^2) &=& \lp u- \bar{u} + d -\bar{d} -2s + 2\bar{s} \rp(x,Q_0^2), \nonumber 
\eea
and of course the gluon.
Here, as in previous NNPDF fits, we do not introduce an independent
parametrization for the charm and anticharm PDFs (intrinsic charm). However we 
do plan to do it in the near future.

As in all previous NNPDF fits, each basis PDF at the reference scale
is parametrized in terms of
a neural network (specifically
a multi-layer feed-forward perceptron) times a preprocessing factor:
\begin{equation}
\label{eq:preproc}
f_i(x,Q_0) = A_i \hat f_i(x,Q_0);\quad   \hat f_i(x,Q_0)=\,x^{-\alpha_i} (1-x)^{\beta_i} \,\textrm{NN}_i(x)
\end{equation}
where $A_i$ is an overall normalization constant, 
and 
$f_i$ and $\hat f_i$ denote the normalized and un-normalized PDF respectively.  
The preprocessing term $x^{-\alpha_i} (1-x)^{\beta_i}$ is simply there to 
speed up the minimization, without biasing the fit. We now discuss the 
overall normalizations $A_i$, while the 
 preprocessing will be addressed in Sect.~\ref{sec:preproc} below.

Out of the seven normalization constants, $A_i$ in  Eq.~(\ref{eq:preproc}),
three can be
constrained by  the valence sum rules (for up, down and
strange quarks) and another by the momentum sum rule.
Which
particular combinations depends of course of the choice of basis.
With the default NNPDF3.0 basis, Eq.~(\ref{eq:nnpdf30basis}), these constraints lead
to
\begin{align}
&A_g = \frac{1-\int_0^1 dx  x \Sigma(x,Q_0)}{\int_0^1 dx \; x\,\hat{g}(x,Q_0)};\quad
A_{V} = \frac{3}{\int_0^1 dx \, \hat{V}(x,Q_0)}; \label{eq:sumrules}\\
&A_{V_3} = \frac{1}{\int_0^1 dx \, \hat{V}_3(x,Q_0)};\quad
A_{V_8} = \frac{3}{\int_0^1 dx \, \hat{V}_8(x,Q_0)}.\nonumber
\end{align} 
The other normalization constants can be set arbitrarily to unity,
that is $A_{\Sigma}=A_{T_3}=A_{T_8}=1$: the overall size of these PDFs is then 
determined by the size of the fitted network.
The finiteness of sum rule integrals  Eq.~(\ref{eq:sumrules}) is
enforced by discarding during the genetic algorithm minimization (see
Sect.~\ref{subsec:ga} below) any mutation for which the integrals
would diverge.
This condition, in particular, takes care of those NN configurations that
lead to a too singular behavior at small-$x$.

Thanks to the flexibility of the
new C{}\verb!++! fitting code, in NNPDF3.0 we support the option of using any
arbitrary basis for the neural network parametrization of PDFs. 
This will allow us in
particular to check explicitly basis independence. 
However, whenever
we use a basis which differs from
the evolution basis  Eq.~(\ref{eq:nnpdf30basis}), the PDFs will be
transformed back to the evolution basis before preprocessing and
normalization are applied. 
This has the advantage of ensuring that the
finite integrals Eq.~(\ref{eq:sumrules}) do not have to be constructed
as the difference of divergent integrals, which would require large
numerical cancellations, thereby potentially leading to numerical
instabilities.

\subsubsection{Effective preprocessing exponents}
\label{sec:preproc}

As mentioned, when parametrizing PDFs in terms of
neural networks, Eq.~(\ref{eq:preproc}),
 preprocessing is introduced as a means to speed up the
minimization, by absorbing in a prefactor the bulk of the fitted
behaviour so that the neural net only has to fit deviations from
it. However, we must make sure that the choice of preprocessing
function does not bias the result.
 As in previous NNPDF fits, starting
with NNPDF1.2~\cite{Ball:2009mk} onwards, this is
done by randomizing the preprocessing exponents, i.e. by choosing a
different value for each replica within a suitable 
range. Unlike in previous NNPDF fits, where this range was determined
based on a stability analysis of the results, we now determine the
range self-consistently in a completely automatic way (the same
methodology was already used in the NNPDFpol family of
 polarized  PDF
determinations~\cite{Ball:2013lla,Nocera:2014gqa}). 

This is done in the following way. First of all, we define effective
asymptotic exponents as follows:
\begin{align}\label{effalpha}
\alpha_{\textrm{eff},i}(x) &= \frac{\ln f_i(x)}{\ln 1/x}\\ 
\beta_{\textrm{eff},i}(x) &= \frac{\ln f_i(x)}{\ln (1-x)}.
\label{effbeta}
\end{align}
Other definitions would be possible, such as 
\be
\alpha_{\textrm{eff},i}(x) = \frac{d \ln f_i(x)}{d \ln 1/x}\quad \, \quad
\beta_{\textrm{eff},i}(x) = \frac{d \ln f_i(x)}{d \ln(1-x)}.
\ee
We have checked that their use would not  modify qualitatively our results.
We choose a wide starting range for the preprocessing exponents for each
PDF, and perform a fit. The effective exponents
Eq.~(\ref{effalpha}-\ref{effbeta}) are then computed for all replicas 
at $x=10^{-6}$ and $10^{-3}$ for the low-$x$ exponent $\alpha_i$   
and at $x=0.95$ and  $0.65$ for the large-$x$ exponent $\beta_i$, for all 
PDFs (except for the gluon and singlet small-$x$ exponent, $\alpha_i$, which
is computed at $x=10^{-6}$). 
 The fit is then repeated by taking as new range for
each exponent the envelope of twice the 68\% 
confidence interval for each $x$ value. The process is then iterated
until convergence, i.e., until the output preprocessing exponents stop
changing. 
Reassuringly, convergence is typically very fast: even in the cases where
the fitted dataset is varied significantly, only one iteration
is needed to achieve stability.

This procedure ensures that the final effective exponents are well
within the range of variation both in the region of the smallest and
largest $x$ data points, and in the asymptotic region (these two regions
coincide for the gluon and singlet at small $x$), thereby ensuring
that the allowed range of effective exponents is not artificially
reduced by the preprocessing, either asymptotically or
sub-asymptotically. 

The final output values of the preprocessing
exponents for the central NLO and NNLO NNPDF3.0 fits are listed
 in Tab.~\ref{tab:preproc}. 
These ranges have been redetermined
 self-consistently for different fits: for example, for fits to
 reduced datasets, wider ranges are obtained due to the 
 experimental information being less contraining.

\begin{table}
\centering
\begin{tabular}{|c||l|l||l|l|}
\hline
 & \multicolumn{2}{c||}{NLO} & \multicolumn{2}{c|}{NNLO} \\
\cline{2-5}
PDF & [$\alpha_{\textrm{min}},\alpha_{\textrm{max}}$] & [$\beta_{\textrm{min}},\beta_{\textrm{max}}$] & [$\alpha_{\textrm{min}},\alpha_{\textrm{max}}$] & [$\beta_{\textrm{min}},\beta_{\textrm{max}}$] \\
\hline
\hline
$\Sigma$ & [1.06, 1.22] & [1.31, 2.68] & [1.02, 1.33] & [1.31, 2.74] \\
\hline
$g$ & [0.96, 1.37] & [0.28, 5.45] & [1.05, 1.53] & [0.85, 5.20] \\
\hline
$V$ & [0.54, 0.70] & [1.20, 2.91] & [0.54, 0.70] & [1.18, 2.80] \\
\hline
$V_3$ & [0.29, 0.58] & [1.31, 3.42] & [0.29, 0.61] & [1.36, 3.73] \\
\hline
$V_8$ & [0.54, 0.73] & [0.80, 3.09] & [0.55, 0.72] & [1.06, 3.07] \\
\hline
$T_3$ & [-0.17, 1.36] & [1.58, 3.14] & [-0.25, 1.41] & [1.64, 3.20] \\
\hline
$T_8$ & [0.54, 1.25] & [1.30, 3.42] & [0.54, 1.27] & [1.33, 3.23] \\
\hline
\end{tabular}

\caption{\small Ranges from which the
small- and large-$x$ preprocessing exponents in Eq.~\ref{eq:preproc}
 are randomly chosen for each PDF.
For each replica, a value is chosen from these ranges assuming
a flat probability distribution.
We provide the results for the global NLO and NNLO NNPDF3.0 fits.
The two sets of ranges,
obtained at each perturbative order,
are determined independently using an iterative procedure, as 
explained in the text.
\label{tab:preproc}
}
\end{table}

\subsubsection{Positivity constraints}
\label{sec:positivity}

As is well known~\cite{Altarelli:1998gn}, beyond leading-order parton
distributions do not need to be positive definite. However, the
requirement that measurable physical observables be positive still imposes
a generalized positivity constraint on the PDFs. 
In previous NNPDF fits these constraints were enforced by imposing positivity
of the deep-inelastic structure functions $F_L$, $F_2^c$ and of the
neutrino charm production (``dimuon") cross-section at
a scale of $Q^2_{\rm pos}=5$ GeV$^2$ 
in the range $x \in \lc 10^{-5},1 \rc$ (see in particular Sect.~4.5
and~5.5 of Ref.~\cite{Ball:2010de}). 
However, while these conditions
were sufficient to guarantee the positivity of most physical
observables in previous NNPDF fits,
in order to ensure positivity of all observables, the number
of independent positivity constraints must be at least equal to the number of
independently parametrized PDFs.

In order to guarantee full positivity of physical observables we have
thus enlarged the set of positivity constraints. In particular, we
have chosen to impose positivity of some pseudo-observables which
must respect positivity for reasons of principle, but which are not
measurable in practice. We choose the three tagged deep-inelastic
structure functions  $F_2^u$, $F_2^d$ and $F_2^s$, 
and the  three
flavor Drell-Yan rapidity distributions,
$d\sigma_{u\bar{u}}^{\rm DY}/dy$ , $d\sigma_{d\bar{d}}^{\rm DY}/dy$ and
$d\sigma_{s\bar{s}}^{\rm DY}/dy$, 
defined
respectively
as the
contribution to the structure function or the Drell-Yan rapidity
distribution which is obtained when all quark
electric charges are set to zero except that of the up, down or strange
quark flavor, respectively. 
These six constraints enforce generalized positivity of
the quark and antiquark PDFs. 
As in previous fits, our positivity conditions are imposed
at the low scale of $Q^2_{\rm pos}=5$ GeV$^2$, and will then be also
satisfied once the PDFs are evolved upwards in $Q^2$.

Generalized positivity of the gluon is
enforced by requiring positivity 
of the light component of the longitudinal
structure function  $F_L^l$, defined as the contribution to the
structure function $F_L$ when all quark electric charges are set to
zero but those of the three lightest flavors.
The use of $F_L^l$ allows us to impose gluon positivity without having
to make specific choices for the treatment of heavy quarks at the low
scale where these conditions are imposed.
This constraint is supplemented 
with that from the rapidity distribution
$d\sigma^H_{gg}/dy$  for  the production in gluon-gluon fusion of a
Higgs-like scalar with mass   $m_H^2=5$ GeV$^2$ (such a constraint being 
much more stringent than that from prodiuction of a heavier Higgs). 
The use of these two
observables ensures  positivity of the gluon
PDF both at small and at large $x$ values.

In summary, the pseudo-observables used to enforce generalized
PDF positivity, expressed schematically in terms of their underlying
parton content, the following:
\bea
F_2^u(x,Q^2) &\propto &  \lp u(x,Q^2) + \bar{u}(x,Q^2)\rp + \mathcal{O}\lp \alpha_s\rp
\nonumber
\\
F_2^d(x,Q^2) &\propto&  \lp d(x,Q^2) + \bar{d}(x,Q^2)\rp + \mathcal{O}\lp \alpha_s\rp \nonumber\\
F_2^s(x,Q^2) &\propto&  \lp s(x,Q^2) + \bar{s}(x,Q^2)\rp + \mathcal{O}\lp \alpha_s\rp \nonumber\\
\frac{d^2\sigma^{\rm DY}_{u\bar{u}}}{d M^2\,dy } &\propto& \, u(x_1,Q^2) \bar{u}(x_2,Q^2)
+ \mathcal{O}\lp \alpha_s\rp \label{eq:posobs}\\
\frac{d^2\sigma^{\rm DY}_{d\bar{d}}}{d M^2\,dy } &\propto& \, d(x_1,Q^2) \bar{d}(x_2,Q^2) 
+ \mathcal{O}\lp \alpha_s\rp \nonumber\\
\frac{d^2\sigma^{\rm DY}_{s\bar{s}}}{d M^2\,dy } &\propto& \, s(x_1,Q^2) \bar{s}(x_2,Q^2) 
+ \mathcal{O}\lp \alpha_s\rp \nonumber\\
F_L^l(x,Q^2) &\propto& C_g \, \otimes \, g(x,Q^2) \, + \,  C_q \, \otimes \, q(x,Q^2) + \mathcal{O}\lp \alpha_s^2\rp \nonumber\\
\frac{d \sigma^{H}_{gg}}{d y} &\propto& g(x_1,M_H^2)g(x_2,M_H^2) + \mathcal{O}\lp \alpha_s^3\rp \quad M_H \, \equiv \, Q_{\rm pos} \nonumber
\eea
All these positivity constraints are imposed at $Q^2_{\rm pos}=5$ GeV$^2$, and for
$x \in \lc 10^{-7},1 \rc$, which, because of the
structure of QCD evolution, ensures positivity at all higher scales
(and explains our choice for the mass of the Higgs-like scalar).
In practice we compute the observables at 20 points in the given $x$
range, equally spaced on a log scale for $x<0.1$ (ten points) and on a linear
scale for $x\ge 0.1$. 

During the minimization, the positivity
constraints are imposed by adding a Lagrange
multiplier, and then further discarding replicas for which any of the
pseudo-observables is negative by more than 25\% of its absolute
value computed with a fixed reference PDF set (typically, the outcome
of a previous fit). 
The latter condition is necessary for cross sections which are
very close to zero (e.g. close to kinematic boundaries, like
the rapidity tails of Drell-Yan
distributions) where the
Lagrange multiplier strategy is not effective.
For DIS structure functions and Drell-Yan distributions, the
pseudo-observables are computed using the
internal NNPDF  {\sc\small FastKernel} implementation, while the Higgs-like
scalar production cross section is computed 
using
 {\sc\small MadGraph5\_aMC@NLO}~\cite{Alwall:2014hca} interfaced to
{\sc\small aMCfast}~\cite{aMCfast}, using a tailored model which goes
beyond the effective theory
approximation for top quark mass 
effects~\cite{Artoisenet:2013puc,Demartin:2014fia}.

The strategy outlined above has been used to enforce 
 generalized positivity 
both in the NLO and NNLO fits.
The only difference is that in 
the NNLO fit we compute the pseudo-observables at NLO (with the PDFs evolved also at
NLO for consistency) because  at the low $Q^2$ values at which the
positivity pseudo-observables are computed, large unresummed NNLO
corrections lead to perturbatively unstable
predictions at large and small $x$. In addition, there
is some evidence that the resummed result is closer to NLO than
to NNLO, see for example Ref.~\cite{Altarelli:2008aj} 
for the case of deep-inelastic structure
functions.
The impact of the positivity constraints on the final PDFs is quite substantial, 
especially in the extrapolation regions, as one could expect.
This statement will be quantified and discussed in Sect.~\ref{sec:stability}, where
we will compare two NNPDF3.0 NLO fits with and without the positivity
constraints, and we will discuss further a posteriori checks of 
the implementation of the positivity conditions.
In the results section, we also explore the impact of the improved 
positivity constraints on searches for high-mass new physics.

In the leading order fits, where PDFs should be
strictly positive-definite, exactly the same strategy
as in the NLO and NNLO fits is used (with pseudo-observables
now computed at LO): we have verified
that, as one could expect, using the conditions
Eq.~(\ref{eq:posobs}) expressed at LO leads to effectively
positive-definite PDFs.
To avoid the problem of the occasional replica of the LO fit
that might become slightly negative, the {\sc\small LHAPDF6} NNPDF3.0LO
grids provide by construction a positive-definite output.
We have verified the robustness of our
LO positivity implementation by comparing it to alternative strategies,
such as using a neural network where the output of the last layer
is squared.

\subsection{Minimization algorithm}
\label{sec:minim}

As in previous NNPDF fits, minimization is performed using genetic
algorithms, which are especially suitable for dealing with very large
parameter spaces. Because of the extreme flexibility of the fitting
functions and the large number of parameters, the optimal fit is not
necessarily the absolute minimum of the $\chi^2$ (see in particular
the discussion in Sect.~4 of Ref.~\cite{Ball:2008by}), which might
correspond to an `overfit' in which not only the desired best fit is
reproduced, but 
also statistical fluctuation about it. As a consequence, a stopping
criterion is needed on top of the minimization method.
  In NNPDF3.0 we have improved both the minimization strategy and the stopping
criterion.

\subsubsection{Genetic Algorithms}
\label{subsec:ga}

We have performed a comprehensive re-examination of the genetic
algorithm minimization procedure utilized in previous NNPDF
determinations. 
Our approach has been to take a minimal starting methodology with 
only a few basic features. New features were added in turn and 
only retained if they resulted in faster fitting. 

The algorithm consists of three main steps: mutation, evaluation and selection. 
Firstly, a large number of mutant PDF sets are generated based on a parent set 
from the previous generation. The goodness of fit to the data for each mutant is then calculated. The best fit mutant is identified and passed on to the next generation, 
while the rest are discarded. The algorithm is then iterated until a set of stopping criteria are satisfied. 

The number of mutants tested each generation is now set to 80 
for all generations, removing
the two GA `epochs' used in previous determinations. However, this number is somewhat arbitrary, as the more significant quantity 
in terms of fit quality is the total number of mutants produced during the fit, i.e.\ the number of mutants 
multiplied by the number of generations. If one is increased while the other is decreased by the same factor the fit results 
are largely unchanged. All mutants are generated from the single best mutant from the previous generation.  

To generate each mutant, the weights of the neural networks from the parent PDF 
set are altered by mutations. 
In previous NNPDF fits the mutations have consisted of point changes, where 
individual weights or thresholds in the networks were mutated at random. 
However, investigations of strategies for training neural
networks~\cite{Montana:1989}  
have found that employing coherent mutations across the whole network
 architecture instead leads to improved fitting performance.
The general principle that explains this
 is that of changing multiple weights which are related by the structure 
of the network, leading to improvements in both the speed and quality of the training.

The neural networks used in NNPDF fits consist of connected nodes 
organized in layers. 
To get a value from the network, the nodes in the input layer are set with the required $x$ 
and $\log x$ value and then the activations of nodes in successive layers are 
calculated according to
\begin{align}
\xi_i^{(l)} &= g\left(\sum\limits_j w_{ij}^{(l)} \xi_j^{(l-1)} + \theta_i^{l}\right) \\
\label{eq:gfun}
g(a) &= \frac{1}{1+e^{-a}}
\end{align}
where $\xi_i^{(l)}$ is the activation of the $i$th node in the $l$-th layer of the network, 
$w_{ij}^{(l)}$ are the weights from that node to the nodes in the previous layer and 
$\theta_i^{l}$ is the threshold for that node. 
The weights and the thresholds are the parameters in the fit, and so are the objects 
which are changed in the mutation. 
The exception to Eq.~(\ref{eq:gfun}) is the last layer, where in order
to allow for an unbounded output a linear activation function $g(a)=a$ is
used instead. The flexibility of the fitting code allows us to 
easily explore other choices, for instance a quadratic output of the last layer,
$g(a)=a^2$, has been used in studies of the PDF positivity in leading order
fits.

In the NNPDF3.0 fits we use
a nodal mutation algorithm, which gives for each
node in each network an  
independent probability of being mutated. 
If a node is selected, its threshold and all of the weights are mutated 
according to 
\begin{equation}
w \rightarrow w + \frac{\eta r_{\delta}}{N_{\rm ite}^{r_{\rm ite}}} \, ,
\end{equation} 
where $\eta$ is the baseline mutation size, $r_{\delta}$ is a uniform random number between 
$-1$ and $1$, different for each weight, $N_{\rm ite}$ is the number of generations elapsed 
and $r_{\rm ite}$ is a second uniform random number between $0$ and $1$ shared by all of the weights. 
An investigation performed on closure test fits found that the best value for $\eta$ is 15, while for 
the mutation probability the optimal value turns out to be around 5\%, which corresponds to an 
average of 3.15 nodal mutations per mutant PDF set. 

As with the removal of the fast- and slow-epochs and their replacement with a single set of GA parameters,
the Targeted Weighted Training (TWT) procedure adopted in previous fits has also been dropped. This was originally introduced in order to avoid imbalanced training between datasets. With the considerably larger dataset
of NNPDF3.0 along with numerous methodological improvements, such an imbalance is no longer observed even in fits without weighted
training. Whereas previously the minimization was initiated with a TWT epoch in which the fit quality to individual datasets was minimized
neglecting their cross-correlations, in NNPDF3.0 the minimization always includes all available cross-correlations between experimental datasets. 

With the removal of the TWT mechanism, along with the consolidation of GA training into a single epoch with a unified set of mutation probabilities and sizes, the number
of free parameters in the NNPDF minimization has been considerably reduced. In Tab.~\ref{tab:gapars} we provide a comparison summarizing the relevant parameters in the NNPDF2.3 and NNPDF3.0 determinations.

\begin{table}[h]
\begin{center}
  \begin{tabular}{|c||c|c|c|c|c|c|}
    \hline 
 &   $N_{\rm gen}^{\rm wt}$ & $N_{\rm gen}^{\rm mut}$
&   $N_{\rm gen}^{\rm max}$ & $E^{\mathrm{sw}}$ & $N_{\rm mut}^a$ 
&  $N_{\rm mut}^b $\\
    \hline
NNPDF 2.3 &    $10000$ & 2500 & 50000 & 2.3 & 80 & 30\\
\hline
NNPDF 3.0 &    $ - $ & - & 30000 & - & 80 & - \\
    \hline
  \end{tabular}\\

\bigskip

  \begin{tabular}{|c||c|c|}
\hline
    \multicolumn{3}{|c|}{NNPDF2.3}    \\ \hline
    \multicolumn{3}{|c|}{Single Parameter Mutation}    \\
    \hline 
PDF &   $N_{\rm mut}$ &  $\eta$  \\
    \hline
\hline 
$\Sigma(x)$    & 2 & 10, 1 \\
$g(x)$  & 3 & 10, 3, 0.4 \\
$T_3(x)$   & 2 &  1, 0.1 \\
$V(x)$   & 3 &  8, 1, 0.1\\
$\Delta_S(x)$   &3 & 5, 1, 0.1 \\
$s^+(x)$  &  2 & 5, 0.5 \\
$s^-(x)$  &  2 & 1, 0.1\\
\hline 
  \end{tabular}   \hskip10pt \begin{tabular}{|c||c|c|}
\hline
    \multicolumn{3}{|c|}{NNPDF3.0}    \\ \hline
    \multicolumn{3}{|c|}{Nodal Mutation}    \\
    \hline 
PDF &   $P_{\rm mut}$ &  $\eta$  \\
    \hline
\hline 
$\Sigma(x)$    & 5\% per node & 15 \\
$g(x)$  & 5\% per node & 15 \\
$V(x)$   & 5\% per node &  15 \\
$V_3(x)$   & 5\% per node &  15\\
$V_8(x)$   &5\% per node & 15 \\
$T_3(x)$  &  5\% per node & 15 \\
$T_8(x)$  &  5\% per node & 15\\
\hline 
  \end{tabular}
  \end{center}
  \caption{\small  Comparison of genetic algorithm parameter between the NNPDF2.3 and NNPDF3.0 fits. In the top table, parameters controlling the maximum fit length, number of mutants, and (for NNPDF2.3) target weighted training settings are shown. In the tables below, the mutation parameters are shown for the two determinations in terms of their respective fitting bases. For the NNPDF3.0 fit the mutation probability is now set at 5\% per network node, and the mutation size is set to a consistent $\eta=15$.}
  \label{tab:gapars}
\end{table}

\subsubsection{Stopping criterion}
\label{subsec:cv}

As in previous NNPDF fits, the optimal fit is determined using a
cross-validation method. This is based on the idea of separating the data in
two sets, a training set, which is fitted, and a validation set, which
is not fitted. The genetic algorithm is used in order to minimize
the $\chi^2$ (or other figure of merit) of the training set, while the
$\chi^2$ of the validation set is monitored along the minimization, and
the optimal fit is achieved when the validation $\chi^2$ stops
improving (see Ref.~\cite{Ball:2008by} for a more detailed
discussion).

In previous NNPDF fits this stopping criterion was implemented by monitoring a
moving average of the training and validation $\chi^2$, and stopping
when the validation moving average increased while the training moving
average decreased by an amount which exceeded suitably chosen
threshold values. The use of a moving average and of threshold values
was necessary in order to prevent the fit from stopping  due to a
statistical fluctuation, but introduced a certain arbitrariness since 
the value of these three parameters (the length of the moving
average and the two thresholds) had to be tuned.

In NNPDF3.0 we have improved on this: we now simply stop all fits at
the point in which the fit reaches the absolute minimum of the
validation $\chi^2$ within the maximum number of generations $N^{\rm
  max}_{\rm gen}$. 
In practice this is done in the following way. All
replicas are minimized for $N^{\rm max}_{\rm gen}$ generations. In the 
beginning of the fit, both the
validation $\chi^2$ and the PDF configurations are stored. 
At the end of the each generation, the validation $\chi^2$ is computed
again, and if it is lower than the previous stored values, 
the PDFs are replaced with the current ones, if not the fit proceeds
to the next iteration. 
At the end, the stored PDFs  
have the the lowest validation $\chi^2$ seen 
during the fit, and they are taken as 
the global best fit. We call this the 'look-back' method, and it is
clearly completely objective. The price to pay for this is that now
all replicas have to be run up to $N^{\rm max}_{\rm gen}$.

In order for the look-back method to be effective, the value of  $N^{\rm max}_{\rm gen}$ 
must be large enough that the PDFs at the minimum do
not depend on it, i.e., such that the choice of a larger value does
not lead to a different minimum. We have verified this explicitly by
checking that results are unchanged if the value of $N^{\rm
  max}_{\rm gen}$ is increased, see Sect.~\ref{sec-longer} below. We have
furthermore checked that  results are stable upon small variations of
the position of the  minimum: this guarantees that the choice of the
absolute minimum (which could correspond to a local fluctuation) does
not bias the result in any way. For this purpose, we have changed the
look-back algorithm, by only updating the PDFs when the $\chi^2$
decreases by more than some threshold: we tried increasingly large
threshold values (0.1, 1, and 10 on the total $\chi^2$ not divided by
the number of data points) and verified that even though, of course,
the stopping point changes, and happens at an earlier stage for large
values of the threshold, the resulting PDFs are indistinguishable. This
explicitly verifies
that local minima whose validation $\chi^2$ values differ from 
the absolute minimum
by a small amount correspond to essentially the same PDFs, as one
might expect.

Even so, there is still the possibility that occasionally for a particular
replica the training length required to reach the global minimum
would be exceptionally long, either because of an unusual fluctuation
of the pseudodata, or because of fluctuations in  the minimization. In
that case, the look-back method would not be effective, because the
best $\chi^2$ within the maximal training length would still be far
from the absolute minimum. In order to safeguard against this
possibility, we consider the PDF arc-length defined as
\begin{equation}
\label{eq:arclength}
L = \int \sqrt{1+\left(\frac{df}{dx}\right)^2} dx, 
\end{equation}
suitably discretised. 
Use of an the arc-length penalty has been previously suggested
as a way of penalizing
functional forms that are too wiggly, see e.g.
example the studies in Ref.~\cite{Glazov:2010bw}.

We then take the value of $L$ as an indicator of convergence: a PDF
replica with a very unlikely value of $L$ is assumed not to have
converged, and thus discarded. This is implemented through an  arc-length veto.
The arc-length is computed for each PDF replica, and the average length
and its standard deviation over the replica sample are determined for
each PDF. We then discard PDF replicas for which at least one PDF
the arc-length exceeds by more than four sigma the mean arc-length for
that PDF.

Finally, at the end of the minimization, an a posteriori
quality check on the resulting sample of Monte Carlo replicas is performed
for each fit.
These quality tests verify that the PDF generalized positivity has been
successfully implemented (see discussion in Sect.~\ref{sec:positivity}),
that no replica has a too unlikely arc-length, Eq.~(\ref{eq:arclength}),
and that likewise no replica has a too unlikely value of the $\chi^2$.
As in the case of arc-length, if any given replica has a $\chi^2$ whose
value is more that four-sigma away the mean $\chi^2$, it is automatically 
replaced by another replica that instead satisfies this condition.

\section{Closure testing}
\label{sec:closure}

In this section we describe in detail the strategy that has
been used to validate
the new fitting methodology described in the previous section:
closure testing.
The benchmarking of fitting  methodology is especially important
due to the substantial increase in experimental data included
in NNPDF3.0, and the increased precision of the resulting PDFs: as
data become more precise and their kinematic coverage increases, it becomes
more and more important to eliminate methodological uncertainties, so that
the only uncertainties in our PDFs are experimental and theoretical.
For example a basic requirement for a successful methodology is to be
able to fit
widely different datasets with the same methodology, and without having to
tune it or modify it according  to the dataset, or to the theory which
is used to describe it (for example at different orders in perturbative QCD). As we will show
below, these requirements can indeed be achieved in the NNPDF3.0 framework.
NNPDF3.0 is in fact the first PDF determination for
which the complete fitting methodology has been thoroughly tested
and tuned in closure tests based
on pseudo-data that have the same kinematical coverage and statistical
properties as the experimental data included in the fit. The
idea of using perfect pseudo-data to validate some specific aspects of a PDF
fitting methodology has been previously explored in Ref.~\cite{Watt:2012tq}.

The basic idea of the closure test is simple~\cite{demortier}:
we take a given assumed form for the PDFs (for example MSTW08), a given
theoretical model (for example NLO pQCD), and with them generate a set of
global pseudo-data with known but realistic
statistical properties (by using the covariance
matrices of the real datasets that together
make up, for example, the NNPDF3.0 dataset). These pseudo-data are
then `perfect', in the sense
that they have known statistical properties, no internal inconsistencies,
and are also entirely consistent with the theoretical model used to
produce them. Thus if we then use our fitting methodology to perform a
fit to these pseudo-data, we
should reproduce the assumed underlying PDF, within the correct uncertainties.
This latter point can be explored in some depth by changing by hand
the level of uncertainties incorporated within the pseudo-data.

We will first introduce the idea of closure testing in the context
of the NNPDF fits, and discuss its practical implementation.
Then we will quantify the efficiency of our neural network training methodology
by performing closure tests to perfect pseudo-data, without any fluctuations,
so that the fit quality can be arbitrarily good.
We then perform fits where the pseudo-data is supplemented with different
levels of statistical and systematic fluctuations.
Finally, in the last part of this section we use closure tests to assess
the robustness of our methodology against variations of
some of its ingredients, like the choice of PDF fitting basis or the dependence
on the size of the neural network.
Some of these last checks have also been performed on real
data, as will be reported in Sect.~\ref{sec:results}.

\subsection{NNPDF closure testing}

The  framework used in NNPDF3.0
for the computation of observables, as presented
in Sect.~\ref{sec:fk}, provides us with the ideal tool to successfully
implement closure tests. In particular, the clean separation between
theoretical assumptions and input PDFs allows us to generate pseudo-data
using a given set of PDFs and the experimental covariance matrix
as an input, and to perform a fit to this
pseudo-data using exactly the same theoretical settings (encoded in
the {\sc\small  FK} tables) that were used for generating them.

Throughout this section we shall refer to the parton distributions used to
generate the pseudo-data as the {\em input} PDFs,
and denote them by $f_{\rm in}$. Any PDF set available through
the {\sc\small LHAPDF} interface can be used as an input
set to generate the pseudo-data: most of the closure tests described
here will be performed using MSTW08, though we will also later
describe results using other input PDFs. We denote
the set of pseudo-data by $\mathcal{D} = \{D_I\}$: the dependence of the
pseudo-data on the input PDF  $f_{\rm in}$ and experimental covariance
matrix will be left implicit.

The outcome of the closure test fits
is then a set of fitted PDFs, $f_{\rm fit}$, which we
will compare to the input PDFs in order to study the statistical
precision and possible systematic biases in the fitting methodology.
For any PDF set $f$, whether input or fitted,
the {\sc\small FastKernel} framework delivers a set of
theoretical predictions, $\mathcal{T}[f] = \{T_I[f]\}$, based on
a particular theoretical model, which for present purposes we take to
be NLO perturbative QCD, precisely as implemented in the NNPDF3.0 fits
to real data, with the same parameter choices and so on.
The one exception to this statement is that for closure tests to MSTW08
we drop the generalized positivity constraints introduced
in Sect.~\ref{sec:positivity}, since positivity was imposed differently
by MSTW. We have checked however, by performing closure tests based on NNPDF3.0
as input PDFs, that none of the conclusions drawn in this
section is affected once the generalized positivity
is included in the fit.

Fitting is performed by minimizing a $\chi^2$ function (denoted during
the fitting as the `error function'): this is also used to assess the
quality of the resulting fit.
Such a $\chi^2$ depends on the dataset, $\mathcal{D}$, and on the
theoretical predictions of the
PDFs $f$ being fitted, $\mathcal{T}[f]$:
\begin{equation}
  \label{eq:chi2not}
  \chi^2[\mathcal{T}[f],\mathcal{D}] = \frac{1}{N_{\mathcal{D}}}
  \sum_{I,J} (T_{I}[f] - D_{I}) \, C^{-1}_{IJ} \,
  (T_{J}[f] - D_{J} )\, .
\end{equation}
In this expression, $C_{IJ}$ is the covariance matrix of the data (here we
always use the $t_0$-covariance matrix in order to avoid bias in the
inclusion of multiplicative uncertainties, see Sect.~\ref{sec:chi2definition}),
and $N_{\mathcal{D}}$ is the total number of data points of the dataset.
Note that when fitting the pseudo-data
we thus use exactly the same procedure (in fact the same code) as we use
in a fit to real data: we use
the same error function, evaluated with the real data replaced by pseudo-data,
the $t_0$ covariance matrix, and then the same fitting methodology
(genetic algorithm, stopping criterion, etc). Since in the closure test
fits the ``correct'' solution is known -- it is
given by the PDFs $f_{\rm in}$ used
as an input -- the result $f_{\rm fit}$ of the fit should then
reproduce the input PDFs within the statistical uncertainties of
$f_{\rm fit}$ as determined by the fit.

For the purposes of the studies that will be performed in
this paper, we will introduce three distinct categories (levels) of
closure tests depending
on the amount of stochastic noise added to the
pseudo-data points generated from the initial PDFs.
In order to make these tests as realistic as possible,
this stochastic noise is generated using the complete information
in the experimental covariance matrix, so that the fluctuations
and correlations of the pseudo-data reproduce precisely those
of the real experimental data.
For the baseline closure tests presented in this section,
the pseudo-data is in one-to-one correspondence with the experimental data used in the global
fit, that is we have generated pseudo-data for every point in the
NNPDF3.0 global dataset described
in Tabs.~\ref{tab:completedataset} and~\ref{tab:completedataset2}.

The three levels of closure test that we will study,
which we call {\bf Level~0},  {\bf Level~1}
and  {\bf Level~2} for reasons that will soon become apparent,
are set up as follows:

\begin{itemize}
\item {\bf Level~0}.

Using a given set of input PDFs,
pseudo-data $\mathcal{D}_0 = \{D_I^{0}\}$ are generated
using the {\sc\small FastKernel} convolution,
Eqs.~(\ref{eq:disFK},\ref{eq:colFK}).
In these Level~0 fits, no stochastic noise is added to
the pseudo-data.
Then we perform $N_{\rm rep}$ fits, each to exactly the same set of
pseudo-data, minimizing the error function (which here is the
same as the $\chi^2$ per degree of freedom, i.e.\
$\chi^2[\mathcal{T}[f],\mathcal{D}_0]$), but using different
seeds for the initialization of the
random numbers used in the minimization. This yields an ensemble of PDF
replicas $\{f_{\rm fit}^k\}$, where $k=1,\ldots,N_{\rm rep}$.

Note that the error function which we minimize is still computed using
the covariance matrix of the data, even though now the pseudo-data have
zero uncertainty. While of course the overall normalization of the
error function is immaterial (as the minimum is at zero), this has the
advantage of reproducing the correlations in the underlying dataset,
which means that the total amount of independent experimental
information (the number of independent data points) is the same as in
the original dataset.

It should be clear from its definition that in Level~0 closure tests, the
fit quality can be arbitrarily good, provided we use a sufficiently flexible
PDF parametrization and a sufficiently efficient minimization algorithm.
Indeed, since, by construction, the pseudo-data does not have any stochastic
noise, and there are no inconsistencies, there exist {\it  perfect}\ fits
to the Level~0 pseudo-data that have a vanishing $\chi^2$. We use the
plural here because there clearly exists an infinity of fits which
lead to vanishing $\chi^2$ by going through all data points, but differ
in the way they interpolate between data points.
These optimal solutions to the minimization problem reproduce precisely the
predictions of the set of PDFs used as input in the generation of
the pseudo-data at each of the experimental data points.
The Level~0 closure test is thus a highly non-trivial test of the
efficiency of the minimization procedure: at Level~0
the value of the error function (i.e., the
$\chi^2$ evaluated for each replica)
should  decrease monotonically towards zero
as the fit proceeds, if the functional form used for
parametrizing the fitted PDFs is flexible enough. Consequently, the
best-fit $\chi^2$ (i.e. the $\chi^2$ evaluated for the average of all
replicas) should also go to zero.

\item {\bf Level~1}.

Now we add stochastic fluctuations on top of the pseudo-data
generated for the Level~0 closure tests, $D_I^{0}$, as follows:
  \begin{equation}
    \label{eq:L0shift}
    D_{I}^{1} = \left(1 + r_I^{\rm nor} \sigma_I^{\rm nor}\right)\,
    \Big(
      D_I^{0} + \sum_{p=1}^{N_{\rm sys}} r_{I,p}^{\rm sys} \sigma^{\rm sys}_{I,p} + r_I^{\rm stat} \sigma^{\rm stat}_I
    \Big)\, ,
  \end{equation}
where, as explained in~\cite{Ball:2008by}, $\sigma^{\rm stat}_I$,
$\sigma^{\rm sys}_{I,p}$ and $\sigma_I^{\rm nor}$ are the statistic,
systematic and normalization uncertainties for each dataset,
and the random numbers $r_I^{\rm nor}$, $r_{I,p}^{\rm sys}$ and
$r_I^{\rm stat}$ are generated with the appropriate
distribution to reproduce
the experimental covariance matrix.
These shifted data points represent the central values of a hypothetical experiment,
for which the size of the statistical, systematic and
normalization uncertainties are given in Eq.~(\ref{eq:L0shift}).

In Level~1 fits, the same underlying pseudo-data, generated
by the random numbers in Eq.~(\ref{eq:L0shift}), are
used for the fit of all the $N_{\rm rep}$ replicas, but as at Level~0
a different random seed is used to initialize the
minimization of each replica.
Therefore no additional stochastic fluctuations (``Monte Carlo
replicas") are added to Eq.~(\ref{eq:L0shift}), and thus, as we will
confirm later, in a Level~1 fit the experimental uncertainties are not
propagated into the uncertainties of the fitted PDFs. The ensemble of PDF
replicas $\{f_{\rm fit}^k\}$ resulting from the Level~1 fits is thus
expected to underestimate the PDF uncertainties.

From its definition, given that  the pseudo-data have
fluctuated on average by one standard deviation away from the Level~0
value,  we expect that in Level~1 closure tests the error function
(which as at Level~0 coincides with the $t_0$ $\chi^2$ per degree of
freedom, i.e.\ $\chi^2[\mathcal{T}[f],\mathcal{D}_1]$) of the best fit
will be around one.
Moreover, also here there exist {\it perfect} solutions for
the minimization, for which the $\chi^2$ of the fitted PDFs
coincides exactly with the $\chi^2$ computed with the
input PDFs used for the generation of the pseudo-data, i.e.\ such that
$\chi^2[\mathcal{T}[f_{\rm fit}],\mathcal{D}_1] =
\chi^2[\mathcal{T}[f_{\rm in}],\mathcal{D}_1]$. Nevertheless, not all PDFs
produced by the fitting will be perfect, and the distribution of the fitted
PDFs will be too narrow.

\item {\bf Level~2}.

Now, starting from the shifted pseudo-data in  Eq.~(\ref{eq:L0shift}),
we generate $N_{\rm rep}$ Monte Carlo replicas as in the standard
NNPDF procedure.
Schematically, this means that we will have, for each replica $k$, a set
of pseudo-data $\mathcal{D}_2^k=\{D_{I}^{2,k}\}$ with
 \begin{equation}
    \label{eq:L2shift}
    D_{I}^{2,k} = \left(1 + r_I^{{\rm nor},k} \sigma_I^{\rm nor}\right)\,
    \Big(
      D_I^{1} + \sum_{p=1}^{N_{\rm sys}} r_{I,p}^{{\rm sys},k} \sigma^{\rm sys}_{I,p} + r_I^{{\rm stat},k} \sigma^{\rm stat}_I
    \Big)\, ,
  \end{equation}
for $k=1,\ldots,N_{\rm rep}$, and the set of random numbers
is different replica by replica.
From the practical point of view, once we have a set of Level~1
pseudo-data Eq.~(\ref{eq:L0shift}), the Level~2 $N_{\rm rep}$ Monte Carlo
pseudo-data replicas Eq.~(\ref{eq:L2shift}) are obtained using
exactly the same code as is used for the fits to real data. Fits to the
Level~2 pseudo-data replicas are also performed in exactly the same way
as to real data replicas, using the same error function, and with different
random seeds to initialize the minimization fitting algorithm for each replica.

In Level~2 fits, each Monte Carlo replica represents a fluctuation around
the Level~1 pseudo-data, and the procedure should correctly propagate
the fluctuations
in the pseudo-data, due to experimental statistical, systematic and
normalization uncertainties, into the fitted PDFs. The fit to each data
replica yields a PDF replica $f_{\rm fit}^k$, and the ensemble of PDF
replicas then contains all the information on PDF uncertainties
and correlations.
Hence we expect the final error function of a Level~2 fit (taken as
$\chi^2[\mathcal{T}[f_{\rm fit}^k],\mathcal{D}_2^k]$ just like
in a fit to real data)
to be close to two (since,
in a sense, each replica contains two fluctuations), while the $\chi^2$ per
degree of freedom of the replica PDFs to the pseudo-data (i.e.\
$\chi^2[\mathcal{T}[f_{\rm fit}^k],\mathcal{D}_1]$) will be close to one.
Moreover we expect the input PDFs $f_{\rm in}$ to lie within the one-sigma
band of the fitted PDFs with a probability of around 68\%.
\end{itemize}

\begin{table}[t]
\small
  \centering
  \begin{tabular}{|l|l|c|l|c|l|}
    \hline
    tag & $f_\mathrm{in}$ & level & stopping & training fraction & details\\
    \hline
    C1 & MSTW08NLO & 0 & fixed length 1k & 100\% & \\
    C2 & MSTW08NLO & 0 & fixed length 3k & 100\% & \\
    C3 & MSTW08NLO & 0 & fixed length 5k & 100\% & \\
    C4 & MSTW08NLO & 0 & fixed length 10k & 100\% & \\
    C5 & MSTW08NLO & 0 & fixed length 30k & 100\% & \\
    C6 & MSTW08NLO & 0 & fixed length 50k & 100\% & \\
    C7 & MSTW08NLO & 0 & fixed length 100k & 100\% &\\
    C8  & MSTW08NLO & 1 & look-back 30k & 50\% & \\
    C9    & MSTW08NLO & 2 & look-back 30k & 50\% & \\
    C10    & CT10 & 2 & look-back 30k & 50\% & \\
    C11  & MSTW08NLO & 2 & look-back 80k & 50\% & \\
    C12 & MSTW08NLO & 2 & look-back 30k & 50\% & NNPDF2.3 basis\\
    C13 & MSTW08NLO & 2 & look-back 30k & 25\% & \\
    C14& MSTW08NLO & 2 & look-back 30k & 75\% & \\
    C15 & MSTW08NLO & 2 & look-back 30k & 50\% & 2-20-15-1 NN arch \\
    C16 & NNPDF3.0 & 2 & look-back 30k & 50\% & \\
    C17 & NNPDF3.0 & 2 & look-back 30k & 50\% & with positivity\\
    \hline
  \end{tabular}
  \caption{\small List of closure fits used in this section. The tag in the first column is used to identify the
    fits in the text. The second column reports the choice of $f_\mathrm{in}$ for each fit. The level of the closure
    fit can be found in the third column, while the fourth column specifies the stopping method and the length
    of the fit, and the last columns provides additional relevant information. }
  \label{tab:CTfits}
\end{table}

In the following sections, we discuss in detail the
results of these closure tests, starting
from Level~0 and then moving to Level~1 and Level~2, including
variations of some of the fitting methodology settings such
as the stopping or the fraction of trained data.
For ease of reference, the closure fits discussed in this section
have been summarized in Table~\ref{tab:CTfits}, where we collect
the tag of the fit, the set of input PDFs used for the pseudo-data
generation, the level of closure fit, the settings for the stopping
and data training fraction, and other relevant information.
All these fits are done using NLO theory, though of course given
the use of pseudo-data this choice is immaterial in the interpretation
of the closure tests.

\subsection{Validation of the training efficiency: Level~0 closure tests}
\label{sec:level0cl}

We present the results of a number of Level~0 closure tests,
using them to assess the training efficiency of the NNPDF3.0
minimization.
Indeed, in Level~0 fits there exists in principle an optimal
solution for the minimization: that which reproduces exactly the input PDFs
used for the pseudo-data generation.
Therefore, we can study which minimization strategy gets closest
to this optimal solution, with the smallest computational effort.

\subsubsection{Training methodology efficiency}

The two main ingredients of our fitting methodology that can
be tested in Level~0 closure tests are the
adequacy of the neural network architecture, ensuring that it
provides a flexible enough parametrization to reproduce
any kind of input PDFs, and the efficiency of the genetic
algorithm minimization, by comparing different options and verifying which
one gets to the optimal solution quickest.
Since in this case no stochastic fluctuations are added in the generation of
the pseudo-data, when the fitted PDFs are equal to the input ones,
the $\chi^2$ of the fit, as already mentioned, will exactly vanish:
$\chi^2[\mathcal{T}[f_{\rm fit}],\mathcal{D}_0] = 0$.
As the length of the training is increased, the fitted PDFs are expected to
get closer and closer to the input ones.

In order to verify this feature, we have performed
a number of fixed length fits to the full dataset, and studied the
dependence on the training length of the $\chi^2$ of the best-fit PDF, i.e.
$\chi^2[\mathcal{T}[f_{\rm fit}],\mathcal{D}_0]$ evaluated for the
average of all replicas.
The fits used for this study are the fits C1 to C7 in Table~\ref{tab:CTfits}.
All these fits use identical settings with the only difference being
the training length, that ranges from short fits
of 1k to very long fits of 100k genetic algorithm generations.
In all these Level~0 closure tests, the input PDF set is taken
to be MSTW08 NLO.

\begin{figure}[t]
  \centering
  \epsfig{width=0.90\textwidth,figure=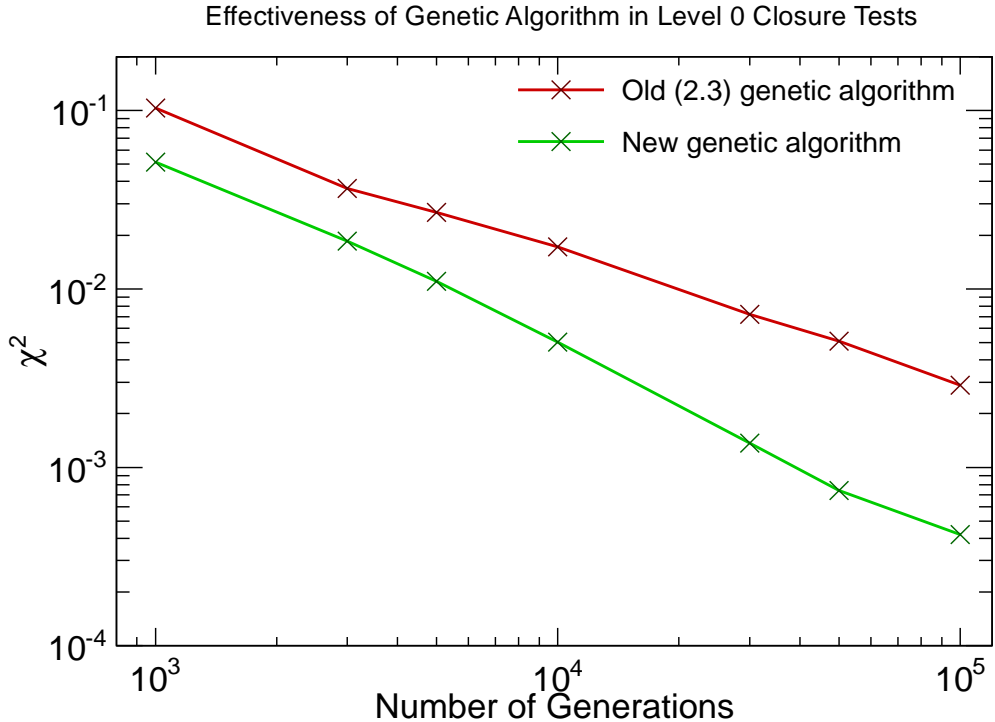}
  \caption{\small The normalized $\chi^2$ per degree of freedom of Level~0
  closure tests, Eq.~(\ref{eq:chi2not}) computed for the central
 PDF (average over replicas), as a function of the length of the genetic algorithms
minimization.
We compare the results for the GA settings used in NNPDF3.0,
based in particular on the nodal mutations strategy,
with the corresponding GA settings used in the NNPDF2.3 fit.
The results for each closure test are marked with crosses;
lines joining these points are added to guide the eye.
}
  \label{fig:chi2vslenL0}
\end{figure}

The dependence of the $\chi^2$ per degree of freedom on the training length
for these Level~0 closure tests
 is plotted in Fig.~\ref{fig:chi2vslenL0}.
We compare the results for the genetic algorithm used in NNPDF3.0,
based in particular on the nodal mutations strategy discussed in
Sect.~\ref{subsec:ga},
with the corresponding genetic algorithm used in the NNPDF2.3 fit.
The results for each closure test are marked with crosses,
with auxiliary lines joining these points to guide the eye.
It is clear from the figure that as expected the $\chi^2$ of the fit keeps
 decreasing as the fit length is increased, with a behaviour that is
approximately described by a power law.

It is important to recall that the normalization of the $\chi^2$
Eq.~(\ref{eq:chi2not}) for Level~0 fits is arbitrary: the uncertainty
in the data being fitted vanishes, and thus the inverse of its covariance matrix is singular. As explained above, we have chosen
to normalize the $\chi^2$ using the covariance matrix of the original experimental
data, even though at Level~0 this
plays no role neither in the data generation nor in the fitting.
Thus the absolute numerical value of the $\chi^2$ simply tells us what
is the average distance of the best fit to the data on the scale of
these experimental uncertainties. Thus  Fig.~\ref{fig:chi2vslenL0} shows that after
$30k$ GA generation each fitted point differs from the data by about
$0.03\sigma$. This fit quality is very uniform across
replicas. For instance, at $30k$ GA, where the $\chi^2$ of the central PDF is
$0.0014$ (as shown in Fig.~\ref{fig:chi2vslenL0}),
the average over replicas is $\langle\chi^2\rangle = 0.010\pm0.008$:
the quality of the fit of individual
replicas is about one order of magnitude worse that that of the
average over replicas.

Therefore, it is possible to achieve an arbitrarily small value of
$\chi^2$ by increasing the training length, though as we approach
the minimum the number of generations needed will grow exponentially.\footnote{
This is a well known behavior of genetic algorithms. In this particular case,
once we are close enough to the absolute minimum, it might be more useful
to switch to other strategies like steepest descent, though of course this
is purely an academic issue.}

It is also clear from Fig.~\ref{fig:chi2vslenL0} that the NNPDF3.0
minimization, based on the nodal genetic algorithm, is substantially
more efficient than the NNPDF2.3 one, reaching values of $\chi^2$ which
are up to one order of magnitude smaller for a given fixed training length.
This of course will be of great importance when we come to the fits to
experimental data, allowing a more efficient exploration of the minima of
the parameter space.

\begin{figure}
  \centering
  \epsfig{width=0.45\textwidth,figure=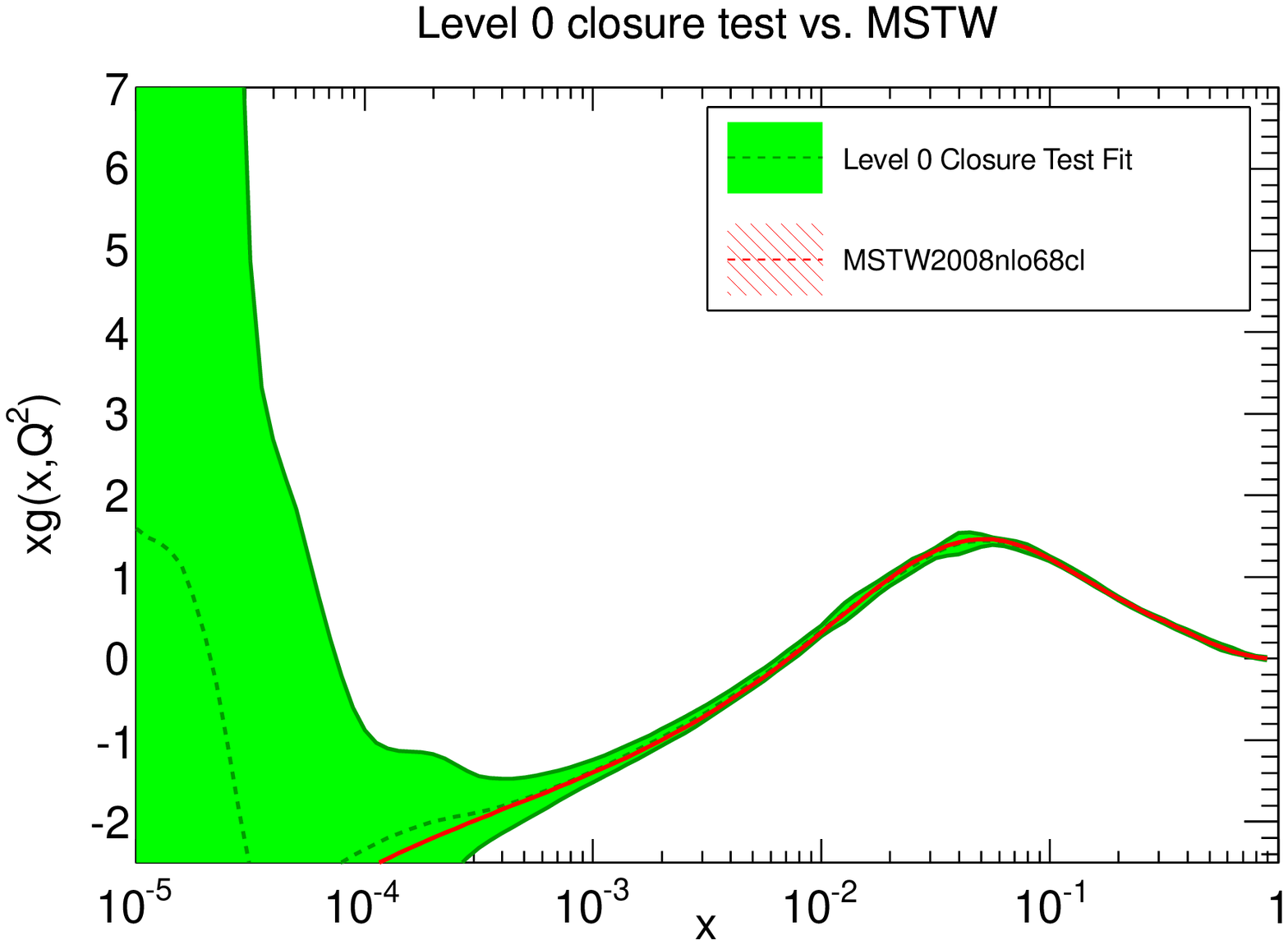}
  \epsfig{width=0.45\textwidth,figure=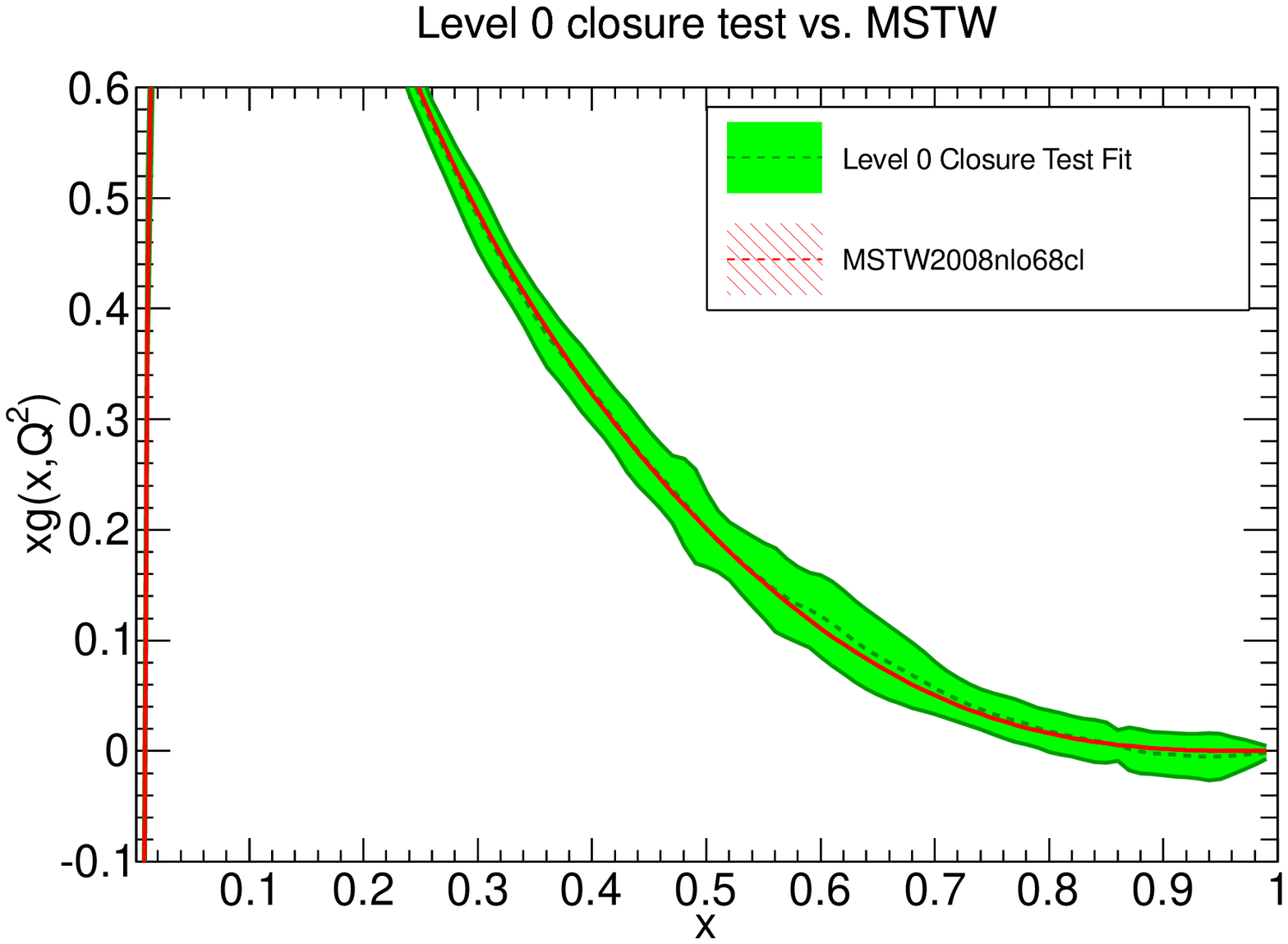}
  \epsfig{width=0.45\textwidth,figure=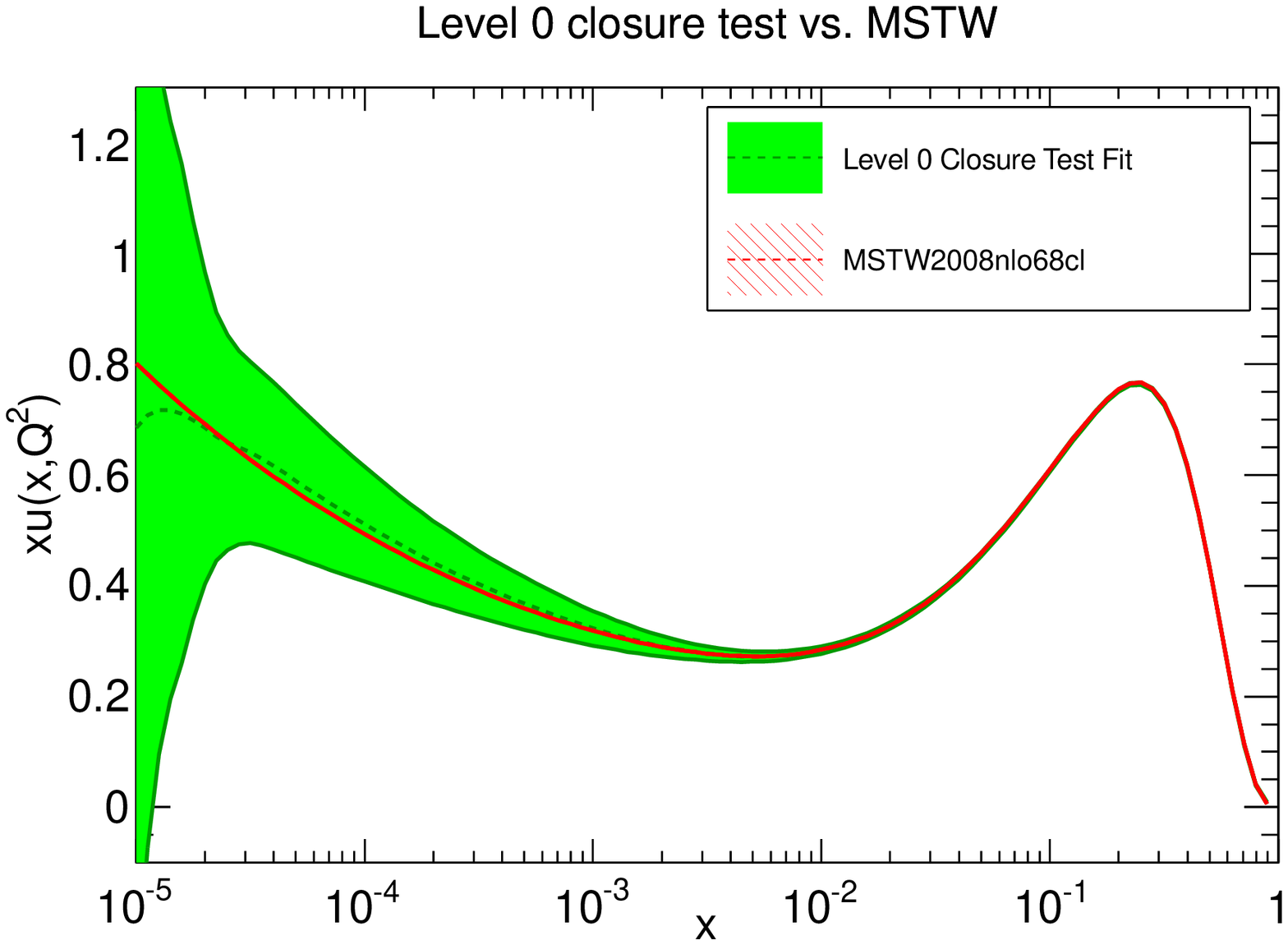}
  \epsfig{width=0.45\textwidth,figure=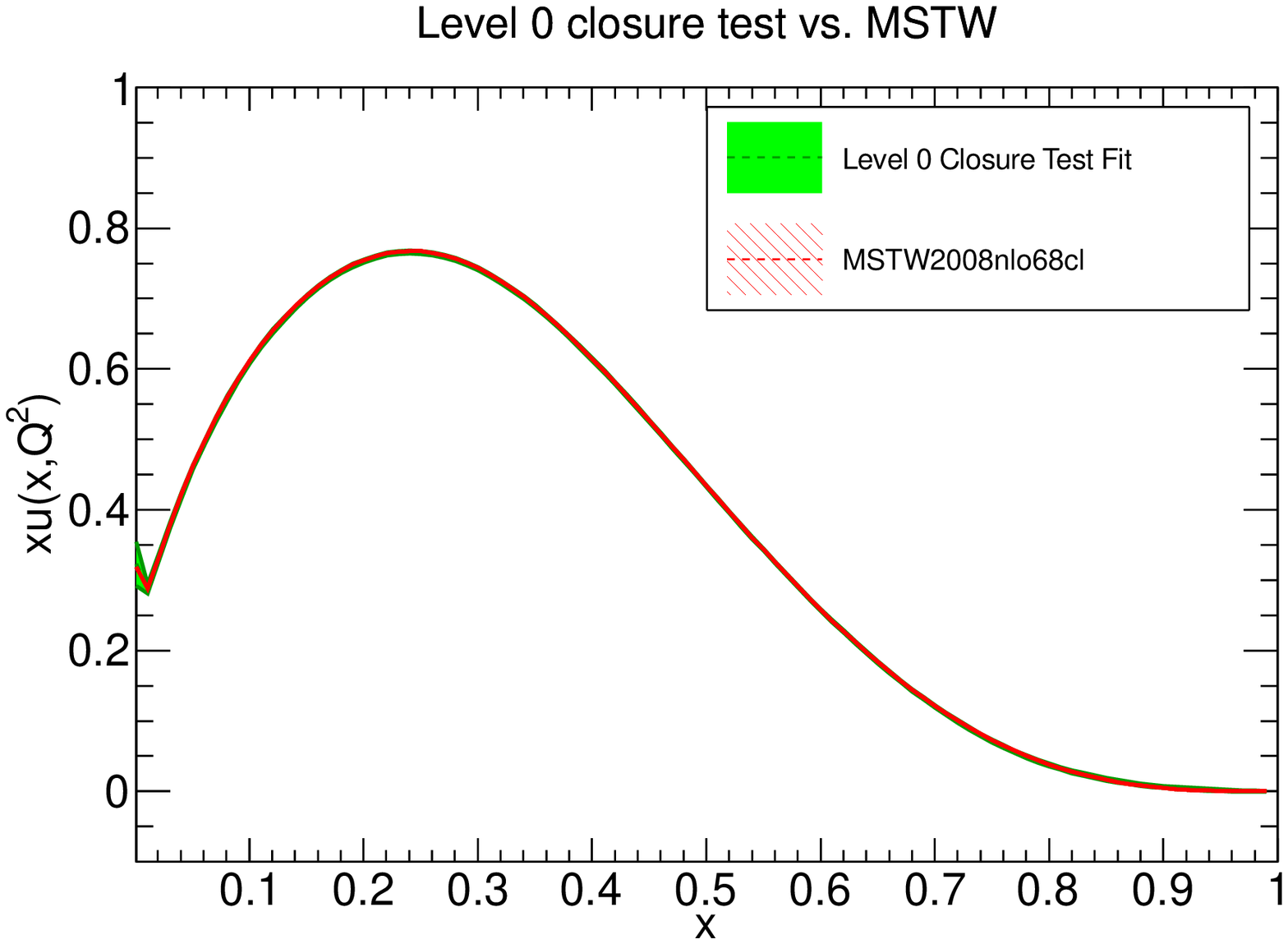}
  \epsfig{width=0.45\textwidth,figure=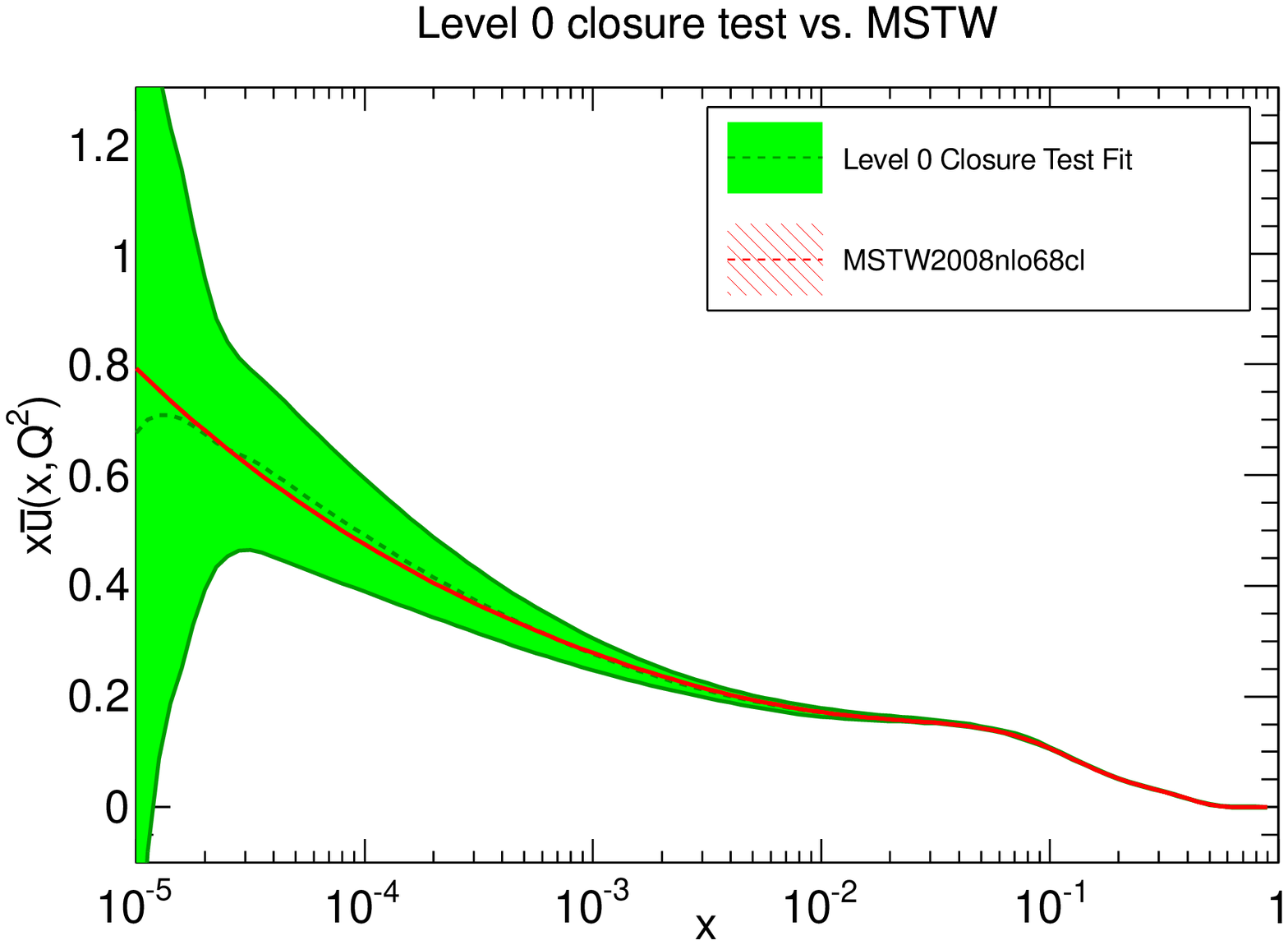}
  \epsfig{width=0.45\textwidth,figure=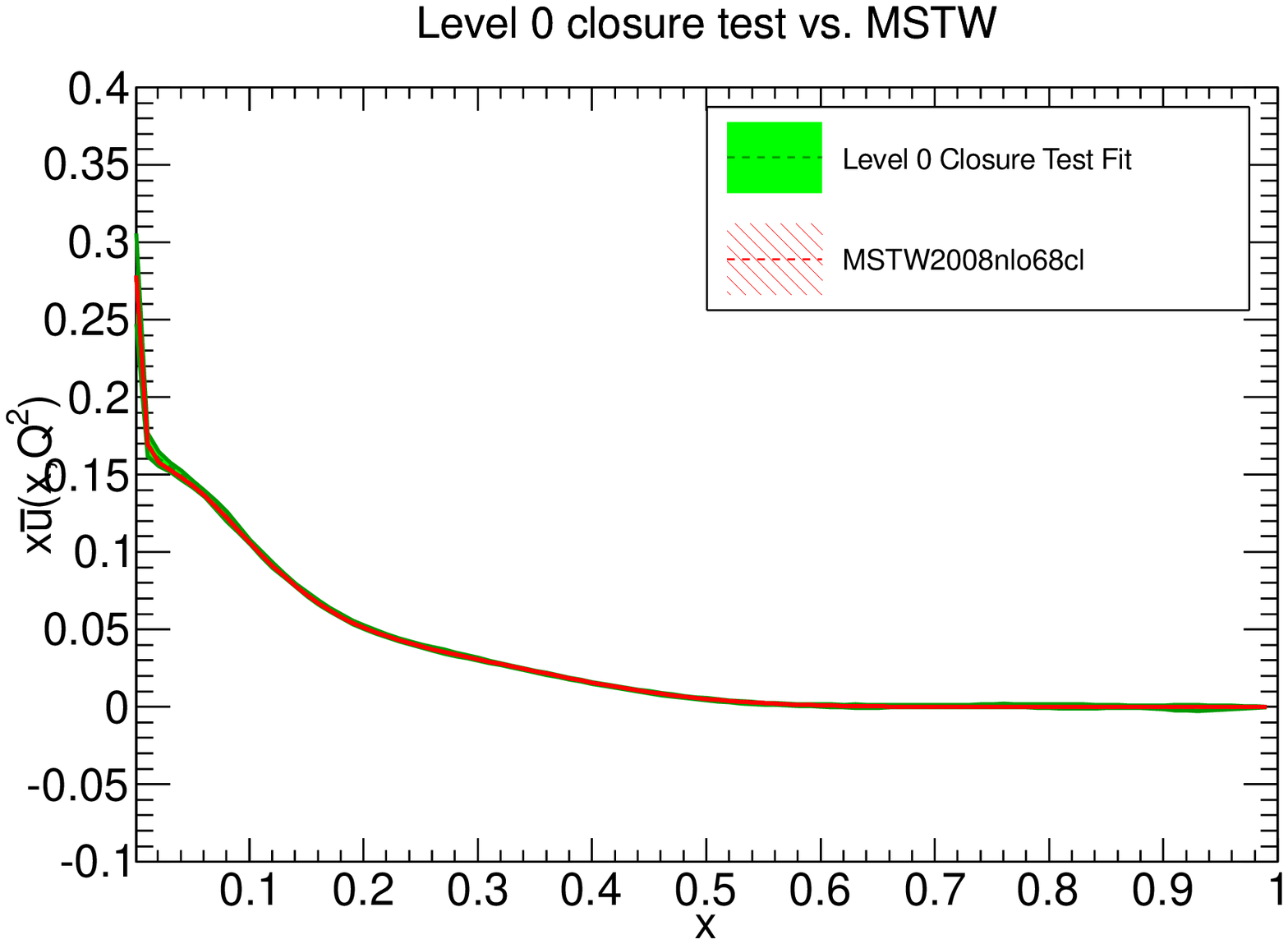}
  \epsfig{width=0.45\textwidth,figure=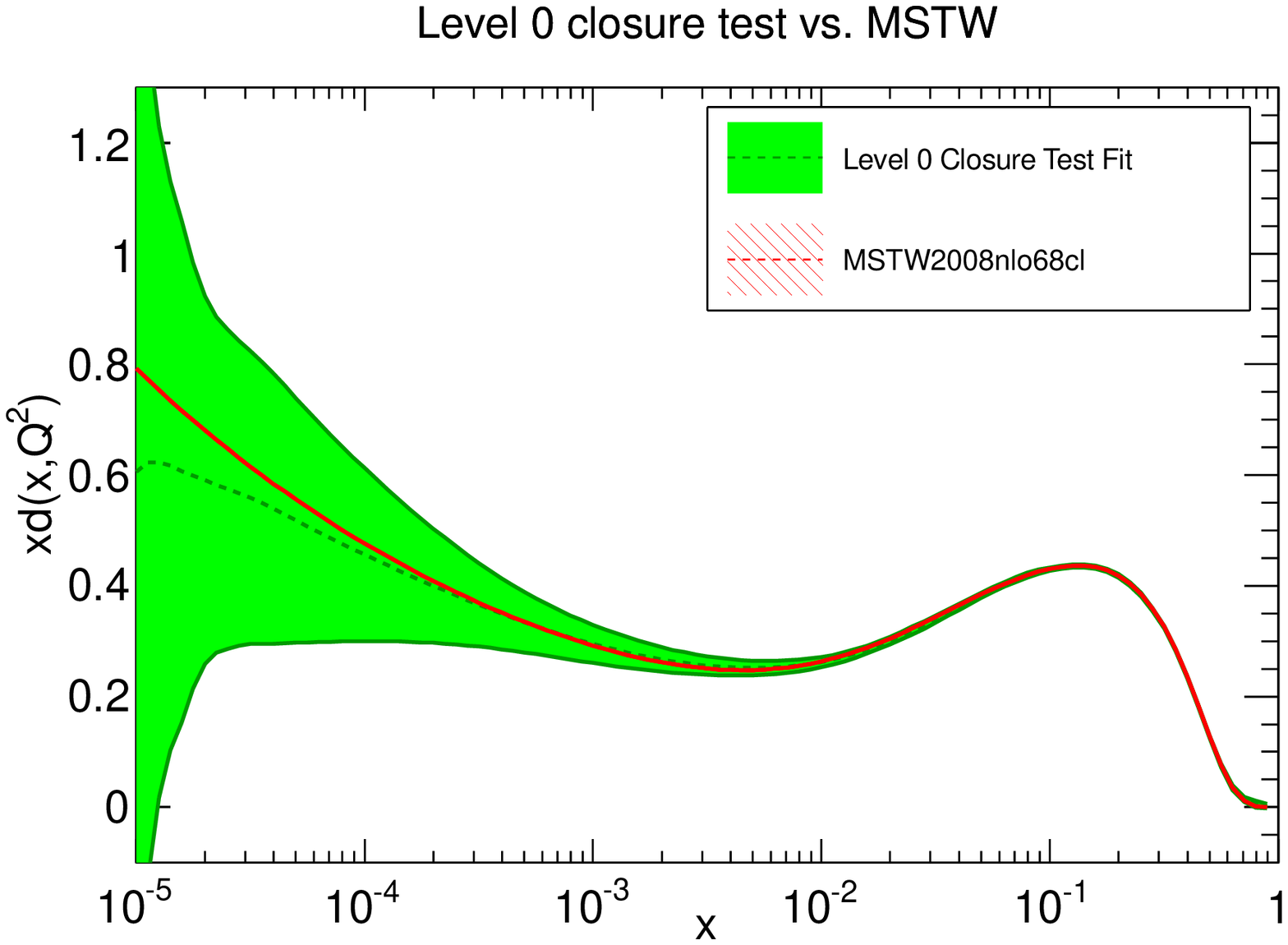}
  \epsfig{width=0.45\textwidth,figure=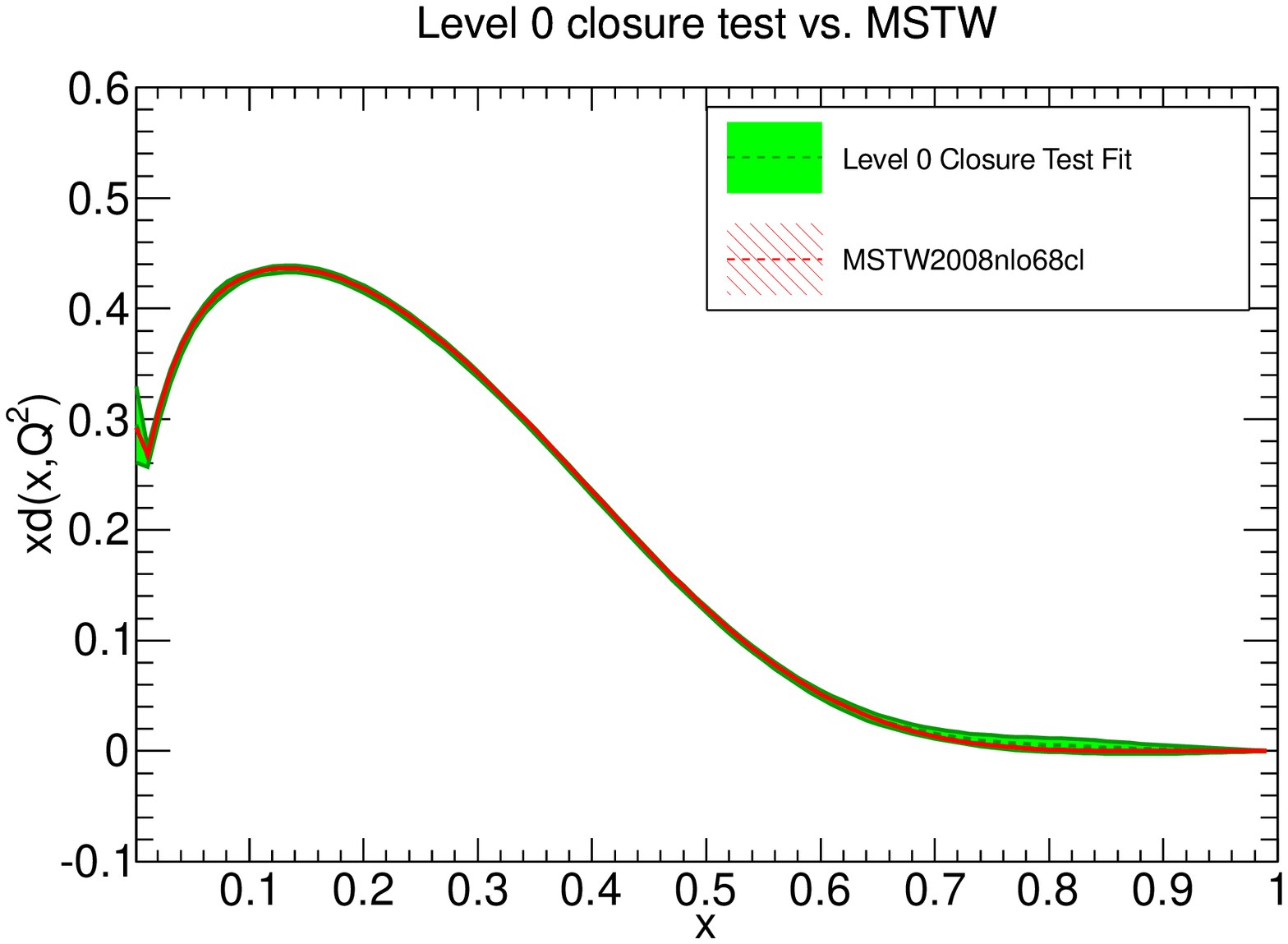}
  \caption{\small Comparison between
the results of the Level~0 closure fit with 100K GA generations
(fit C7) and the corresponding input PDF set, the
 central value of MSTW08 NLO PDF set.
  The green band shows the one-sigma interval
computed over the sample of $N_{\rm rep}=100$ replicas, with the
the green dotted line showing the mean value.
  The plots show the gluon, $u$, $\bar{u}$ and $d$ PDFs on both linear (right hand side) and
  logarithmic (left) scales in $x$, at the scale $Q^2 = 1~\textrm{GeV}^2$
where the PDFs are parametrized.
}

  \label{fig:l0pdfplot}
\end{figure}

\begin{figure}
  \centering
  \epsfig{width=0.45\textwidth,figure=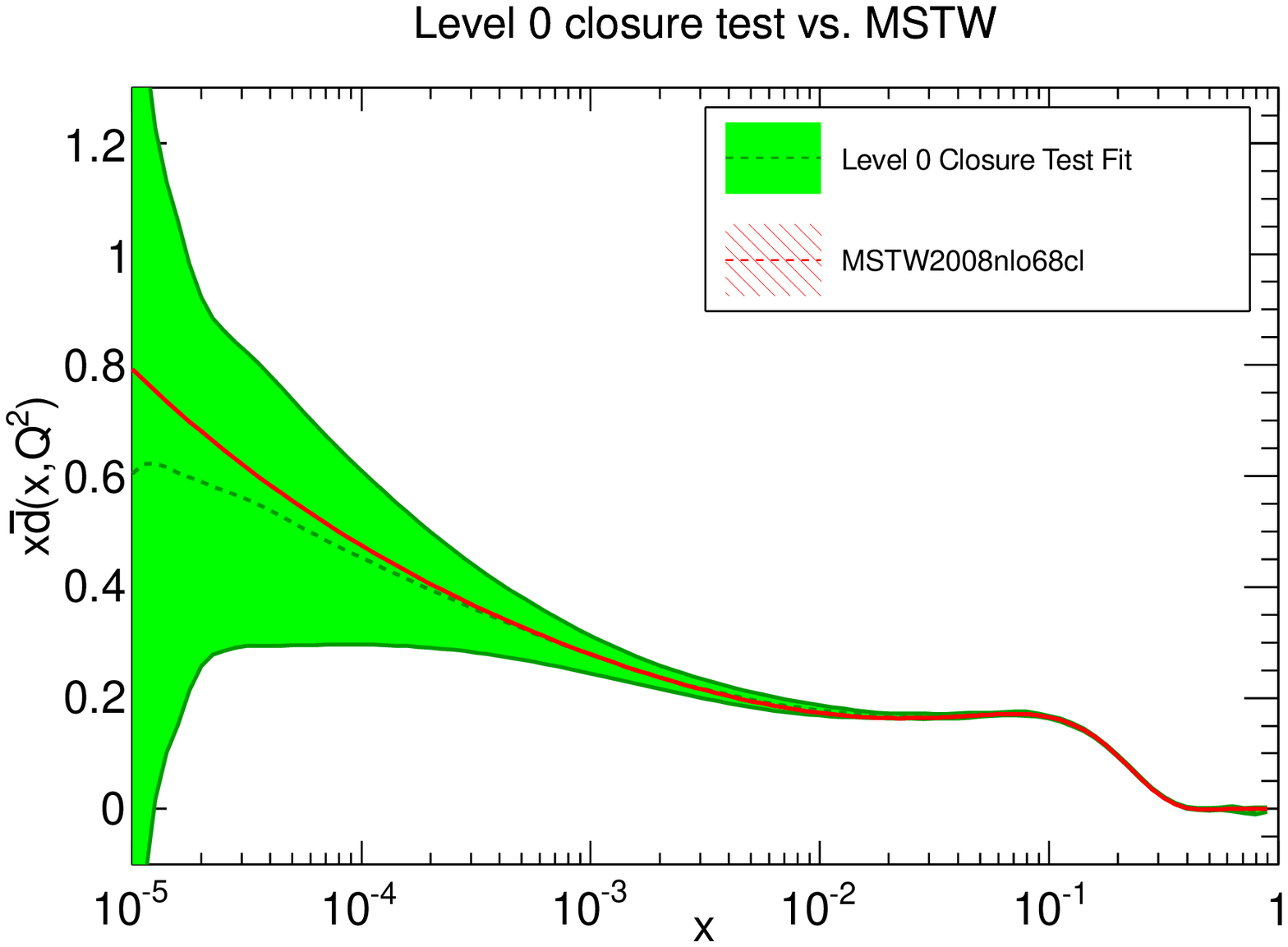}
  \epsfig{width=0.45\textwidth,figure=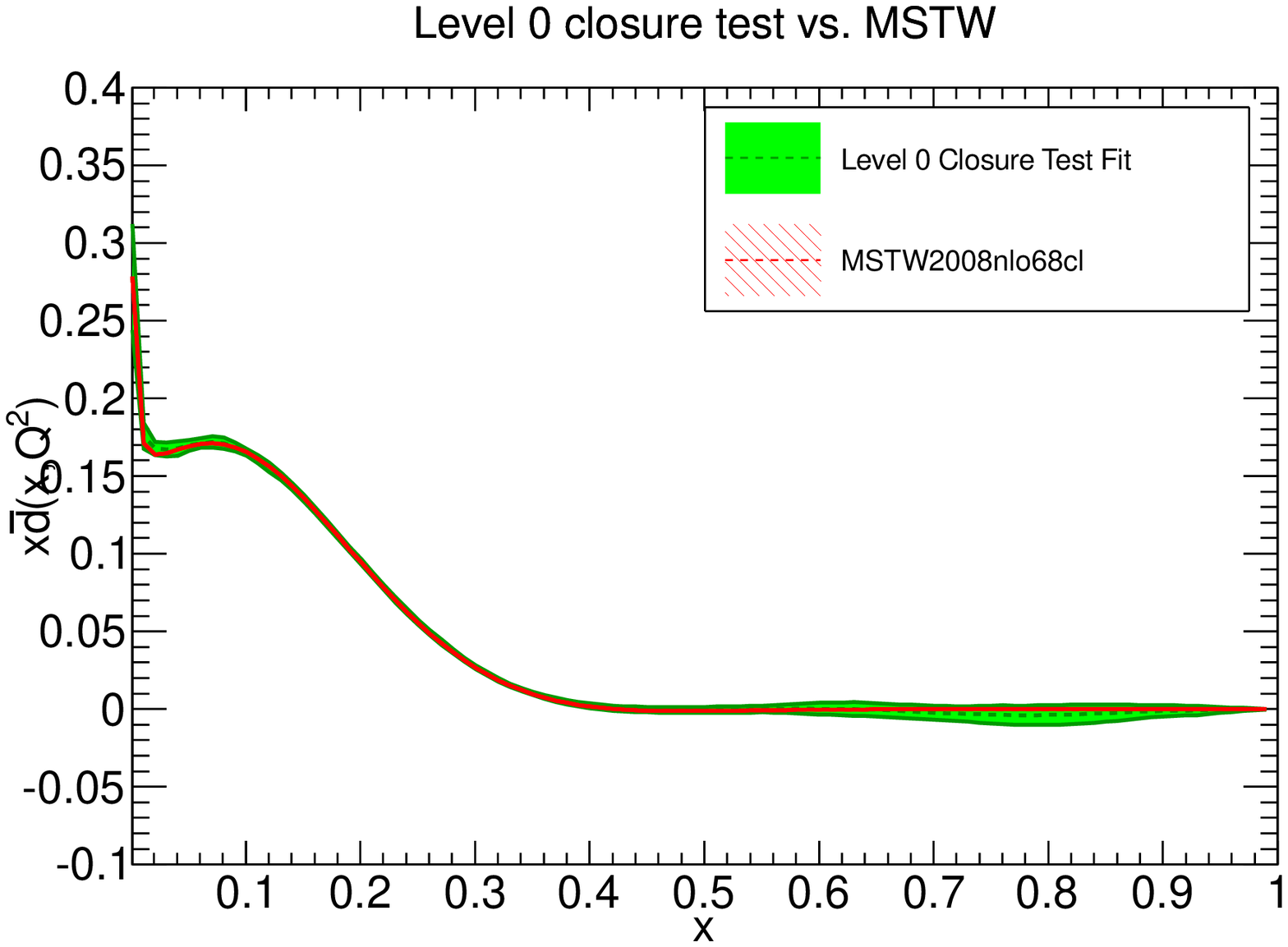}\\
  \epsfig{width=0.45\textwidth,figure=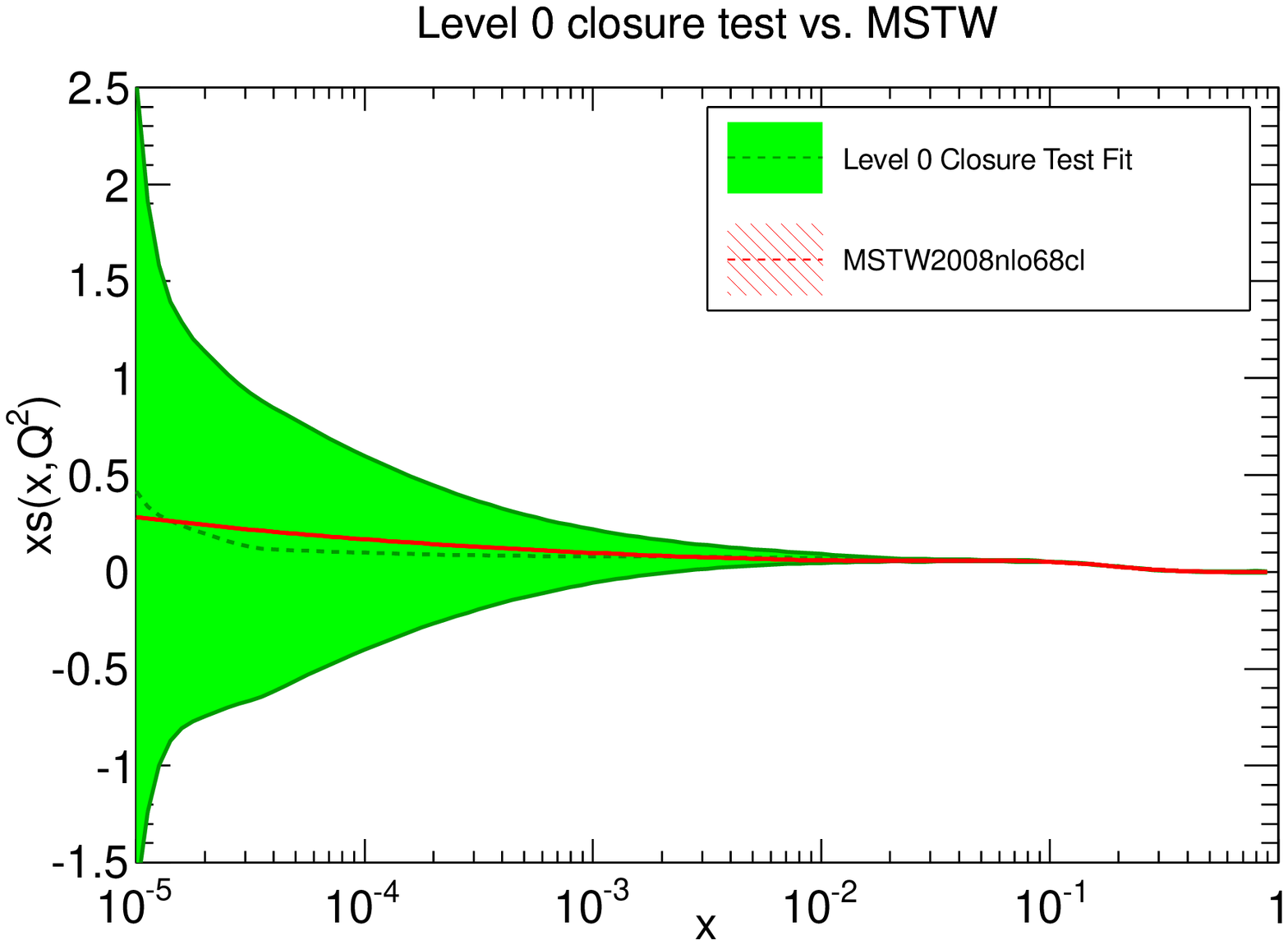}
  \epsfig{width=0.45\textwidth,figure=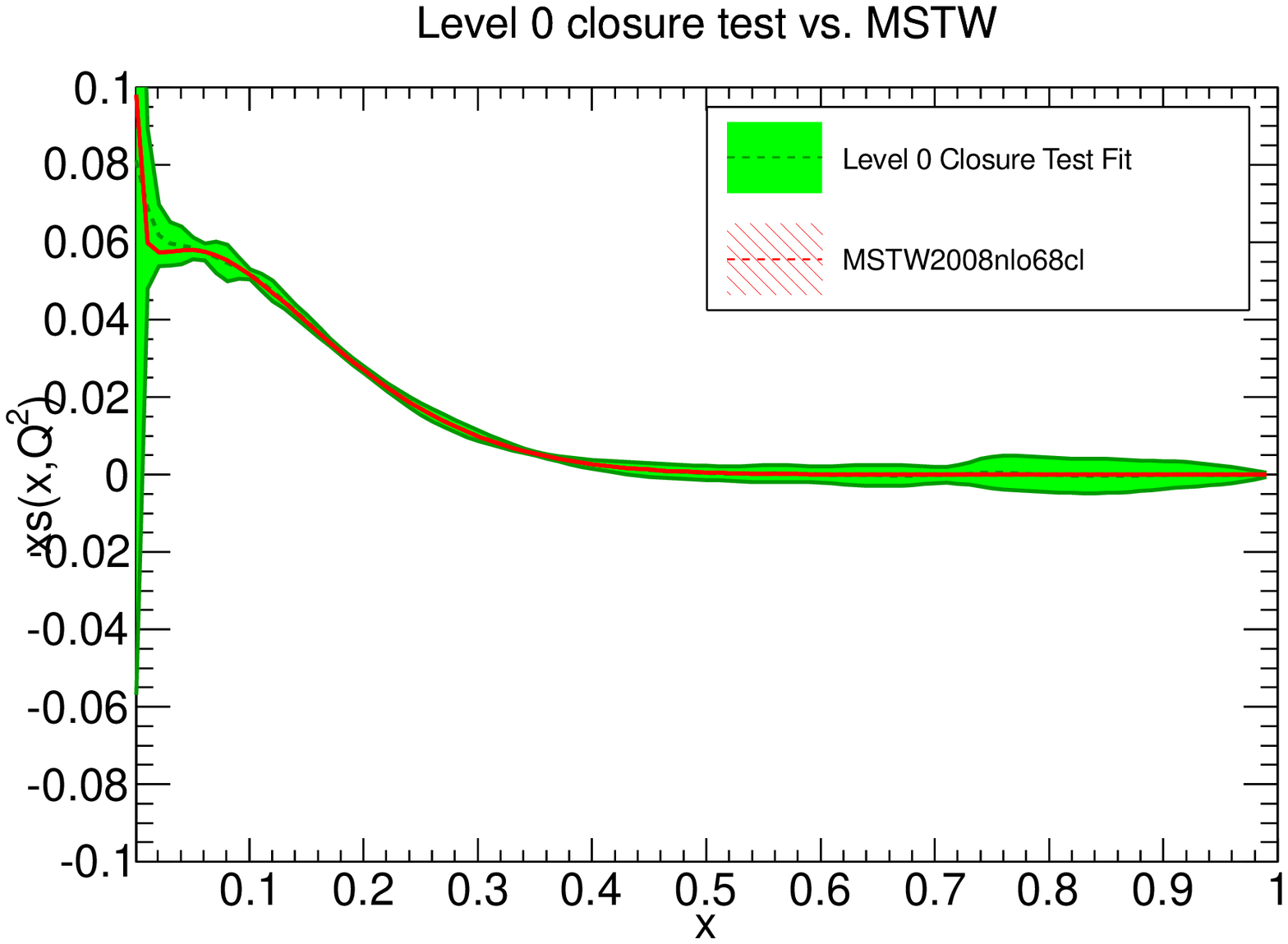}\\
  \epsfig{width=0.45\textwidth,figure=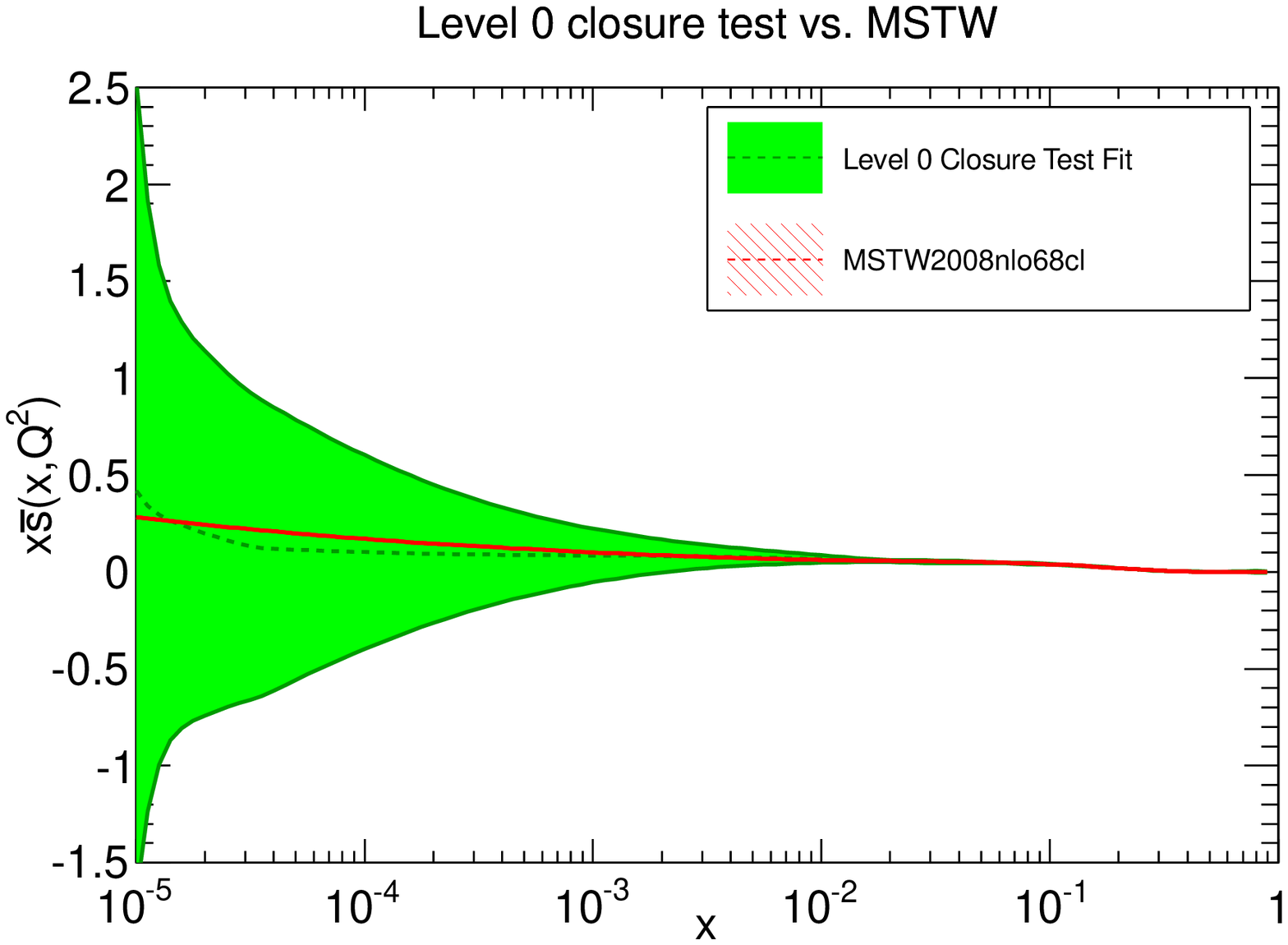}
  \epsfig{width=0.45\textwidth,figure=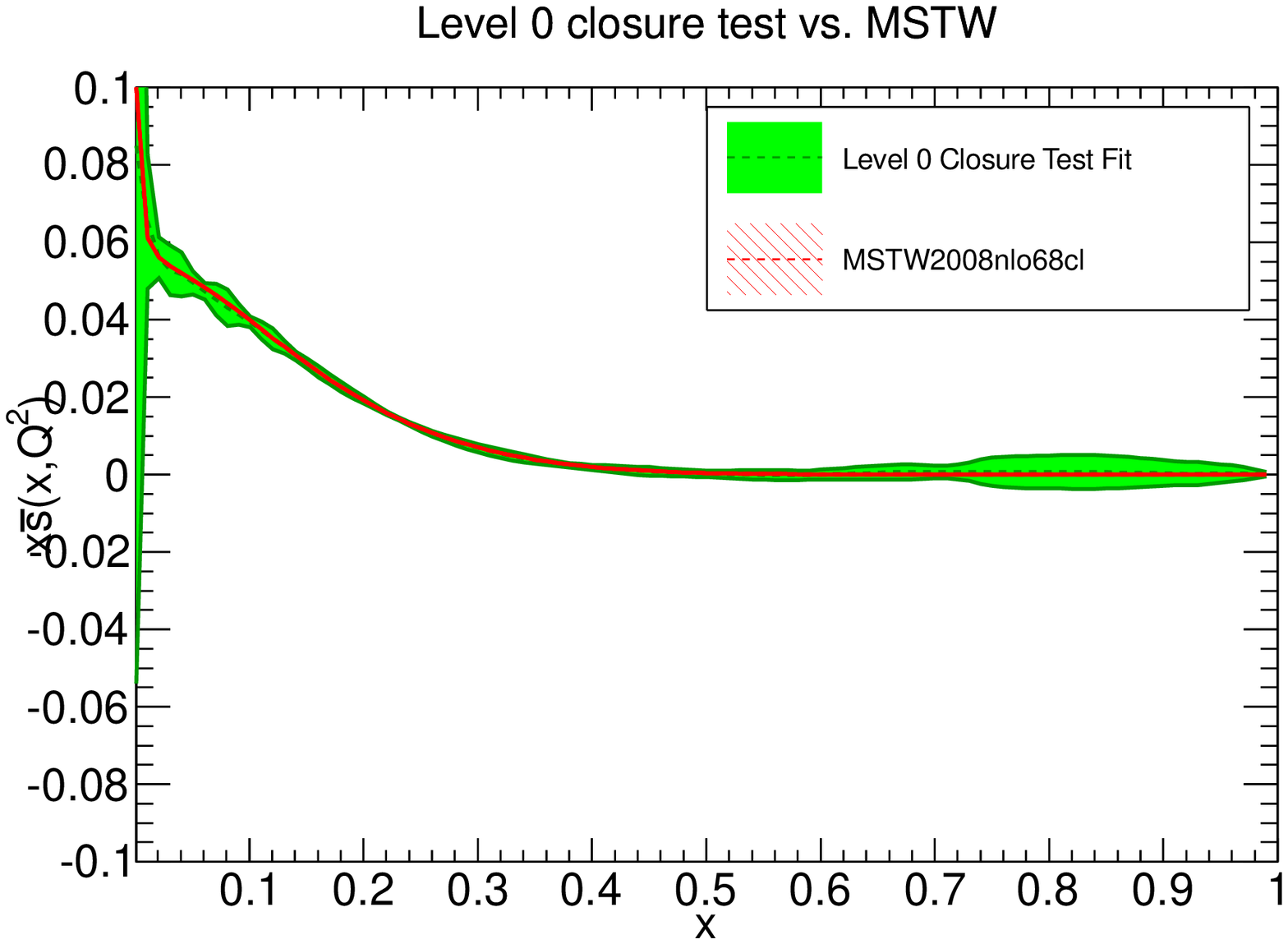}\\
\epsfig{width=0.45\textwidth,figure=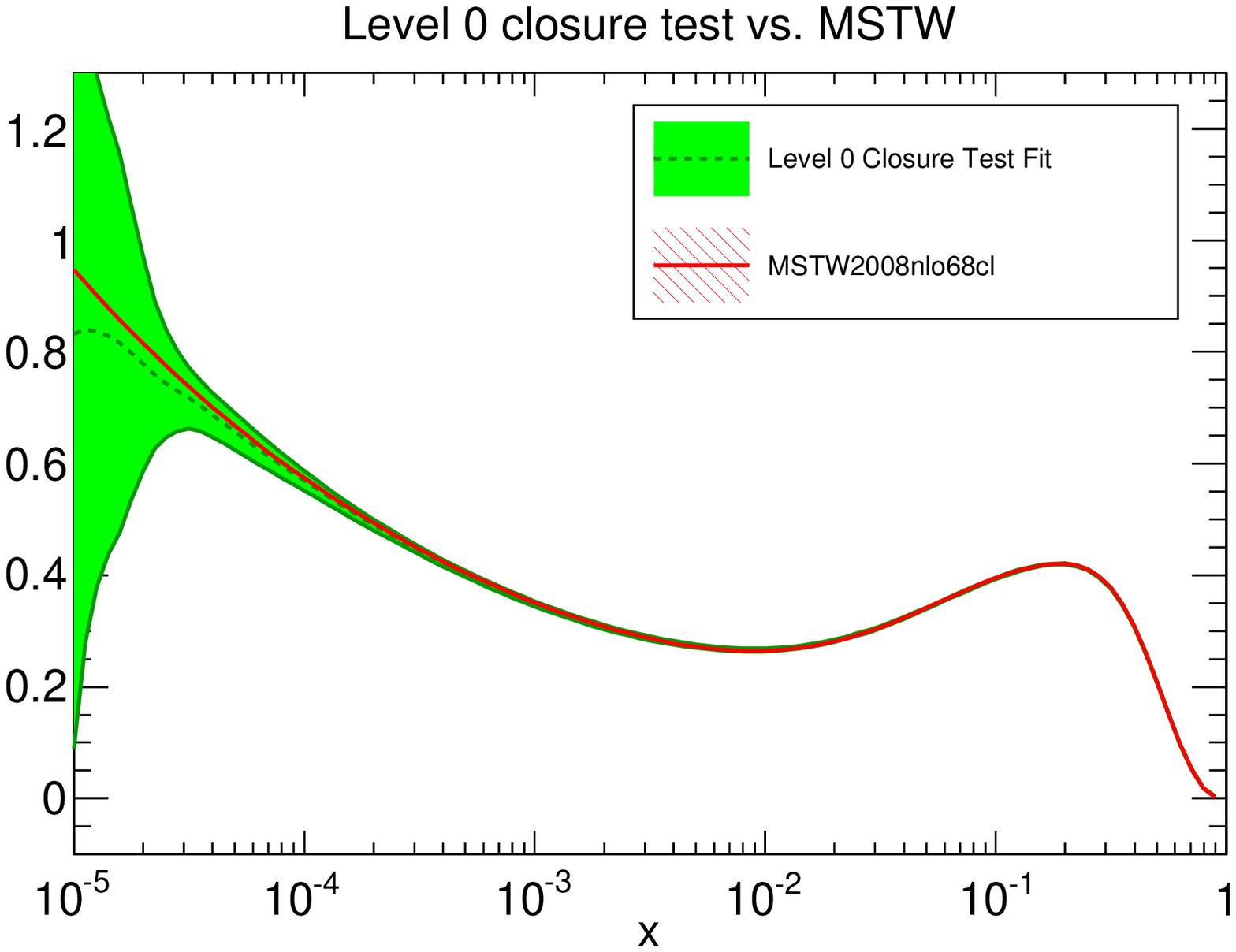}
  \epsfig{width=0.45\textwidth,figure=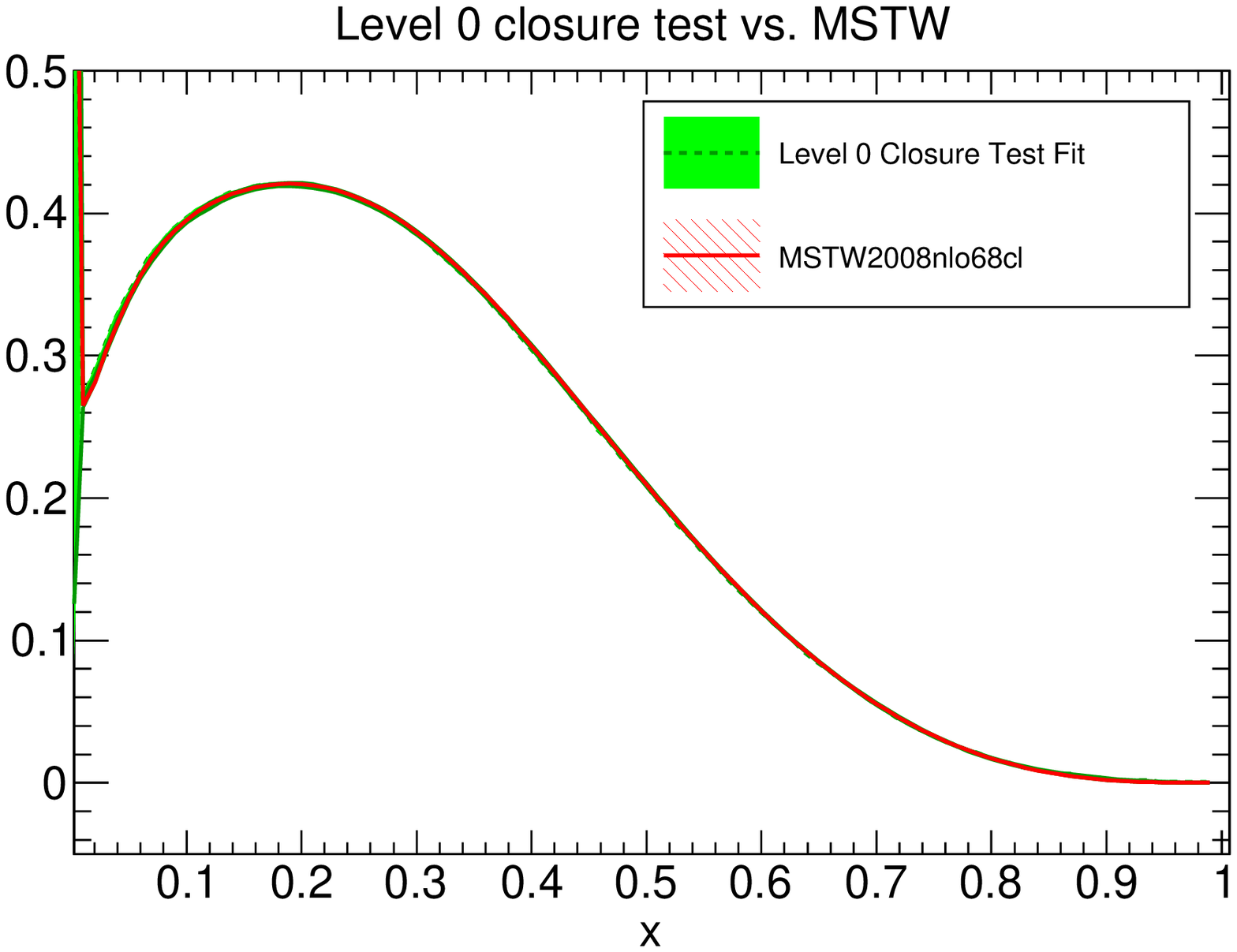}
  \caption{\small Same as Fig.~\ref{fig:l0pdfplot}
for the $\bar{d}$, s and $\bar{s}$ PDFs, and for the combination of
PDFs which corresponds to the leading-order expression of $F_2^p$.}
  \label{fig:l0pdfplot2}
\end{figure}

Given that the $\chi^2$ is essentially vanishing, we expect
almost perfect agreement between the fitted and input
PDFs.
Indeed such an agreement  can be found in the corresponding comparison shown
in Figs.~\ref{fig:l0pdfplot} and~\ref{fig:l0pdfplot2}.
In these plots we show, for the Level~0 fit with a training length of
100K (fit C7), the results of the fit and the input PDFs, in this
case MSTW08. The central value of our fitted PDFs is computed as the average over replicas in the usual way:
\begin{equation}
\label{eq:fbarfit}
\langle {f}_{\rm fit}\rangle = \frac{1}{N_{\rm rep}}
\sum_{k=1}^{N_{\rm rep}} f_{\rm fit}^{k}\, ,
\end{equation}
the angled brackets denoting the average over replicas, and the
variance as,
\begin{equation}
\label{eq:sigmafit}
\sigma_{\rm fit}^2 = \langle({f}_{\rm fit}^2 - \langle{f}_{\rm fit}\rangle)^2\rangle
= \frac{1}{N_{\rm rep}}
\sum_{k=1}^{N_{\rm rep}} \left(f_{\rm fit}^{k}-\langle{f}_{\rm fit}\rangle\right)^2\, .
\end{equation}
It is clear from these plots that the neural network
architecture is flexible enough to reproduce the input PDFs,
and that the genetic algorithm minimization is working as expected. It is interesting to observe
that PDFs for which experimental information is very dense, such as for
example the up quark in the valence region, are perfectly reproduced
with essentially zero uncertainty. PDFs for which information is more sparse or
indirect, such as for example the gluon, have an uncertainty even when
the $\chi^2$ at the data points is essentially zero, because there is
still a certain freedom in interpolating between data points.
Indeed, if we look at the particular combination of PDFs which
corresponds to the leading-order expression of the structure function
$F_2^p$, namely $\frac{4}{9}\left(u+\bar
u+c+\bar c\right)+\frac{1}{9}\left(d+\bar d+s+\bar s\right)$, which is
directly probed by the HERA data, the uncertainty on it at small $x$
in the HERA data region $10^{-4}\lsim x\lsim10^{-3}$ is significantly
smaller than that on each individual PDF entering this combination
(see the bottom row of Fig.~\ref{fig:l0pdfplot2}).


Additional interesting  information can be extracted from Level~0 fits
by looking at the PDF uncertainties
of the resulting fit, computed as usual as the standard
deviation over the sample of $N_{\rm rep}=100$ fitted replicas, either
at the level of parton distributions or at the level of physical
observables.
Recall that at Level~0 the only difference between each
of the replicas is the random seed used to initialize the minimization,
and therefore, their spread quantifies different possibilities
of approaching the underlying input PDFs.
Given
that the Level~0 input pseudo-data do not fluctuate, the cross-sections
computed from the fitted PDFs must converge to these input  values
for each replica as the training length is increased; i.e.\ the
uncertainty on the predicted value for all the observables included in the
fit must go to zero.

\begin{figure}[t]
  \centering
  \epsfig{width=0.80\textwidth,figure=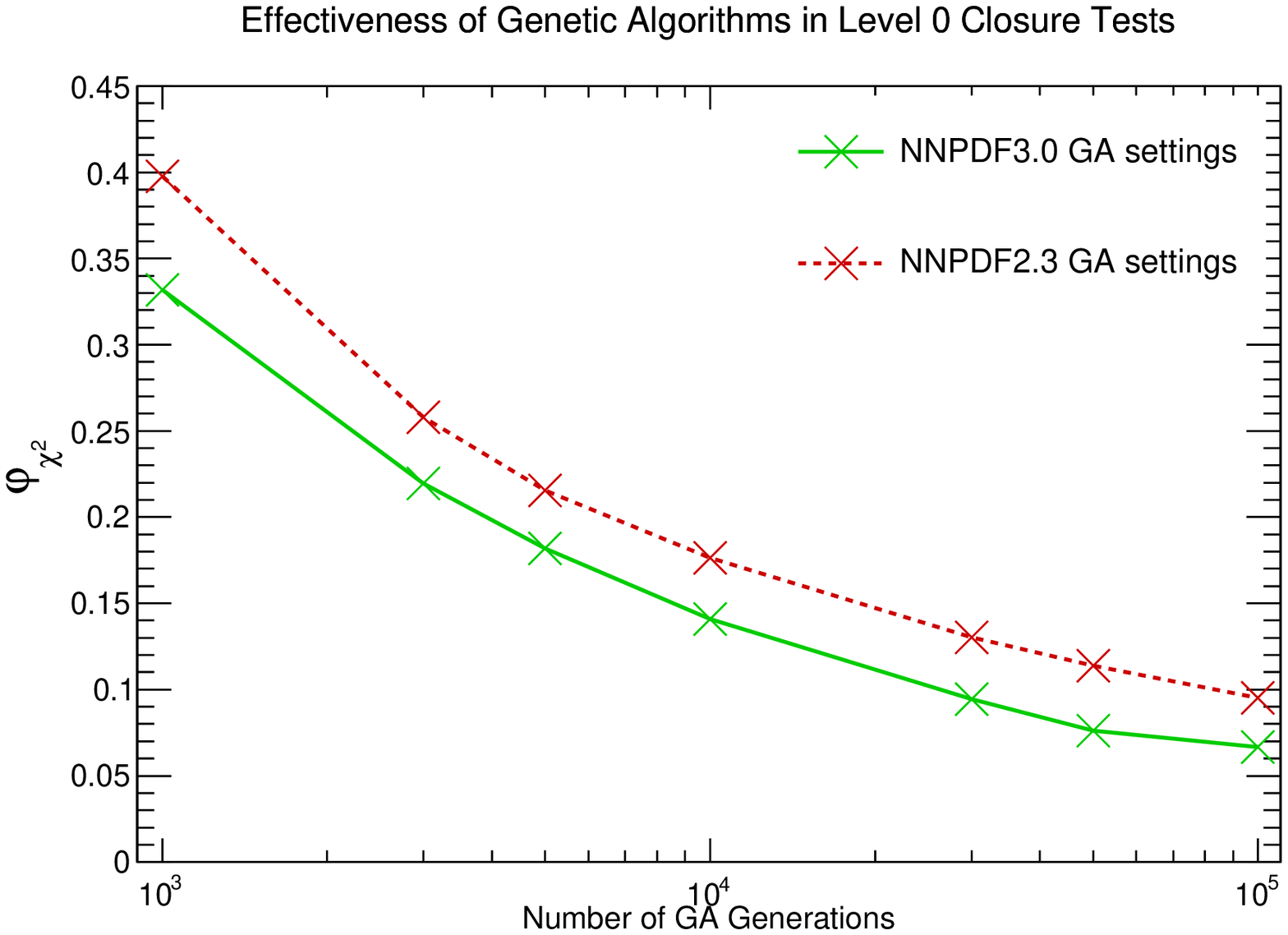}
  \caption{\small The estimator $\varphi_{\chi^2}$ Eq.~(\ref{eq:frdbdef}),
as a function of the length of the genetic algorithms
minimization.
We compare the results for the GA settings used in NNPDF3.0,
with the corresponding GA settings used in the NNPDF2.3 fit.
The results for each closure test are marked with crosses,
with auxiliary lines joining these points.
}\label{fig:level0_fchi2}
\end{figure}

To verify this expectation, it is convenient to define an indicator
which measures the standard deviation over the replica sample in
units of the data uncertainty. This can be defined as follows:
\begin{equation}
\varphi_{\chi^2} \equiv \sqrt{\langle \chi^2[\mathcal{T}[f_{\rm fit}],\mathcal{D}_0]\rangle - \chi^2[\langle\mathcal{T}[f_{\rm fit}]\rangle,\mathcal{D}_0]}\, .
\label{eq:frdbdef}
\end{equation}
To see that this does the job,
consider the first term in Eq.~(\ref{eq:frdbdef}) in more detail:
using the definition of the $\chi^2$ in Eq.~(\ref{eq:chi2not}), we find that
\bea
  &&N_{\mathcal{D}} \langle \chi^2[\mathcal{T}[f],\mathcal{D}]\rangle
  = \big\langle
  \sum_{I,J} (T_{I}[f]  -D_{I}) \, C^{-1}_{IJ} \,
  (T_{J}[f] - D_{J}) \big\rangle
\nonumber
 \\
  \nonumber&=& \sum_{I,J}\langle
  T_{I}[f] \, C^{-1}_{IJ} \, T_{J}[f] \rangle
   -  \sum_{I,J}\langle T_{I}[f]\rangle \, C^{-1}_{IJ} \, D_{J}
-  \sum_{I,J}D_{I} \, C^{-1}_{IJ} \, \langle T_{J}[f]\rangle
+  \sum_{I,J}D_{I}\, C^{-1}_{IJ} \, D_{J}   \, ,
\eea
so that
\beq
\langle \chi^2[\mathcal{T}[f],\mathcal{D}]\rangle -
\chi^2[\langle\mathcal{T}[f]\rangle,\mathcal{D}] =
\frac{1}{N_{\mathcal{D}}} \sum_{I,J}
 \big( \langle T_{I}[f] C^{-1}_{IJ} T_{J}[f]\rangle -  \langle T_{I}[f]\rangle C^{-1}_{IJ} \langle T_{J}[f]\rangle\big).
\eeq

Thus in terms of the covariance matrix of the theoretical predictions
\beq
T_{IJ}\equiv   ( \langle T_{I}[f] T_{J}[f]\rangle -  \langle T_{I}[f]\rangle\langle T_{J}[f]\rangle),
\eeq
we have
\begin{equation}
\label{eq:varphiinterp}
\varphi_{\chi^2}^2 \equiv
\frac{1}{N_{\mathcal{D}}} \sum_{I,J} C_{IJ}^{-1}T_{JI},
\end{equation}
i.e. the average over all the data points of the uncertainties and
correlations of the theoretical predictions, $T_{IJ}$, normalized according
to the corresponding uncertainties and correlations of the data as
expressed through the covariance matrix $C_{IJ}$. If the covariance
matrix was diagonal, i.e. in the absence of correlations, this would
just be the variance of the predictions, divided by the experimental
variance, averaged over data points; $\varphi_{\chi^2}^2$ is thus
recognized to provide the generalization of this in the presence of
correlations.

In Fig.~\ref{fig:level0_fchi2}
we show $\varphi_{\chi^2}$, Eq.~(\ref{eq:frdbdef}),
for the Level~0 fits as a function of the length of the genetic algorithms
minimization.
As before, we compare the results for the GA settings used in NNPDF3.0,
with the corresponding GA settings used in the NNPDF2.3 fit.
We can see that indeed, as we increase the training length, the
spread of the theoretical predictions at the data points for different
replicas at the level of fitted cross-sections decreases monotonically,
and again here we observe the improvement from the more efficient
minimization strategy in NNPDF3.0. Specifically, for the longest
training length we find that standard deviation of the theoretical
predictions is on average
almost by a factor 20 smaller than the nominal standard
deviation of the data.

On the other hand, the fitted PDFs themselves need not
become identical replica by replica, even at Level~0, since
different functional forms for the PDFs can yield the same
predictions for the observables.
Of course, in
the kinematic regions where data are available, the fluctuations of
the PDFs are limited by the fact that the neural networks provide a
smooth interpolation between points that are constrained by data.
In this respect, in regions where the PDFs are well constrained
by experimental data, PDF uncertainties should be very small, and
indeed they are.

This is however not necessarily the case in the extrapolation regions,
where we expect large PDF uncertainties, which moreover are essentially
independent of the training length. This is due to the fact that there the
functional  forms can vary substantially without affecting the fitted figure of
merit, $\chi^2[\mathcal{T}[f_{\rm fit}],\mathcal{D}_0]$.
These two effects, very small PDF uncertainties in the data region,
and large PDF uncertainties in the extrapolation regions,
in particular at small and large $x$ are still clearly visible
in the plots in Fig.~\ref{fig:l0pdfplot}, even at the end of a
100k-generation training.
These results provide a way of quantifying the {\em extrapolation}
uncertainty on the PDFs due to the lack of direct constraints
in these regions: this is an irreducible source of PDF uncertainty that
can only be reduced if new data is provided, and that accounts
for most of the PDF uncertainties in the fits to real data
in the extrapolation regions.

\subsubsection{Effective preprocessing exponents}

When using as an input
to the closure test
a PDF set based on a relatively
simple functional form, such as for example
MSTW08~\cite{Martin:2009iq} or CT10~\cite{Lai:2010vv,Gao:2013xoa},
the exponents that characterize the
asymptotic small-$x$ and large-$x$
behavior of the individual PDFs are known, since they are part
 of the set of parameters defining the PDF parametrization
 at the initial scale.
For instance, in the MSTW08 fit, the PDFs are parametrized at
the initial scale of $Q_0^2=$ 1 GeV$^2$ using a generic functional form
$A x^{-\alpha} (1-x)^\beta \left(1+\epsilon x^{0.5} + \gamma x\right)$
and therefore, the asymptotic small-$x$ and large-$x$ behaviors will both be powerlike, parametrized by the exponents $\alpha$ and $\beta$.
Note that in general one has to be careful with this naive interpretation since
sub-asymptotic corrections can be numerically large.

In the context of NNPDF fits, no explicit assumption
is made on the functional form of PDFs, which are
instead parametrized using  neural networks. On the other hand,
preprocessing is introduced in order to
absorb the dominant behavior of the fitted PDFs at small- and large-$x$ and
speed up the fitting. As explained in Sect~\ref{sec:preproc},
when fitting real data, the range in which preprocessing is varied for
each PDF is determined dynamically through an iterative procedure.

\begin{figure}[h!]
  \centering
  \epsfig{width=0.49\textwidth,figure=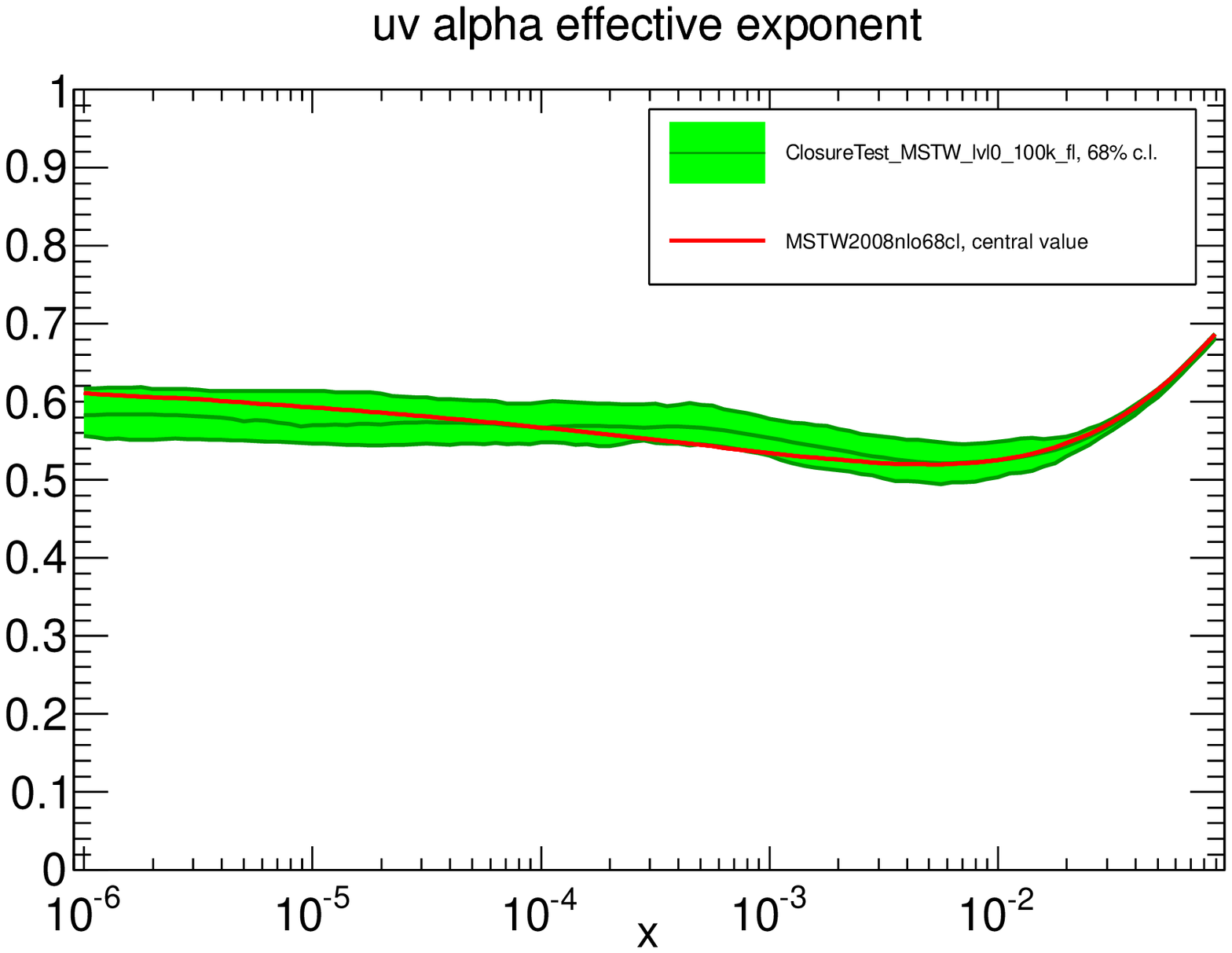}
  \epsfig{width=0.49\textwidth,figure=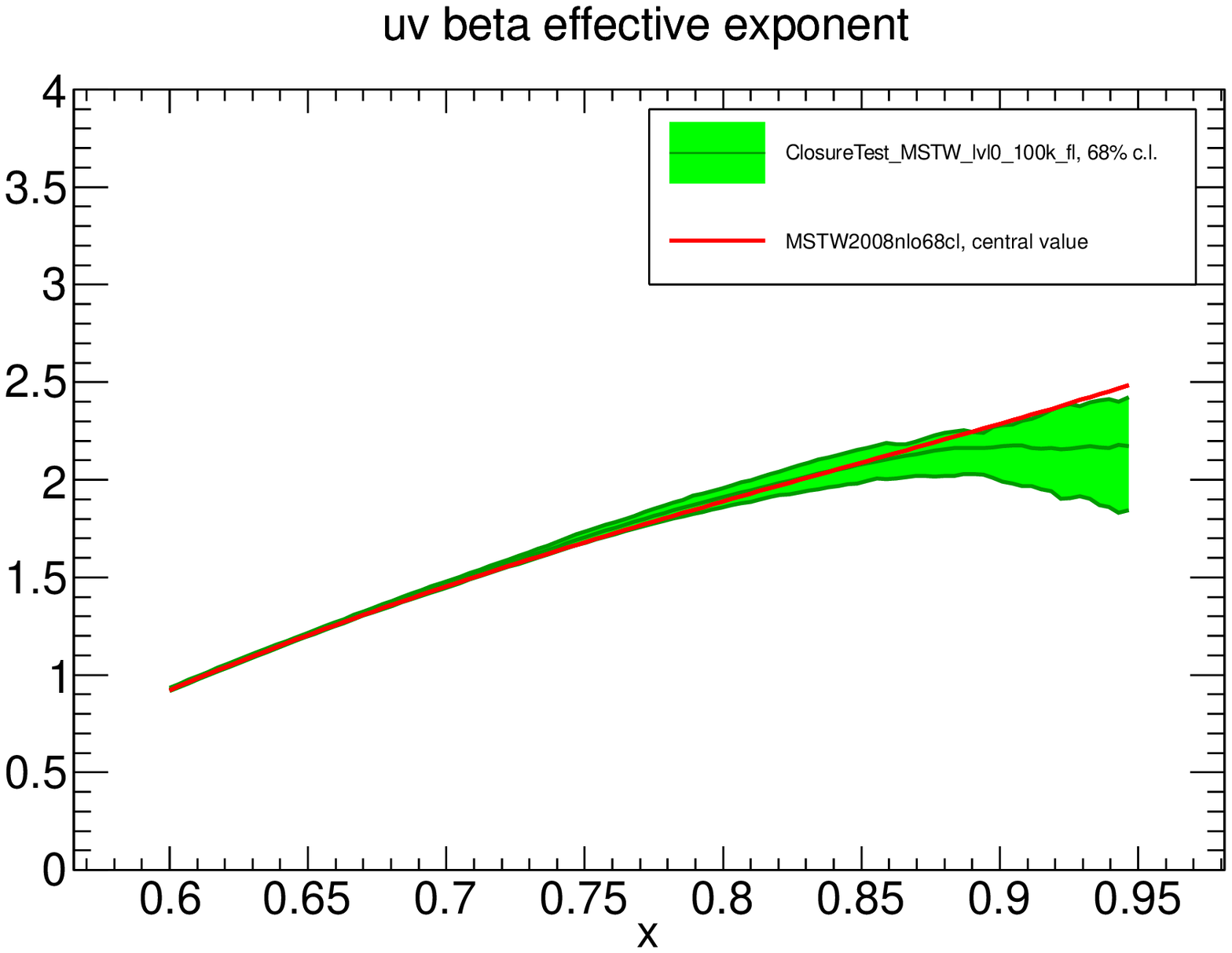}
  \caption{\small Small-$x$ (left) and large-$x$ (right) behaviour of the fitted PDF for $u_v$
    compared to the behaviour of the MSTW08 input functions. In both cases
    the values are obtained using Eq.~(\ref{effalpha}-\ref{effbeta}).}
  \label{fig:exp-est}
\end{figure}

In the context of a Level~0 closure test,
when a PDF set based on a  known functional form
is used as input, it is interesting to verify that the small-$x$
and large-$x$ behavior of the fitted PDFs reproduce the ones of the input PDFs.
This can be achieved by comparing the effective preprocessing exponents
between the fitted and input PDFs.
These effective exponents can be computed as in
Eq.~(\ref{effalpha}-\ref{effbeta}). The results of this comparison are
shown in Fig.~\ref{fig:exp-est},
for the Level~0 closure test with 100K generations (fit C7
in Table~\ref{tab:CTfits}).
A beautiful agreement is found, with the
MSTW08 value always within in the one-sigma band of the fitted
PDFs, thus showing that our methodology is capable
of precise quantitative predictions on  the behavior of the
functions that we are trying to fit. Similar results are obtained for
all the other PDF combinations.

\subsection{PDF uncertainties in closure tests}
\label{sec:uncert}

In the previous section we studied Level~0 closure tests,
in which the fit quality can become arbitrarily good, and the PDF
uncertainties arise purely due to the fact that the experimental
data used in the fit has finite kinematical coverage.
Now we turn to Level~1 and Level~2 closure tests, and
in doing so we shed some light, in the cleanly controlled
environment of closure testing, on the  various different origins of
PDF uncertainties: specifically those
due to the fluctuations of the experimental data,
to the choice of functional form, and to the extrapolation
uncertainties due to the finite coverage of the data.
Following this, we go on to quantify how well the closure test is
passed,
both in terms of central values
and uncertainties, using various statistical estimators.

\subsubsection{PDF uncertainties: data, functional and extrapolation
components}
\label{sec:components}

A more sophisticated understanding of the various sources
that form the total PDF uncertainties
can be obtained in the context of closure tests by
comparing Level~0, Level~1, and Level~2 fits.
Indeed, in each of these fits the PDF uncertainty band has different
components.
In Level~0 fits, the only significant component is the interpolation
and extrapolation
uncertainty (which we will refer to as extrapolation uncertainty for
short);
in Level~1, one also has the uncertainty due to the choice
of functional form; and in Level~2 finally one also adds the uncertainties
due to the fluctuations of the experimental data.
Therefore, by comparing Level~0, Level~1 and Level~2 closure fits
we can analyse how the total PDF uncertainties
are decomposed into data, functional and extrapolation
uncertainties.

Let us begin with the extrapolation uncertainty.
As discussed in the previous section, in a Level~0
closure test, the PDF uncertainty at the fitted data points
should go to zero as the training length is increased.
This implies that, in the experimental data region, PDF
uncertainties should also decrease monotonically as a function
of the training length wherever data are available.
However, in between data (interpolation) and outside the data region
(extrapolation) PDFs can fluctuate.
This residual uncertainty, which remains even at infinite  Level~0
training length,  we refer to as the {\em extrapolation}\ uncertainty.

Note that, given the highly non-trivial dependence of
PDFs on the measured cross-sections, and the wide
range of observables included in the fit, it is very difficult
to determine precisely how this extrapolation region
is defined: while  a non-negligible
extrapolation component is expected for all PDFs at small enough
and large enough values of $x$, significant uncertainties due to
interpolation could also be present at intermediate $x$.
In fact,  this also accounts for possible
degeneracies between PDFs: for example, even in Level~0 closure
fits some PDFs can compensate  each other to produce exactly
the same cross-section, and this effect will also be included
in the extrapolation uncertainty.

\begin{figure}[h!]
  \centering
  \epsfig{width=0.46\textwidth,figure=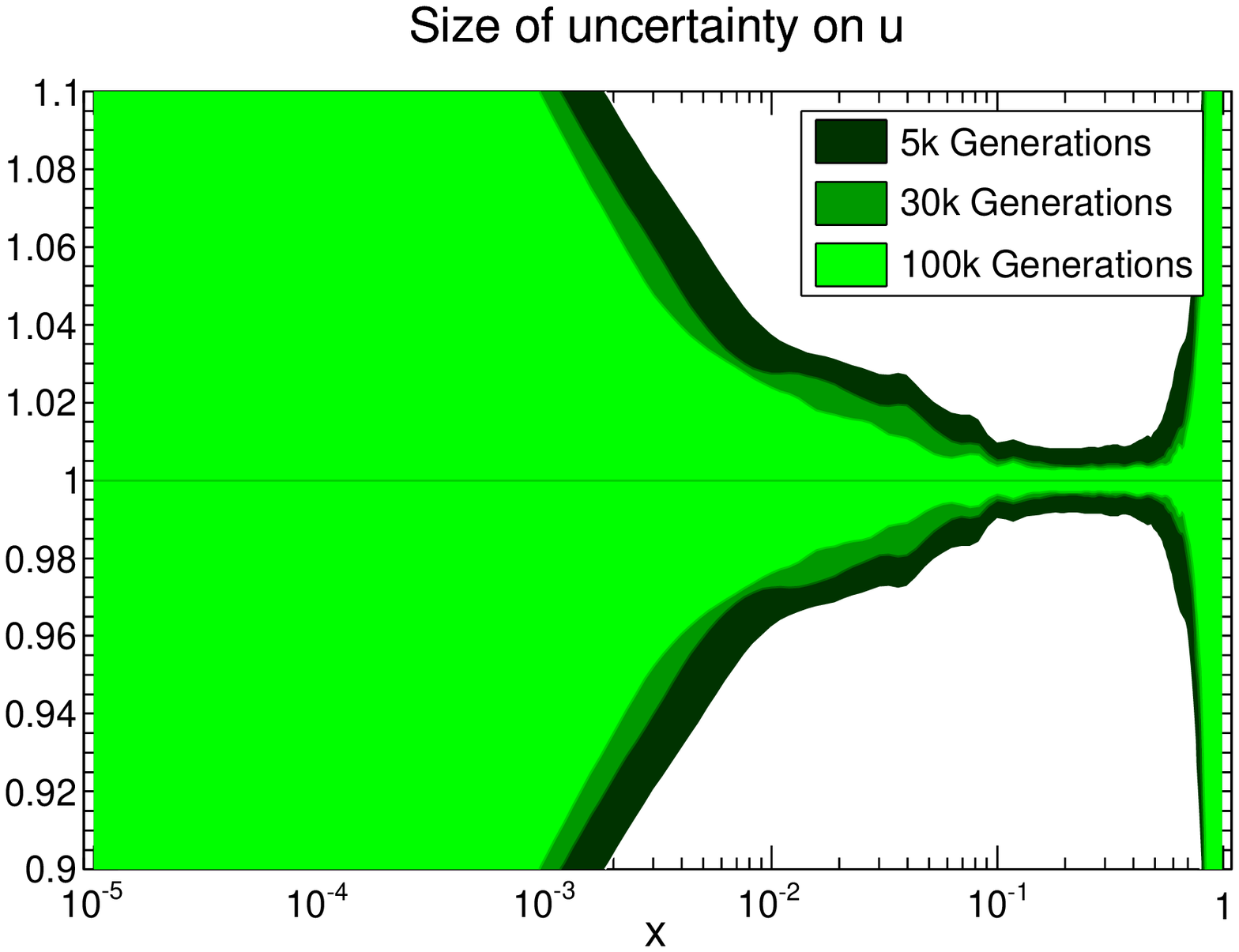}
  \epsfig{width=0.46\textwidth,figure=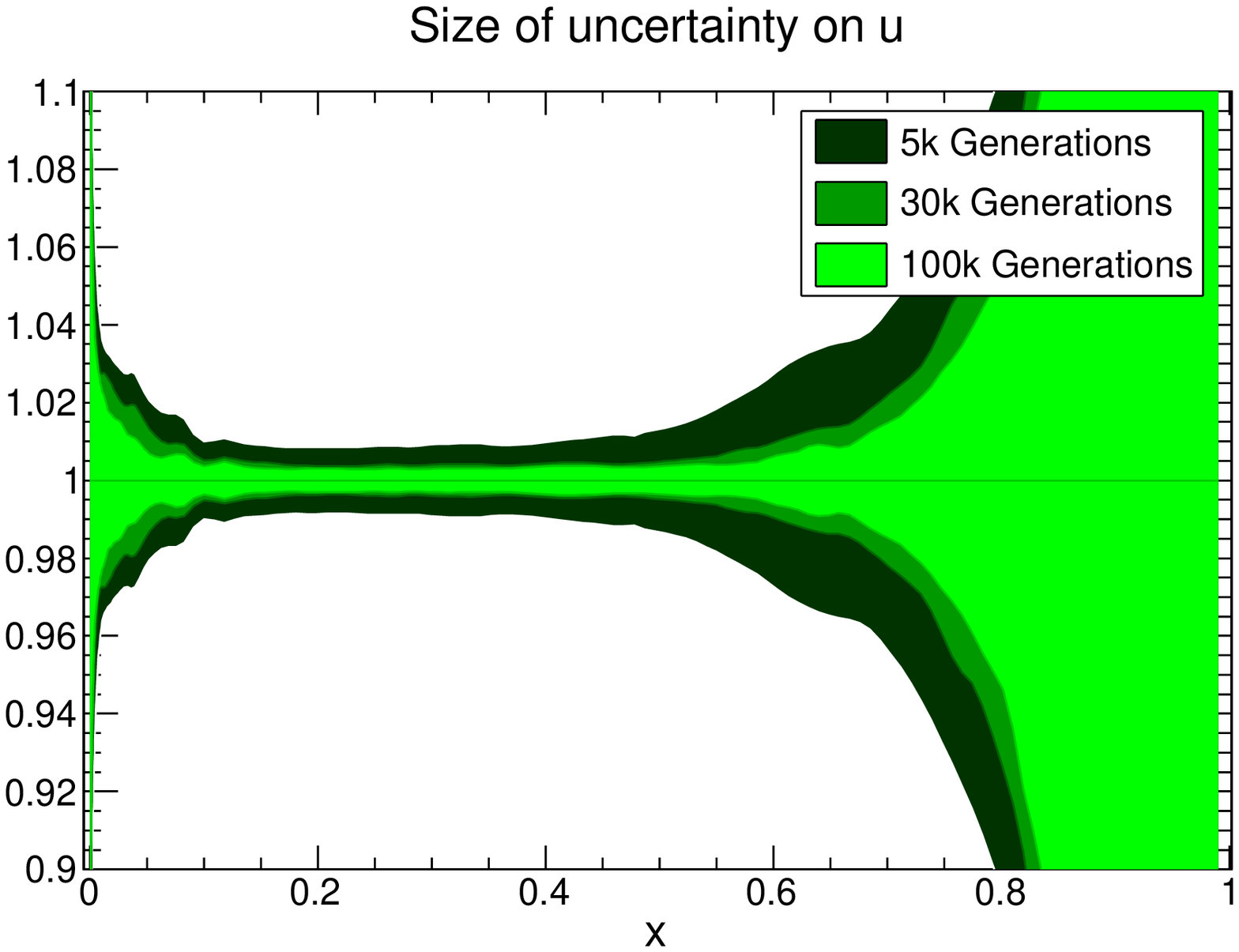}
  \epsfig{width=0.46\textwidth,figure=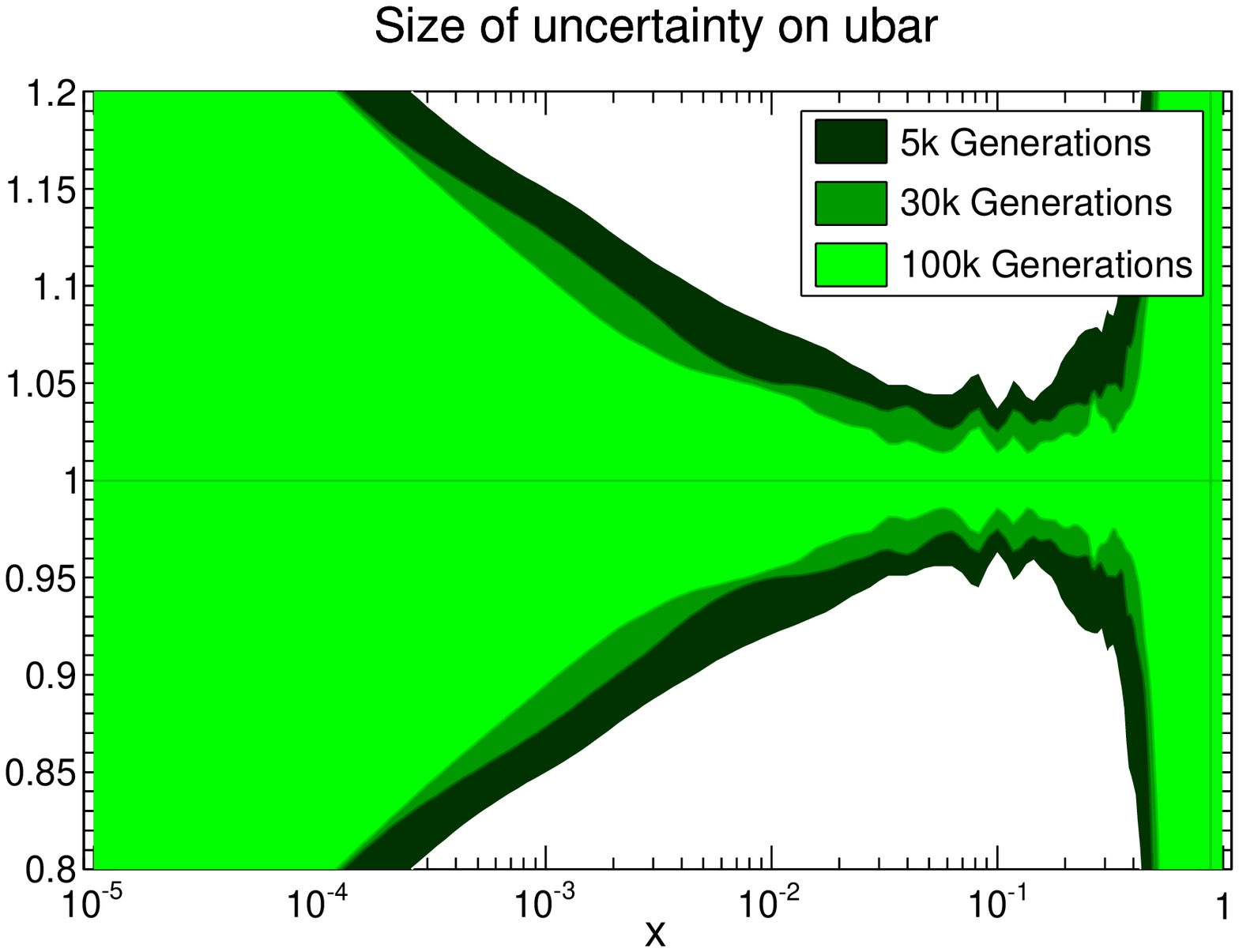}
  \epsfig{width=0.46\textwidth,figure=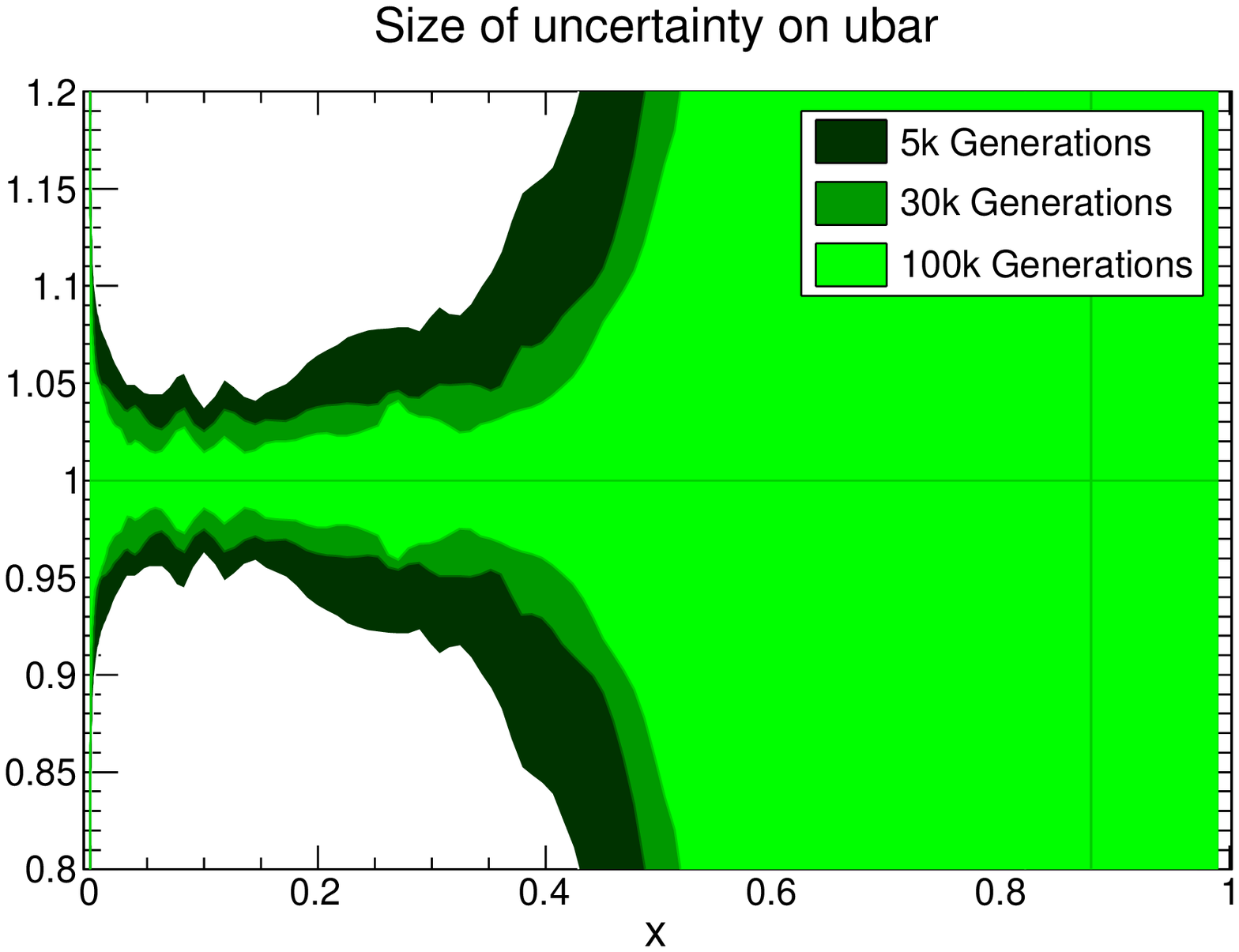}
  \epsfig{width=0.46\textwidth,figure=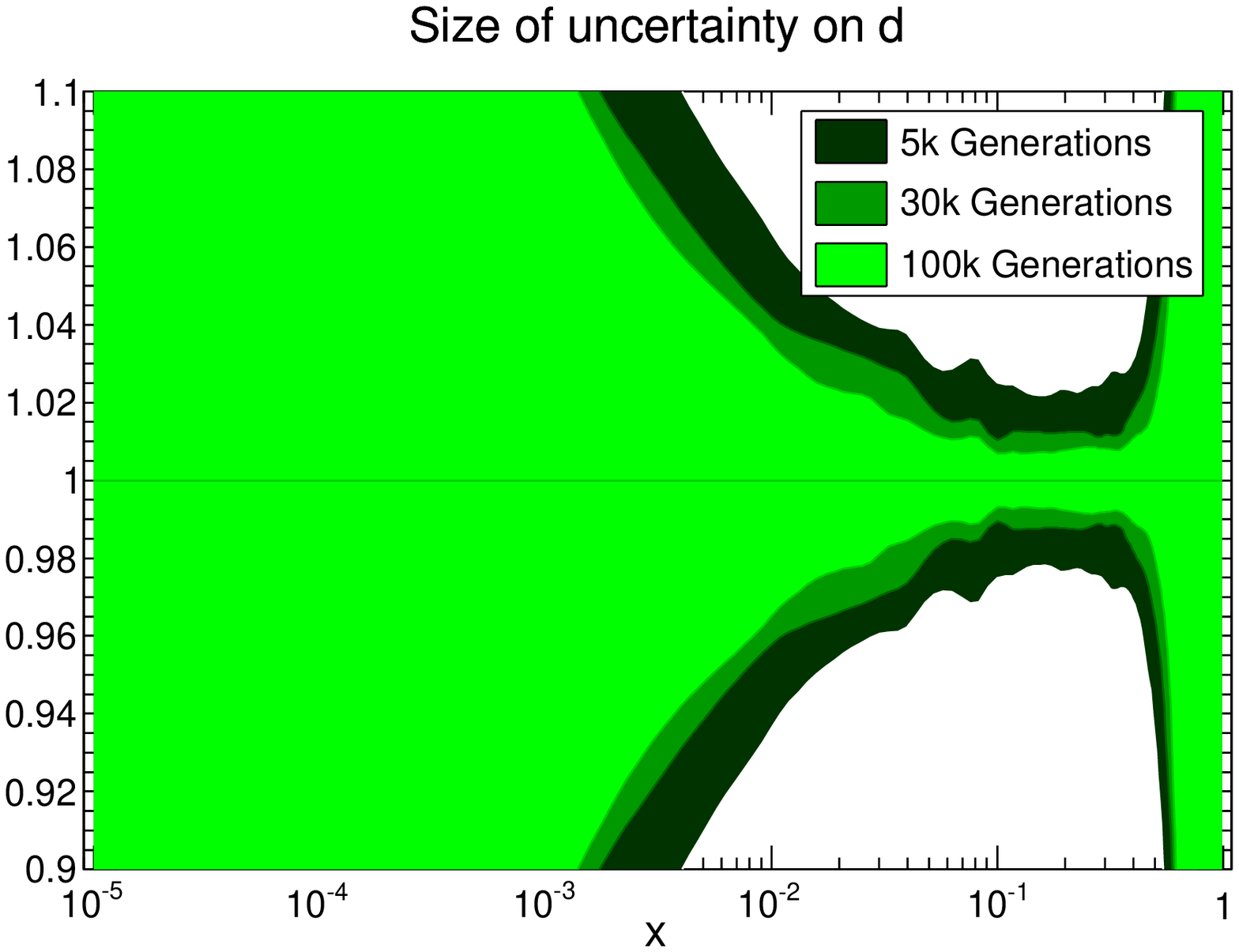}
  \epsfig{width=0.46\textwidth,figure=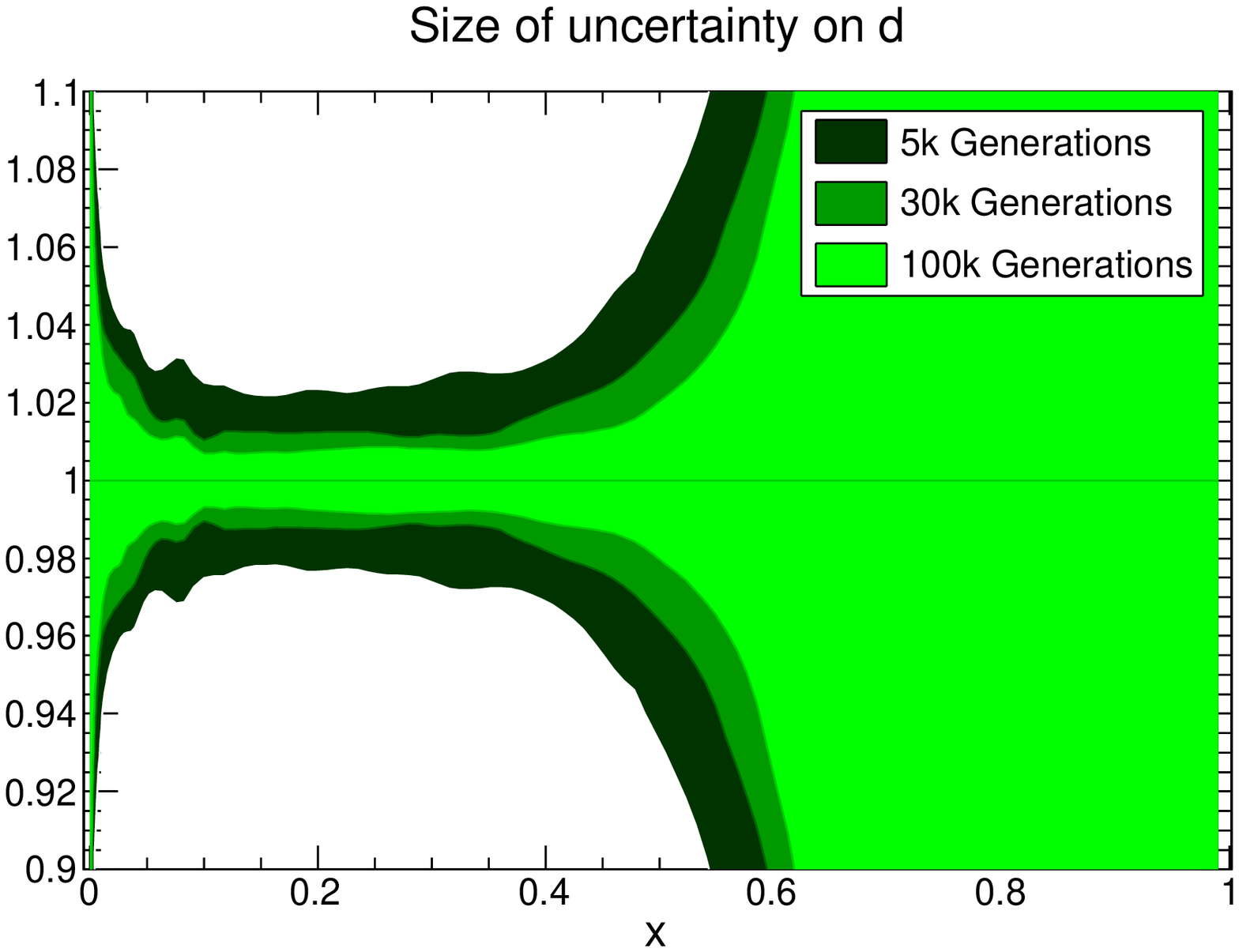}
  \epsfig{width=0.46\textwidth,figure=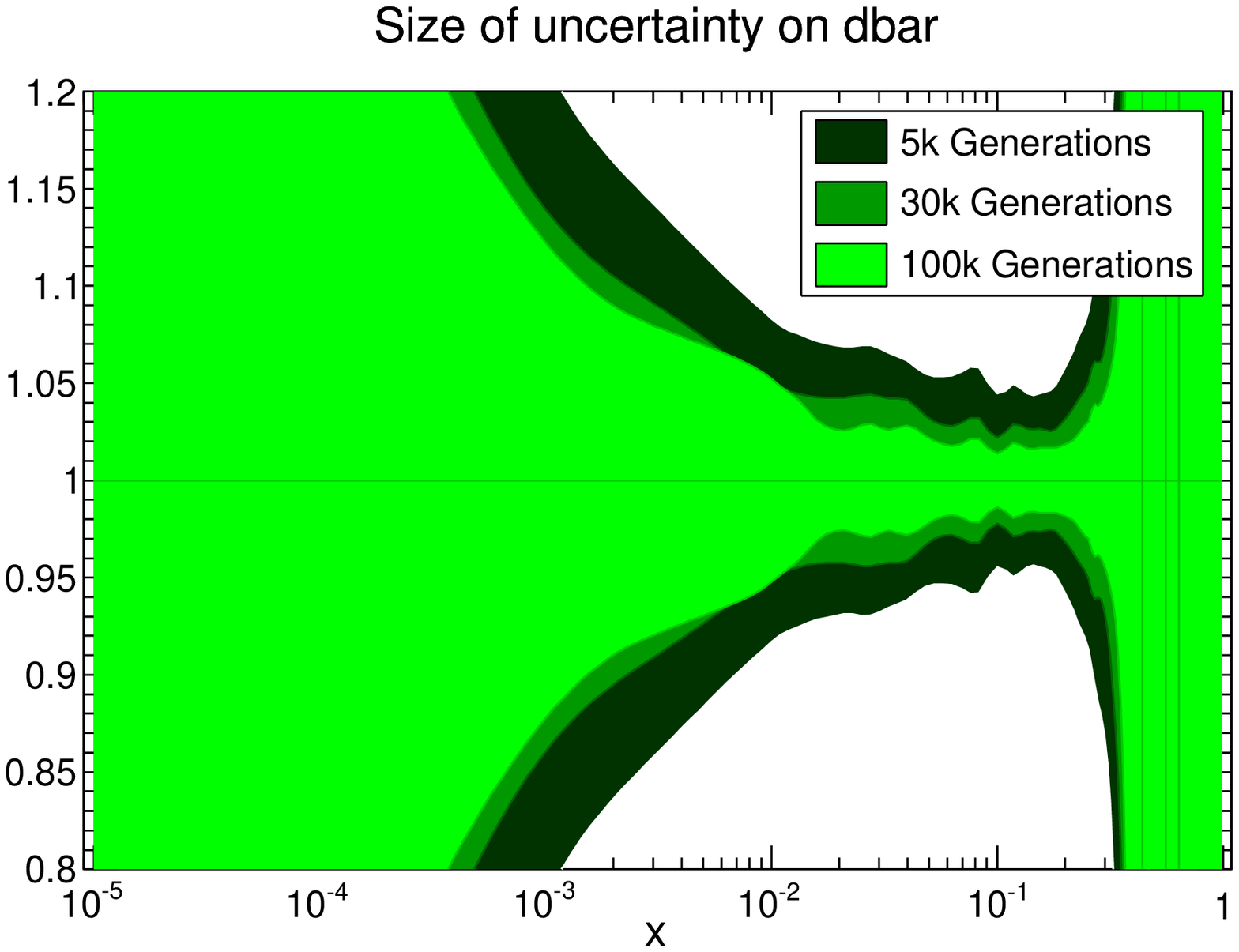}
  \epsfig{width=0.46\textwidth,figure=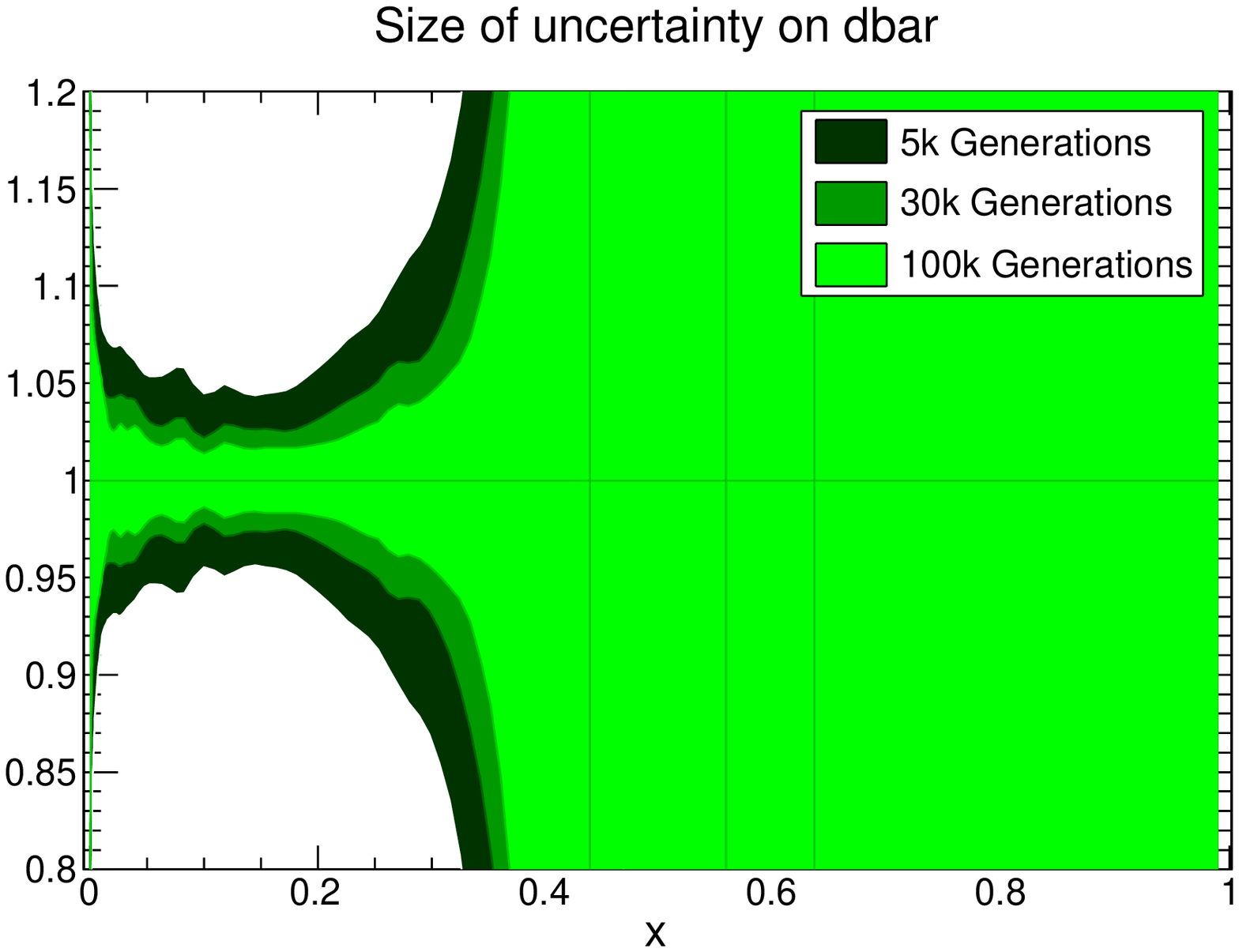}
  \caption{\small
Comparison of relative
PDF uncertainties obtained
  from Level~0 closure test fits with MSTW2008 NLO as input set and
three different training
  lengths. The black band shows the results with 5k GA generations,
the dark green band with 30k generations, and the pale green band
with 100k generations.
Results for the
$u,\bar{u},d$ and $\bar{d}$ PDFs are shown at the input
parametrization scale of $Q^2 = 1~\textrm{GeV}^2$,  both
in a logarithmic (left) and in a linear (right)
scales.
  \label{fig:5v100-unc}}
\end{figure}

To illustrate this point, in Fig.~\ref{fig:5v100-unc} we show the
results of  Level~0 closure tests for all PDFs, for three  fits with identical
settings but different training length, 5k, 30k and 100k GA generations
respectively.
We can see that while the uncertainties at 100k  generations are
smaller than at 5k, they do not go to zero, and in fact for
very small or very large values of $x$, where there is no experimental information, they
remain very large.
This is a direct consequence of the lack of experimental data
in this region, and illustrates this irreducible component of
the PDF uncertainties that we denote as extrapolation
uncertainty. Furthermore, in most of the data region (with the possible
exception of a very small region around the valence peak
$x\sim0.2-0.3$ for the up and down distributions) the uncertainty at
30k generations and 100k generations are very close, and in particular
rather closer than the corresponding  $\varphi_{\chi^2}$ values of
Fig.~\ref{fig:level0_fchi2}. This is especially noticeable for the up
and gluon distributions, and it means that, even in the data region,
there is an irreducible component of the uncertainty which does not go
to zero even for extremely long training length, and can only be
reduced with more data, as it is due to the interpolation between data
points. Otherwise stated, at 30k, the uncertainty of the Level~0 fit
shown in Fig.~\ref{fig:level0_fchi2}, which is evaluated only at
the data points and is thus due to fitting inefficiency, is already subdominant
in comparison to the extrapolation uncertainty almost
everywhere.

\renewcommand{\thefigure}{\arabic{figure} (Cont.)}
\addtocounter{figure}{-1}
\begin{figure}[ht]
  \centering
  \epsfig{width=0.46\textwidth,figure=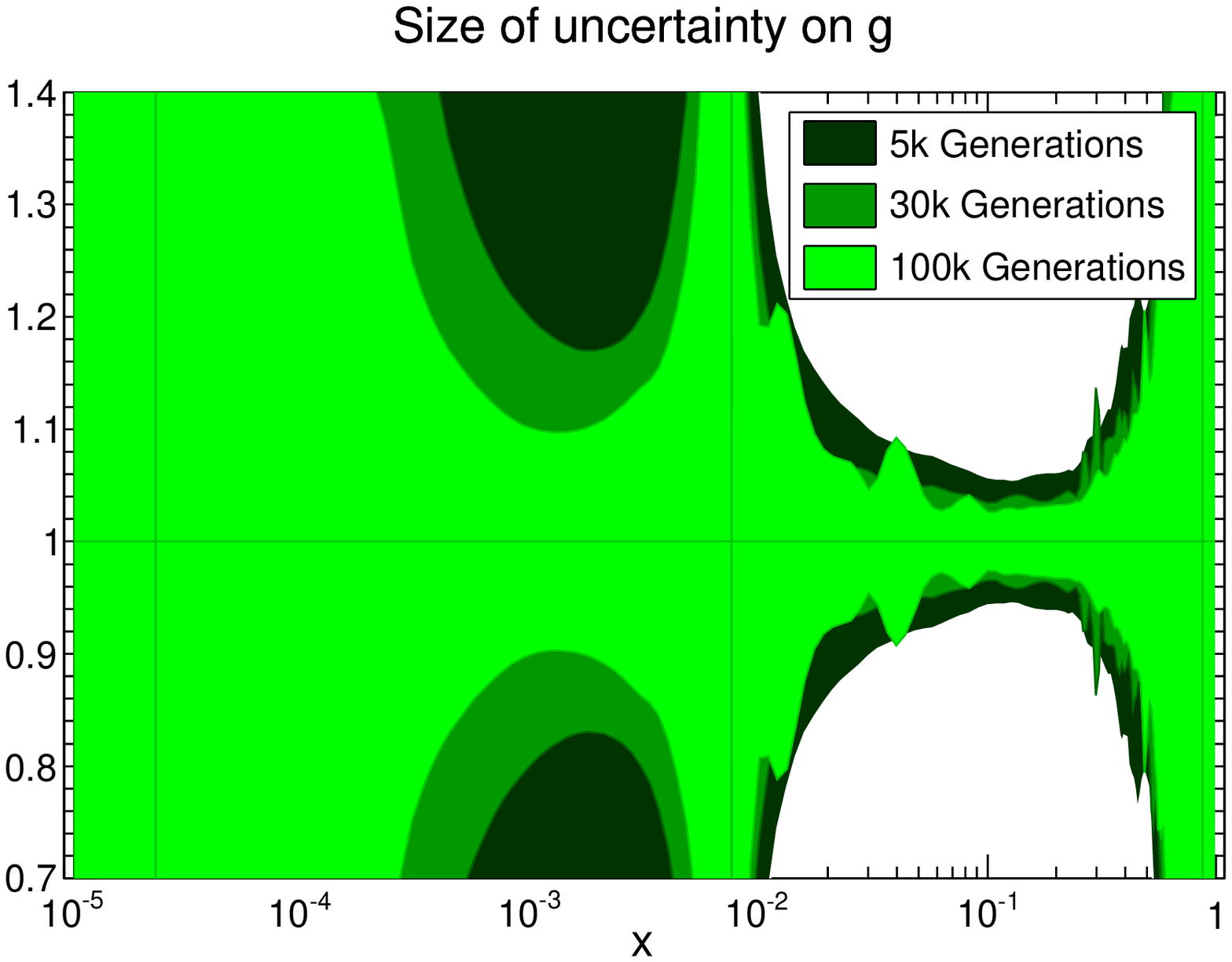}
  \epsfig{width=0.46\textwidth,figure=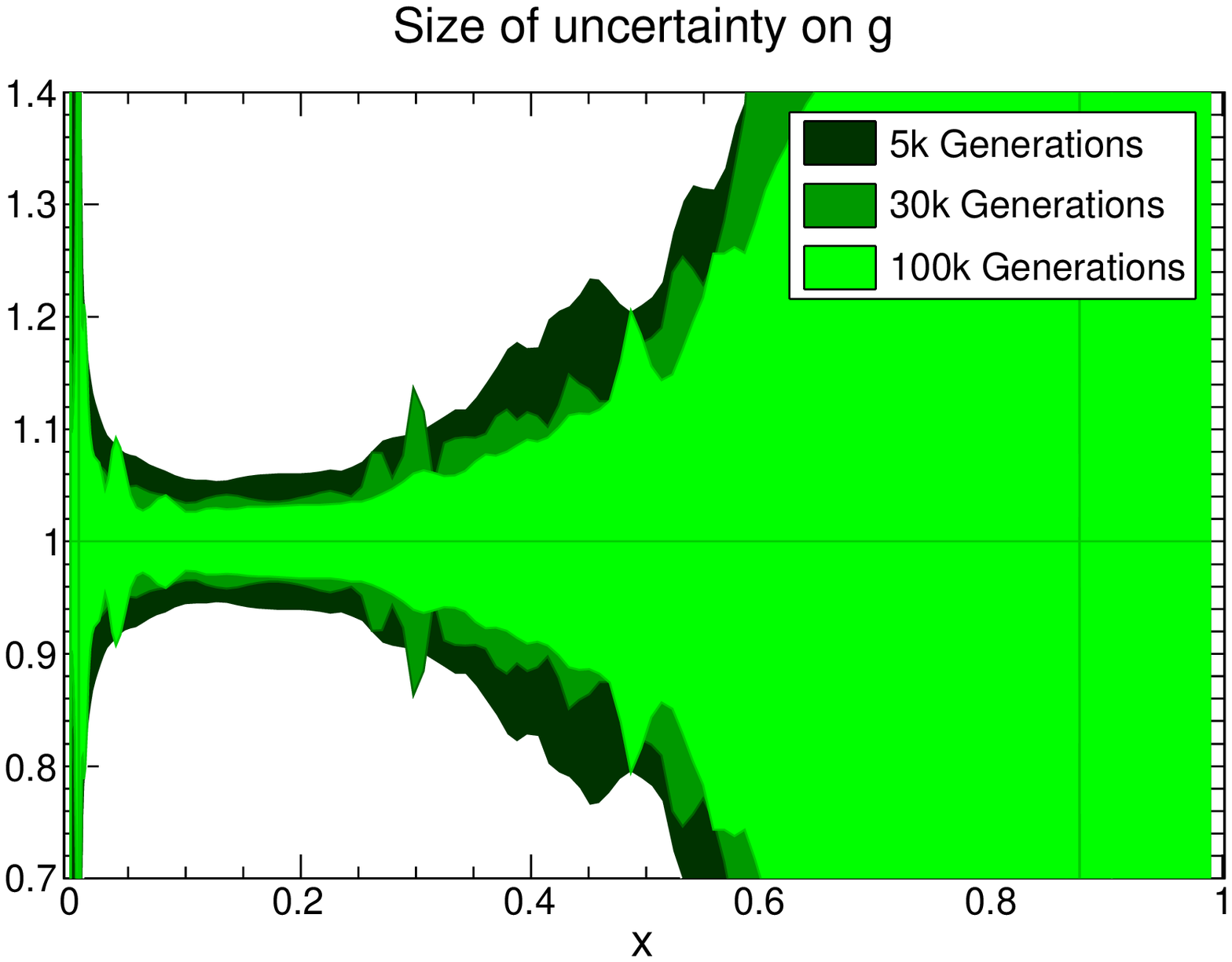}
  \epsfig{width=0.46\textwidth,figure=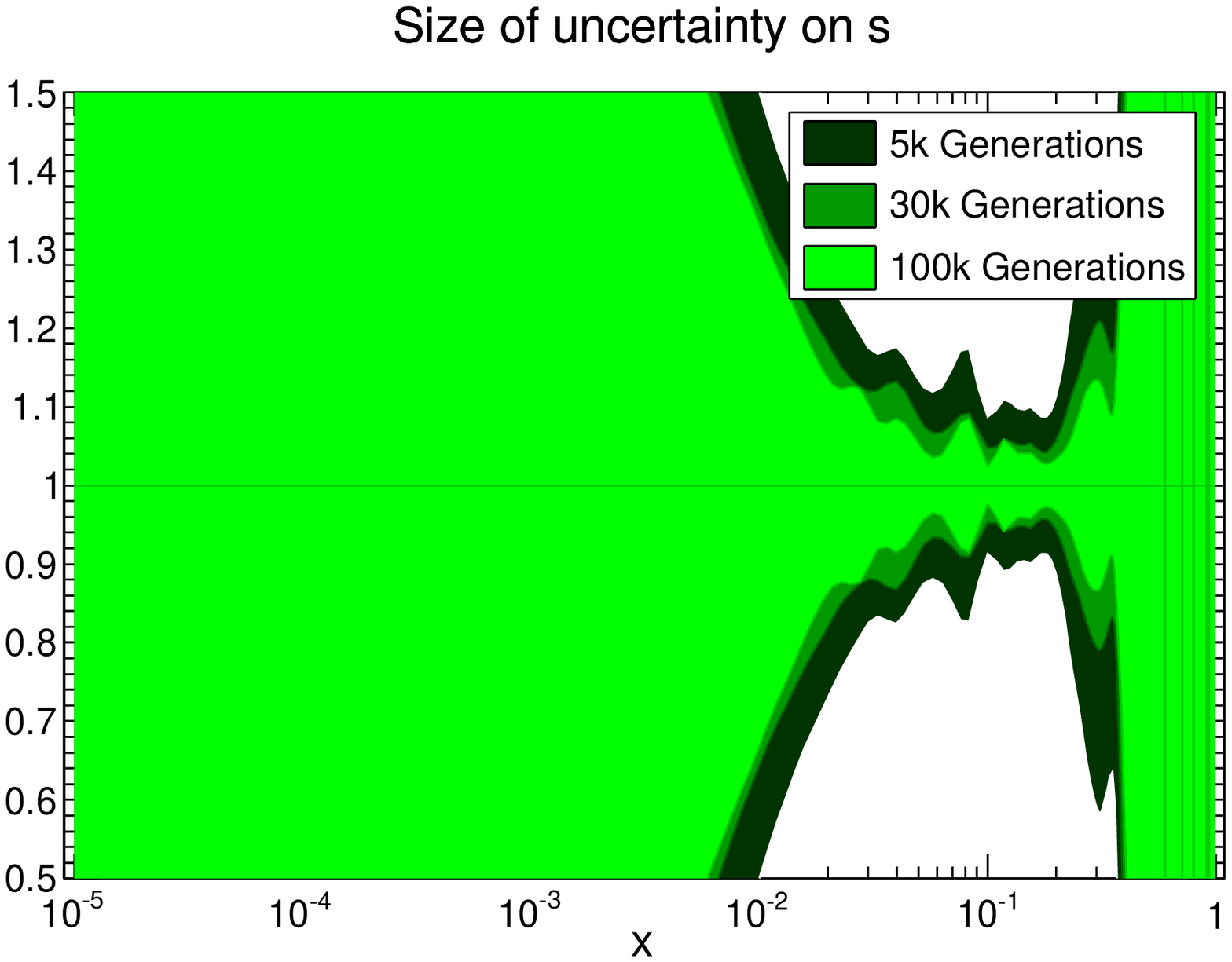}
  \epsfig{width=0.46\textwidth,figure=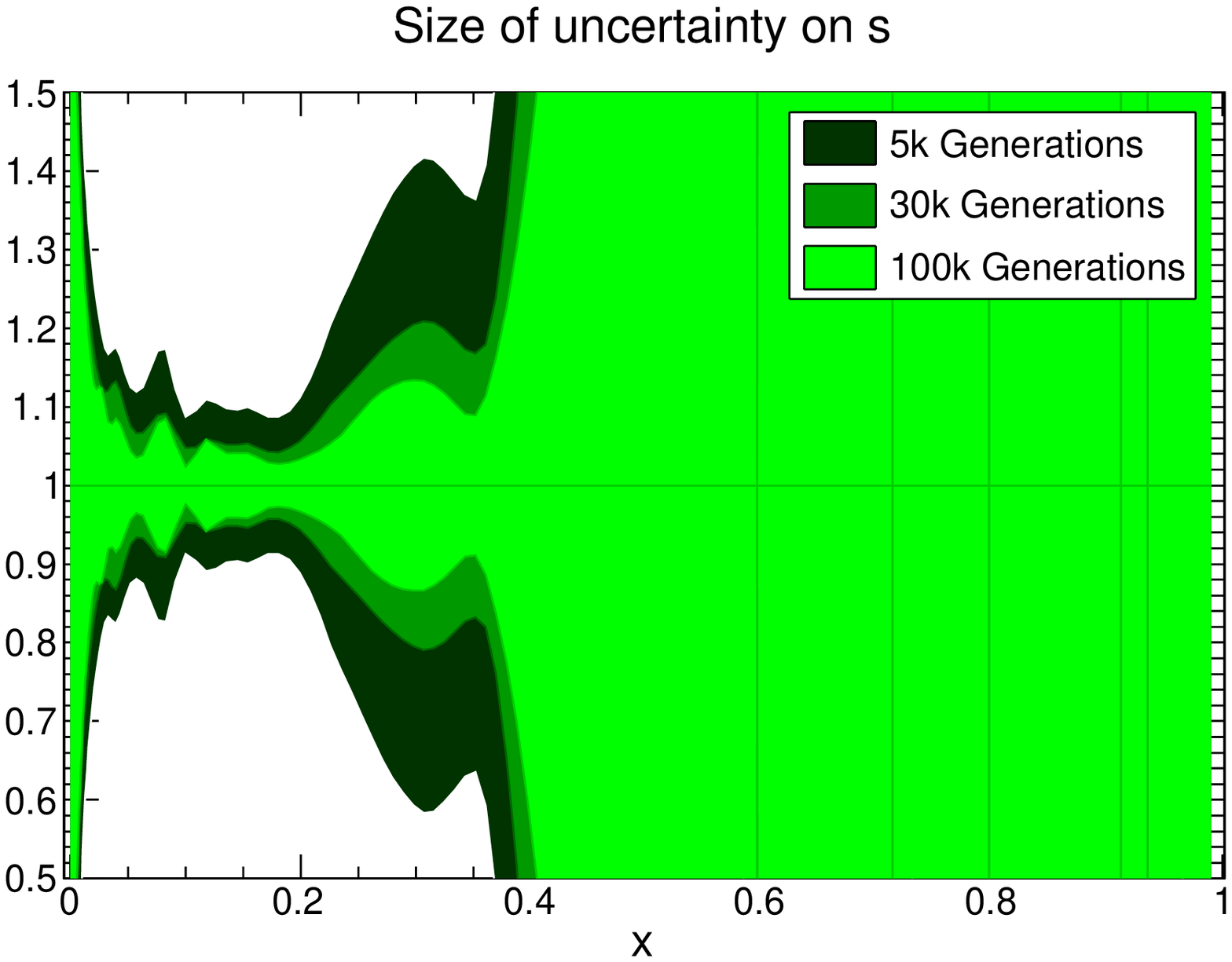}
  \epsfig{width=0.46\textwidth,figure=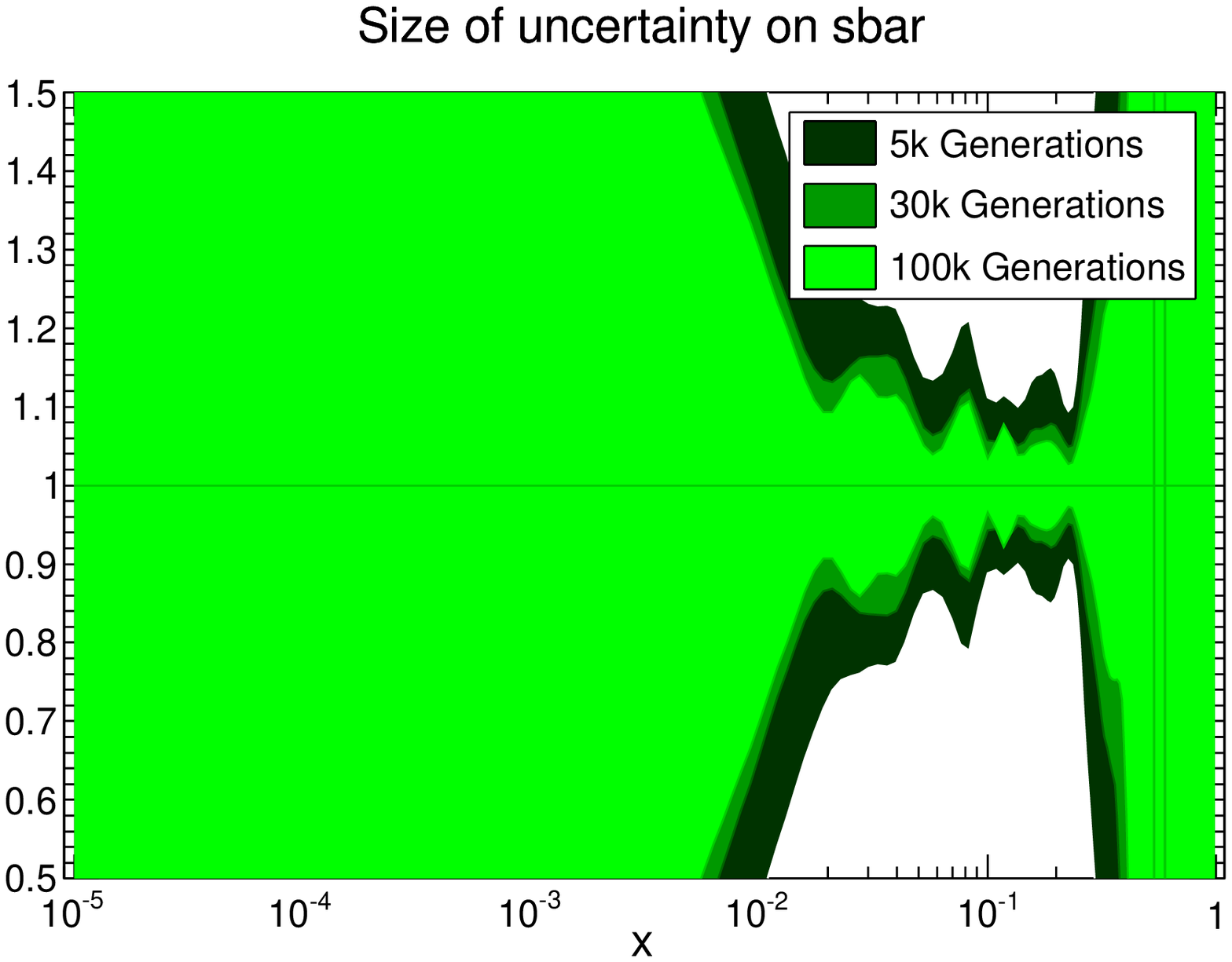}
  \epsfig{width=0.46\textwidth,figure=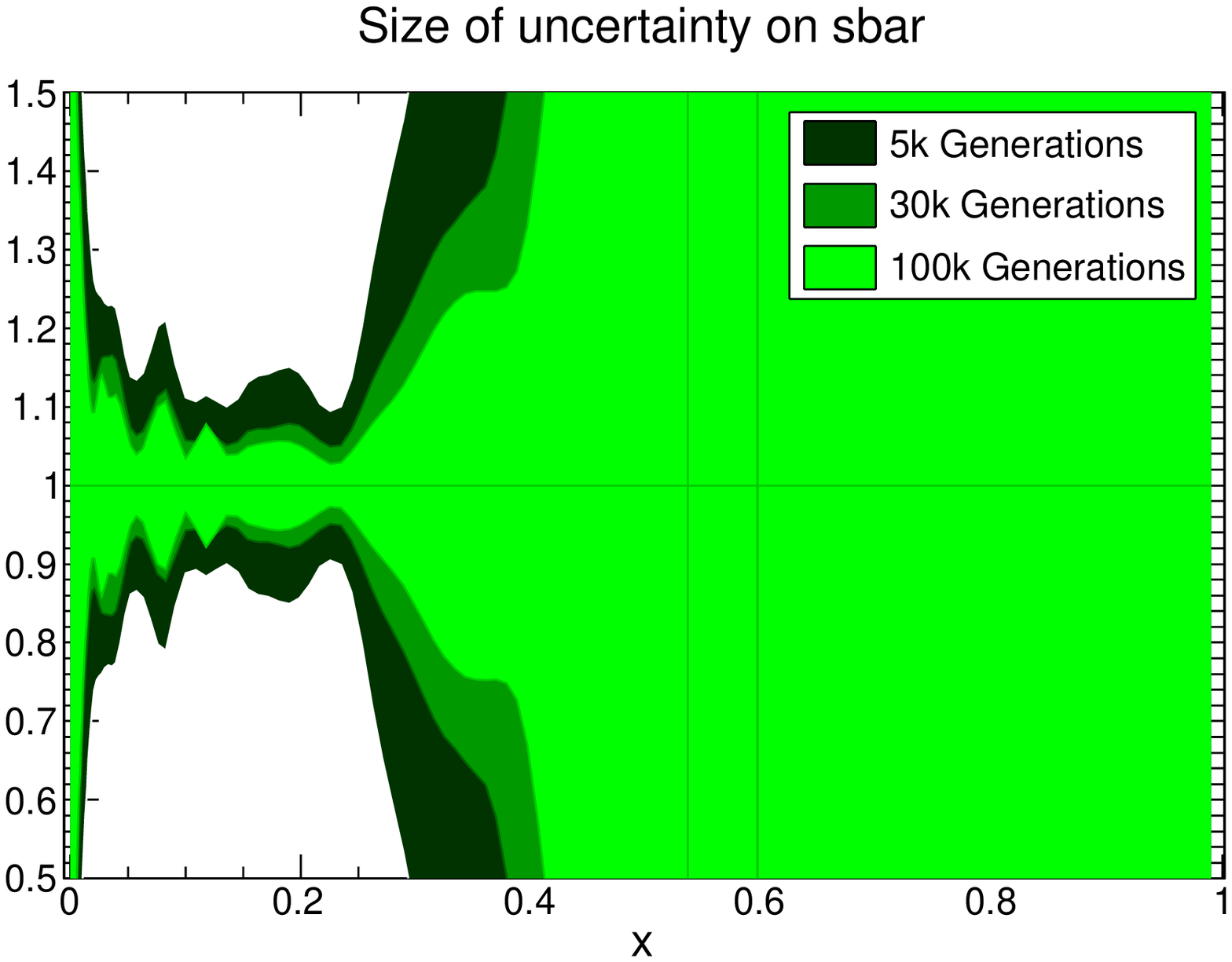}
  \caption{\small Same as above for the $g,s$ and $\bar{s}$ PDFs.
The spike on the gluon PDF at $x\sim 0.005$ is caused by the fact
that the gluon from the input PDF set, MSTW08, has a node
in this region, and thus the relative PDF uncertainty blows up.
 }
  \label{fig:5v100-unc1}
\end{figure}
\renewcommand{\thefigure}{\arabic{figure}}

\begin{figure}[ht!]
  \centering
  \epsfig{width=0.46\textwidth,figure=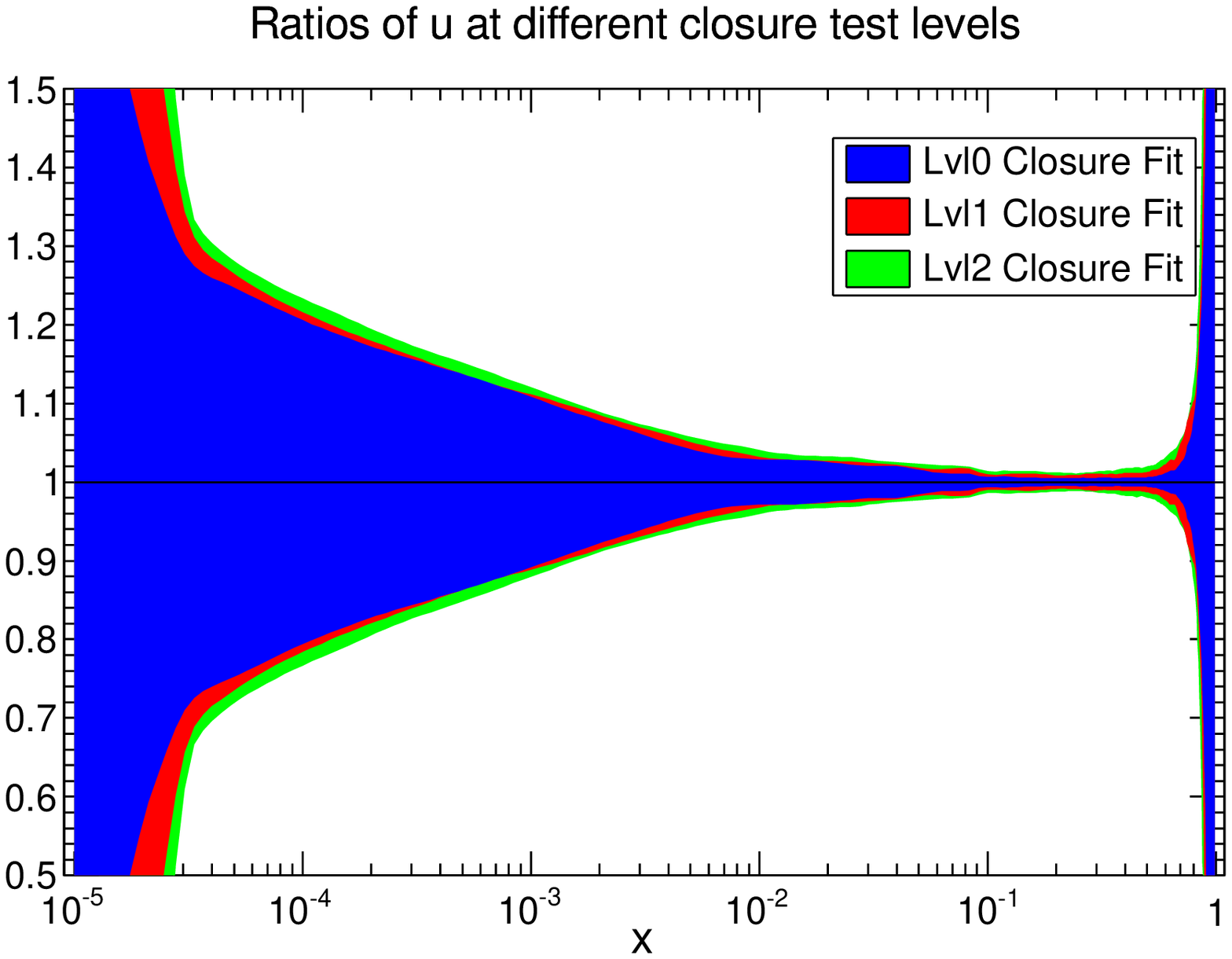}
  \epsfig{width=0.46\textwidth,figure=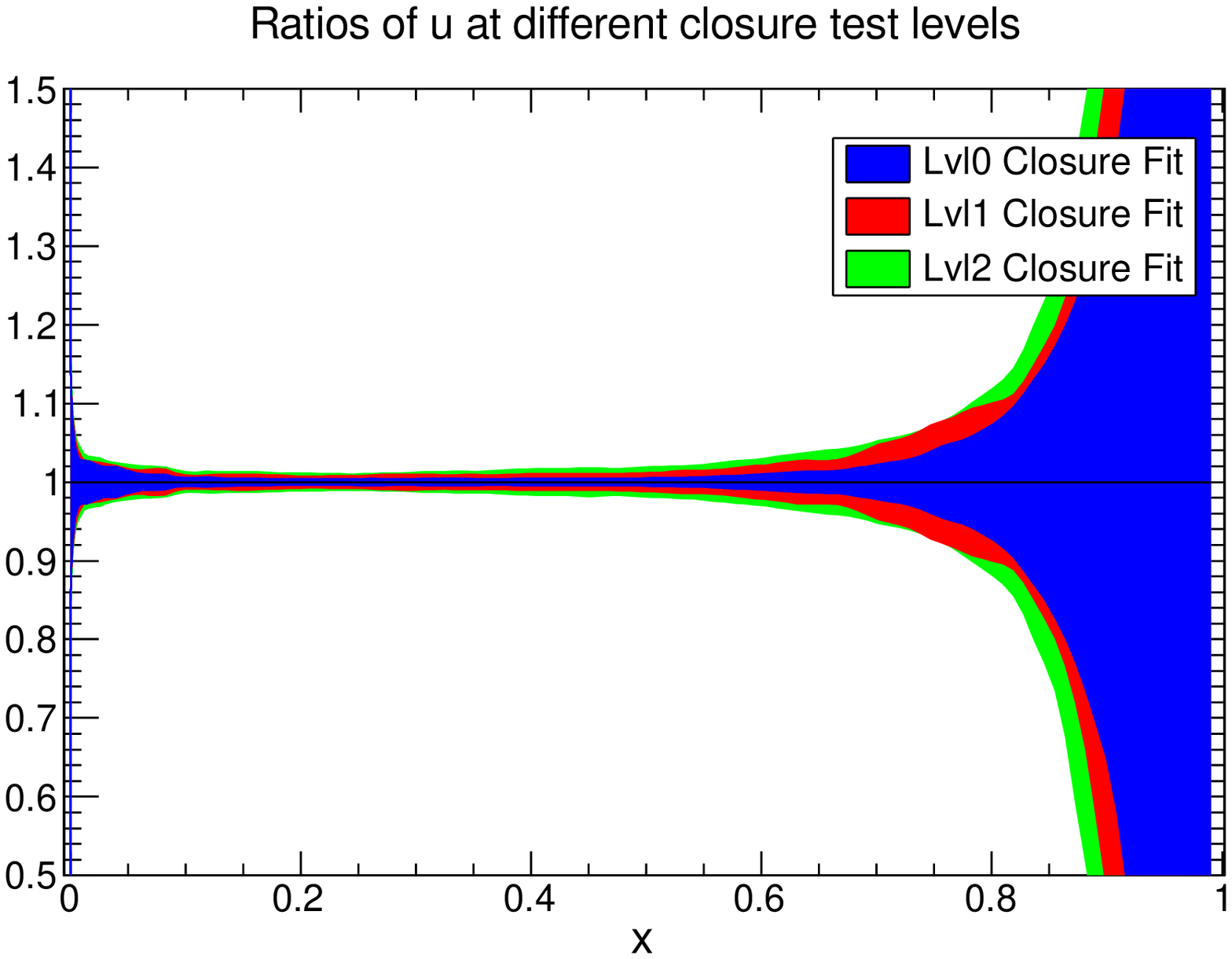}
  \epsfig{width=0.46\textwidth,figure=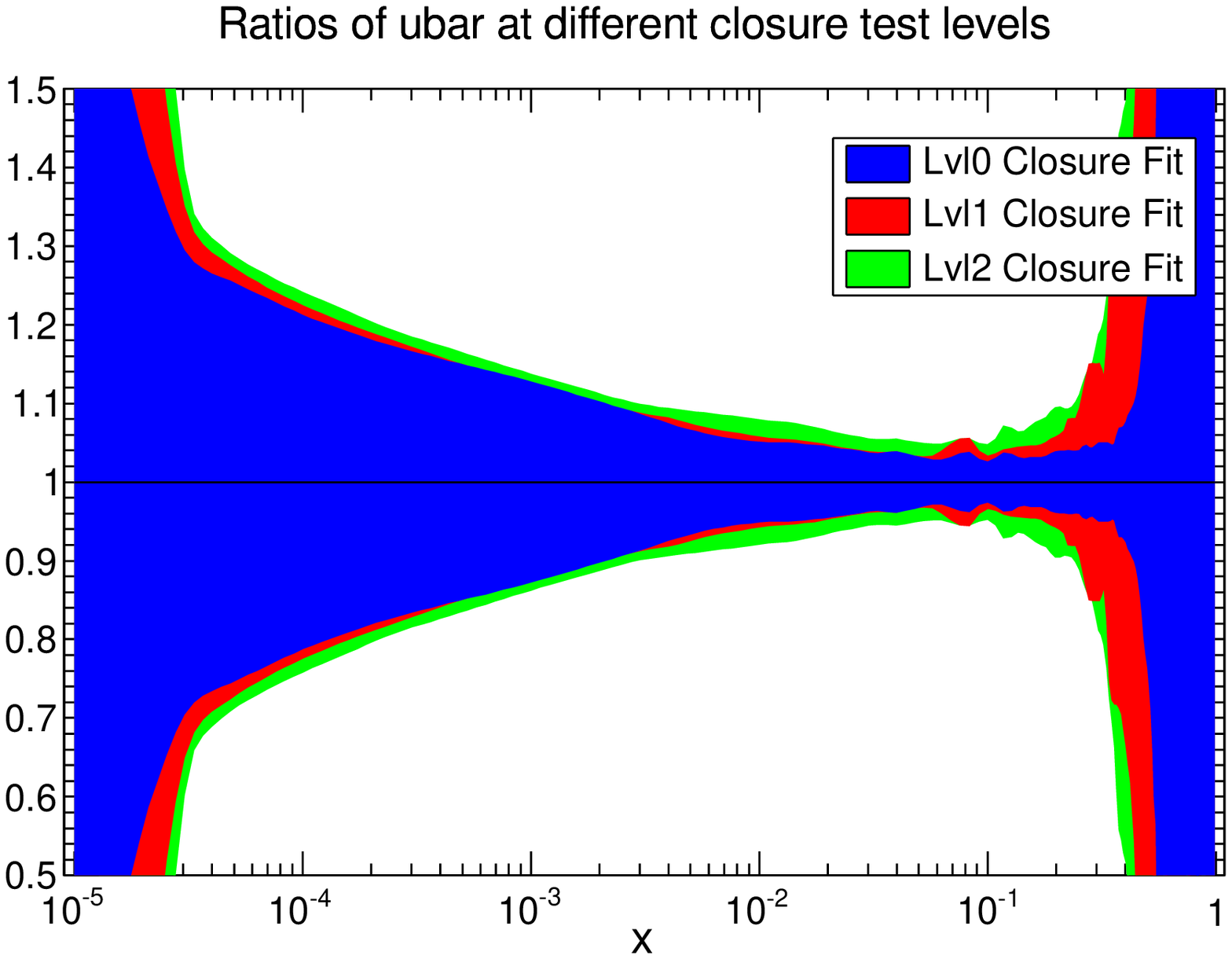}
  \epsfig{width=0.46\textwidth,figure=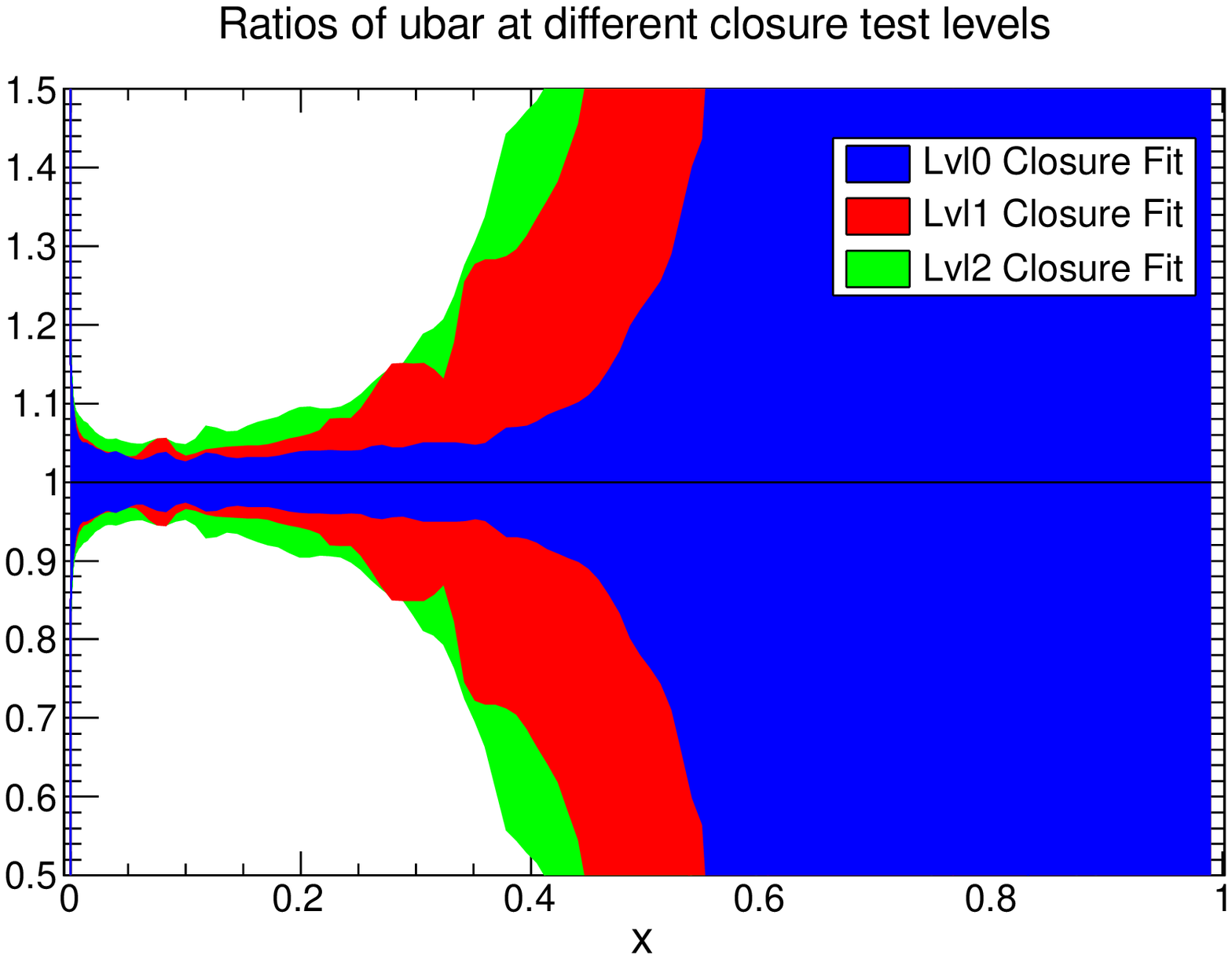}
  \epsfig{width=0.46\textwidth,figure=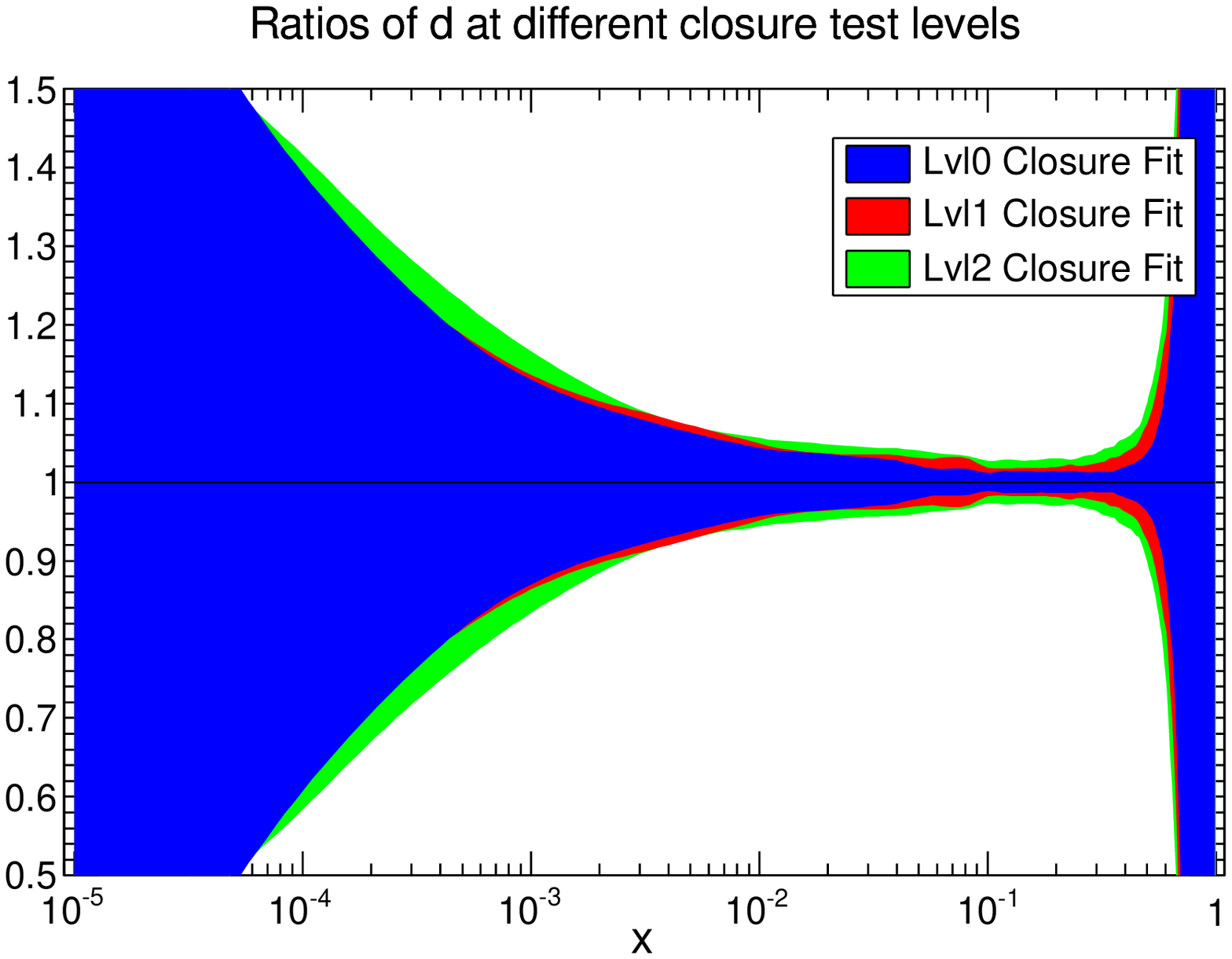}
  \epsfig{width=0.46\textwidth,figure=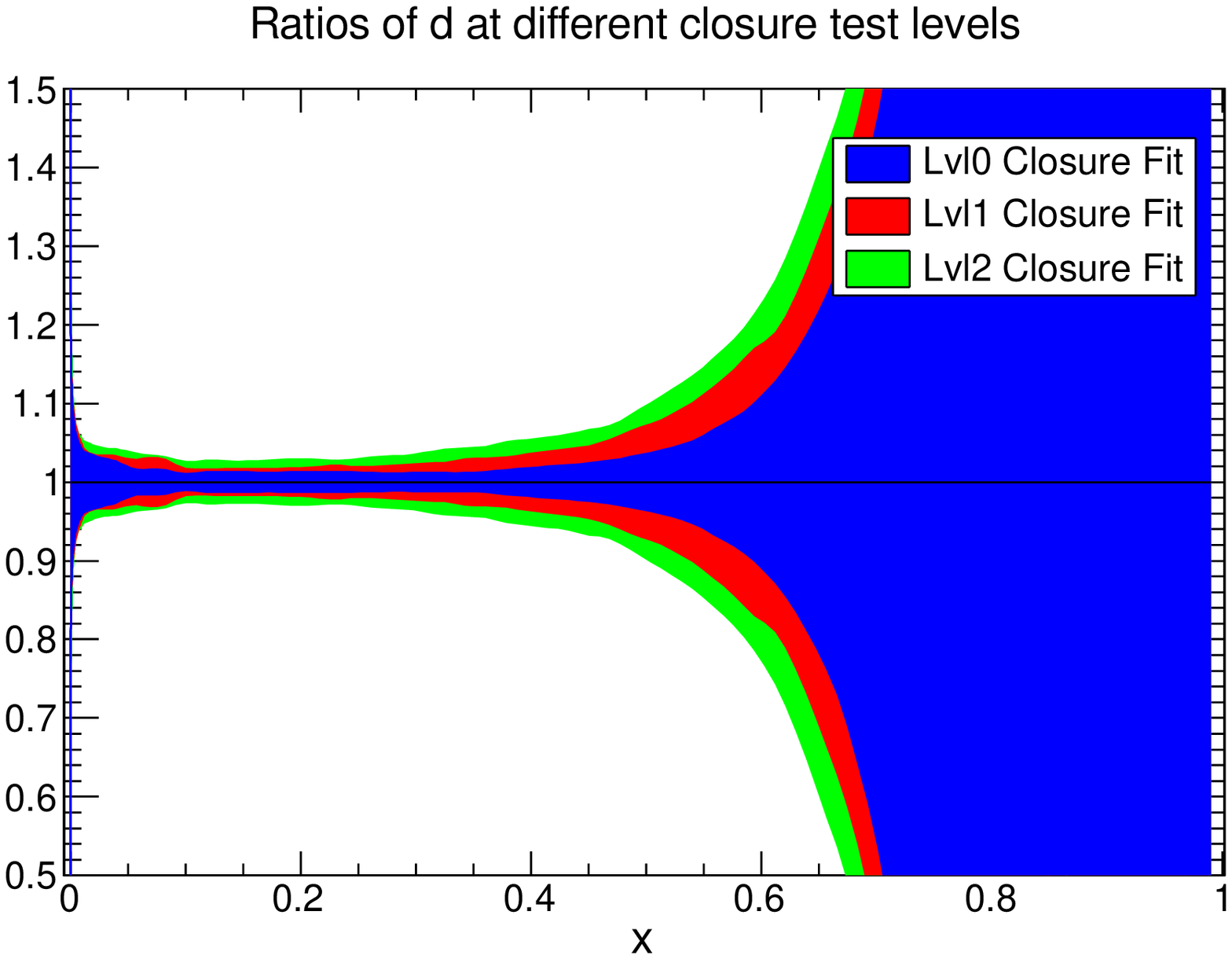}
  \epsfig{width=0.46\textwidth,figure=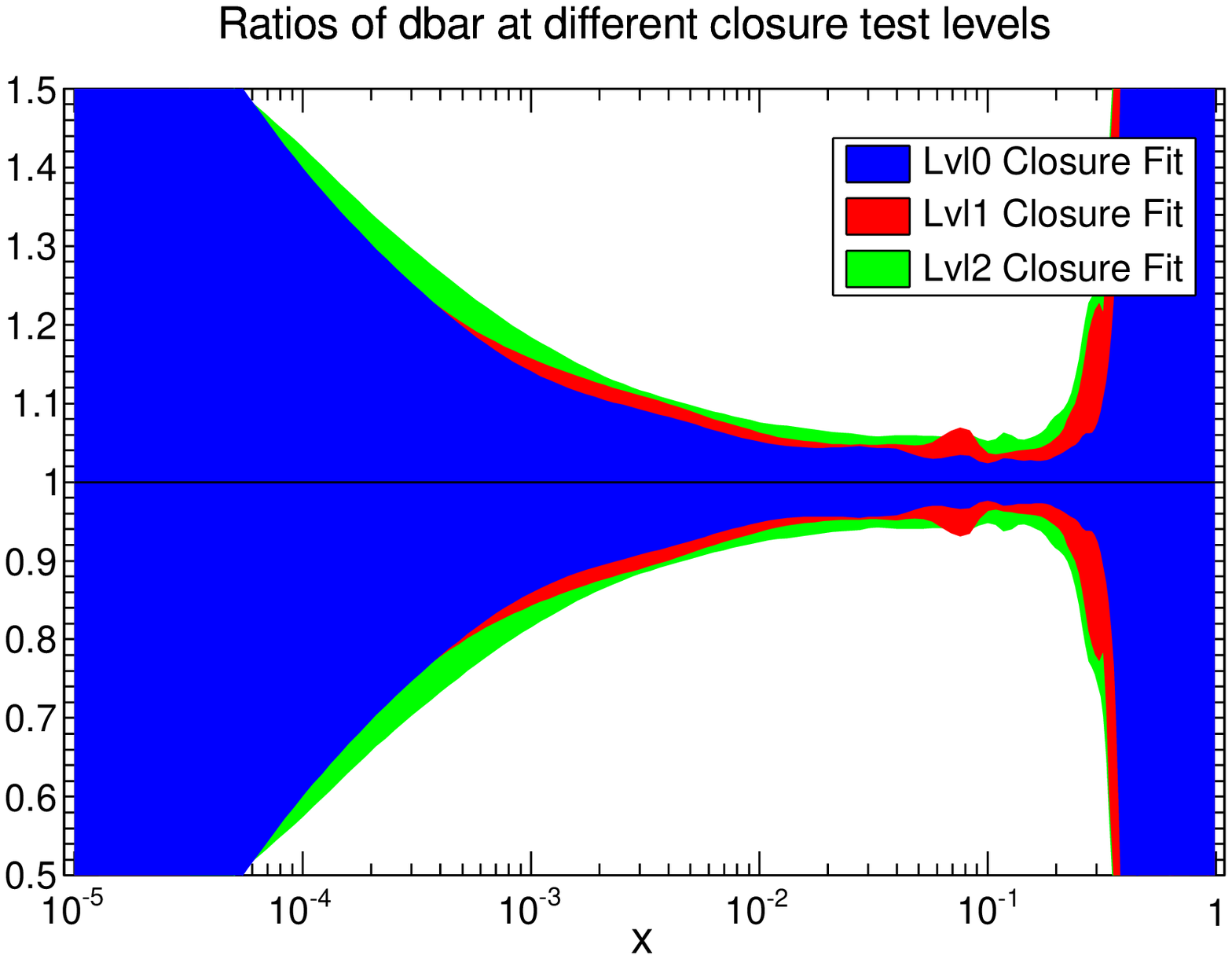}
  \epsfig{width=0.46\textwidth,figure=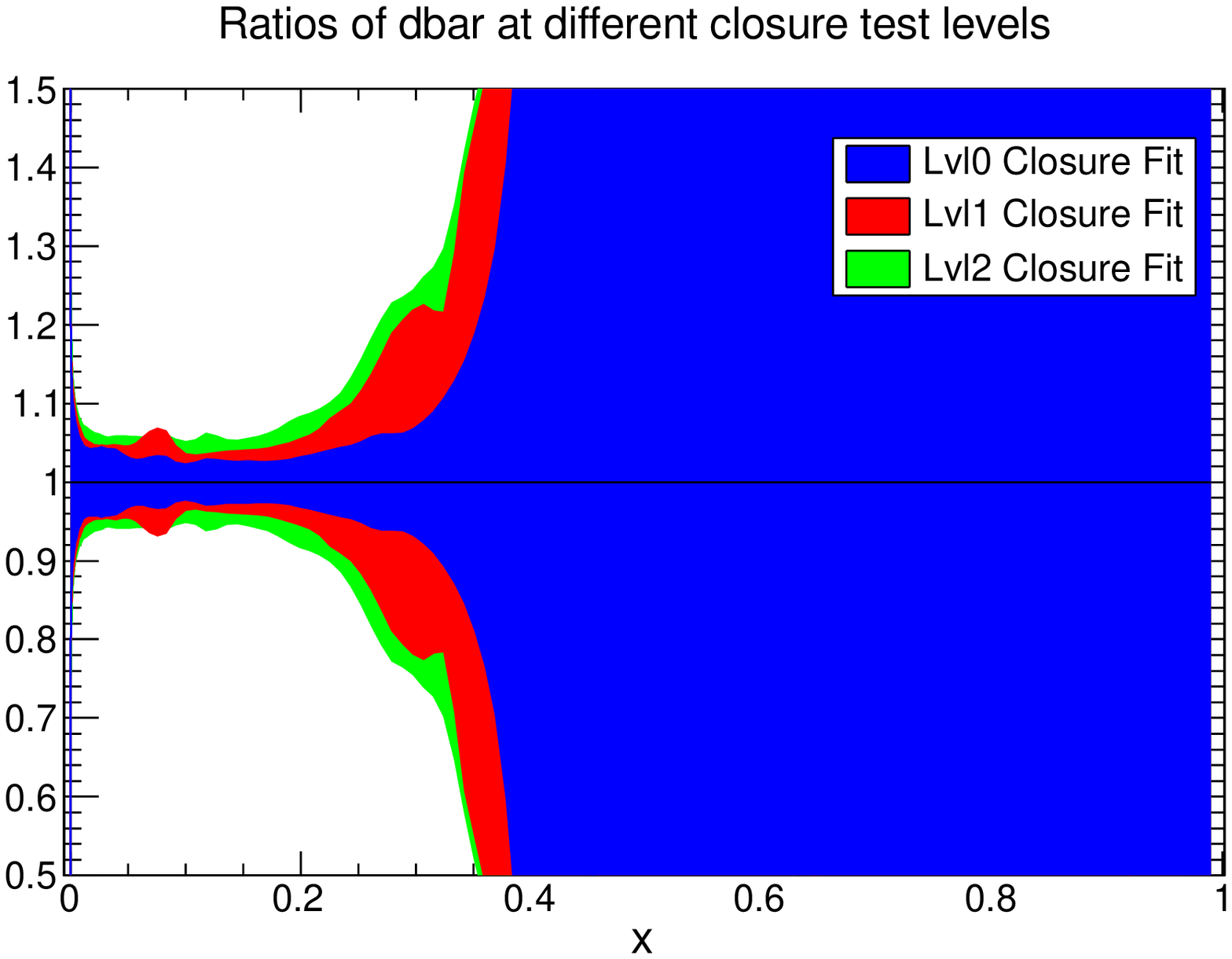}
  \caption{\small Same as Fig.~\ref{fig:5v100-unc}, but now comparing
    PDF uncertainties obtained
  from Level~0, Level~1 and Level~2 closure test fits.
The plots show the 68\% confidence level PDF
uncertainty band for each of the fits, normalized
to the corresponding central value of each fit.
}
  \label{fig:ratiofit1}
\end{figure}

Now we can see how the total PDF uncertainty also contains
functional and data uncertainties
by comparing the results of Level~0, Level~1 and Level~2 closure tests
with otherwise all other fit settings identical.
Figure~\ref{fig:ratiofit1}  shows the ratios of the
uncertainty of the fitted PDFs to the respective central values
in each case, for the
C5, C8 and C9 closure test fits, see Table~\ref{tab:CTfits}, corresponding
to Level~0, Level~1 and Level~2 closure tests respectively.
The MSTW08 NLO set is used as input PDF $f_{\rm in}$,
and a total training length of  30k generations is
used in all cases.
For the three fits, the PDF uncertainty bands are defined
as the 68\% confidence interval
from the sample of $N_{\rm rep}=100$ fitted replicas.
Results are provided for the PDFs in the flavor basis
at the input parametrization scale of
 $Q^2 = 1~\textrm{GeV}^2$, though the qualitative interpretation
is the same in any other basis.

In order to understand these results, it is useful to review our
expectations for the results of the Level~1 and Level~2 closure
fits, as we have done above for Level~0.
In a Level~1 fit, the central values of the data have been
fluctuated around the theoretical prediction, and
therefore $f_{\rm fit}=f_{\rm in}$ no longer
provides an absolute minimum for the $\chi^2$.
Indeed, provided the PDF parametrization
is flexible enough, the minimization algorithm should find a large
number of different
functional forms that yield an equally good
$\chi^2[\mathcal{T}[f_{\rm fit}],\mathcal{D}_1]\approx 1$.
Therefore, in Level~1 closure fits, on top of the extrapolation
component, the total PDF uncertainty will include a new component,
 which we refer to as {\em functional}\ uncertainty.

This functional uncertainty is a
consequence of the fact that, as discussed in Sect.~\ref{subsec:cv},
the optimal $\chi^2$ in the presence of data fluctuations is not the
absolute minimum of the $\chi^2$. In a closure test, this is obvious:
the optimal result corresponds to the true underlying functional
form, and thus the optimal $\chi^2$ is the one of the Level~1
pseudo-data, whose value is of order one (and tends to one in the limit
of infinite size of the data sample). For an infinite-dimensional
space of functions, this $\chi^2$ value can be obtained in an infinity
of different ways, whose spread provides the functional uncertainty.
The look-back method discussed  in Sect.~\ref{subsec:cv}
ensures that regardless of the length of the fit,
the final $\chi^2$ does not decrease beyond the overlearning point,
and thus again this functional uncertainty will survive even for
infinite training length.

In a Level~2 fit, the starting point
is again the Level~1 pseudo-data generated by
adding a Gaussian fluctuation over the predictions obtained from the input PDFs.
Now however there is a second step, which reproduces the actual
fitting strategy used in the NNPDF framework:
starting from these pseudo-data, a set of
$N_{\mathrm{rep}}$ Monte Carlo replicas is generated that reflects the
statistical
and systematic errors given by the experimental measurements.
Because each replica fluctuates around Level~1 data,
the expectation for the minimized figure of merit for each replica in Level~2 fits
is $\chi^2[\mathcal{T}[f_{\rm fit}],\mathcal{D}^k_2]\approx 2$, although we still expect
$\chi^2[\mathcal{T}[f_{\rm fit}],\mathcal{D}_1]\approx 1$.
In Level~2 closure tests,
each replica is fitted with exactly the same algorithm, yielding an
ensemble of fitted PDFs $\{f^k_{\rm fit}\}$ whose statistical
properties are a faithful propagation of the fluctuations in the
underlying dataset.
The increase in the uncertainty from Level~1 to Level~2 fits
is the genuine {\em data}\ uncertainty.

Following these considerations, it is possible to understand
quantitatively the features that are
observed in Fig.~\ref{fig:ratiofit1}.
Firstly, we see that Level~0 uncertainties are smaller
than Level~1 and in turn these are smaller than those at
Level~2: this confirms the expectation that at each Level
we are adding a new component of the total PDF uncertainty,
extrapolation, functional and data components, respectively.
We also observe that in the small-$x$ and
large-$x$ regions it is the extrapolation uncertainty that dominates,
that is, the Level~2 PDF uncertainties are already reasonably reproduced
by those of Level~0 closure fits.

However, as already noticed from Fig.~\ref{fig:5v100-unc},  the
Level~0 extrapolation
component can be significant also in some regions
with experimental data, like medium $x$, due to interpolation and degeneracies.
Indeed, it is clear from Fig.~\ref{fig:ratiofit1} that the Level~0
uncertainty in
the data region is typically of order of 5\% or more, and only goes down to
about 1\% in a very small region close to the valence peak for the up
and down quark distributions. But we know from
Fig.~\ref{fig:level0_fchi2} that on average the uncertainty at the
data points is by about a factor $\varphi_{\chi^2}\approx0.09$ (for a 30k fit)
smaller than the uncertainty of the original experimental data
$\langle\sigma_{\rm dat}\rangle\approx 18\%$,
i.e. it is of order of 1-2\%. This uncertainty is due to fitting inefficiency,
but anything on top of that, i.e. the Level~0 uncertainty seen in
Fig.~\ref{fig:ratiofit1} except for the up and down quarks at the
valence peak, is a genuine extrapolation uncertainty.

By comparing to the Level~1
results,  we see that the functional uncertainty is generally sizable,
especially
in regions at the boundary between data and extrapolation,
in particular at large $x$.
Interestingly, in regions where we have
a rather reasonable coverage from available data, the three components
are roughly of similar size.
Take for example the gluon around $x\sim 10^{-3}$, which is
well constrained by the high-precision HERA measurements.
We see that the extrapolation, functional and
data uncertainties are all of similar size,
and thus a  proper estimate of the total
PDF uncertainty must include all three components.
The same applies for other PDF flavors, such as for example
strangeness for $x \gsim 0.01$ (with abundant constraints
from neutrino DIS and LHC data) or the up and down quarks
at medium and large-$x$ (with many DIS and LHC datasets
providing information on them).

An important general conclusion is that data uncertainties are
not dominant  anywhere, and thus a  PDF determination that does
not include the extrapolation and functional components
will underestimate the overall PDF uncertainty.
This conclusions is consistent with that
of  previous rather less sophisticated,
NNPDF studies such as those of Ref.~\cite{Ball:2011eq}.
It is natural to conjecture that the tolerance
method~\cite{Pumplin:2009bb} which is used
in Hessian fits, provides an alternative way of supplementing the data
uncertainty with these extra necessary
components.
\renewcommand{\thefigure}{\arabic{figure} (Cont.)}
\addtocounter{figure}{-1}
\begin{figure}[h!]
  \centering
  \epsfig{width=0.46\textwidth,figure=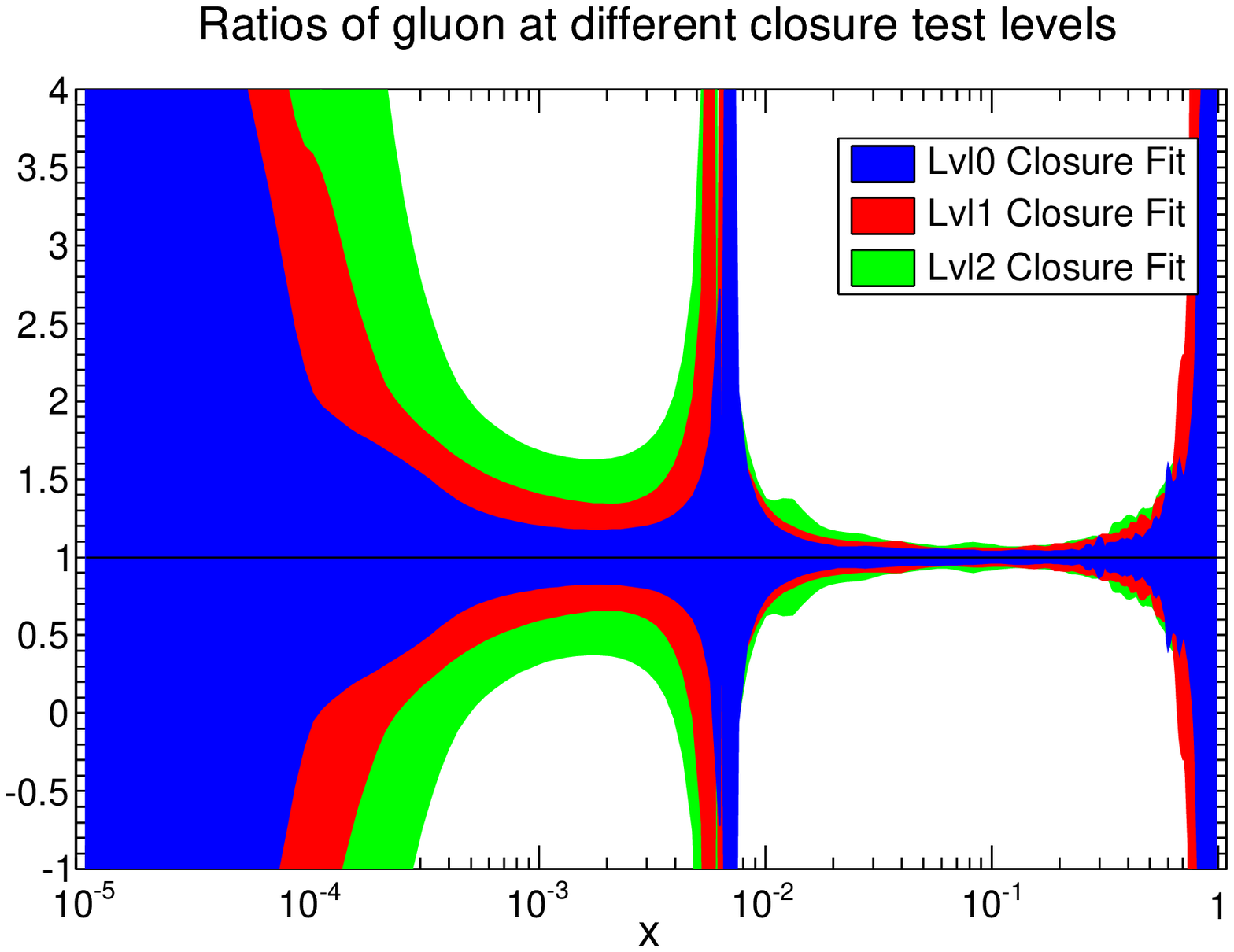}
  \epsfig{width=0.46\textwidth,figure=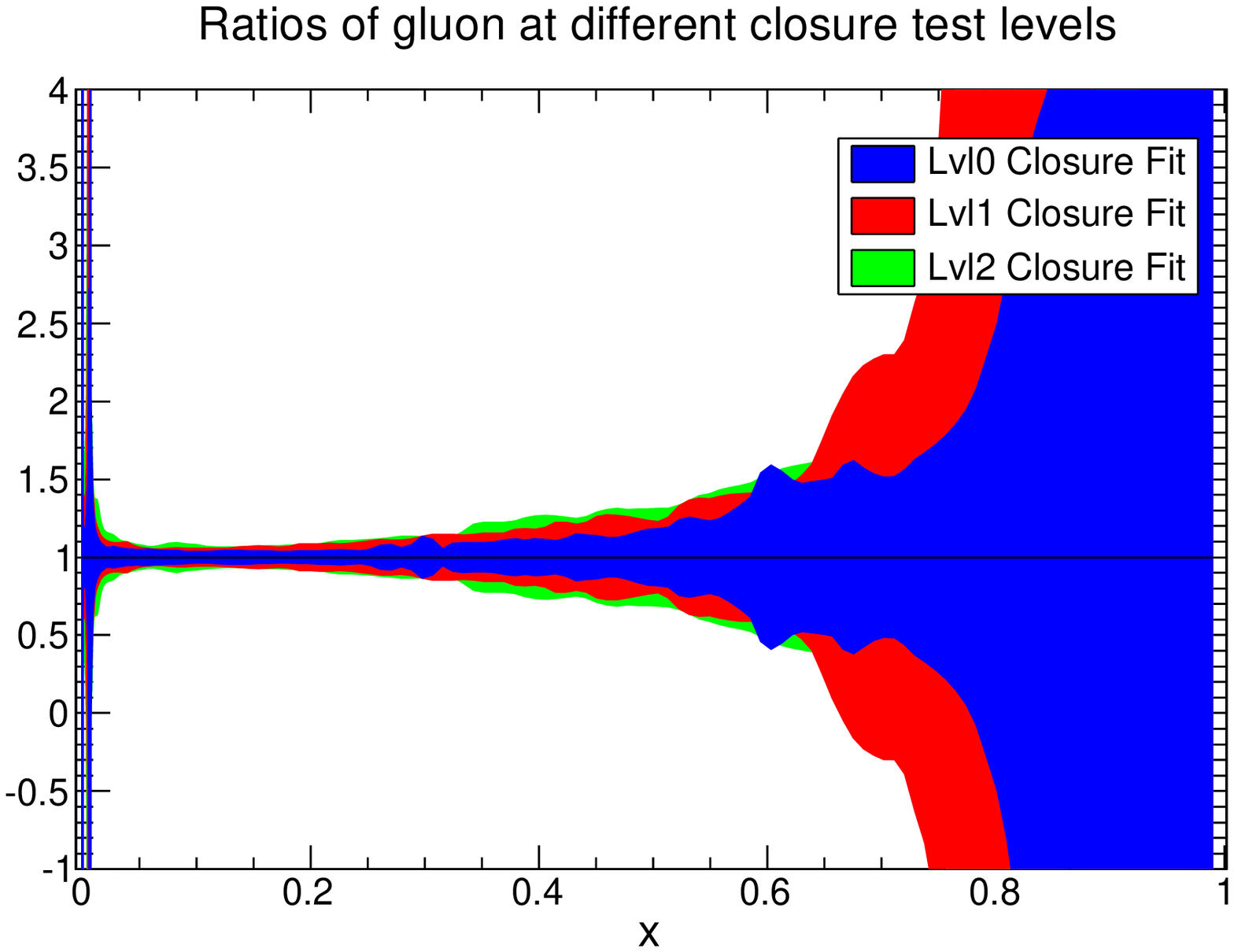}
  \epsfig{width=0.46\textwidth,figure=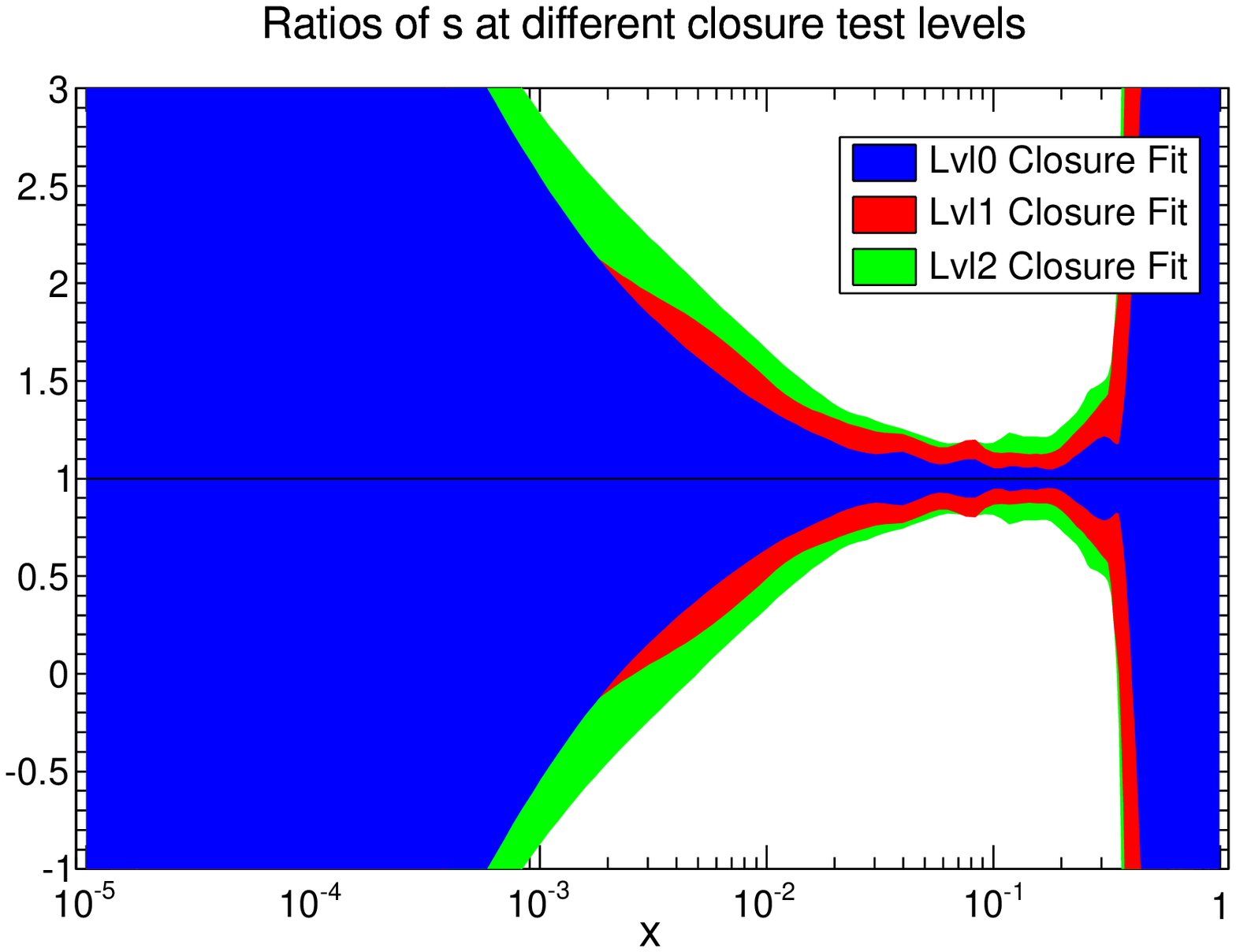}
  \epsfig{width=0.46\textwidth,figure=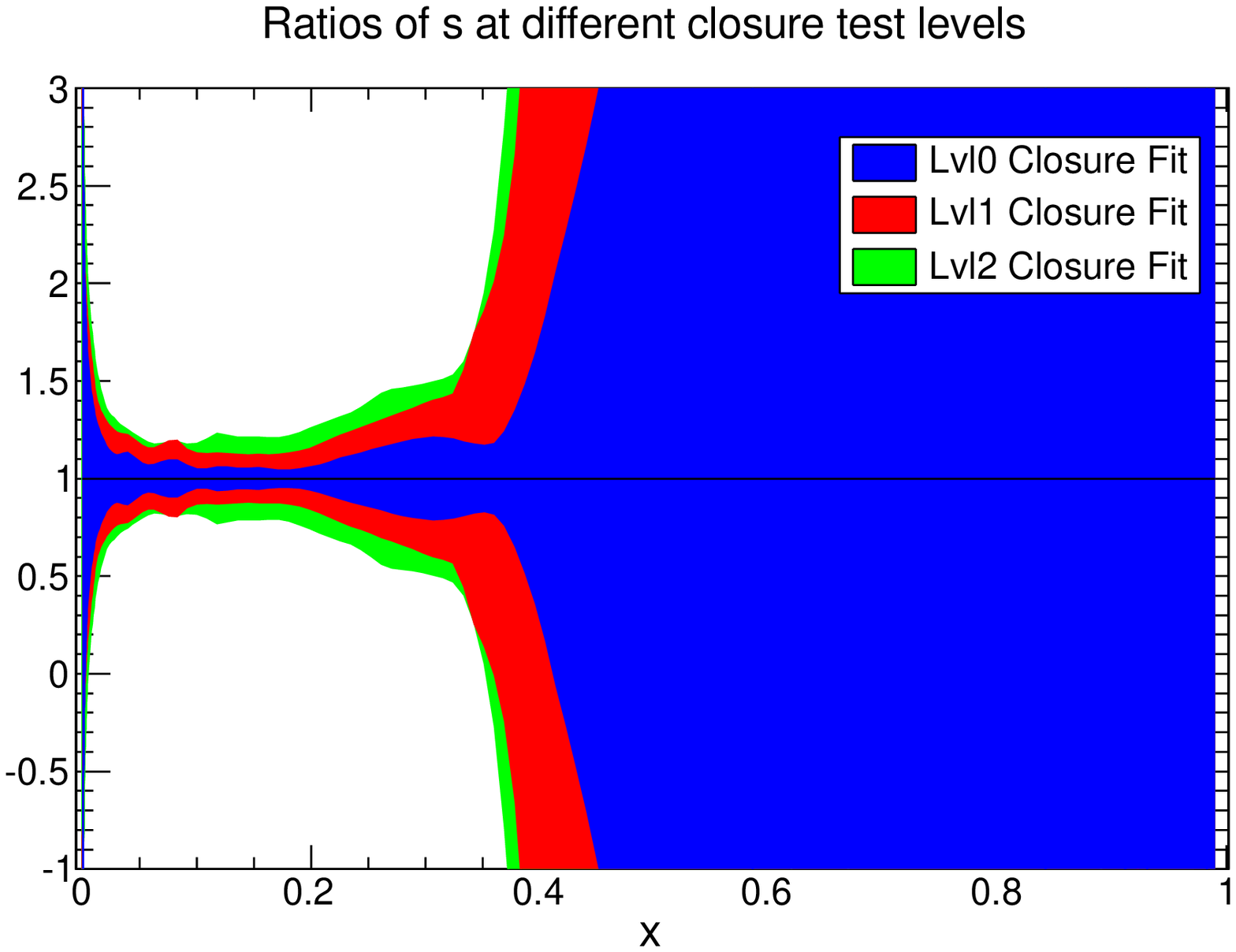}
  \epsfig{width=0.46\textwidth,figure=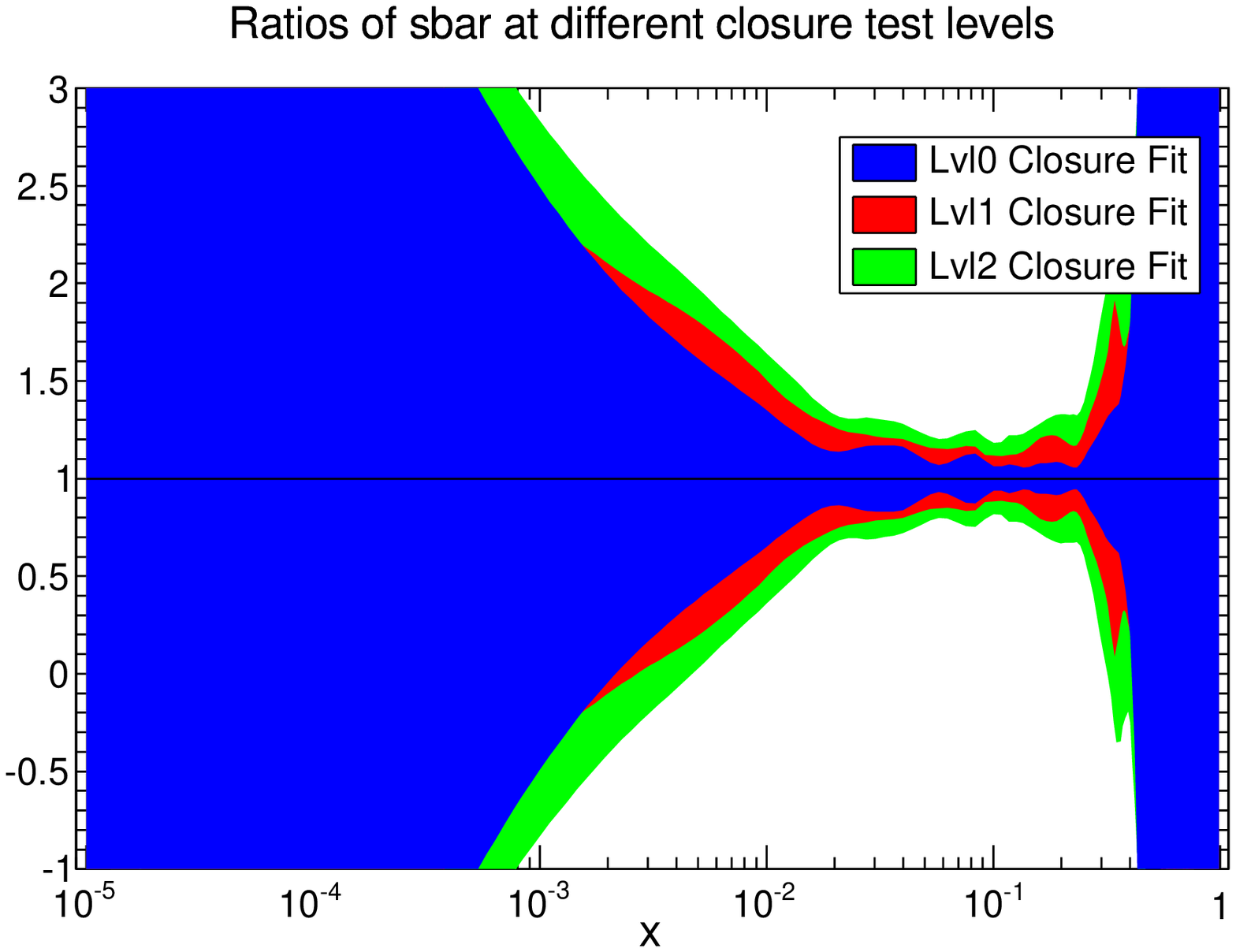}
  \epsfig{width=0.46\textwidth,figure=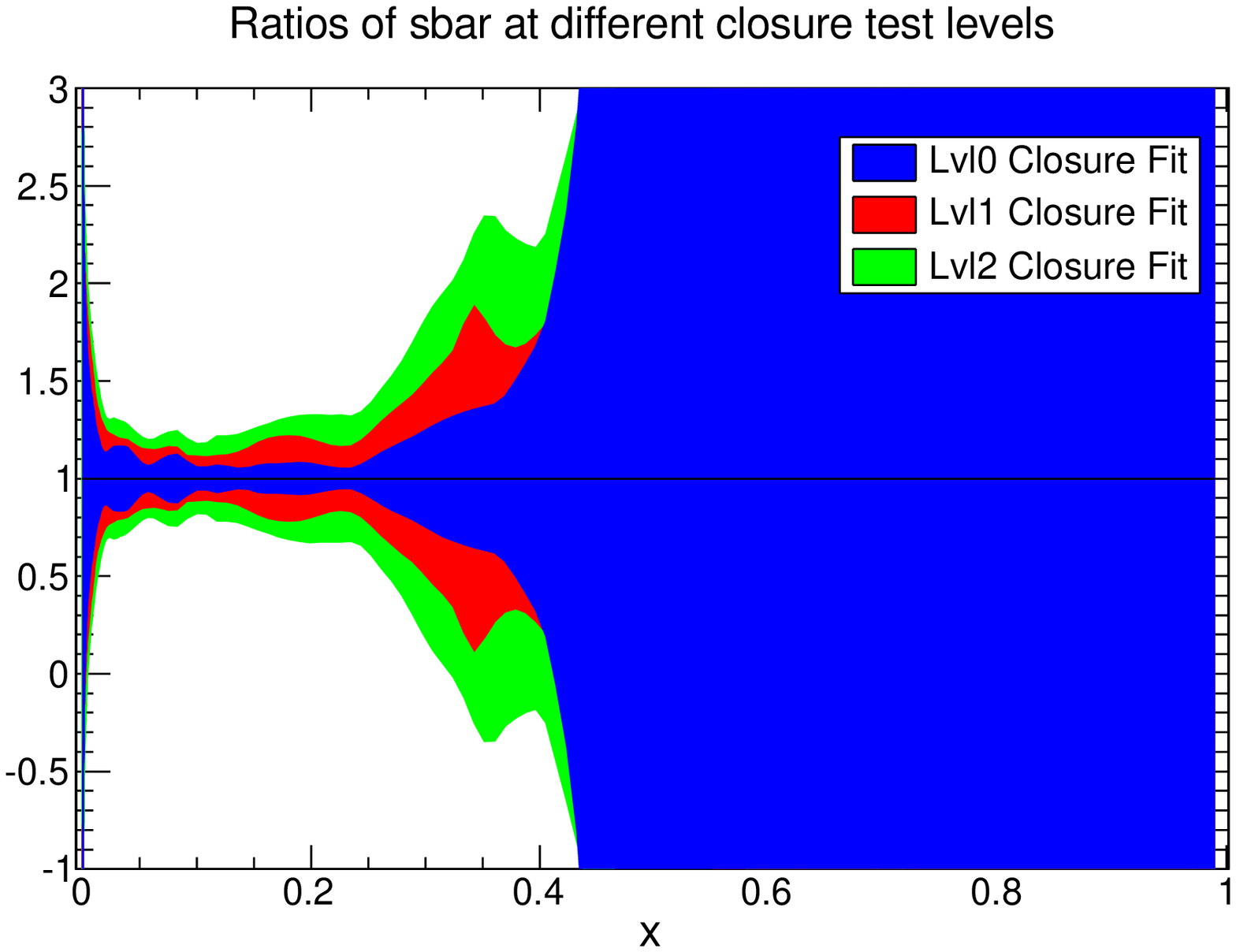}
  \caption{\small}
  \label{fig:ratiofit2}
\end{figure}
\renewcommand{\thefigure}{\arabic{figure}}
\clearpage

\subsubsection{Quantitative validation of PDF uncertainties in closure tests}
\label{sec:closureestimators}

In the previous section we discussed the various components
of the total PDF uncertainty in a purely qualitative way,
by comparing the results of Level~0, Level~1 and Level~2
closure tests.
In this section we provide some quantitative estimators of the
successfulness and effectiveness of the closure tests, which will be
used in Sect.~\ref{sec:ctresults} to validate the closure test fits.

In Level~1 and Level~2 closure fits, we expect that the
central $\chi^2$ obtained
from the average of the fitted PDFs should reproduce the one computed using
the input PDFs, that is $\chi^2[\langle\mathcal{T}[f]\rangle,\mathcal{D}_1]
\approx \chi^2[\mathcal{T}[f_{\rm in}],\mathcal{D}_1]$
where by $\mathcal{D}_1$ we indicate that
that we use the Level~1 pseudo-data, which include
the fluctuations.

We can test for a particular fit whether this
is the case by defining the following
statistical estimator
\begin{equation}
\label{eq:dchi2}
\Delta_{\chi^2}= \frac{\chi^2[\langle\mathcal{T}[f]\rangle,\mathcal{D}_1]
- \chi^2[\mathcal{T}[f_{\rm in}],\mathcal{D}_1]}{\chi^2[\mathcal{T}[f_{\rm in}],\mathcal{D}_1]}\, ,
\end{equation}
that is, the difference between the central $\chi^2$ of the closure test
fit, computed with an average over replica PDFs, and the $\chi^2$
of the input PDF set, both  computed
with respect to the same closure test dataset.
This estimator
is therefore a measure of how close the closure test fit reproduces
the theoretical predictions of the input PDF, and therefore
it provides
a quantitative measure of the
success of the test.

In particular, noting that in genetic algorithm minimization $\chi^2$
is a decreasing function along the training,
$\Delta_{\chi^2}>0$ corresponds to
underlearning (the optimal $\chi^2$ has not been reached yet),
$\Delta_{\chi^2}=0$ corresponds to perfect learning of
the underlying law, while  in fixed-length genetic minimization
 $\Delta_{\chi^2}<0$ would correspond to overlearning, i.e., the fit
is learning the noise in the data. Of course in a closure test the
true underlying $\chi^2$ is known and thus one could stop at the
optimal point by comparing to it, which of course is not possible in a
realistic situation. The effectiveness of the method for determining the optimal stopping point
that is actually used --- in our case, the look-back method described in
Sect.~\ref{subsec:ga} --- can then be measured by simply evaluating how
close $\Delta_{\chi^2}$ Eq.~(\ref{eq:dchi2}) is to zero
at the actual stopping point.
The estimator Eq.~(\ref{eq:dchi2}) thus quantifies how successful
the closure test fit is in terms of reproducing the
central values of the input PDFs.

We now introduce an estimator which allows for an assessment of the
accuracy with which PDF uncertainties are reproduced. To this
purpose, we first recall that for an unbiased estimator, assuming
gaussianity,
the $n$-sigma intervals about the prediction can be interpreted as
confidence levels for the true value (see e.g. Ref.~\cite{cowan}):
this means that the true value must fall on average within the one sigma
uncertainty band in 68.3\% of cases, within the two-sigma band in 95.5\%
of cases, and so on.

We thus define the estimator
\begin{equation}
  \label{eq:L268CL}
  \xi_{n\sigma} = \frac{1}{N_{\mathrm{PDF}}}\frac{1}{N_{x}}\frac{1}{N_{\mathrm{fits}}}\sum_{i=1}^{N_{\mathrm{PDF}}}\sum_{j=1}^{N_{x}}\sum_{l=1}^{N_{\mathrm{fits}}}
  I_{[-n\sigma^{i(l)}_{\mathrm{fit}}(x_j),n\sigma^{i(l)}_{\mathrm{fit}}(x_j)]}\left( \langle{f}^{i(l)}_{\mathrm{fit}}(x_j)\rangle
    - f_{\mathrm{in}}^{i}(x_j) \right)\, ,
\end{equation}
where $n$ is a positive integer, and
$N_{\mathrm{PDF}}$, $N_{x}$ and $N_{\mathrm{fits}}$ are the number of PDF
flavors, $x$ values and fits respectively, over which averages are performed.
For the sampling of the PDFs in $x$, we use 20 points between
$10^{-5}$ and 1, half of them log spaced below 0.1 and
the rest linearly spaced.
In Eq.~(\ref{eq:L268CL}),
$I_A(x)$ denotes the indicator function of the interval $A$:
it is only non-zero, and actually equal to one, if its argument lies in the interval $A$,
while it vanishes for all other values of its argument.
Finally,
$\langle{f}^{i(l)}_{\mathrm{fit}}\rangle$ and $\sigma^{i(l)}_{\mathrm{fit}}$
are the average PDFs and the corresponding standard deviation of
the $i$ PDF flavor for fit $l$, computed over the sample of
$N_{\rm rep}=100$ replicas of fit $l$.
The estimators $\xi_{1\sigma},\, \xi_{2\sigma},\, \ldots$ provide
the fraction of those fits for which the input
PDF falls within one sigma, two sigma, etc of the central
PDF $\bar{f}_{\rm fit}^{i(l)}$, averaged over PDF flavors and values of $x$.
In a successful closure test we should thus find that
$\xi_{1\sigma}\approx0.68$, $\xi_{2\sigma}\approx0.95$, etc.

In principle, in order to calculate $\xi_{1\sigma}$, Eq.~(\ref{eq:L268CL}),
 we would need to, for instance, generate 100 closure
test fits each one with
$N_{\rm rep}=100$ replicas, following
the procedure explained
above.
However performing this very large number of fits
is very computationally expensive, and
we would instead like to obtain an estimate
of $\xi_{1\sigma}$ which involves fewer fits.
 To achieve this, we can approximate the
mean of each fit, $\langle{f}_{\rm fit}^{i(l)}\rangle$, by
fitting a single replica to each set of closure test
data at Level~1, i.e.\ without additional replica fluctuations.
We can also replace the individual values of $\sigma^{i(l)}$ in
Eq.~(\ref{eq:L268CL})
with the corresponding values taken from a single 100 replica fit,
making use of the
fact that the variation in the PDF uncertainties
between different closure test fits is small.

\subsection{Validation of the closure test fits}
\label{sec:ctresults}

Using the statistical estimators
introduced in the previous section, we now move to validate
quantitatively the results of the closure fits.
First,  we show how close the central values and $\chi^2$ values
of the input and fitted PDFs are to each other,
both for the total dataset and for
individual experiments.
Then we discuss a quantitative validation of
the PDF uncertainties obtained in the closure tests,
using the estimators defined in Sect.~\ref{sec:closureestimators} for
this purpose.
Finally in this section we will show how one can also use the Bayesian
reweighting procedure~\cite{Ball:2010gb,Ball:2011gg} to provide further
evidence that the closure tests
are working as expected.
Indeed, it turns out that reweighting provides the most stringent validation
test, since for it to be successful it is necessary to reproduce
not only central values and uncertainties, but also
higher moments and correlations.

\subsubsection{Central values}
\label{subsec:closureCV}

A first indicator of the quality of a closure test is provided by the values of the central $\chi^2$
obtained using the fit, $\chi^2[\langle\mathcal{T}[f_\mathrm{fit}]\rangle,
\mathcal{D}_1]$.
This should reproduce the values of
the $\chi^2$ obtained using the generating PDFs, that is
$\chi^2[\mathcal{T}[f_\mathrm{in}],
\mathcal{D}_1]$.
For our baseline Level~2 closure test, using pseudo-data based on MSTW08, and
performed with the look-back stopping criterion rather than at fixed length
(fit C9 in Table~\ref{tab:CTfits}), we obtain the following
result for Eq.~(\ref{eq:dchi2}),
\be
\Delta_{\chi^2} = -0.011 \, ,
\ee
which shows that the fitted PDFs reproduce the $\chi^2$ of the generating
PDFs at the 1\% level, with a small amount of overlearning, which must
be viewed as an inefficiency of the look-back method.
The result for the corresponding Level~1 fit (C8) is very similar: $\Delta_{\chi^2} = -0.015$.

\begin{figure}[t]
  \centering
  \epsfig{width=0.90\textwidth,figure=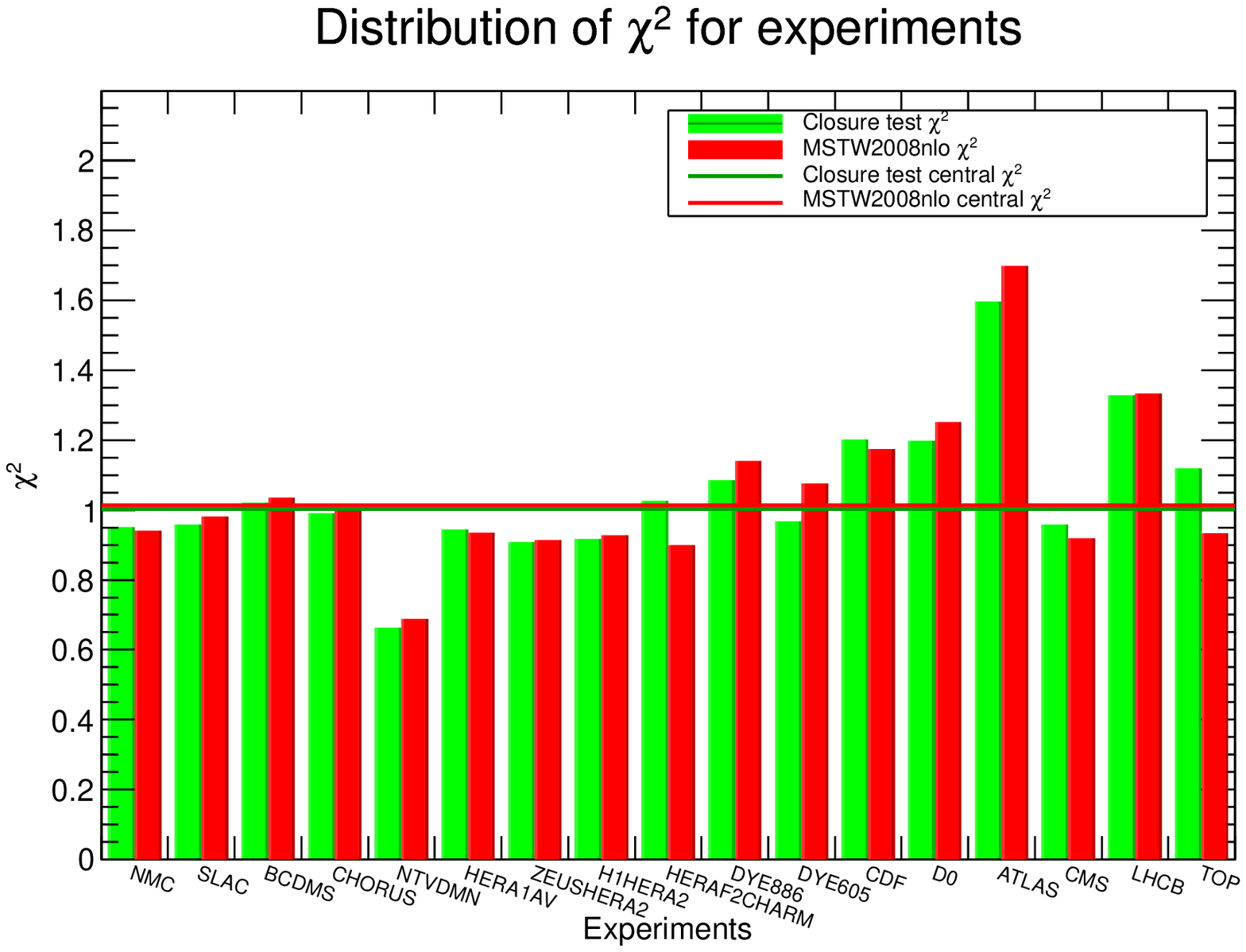}
  \caption{\small Comparison of the $\chi^2$ to the closure test data,
 $\chi^2[\mathcal{D}_1]$, obtained with the input
 (red) and with the fitted (green)
  PDFs, for the Level~2 closure test fit
 based on
MSTW08 pseudo-data,
fit C9 in Table~\ref{tab:CTfits}.
The horizontal bars show the total
 $\chi^2$ for the two PDF sets, and are computed as
the weighted average of the $\chi^2$ of each individual
dataset.
The datasets shown are the same used in the baseline
NNPDF3.0 global fit, see Tables~\ref{tab:completedataset} and
  \ref{tab:completedataset2} in Sect.~\ref{sec:expdata}.}
  \label{fig:chi2-L2-expt}
\end{figure}

This level of agreement is achieved not only for the total
$\chi^2$, but also for the individual experiments included
in the global fit.
This is important to test, since each experiment fluctuates by a different
amount, and we want the closure test to reproduce these fluctuations.
Figure~\ref{fig:chi2-L2-expt} shows the contributions in the Level~2 fit C9 to
$\chi^2[\langle\mathcal{T}[f_{\rm fit}]\rangle,\mathcal{D}_1]$ from
the pseudo-data generated for each individual experiment, compared to
the corresponding contributions to
$\chi^2[\mathcal{T}[f_{\rm in}],\mathcal{D}_1]$.
The horizontal bars show the total
 $\chi^2$ for the two PDF sets, and are computed as
the weighted average of the $\chi^2$ of each individual
dataset.
The datasets shown are the same used in the baseline
NNPDF3.0 global fit (see Tables~\ref{tab:completedataset} and
  \ref{tab:completedataset2} in Sect.~\ref{sec:expdata}).

We see from Fig.~\ref{fig:chi2-L2-expt} that the closure test
successfully reproduces the $\chi^2$ of the input PDFs, MSTW08,
not only for the total dataset but also experiment by experiment.
Note that especially for experiments with only a small number of points
the fluctuations added to the pseudo-data can lead to a
$\chi^2$ quite different from one, and this is exactly reproduced
by the closure test results.
Figure~\ref{fig:chi2-L2-expt} is thus a strong check that,
at least at the level of central values, the Level~2 closure fits
are successful.

\subsubsection{PDF uncertainties: qualitative validation}
\label{subsec:qual}

We can get a first assessment of the uncertainty of the
Level~2 closure fit by means  of the estimator $\varphi_{\chi^2}$
Eq.~(\ref{eq:frdbdef}).
For  the  Level~2 closure fit C9 we get
\be
\varphi_{\chi^2}=0.254 \, ,
\ee
a result that indicates that the PDF uncertainty
on the data points from the Level~2 closure fits
is roughly a factor of four smaller than that the original
experimental uncertainty, due to the combination of all the data
into the PDFs. For the Level~1 fit C8 this number is rather smaller:
$\varphi_{\chi^2}=0.173$, confirming that at Level~1 the data uncertainty is
still missing.
Since contributions to $\varphi_{\chi^2}$ are entirely
from the data points, there is no extrapolation uncertainty; however,
the
Level~0 uncertainty, though small, is not exactly zero because this
would require infinite training length. From
Fig.~\ref{fig:level0_fchi2} we may read off that the Level~0
contribution to  $\varphi_{\chi^2}$ is of order of  $\varphi_{\chi^2}\approx 0.09$.
Hence, combining uncertainties in quadrature, we get that the
contribution of the functional uncertainty to $\varphi_{\chi^2}$
is about $0.15$, and the data uncertainty contribution
is about $0.21$, so,  at least on average, the data and
functional uncertainties are of comparable size.

Recalling that $\varphi_{\chi^2}$ is essentially the ratio of the
uncertainty of the fitted PDFs at the data points to that of the
original data, we  conclude that the fitting procedure leads to an
error reduction by a factor of four or so.
However, we must now verify that this error reduction is real, namely
that uncertainties are correctly estimated by the closure test. We do
this first in a more qualitative way, and then more quantitatively
in Sects.~\ref{subsec:quant} and \ref{sec:bayes}  below.

To this purpose, we first look at the distance between the fitted PDFs
and the underlying ``truth'' in units of the standard deviation.
This is most efficiently done  using the
distance estimators, as defined in
Appendix~\ref{app:distances}, between the closure test
fit PDFs and the input MSTW08 PDFs.
The distances between fitted and input PDFs in
closure tests are normalized such that $d(x,Q)\sim 1$ corresponds
to agreement at the one-sigma level (in units of the
uncertainties of the closure test fitted PDF), $d(x,Q)\sim 2$ corresponds
to agreement at the two-sigma level and so on, that is,
\begin{equation}
\label{eq:distancesclosure2}
  d_{\sigma}\left[f_{i,\mathrm{fit}},f_{i,\mathrm{in}}\right](x,Q) \equiv
  \sqrt{\frac{\left(\bar{f}_{i,\mathrm{fit}}(x,Q) - f_{i,\mathrm{in}}(x,Q) \right)^2}
  {\sigma^2 \left[ f_{i,\mathrm{fit}} \right](x,Q)}}\, ,
\end{equation}
where $i$ stands for the PDF flavor.
In the following, distances are computed at the initial parametrization scale
of $Q^2=1~\GeV^2$.

\begin{figure}[h!]
  \centering
  \epsfig{width=0.99\textwidth,figure=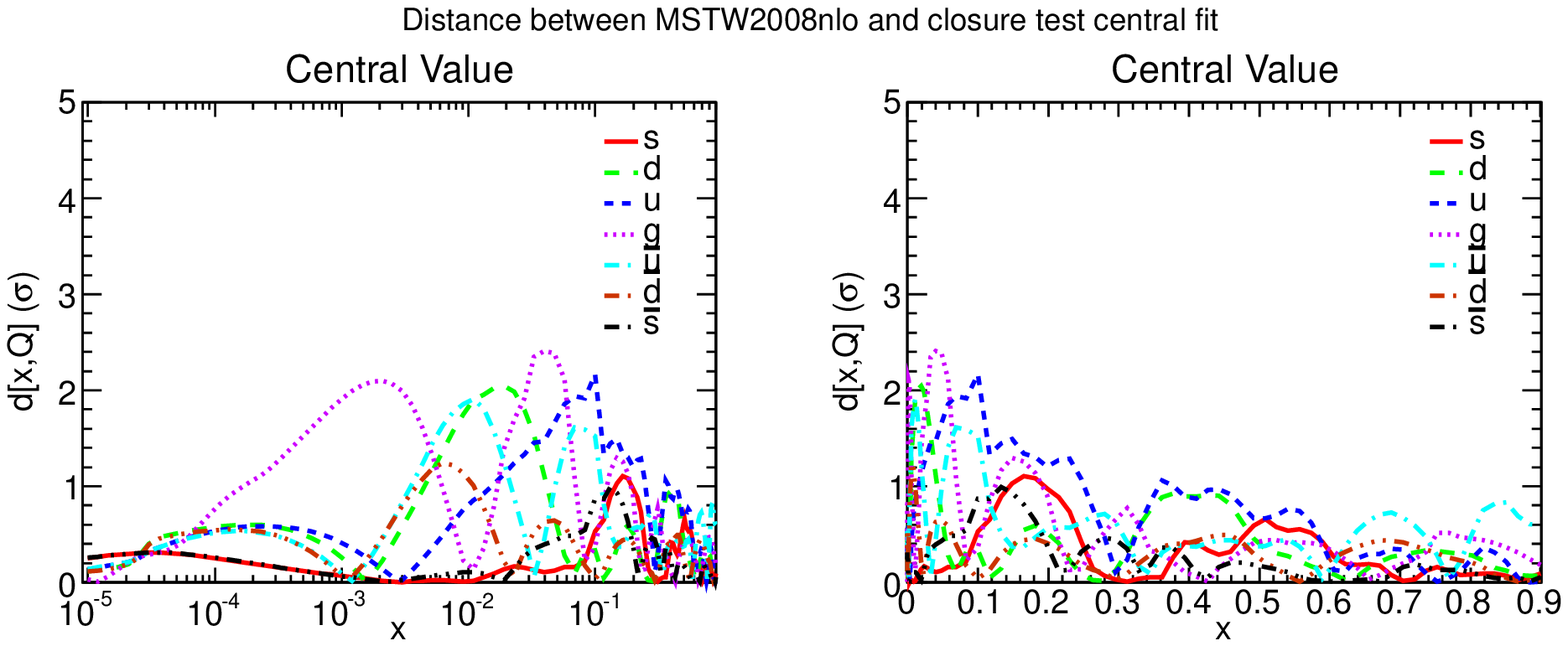}
  \caption{\small Distances,
Eq.~(\ref{eq:distancesclosure2}),
 between the central values of the fitted PDFs from the
Level~2 closure test C9 and the MSTW08 PDFs, which
are used as input to generate the theory predictions of the closure test.
These distances are computed using Eq.~(\ref{eq:distancesclosure2}), that
is, normalized to the standard deviation of the fitted PDFs.
Distances are computed at the input parametrization scale
of $Q^2=$ 1 GeV$^2$, see text for more details.}
  \label{fig:L2-MSTW-dist}
\end{figure}

Using the definition Eq.~(\ref{eq:distancesclosure2}), the distances between the
central values of the fitted and the input PDFs (MSTW08) in the Level~2
closure test C9 are shown in Fig.~\ref{fig:L2-MSTW-dist}.
The distances in Fig.~\ref{fig:L2-MSTW-dist} show that the fitted
and input PDFs are in good agreement, at the level of
one sigma or better (in units of the
uncertainty of the fitted PDF), with some PDFs for some points in $x$
differing by an amount between one and two sigma, as one would
expect if the underlying distribution was roughly Gaussian.
In the extrapolation regions, at small and large $x$, the distances
between input and fitted PDFs become smaller because of the
large PDF uncertainties in these regions.

\begin{figure}[h!]
  \centering
  \epsfig{width=0.45\textwidth,figure=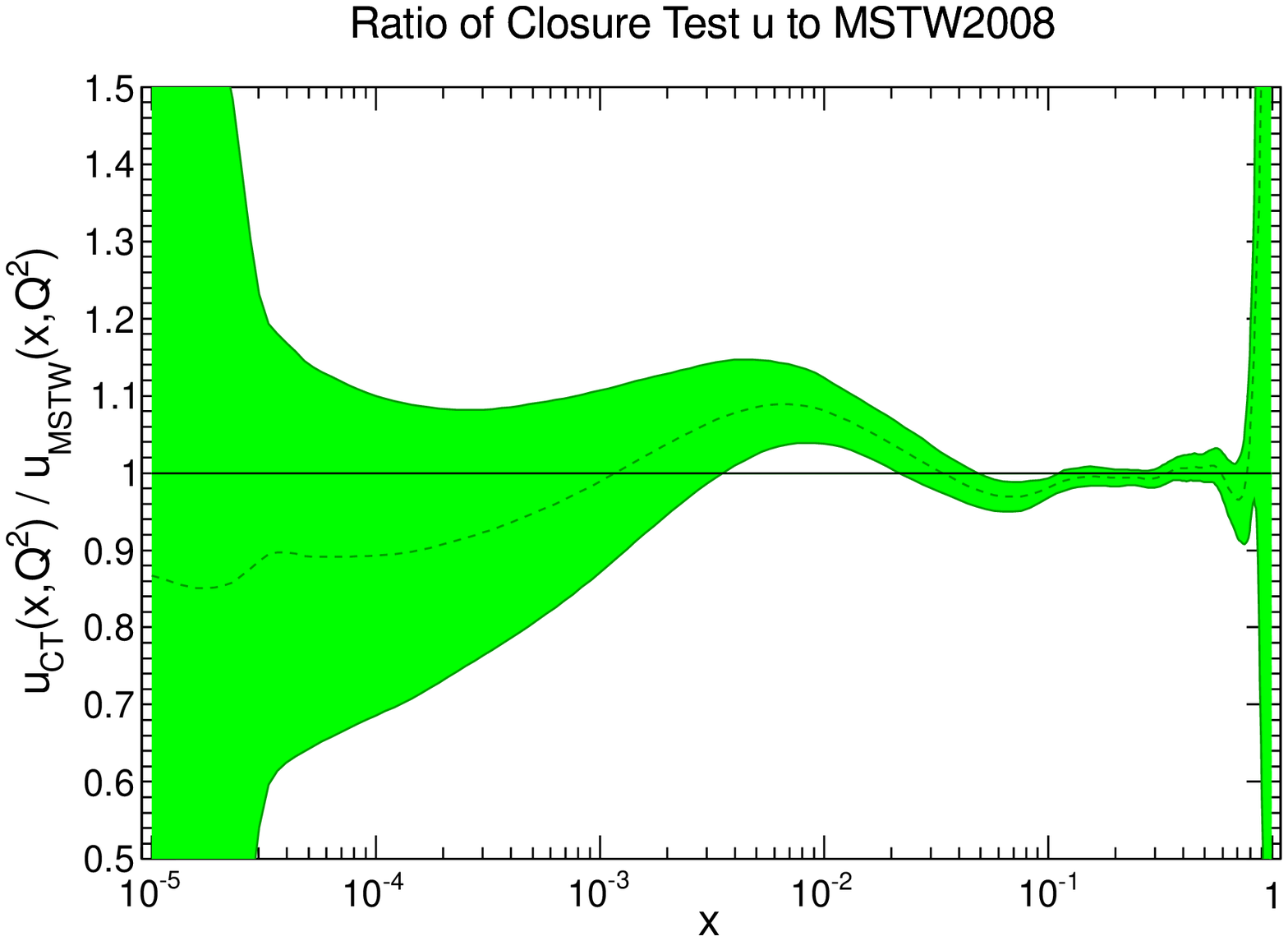}
  \epsfig{width=0.45\textwidth,figure=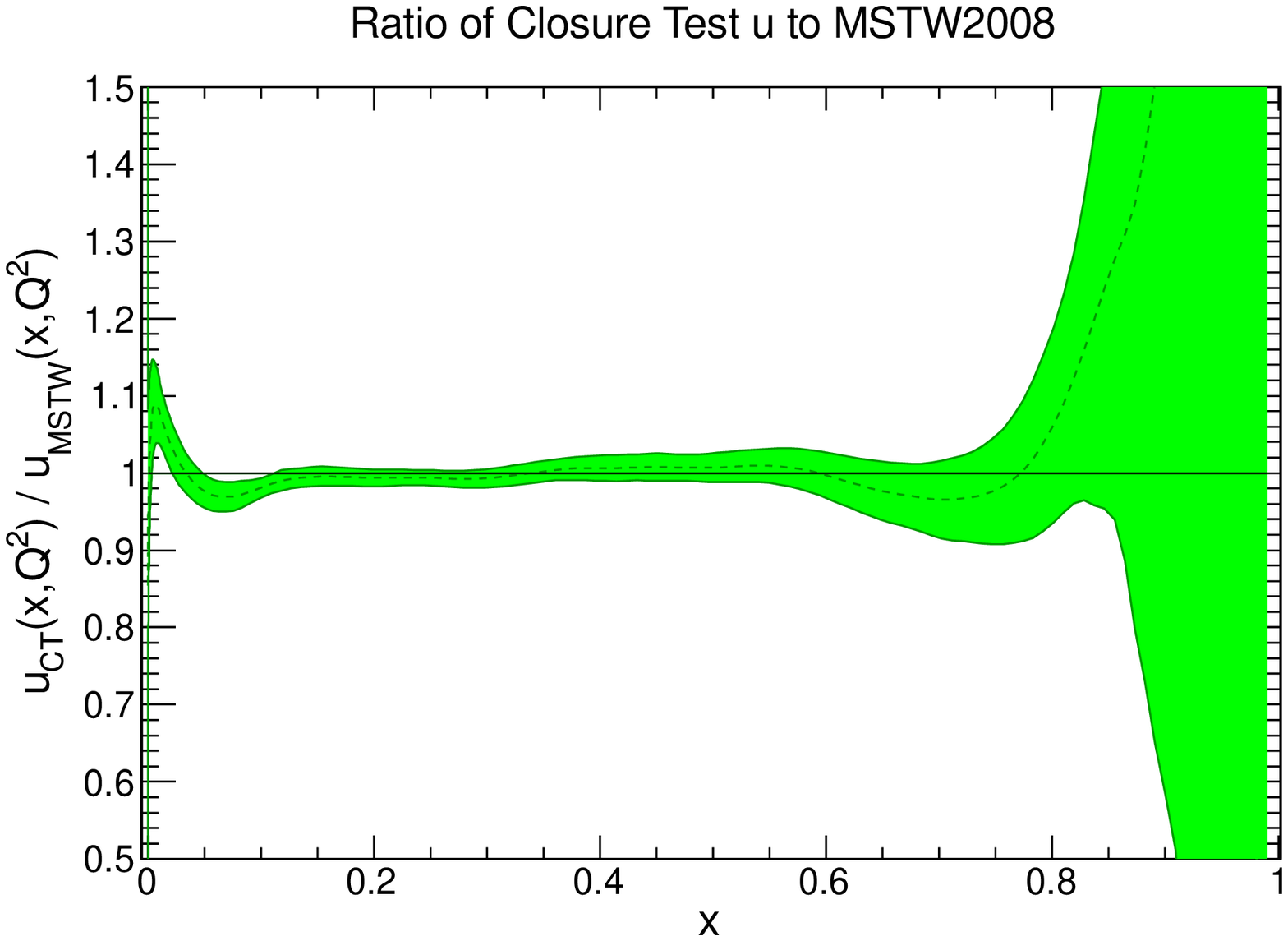}
  \epsfig{width=0.45\textwidth,figure=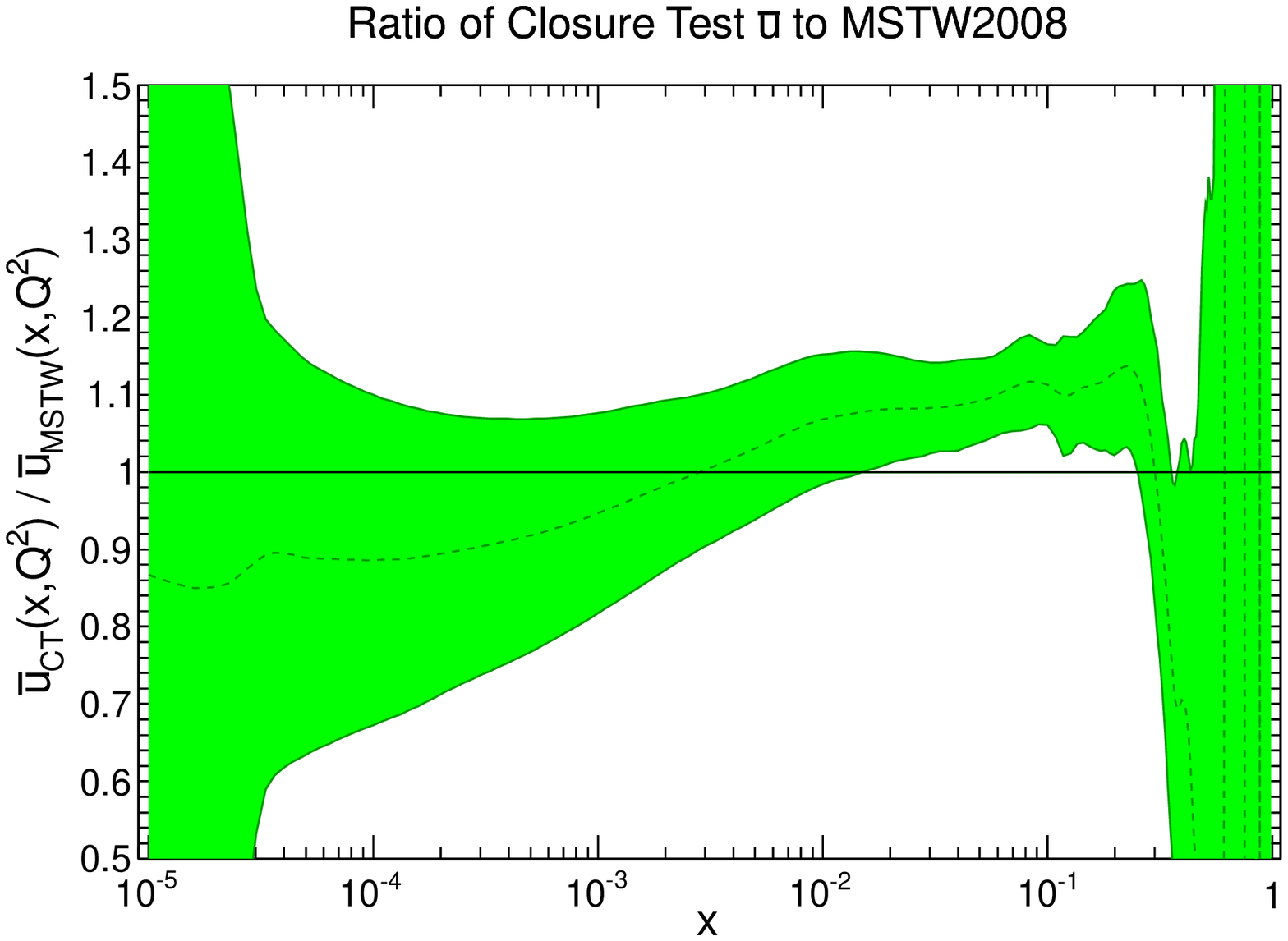}
  \epsfig{width=0.45\textwidth,figure=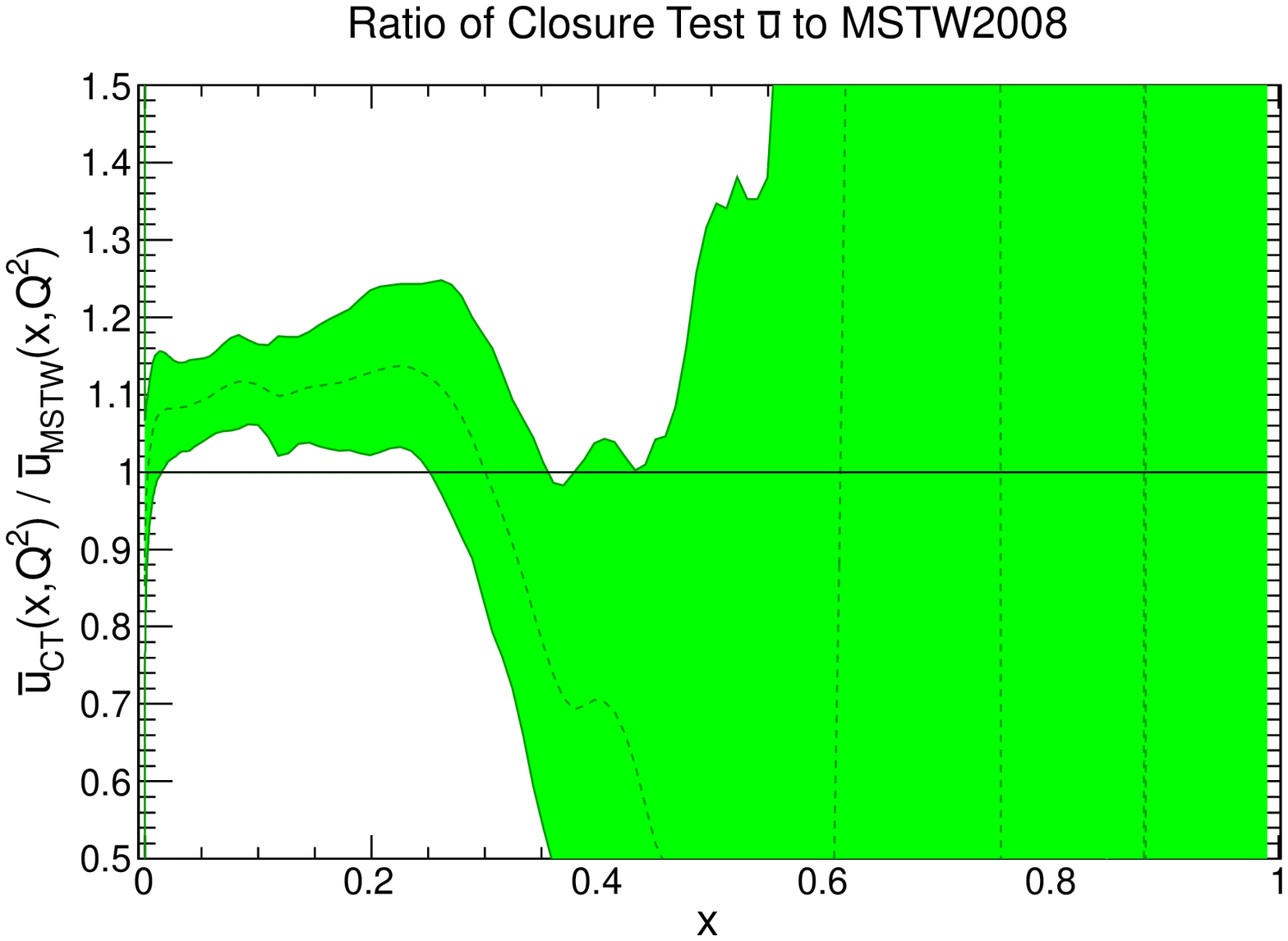}
  \epsfig{width=0.45\textwidth,figure=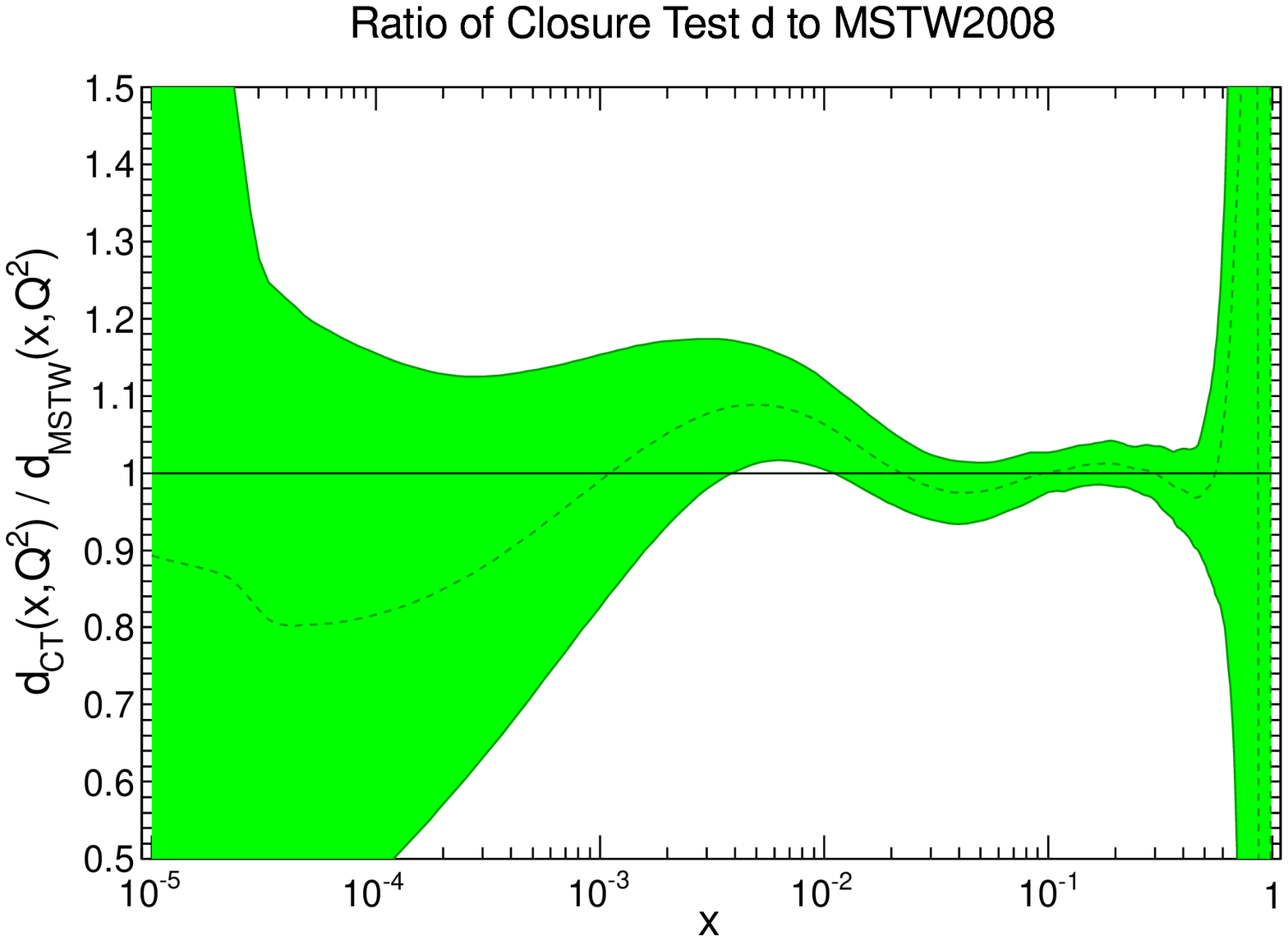}
  \epsfig{width=0.45\textwidth,figure=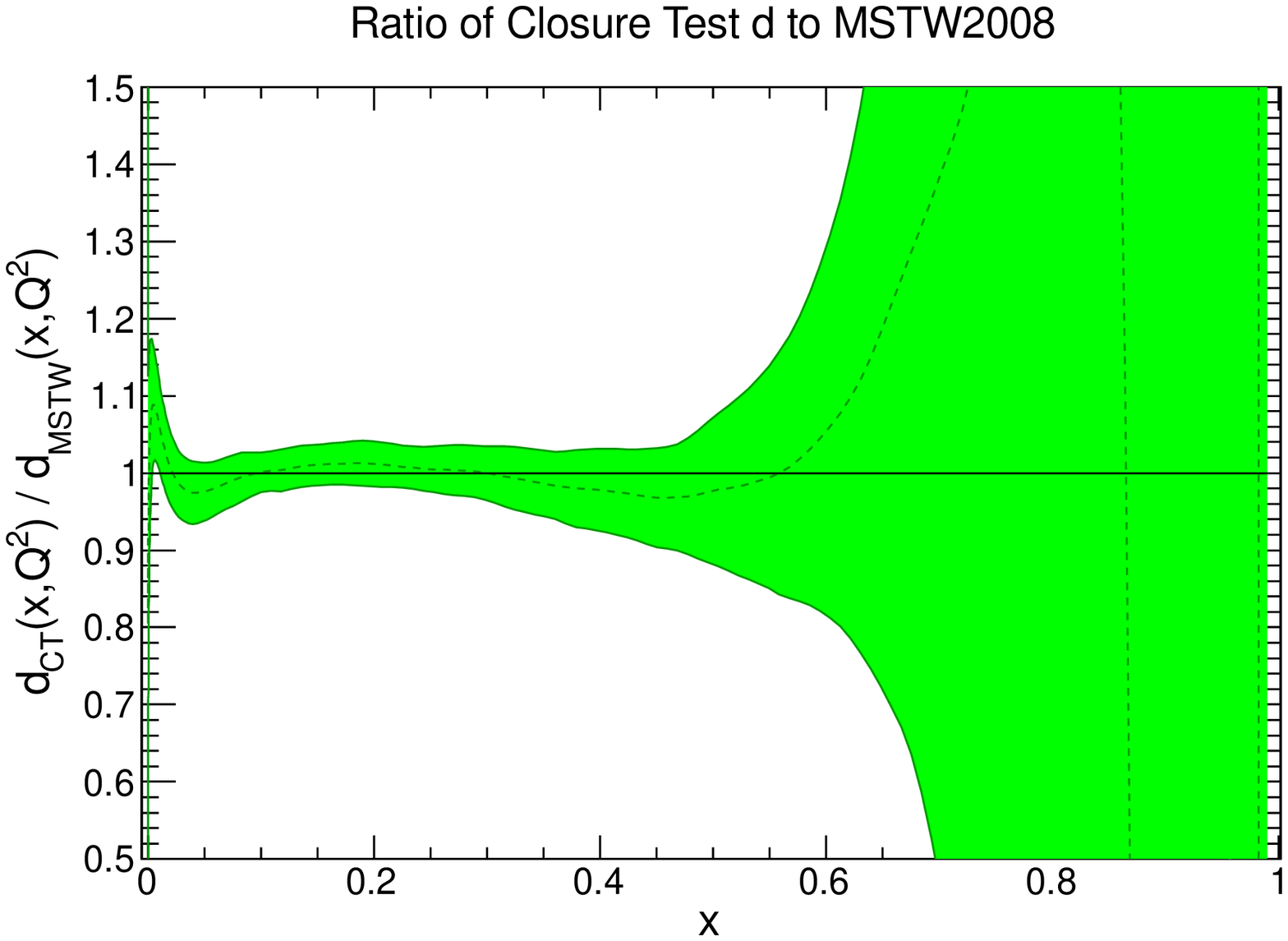}
  \epsfig{width=0.45\textwidth,figure=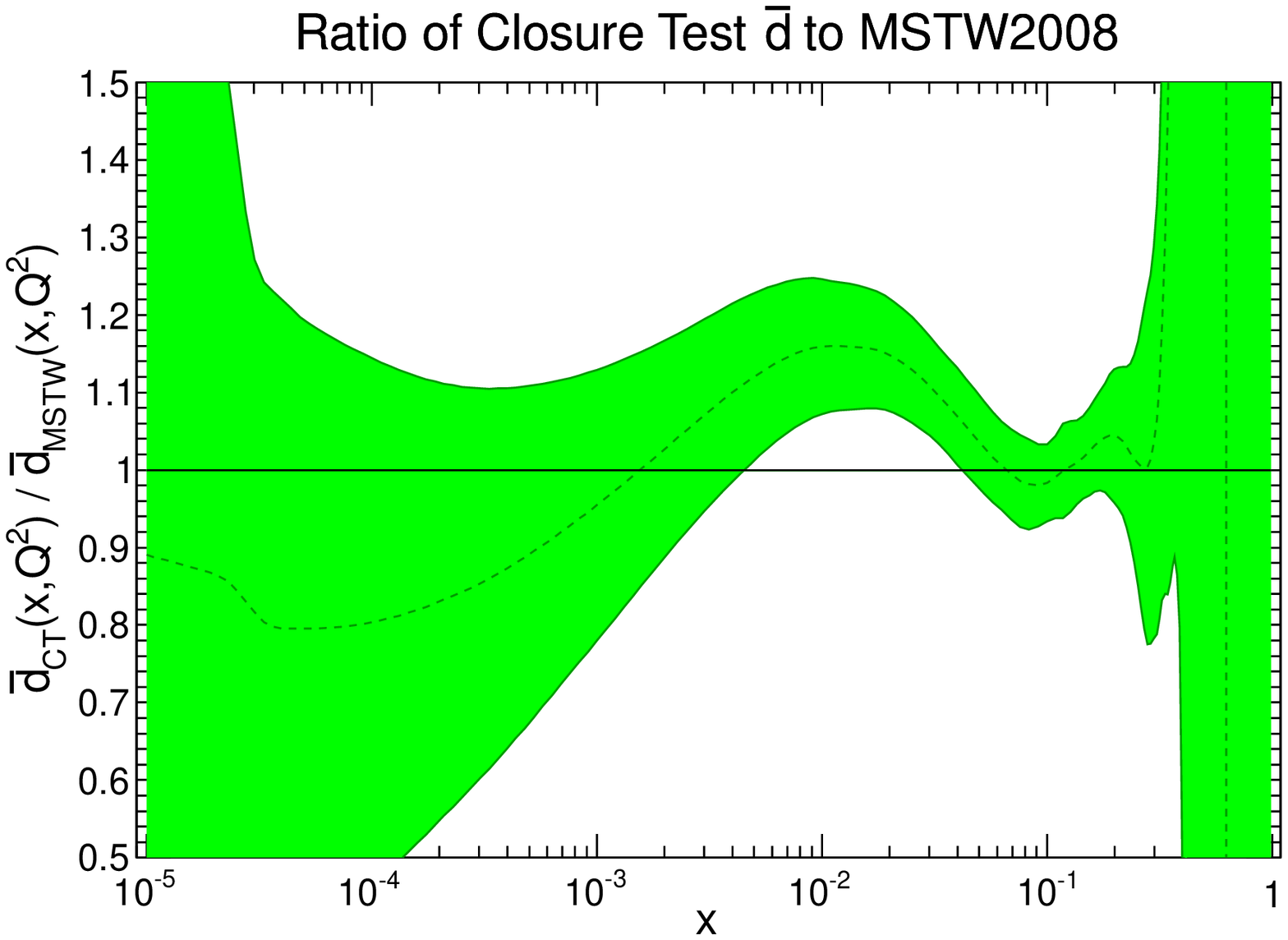}
  \epsfig{width=0.45\textwidth,figure=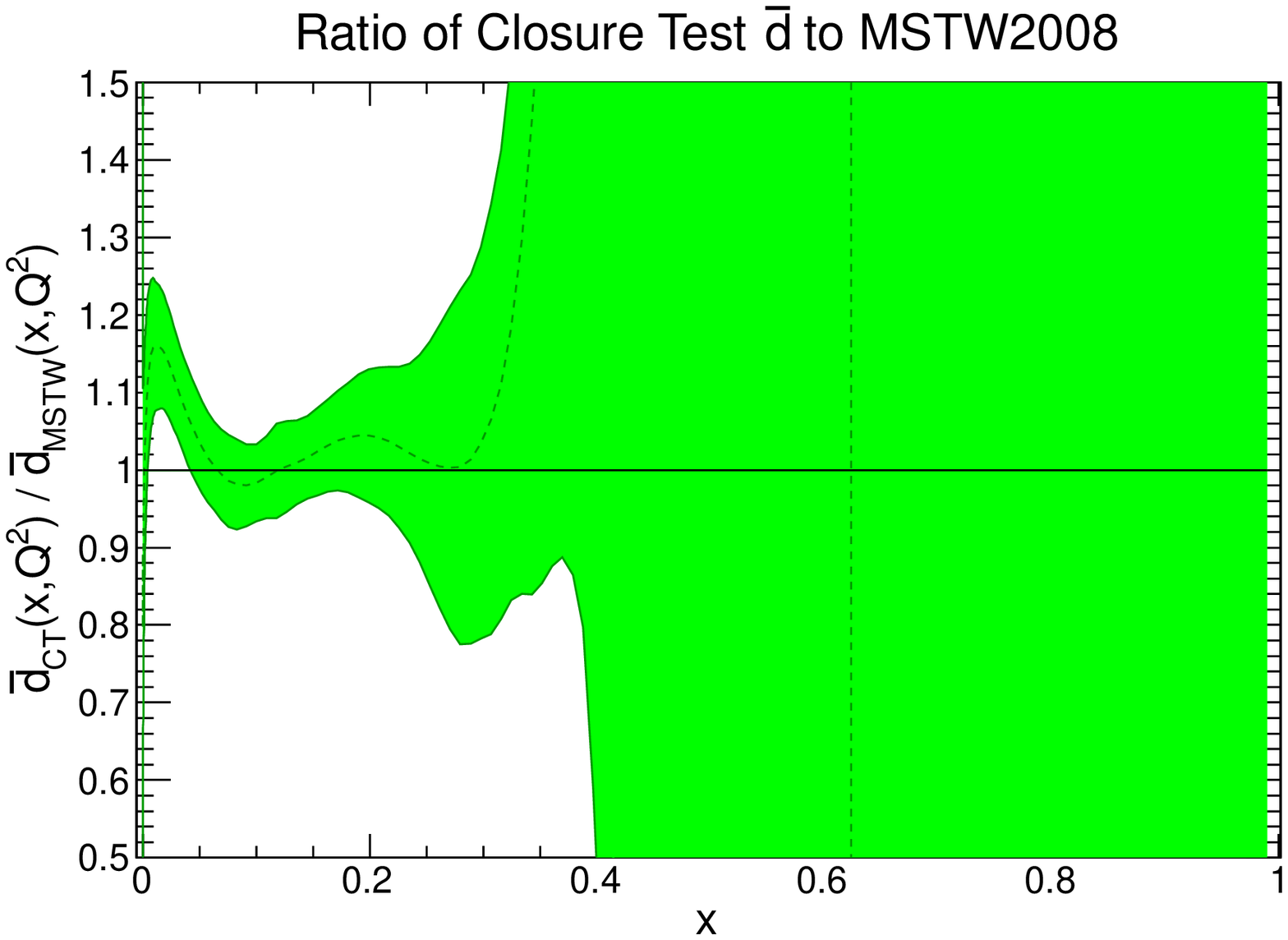}
  \caption{\small Ratio of the PDFs obtained from the Level~2 closure test
C9, which uses MSTW08 as input, with respect to the  input MSTW2008 PDFs
themselves.
The green  band shows the one sigma interval of the
fitted PDFs, while the green dotted line is the
corresponding mean.
  The plots show the up and down PDFs on both linear (right hand side) and
  logarithmic (left) scales in $x$.
The corresponding comparison in terms of distances, Eq.~(\ref{eq:distancesclosure2}),
 is shown in Fig.~\ref{fig:L2-MSTW-dist}.
The comparison is performed at the fitting scale of $Q^2 = 1~\textrm{GeV}^2$.}
  \label{fig:L2-MSTW-ratio1}
\end{figure}

From the distance comparisons of Fig.~\ref{fig:L2-MSTW-dist}
we can see that, at the qualitative level, the closure test is successful
since the fitted PDFs fluctuate around the truth by an amount which
is roughly compatible with statistical expectations.
More insight on this comparison is provided by plotting, for
all PDF flavors, the ratio ${f}_\mathrm{fit} /f_\mathrm{in}$
between the fitted and input PDFs, including both the fitted PDF
central values and uncertainties.
This comparison is shown in Fig.~\ref{fig:L2-MSTW-ratio1} for each parton flavor.
It is clear
from these plots that the NNPDF methodology reproduces successfully the input PDFs, with deviations
from the input functions usually by two standard deviations at most.
This comparison provides
further evidence that PDF uncertainties are properly estimated
in Level~2 closure tests, in that the central value of the fitted
PDFs fluctuates around the truth by an amount which is
consistent with the size of the PDF errors.
%

\renewcommand{\thefigure}{\arabic{figure} (Cont.)}
\addtocounter{figure}{-1}
\begin{figure}[h!]
  \centering
  \epsfig{width=0.45\textwidth,figure=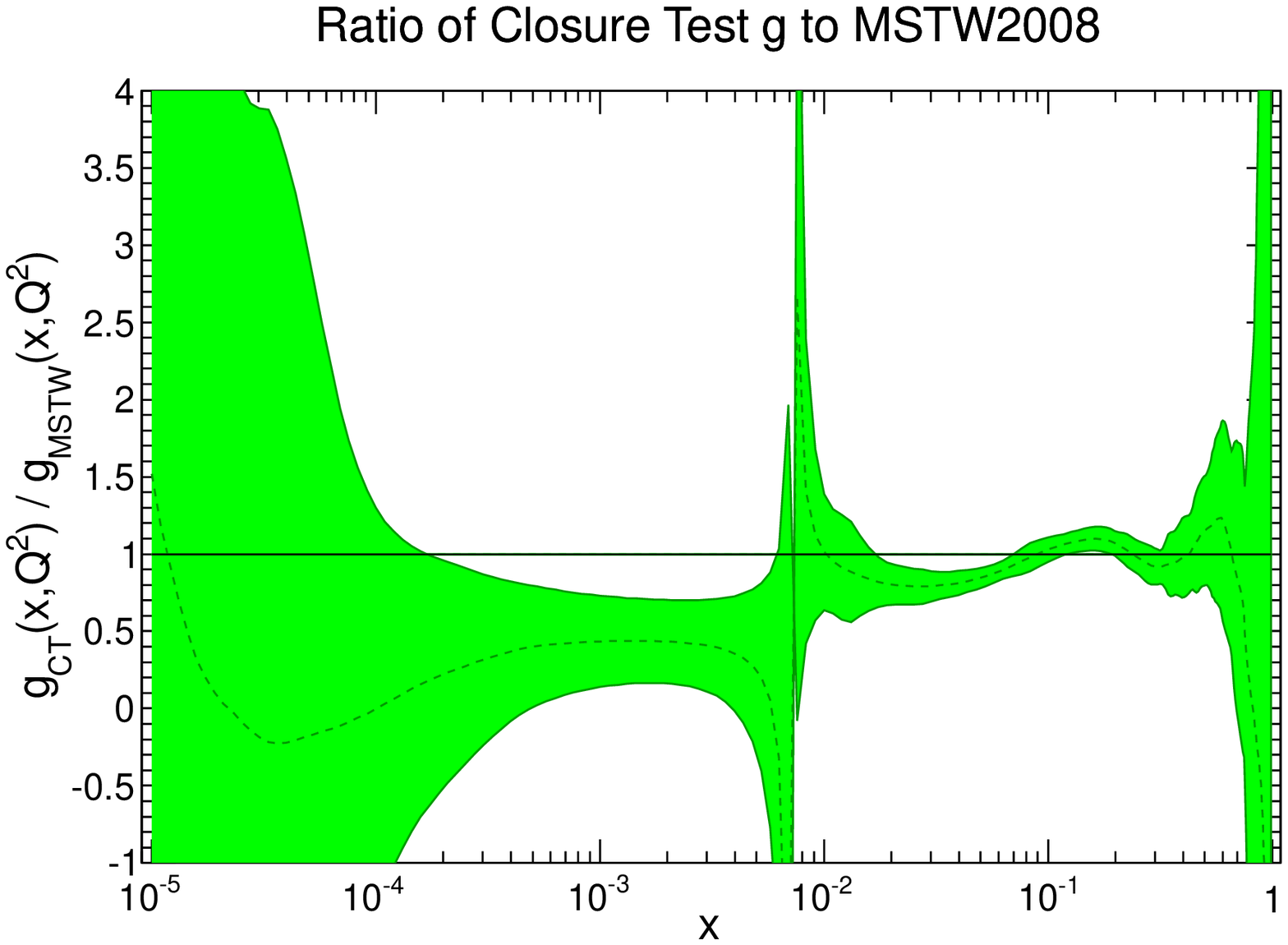}
  \epsfig{width=0.45\textwidth,figure=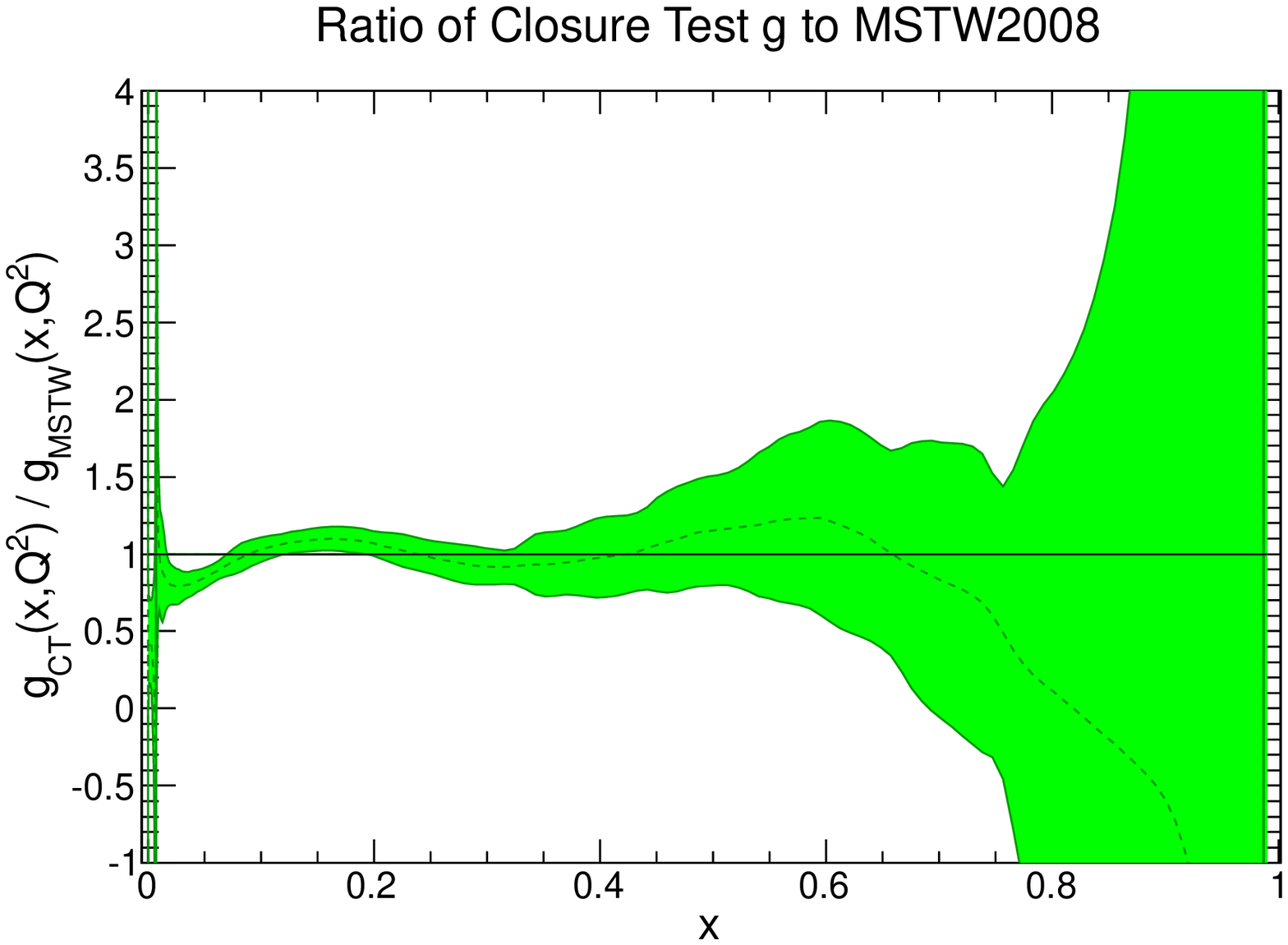}
  \epsfig{width=0.45\textwidth,figure=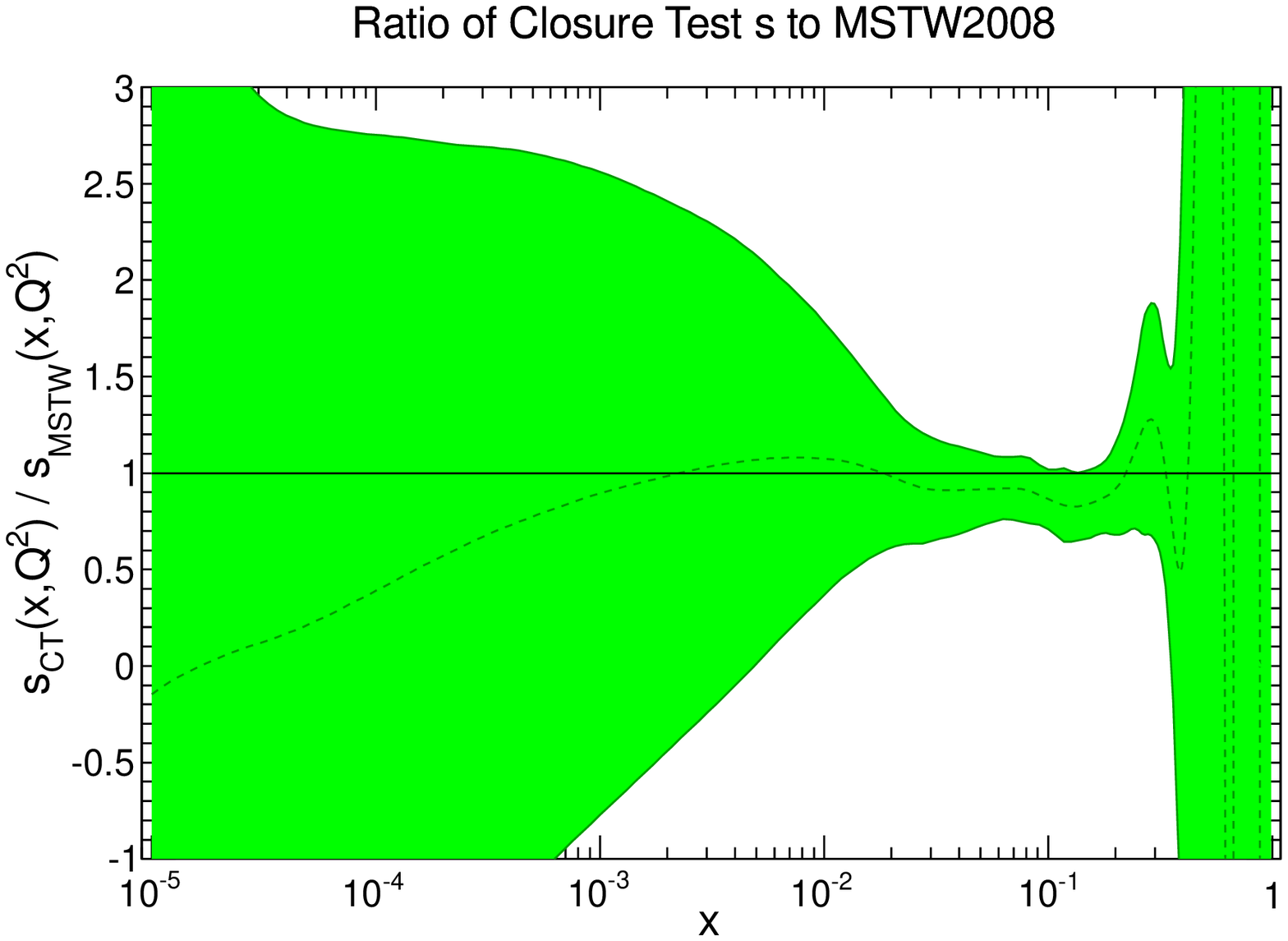}
  \epsfig{width=0.45\textwidth,figure=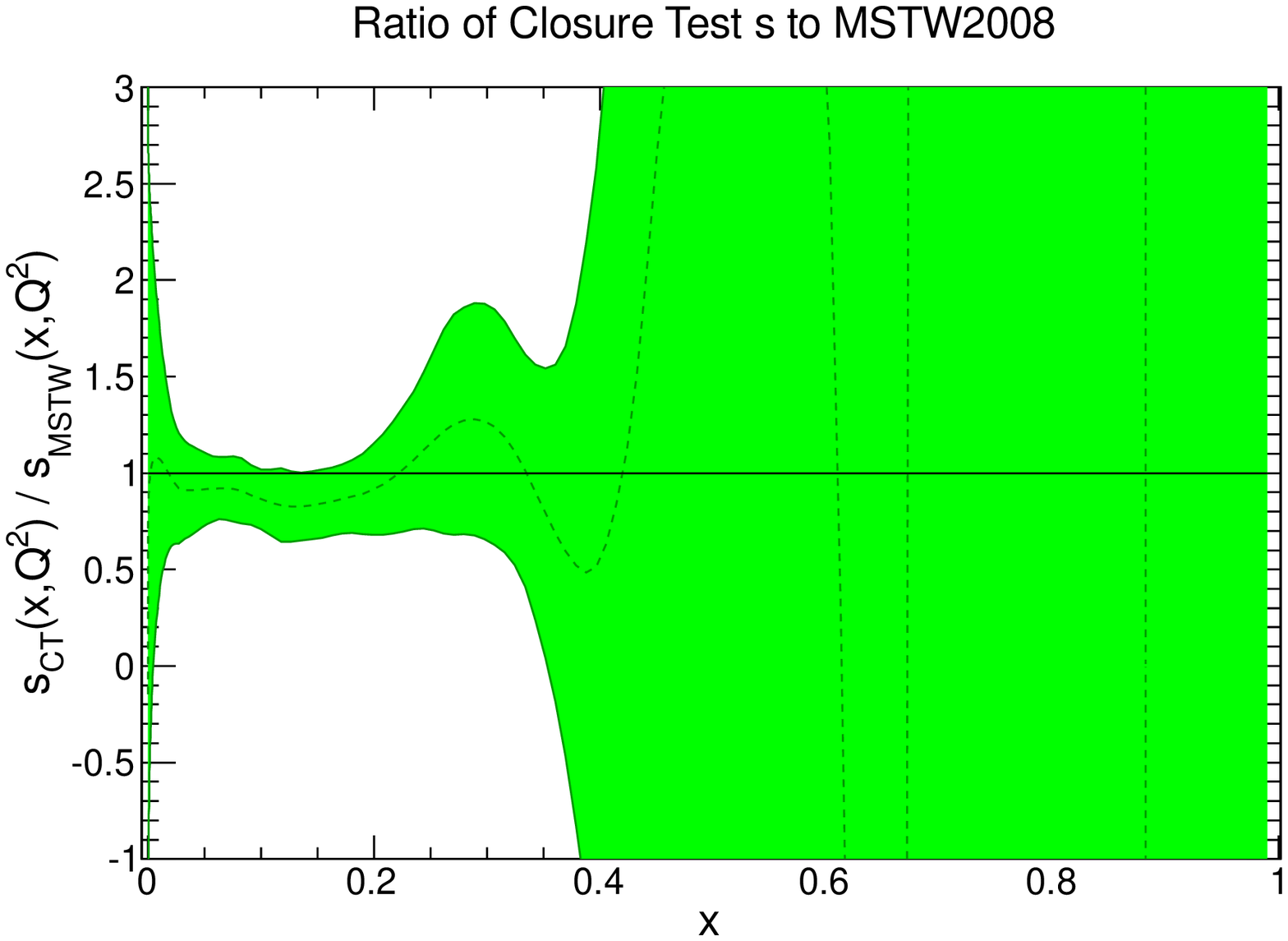}
  \epsfig{width=0.45\textwidth,figure=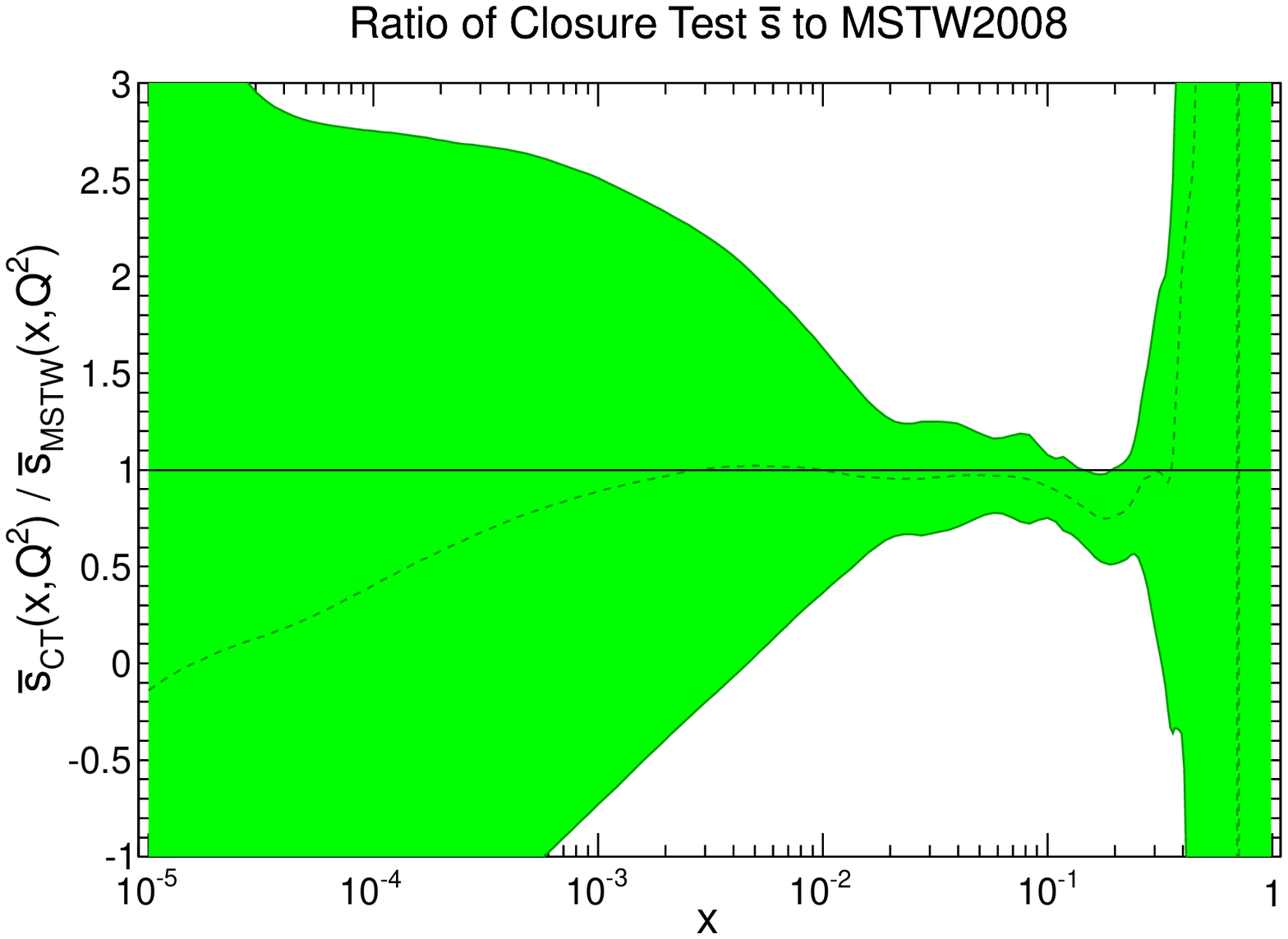}
  \epsfig{width=0.45\textwidth,figure=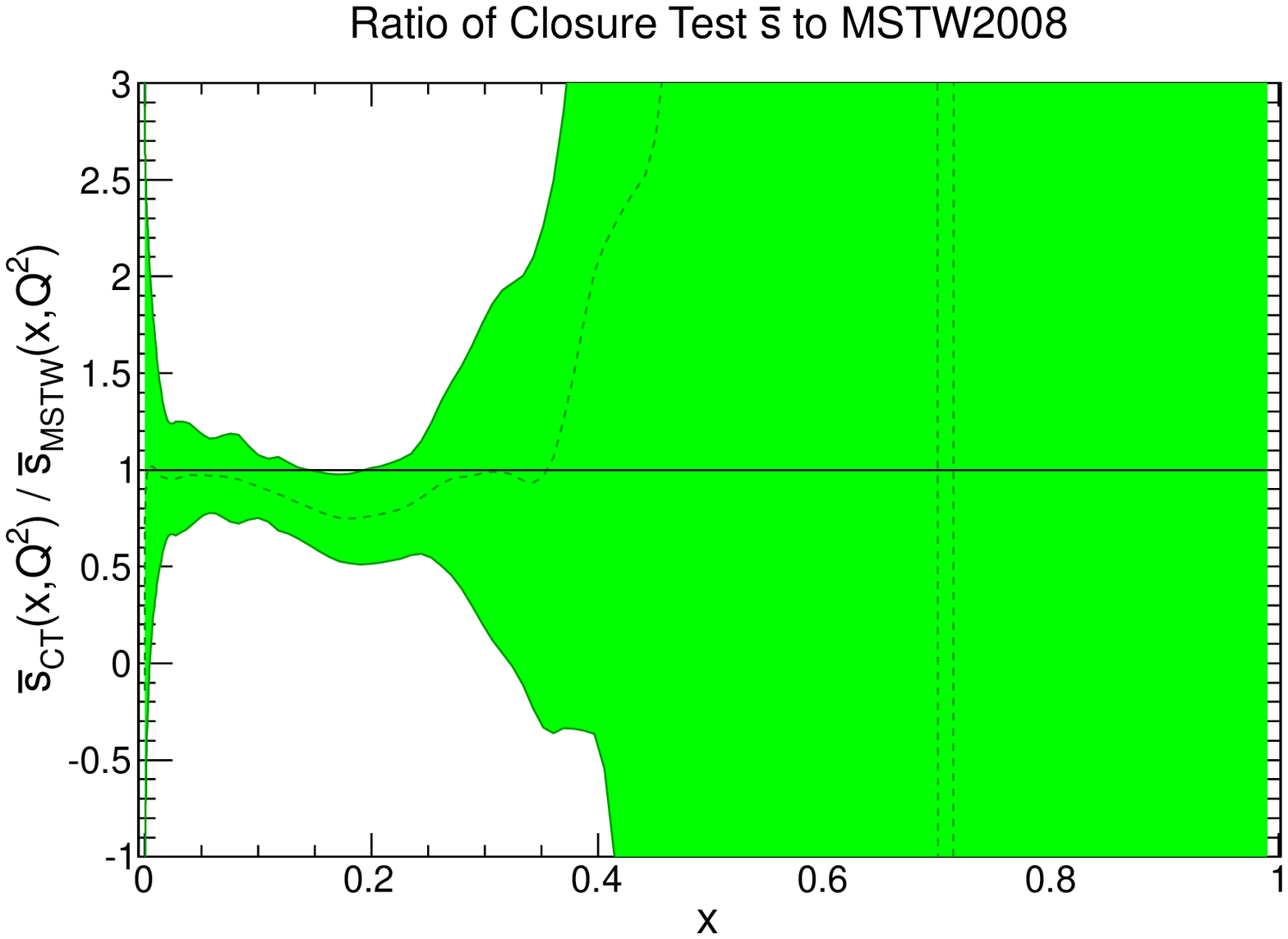}
  \caption{\small Same as above for gluon and strange PDFs.}
  \label{fig:L2-MSTW-ratio2}
\end{figure}
\renewcommand{\thefigure}{\arabic{figure}}

A final qualitative comparison can be performed by using the arc-length,
which, as discussed in Sect.~\ref{subsec:cv}, provides a measure of
the complexity of a
function defined over a finite interval.
A function that has a more
complicated structure is expected to have a larger arc-length,
and vice-versa.
In the context of our fits, PDFs with too large arc-lengths
might be an indication of over-fitting, with a PDF trying to use a contrived
shape in order to fit statistical fluctuations.
For this reason, as explained in Sect.~\ref{subsec:cv},
replicas that at the end of the fit have some PDFs with unnaturally
large arc-length are discarded from the final sample.

\begin{figure}[h!]
  \centering
 \epsfig{width=0.90\textwidth,figure=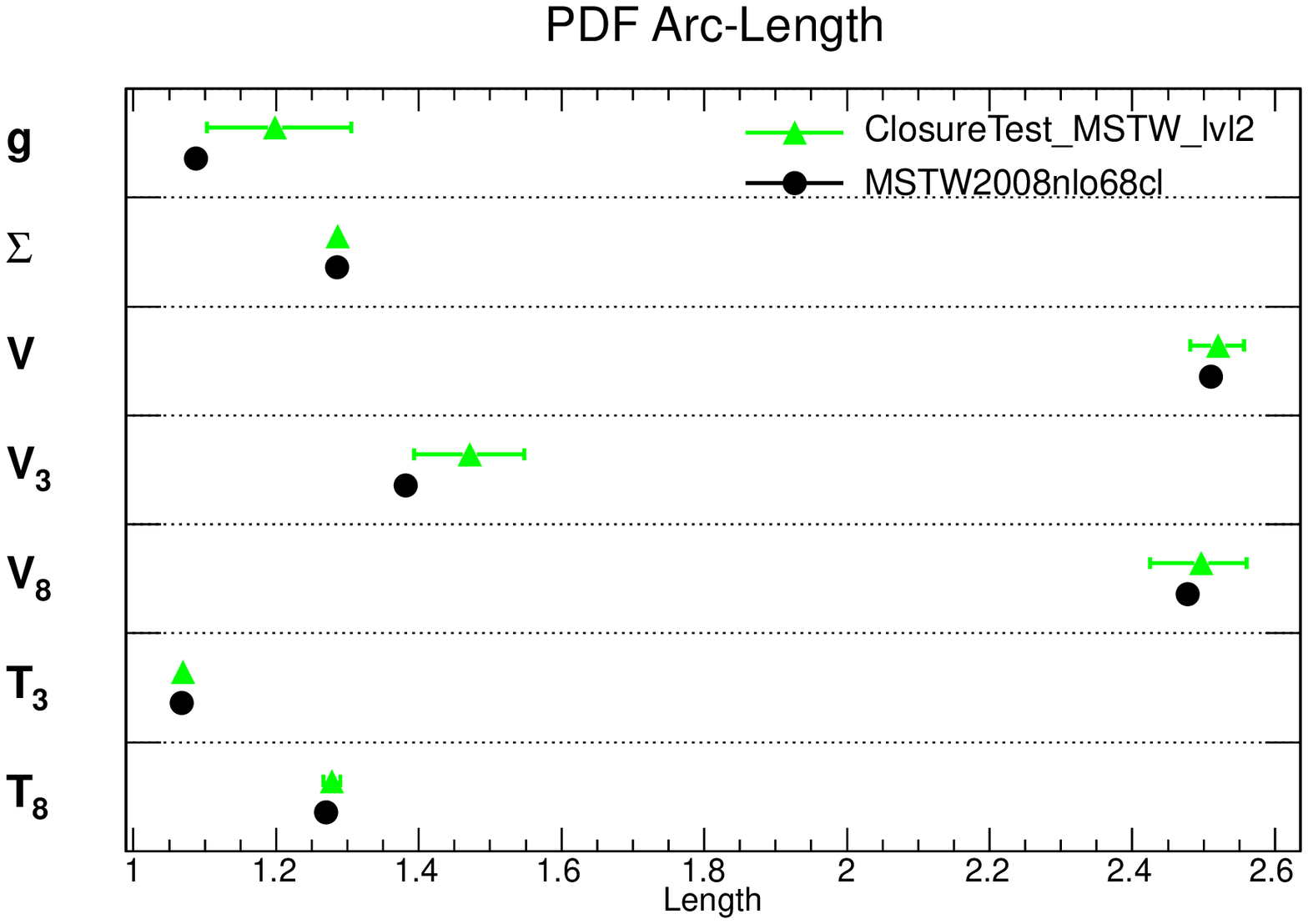}
  \caption{\small Comparison
of the arc-lengths of the fitted and input PDFs, calculated using
  Eq.~(\ref{eq:arclength}) for the C9 Level~2 closure test,
using MSTW08 NLO as input.
Arc-lengths are computed at the input parametrization scale of
$Q^2=1$~GeV$^2$.
For the fitted PDFs, we show the central value and the  68\% confidence interval.}
  \label{fig:arc-le}
\end{figure}

In the context of closure tests, comparing the arc-lengths of the
fitted PDFs with those of the input PDFs provide an integrated
comparison, rather than point by point, that the fitted PDFs
reproduce the input ones.
The arc-lengths of
the Level~2 fit C9, using MSTW08 as input, are shown in Fig.~\ref{fig:arc-le}.
Arc-lengths are computed at the input parametrization scale of
$Q^2=1$~GeV$^2$.
As expected from the comparison at the level of PDFs,
there is also good agreement between the input and
fitted PDFs at the level of arc-lengths, the arc-length of the
input PDF typically lying within the 68\% confidence
level of the fitted PDFs.

\clearpage

\subsubsection{PDF uncertainties: quantitative validation}
\label{subsec:quant}

The results of Sects.~\ref{subsec:closureCV} and~\ref{subsec:qual} provide
evidence  the
Level~2 closure tests are successful, both in terms
of central values and of PDF uncertainties.
We now provide a more quantitative assessment of this success using the
indicators described in Sect.~\ref{sec:closureestimators}.

In order to validate quantitatively the agreement in central values
we  compute the $\xi_{n\sigma}$  estimators Eq.~(\ref{eq:L268CL}) for Level~2 fit C9.
We obtain the following results:
\be
\xi_{1\sigma}^{\rm (l2)} = 0.699\, , \qquad \xi^{\rm (l2)}_{2\sigma} = 0.948\, ,
\ee
to be compared with the theoretical expectations of  0.68 and 0.95.
This excellent agreement confirms that the PDF replicas obtained by our
fitting methodology can indeed be
interpreted as a representation of the probability distribution for
the PDFs given the data used in the fit.

To verify that this agreement is not accidental,
or a fluke of the definition of the estimator $\xi_{n\sigma}$,
but rather a robust feature of our analysis,
we have redone the whole procedure but this time based
on Level~1 closure fits, set up in the same way as C8,
and computed again the $\xi_{n\sigma}$  estimators.
We already know from the qualitative comparisons
between Level~1 and Level~2 uncertainties
from Fig.~\ref{fig:ratiofit1} that in Level~1 closure tests
PDF uncertainties are underestimated, and therefore in this case
the central value of the fitted PDF will fluctuate more
than its estimated uncertainty would suggest.

This implies that in Level~1 closure tests we expect
the $\xi_{n\sigma}$  estimators to be smaller than the
theoretical
expectations above.
Indeed, computing $\xi_{1\sigma}$ and $\xi_{2\sigma}$ at Level~1 this is precisely what we find:
\be
\xi_{1\sigma}^{\rm (l1)} = 0.512\, , \qquad \xi_{2\sigma}^{\rm (l1)} = 0.836\, ,
\ee
which shows that indeed the Level~1 closure tests fail, in the sense that
Level~1 fits underestimate PDF uncertainties. This confirms that the Level~2
step, that is the generation of $N_{\rm rep}$ Monte Carlo
replicas on top of the pseudo-data with the fluctuations, is essential to obtain
the correct PDF uncertainties.

\begin{figure}[H]
  \centering
  \epsfig{width=0.49\textwidth,figure=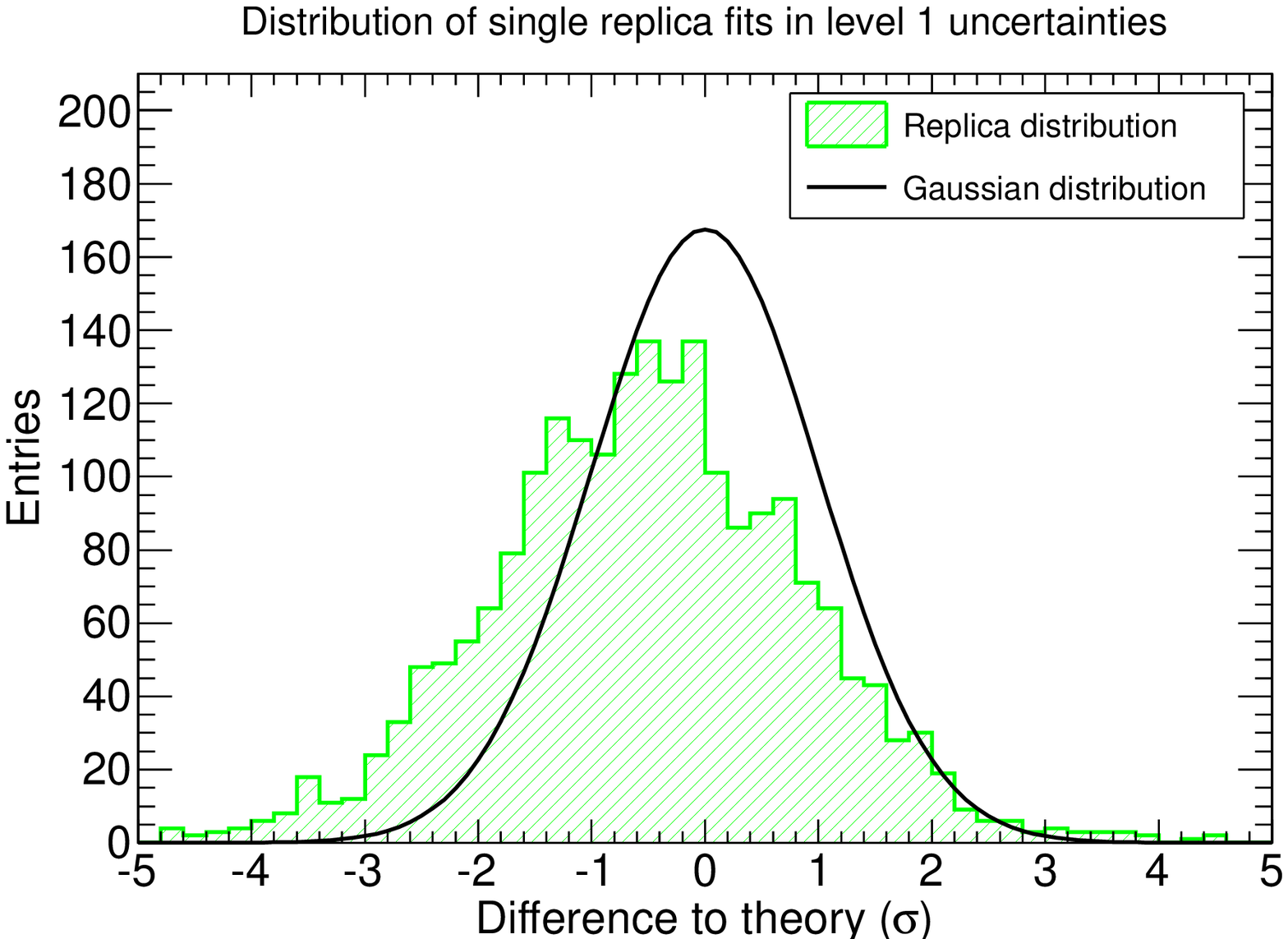}
  \epsfig{width=0.49\textwidth,figure=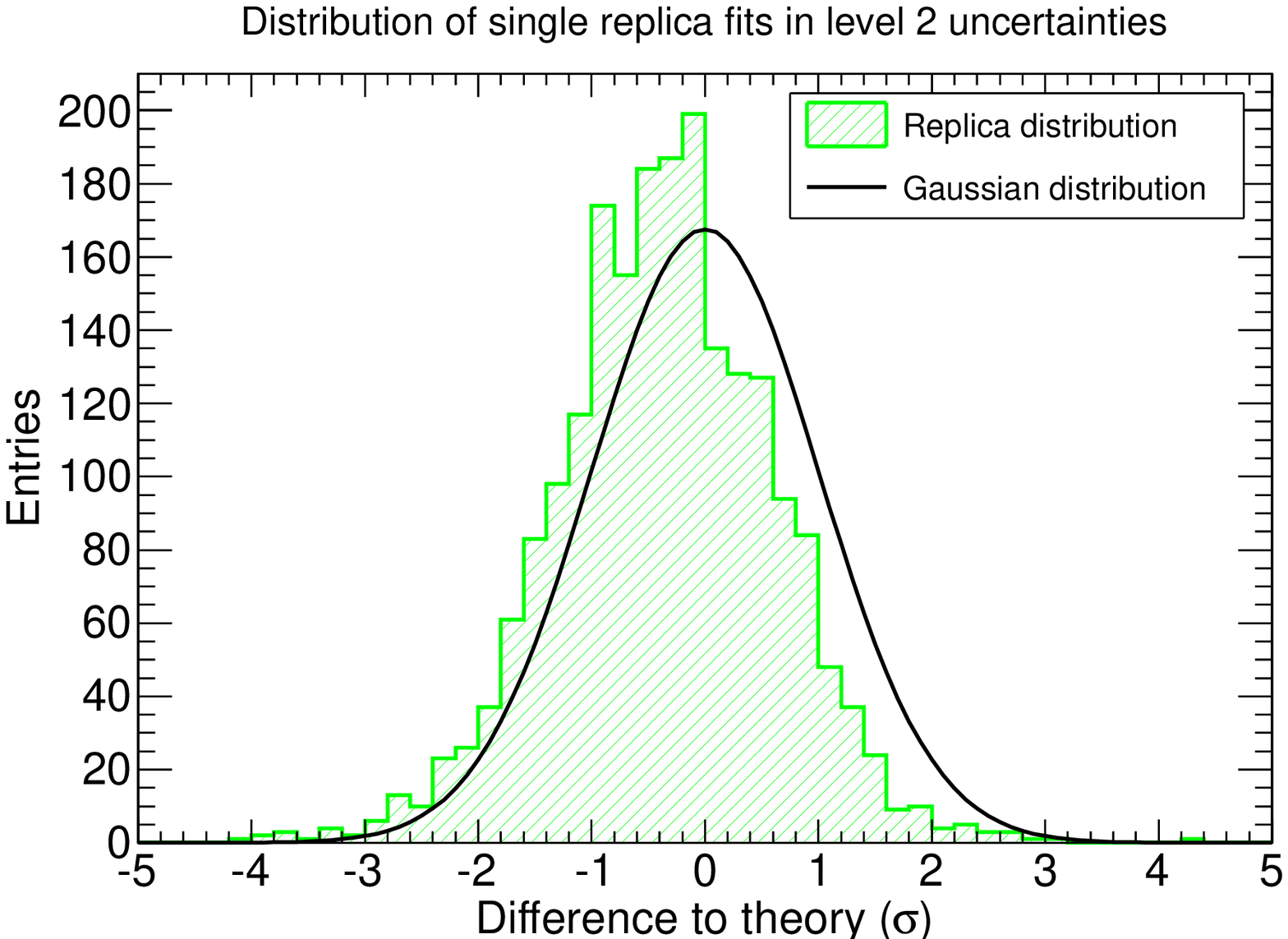}
  \caption{\small Histograms for the difference between the input PDF and
  the single replica fits obtained from
  different sets of closure test data, in units of the
  standard deviation of standard Level~1 (left) and Level~2 (right) fits.
  An appropriately scaled Gaussian distribution is shown for comparison.}
  \label{fig:dist-histo}
\end{figure}

The fact that the Level~2 closure test leads to a correct estimate of the PDF uncertainties,
while the Level~1 fails, can be tested in more detail by looking at the distribution of the
mean of our fit for different closure test datasets: this tests not only the one- and two-sigma
confidence intervals, but the shape of the whole distribution of deviations between the
prediction and the truth.
Figure~\ref{fig:dist-histo} shows the
histograms of the differences between $\langle f_{\rm fit}\rangle$ obtained
using different closure test datasets (that is, pseudo-data generated
with different random seeds) and the central value $f_{\rm in}$ of the
MSTW input PDFs, in units of
the standard deviation of a standard Level~1 or Level~2 closure test.
The histogram is generated using the
values at $x = 0.05$, 0.1 and 0.2 for each PDF, as a representative
sampling.
The resulting distribution is very close
to a Gaussian distribution with standard deviation 1
when scaled with the Level~2 uncertainties,
and is considerably wider using the Level~1 uncertainties.
This is consistent with the results found above that Level~2 uncertainties
reproduce the correct fluctuations of the central values while Level~1
underestimates them.

\subsubsection{Closure test validation using Bayesian reweighting}
\label{sec:bayes}

The results above show that our
Level~2 closure tests are successful in terms
of both central values and uncertainties,
both at the qualitative and quantitative levels.
However, it is still conceivable that they could fail
for higher moments of the fitted distribution, including
the correlations.
In this respect, the most stringent validation of
our fitting methodology, also in the context
of closure tests, is provided by exploiting the
Bayesian reweighting method~\cite{Ball:2011gg}. Reweighting allows one
to determine a new set of PDFs from a prior set when new data are added. Since
reweighting is analytic, in the sense that the weights are computed using the
$\chi^2$ to the new data without the need for refitting, by comparing the
reweighted distribution to a refitted distribution in a closure test, one
can test rather precisely the fitting methodology.

With this motivation,
we have performed, based on closure tests,
two reweighting analyses, one at Level~1 and
the other at Level~2.
The priors consist in each case of sets of $1000$ PDF replicas,
produced using the same dataset as was used in the NNPDF2.3-like fits described
in Sect.~\ref{sec:results}, in particular without the
inclusive jet production data (which in NNPDF2.3 are from CDF and
ATLAS).
The pseudo-data were generated using
MSTW08 as the input PDF, NLO QCD, and so on,
just as for the C8 and C9 closure tests. They were then fitted in the usual way.
The jet data were then included by reweighting the prior
in the usual way~\cite{Ball:2011gg}, and the results compared
with the corresponding closure test PDFs in which instead the
jet data were included by generating pseudo-data,
adding them to the prior pseudo-data and then refitting.

\begin{figure}[H]
  \centering
 \epsfig{width=0.49\textwidth,figure=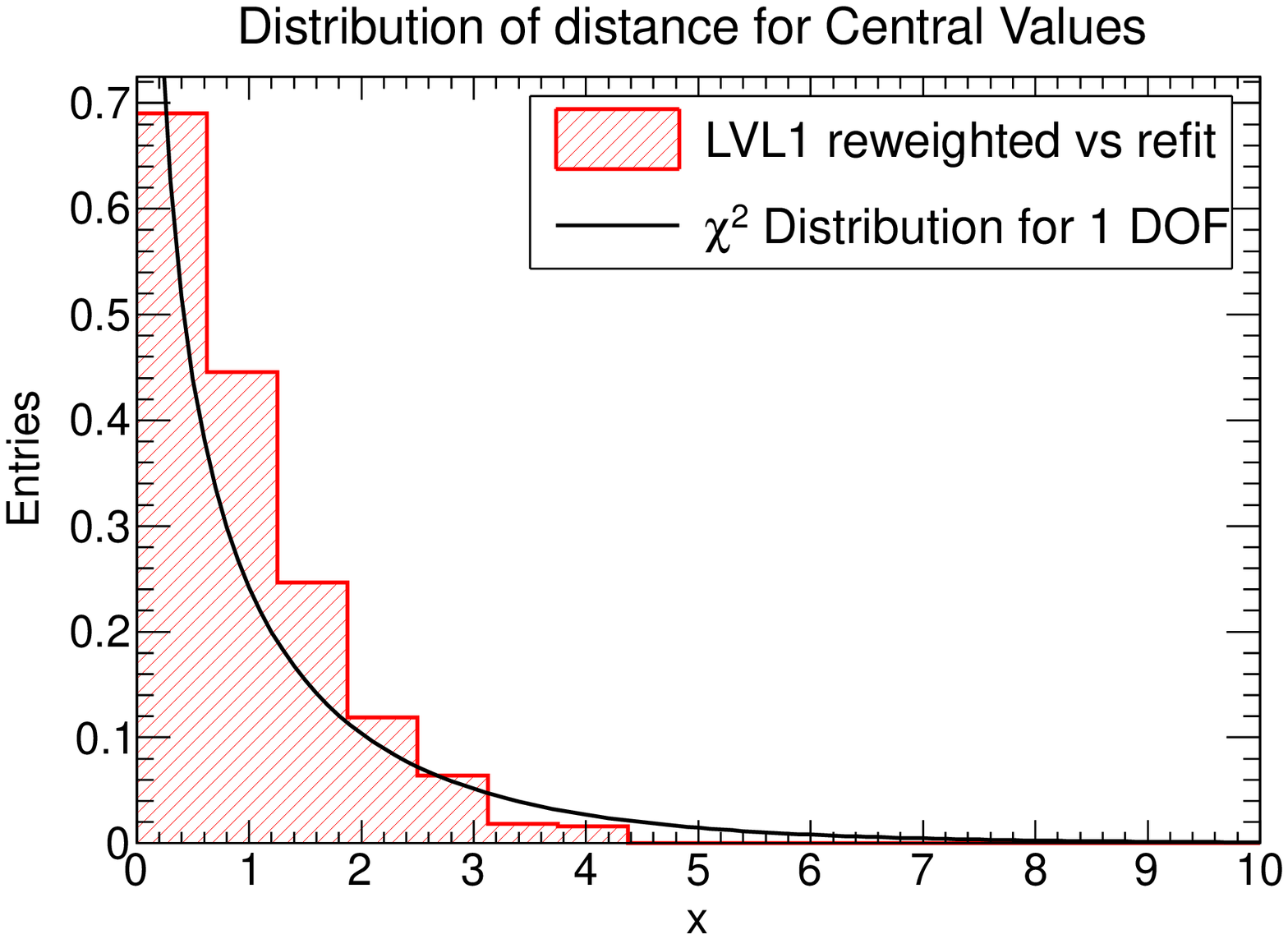}
 \epsfig{width=0.49\textwidth,figure=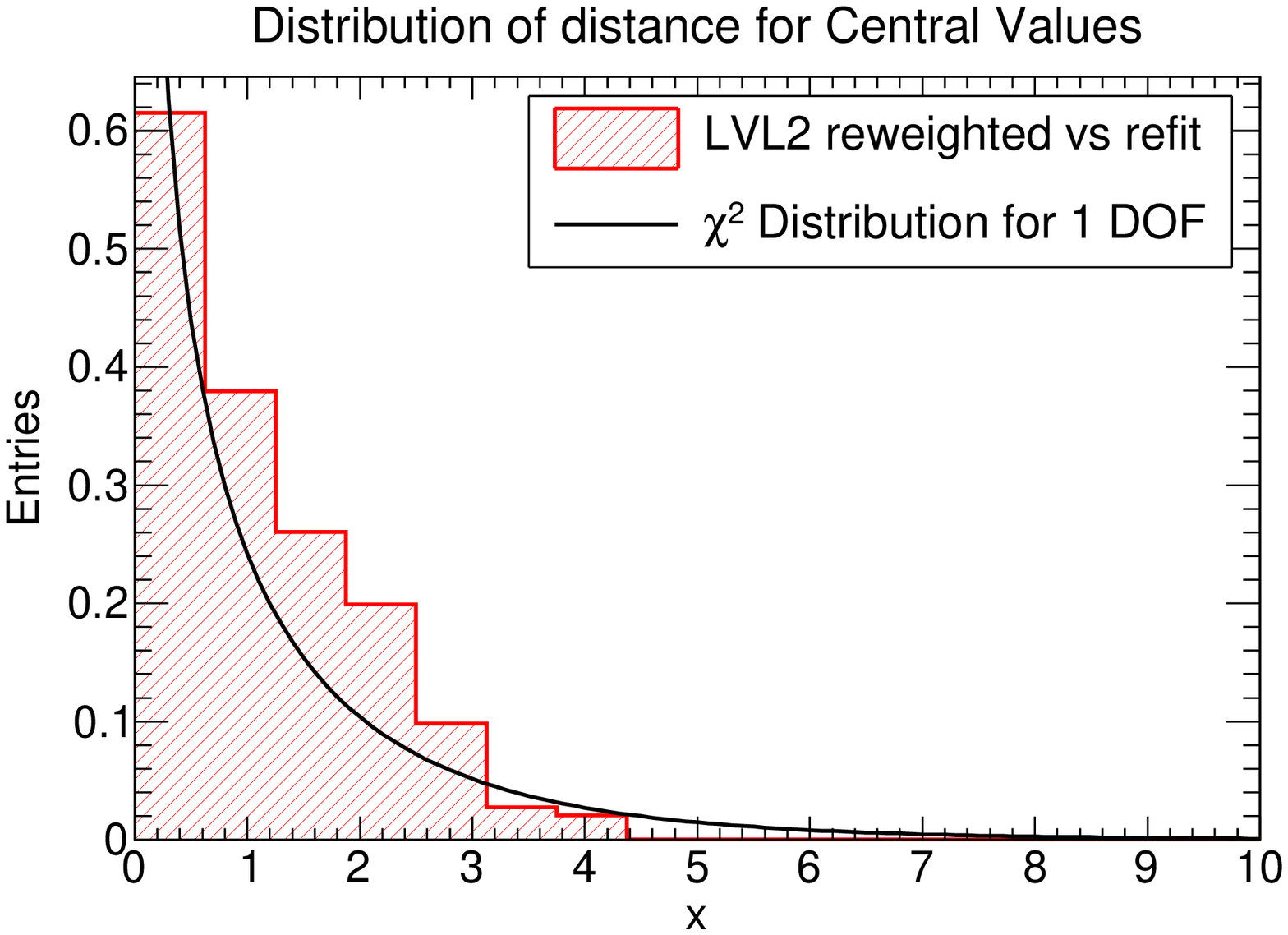}
  \caption{\small Distribution of
the distances, for the central values, between the reweighted and the
 refitted PDFs for the
  Level~1 (left) and the  Level~2 (right) closure test fits.
See Appendix~\ref{app:distances} for the definition
of the distances. Also
  shown is a $\chi^2$ distribution with one degree of freedom.}
  \label{fig:CTL2-rwt-hist}
\end{figure}

First, we compute the distance
between the central values of the reweighted and refitted PDFs.
The  distribution of these distances, in both the Level~1
and Level~2 cases, is shown in Fig.~\ref{fig:CTL2-rwt-hist}.
These distributions are obtained from a sampling
 produced using
100 points in $x$ for each PDF in the flavor basis,
in the range between $10^{-5}$ and 1.
We expect this distribution of distances, for a successful
reweighting test, to follow a $\chi^2$ distribution with
one degree of freedom.

From Fig.~\ref{fig:CTL2-rwt-hist} we see that
the distribution of
distances follows the statistical expectation of
a $\chi^2$ distribution with one degree of freedom, as seen
by comparing with the superimposed curve representing the latter, both
in Level~1 and Level~2. This means that both Level~1 and Level~2
closure tests faithfully reproduce central values.

\begin{figure}[h!]
  \centering
 \epsfig{width=0.49\textwidth,figure=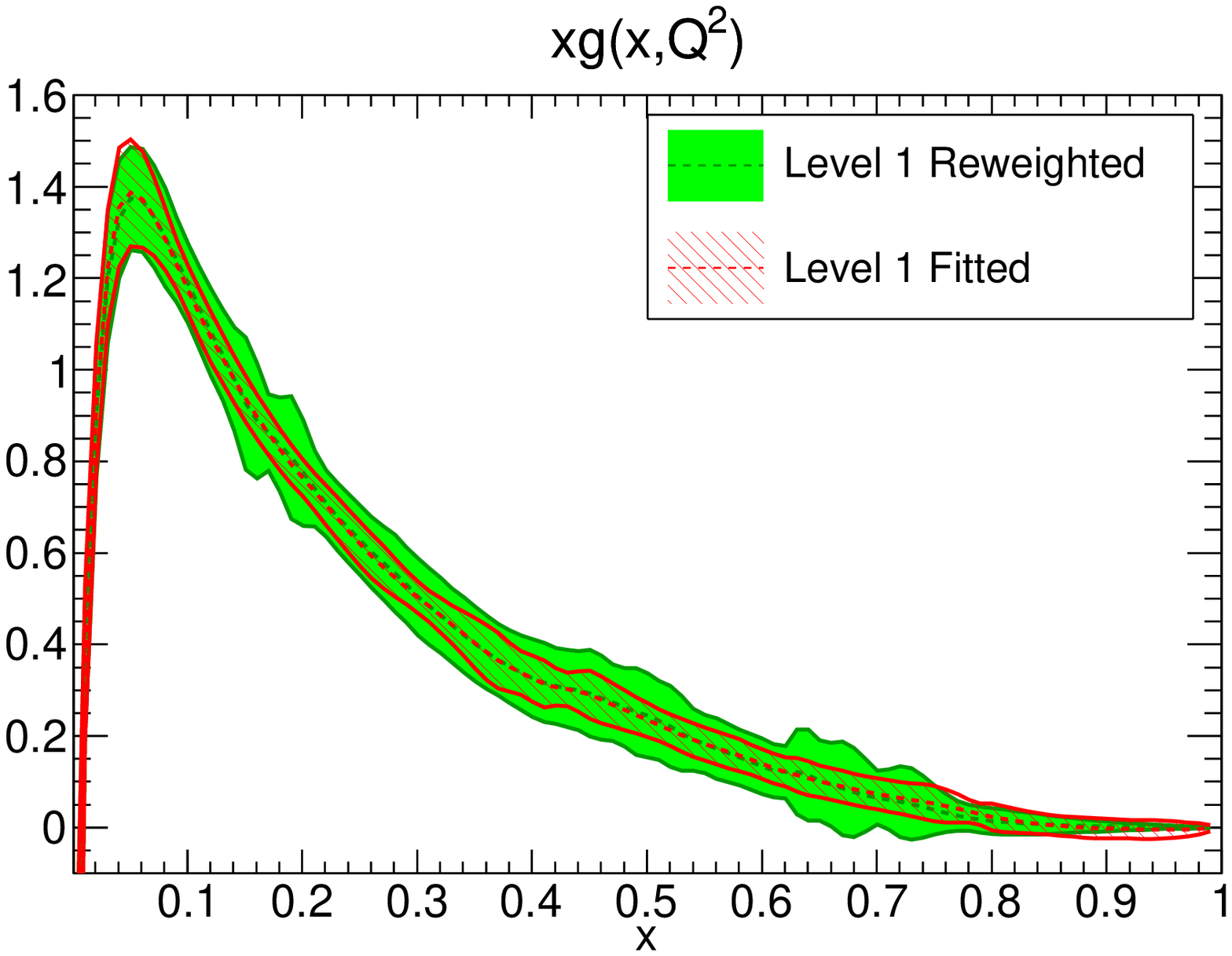}
 \epsfig{width=0.49\textwidth,figure=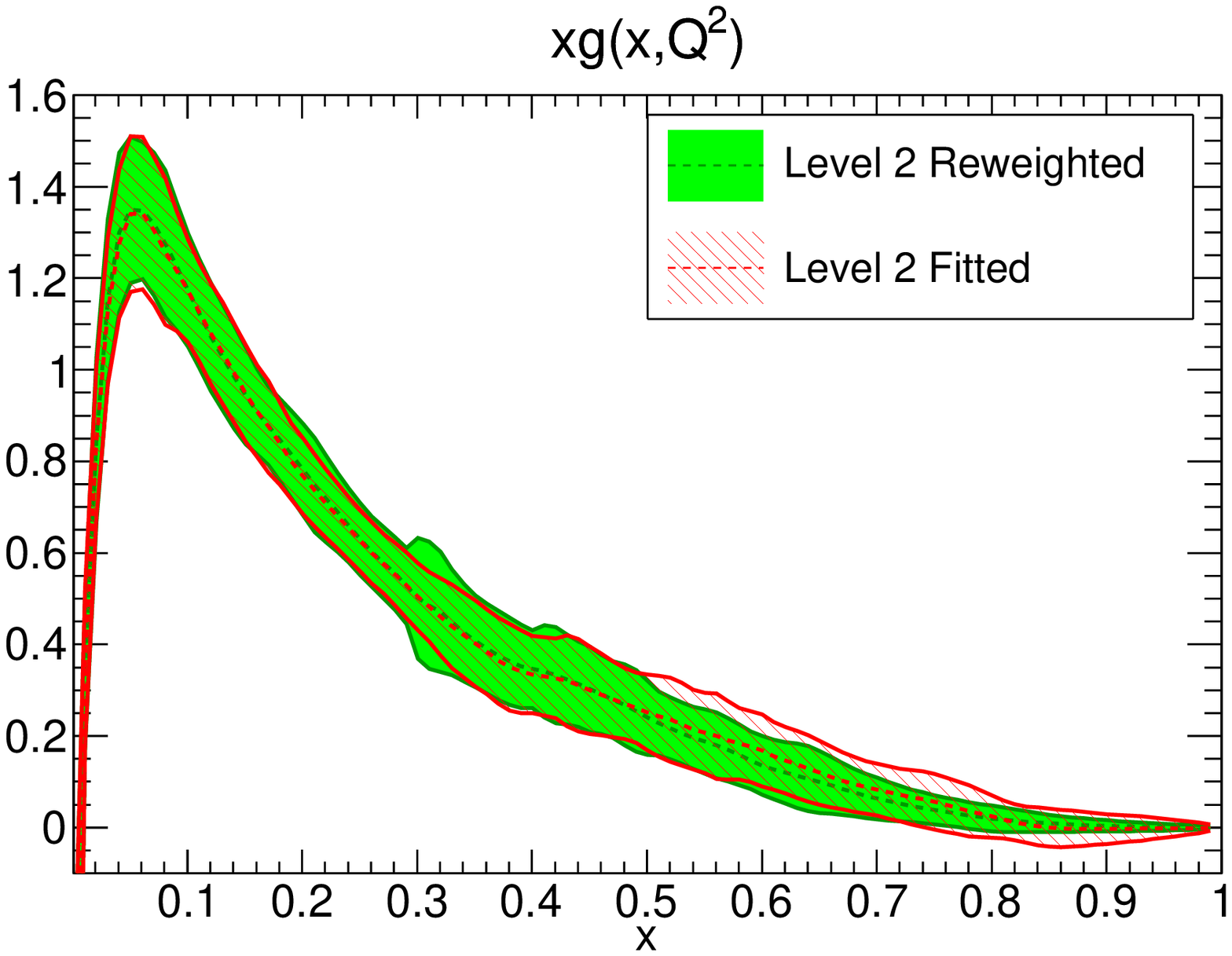}
  \caption{\small The large-$x$ gluon PDF at the
initial parametrization scale of $Q^2 = 1~\textrm{GeV}^2$, comparing
the results of adding the inclusive
jet data by reweighting (green solid band) and by refitting (red
hatched band).
Results are shown for a Level~1 closure test fit (left)
and for a Level~2 closure test fit (right).
See text for more details.
}
  \label{fig:CTL2-rwt}
\end{figure}

A more stringent test is obtained when computing uncertainties.
The comparison between the reweighted and refitted results including uncertainties
for Level~1 and Level~2 closure tests, are shown in
Fig.~\ref{fig:CTL2-rwt}.
Several interesting conclusions can be drawn from this
comparison. First, it is clear that at Level~2 the reweighted and
refitted results are in perfect agreement, thus validating the
procedure.
Second, at Level~1 the refitted uncertainties are rather smaller than
the reweighted ones, thus signaling a failure of the procedure.

The more interesting observation however is that in fact the Level~1
reweighted and Level~2 reweighted results essentially coincide. This,
upon reflection, is to be expected: indeed, if the new data bring in
sufficient new information, Bayesian reweighting produces results
which are largely independent of the prior. Because the jet data are
controlling the large-$x$ gluon uncertainty, and these are introduced by
reweighting in both cases, this uncertainty comes out to be the
same even when the uncertainties in the prior are not accurately
estimated, as in the Level~1 fit. If instead the jet
data are introduced by Level~1 refitting, the ensuing uncertainty
comes out to be too small, because Level~1 does not properly account
for the data uncertainty as discussed in Sect.~\ref{sec:components}.

\clearpage

\subsection{Robustness of the fitting methodology}
\label{sec:robustness}

We now turn to study the robustness of the fitting methodology against
changes in some of its parameters, in the context of the closure test fits.
The main advantage of using closure tests, as compared to attempting to address
these issues in fits to real experimental data, is that any deviation from expectation
must be attributed uniquely to deficiencies in the methodology, rather than blamed
on spurious reasons like inconsistencies in the data, or failings of the theoretical
description: in a closure test data and theory are by construction `perfect'.

First, we study how the fit results change when the training length
is modified, then when the fraction of data included in the training set
is varied, then we examine the effects of modifying the basis in which the PDFs
are parametrized, and finally we verify that our baseline neural network
architecture is redundant enough by comparing with the results
of a fit with a huge neural network.
Most of the closure tests in this section are performed using MSTW08
as input PDF, but we finally also explore the robustness of the
closure testing against variations of this choice, in particular
using CT10 and NNPDF3.0 itself as input PDFs.

\subsubsection{Independence of the maximum number of GA generations}
\label{sec-longer}

In NNPDF3.0, the cross-validation strategy that determines the optimal stopping
point
for the neural network training is based on the look-back method, described
in Sect.~\ref{subsec:cv}.
Provided that the maximum number of genetic algorithm generations
$N_{\rm gen}$ is large enough,
we expect results based on the look-back method to be independent of
the value of $N_{\rm gen}$.
The condition for this to take place is that $N_{\rm gen}$ must be sufficiently
large so that the look-back algorithm is able to the identify
an optimal stopping point by finding somewhere close to the absolute
minimum of the validation $\chi^2$.

\begin{figure}[t]
  \centering
  \epsfig{width=0.99\textwidth,figure=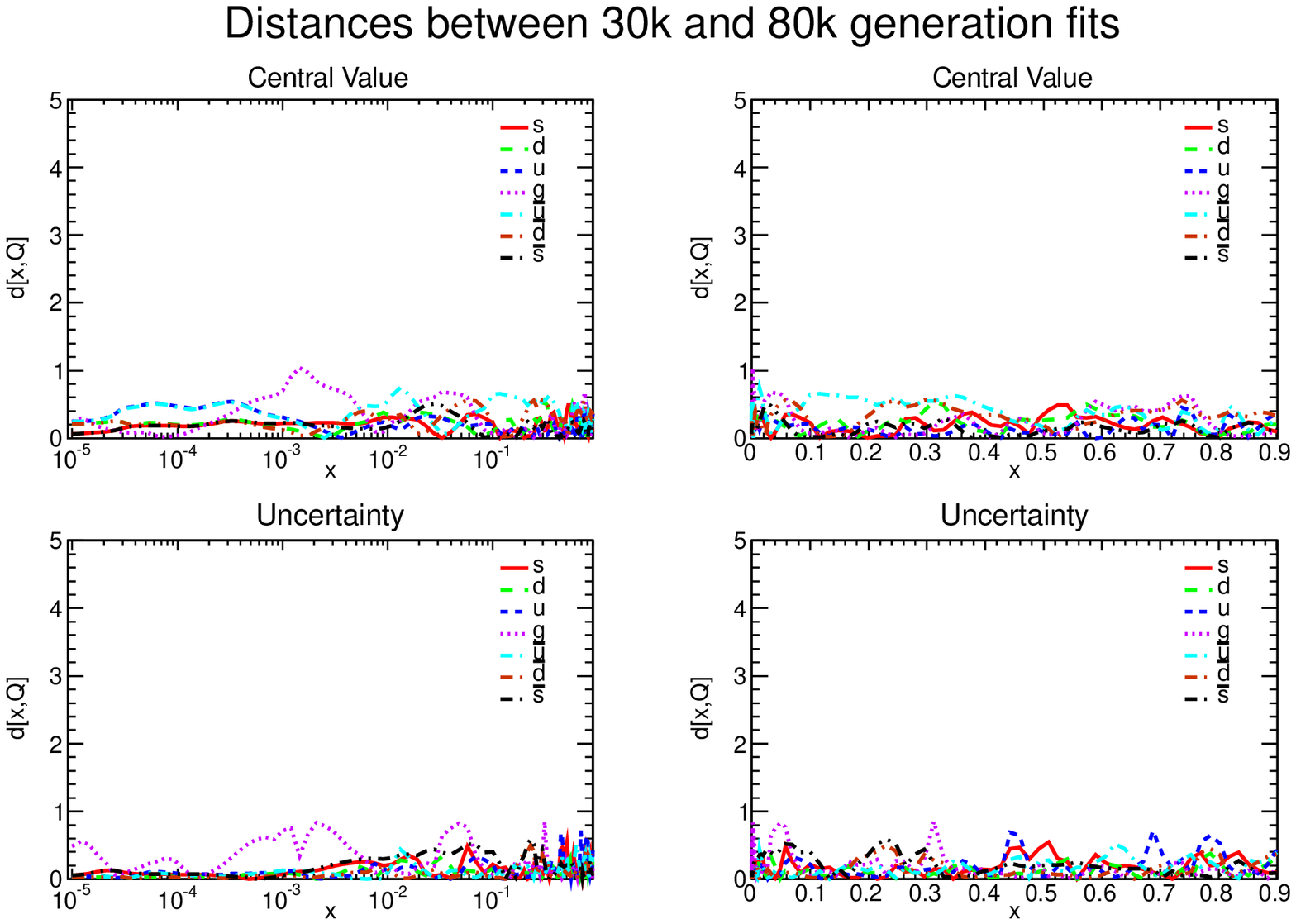}
  \caption{\small Distances between central values and uncertainties of two Level~2 closure test fits with maximum numbers of generations set to 30k and 80k,  fits C9
and C11 in  Table~\ref{tab:CTfits}, with
all other fit setting identical.
Distances are calculated at the input parametrization
 scale of $Q^2 = 1~\textrm{GeV}^2$. Left plots are in a log scale while right
plots are in a linear scale.}
  \label{fig:long-tr}
\end{figure}

To verify this expectation,  in Fig.~\ref{fig:long-tr}
we show the distances for two Level~2 closure test fits with the
  maximum numbers of generations set to 30k and 80k (fits C9
and C11 in  Table~\ref{tab:CTfits}), all other settings being
identical.
 Both fits use the look-back procedure to determine the ideal stopping point,
and only differ in the maximum number of generations that are analysed by the
look-back algorithm.
Since  the random seed used for the fit is the
same in the two cases, this is completely equivalent to looking
at the same fit had it been left to run for a larger number of generations.

Both the distances between central values, and the distances between uncertainties,
are smaller than one for all the PDFs in the flavor basis over the whole range of $x$ values,
showing that the two fits are statistically indistinguishable.
 We can therefore rule out any
sizable dependence on the total training length in our current results, and
we can stick to a baseline maximum number of generations of
30k in our fits to real data.

\subsubsection{Dependence on the training fraction}

As discussed in Sect.~\ref{subsec:cv}, the look-back cross-validation
algorithm requires us to separate the fitted dataset into two disjoint subsets,
the training set and the validation set, and in the standard fits each contains
50\% of the total dataset.
During the minimization, only the training set is seen by the neural
network, with the validation being used only to determine the
optimal stopping point.
While the exact value of the training fraction would be irrelevant
in the limit of infinite statistics, one could argue that
for our large, but finite, dataset, results might change substantially
if this 50\% is varied to some other value.
In particular, we must be sure that a training fraction of 50\% is
enough to retain all the
relevant information contained in the original dataset.

\begin{figure}[t]
  \centering
  \epsfig{width=0.99\textwidth,figure=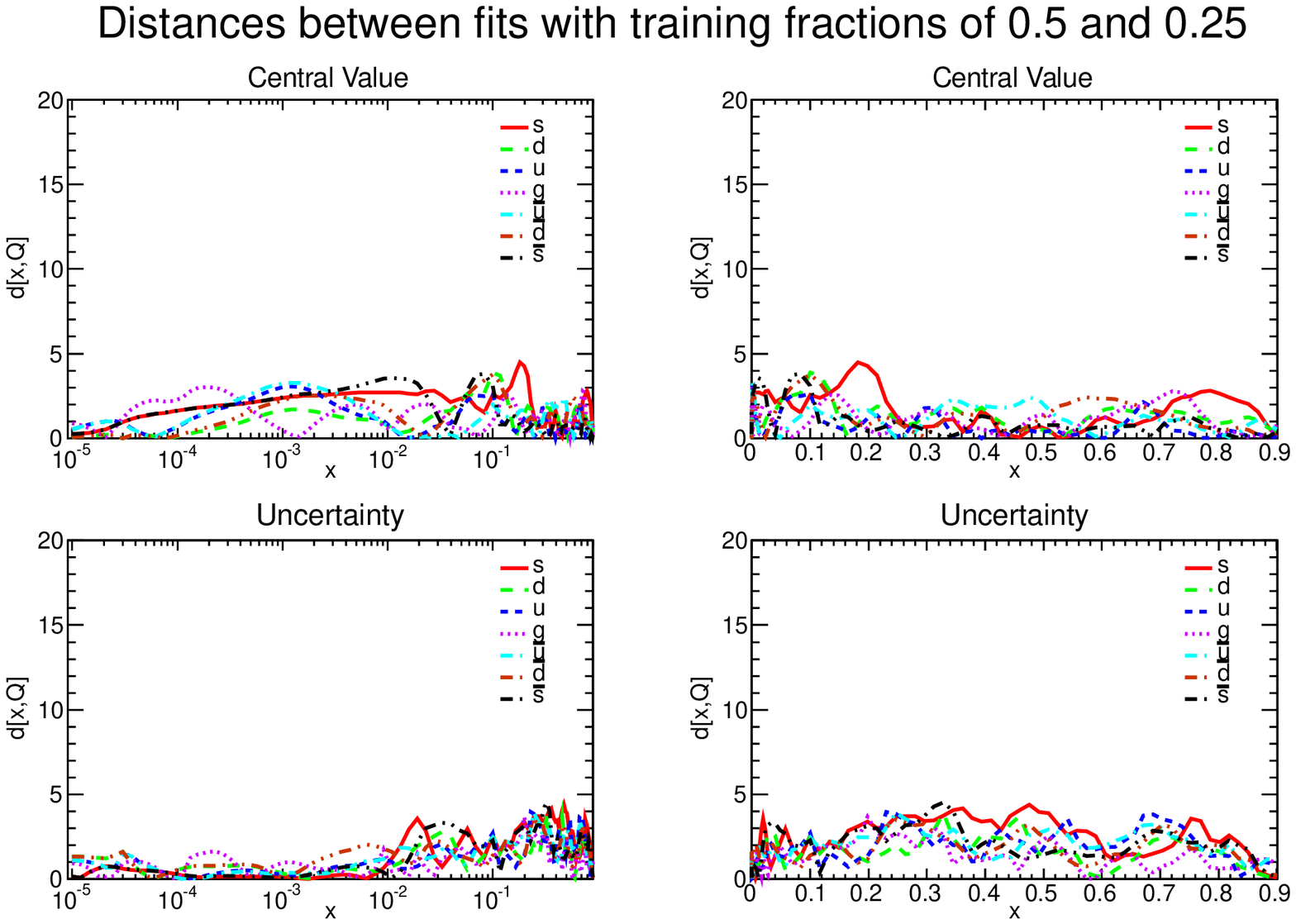}
  \caption{\small Same as Fig.~\ref{fig:long-tr} for the
Level~2 closure fits based on training fractions of 25\% and 50\%,
fits C9 and C13 in Table~\ref{tab:CTfits}. }
  \label{fig:dist-diff-tf25}
\end{figure}

\begin{figure}[t]
  \centering
  \epsfig{width=0.99\textwidth,figure=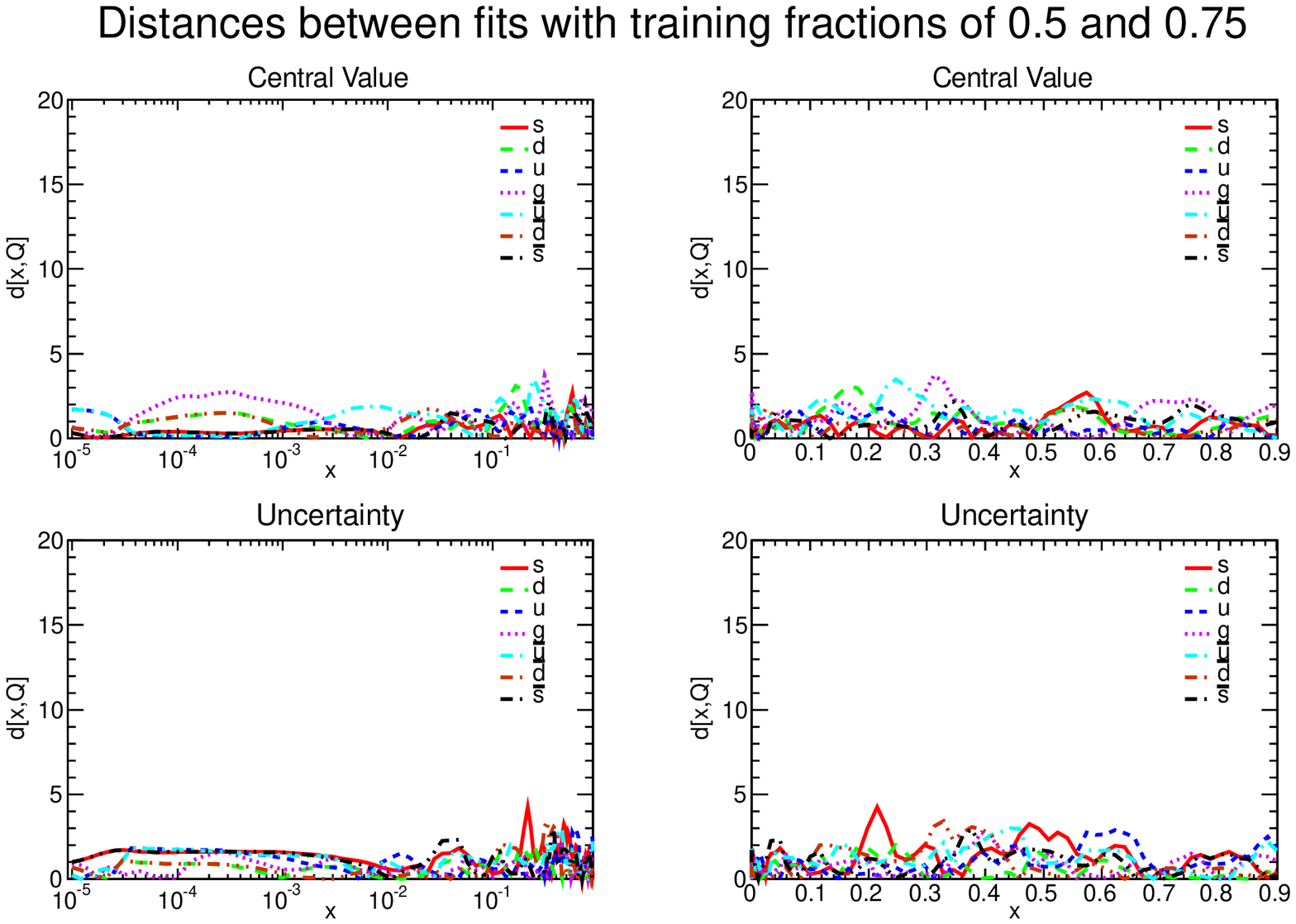}
  \caption{\small Same as Fig.~\ref{fig:long-tr} for the
Level~2 closure fits based on training fractions of 50\% and 75\%,
fits C9 and C14 in Table~\ref{tab:CTfits}.   }
  \label{fig:dist-diff-tf75}
\end{figure}

\begin{figure}[t]
  \centering
  \epsfig{width=0.49\textwidth,figure=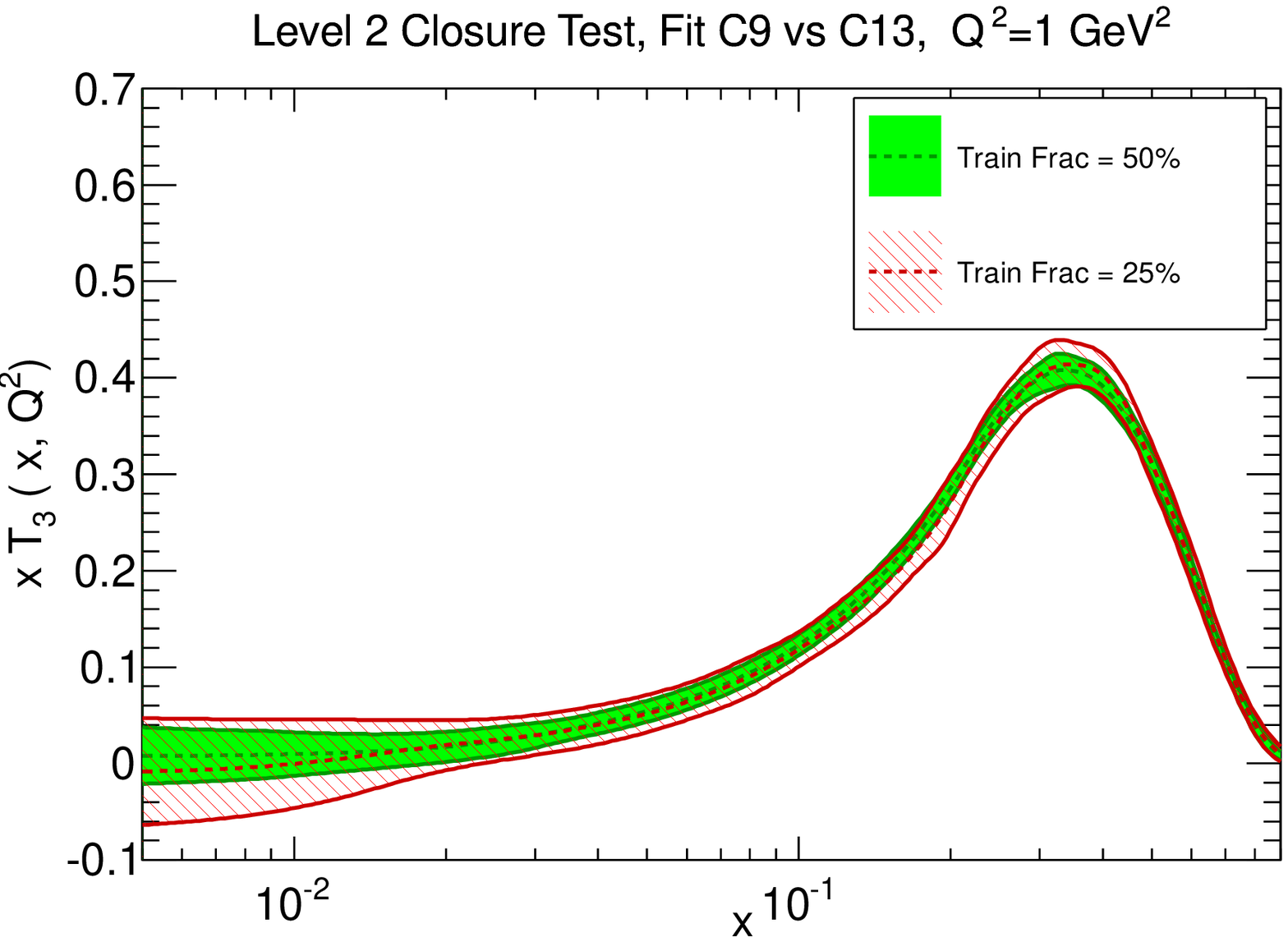}
 \epsfig{width=0.49\textwidth,figure=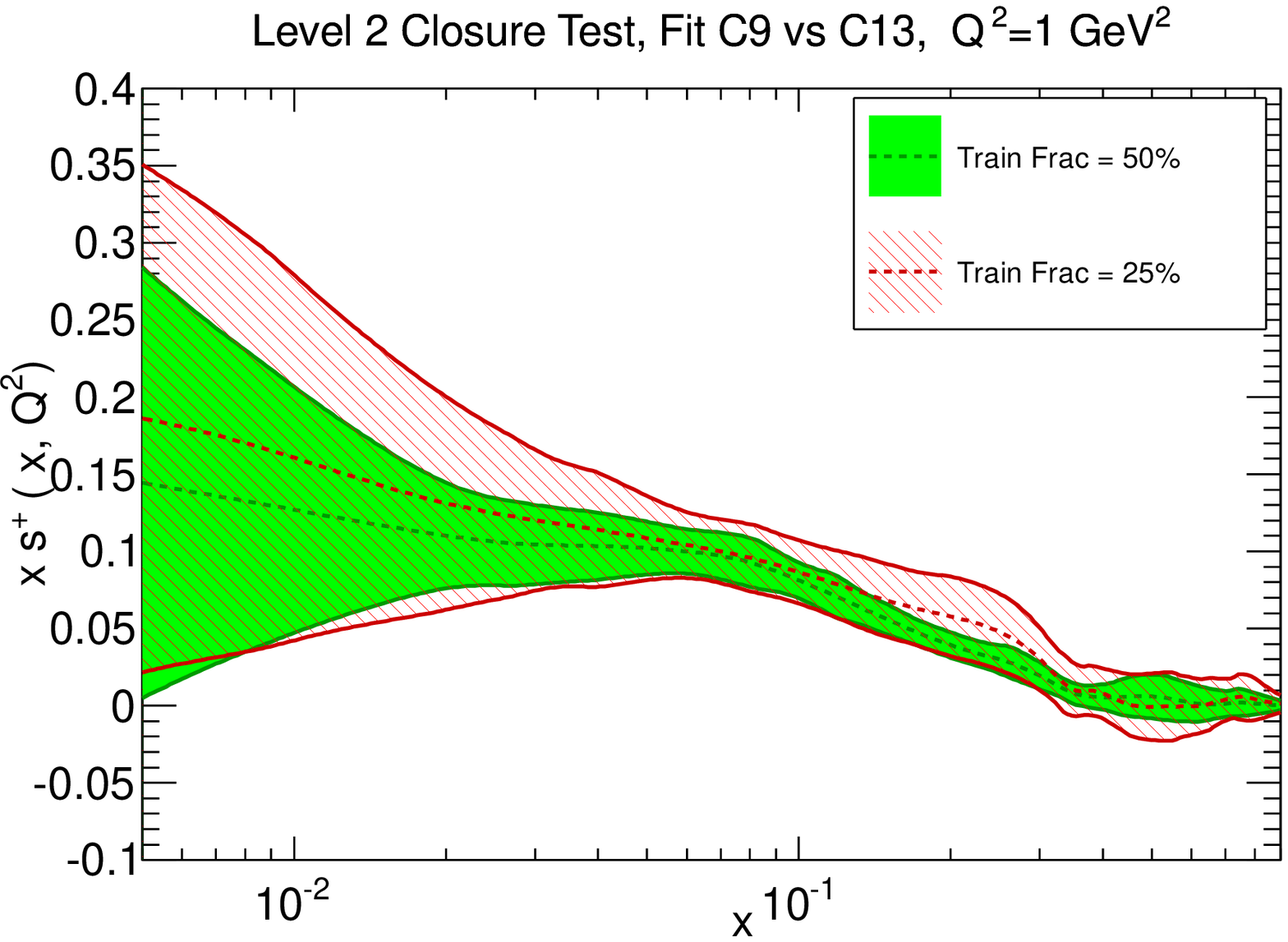}
  \caption{\small Comparison between PDFs
 in the
Level~2 closure fits based on two different
training fractions, 50\% and 25\%, corresponding to
fits C9 and C13 in Table~\ref{tab:CTfits}.
We show the quark triplet $xT_3(x,Q_0^2)$ (left plot)
and the total strangeness $xs^+(x,Q_0^2)$ (right plot)
at the initial parametrization scale of $Q_0^2=$ 1 GeV$^2$.
The increase in PDF uncertainties when the training fraction is decreased
to 25\% is a result of the loss of non-redundant information
as compared to the baseline fit with 50\% training fraction
  }
  \label{fig:dist-diff-tf75-pdfs}
\end{figure}

In order to study the impact on the fitted PDFs of the use of
a different training fraction, we have produced Level~2 closure fits
with identical settings, the only difference being the value
of this training fraction.
Compared to the baseline fit with a training fraction of 50\%
(fit C9 in Table~\ref{tab:CTfits}), we have produced a
fit with a smaller
training fraction, 25\% (fit C13), and another one with a larger
training fraction, 75\% (fit C14).
The comparison between these three values of the training fraction,
quantified by the distance plots of Fig.~\ref{fig:dist-diff-tf25}
and  Fig.~\ref{fig:dist-diff-tf75},
indicate that the resulting uncertainties on the fitted PDFs increase when
the training fraction is reduced to 25\%, showing that, for the
NNPDF3.0 dataset, some of the information which is lost when using the smaller
training fraction is not redundant.
This effect can be quantified more directly by looking at the PDFs
in the two fits, shown in Fig.~\ref{fig:dist-diff-tf75-pdfs}.
Is clear that the reduction of the training fractions entails a considerable
increase of the PDF uncertainties.


On the other hand, as can be seen from
the distances in  Fig.~\ref{fig:dist-diff-tf75}, the fits with
training fractions of 50\% and 75\% are statistically indistinguishable.
We thus conclude that the loss of information due to the
training-validation
splitting of the dataset required by the cross-validation procedure
is small provided the training fraction is above 50\%, but
would be problematic for smaller training fractions.

\subsubsection{Parametrization basis independence}
\label{sec:basis}

One of the most attractive features of the new NNPDF3.0
methodology is that thanks to the new flexible C++ fitting code,
it is possible to change the basis for the PDF parametrization rather easily,
as discussed in Sect.~\ref{sec:inputbasis}.
In turn, this allows us to perform detailed tests of the robustness
of the whole fitting procedure: indeed, for a truly unbiased fit,
results should be independent of the parametrization basis,
since they are all related to each other by linear transformation.
Note that in the standard Hessian approach to PDF fitting even adding a few extra
parameters to the input parametrization can be quite complicated,
let alone changing  the parametrization basis altogether.

\begin{figure}[t]
  \centering
  \epsfig{width=0.99\textwidth,figure=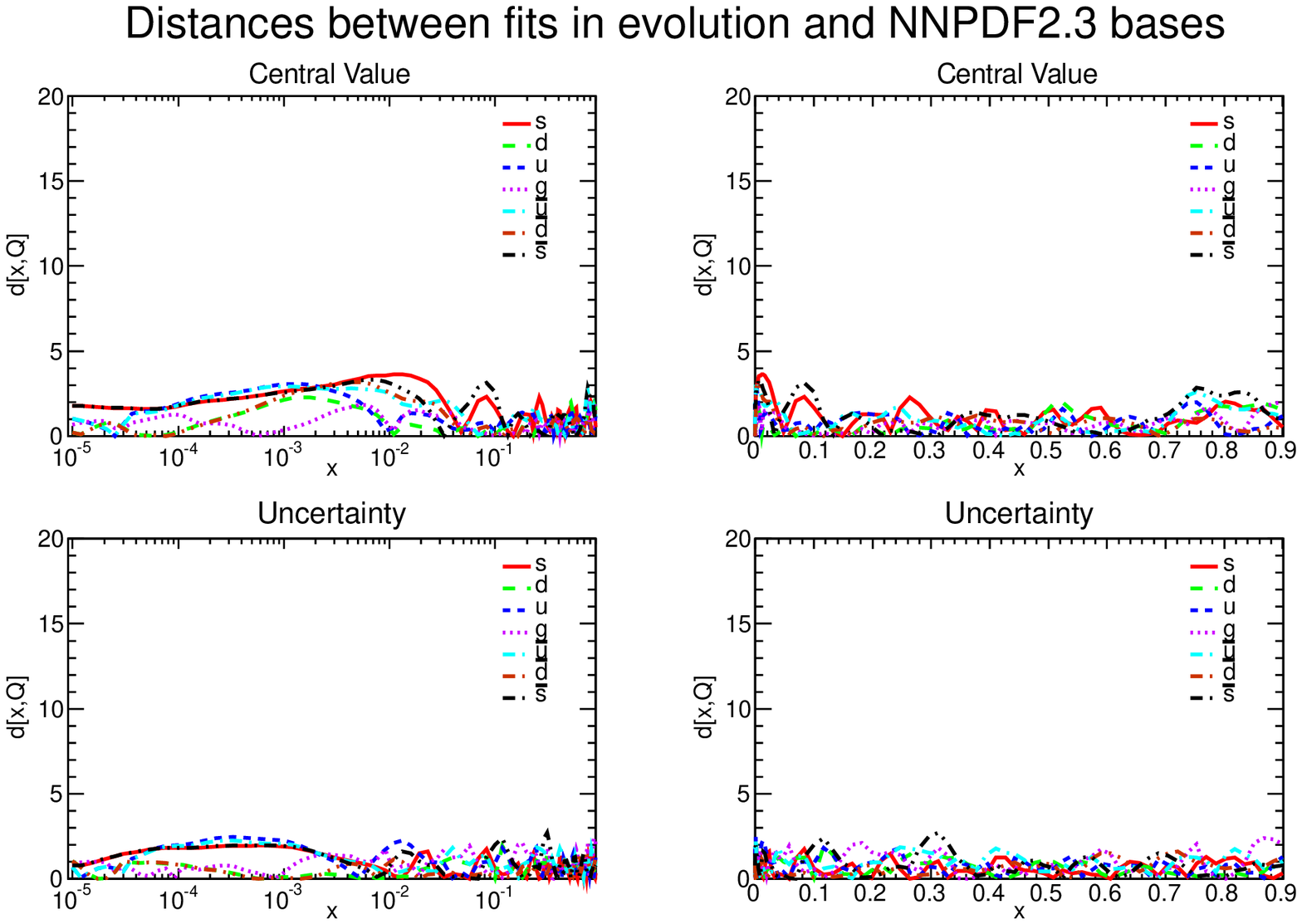}
  \caption{\small Same as Fig.~\ref{fig:long-tr} for the
Level~2 closure fits based on the NNPDF3.0 and
NNPDF2.3 input parametrization basis,
fits C9 and C112 in Table~\ref{tab:CTfits}.  }
  \label{fig:dist-diff-basis}
\end{figure}

In order to test for this,  we have produced a closure test Level~2 fit,
C12 in Table~\ref{tab:CTfits}, which is otherwise identical to
fit C9 but uses the NNPDF2.3 parametrization basis,
Eq.~(\ref{eq:nnpdf23basis}), rather than the new NNPDF3.0
parametrization basis, Eq.~(\ref{eq:nnpdf30basis}).
The distances between fits C9 and C12, that is, between
fits in the NNPDF3.0 and NNPDF2.3 parametrization basis, are shown in
Fig.~\ref{fig:dist-diff-basis}.
We see that distances are quite small, though slightly larger than
for statistically indistinguishable fits.

Distances are typically between $d\sim 1$ and $d\sim 3$, so some
slight differences due to the choice of basis are found, which are
however much smaller than the corresponding PDF
uncertainties: this means that basis independence is satisfied up to
inefficiencies of the algorithm at the quarter-sigma level.
Of course, this does not fully test for basis independence, as three
of the basis combinations (singlet, triplet, and valence) are the same
in the two bases which are being compared. However, it is worth
noticing that PDFs with a relatively simple shape (with a single
maximum) such as the strange combinations in one basis, are obtained
as linear combinations of PDFs with a more complex dip-bump shape in
the other basis: the stability of results then provides a strong check
that the final PDF shape is independent of a bias due to the form
of the parametrizing function.
We can conclude that our results
are independent for all practical purposes of the choice
of fitting basis.
As will be shown in Sect.~\ref{sec:bastability}, similar conclusions will
be derived of the corresponding tests in the fits
to real experimental data.

\subsubsection{Redundancy of the neural network architecture}
\label{sec:huge}

To be used as reliable unbiased interpolants,  neural networks
have to be characterized by an architecture redundant enough for the
problem at hand: adding or removing nodes, even a substantial number of
them,
should leave the final results unchanged.
The choice of a 2-5-3-1 architecture as baseline for the NNPDF fits
can be justified precisely by showing its redundancy, as
we did in previous work~\cite{Ball:2008by}.
Using the closure testing technology, we have revisited for
the current fits the  stability of the fit results
with respect to this choice of
the architecture of the neural networks.

Starting from our baseline architecture, 2-5-3-1, which has 37 free
parameters for each PDF, or 259 in total, we have performed a
closure fit with instead a huge neural network, with architecture
2-20-15-1, therefore increasing by more than a factor 10 the
number of free parameters in the fit.
Other than this modification, exactly the same fit settings were used
in the two cases.

\begin{figure}[t]
  \centering
  \epsfig{width=0.99\textwidth,figure=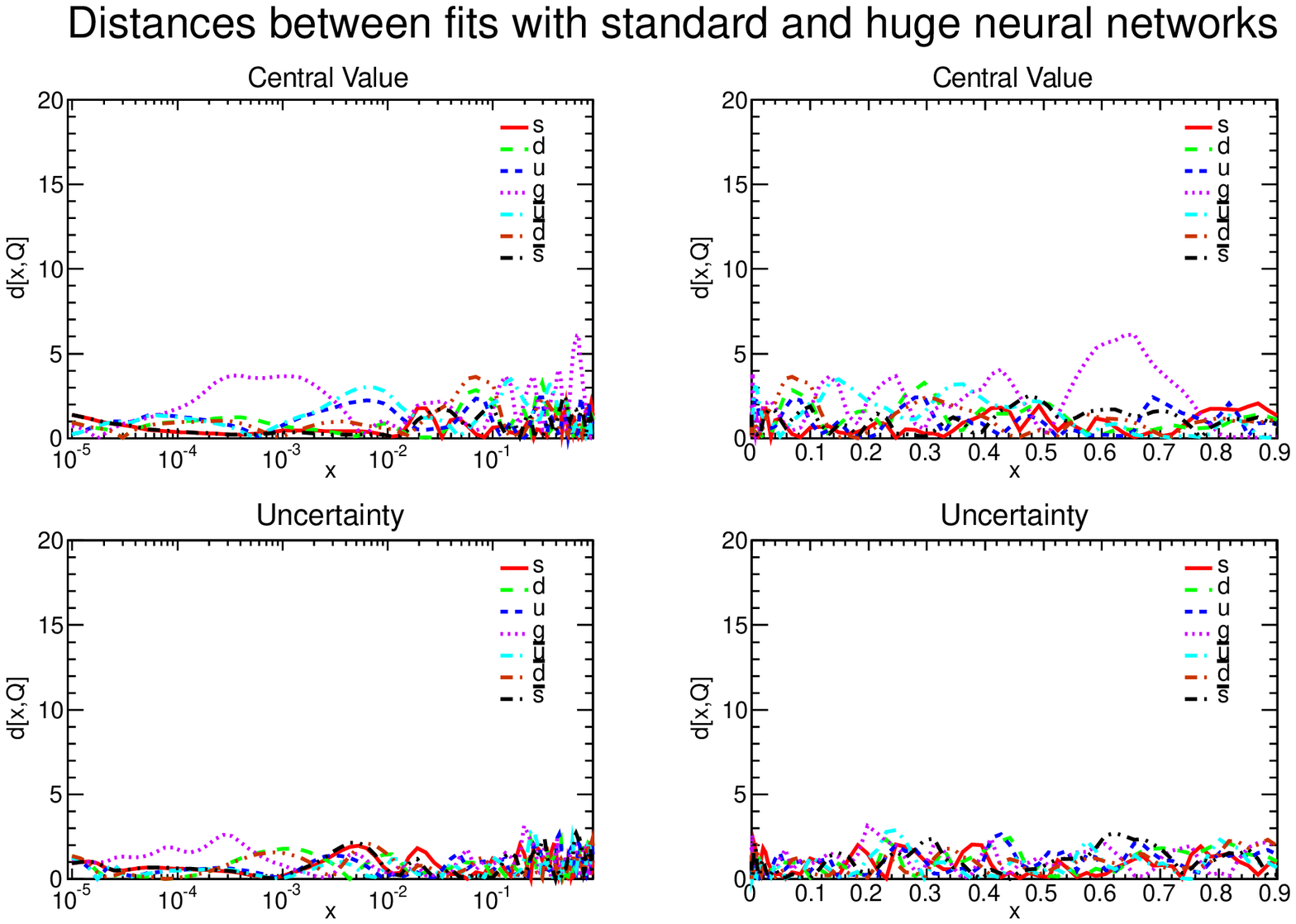}
  \caption{\small Same as Fig.~\ref{fig:long-tr} for the
Level~2 closure fits based on 2-5-3-1 and 2-20-15-1
architectures,
fits C9 and C15 in Table~\ref{tab:CTfits}. }
  \label{fig:dist-diff-arch}
\end{figure}

 The distances between the two fits can be seen in
Fig.~\ref{fig:dist-diff-arch}.
Remarkably, even increasing the architecture to have a factor 10 more
free parameters,
the fit using the huge networks is still quite close to the
fit with standard sized networks.
Again, we do not have perfect statistical equivalence, but
inefficiencies are well below the half-sigma level.

This test also provides another demonstration of the resilience to
over-fitting of our fitting methodology.
Indeed, one might fear that such a large increase in architecture
would lead to massive overlearning, while in actual fact results are
virtually unchanged.
This is thus a significant test of the effectiveness of the look-back method
in determining the optimal stopping point.

Interestingly, we also observe a very moderate increase in
CPU time when using the huge network as compared to our
baseline architecture. This is because when
a network is very redundant, it often makes little difference
which parameters the algorithm changes.

Finally, let us mention that
several other tests of neural network architecture were also
performed, including changing the number of layers in the networks and
the number of inputs.
In most cases the results were either very similar to those from the
standard structure or (for some smaller networks) noticeably worse.

\subsubsection{Robustness with respect to the choice of input PDF set}

So far in this section we have shown results fitted to the
pseudo-data generated using MSTW08 as input PDF,
and studied
 the effects of changing some of the settings of the closure test fits.
However, it is important to verify that there is nothing special in
using pseudo-data generated with MSTW08, and
that our methodology is flexible enough so that
similar successful closure fits are achieved if other PDFs
are used as input.
In particular, we want to explicitly verify that self-closure is successful:
using NNPDF3.0 as input PDF, and checking
 that it is correctly reproduced
by the closure fit.
Note that the MSTW08 and CT10 parametrizations are relatively simple, so
we need to verify that everything works fine even when a very flexible
PDF parametrization is used as input in the closure test.

As a preliminary test, we have verified  that Level~0 tests works
regardless of the input PDFs, by
producing a number of
fixed length Level~0 fits for different input PDFs, and checking that all the
conclusions of Sect.~\ref{sec:level0cl} are unchanged if input PDFs other than MSTW08
are used to generate the pseudo-data.
We have then moved to Level~2 tests:
we have performed closure test fits using exactly the same methodology as those
based on MSTW08,
but now using pseudo-data generated using different input PDF sets, in
particular with the
CT10 and NNPDF3.0 NLO sets.
In the latter case, we have performed closure tests
with and without the inclusion of the generalized positivity constraints
described in Sect.~\ref{sec:positivity}.
Given that NNPDF3.0 satisfies these constraints, this is a useful
cross-check.
These closure tests are labeled in Table~\ref{tab:CTfits}
as C10 (for CT10) and C16 and C17 (for NNPDF3.0
without and with positivity constraints).

\begin{figure}[t]
  \centering
  \epsfig{width=0.95\textwidth,figure=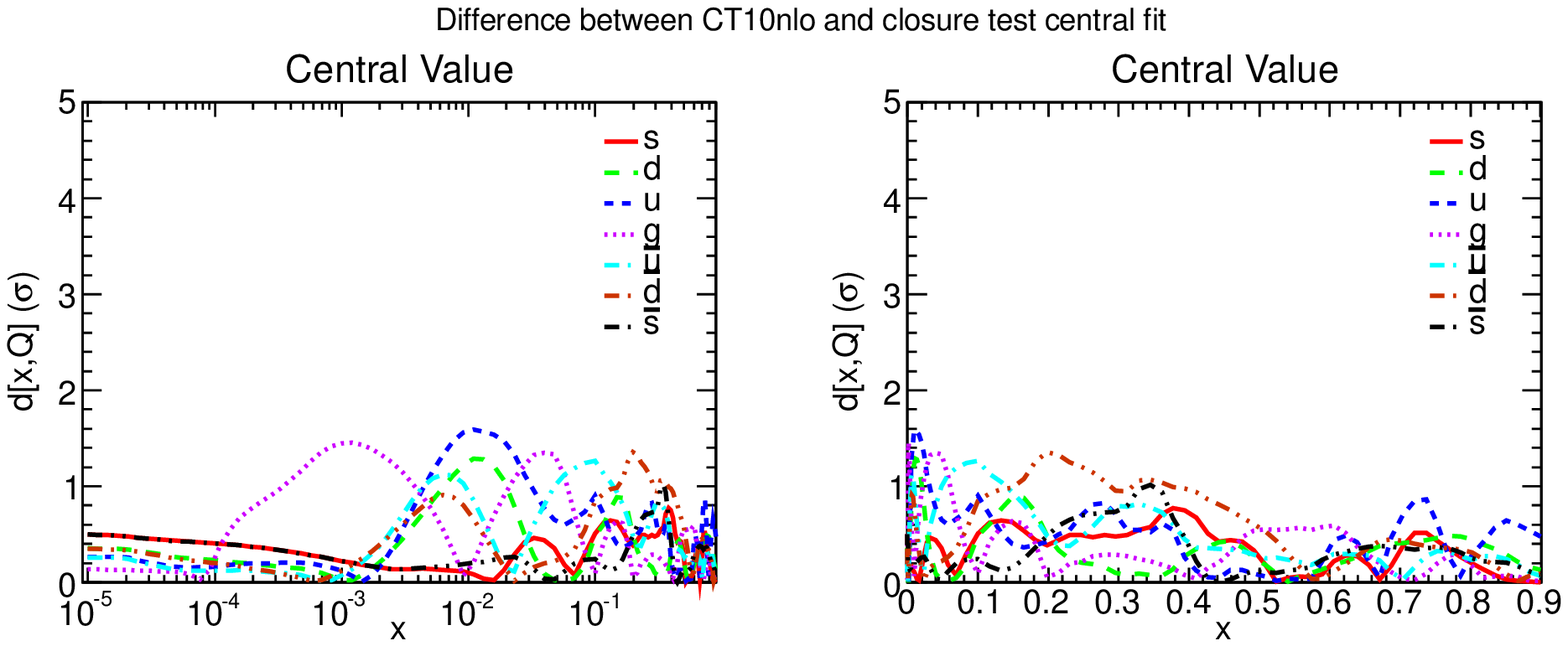}
  \caption{\small Distances,
Eq.~(\ref{eq:distancesclosure2}), between the central values of the PDFs from the
closure test fit and the CT10 PDFs, in units of the standard deviation of the fit PDFs.}
  \label{fig:dist-diff-ct10}
\end{figure}

First,  Fig.~\ref{fig:dist-diff-ct10} shows the
distances between the fitted and the input PDFs,
Eq.~(\ref{eq:distancesclosure2}), for the closure test fit which uses the
CT10 NLO PDF set to generate the data.
Let us recall, as discussed above, that
these distances are normalized
to the standard deviation of the fitted PDFs, and thus a value
$d_{\sigma}\sim 1$ indicates that input and fitted PDFs agree at the one sigma
level, and so on.
These distances should be compared with the corresponding results
obtained using MSTW08 as input, Fig.~\ref{fig:L2-MSTW-dist}.
We observe that, just as with MSTW08,
the fitted PDFs are mostly within one sigma of the input PDFs, and never
more than around two sigma away.
In this respect, the closure test based on CT10 is as successful as that based
on MSTW08.
We have verified that this is also the case using some of the other
estimators considered in this section.

\begin{figure}[t]
  \centering
  \epsfig{width=0.95\textwidth,figure=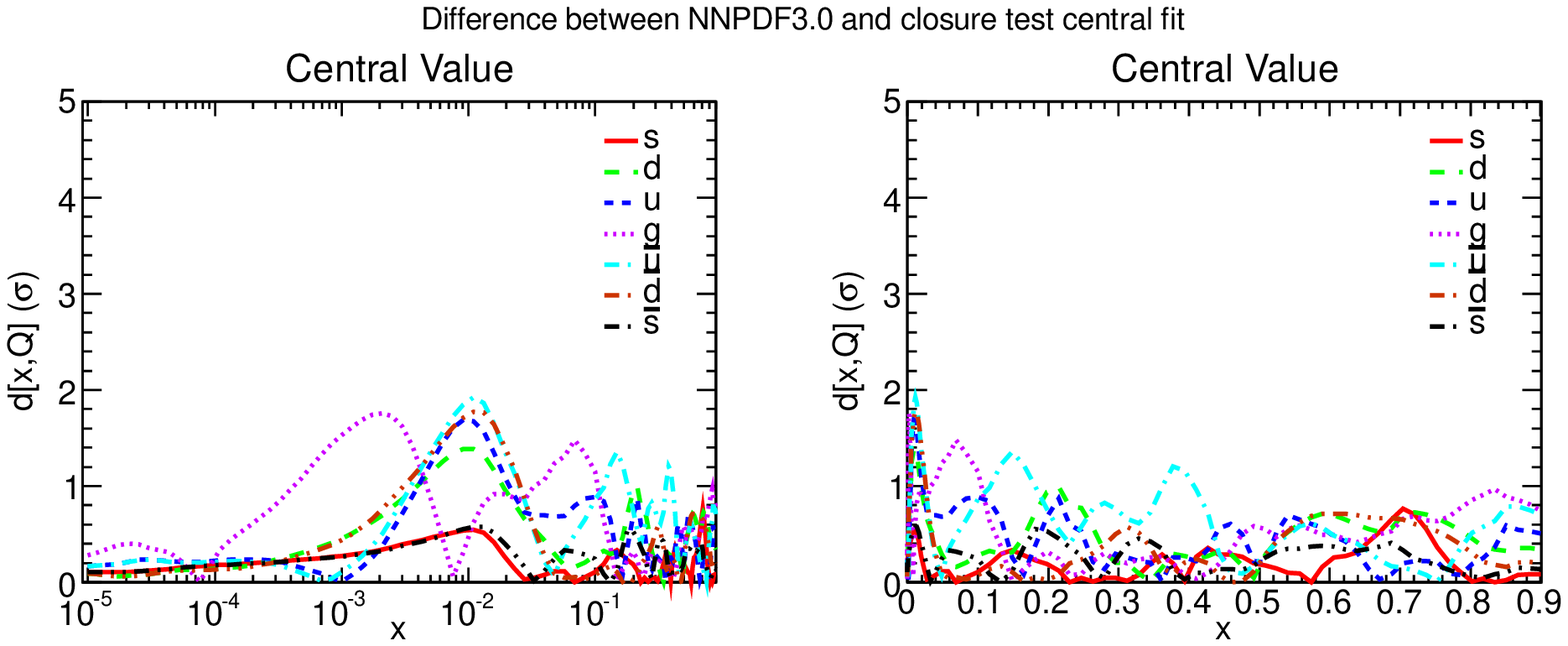}
  \caption{\small Same as Fig.~\ref{fig:dist-diff-ct10} for the closure test
based on NNPDF3.0 as input PDFs, without positivity constraints.}
  \label{fig:dist-nn30}
\end{figure}

The distances Eq.~(\ref{eq:distancesclosure2}) are shown again in
Fig.~\ref{fig:dist-nn30} now for a closure test fit using as input
the NNPDF3.0 NLO PDFs.
As in other closure tests, no generalized positivity is required here.
This fit is therefore the ultimate closure test,
providing evidence of self-closure,
where the PDF set used to generate the pseudo-data
has been determined using the same methodology as the closure test fit.
Again, the agreement using NNPDF3.0 as input is as good as that obtained
using other PDFs with a less flexible parametrization, showing
that the closure tests works also in the case of a initial condition
with substantial structure.

\begin{figure}[t]
  \centering
  \epsfig{width=0.95\textwidth,figure=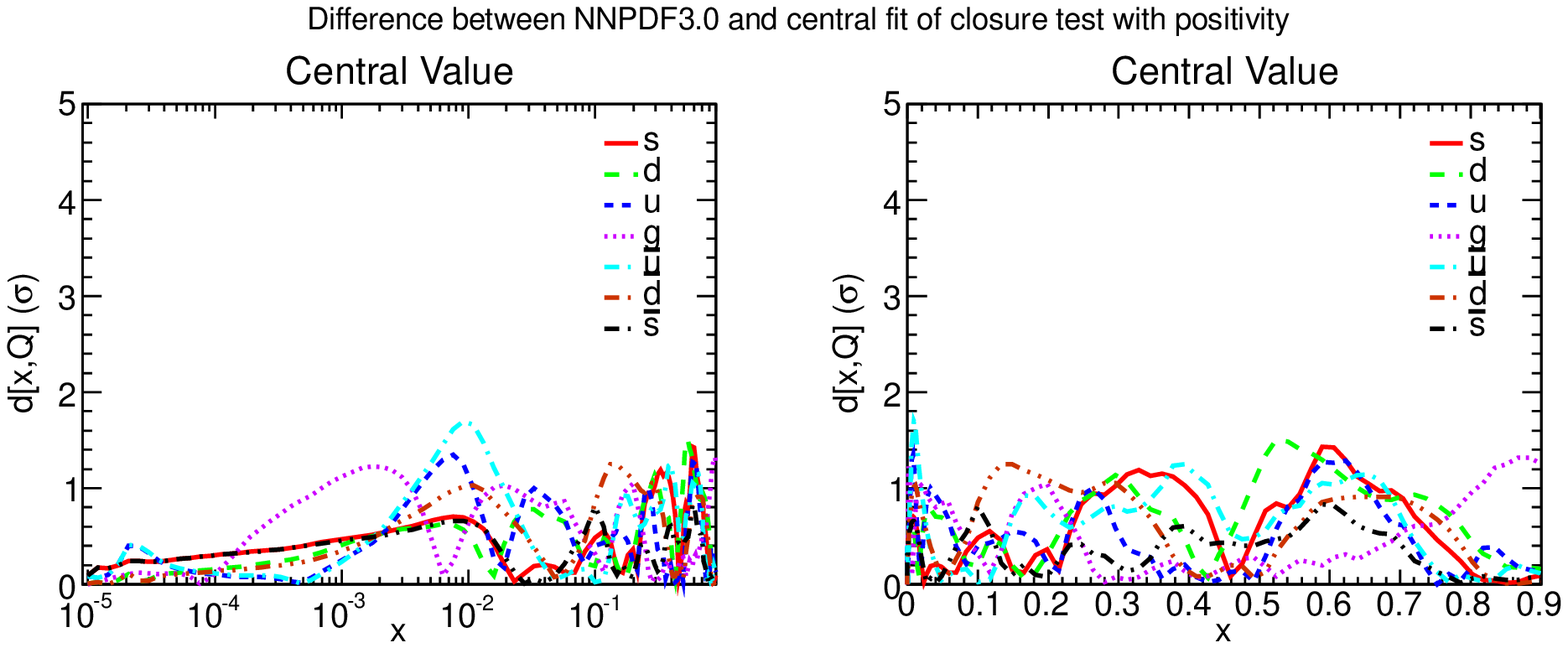}
  \caption{\small Same as Fig.~\ref{fig:dist-nn30}, this time
including the generalized positivity constraints in the closure
test fit.
See text for more details.
}
  \label{fig:dist-diff-nn30p}
\end{figure}

All of the closure test fits shown so far have been performed without
the positivity constraints used in the fits to real data,
described in Sect.~\ref{sec:positivity}.
The motivation for this is that for some of the input PDFs used in the closure
tests, in particular MSTW08, generalized positivity is not satisfied
 and therefore
including such constraints
would result in inconsistencies
with the generated pseudo-data in the closure test, potentially biasing
the results.
On the other hand, as shown in Sect.~\ref{sec:positivityresults},
the NNPDF3.0 fits satisfy the generalized positivity constraints by construction,
and therefore if we use NNPDF3.0 as input PDF we can include generalized
positivity in the closure test, expecting to find no differences
with respect to the previous case.
Fig.~\ref{fig:dist-diff-nn30p} shows the
distances, Eq.~(\ref{eq:distancesclosure2}), for the closure test
using as input the NNPDF3.0 NLO PDFs and now also with
the generalized positivity constraints imposed during the closure test fit.
We indeed find that the level of agreement is similar as that of the
closure test
when no positivity constraints were imposed during the minimization.
%


\section{The NNPDF3.0 PDF set}
\label{sec:results}

In this section we present the NNPDF3.0 LO, NLO and NNLO global fits.
First, we discuss  fit quality and the  dependence of the
$\chi^2$ on its exact definition and on the treatment of
systematic and normalization uncertainties.
Then we show results for PDFs, we compare
the new sets with NNPDF2.3 and with other existing PDF sets.

Next, we explore  the dependence
of the NNPDF3.0 partons on the choice of dataset.
We propose a new definition of conservative PDF set,
based on an estimator which allows for an assessment of the
consistency of an individual dataset with the global fit,
 and compare these conservative
partons with the global fit results.
We then study a wide range of variations of the fitted dataset, including
fits without LHC data, fits with only HERA and LHC data, and fits
based on the NNPDF2.3 dataset.
Fits to reduced datasets are also used to study
the impact of jet data  on the global fit and their stability,
and the size of the strange PDF, which has been the object of various
recent studies.

We then turn to an assessment of the stability of the NNPDF3.0 results
upon variations in the fitting methodology.
These include the impact of positivity constraints, the stability
upon change of fitting basis, and the dependence on the
additive or multiplicative (recall Sect.~\ref{sec:chi2definition})
treatment of systematic errors.

Finally, we study the implications of NNPDF3.0
for LHC phenomenology.
We compare NLO and NNLO PDF luminosities at $\sqrt{s}=$13 TeV
with NNPDF2.3 and with CT10 and MMHT.
We then provide predictions for  the LHC
at 13 TeV, using the {\sc\small MadGraph5\_aMC@NLO} program~\cite{Alwall:2014hca}, for a number of representative
processes like vector boson production, top quark production
and Higgs production.
In addition, we also  discuss,
using the {\sc\small iHixs}~\cite{Anastasiou:2011pi} code, the implications
for the dominant Higgs production channel at the LHC,
gluon-fusion, and compute NNLO cross-sections.
In this case we also study the dependence of $\sigma(gg\to h)$
on the choice of fitted dataset.
To complete our brief exploration of
the implications of NNPDF3.0 at the LHC, we study the production
of systems of very high invariant masses, close to the kinematic
threshold, which are relevant for searches of massive New Physics
at the energy frontier.

\subsection{The NNPDF3.0 set of parton distributions}

In this section first of all
we discuss the quality of the LO, NLO and NNLO
fits and the  dependence of the $\chi^2$ on its definition,
and then present
the NNPDF3.0 PDFs and uncertainties and compare this PDF set
to the previous NNPDF2.3 set.
We then turn to a discussion on the way theoretical uncertainties on PDFs
could be estimated by comparing PDFs at different perturbative
orders. We finally briefly discuss other sources of uncertainties
which are currently not included in the total PDF uncertainty.

\subsubsection{Fit quality}
\label{sec:quality}

In Tab.~\ref{tab:chi2tab_exp_vs_t0} we present
the results for the fit quality of the global LO, NLO and NNLO
sets.
The comparison is performed for a common value of
$\alpha_s(M_Z)=0.118$ (but PDFs  in a wide range of
values of $\alpha_s(M_Z)$ are also available, see Sect.~\ref{sec:delivery}).
We show the results obtained  both using the experimental
and the $t_0$ $\chi^2$  definition (see Refs.~\cite{Ball:2009qv,Ball:2012wy}).
Note that the $t_0$ definition varies with the perturbative order, as  it
depends on the theoretical values of the
cross-sections included in the fit.
The $t_0$
$\chi^2$ is used for minimization as it corresponds to an unbiased
maximum-likelihood estimator even in the presence of multiplicative
uncertainties~\cite{Ball:2009qv} (see Sect.~\ref{sec:chi2definition}).
The experimental definition, which is based on the
experimental covariance matrix, cannot be used for minimization as it
would lead to biased results, but it is best suited for benchmarking as
it only depends on publicly available results (the final PDFs and the
experimental covariance matrix).

\begin{table}[t]
\footnotesize
\begin{center}
\begin{tabular}{c||ccc|ccc|ccc}
\hline
 & \multicolumn{3}{c}{LO} & \multicolumn{3}{c}{NLO} & \multicolumn{3}{c}{NNLO} \\
\hline
 & $N_{\rm dat}$  & $\chi^{2}_{\rm exp}$ & $\chi^{2}_{\rm t_0}$  & $N_{\rm dat}$  & $\chi^{2}_{\rm exp}$ & $\chi^{2}_{\rm t_0}$ & $N_{\rm dat}$  & $\chi^{2}_{\rm exp}$ & $\chi^{2}_{\rm t_0}$ \\
\hline
\hline
Total & 4258 	& 2.42 & 2.17  & 4276 	& 1.23  & 1.25 & 4078 	& 1.29 & 1.27 \\
\hline
NMC $d/p$ & 132&  1.41 & 1.09 & 132 & 0.92 & 0.92 & 132 & 0.93 & 0.93\\
NMC  & 224 & 2.83 & 3.3 & 224 & 1.63 & 1.66 & 224 & 1.52 & 1.55\\
SLAC  & 74 & 3.29 & 2.96 & 74 & 1.59 & 1.62 & 74 & 1.13 & 1.17\\
BCDMS & 581 & 1.78 & 1.78 & 581 & 1.22 & 1.27 & 581 & 1.29 & 1.35\\
CHORUS  & 862 & 1.55 & 1.16 & 862 & 1.11 & 1.15 & 862 & 1.09 & 1.13\\
NuTeV  & 79 & 0.97 & 1.03 & 79 & 0.70 & 0.66 & 79 & 0.86 & 0.81\\
HERA-I  & 592 & 1.75 & 1.51 & 592 & 1.05 & 1.16 & 592 & 1.04 & 1.12\\
ZEUS HERA-II  & 252 & 1.94 & 1.44 & 252 & 1.40 & 1.49 & 252 & 1.48 & 1.52\\
H1 HERA-II  & 511 & 3.28 & 2.09 & 511 & 1.65 & 1.65 & 511 & 1.79 & 1.76\\
HERA $\sigma_{\rm NC}^{c}$  & 38 & 1.80 & 2.69 & 47 & 1.27 & 1.12 & 47 & 1.28 & 1.20\\
\hline
E886 $d/p$  & 15 & 2.04 & 1.10 & 15 & 0.53 & 0.54 & 15 & 0.48 & 0.48\\
E886 $p$  & 184 & 0.98 & 1.64 & 184 & 1.19 & 1.11 & 184 & 1.55 & 1.17\\
E605  &  119 &0.67 & 1.07 & 119 & 0.78 & 0.79 & 119 & 0.90 & 0.72\\
CDF $Z$ rapidity  & 29 & 2.02 & 3.88 & 29 & 1.33 & 1.55 & 29 & 1.53 & 1.62\\
CDF Run-II $k_t$ jets  & 76 &  1.51 & 2.12 & 76 & 0.96 & 1.05 & 52 & 1.80 & 1.20\\
D0  $Z$ rapidity & 28 & 1.35 & 2.48 & 28 & 0.57 & 0.68 & 28 & 0.61 & 0.65\\
\hline
ATLAS $W,Z$ 2010  & 30 & 5.94 & 3.20 & 30 & 1.19 & 1.25 & 30 & 1.23 & 1.18\\
ATLAS 7 TeV jets 2010   & 90 & 2.31 & 0.62 & 90 & 1.07 & 0.52 & 9 & 1.36 & 0.85\\
ATLAS 2.76 TeV jets  & 59 & 3.88 & 0.61 & 59 & 1.29 & 0.65 & 3 & 0.33 & 0.33\\
ATLAS high-mass DY  & 5 & 13.0 & 15.6 & 5 & 2.06 & 2.84 & 5 & 1.45 & 1.81\\
ATLAS $W$ $p_T$  &  -  & - & - & 9 & 1.13 & 1.28 & - & - & - \\
CMS $W$ electron asy  & 11 & 10.9 & 0.95 & 11 & 0.87 & 0.79 & 11 & 0.73 & 0.70\\
CMS $W$ muon asy   &11 & 76.8 & 2.25 & 11 & 1.81 & 1.80 & 11 & 1.72 & 1.72\\
CMS jets 2011  & 133 & 1.83 & 1.74 & 133 & 0.96 & 0.91 & 83 & 1.9 & 1.07\\
CMS $W+c$ total  & 5 & 11.2 & 25.8 & 5 & 0.96 & 1.30 & 5 & 0.84 & 1.11\\
CMS $W+c$ ratio  & 5 & 2.04 & 2.17 & 5 & 2.02 & 2.02 & 5 & 1.77 & 1.77\\
CMS 2D DY 2011  & 88 & 4.11 & 12.8 & 88 & 1.23 & 1.56 & 110 & 1.36 & 1.59\\
LHCb $W$ rapidity  & 10 & 3.17 & 4.01 & 10 & 0.71 & 0.69 & 10 & 0.72 & 0.63\\
LHCb $Z$ rapidity  & 9 & 5.14 & 6.17 & 9 & 1.10 & 1.34 & 9 & 1.59 & 1.80\\
$\sigma(t\bar{t})$ & 6 & 42.1 & 115 & 6 & 1.43 & 1.68 & 6 & 0.66 & 0.61\\
\hline
\hline
\end{tabular}

\end{center}
\caption{\small The values of the $\chi^2$ per data point for the LO, NLO and NNLO
central fits of the NNPDF3.0 family with $\alpha_s(M_Z)=0.118$, obtained using
both the experimental and the  $t_0$ definitions.
 \label{tab:chi2tab_exp_vs_t0}
}
\end{table}

The overall fit quality is  good, with an  experimental $\chi^2$ value
of 1.23 at NLO and 1.29 at NNLO; the $t_0$ values are very close, 1.25
and 1.27 at NLO and NNLO, respectively (see Table~\ref{tab:chi2tab_exp_vs_t0}).
The LO fit is characterized of course by a much poorer fit quality, due
to the missing NLO corrections.
For some experiments like CHORUS, SLAC, ATLAS high-mass Drell-Yan, the $W$ lepton
asymmetry or top quark pair production, the $\chi^2$ improves when going from
NLO to NNLO: for top, in particular, this is related to the presence
of large NNLO
corrections~\cite{Czakon:2013goa,Czakon:2013tha} (a good $\chi^2$
at  NLO would require unnaturally small values of the top mass, far
from the current PDG value). However,
for most of the experiments it remains either very similar or gets
slightly worse.
This is also the case for the new HERA-II datasets.
For the jet data the fit quality is quite similar at NLO and NNLO using the
$t_0$ definition, but note that the kinematical cuts in the two cases
are different (see Sect.~\ref{sec:exclusion}).
This is also the case for the CMS Drell-Yan data: the $\chi^2$ is slightly
worse at NNLO but only because at NLO we impose kinematical cuts that
remove the region with large NNLO corrections: without such cuts, the
$\chi^2$ is much poorer at NLO.

Another interesting feature that
one can observe from Tab.~\ref{tab:chi2tab_exp_vs_t0} is that, even for
the same underlying PDFs,
the numerical differences between the two definitions of the
$\chi^2$ can be substantial.
This effect is particularly acute
for experiments where systematic uncertainties dominate
over statistical ones, and emphasizes the crucial role
of a careful estimation of systematic errors for PDF fitting.
One such example is provided by the NNLO fit, where for
the CMS inclusive jet data the best-fit $\chi^2$ changes
from 1.90 (experimental definition) to 1.07 ($t_0$ definition).
Reassuringly, as we will show in Sect.~\ref{sec:additive} below, these differences
in the value of the $\chi^2$  do not have a large impact on the PDFs,
which are rather stable upon changes of the $\chi^2$ definition.
The dependence of the $\chi^2$ on its definition is weaker
for fixed
target experiments
and  DIS data, for which statistical uncertainties are dominant.

\subsubsection{Parton distributions}
\label{sec:nnpdf30set}
\begin{figure}[t]
\begin{center}
\epsfig{width=0.73\textwidth,figure=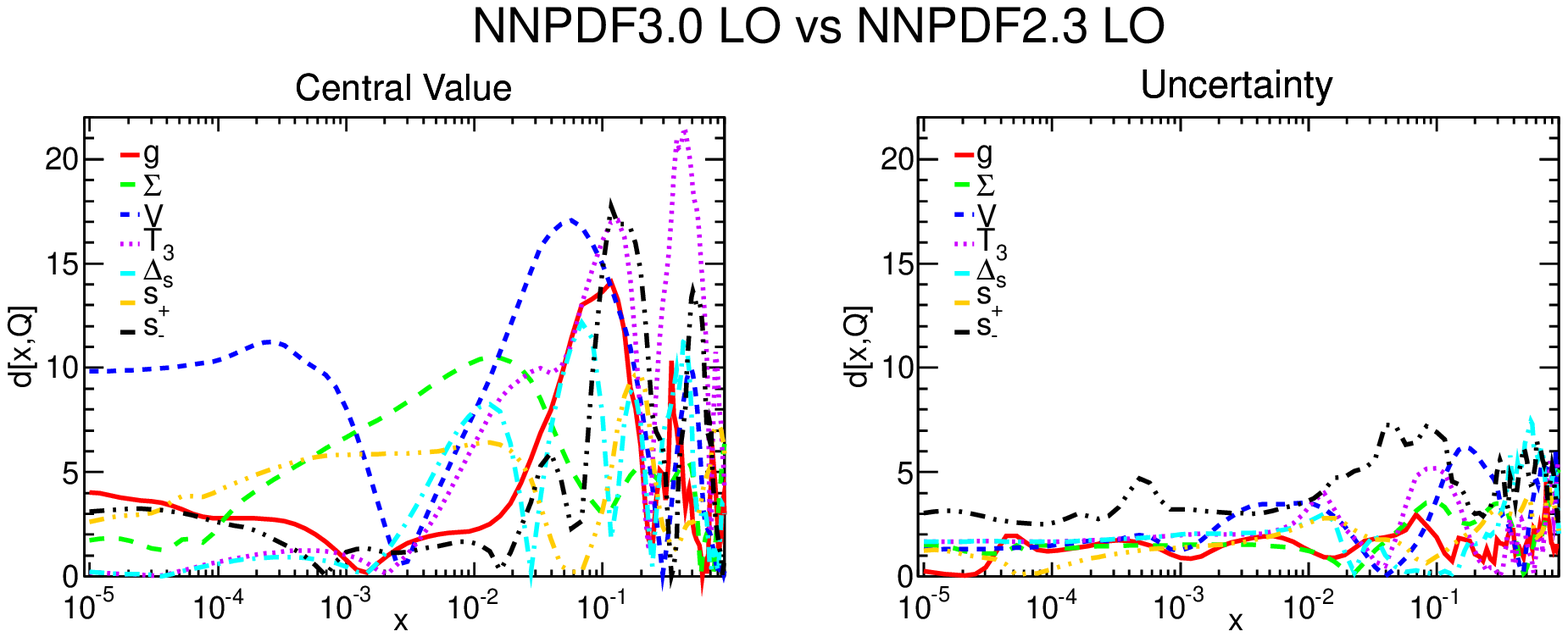}
\epsfig{width=0.73\textwidth,figure=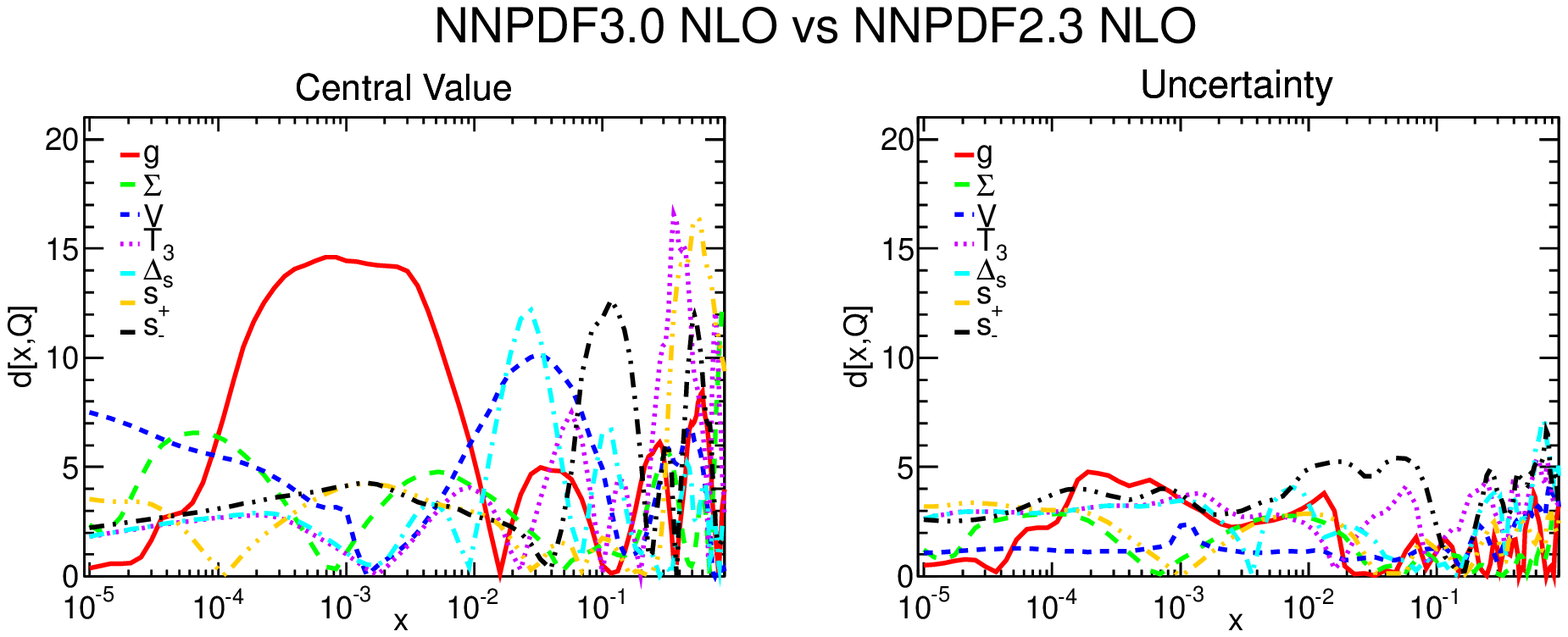}
\epsfig{width=0.73\textwidth,figure=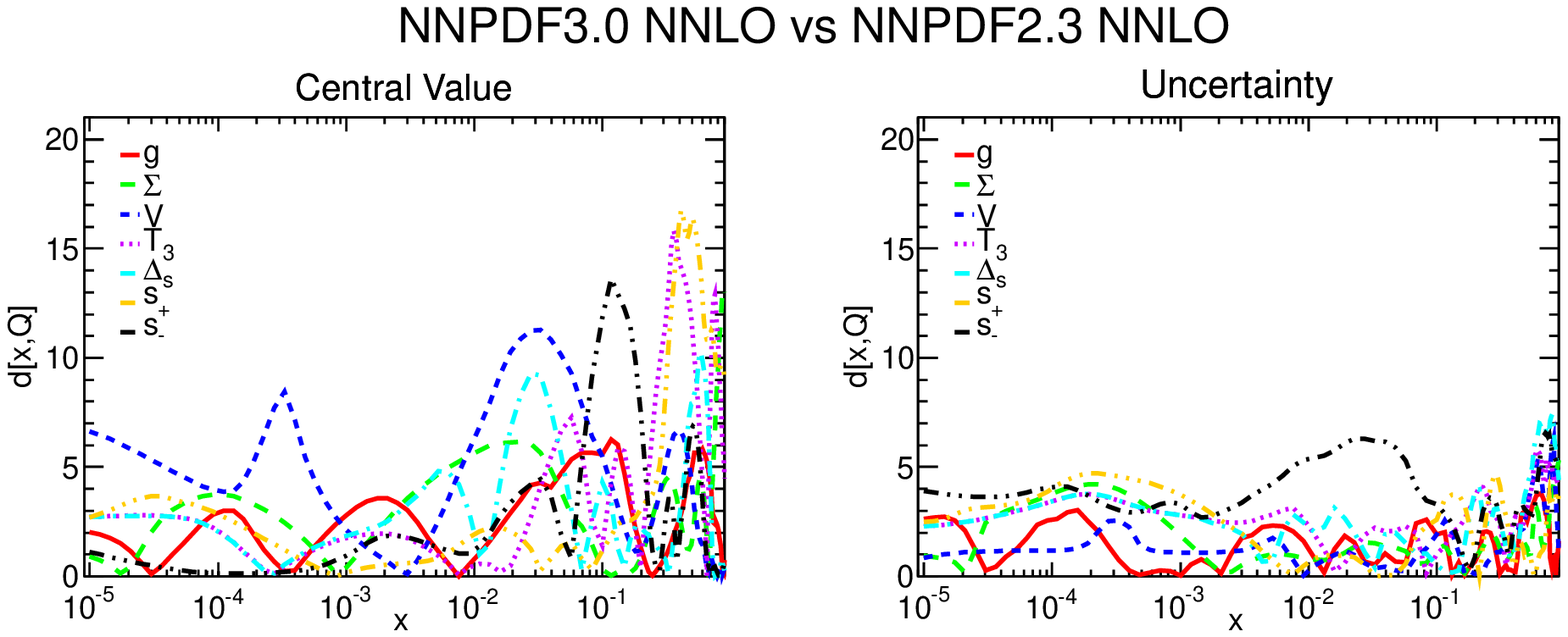}
\caption{\small
Distances  between NNPDF2.3 and NNPDF3.0 at LO (top), NLO
(center)
and  NNLO (bottom) PDFs, with $\alpha_s(M_Z)=0.118$.
 computed between sets
of $N_{\rm rep}=100$ replicas at  $Q^2=2$ GeV$^2$.
Note that at LO, the  NNPDF2.3 set has  $\alpha_s(M_Z)=0.119$.
 \label{fig:distances_30_vs_23_nnlo}}
\end{center}
\end{figure}

We now compare the NNPDF3.0 LO, NLO and NNLO partons, with  $\alpha_s(M_Z)=0.118$,
with the corresponding NNPDF2.3 sets
 and with each other.
In Fig.~\ref{fig:distances_30_vs_23_nnlo} we show the distances
between the parton distributions in the NNPDF3.0 and NNPDF2.3 sets
for the three perturbative orders, LO, NLO and NNLO.
We recall that when comparing two sets of $N_{\rm rep}=100$ replicas,
$d\sim 1$
means that the two sets are statistically equivalent (they cannot be
distinguished from replica sets extracted from the same underlying
probability distribution), while  $d\sim 10$ means that the sets
correspond to PDFs  that agree at the one-sigma level. A
full discussion of distances in given in Appendix~\ref{app:distances};
note that in comparison to previous NNPDF papers we have slightly
changed the algorithm used in computing distances (in particular by
removing an averaging and smoothing procedure), without changing
their statistical interpretation.
Distances are computed at a scale of $Q^2=2$ GeV$^2$.
At LO, when comparing to  NNPDF2.3 we use the set with
$\alpha_s(M_Z)=0.119$, since
$\alpha_s(M_Z)=0.118$ is not available for NNPDF2.3 LO. This has a
minor effect on  the comparison.

The sizes of the distances shown  in Fig.~\ref{fig:distances_30_vs_23_nnlo},
vary significantly with the  perturbative order.
At LO, the gluon in the two sets is in very good agreement for $x \lsim 0.01$.
This suggests that Monte Carlo tunes (which strongly depend on the
small-$x$ gluon) based on NNPDF2.3LO, such as the Monash 2013
tune of {\sc\small Pythia8}~\cite{Skands:2014pea} should also work
reasonably well with NNPDF3.0LO.
On the other hand, larger differences, between one and two-sigma, are found at medium and large $x$,
both for the quarks and the gluon.
Note that, however, at LO theory uncertainties dominate over PDF uncertainties.

At NLO and NNLO, NNPDF2.3 and NNPDF3.0
are typically in agreement at the one-sigma
level, with occasionally somewhat larger distances, of order 1.5--sigma.
In particular, while the total quark singlet PDF is relatively stable, there
are larger
differences for  individual quark flavors, especially at medium
and large-$x$.
Significant differences are also found for the gluon PDF, especially
at NLO, where however it should be kept in mind that NNPDF2.3 used the
FONLL-A treatment of heavy quarks, while NNPDF3.0 uses  FONLL-B (see
Sect.~\ref{sec:hq}).
This comparison also shows that PDF uncertainties change
at the level of one-sigma: this is to be expected, as a consequence of
the constraints coming from new data, and the improved fitting
methodology.

\begin{figure}[t]
\begin{center}
\epsfig{width=0.42\textwidth,figure=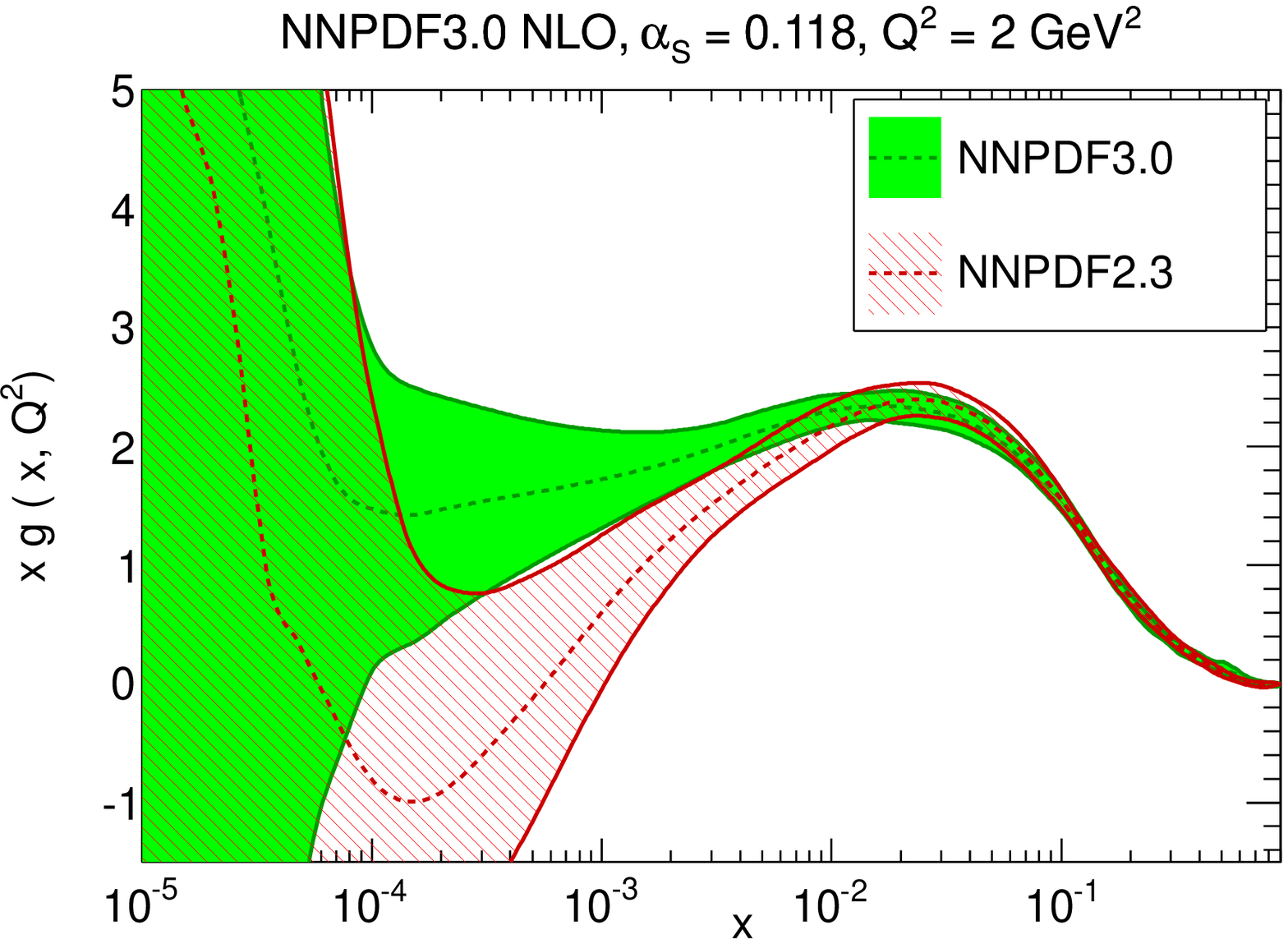}
\epsfig{width=0.42\textwidth,figure=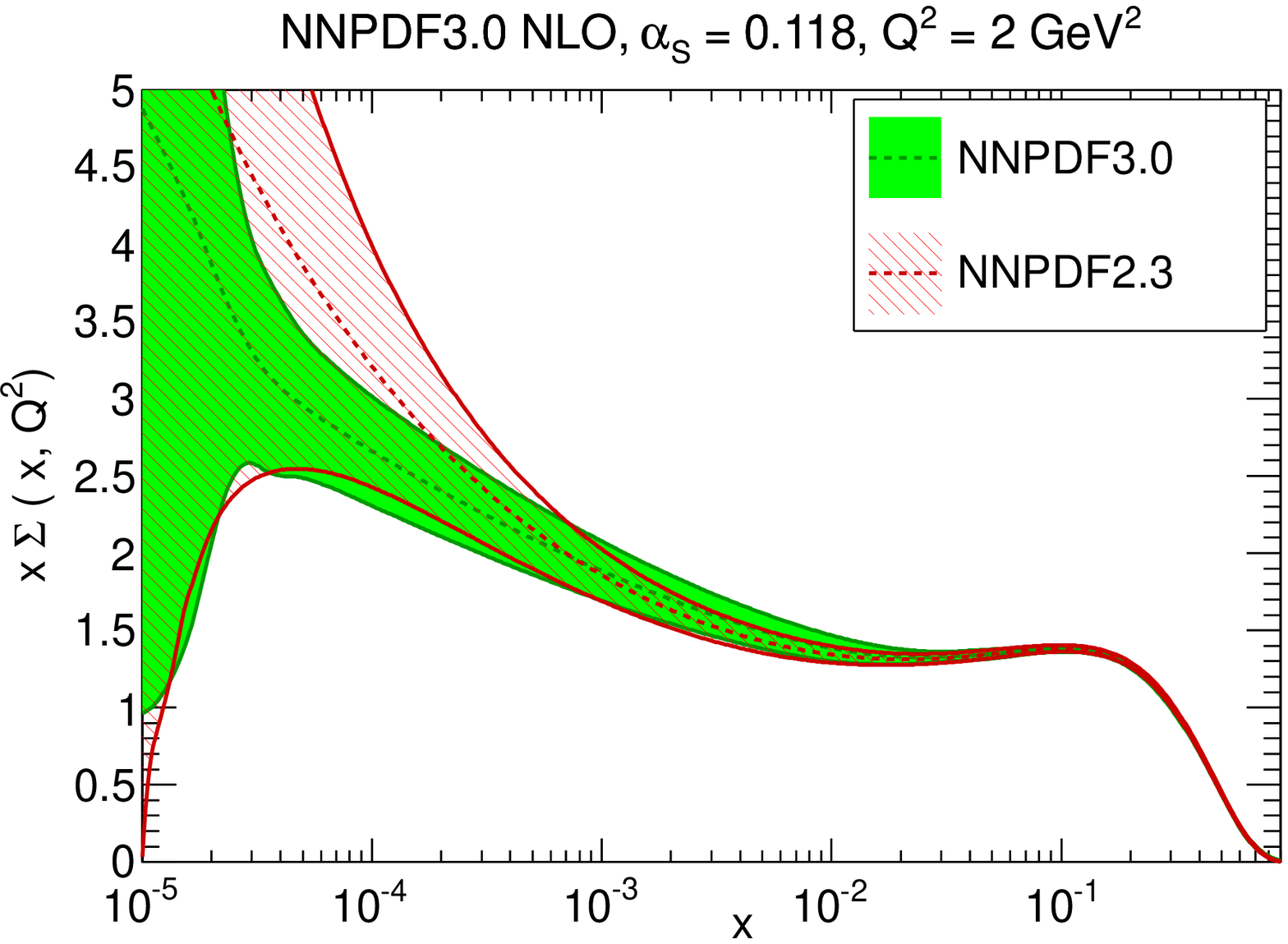}
\epsfig{width=0.42\textwidth,figure=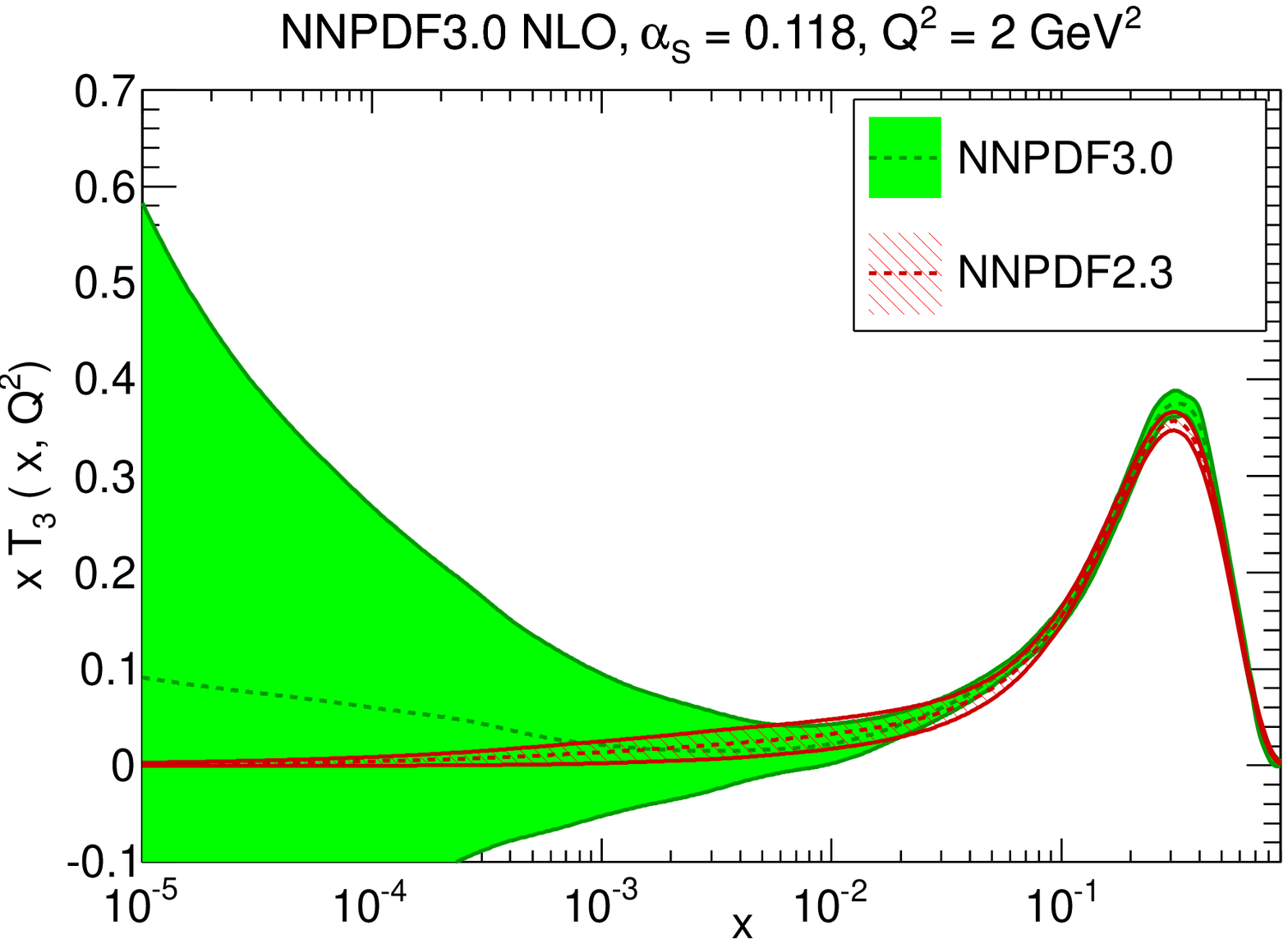}
\epsfig{width=0.42\textwidth,figure=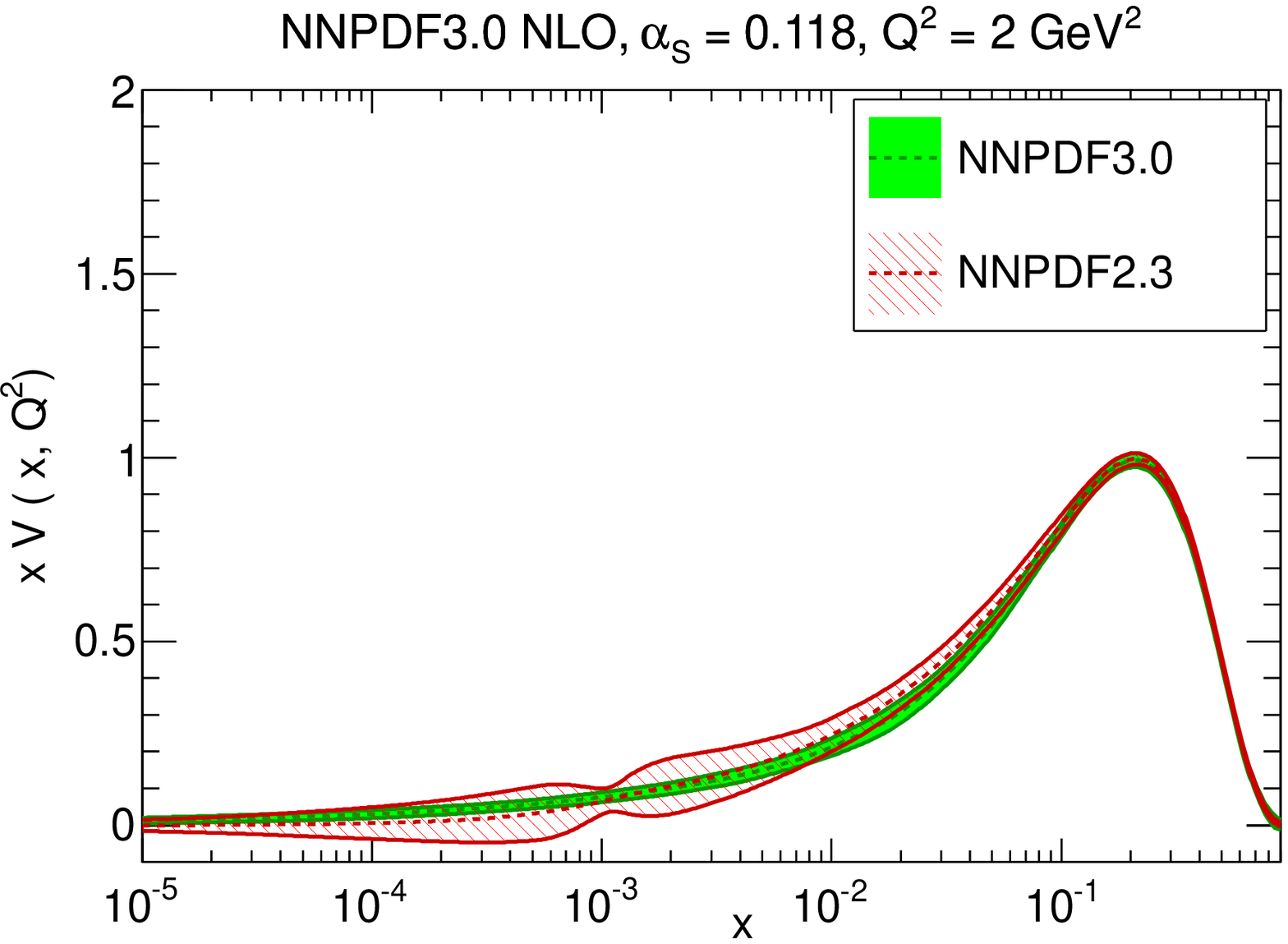}
\caption{\small
Comparison of
NNPDF2.3 and NNPDF3.0 NLO PDFs  at $Q^2=2$ GeV$^2$ with   $\alpha_s(M_Z)=0.118$.
From top to bottom and from left to right the
gluon,  singlet,  isospin triplet and
total valence are shown. \label{fig:30_vs_23_lowscale_nlo}}
\end{center}
\end{figure}

Now we turn to the direct comparison of NNPDF2.3 and NNPDF3.0 NLO PDFs:
in Fig.~\ref{fig:30_vs_23_lowscale_nlo} the gluon, singlet PDF,
isospin triplet and
total valence PDFs are shown, with $\alpha_s(M_Z)=0.118$ at $Q^2=2$ GeV$^2$.
We can see that in the NNPDF3.0 NLO set,
the central value of the gluon never turns negative, even at
small-$x$:
it is flat down to $x\sim 10^4$ and then it begins
to grow, within its large uncertainty, always remaining above its
NNPDF2.3 counterpart. The
 difference can be understood as a consequence  of moving to the
FONLL-B heavy quark scheme, and due
to the more stringent positivity constraints that are imposed now (see
Sect.~\ref{sec:positivity}). For the total  quark singlet there
is good agreement between 2.3 and 3.0.
For the quark triplet we see two interesting features: first at large $x$ the
result in 3.0 is larger than in 2.3, especially in the region where the PDF peaks, and
also the small-$x$ uncertainties are substantially larger, suggesting
that they were somewhat underestimated in NNPDF2.3, presumably due to
the less flexible methodology and less efficient treatment of
preprocessing (see Sect.~\ref{sec:preproc}).
As will be shown below, the small-$x$ uncertainties
in the triplet between NNPDF2.3 and NNPDF3.0 become more similar
when the comparison is performed in terms of 68\%
confidence level intervals, see Fig.~\ref{fig:3068cl}.

\begin{figure}[t]
\begin{center}
\epsfig{width=0.42\textwidth,figure=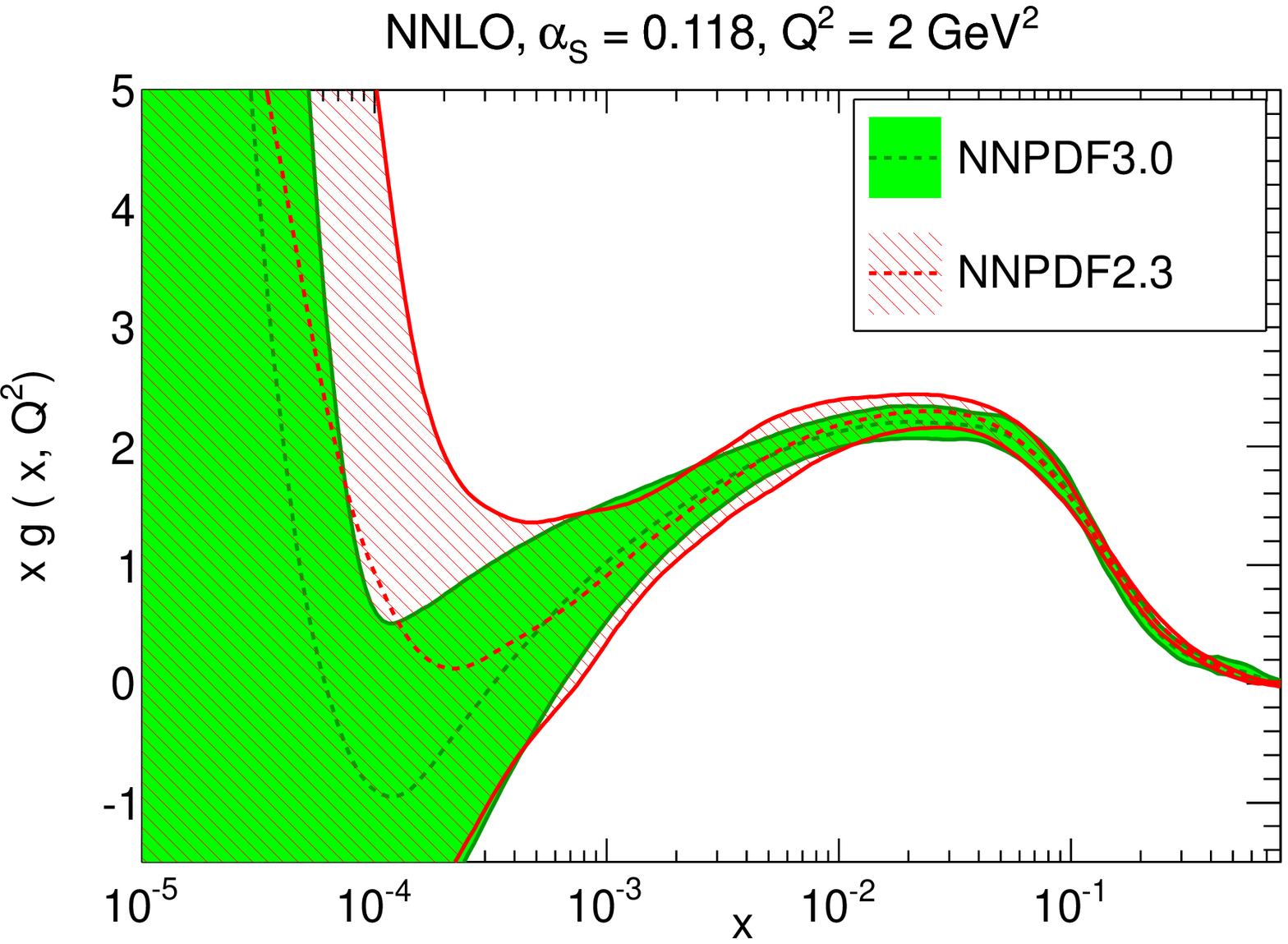}
\epsfig{width=0.42\textwidth,figure=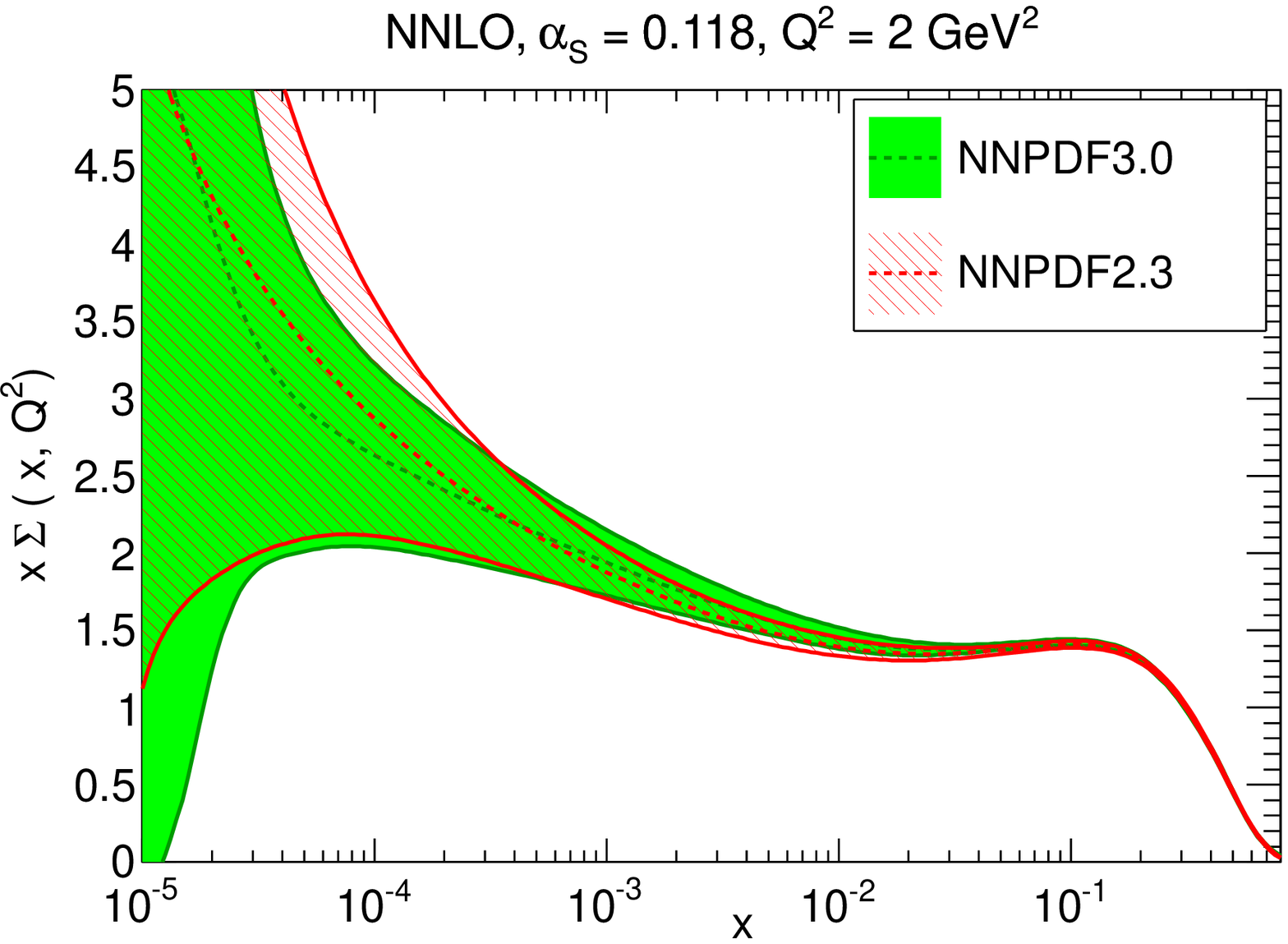}
\epsfig{width=0.42\textwidth,figure=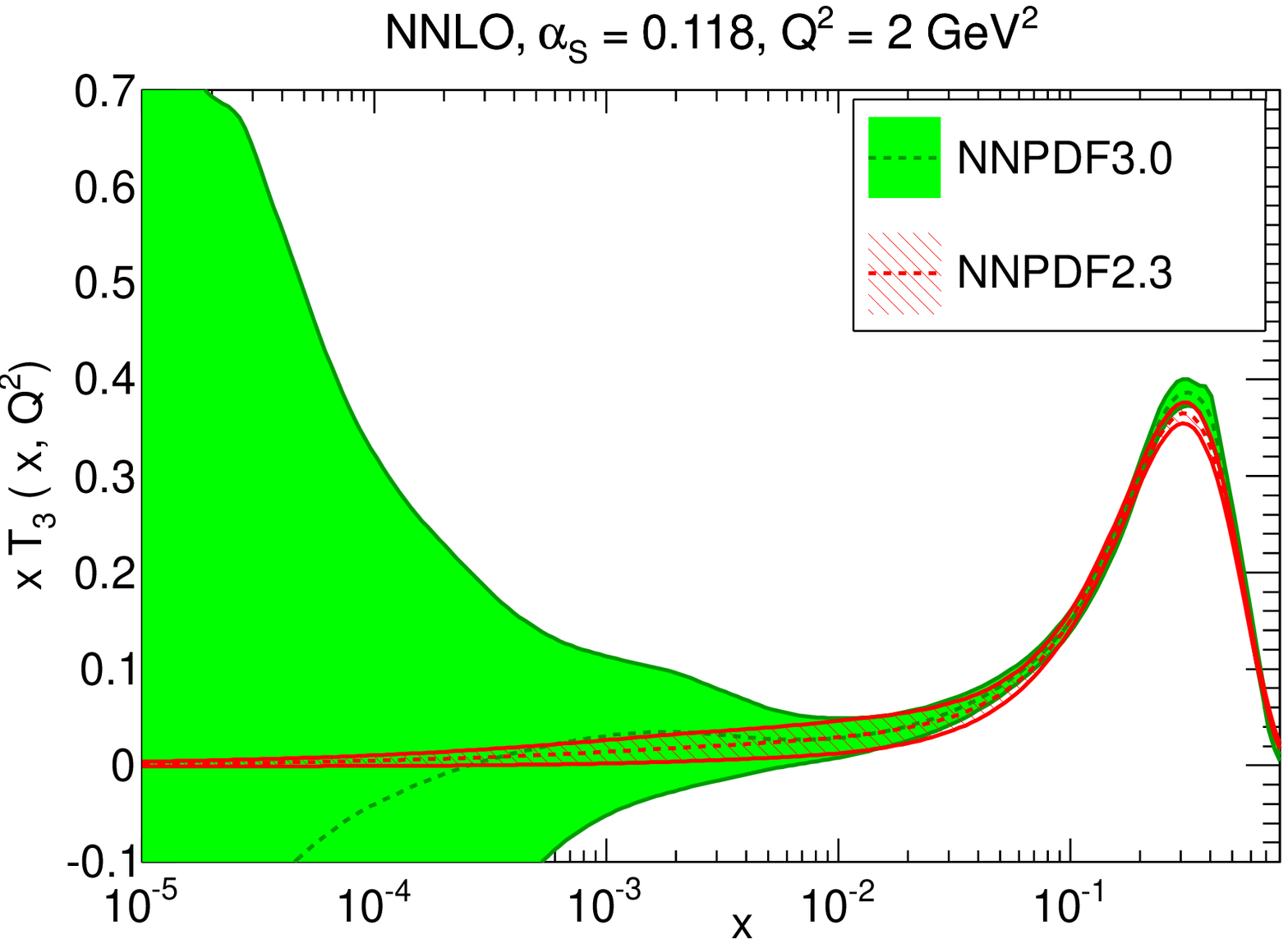}
\epsfig{width=0.42\textwidth,figure=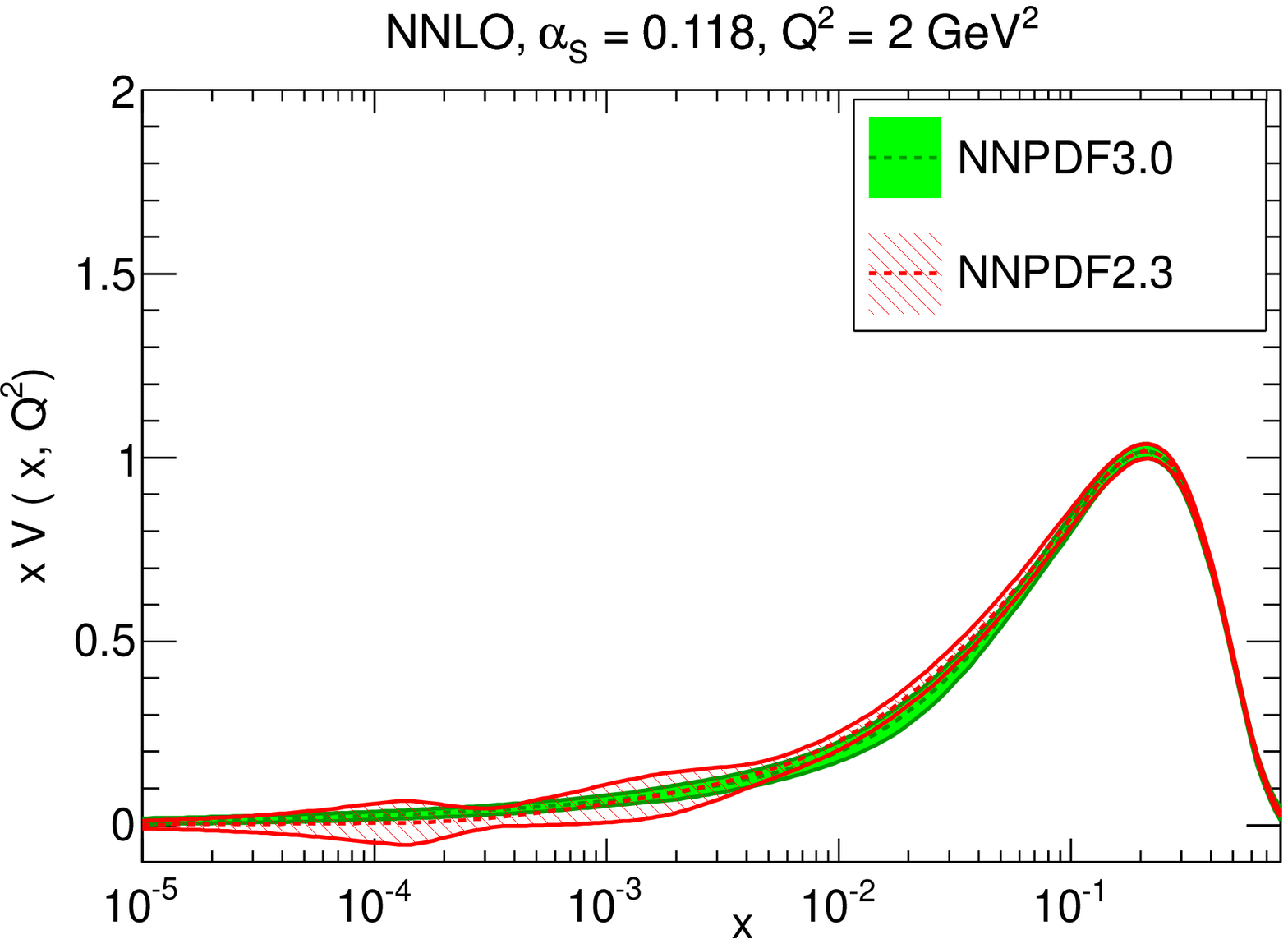}
\caption{\small
Same as Fig.~\ref{fig:30_vs_23_lowscale_nlo}
but at NNLO.
 \label{fig:30_vs_23_lowscale}}
\end{center}
\end{figure}

The same comparison is performed at NNLO in Fig.~\ref{fig:30_vs_23_lowscale}.
In this case, we observe good consistency for the gluon PDF, with a reduction in
the PDF uncertainties at small-$x$.
Note that in both NNLO fits the same FONLL-C GM-VFN scheme is being used.
The agreement is also good for the quark singlet, as it was at NLO.
For the triplet PDF $T_3$, the small-$x$ uncertainties, as at NLO, are
larger in 3.0 than in 2.3, though as
in the case of NLO the two PDF uncertainty bands are in
rather better agreement if  68\% CL contours instead of  one-sigma
uncertainties are compared.

\begin{figure}[t]
\begin{center}
\epsfig{width=0.42\textwidth,figure=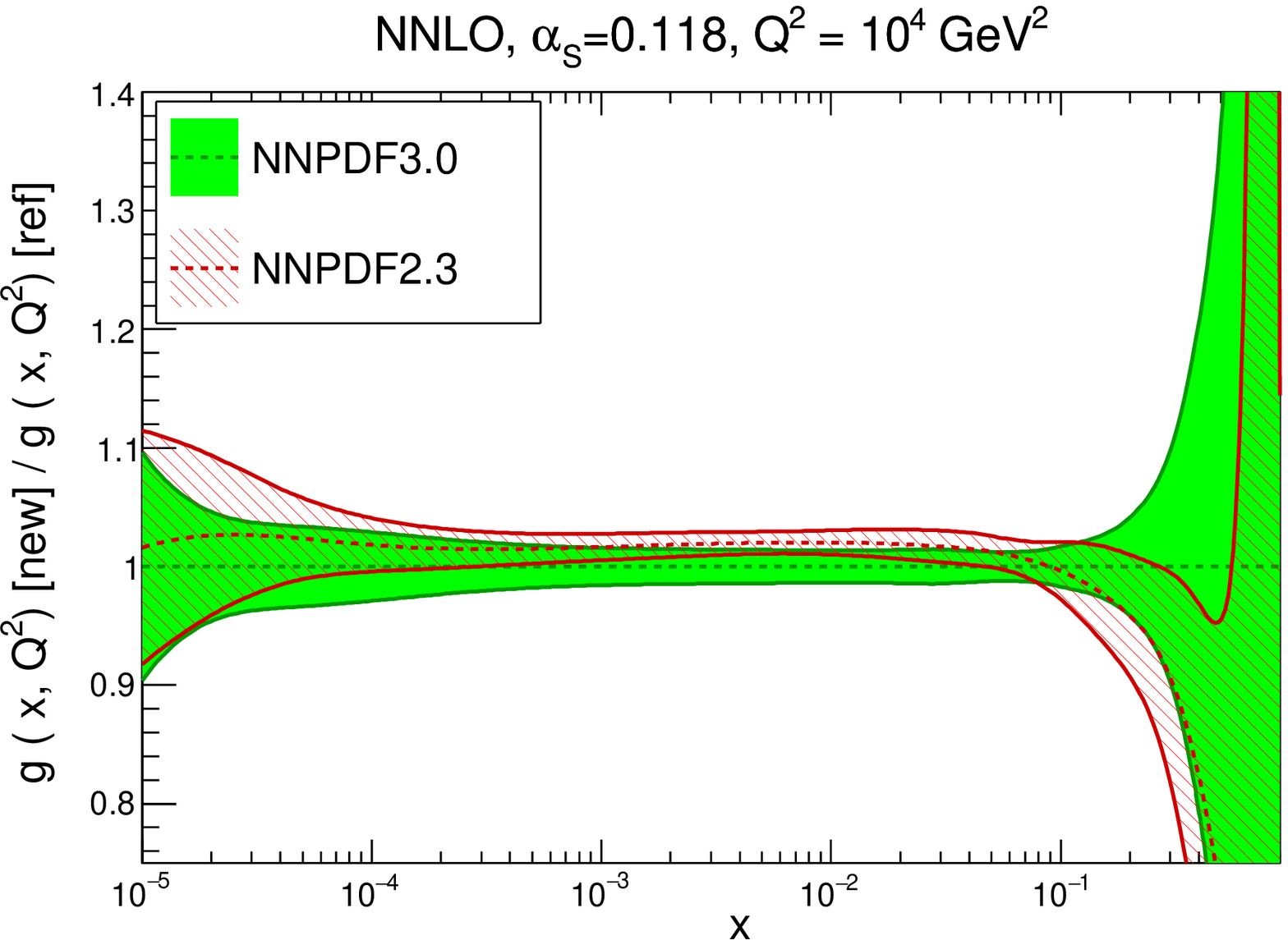}
\epsfig{width=0.42\textwidth,figure=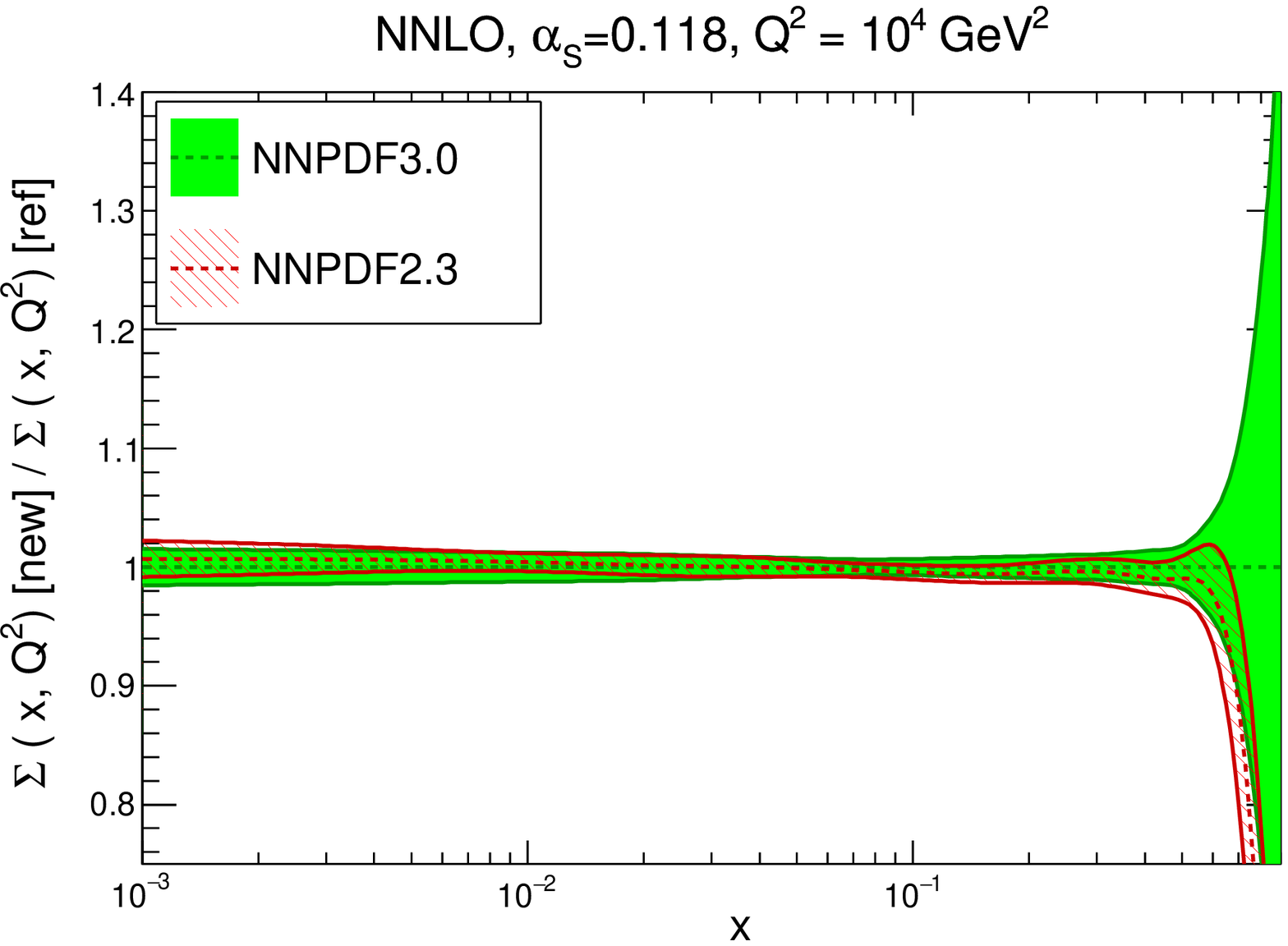}
\epsfig{width=0.42\textwidth,figure=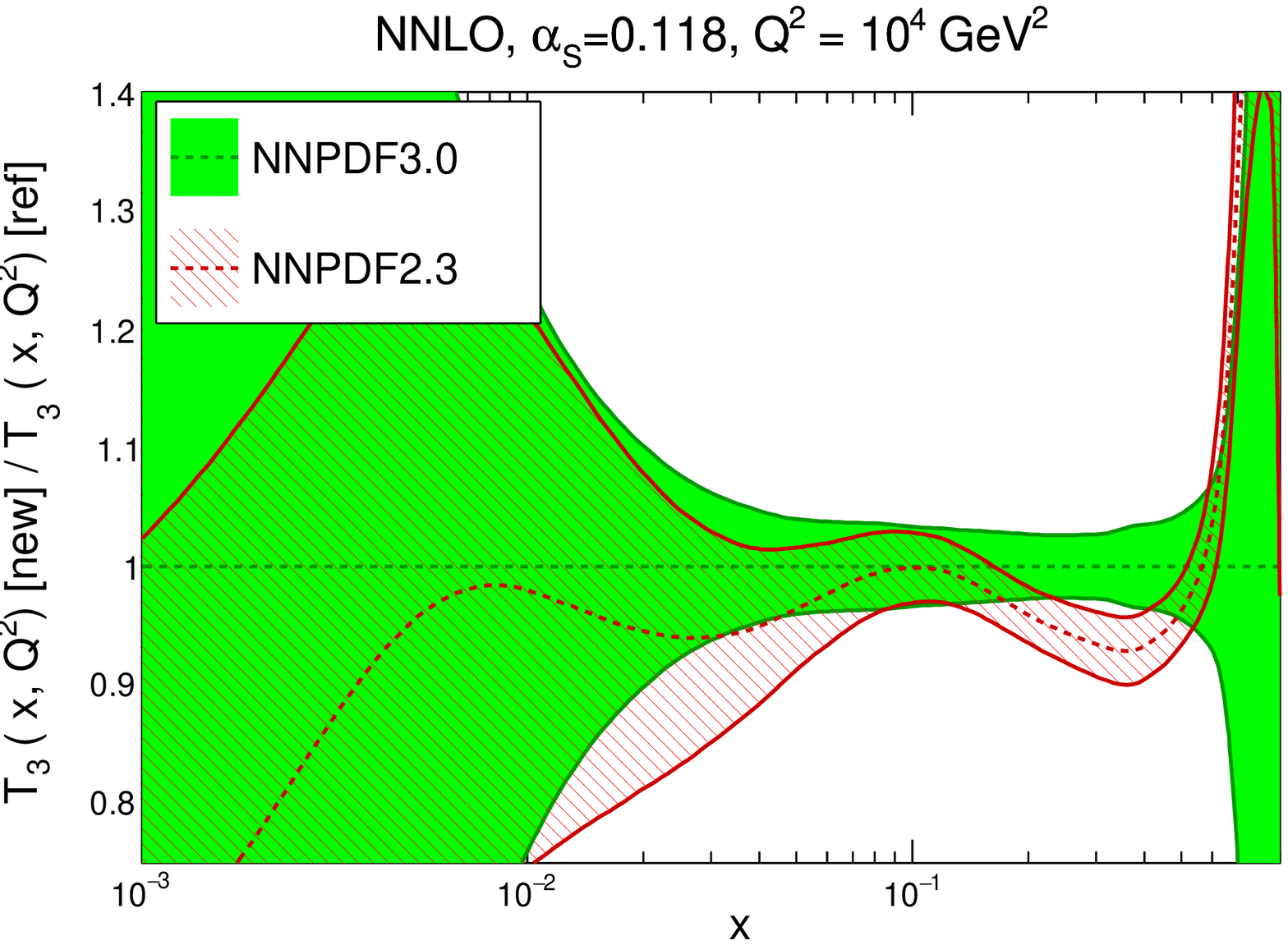}
\epsfig{width=0.42\textwidth,figure=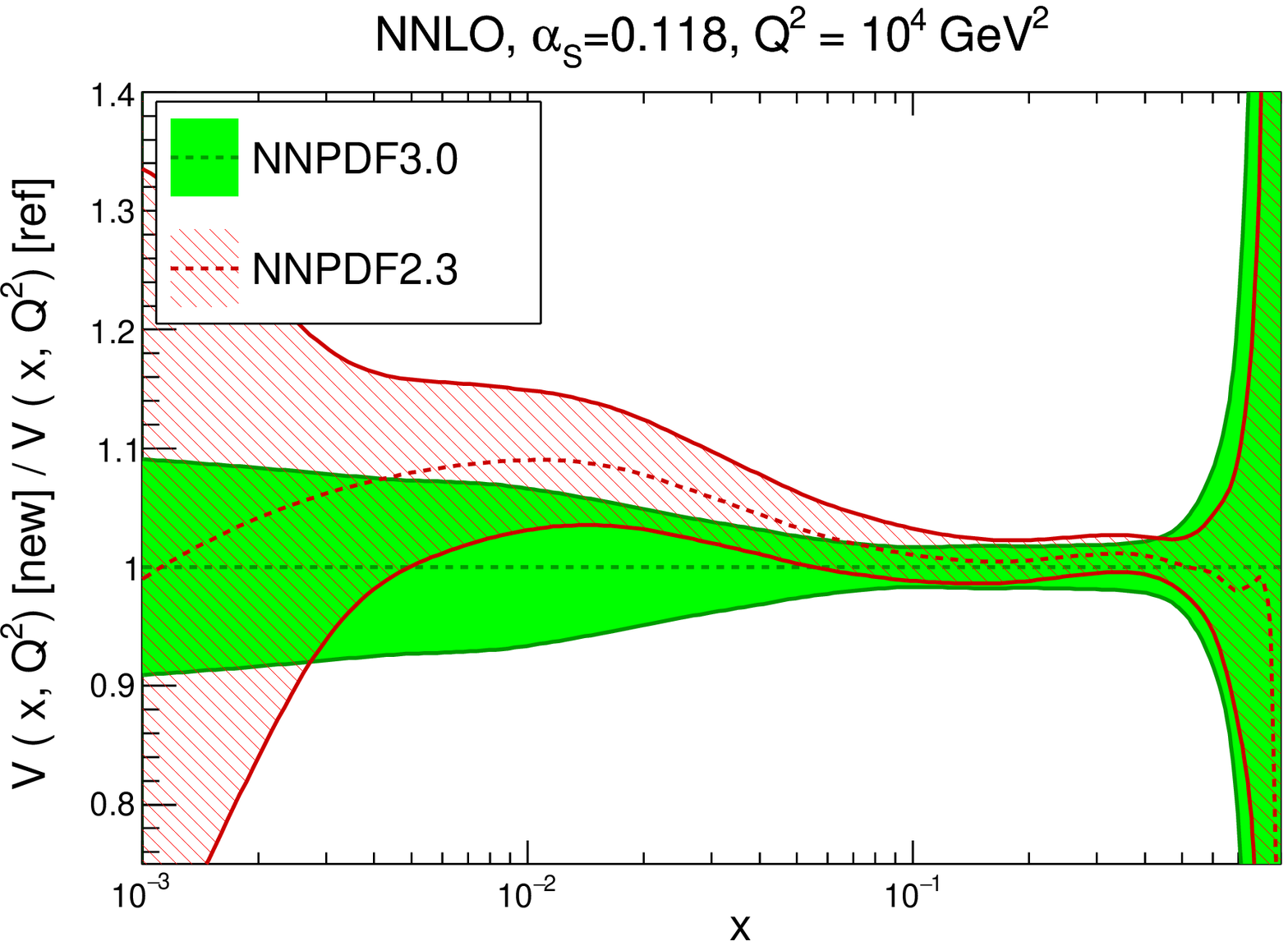}
\caption{\small
Same as Fig.~\ref{fig:30_vs_23_lowscale}, but at $Q^2=10^4$ GeV$^2$,
and with results shown as ratios to the NNPDF3.0 central value.
\label{fig:30_vs_23_highscale}}
\end{center}
\end{figure}

It is also interesting to compare the NNPDF2.3 and 3.0 sets at the
higher scale  $Q^2=10^4$ GeV$^2$, typical of
LHC processes.
Results for this comparison, at NNLO, are shown
Fig.~\ref{fig:30_vs_23_highscale}, as ratios
to the  NNPDF3.0 central value.
We see that the two PDF sets agree typically
at the one-sigma level or better, with some exceptions.
The NNPDF3.0 gluon is somewhat softer than in NNPDF2.3, in
particular in the region around $x\sim 0.01$ which is important
for the Higgs cross-section in gluon fusion.  It is interesting to
observe that this difference is not due to the different value of the
charm mass used in NNPDF2.3 and NNPDF3.0, as the sensitivity of the
gluon to the charm mass is minimal in this region, and becomes
sizable only at very large $x
\gsim0.2$,and to a lesser extent small $x\lsim10^{-5}$, where however
uncertainties are very large (see Ref.~\cite{Ball:2011mu}, Fig.~40,
and also Sect.~\ref{sec:model}, Fig.~\ref{fig:luminnlohq} below).
The total quark singlet is also harder for $x\le 10^{-3}$,
where the two error bands do not overlap, except for
$x \lsim 10^{-4}$ where the two sets agree again at one-sigma.
For the triplet, there is good agreement, except near
$x\sim 0.3$ where the NNPDF2.3 and 3.0 fits disagree at the two-sigma
level; for the total valence PDF there is a reasonable
agreement for all values of $x$.

In order to assess the phenomenological impact of these changes, it is
useful to look at parton luminosities.
Following Ref.~\cite{Campbell:2006wx}, we define
the parton luminosity for the $ij$ initial state as
\be
\Phi_{ij}\lp M_X^2\rp = \frac{1}{s}\int_{\tau}^1
\frac{dx_1}{x_1} f_i\lp x_1,M_X^2\rp f_j\lp \tau/x_1,M_X^2\rp \ ,
\label{eq:lumdef}
\ee
where $f_i(x,M^2_X)$ is the PDF for the $i$-th parton,
$\tau \equiv M_X^2/s$ and
$M_X$ is the invariant mass of the final state.

\begin{figure}
\begin{center}
\epsfig{width=0.42\textwidth,figure=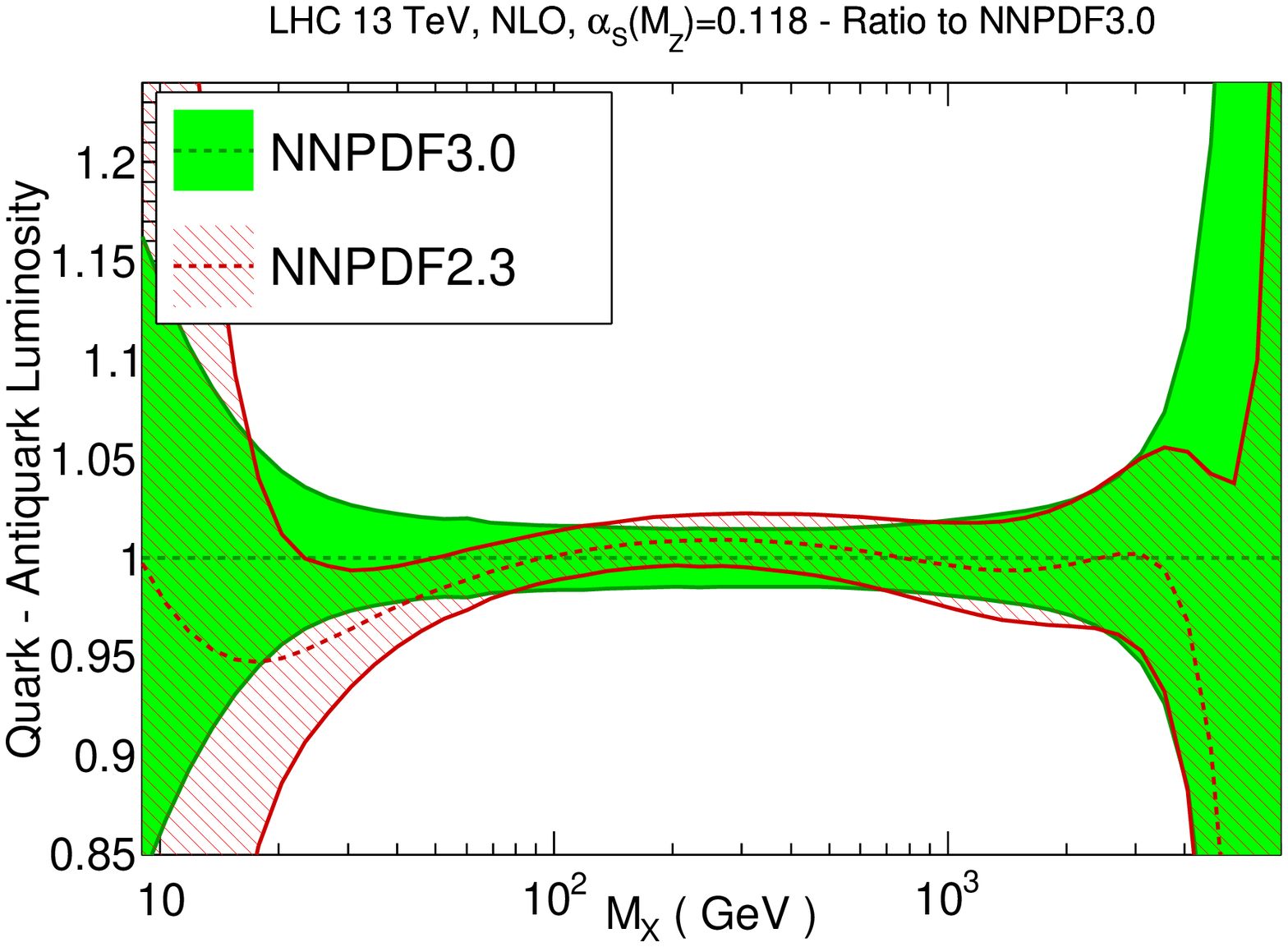}
\epsfig{width=0.42\textwidth,figure=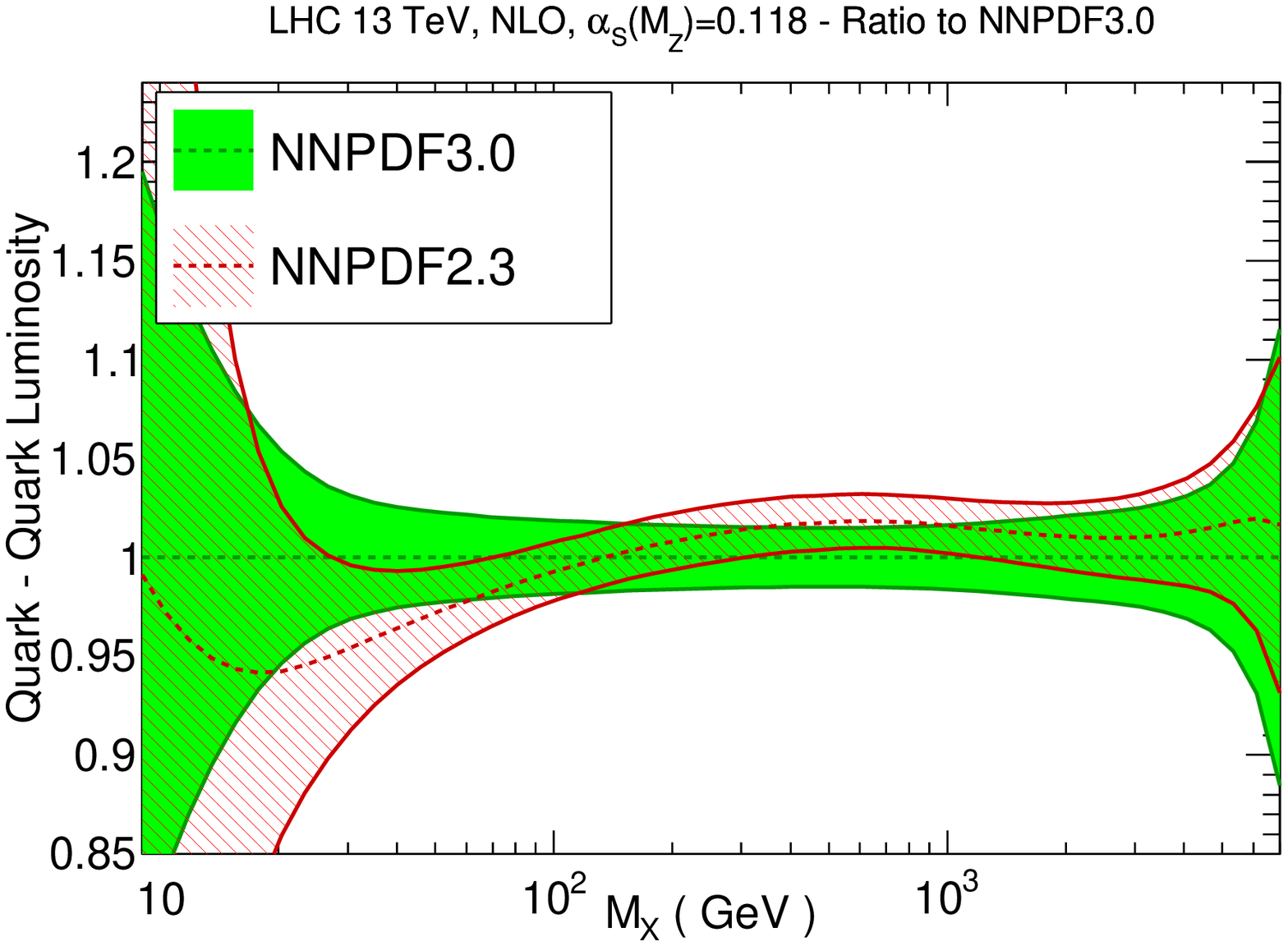}
\epsfig{width=0.42\textwidth,figure=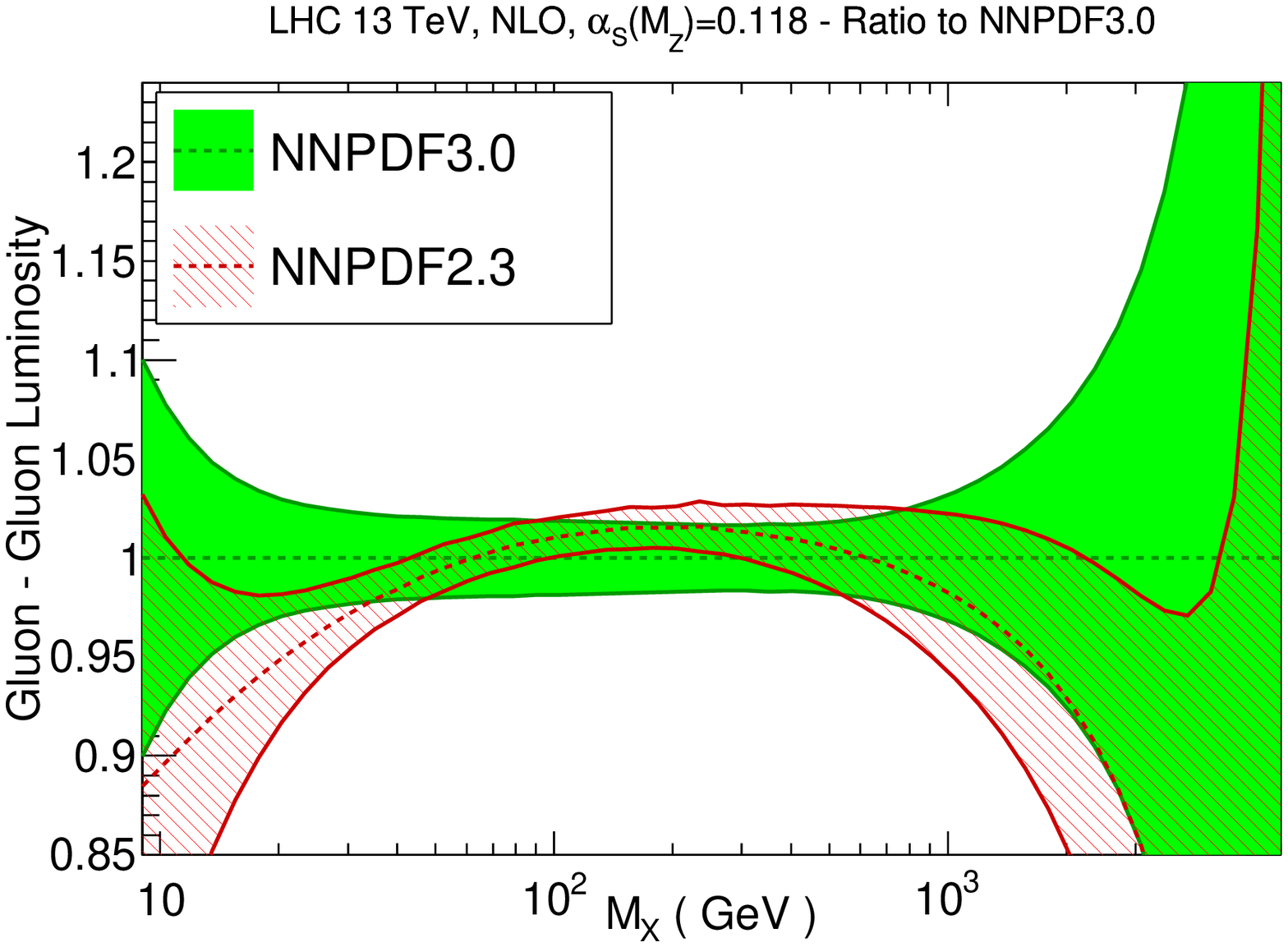}
\epsfig{width=0.42\textwidth,figure=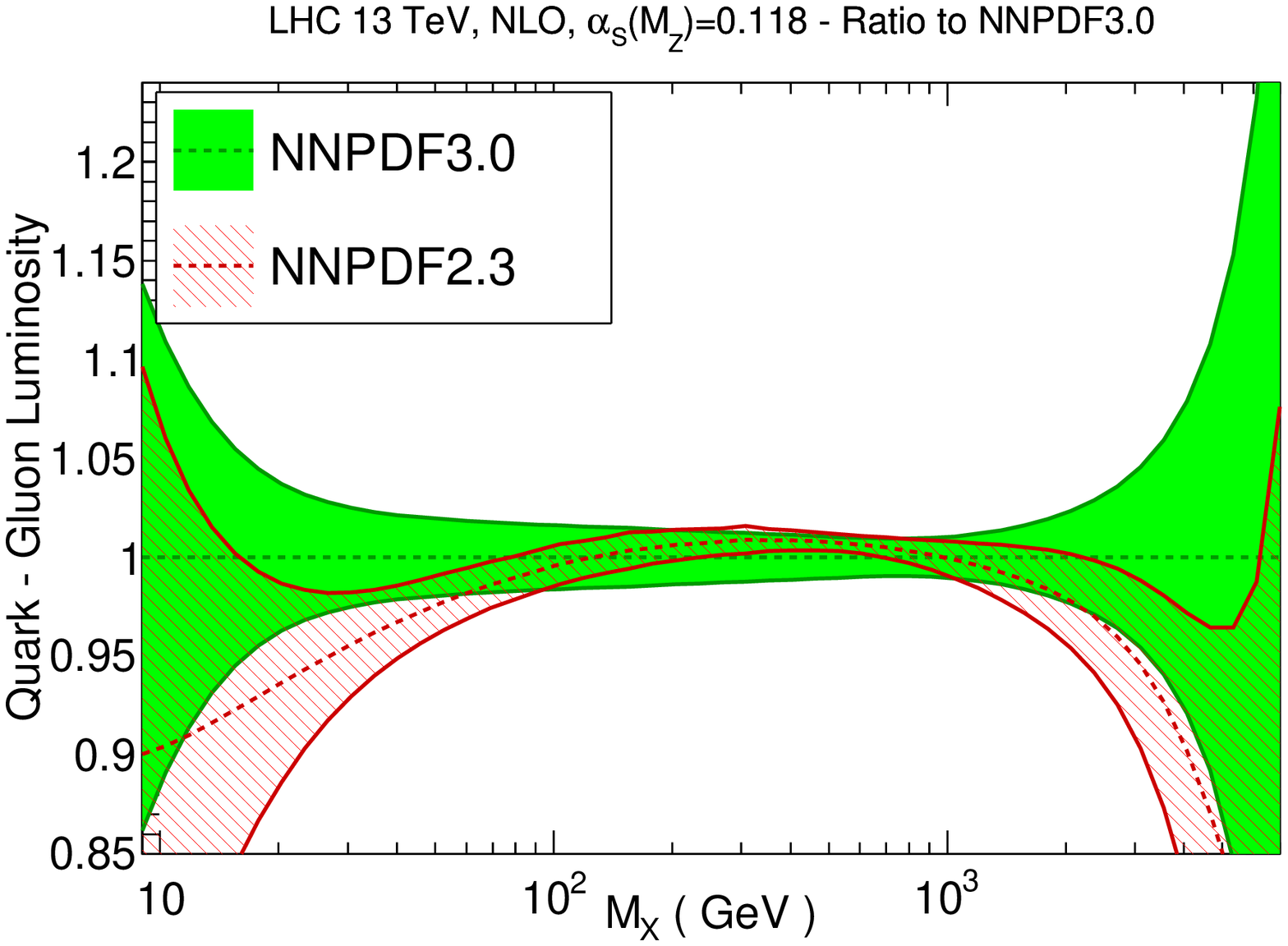}
\caption{\small Parton luminosities, Eq.~(\ref{eq:lumdef}) computed
  using
NNPDF2.3 and NNPDF3.0 NLO PDFs with  $\alpha_s(M_Z)=0.118$, as a function
of the invariant mass of the final state $M_X$. Results are shown as
ratios to NNPDF2.3.
From top to bottom and from left to right the $q\bar q$, $qq$, $qq$
and $qg$ luminosities are shown.
 \label{fig:lumi1nlo}}
\end{center}
\end{figure}

\begin{figure}
\begin{center}
\epsfig{width=0.42\textwidth,figure=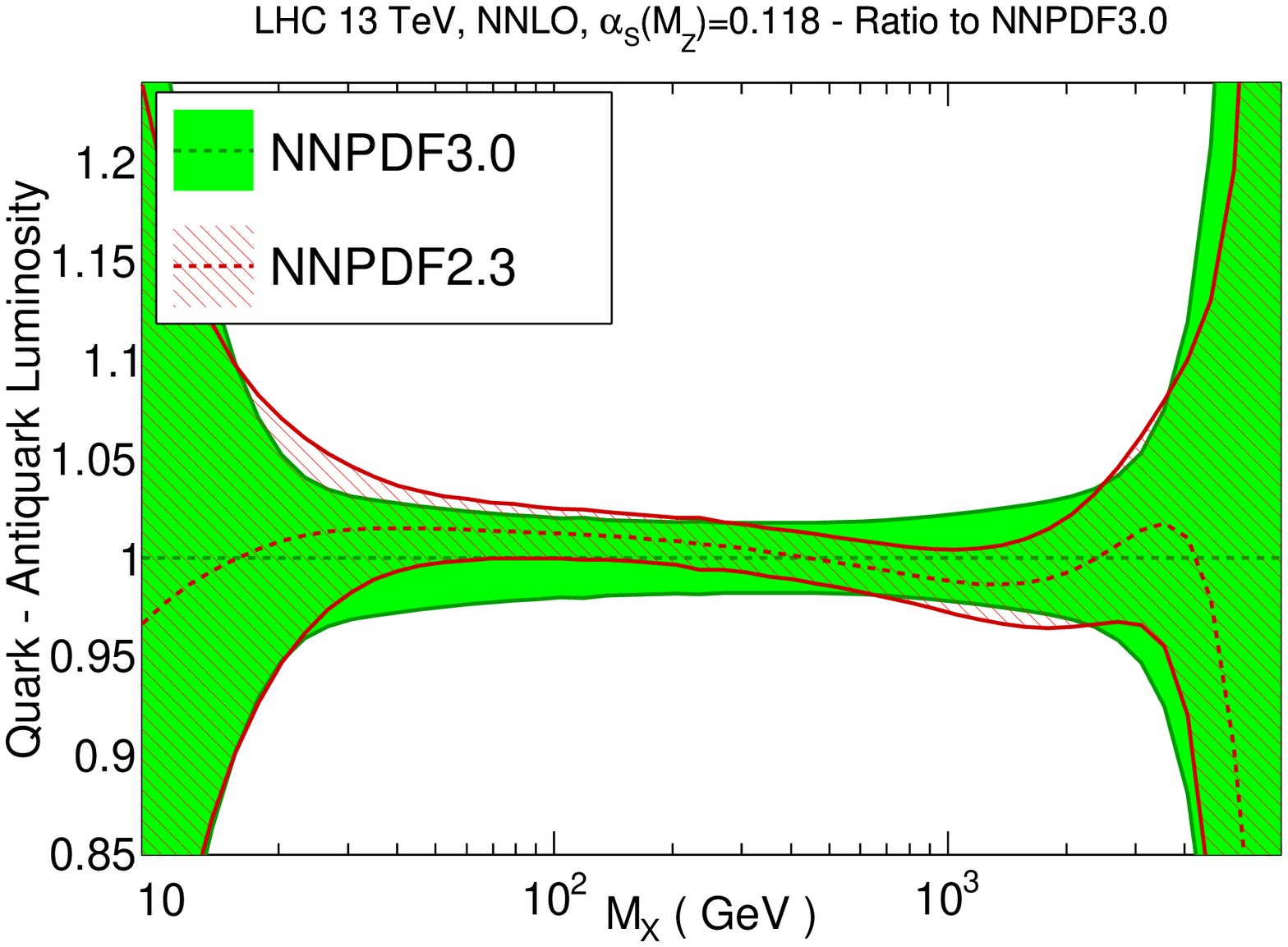}
\epsfig{width=0.42\textwidth,figure=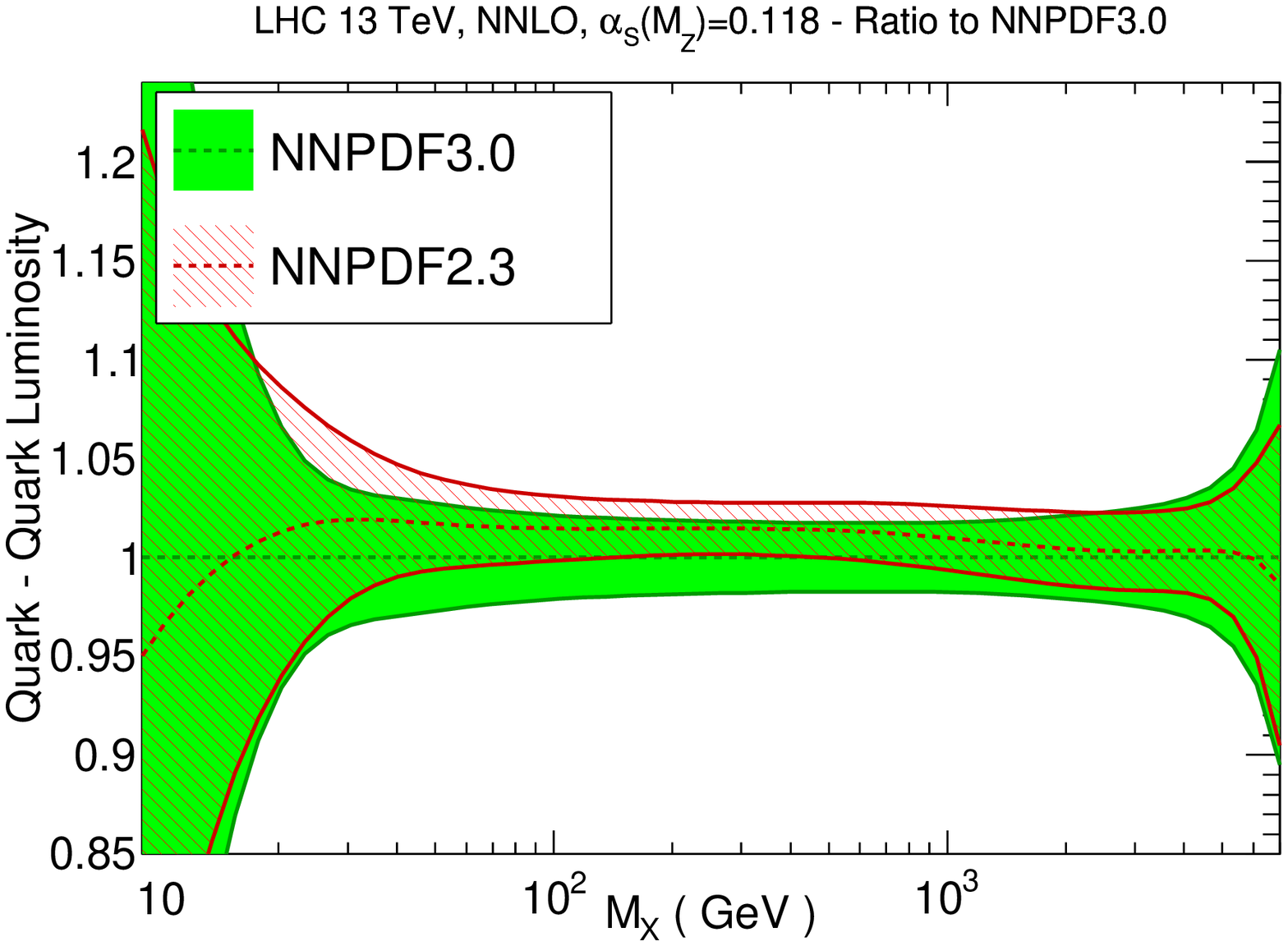}
\epsfig{width=0.42\textwidth,figure=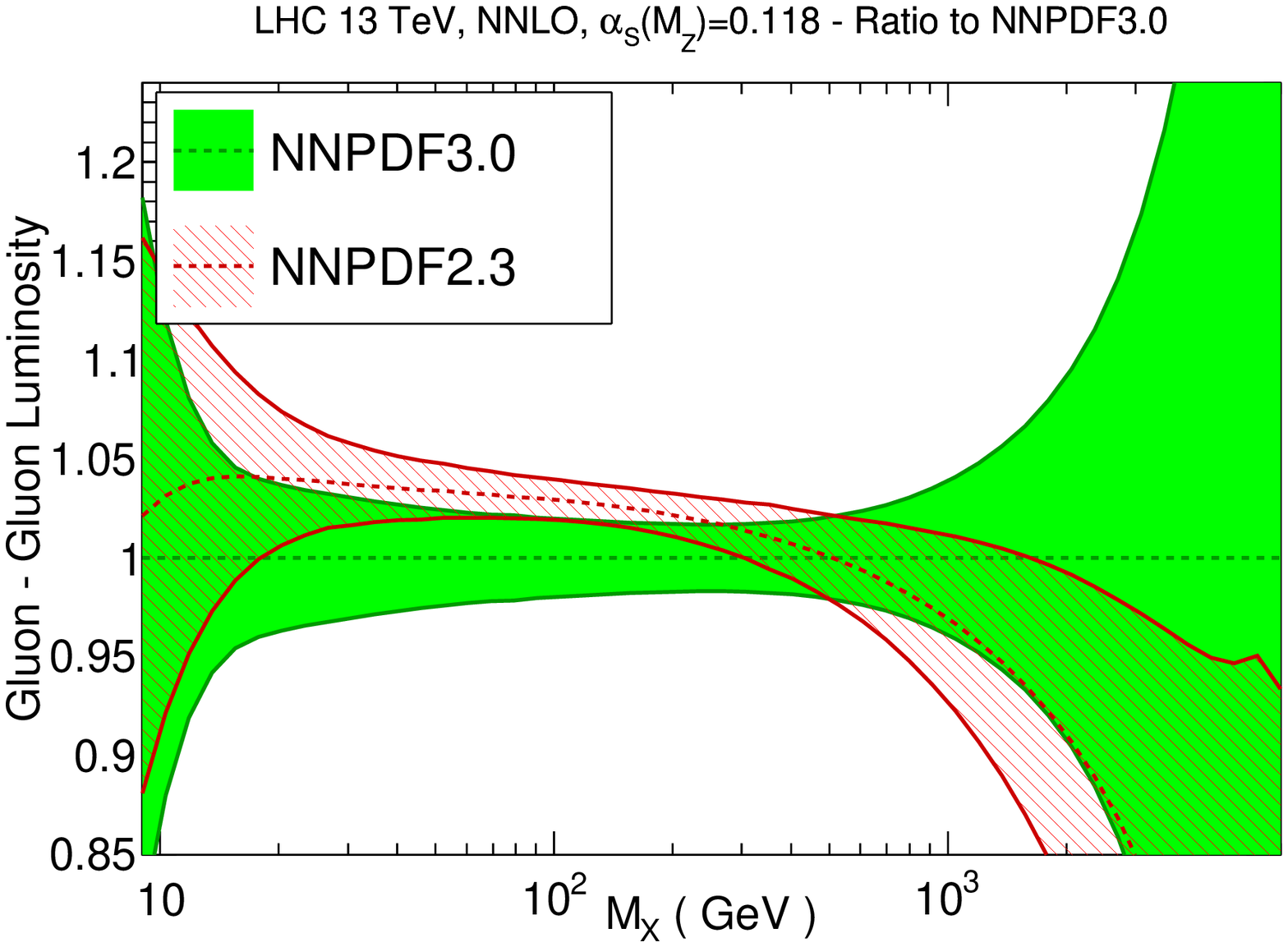}
\epsfig{width=0.42\textwidth,figure=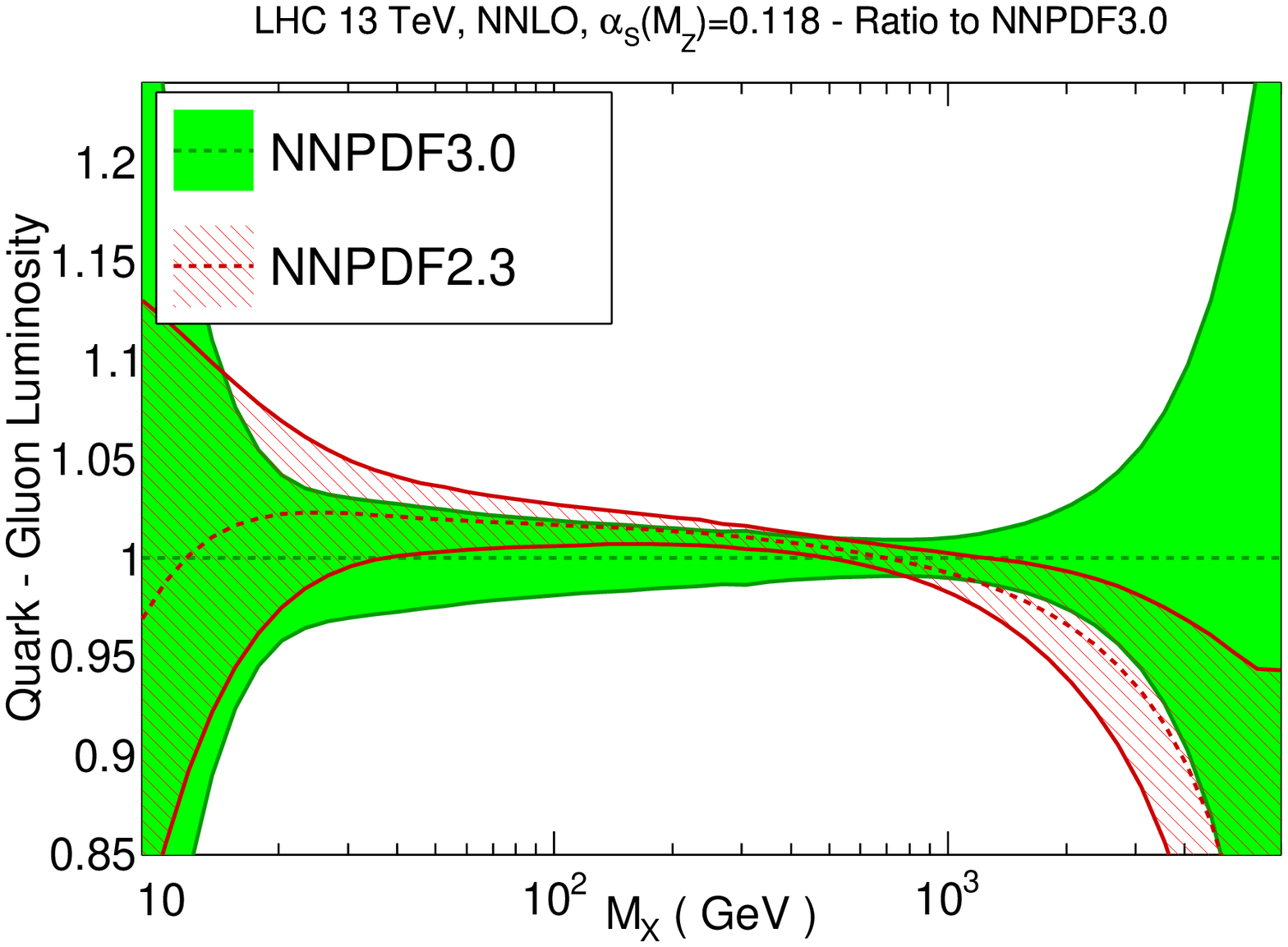}
\caption{\small Same as Fig.~\ref{fig:lumi1nlo} at NNLO.
 \label{fig:lumi1nnlo}}
\end{center}
\end{figure}

\begin{figure}[t]
\begin{center}
\epsfig{width=0.42\textwidth,figure=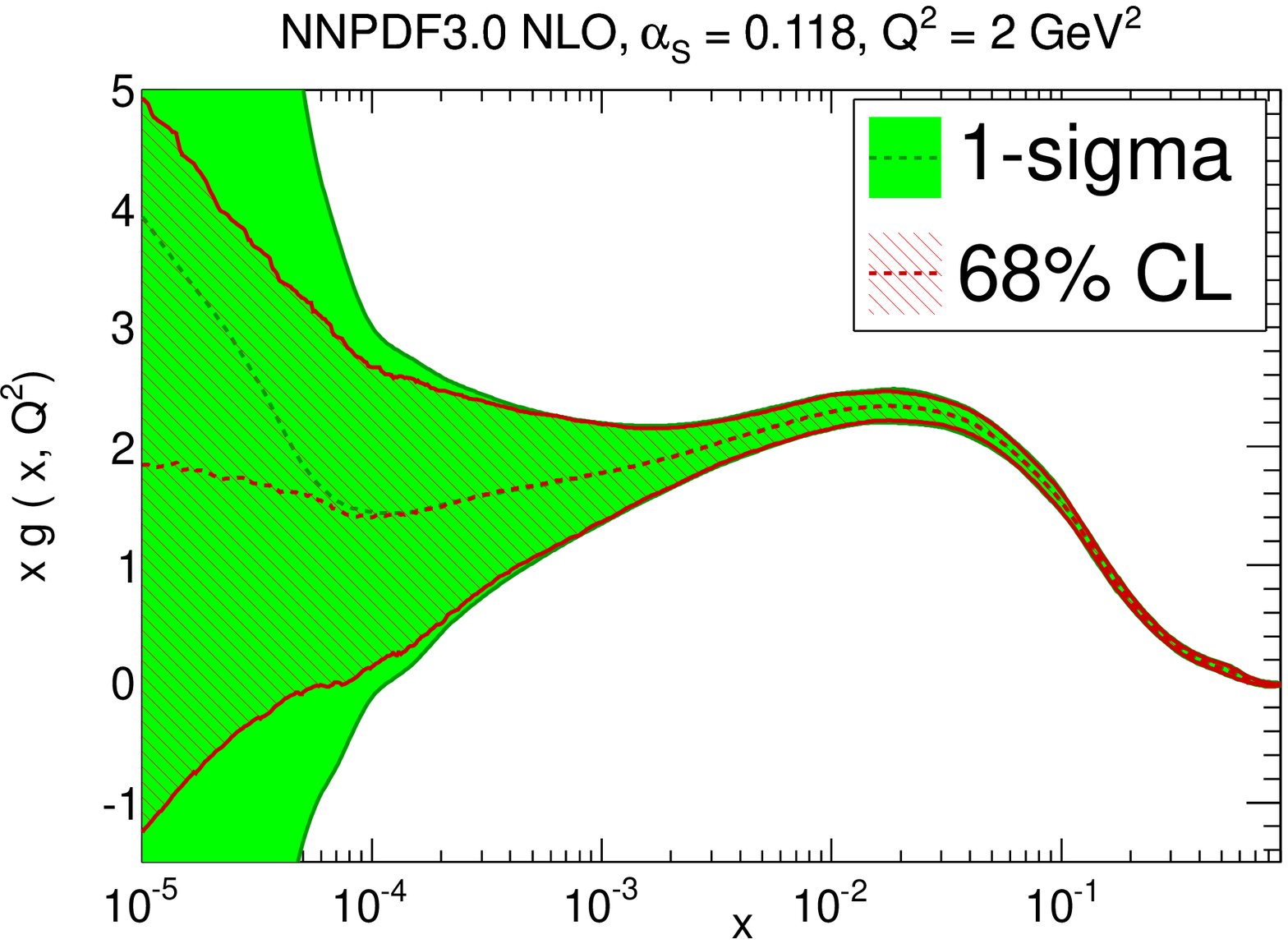}
\epsfig{width=0.42\textwidth,figure=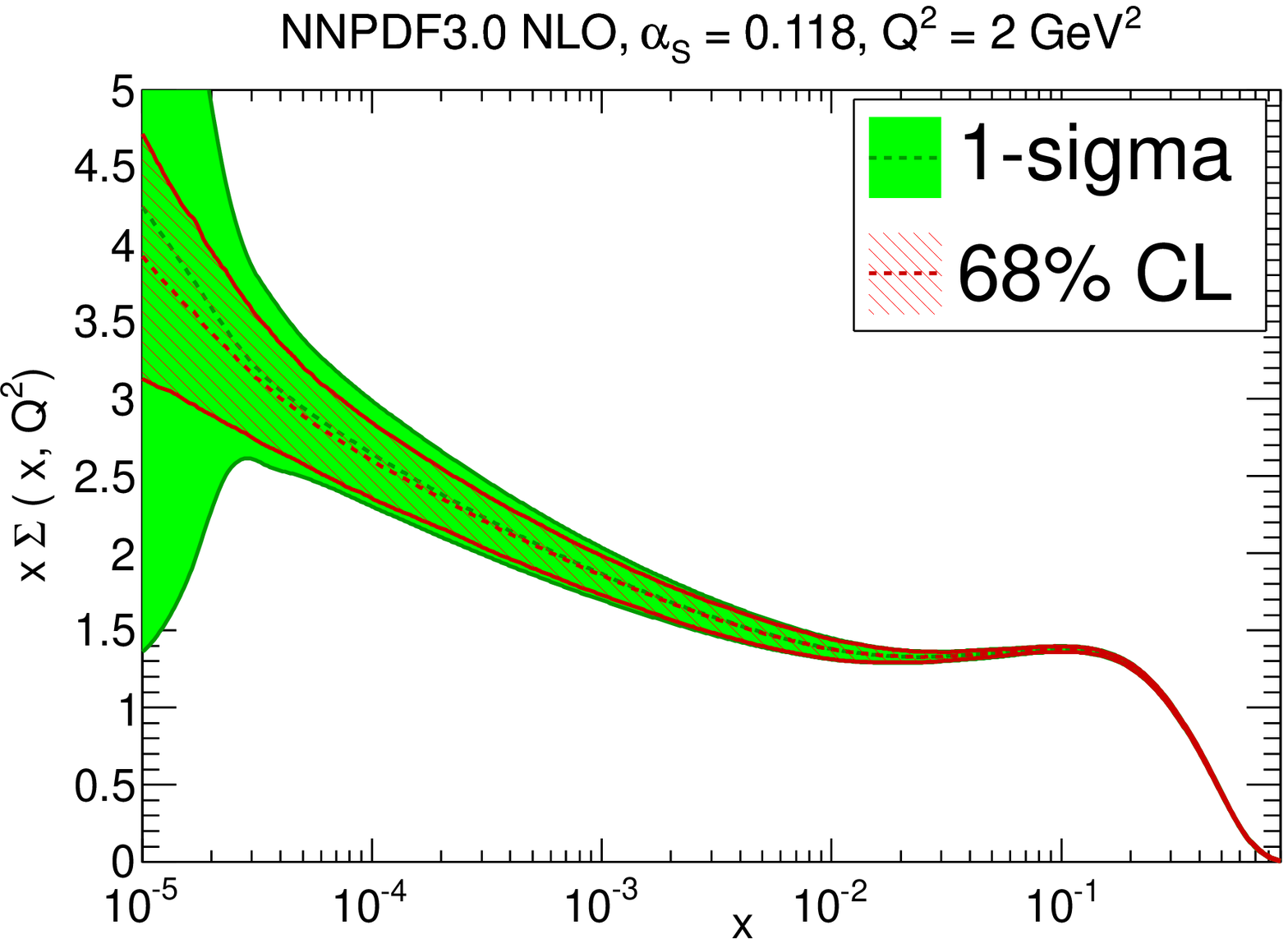}
\epsfig{width=0.42\textwidth,figure=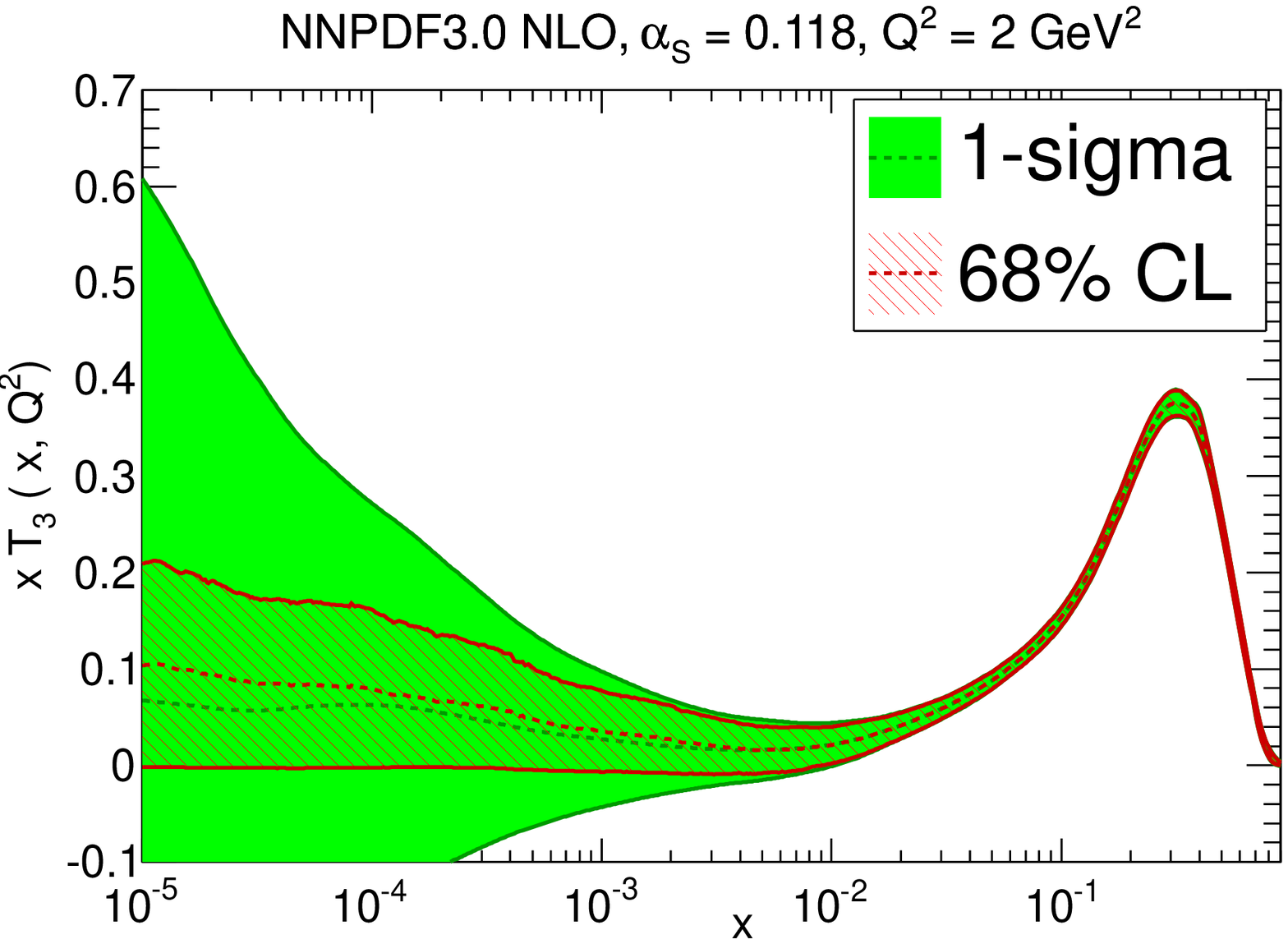}
\epsfig{width=0.42\textwidth,figure=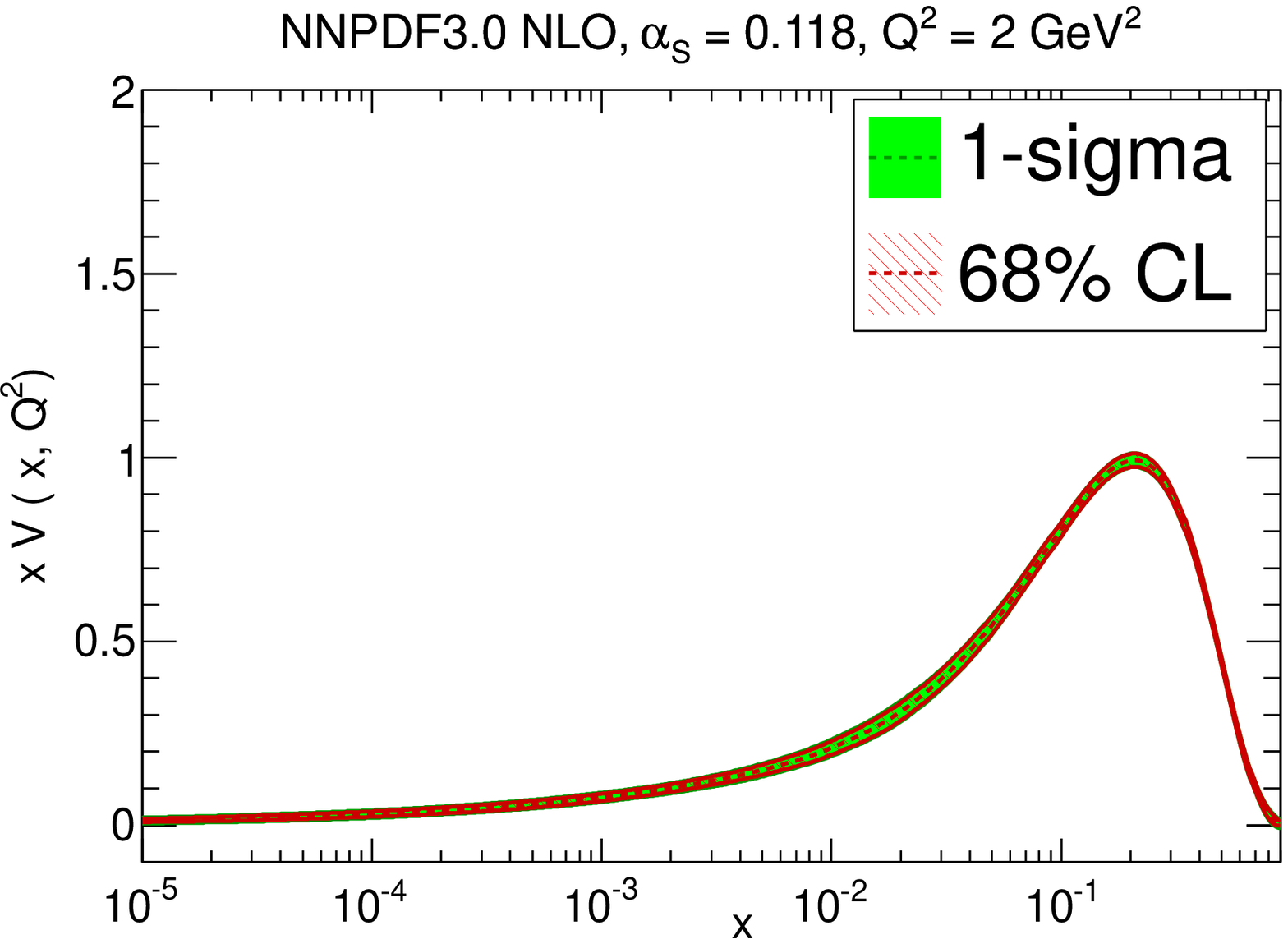}
\caption{\small
Comparison  of one-sigma uncertainty bands and 68\%
confidence level intervals for the NNPDF3.0 NLO set with
$\alpha_s(M_z)=0.119$  at  $Q^2=2$ GeV$^2$.
The set with $N_{\rm rep}=1000$ replicas has been used.
From top to bottom and from left to right the
gluon,  singlet,  isospin triplet and
total valence are shown.
 \label{fig:3068cl}}
\end{center}
\end{figure}

We compare $gg$, $qq$, $q\bar{q}$ and $qg$ luminosities obtained using NLO and
NNLO
PDF sets for  $\sqrt{s}$=13 TeV and $\alpha_s(M_Z)=0.118$
(where for quarks a sum over light flavors is
understood).
The NLO and NNLO comparisons between NNPDF2.3
and NNPDF3.0 luminosities are shown in Figs.~\ref{fig:lumi1nlo} and
\ref{fig:lumi1nnlo} respectively.
At NLO, we generally find agreement  at the
one-sigma level, with some
differences in the $qq$ channel
for $M_X\sim500$~GeV, where the error bands of the two
PDF sets barely overlap, and in the $qg$ channel above
1 TeV, where the luminosity is rather larger in NNPDF3.0 than
in NNPDF2.3.
Note that in the $gg$ channel in the region around 100-200 GeV the
NNPDF3.0 luminosity is somewhat softer than in NNPDF2.3, though
always in agreement within PDF uncertainties.

At NNLO, in the $qq$ and $q\bar q$
channels there is generally good agreement, with
differences well within one sigma: for $q\bar{q}$, the NNPDF3.0 luminosity tends
to be larger at high invariant masses, while for  $qq$ around 500~GeV NNPDF3.0 is
somewhat lower, with barely overlapping error bands.
More significant differences are found in the $gg$  channel,
where the luminosity at medium invariant masses is smaller
by about one sigma in NNPDF3.0 than in NNPDF2.3:
in particular, for 30 GeV $\le M_X \le$ 300 GeV, the $gg$ one sigma bands
barely overlap. This has  important consequences for
gluon-initiated processes such as inclusive Higgs production, see
Sect.~\ref{sec:higgsgg} below.
As discussed in Sect.~\ref{sec:nnpdf30set}, these differences
stem from a combination of the improved fitting methodology and the new
constraints from HERA and LHC data.
%

Finally, we compare one-sigma uncertainty bands to 68\% confidence level intervals.
For this comparison we use the NNPDF3.0 NLO fit with
$N_{\rm rep}=1000$ replicas;  the conclusions would be qualitatively
the same for the NNLO fit.
Of course if the PDF probability distribution is Gaussian the one-sigma and 68\%
intervals coincide. While this is usually the case, for
some PDFs in specific $x$ regions there are significant
deviations from gaussianity that can be quantified by this
type of comparison: typically,
this happens  in extrapolation regions where there
are no direct experimental constraints, especially if positivity
constraints, which are asymmetric, play a significant role.

This comparison is shown in Fig.~\ref{fig:3068cl}:
for most PDFs there is a good agreement,
the only exception being the small-$x$ extrapolation regions,  $x\lsim 10^{-4}$,
where 68\% CL intervals are typically smaller than the one-sigma
bands, indicating the presence of non-gaussian outliers.
For the triplet, differences are already significant for  $x\lsim 10^{-3}$. Note that because
of the generalized
positivity constraints, the lower
limit of 68\% CL interval is typically just above zero.

\subsubsection{Perturbative stability and theoretical uncertainties}
\label{sec:perturbative}

\begin{figure}[tm]
\begin{center}
\epsfig{width=0.84\textwidth,figure=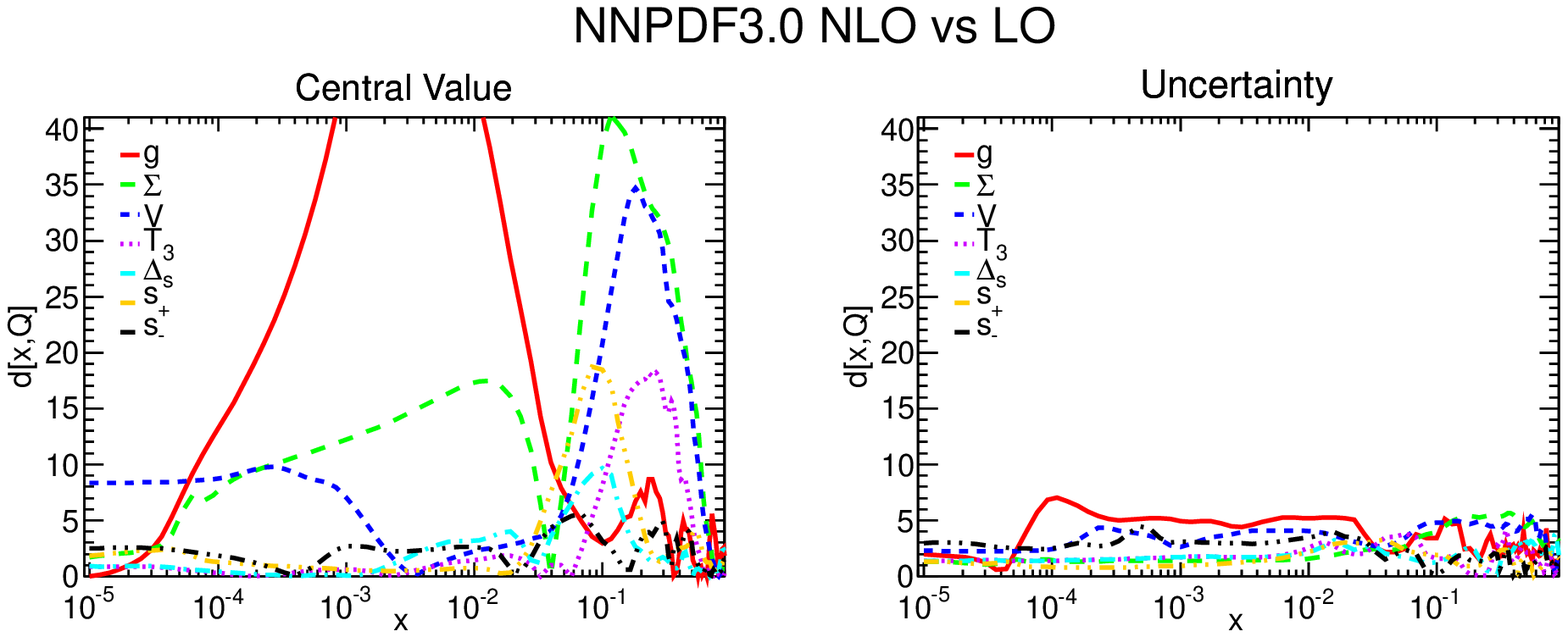}
\epsfig{width=0.84\textwidth,figure=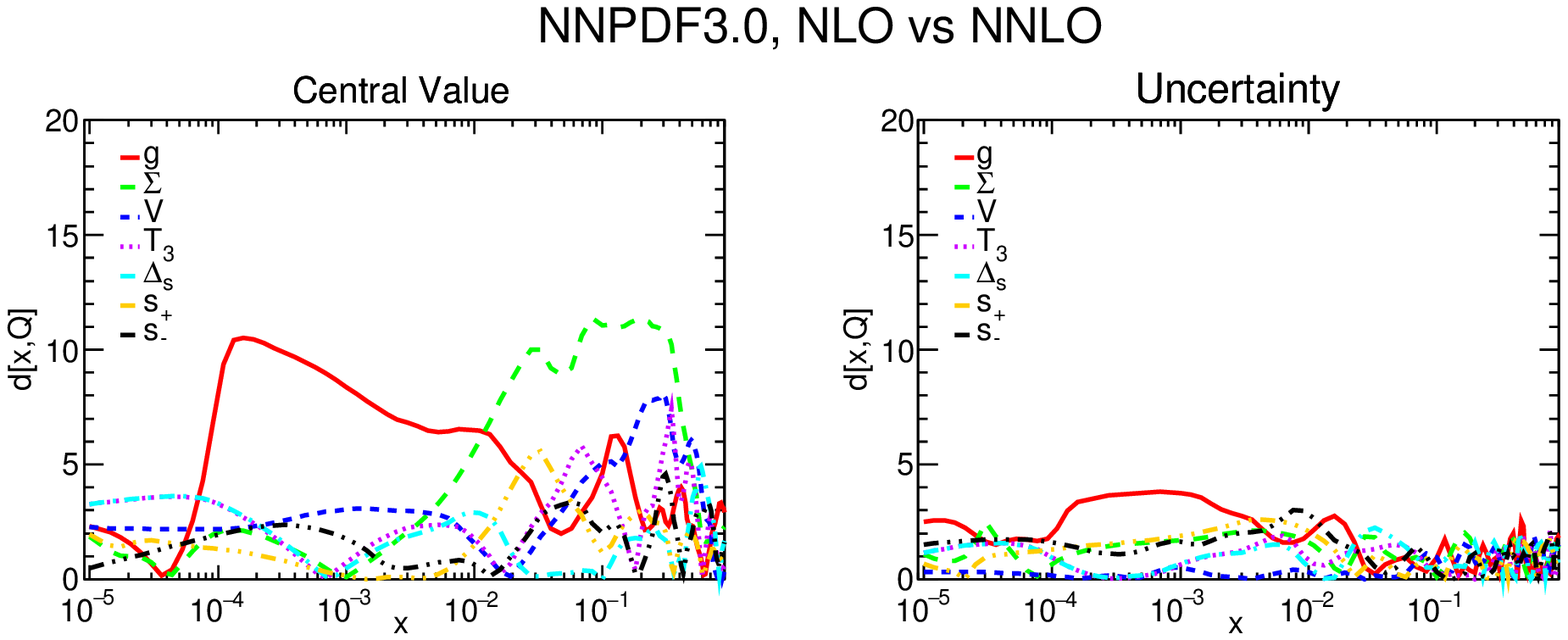}
\caption{\small
Same as Fig.~\ref{fig:distances_30_vs_23_nnlo}, but now comparing
NNPDF3.0 LO vs NLO (top) and NLO vs. NNLO (bottom).
 \label{fig:distances_nnpdf30_nlo_vs_nnlo}}
\end{center}
\end{figure}

We now compare the NNPDF3.0 LO, NLO and NNLO sets: this is a meaningful comparison
because  the same methodology is used at all orders, with only the underlying QCD
theory (and to a small extent the  dataset) changing from one order to the next.
In Fig.~\ref{fig:distances_nnpdf30_nlo_vs_nnlo}
we show the distances between the NNPDF3.0 pairs of fits at two
consecutive orders:
LO vs. NLO, and  NLO vs. NNLO.
In the former case, the main variation
is of course seen in the gluon PDF, which as well known is very
different at LO; there are also significant differences in the large-$x$ quarks.
Of course, at  LO theory uncertainties
completely dominate over the PDF uncertainty, which depends on the
data and is roughly the same at all orders, as this comparison clearly shows.

Distances become much smaller when comparing  NLO to NNLO:
for  central values,
the main differences are in the gluon PDF, both
at small $x$ and at large $x$, and in the medium- and large-$x$
quarks, in particular the total quark singlet.
Uncertainties are again quite stable,  with the
exception on the large-$x$ gluon,
where the PDF uncertainties are larger at NNLO because of
the additional cuts applied to the jet data (recall
Sect.~\ref{sec:thjetdata}).
These differences in the fitted jet dataset also impact the central
values of the two fits.

\begin{figure}[t]
\begin{center}
\epsfig{width=0.42\textwidth,figure=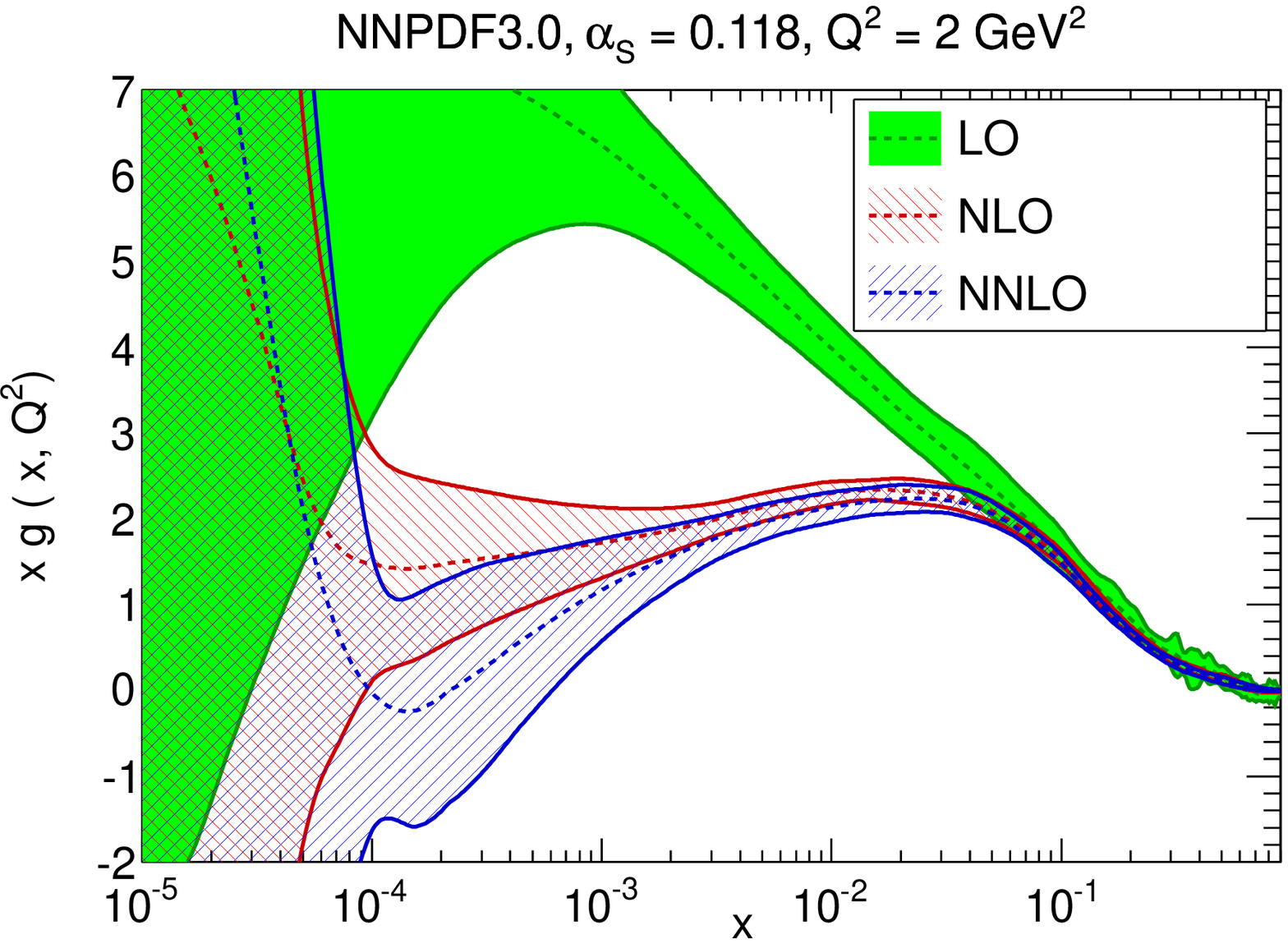}
\epsfig{width=0.42\textwidth,figure=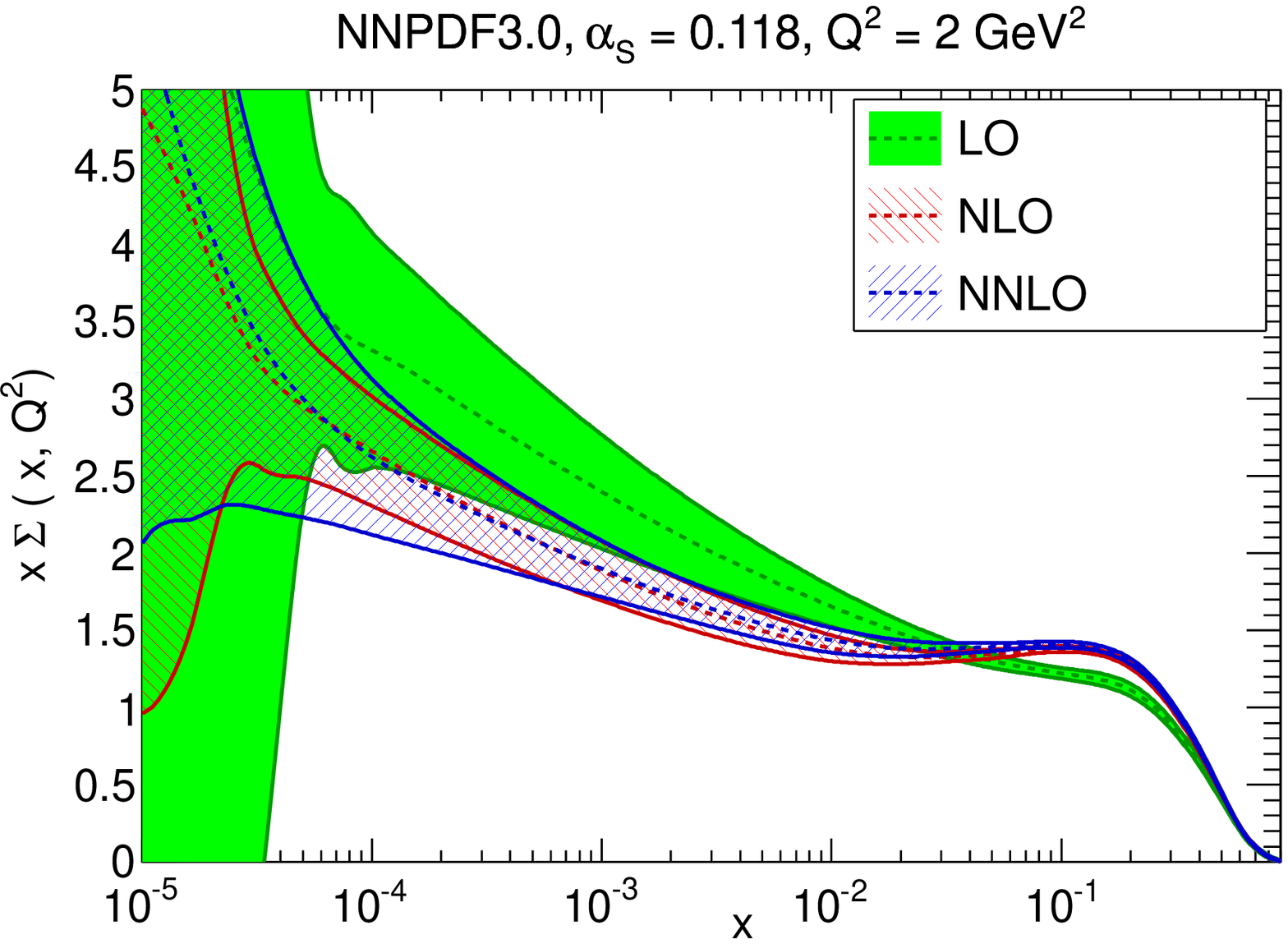}
\epsfig{width=0.42\textwidth,figure=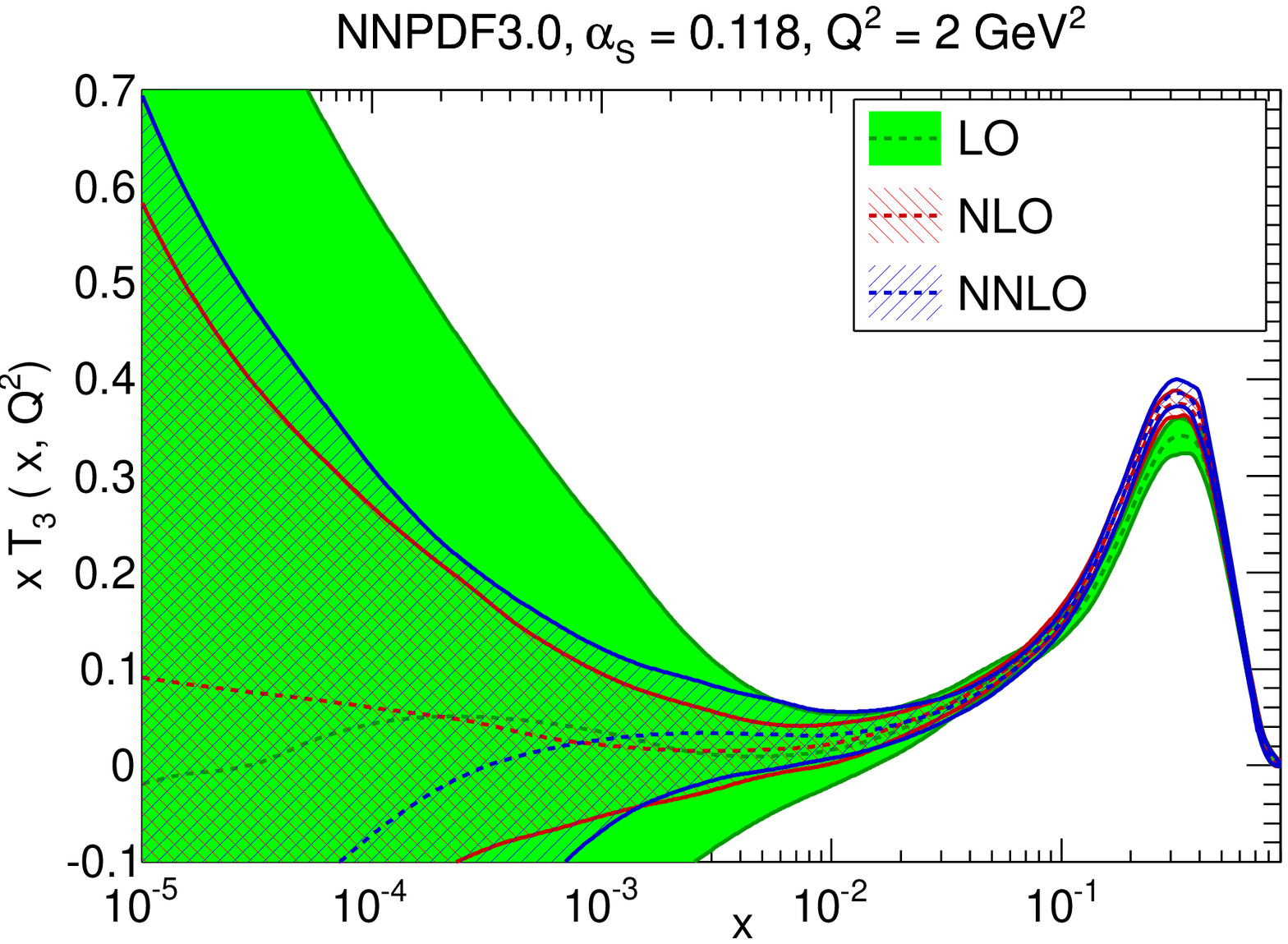}
\epsfig{width=0.42\textwidth,figure=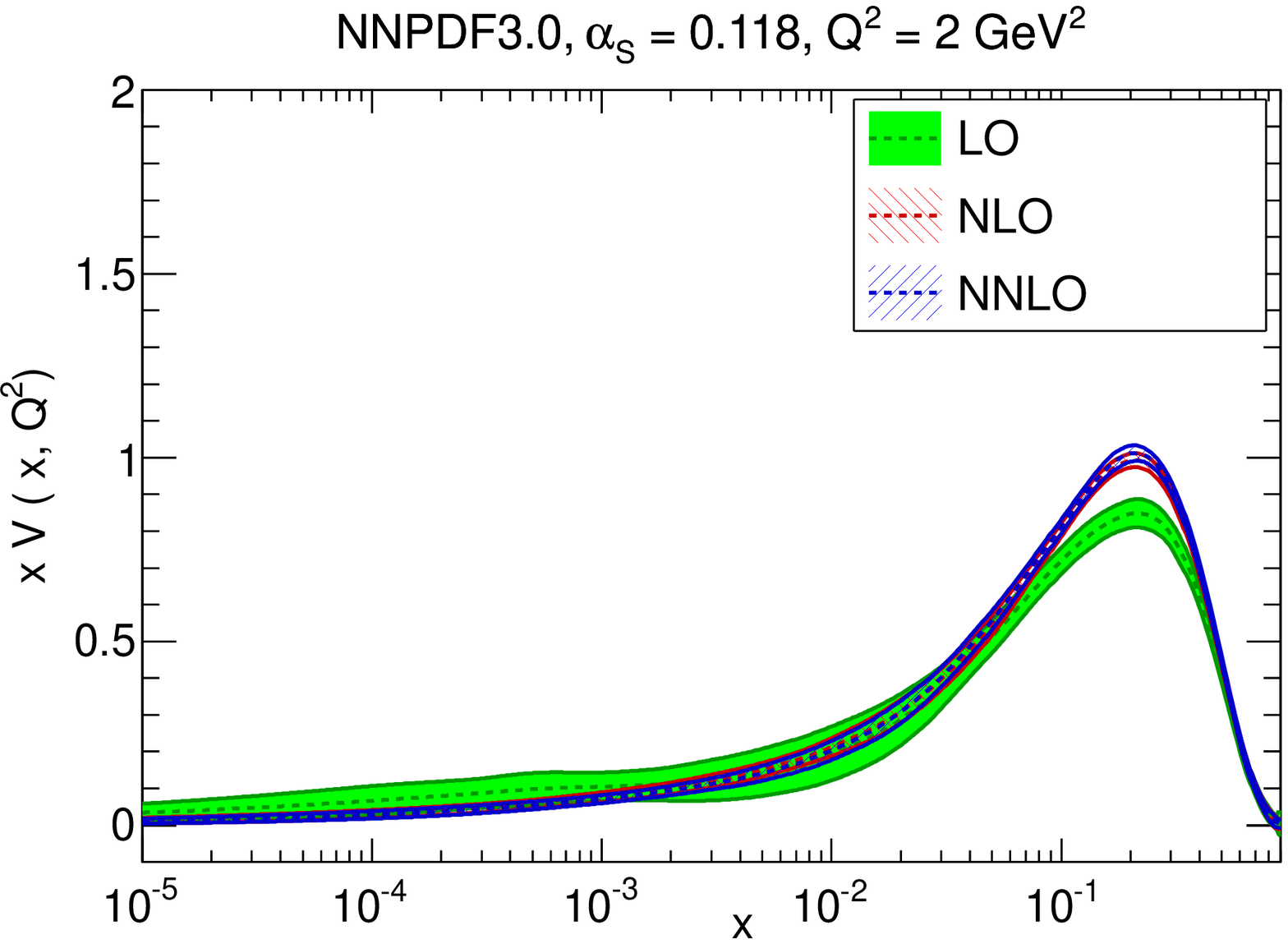}
\caption{\small
Same as Fig.~\ref{fig:30_vs_23_lowscale_nlo}, but now  comparing
NNPDF3.0 LO, NLO and  NNLO PDFs. \label{fig:global_nnlo_vs_nlo}}
\end{center}
\end{figure}

Next, in Fig.~\ref{fig:global_nnlo_vs_nlo} we compare
the LO, NLO and NNLO NNPDF3.0 parton
distributions at $Q^2=2$ GeV$^2$.
The large shift in the gluon between LO and NLO and its subsequent
stability at NNLO is clearly seen.  Specifically, the LO gluon
is very large, compensating for  missing
NLO terms in the DIS splitting functions and anomalous
dimensions. This
is a crucial ingredient for the tunes
to semi-hard and soft data in Monte Carlo parton shower programs (see
e.g.  Ref.~\cite{Skands:2014pea}). However, in the small $x$ region
missing higher order
corrections cause tension between the very accurate HERA data which
results in a bigger uncertainty at NLO than at LO, which then also
propagates onto the singlet quark.

At NLO, the small-$x$ gluon is rather flatter than the NNLO one, which tends
to go negative at small-$x$, being prevented to do so by positivity bounds. This
relatively unstable perturbative  behaviour of the small-$x$ gluon
might be related to unresummed small-$x$ perturbative corrections~\cite{Caola:2010cy}.
Quark PDFs are generally quite stable, with NNLO and NLO always in
agreement at the one-sigma level, and sizable shifts only seen when going from
LO to NLO, especially in the region around $x\sim 0.1$.

The comparison of parton distributions at  different
perturbative orders provides a way of estimating the uncertainty
related to missing  higher-order
corrections in the computations used for PDF determination,
using a method suggested in Refs.~\cite{Cacciari:2011ze,Bagnaschi:2014wea} and applied to
PDFs in Ref.~\cite{Forte:2013mda}.
This source of
theoretical uncertainty is currently not part of the total PDF errors,
which
only includes uncertainties due to the data and the methodology (recall
Sect.~\ref{sec:uncert}, especially Fig.~\ref{fig:ratiofit1}).
As these become increasingly
small, however, an estimate of the theoretical uncertainty related to
missing higher perturbative orders becomes more and more important.

While a full study of these  uncertainties is beyond
the scope of this work, we may provide a first estimate by simply
studying the variation of each individual PDF when going up one
order: this must be taken as an order-of-magnitude estimate for the
time being, as a detailed estimate of the impact of this shift on
the predictions for physical processes would involve a study of the way
perturbative corrections to different processes are
correlated~\cite{Bagnaschi:2014wea,Forte:2013mda}.
In the remainder of this section,
we will refer to the uncertainty related to missing higher
orders as ``theory uncertainty'', while we will call PDF uncertainty
the standard uncertainty as discussed on Sect.~\ref{sec:uncert}, which
only includes the uncertainty from the data and the methodology.

\begin{figure}[t]
\begin{center}
\epsfig{width=0.42\textwidth,figure=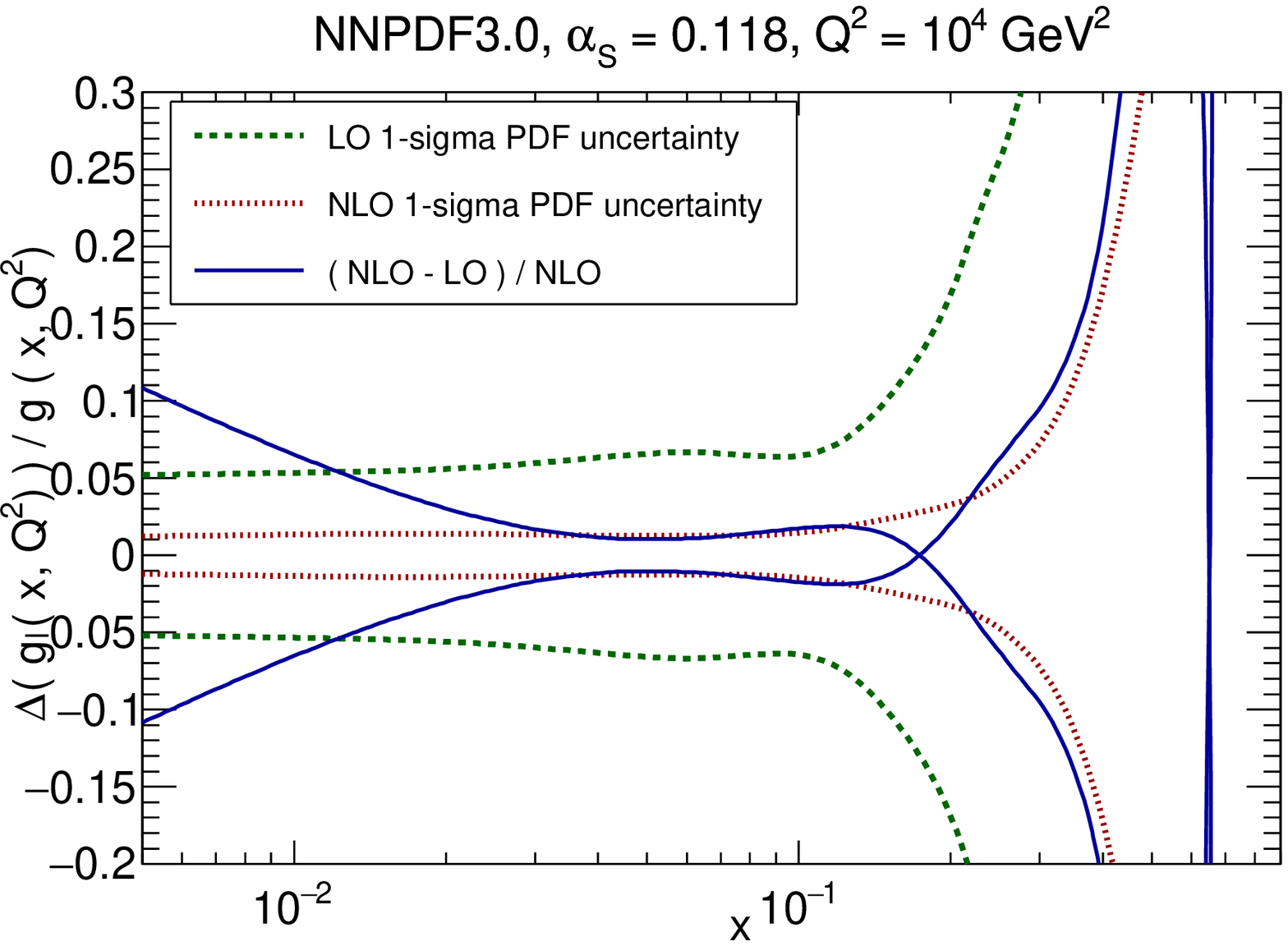}
\epsfig{width=0.42\textwidth,figure=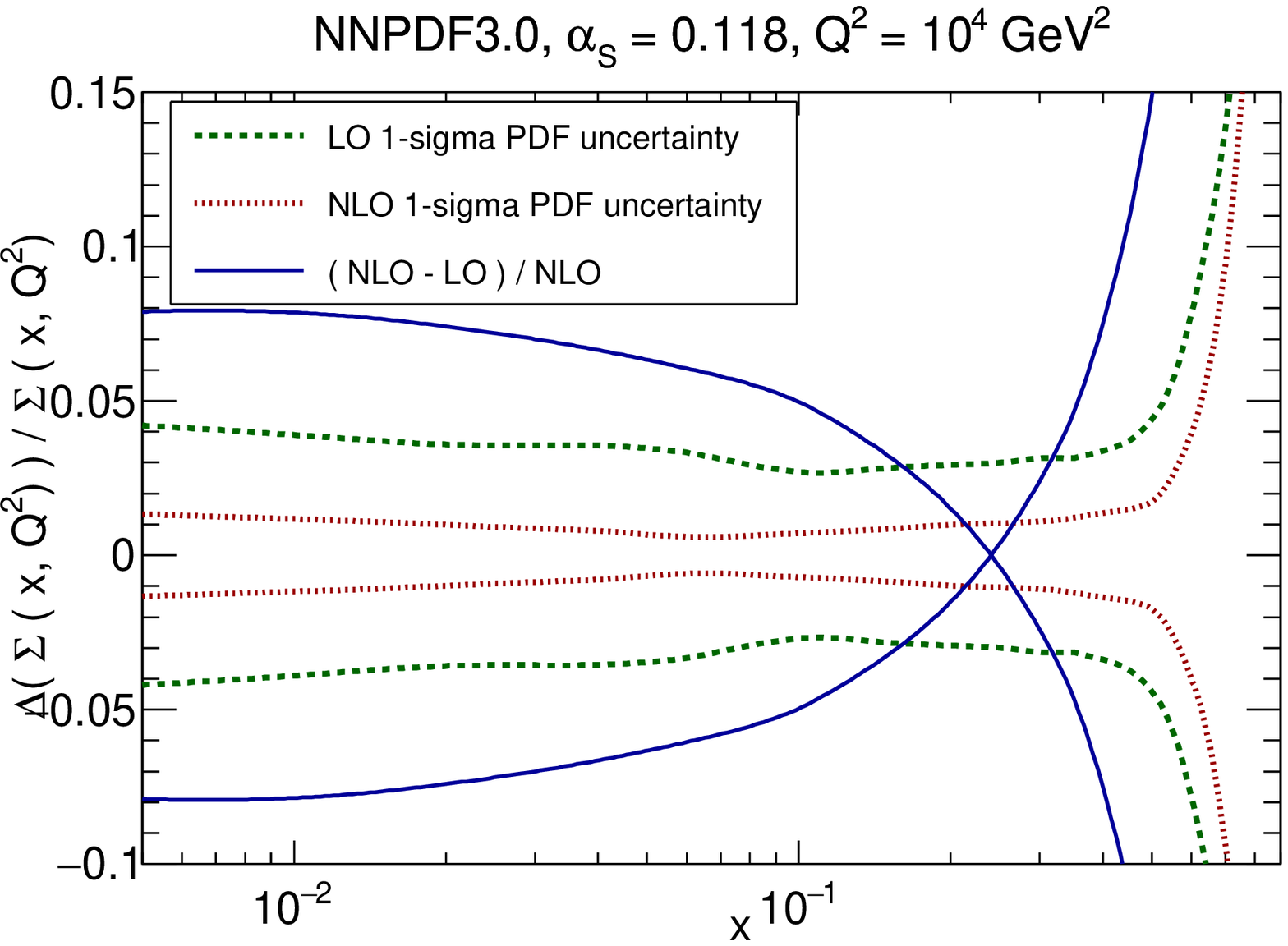}
\epsfig{width=0.42\textwidth,figure=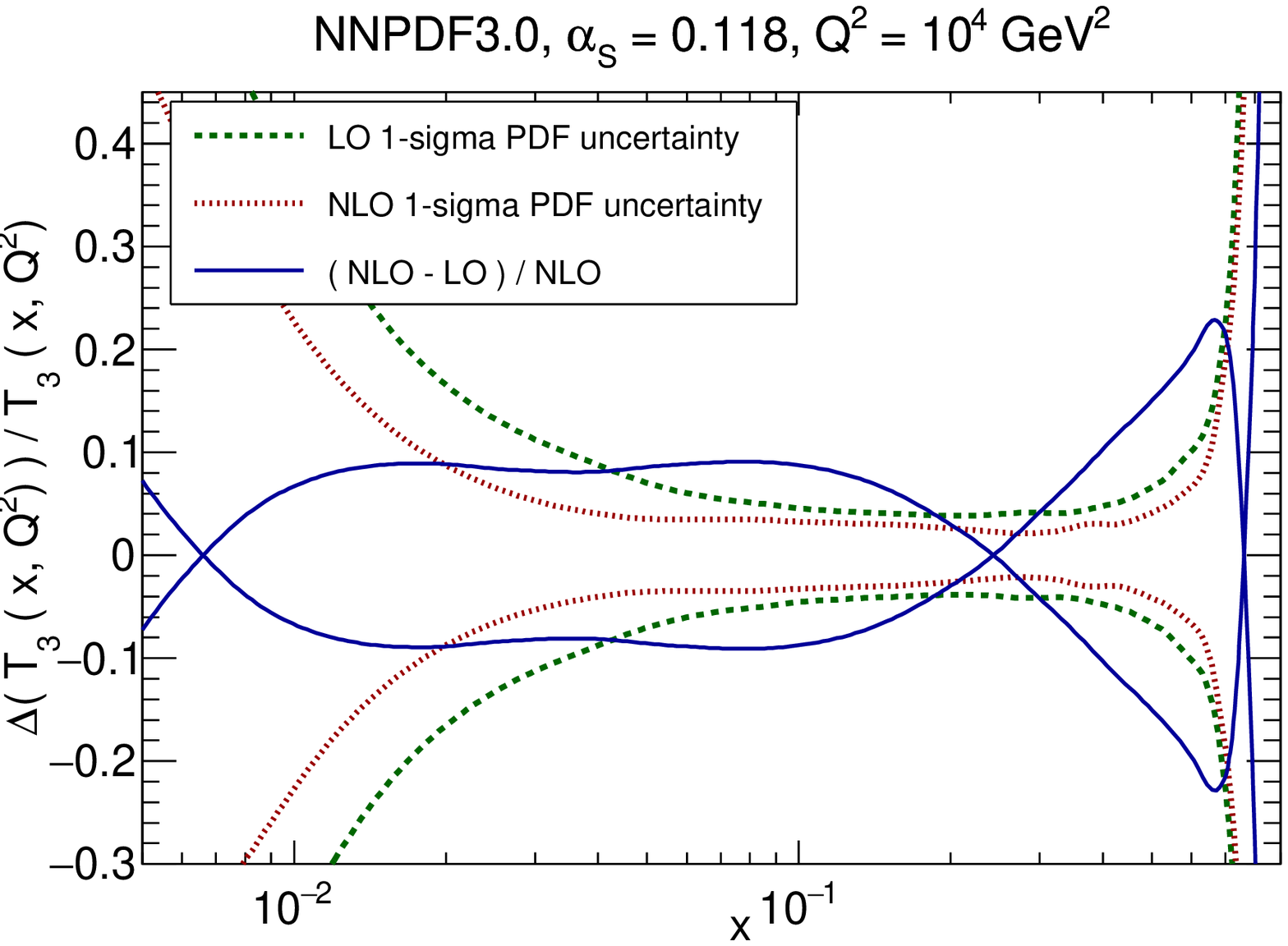}
\epsfig{width=0.42\textwidth,figure=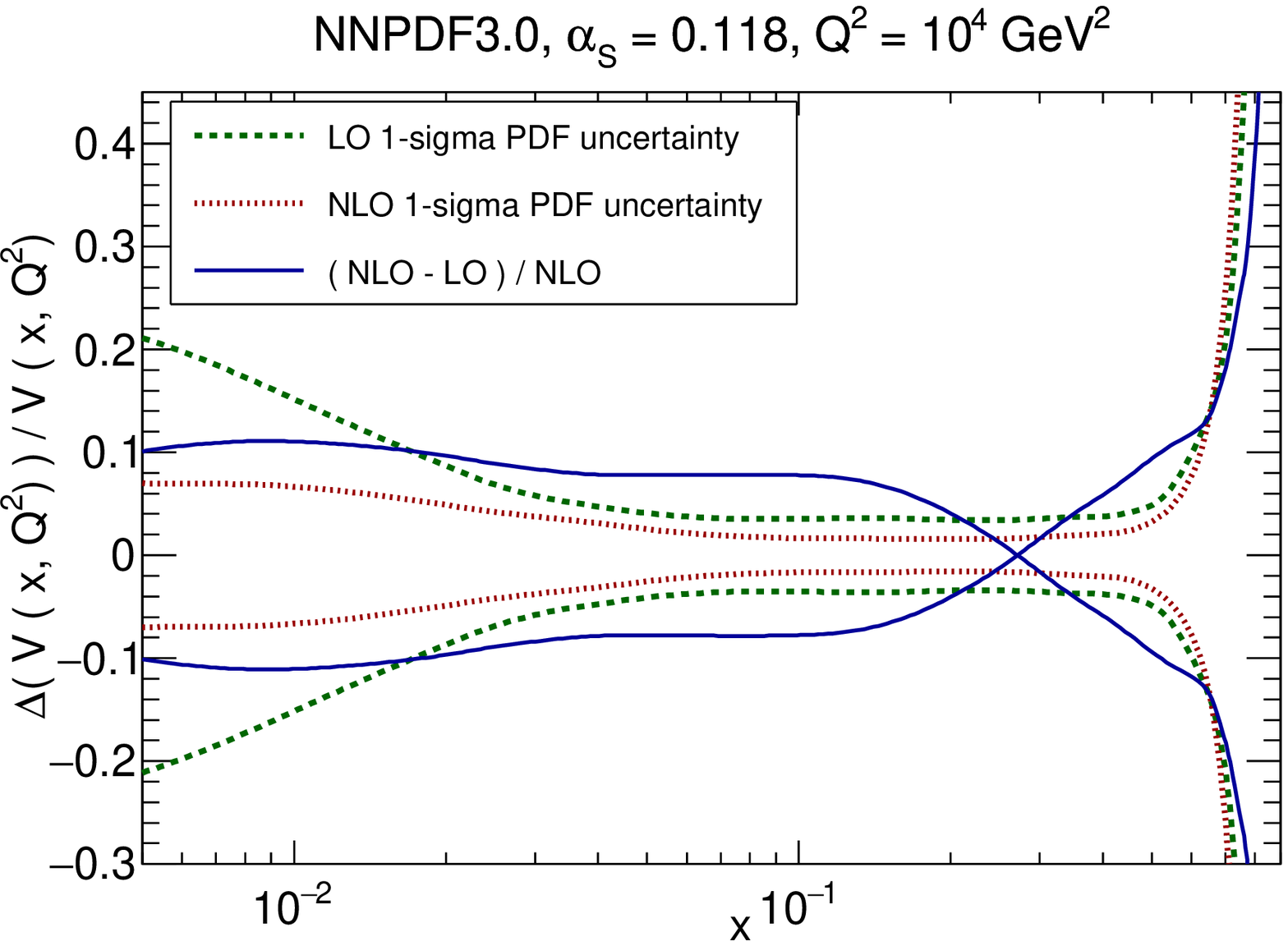}
\caption{\small The shift in NNPDF3.0 PDFs with $\alpha_s(M_Z)=0.118$
at $Q^2=10^4$ GeV$^2$ when going from LO to NLO, normalized to the
NLO central value,
compared to the LO and NLO PDF uncertainties.
From top to bottom and from left to right the
gluon,  singlet,  isospin triplet and
total valence are shown.
 \label{fig:th_unc_lo}}
\end{center}
\end{figure}

In Fig.~\ref{fig:th_unc_lo} the relative shift
in the central values of the NNPDF3.0 PDFs at $Q^2=10^4$ GeV$^2$
when going from LO to NLO (normalized
to the NLO) is compared to
the PDF uncertainty  at the two orders.
This shift may be viewed as
the LO theory uncertainty, and thus likely an upper bound to the NLO theory
uncertainty. It is clear that
at LO theory uncertainties are dominant essentially everywhere,
and especially for the small-$x$ gluon and the medium and small-$x$
quarks, as also apparent from Fig.~\ref{fig:global_nnlo_vs_nlo}.

\begin{figure}[t]
\begin{center}
\epsfig{width=0.42\textwidth,figure=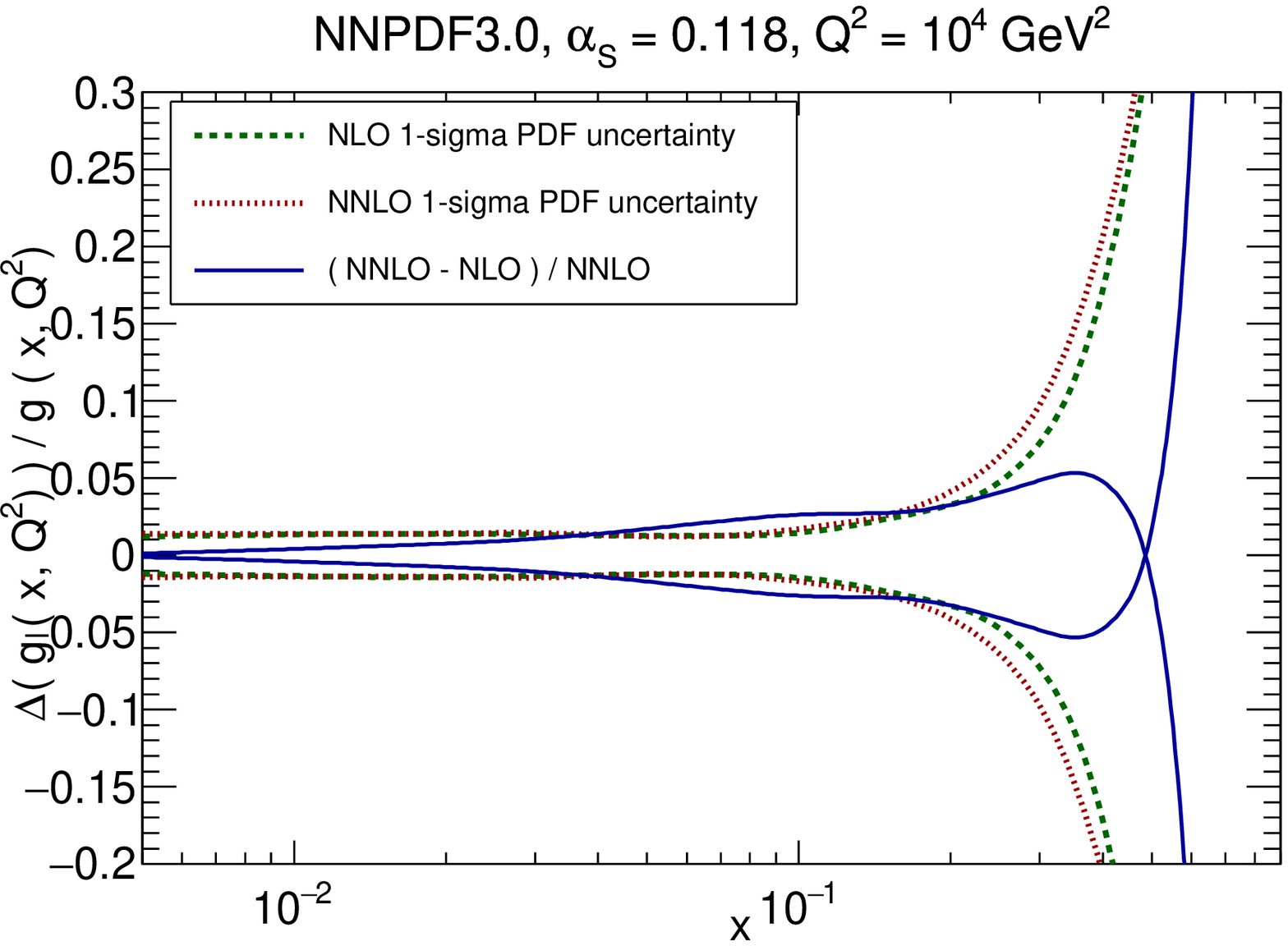}
\epsfig{width=0.42\textwidth,figure=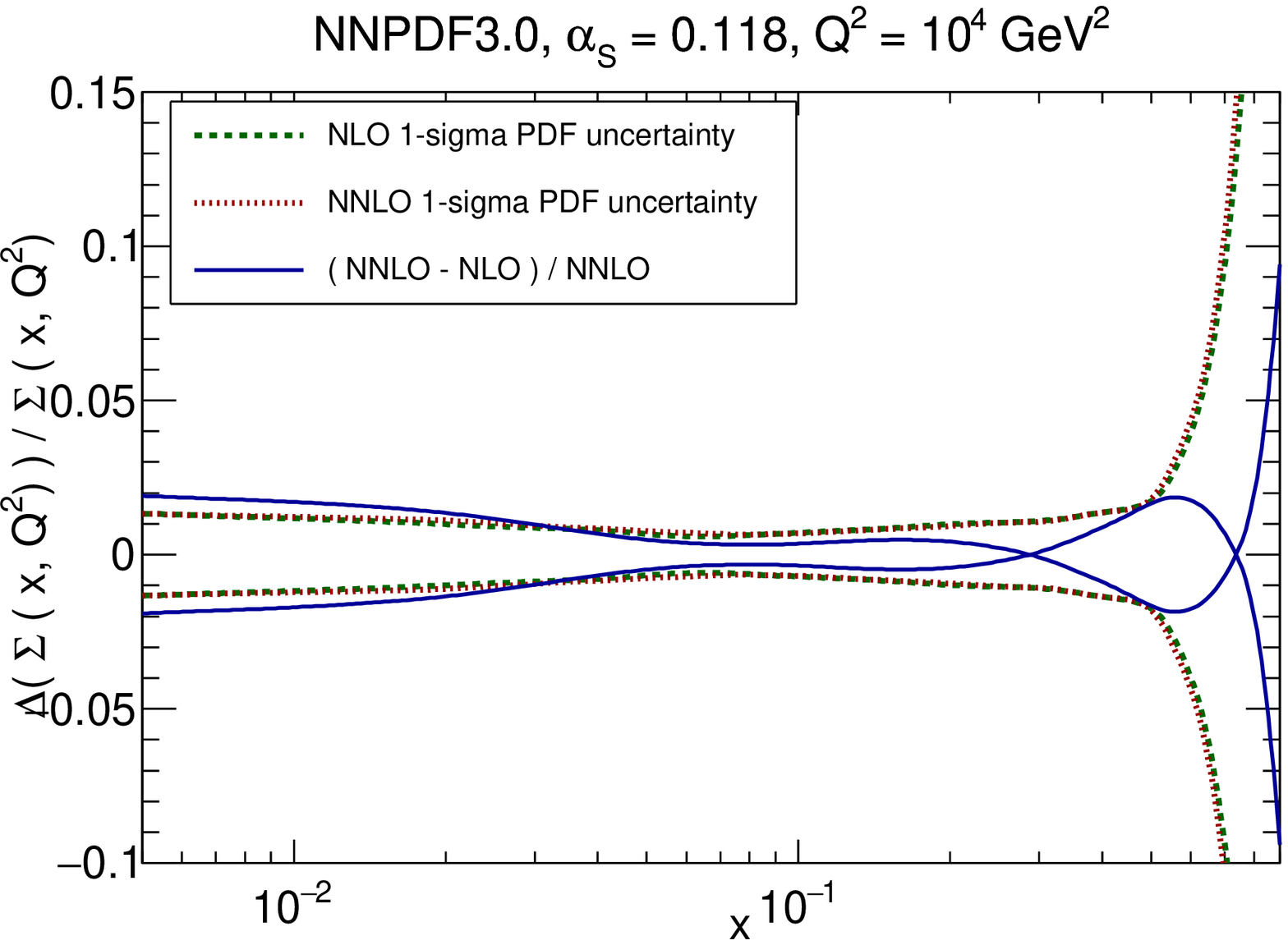}
\epsfig{width=0.42\textwidth,figure=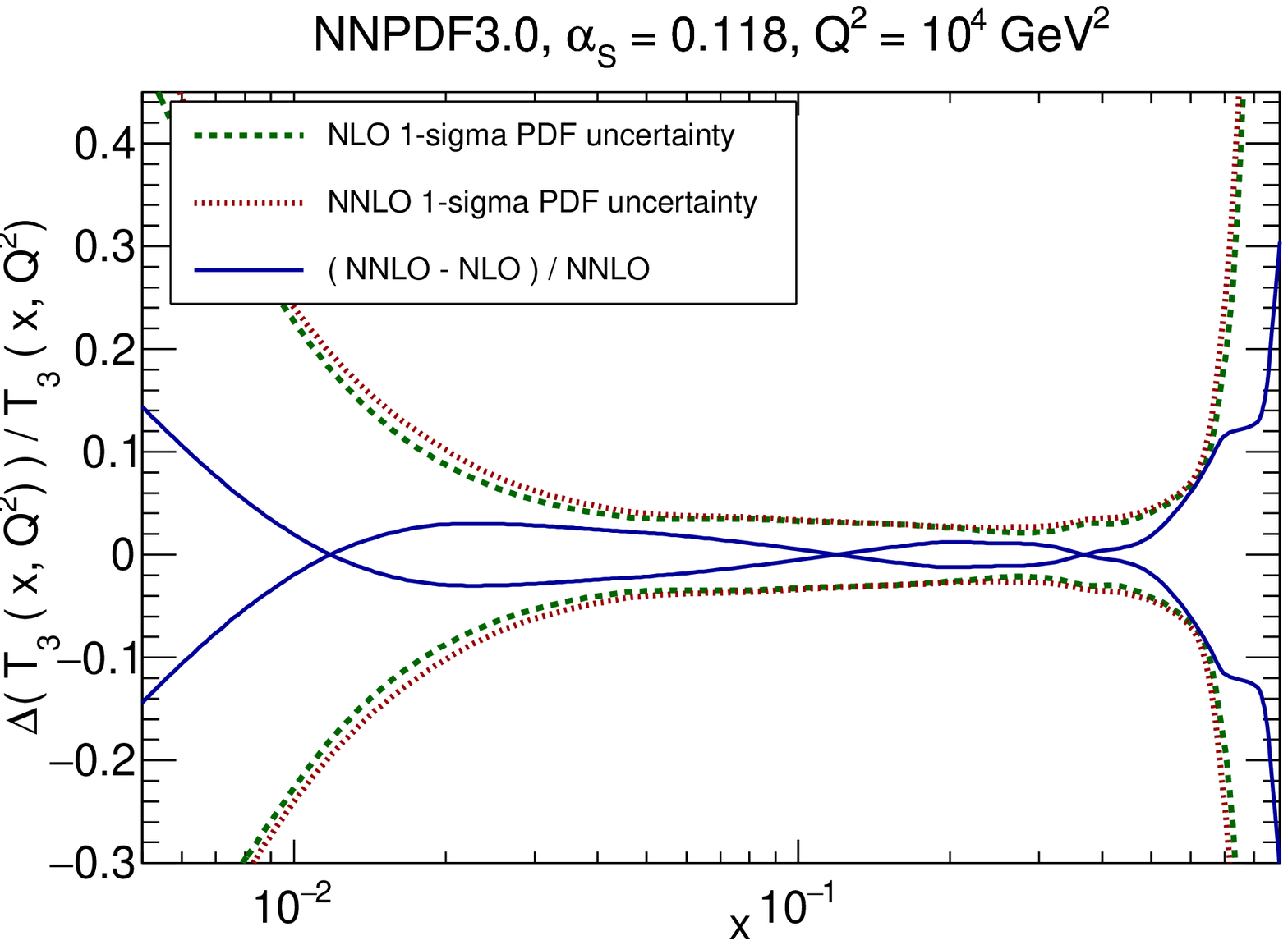}
\epsfig{width=0.42\textwidth,figure=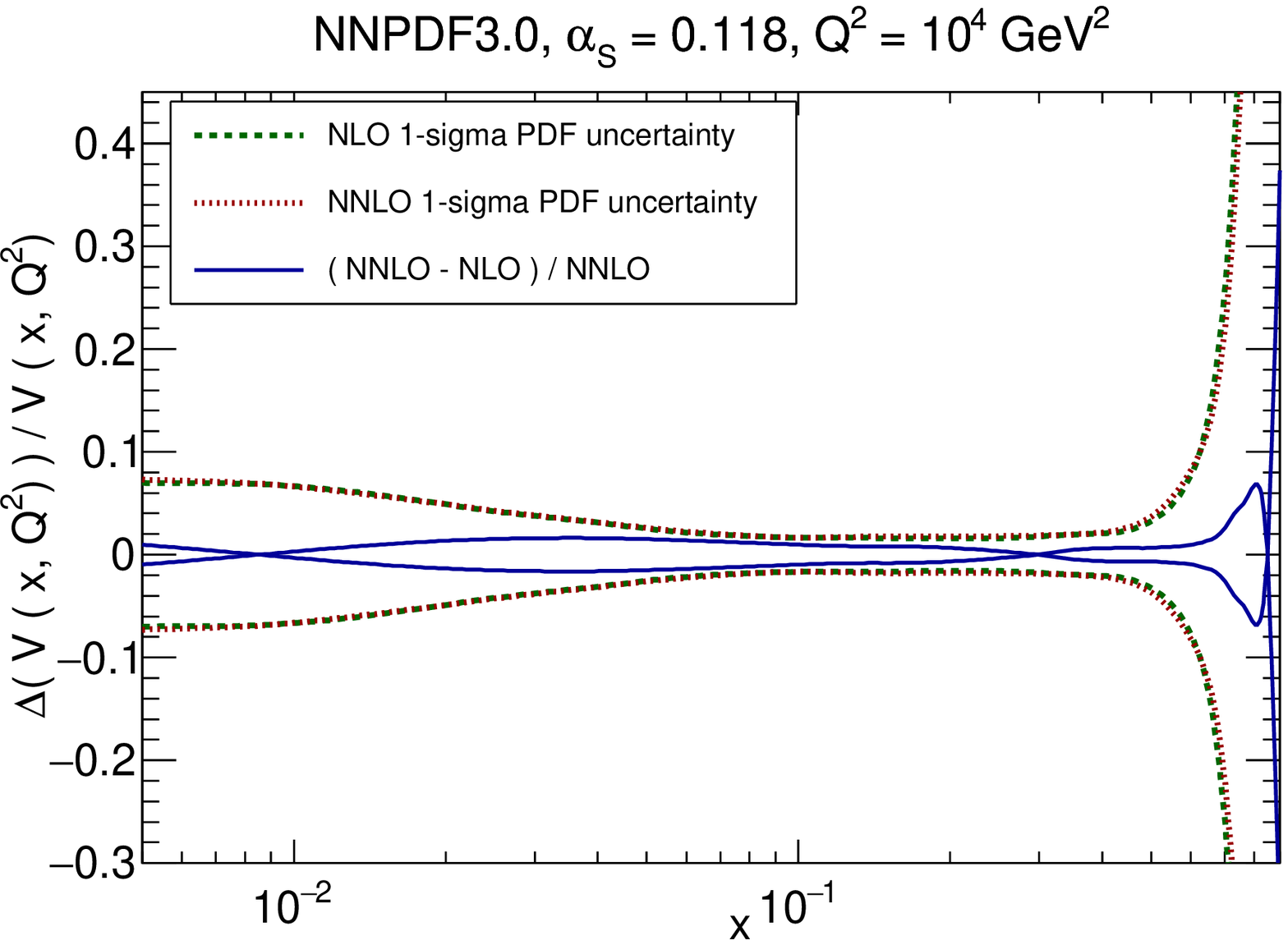}
\caption{\small
Same as Fig.~\ref{fig:th_unc_lo}, but now comparing NLO and NNLO.
 \label{fig:th_unc_nlo}}
\end{center}
\end{figure}

The shifts when going from NLO to NNLO (normalized to the NNLO) are
shown in Fig.~\ref{fig:th_unc_nlo}.
At this order, the theory uncertainty becomes smaller than the PDF
uncertainty,
thereby suggesting that their current neglect is mostly justified.
However, is also apparent that theory and PDF  uncertainties may
become comparable  in some important cases, like
the large-$x$ gluon, relevant for example for top quark pair
production, or the medium-$x$ quark singlet PDF, which is
relevant for LHC electroweak boson production.
Note that for the
gluon at $x\sim 10^{-2}$, relevant for Higgs production
in gluon-fusion, the perturbative convergence is very good,
as already highlighted in Ref.~\cite{Forte:2013mda}.

The results of Figs.~\ref{fig:th_unc_lo}
and~\ref{fig:th_unc_nlo} suggest that as PDF uncertainties
decrease, either by the addition of
new experimental constraints or
by refinements in the fitting methodology,
 a careful estimate of
theory uncertainties will become mandatory.
In the case of NNPDF3.0 this is especially true now
given
that, thanks to the closure
test validation, methodological uncertainties are under full control.

\subsubsection{Model uncertainties}
\label{sec:model}

While uncertainties related to higher order corrections are perhaps
the largest source of uncertainty which is not determined
systematically and thus not included in the standard PDF uncertainty,
there are a few more sources of uncertainty which are also not part of
the current PDF uncertainty and which might become relevant as the
precision of the data increases. These have to do with further
approximations which are made in the theoretical description of the
data, and we generically refer to them as ``model'' uncertainties. We
now discuss the likely dominant sources of model uncertainties,
namely, those related to nuclear corrections and those related to the
treatment of heavy quarks.

\begin{figure}[t]
\begin{center}
\epsfig{width=0.85\textwidth,figure=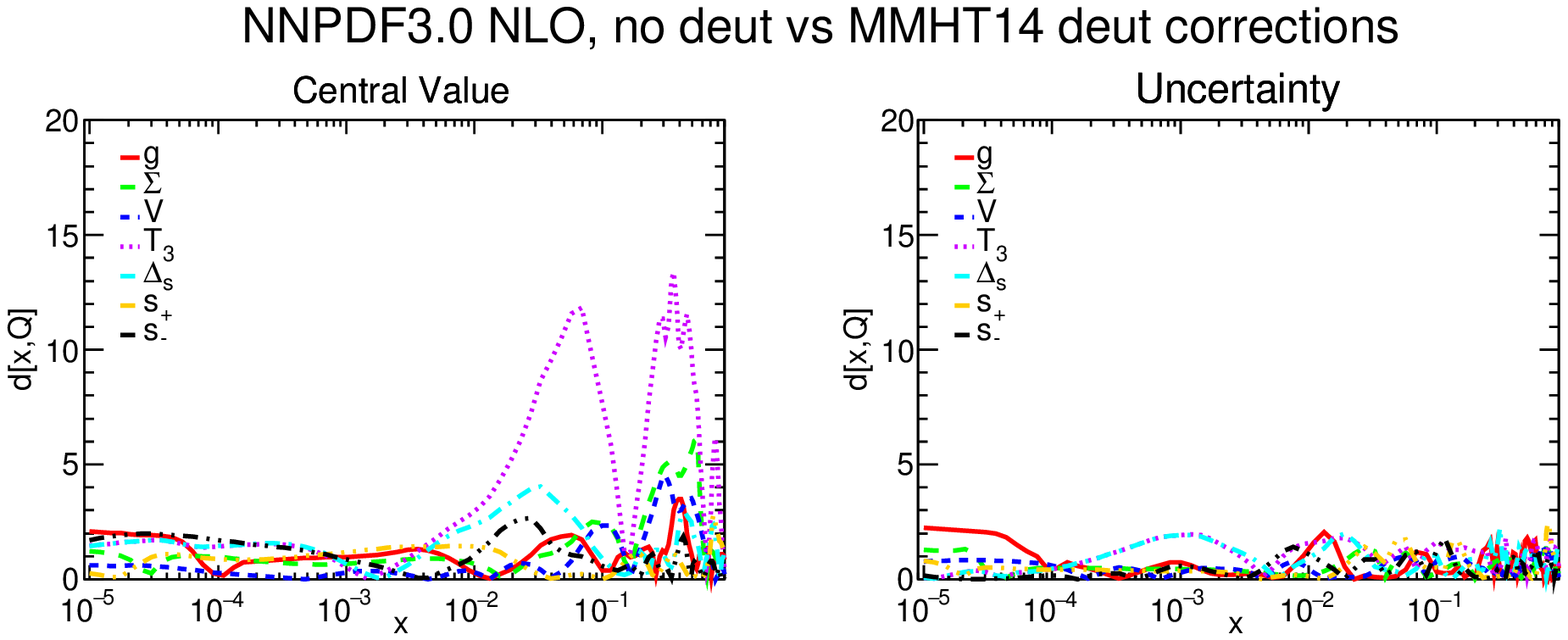}
\caption{\small
Same as Fig.~\ref{fig:distances_30_vs_23_nnlo},
but now comparing the  NNPDF3.0 NLO PDFs
with and without deuterium
nuclear corrections.
\label{fig:nucl30_highscale_nlo_dist}}
\end{center}
\end{figure}
Several fixed-target data included in the NNPDF3.0 PDF determination are taken on nuclear targets.
These include all of the neutrino deep-inelastic scattering data, namely the CHORUS and NuTeV data
sets of Table~\ref{tab:completedataset}, the data for
charged-lepton deep-inelastic scattering from deuteron targets in the NMC,
BCDMS, and SLAC data sets also in Tab.~\ref{tab:completedataset}, and
the data for Drell-Yan production on a deuterium target in the DY~E866
data set in Tab.~\ref{tab:completedataset2}. The impact of nuclear
corrections on the NNPDF2.3 PDF determination  was previously discussed in
Ref.~\cite{Ball:2013gsa}, where the NNPDF2.3 fit was repeated by
introducing deuterium nuclear corrections according to various
models, and found to be non-negligible (up to about one and a half
sigma) but effecting only the down distribution at large $x$.

\begin{figure}[t]
\begin{center}
\epsfig{width=0.42\textwidth,figure=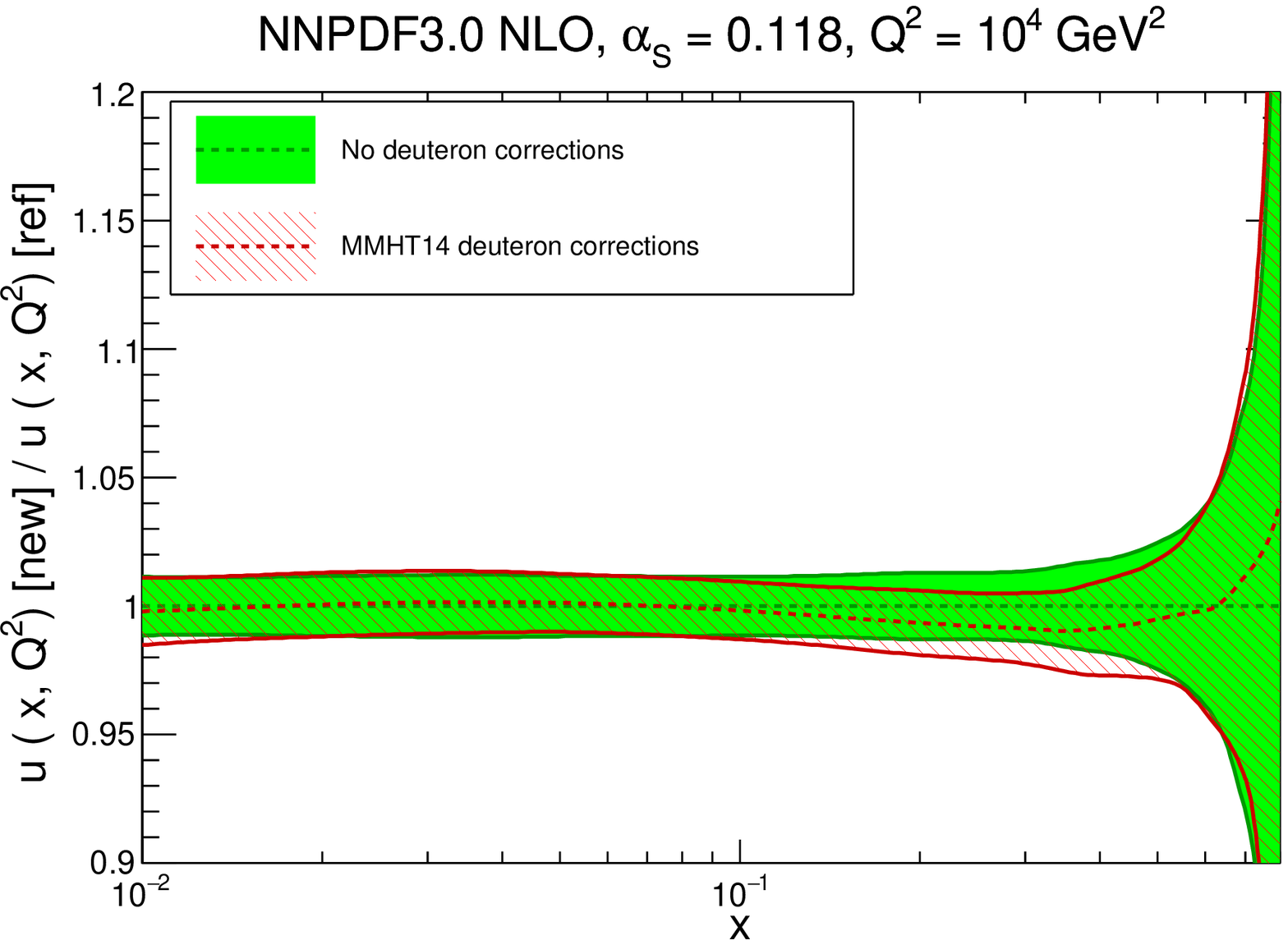}
\epsfig{width=0.42\textwidth,figure=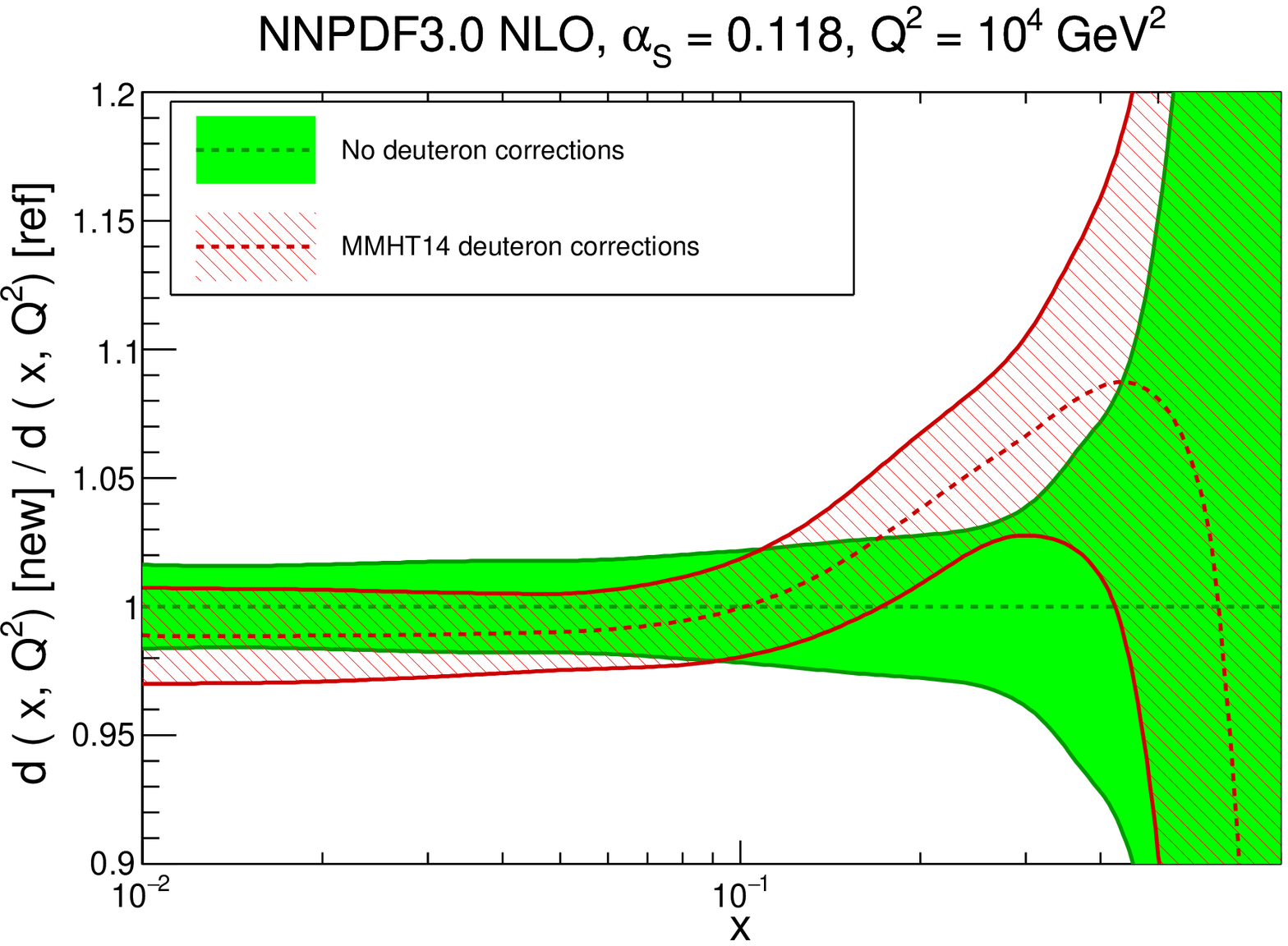}
\caption{\small
Same as Fig.~\ref{fig:30_vs_23_highscale}, but
now comparing the  NNPDF3.0 NLO fit with and without
deuterium
nuclear corrections.
From left to right the up and down quark PDFs are shown. \label{fig:nucl30_highscale_nlo}}
\end{center}
\end{figure}
We have repeated the NNPDF3.0 PDF determination, but now including
deuterium corrections according to the recent model of Ref.~\cite{Harland-Lang:2014zoa},
which supersedes the previous treatment of higher twist corrections of
Ref.~\cite{Martin:2012xx},
considered in Ref.~\cite{Ball:2013gsa} along with other
models. The pattern of deviations between the PDFs determined with and without
nuclear correction is shown in
Fig.~\ref{fig:nucl30_highscale_nlo_dist}, and it
is very similar to that seen in
Ref.~\cite{Ball:2013gsa} when correcting the NNPDF2.3 in an analogous
way,
but with a somewhat more moderate impact, as
one might expect given the fact that the data set used to determine NNPDF3.0 is
bigger than that used for NNPDF2.3. Essentially only the up and down
quark distributions are affected; they are compared in
Fig.~\ref{fig:nucl30_highscale_nlo}, at
$Q^2=10^4$~GeV$^2$:  it is apparent that the effect is always below
one sigma.  In view of the theoretical uncertainty involved in the
modeling of these corrections, we prefer not to include them in the
fit as it is unclear that the uncertainty on them is significantly
smaller than their size. Nuclear corrections to neutrino data are
likely to be yet smaller, with the possible exception of the strange
distribution~\cite{Harland-Lang:2014zoa}.

Another important potential source of theoretical uncertainty is
related to the treatment of heavy quarks. As discussed in
Sect.~\ref{sec:hq}, we use a computational scheme, the FONLL scheme,
which  ensures that all the available perturbative
information is included. However, there are also aspects that go
beyond perturbation theory, namely, the dependence on the quark mass
itself, and the possible presence of an intrinsic heavy quark
component~\cite{Brodsky:1980pb}, namely, the
possibility that the boundary condition for the
evolution of the heavy quark PDFs is not zero, but rather an
independently parametrized PDF~\cite{Pumplin:2007wg}.

 The dependence
of PDFs on the values of the heavy quark masses was previously studied by us in
Ref.~\cite{Ball:2011mu} within the context of the NNPDF2.1 PDF
determination, where the values of $m_c$ and $m_b$ were varied, in the
absence of intrinsic heavy quark PDFs.
The main result of this study
was that the value of the heavy quark  mass mostly affects the heavy quark PDF
by providing the threshold for generating it by perturbative evolution:
 a lower mass value corresponds to a larger PDF
at a given scale, because the evolution length is larger.
For the $b$ quark PDFs, which are expected to be
perturbative quantities, with a negligible intrinsic component, this
dependence on the quark mass value is likely to be a real physical effect.
However since charm has a threshold at the boundary between the perturbative
and nonperturbative regions, charm PDFs might have a significant intrinsic component,
and much of the dependence on the charm mass might be compensated by changes
in this intrinsic PDF.

\begin{figure}
\begin{center}
\epsfig{width=0.32\textwidth,figure=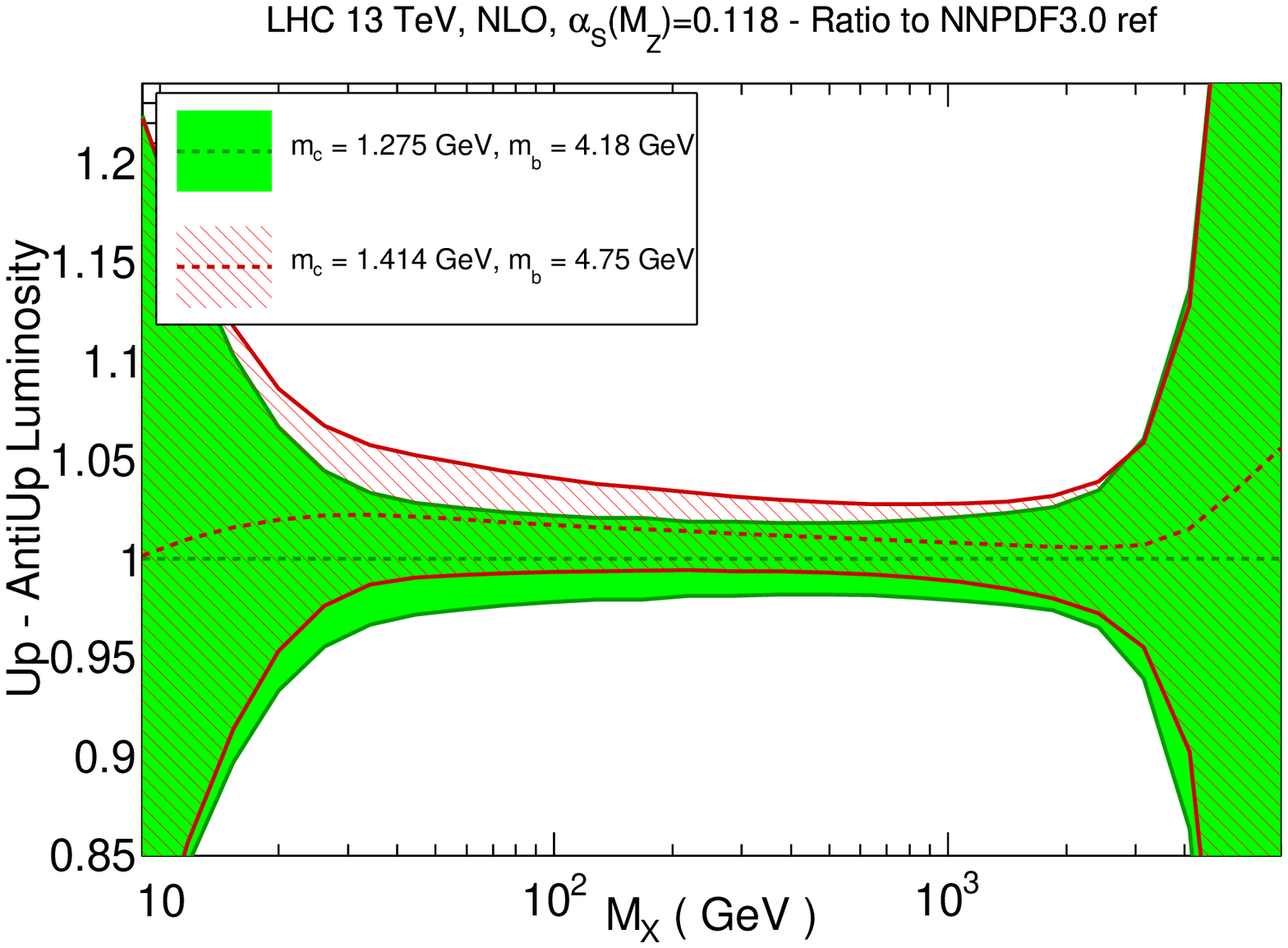}
\epsfig{width=0.32\textwidth,figure=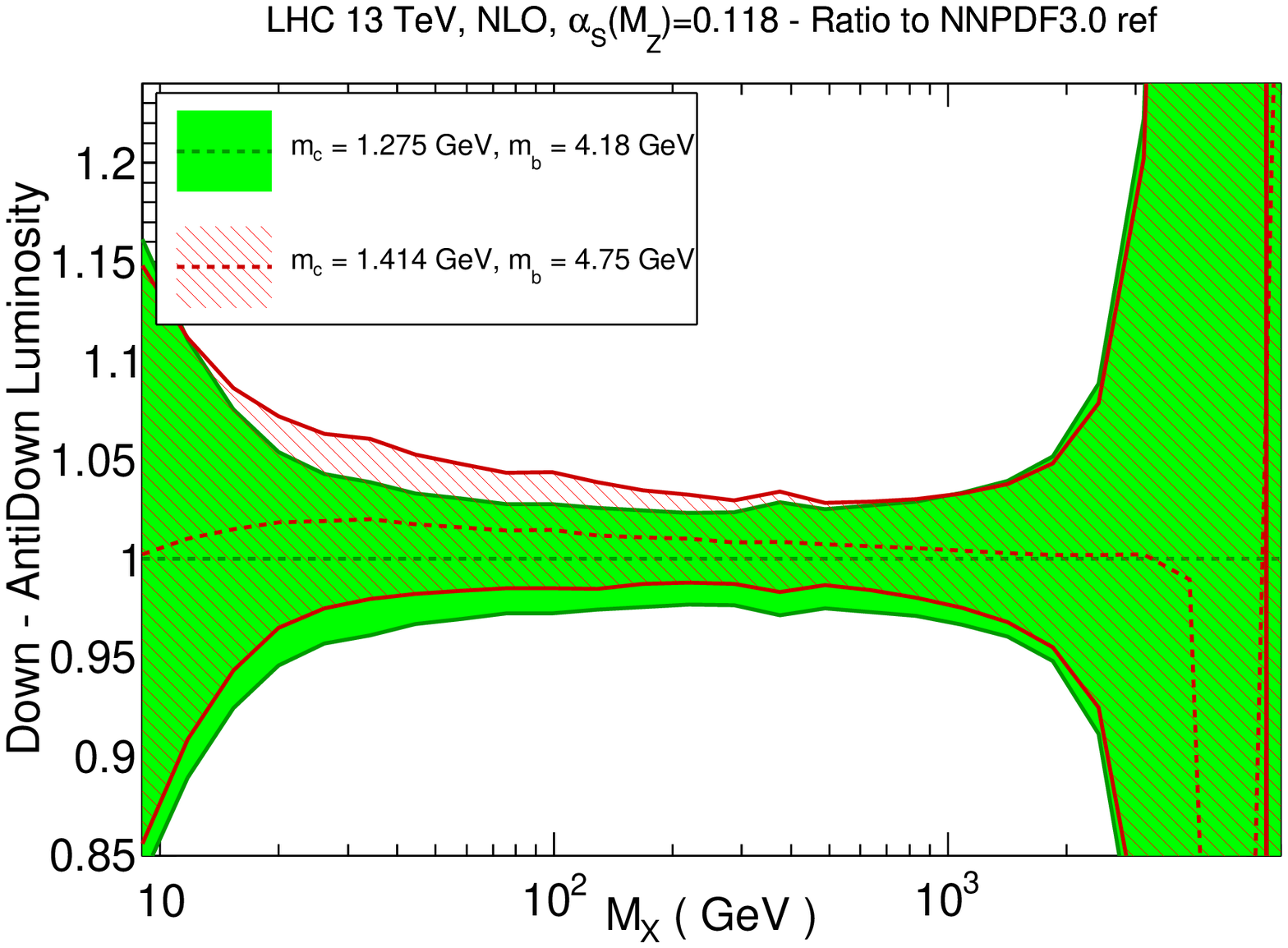}
\epsfig{width=0.32\textwidth,figure=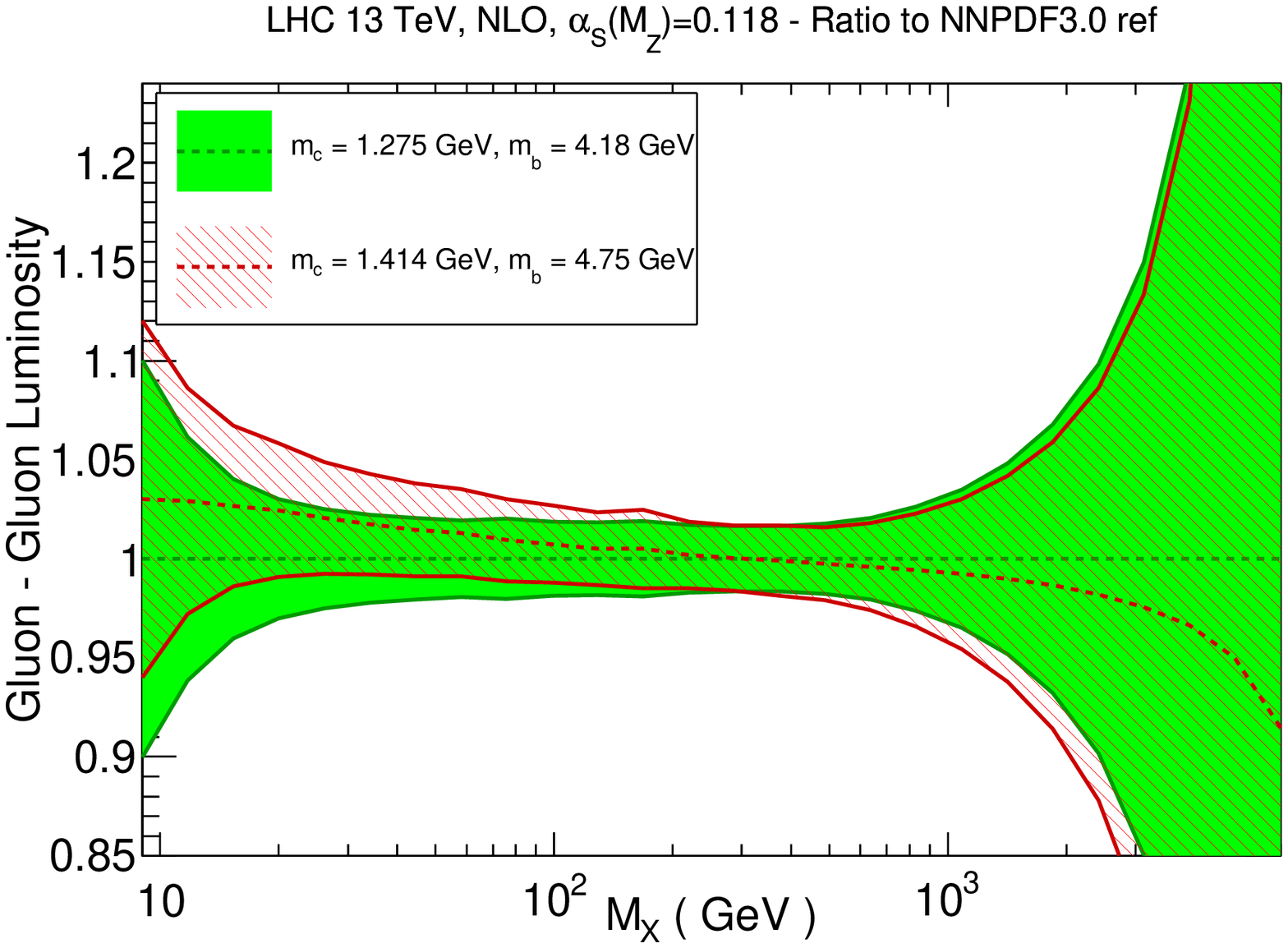}
\caption{\small
 \label{fig:luminnlohq}
Dependence on the value of the heavy quark masses of
 parton luminosities Eq.~(\ref{eq:lumdef}) computed
using NNPDF3.0 NLO PDFs with  $\alpha_s(M_Z)=0.118$. Results are shown as
ratios to the default set. For left to right the up-antiup,
down-antidown and gluon-gluon luminosities are shown.
 }
\end{center}
\end{figure}

As mentioned in Sect.~\ref{sec:hq}, Eq.~\ref{eq:hqmv}, the heavy quark
mass values used in the current NNPDF3.0 PDF determination differ from
the values previously used in the NNPDF 2.3 determination, essentially
because now we use $\overline{\rm MS}$ PDG mass values, while the
previous values were close to pole mass values. The shift is larger
than the current uncertainty on $\overline{\rm MS}$ masses. In order
to assess the impact of this change, and thus also of the dependence
on heavy quark masses, we have repeated the NNPDF3.0 PDF NLO
determination using the heavy quark mass values that were used for the
NNPDF2.3 set. In
Fig.~\ref{fig:luminnlohq} the respective  parton luminosities at
$Q^2=10^4$~GeV$^2$ are compared.

Results are in agreement with the findings of Ref.~\cite{Ball:2011mu},
where a similar effect due to changes of the charm mass was
observed. Note that the impact of changing the bottom mass was found to
be negligible on all luminosities, except the bottom luminosity
itself. The effect is not entirely negligible. However, as mentioned above,
we expect that some of this dependence
might be absorbed into an intrinsic charm PDF. At NLO,
$\overline{\rm MS}$ and pole mass-scheme expressions coincide, with
a small correction at NNLO, hence
it seems more appropriate to use the more accurate $\overline{\rm MS}$ mass value.
The shifts seen in Fig.~\ref{fig:luminnlohq}
should be taken as an upper bound to the size of the uncertainty
related to the charm mass value, whose exact assessment will only be
possible once an intrinsic charm component is introduced in our PDF fits.

We finally mention that further model uncertainties are expected to
come from the treatment of electroweak interactions, both in the
choice of parameters, and the treatment of higher order terms
(including mixed strong-electroweak
corrections~\cite{Barze:2013yca}). These are generally smaller than
the uncertainties discussed here, though they could become significant
in particular kinematic regions or for specific processes, such as for
instance high-mass production of $W$ pairs.

\clearpage

\subsection{Dependence on the dataset}
\label{sec:resdataset}

We now study the  dependence of the NNPDF3.0 PDFs on the choice
of dataset, exploring a wide range of variations as compared to the
dataset used in the reference global fit.
First, we propose a new definition of  conservative
PDFs based on a maximally consistent dataset,
along the lines of previous proposals~\cite{Martin:2003tt}, but now
using an objective criterion, rather than  theoretical
expectations.
Then we explore systematically the impact of the new data in the NNPDF3.0
framework, by comparing fits to various subsets of data both to the
global fit, and to a minimal fit based on HERA data only; in this
context we  also study the impact of the new  HERA-II data.
We also construct sets where the HERA data is supplemented by all
available data by either ATLAS or CMS; these are specially useful
for comparisons with related studies by the LHC collaborations.
We specifically study  in detail the
impact of jet data, the associate theoretical uncertainties, and their
impact on the  uncertainty on the
large-$x$ gluon, and the impact of $W$+$c$ data on strangeness and their
relative weight in comparison to neutrino data.

\subsubsection{Conservative PDFs from a consistent dataset}
\label{sec:conservative}

Inconsistencies between data which enter a global PDF determination can distort the
statistical interpretation of PDF uncertainties. Inconsistency of any individual
dataset with the bulk of the global fit may suggest that its
understanding, either
from the theoretical or experimental point of view, is not complete,
and that its exclusion from the fit might be advantageous.
In order to minimize such inconsistencies, ``conservative'' partons
have been suggested, for example by introducing restrictive kinematic cuts
which remove potentially dangerous regions~\cite{Martin:2003tt}, or
by picking data which one might expect to be more reliable:
for example, the NNPDF2.3 collider-only fit~\cite{Ball:2012cx}, based
on the expectation that  collider data, because of their higher
energy,
should be  more reliable
than fixed-target data.

\begin{table}[t]
\begin{center}
\footnotesize
\begin{tabular}{c||ccc|ccc}
\hline
 & \multicolumn{3}{c|}{NLO global fit} &  \multicolumn{3}{c}{NNLO global fit} \\
\hline
Experiment & mean & mode  & median & mean &  mode & median \\
\hline
NMC $d/p$ & 1.04 &	1.01	&1.03&    1.04&	1.01	&1.03 \\
NMC $\sigma^{\rm NC,p}$ &1.32&	1.31&	1.27 & 1.27&	1.26&	1.27 \\
SLAC &  1.31&	1.27&	1.30 &  1.13&	1.09&	1.12 \\
BCDMS &	1.17&	1.16&	1.17  & 1.20 &	1.19 &	1.20  \\
CHORUS& 1.11 &	1.10 &	1.11 &	1.10 &	1.09 &	1.09 \\
NuTeV & 1.04	&0.90&	0.98&   1.06&	0.92	&1.00 \\
HERA-I &	1.09	&1.09&	1.10&  1.10&	1.09&	1.09 \\
ZEUS HERA-II & 1.23 &	1.22&	1.23&  1.25&	1.24&	1.25 \\
H1 HERA-II  & 	1.30&	1.3&	1.31&  1.35&	1.34&	1.34 \\
HERA $\sigma_{\rm NC}^{c}$ & 1.10&	1.06&	1.09 & 1.14&	1.11&	1.13 \\
\hline
E886 $d/p$ &1.00 &	0.88&	0.96 &  	1.01&	0.88 &	0.96	\\
E886 $p$  &	1.13&	1.11&	1.12 & 1.15&	1.14&	1.15 \\
E605 & 0.97 &	0.94&	0.96 &  0.94 &	0.91 &	0.93 \\
CDF $Z$ rapidity &   1.34 &	1.28&	1.32&  1.39&	1.32&	1.36 \\
CDF Run-II $k_t$ jets & 1.09 &	1.06&	1.08 & 1.15&	1.12&	1.14 \\
D0  $Z$ rapidity & 1.34 &	1.28 &	1.32  & 0.86&	0.82&	0.85 \\
\hline
ATLAS $W,Z$ 2010  &	1.20 &	1.15 &	1.18 & 	1.17	&1.12	&1.15 \\
ATLAS 7 TeV jets 2010 & 0.76 &	0.74&	0.75&  	1.09&	0.92 &	1.02 \\
ATLAS 2.76 TeV jets & 	0.86 &	0.83 &	0.85 &  	1.07 &	0.57 &	0.83 \\
ATLAS high-mass DY  & 2.22&	1.68&	2.03  & 1.82 &	1.34 &	1.63 \\
CMS $W$ electron asy & 1.05&	0.91&	0.99 &	 	1.00 &	0.87 &	0.95 \\
CMS $W$ muon asy   &	1.62&	1.42&	1.54 & 1.60 &	1.40 &	1.53 \\
CMS jets 2011  & 1.01&	0.97 &	0.99 &   1.09 &	1.07 &	1.08 \\
CMS $W$+$c$ total &  1.60 &	1.17 &	1.42 &  1.50 &	1.09	& 1.33 \\
CMS $W$+$c$ ratio &  1.93	& 1.43 &	1.74 &   	1.88 &	1.39 &	1.69 \\
CMS 2D DY 2011  & 1.27 &	1.25 &	1.27  & 1.28	& 1.27 &	1.28 \\
LHCb $W,Z$ rapidity  &1.10 &	1.02 &	1.07 &  1.20 &	1.12 &	1.17 \\
$\sigma(t\bar{t})$  & 1.65 &	1.24 &	1.49  & 1.09 &	0.75 &	0.95  \\
\hline
\end{tabular}
\caption{\small The mean, mode and median of the
$P\lp \alpha\rp$ distributions~\cite{Ball:2010gb,Ball:2011gg} (see text) for
all the experiments in the NNPDF3.0 global fits,
both at NLO (left) and at NNLO (right).
 \label{tab:chi2tab_conservative_v3}
}
\end{center}
\end{table}

\begin{table}[t]
\begin{center}
\footnotesize
\begin{tabular}{c||cc|cc|cc|cc}
\hline
 & \multicolumn{2}{c|}{$\alpha_{\rm max}=1.1$} &  \multicolumn{2}{c|}{$\alpha_{\rm max}=1.2$} & 
 \multicolumn{2}{c|}{$\alpha_{\rm max}=1.3$}  & 
 \multicolumn{2}{c}{Global fit} \\ 
 & $\chi^{2}_{\rm nlo}$ & $\chi^{2}_{\rm nnlo}$  & $\chi^{2}_{\rm nlo}$ & $\chi^{2}_{\rm nnlo}$  & $\chi^{2}_{\rm nlo}$ & $\chi^{2}_{\rm nnlo}$  & $\chi^{2}_{\rm nlo}$ & $\chi^{2}_{\rm nnlo}$ \\
\hline
\hline
Total & 0.96 & 1.01 & 1.06  & 1.10 & 1.12 & 1.16  & 1.23 &  1.29  \\
\hline
NMC $d/p$ & 0.91 & 0.91 & 0.89 & 0.89  &  0.88 & 0.89  & 0.92 & 0.93 \\
NMC $\sigma^{\rm NC,p}$  & -  & - & -  & -  & - & -  & 1.63 & 1.52 \\
SLAC  & -  &  - & -  & -  &  1.77 &  1.19  & 1.59 & 1.13 \\
BCDMS & - & - & 1.11 & 1.15 & 1.12 & 1.16  & 1.22 & 1.29 \\
CHORUS  & -  & -  & 1.06 & 1.02 & 1.09  & 1.07   & 1.11 & 1.09 \\
NuTeV  & 0.35 & 0.34 & 0.62 & 0.64 & 0.70  & 0.70  & 0.70 & 0.86 \\
HERA-I  & 0.97 & 0.98 & 1.02  & 1.00 & 1.02 & 0.99  & 1.05 & 1.04\\
ZEUS HERA-II  & -  & - & -  & -  & 1.41  & 1.48  & 1.40 & 1.48\\
H1 HERA-II  & -  & - & - & - & - & - & 1.65 & 1.79 \\
HERA $\sigma_{\rm NC}^{c}$  & - & -  & 1.21 & 1.32 & 1.20 &  1.31   & 1.27 & 1.28 \\ 
\hline
E886 $d/p$  & 0.30 & 0.30  & 0.43 & 0.40 & 0.44 & 0.46  & 0.53 &  0.48\\
E886 $p$  & - & - &  1.18 & 1.40 &  1.27 &  1.53  & 1.19 & 1.55 \\
E605  & 1.04 & 1.10 & 0.74  & 0.83 & 0.75  & 0.88 & 0.78 & 0.90 \\
CDF $Z$ rapidity  & -  & - & -  & -  & -  & -  &  1.33 & 1.53 \\
CDF Run-II $k_t$ jets  & - & - & 1.01 & 2.01 & 1.04  & 1.84  & 0.96  & 1.80  \\
D0  $Z$ rapidity & 0.56 & 0.61 & 0.62 & 0.71 & 0.60 & 0.69  & 0.57  & 0.61 \\
\hline
ATLAS $W,Z$ 2010  & -  & -  & 1.19 & 1.13 & 1.19 & 1.17  & 1.19 & 1.23 \\
ATLAS 7 TeV jets 2010   & 0.96 & 1.65 & 1.08  & 1.58 &  1.10 & 1.54  & 1.07 & 1.36 \\
ATLAS 2.76 TeV jets  & 1.03 & 0.38 & 1.38 & 0.36 & 1.35  & 0.35 & 1.29 & 0.33 \\
ATLAS high-mass DY  & -  & - & -  & -  & -  & -   & 2.06 & 1.45 \\
ATLAS $W$ $p_T$  & - & -  & -  & -  & -  & -  & 1.13 & -  \\
CMS $W$ electron asy  & 0.98  & 0.84 & 0.82  & 0.72 & 0.85  & 0.73  & 0.87 &  0.73\\
CMS $W$ muon asy   & -  &  -&  - & -  & -  & -  & 1.81  &  1.72\\
CMS jets 2011  & 0.90  & 2.09  & 0.96  & 2.09 & 0.99 & 2.10  & 0.96  & 1.90 \\
CMS $W+c$ total  & - & -  &  - & - & -  & - & 0.96  & 0.84 \\
CMS $W+c$ ratio  & - & -  & -  & -  & - & - & 2.02 & 1.77 \\
CMS 2D DY 2011  & -  & -  & -  & - &  1.20 & 1.30  & 1.23 & 1.36 \\
LHCb $W$ rapidity  & - & - & 0.69 & 0.65 & 0.74 &  0.69  & 0.71 & 0.72 \\
LHCb $Z$ rapidity  & - & -& 1.23 & 1.78  & 1.11 & 1.58  & 1.10 & 1.59 \\
$\sigma(t\bar{t})$  & - & -& - & - & - & - & 1.43 &  0.66 \\
\hline
\end{tabular}

\caption{\small The experimental $\chi^2$ values at  NLO and NNLO
for NNPDF3.0 fits to  conservative datasets corresponding to three different
values of the threshold $\alpha_{\rm max}$ (see text).
In each case, the  $\chi^2$ is shown for the datasets which pass the
conservative cut. The  values
for the global fit (same as in Tab.~\ref{tab:chi2tab_exp_vs_t0}) are
also shown for ease of comparison.
 \label{tab:chi2tab_conservative}
}
\end{center}
\end{table}

%
We propose a new objective  definition of a
 conservative set of PDFs based
on a measure of consistency between
datasets introduced in Ref.~\cite{Ball:2010gb,Ball:2011gg}.
This is
based on observing that lack of compatibility can always be viewed as
an underestimate of the covariance matrix: if the covariance matrix
is inflated by a factor $\alpha^2$, then compatibility can always be
attained if $\alpha^2$ if large enough (crudely speaking, if
uncertainties are all multiplied by a factor $\alpha$).
It is then possible to measure
compatibility by  assuming that the prior knowledge
is given by all experiments in the global dataset but the given one, and
using Bayes' theorem to study how this prior is modified when adding
the experiment whose consistency is under investigation. One may then
compute the a posteriori probability $P\lp \alpha\rp$ that the
 covariance matrix of the given experiment should be rescaled
 by a factor $\alpha$. Compatibility corresponds to the case in which
$P(\alpha)$ peaks around $\alpha\sim1$, while
if the most likely value is at $\alpha_0>1$, this means that
compatibility is only achieved when uncertainties are inflated by
$\alpha_0$ (see Ref.~\cite{Ball:2010gb,Ball:2011gg} for a more
detailed discussion and definition). The $t_0$ definition of the
$\chi^2$, which is used for minimization (see
Sect.~\ref{sec:quality}), is also used in the
determination of $P(\alpha)$.

We then proceed as follows.
We compute the probability distribution
of the rescaling variable $\alpha$,
$P(\alpha)$  for each
dataset included in the global fit, and we determine
the mean, the median and the
mode of the corresponding $P(\alpha)$ distribution.
 We then exclude
from the conservative fit all experiments for which at least two of
these three quantities are above some threshold value,
denoted by $\alpha_{\rm max}$. We discard all datasets for which the
criterion fails either at NLO or at NNLO (or both), which corresponds
to the most conservative choice of only retaining experiments which
are well described at all perturbative orders, and has the obvious
advantage that the dataset does not depend on the perturbative order,
thereby keeping PDF uncertainties separate from the theory
uncertainties discussed in Sect.~\ref{sec:perturbative}.
 In practice, for simplicity we compute
the probability  $P(\alpha)$  without excluding the given experiment
from the global fit: this provides a conservative estimate of the
compatibility (which is clearly increased by including the experiment
under investigation in the prior) without requiring us to construct a
new set of 1000 replicas when each of the experiments is excluded in
turn.
The values of the mean, median and mode thus computed
for all the experiments in the NNPDF3.0 global fits
at NLO and NNLO are  collected in
Tab.~\ref{tab:chi2tab_conservative_v3}.\footnote{We notice that the
ATLAS $W$ $p_T$ data, which are included in the NNPDF3.0 global fit,
are not included in Tab.~\ref{tab:chi2tab_conservative_v2}.
This is due to the fact the we construct
NLO and NNLO conservative PDF sets s based on the same datasets.
Since the ATLAS $W$ $p_T$ data are excluded from the global NNLO fit
because of the lack of availability of a NNLO prediction, they will not be
included in our conservative set.}

\begin{figure}[t]
\begin{center}
\epsfig{width=0.70\textwidth,figure=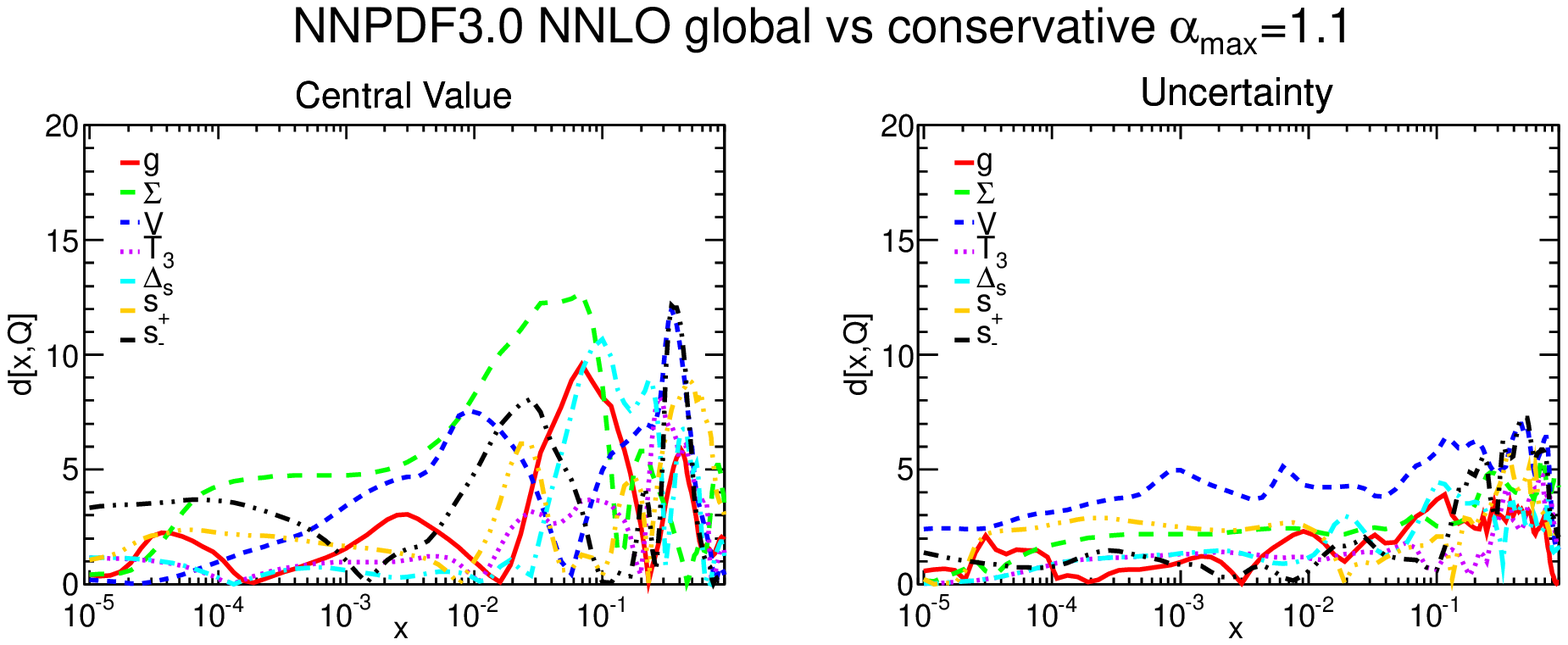}
\epsfig{width=0.70\textwidth,figure=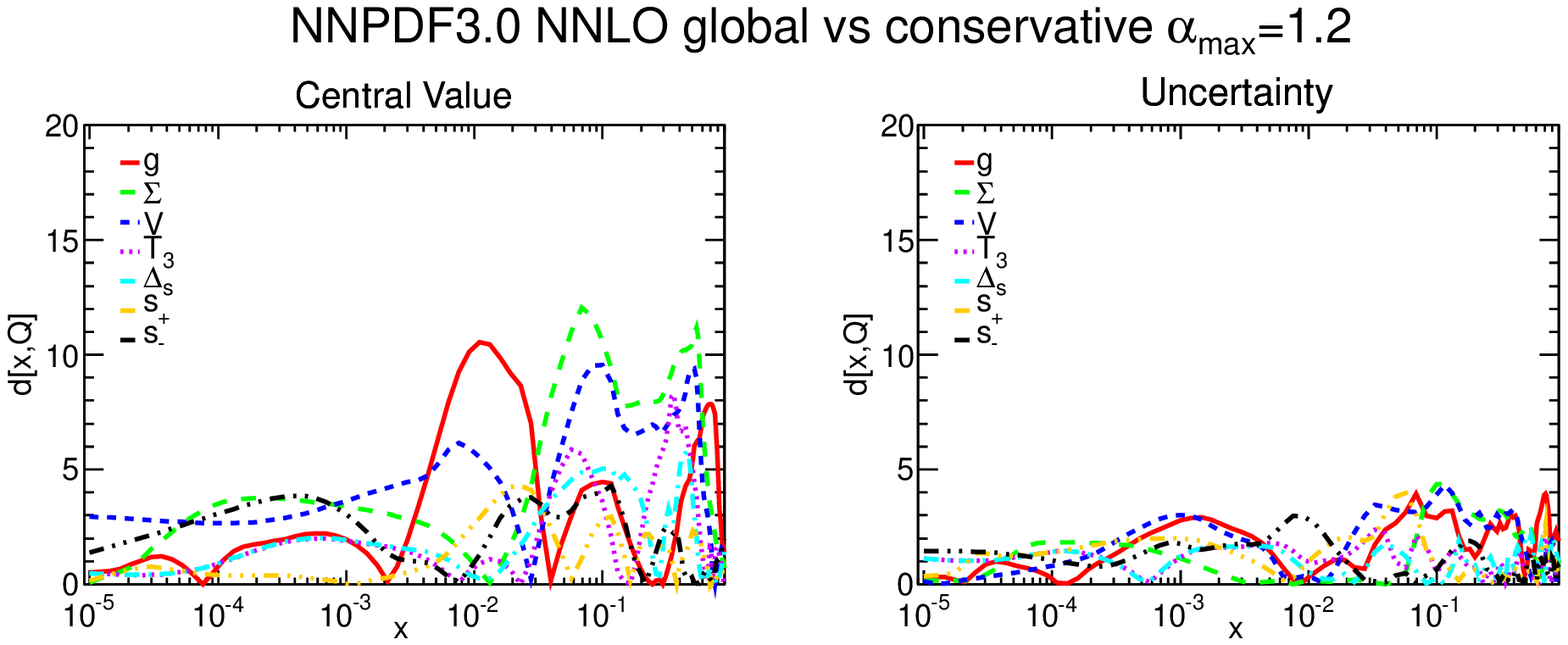}
\epsfig{width=0.70\textwidth,figure=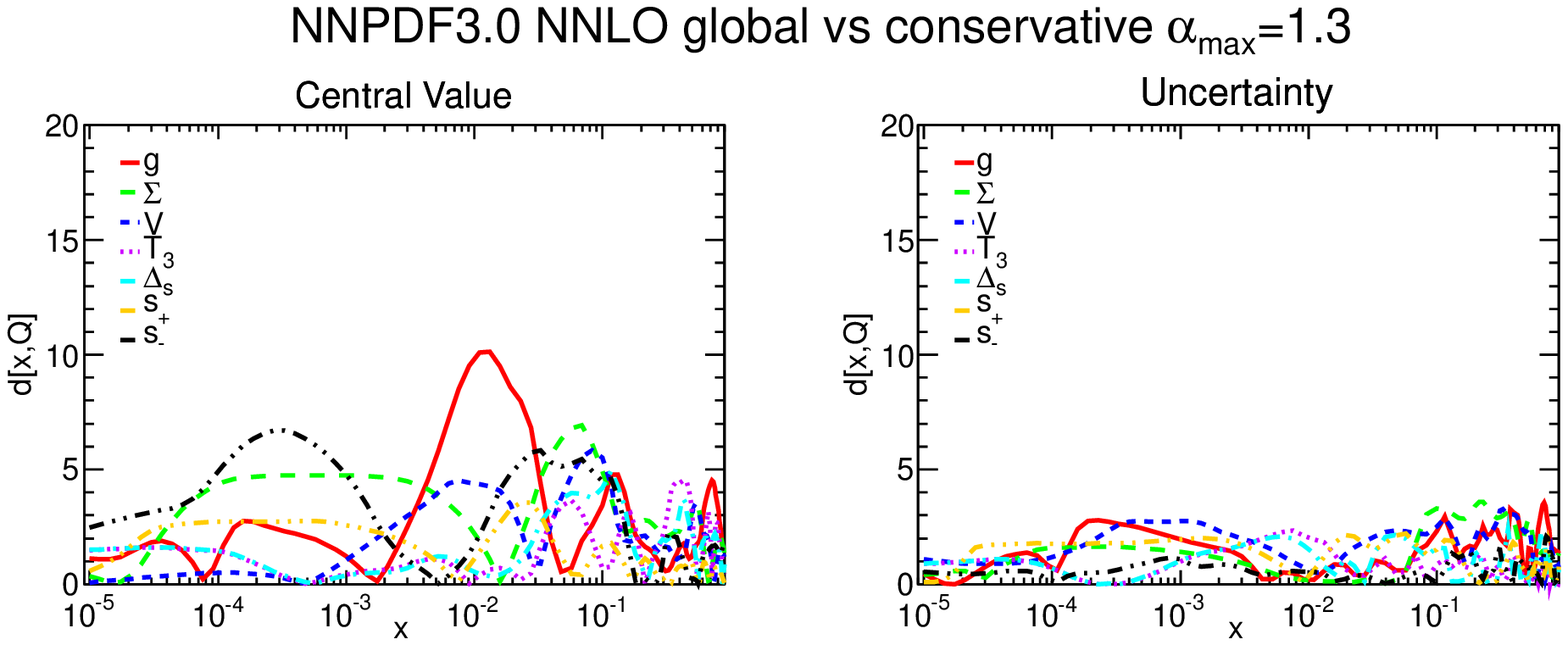}
\caption{\small
Same as Fig.~\ref{fig:distances_30_vs_23_nnlo},
but now comparing the  baseline NNPDF3.0 global
fit to the conservative  fits obtained using  the three values of $\alpha_{\rm max}$ of
Tab.~\ref{tab:chi2tab_conservative}: $\alpha_{\rm max}=1.1$ (top),
$\alpha_{\rm max}=1.2$ (center),  $\alpha_{\rm max}=1.3$ (bottom).
 \label{fig:distances_global_vs_conservative}}
\end{center}
\end{figure}

In the following we present results for
partons obtained by fitting to datasets constructed by
only including data from Tab.~\ref{tab:chi2tab_conservative_v3}
which pass the conservative cuts
 corresponding to three different values
of  $\alpha_{\rm max}$, namely 1.1, 1.2 and 1.3.
In Tab.~\ref{tab:chi2tab_conservative}
we provide the $\chi^2$ (for ease of comparison we show results
obtained using the experimental definition,
see Sect.~\ref{sec:quality}) for the PDF fits to these datasets.
To facilitate the comparison with the global fit, we also provide
its $\chi^2$ values in the same table, taken
from Tab.~\ref{tab:chi2tab_exp_vs_t0}.
%

\begin{table}[t]
\begin{center}
\footnotesize
\begin{tabular}{c||ccc|ccc}
\hline
 & \multicolumn{3}{c|}{NNLO global fit} &  \multicolumn{3}{c}{NNLO cons. fit $\alpha_{\rm max}=1.1$} \\
\hline
Experiment & mean & mode  & median & mean &  mode & median \\
\hline
NMC $\sigma^{\rm NC,p}$  &	1.27 & 1.26  & 1.27  & 1.50 &	1.45 &	1.48 \\
SLAC  &1.13 &	1.09 &	1.12&		1.61&	1.37&	1.48 \\
BCDMS &1.20&	1.19&	1.20  &    2.02&	1.86 &	1.92 \\
CHORUS &      1.10     & 1.09 & 1.09  &        	2.55&	1.69&	2.32 \\
ZEUS HERA-II &  1.25&	1.24  &	1.25 &  1.38&	1.33&	1.36\\
H1 HERA-II  &1.35&	1.34&	1.34&		1.51&	1.47&	1.49 \\
HERA $\sigma_{\rm NC}^{c}$ & 1.14&	1.11&	1.13&	1.13	&1.09 &	1.12 \\
\hline
E886 $p$ & 1.15&	1.14&	1.15	 &	2.18 &	1.62 &	2.03 \\
CDF $Z$ rapidity & 1.39&	1.32 &	1.36  & 1.56 &	1.40 &	1.50 \\
CDF Run-II $k_t$ jets  & 1.15&	1.12 &	1.14  &  1.25&	1.18&	1.22 \\
\hline
ATLAS $W,Z$ 2010 &  1.17&	1.12&	1.15 &	1.38	& 1.25 &	1.32 \\
ATLAS high-mass DY &	1.00&	1.34&	1.63 &	1.63	& 1.19 &	1.45 \\
CMS $W$ muon asy  & 1.60&	1.40&	1.53  & 2.90&	2.48&	2.81 \\
CMS $W$+$c$ total & 1.50&	1.09&	1.33 & 1.85	&1.37&	1.67 \\
CMS $W$+$c$ ratio & 2.00&	1.39&	1.69&	2.12	&1.58&	1.94	\\
CMS 2D DY 2011  &1.28	&1.27&	1.28&		1.29&	1.28&	1.29 \\
LHCb  & 1.20	& 1.12	&1.17	  &	1.58	& 1.22	& 1.48 \\
\hline
\end{tabular}
\caption{\small
The mean, mode and median of the
$P\lp \alpha\rp$ distributions at NNLO
for  the experiments
excluded from the conservative fit with $\alpha_{\rm max}=1.1$, either
when the prior is the global fit (same as
Tab.~\ref{tab:chi2tab_conservative_v3})  or when using
as prior the conservative set itself.
 \label{tab:chi2tab_conservative_v2}
}
\end{center}
\end{table}

The improvement in global fit quality as $\alpha_{\rm max}$ is lowered
is apparent, with the most conservative option leading to an
essentially perfect $\chi^2$ of order one. It is interesting to
observe that  NMC proton data, which are known
to have internal inconsistencies~\cite{Forte:2002fg}, as well as
other datasets such as
the H1 HERA-II data,  the ATLAS high-mass Drell-Yan data,
and the CMS $W$+$c$ data are
excluded even from the least conservative set, the one with
 $\alpha_{\rm max}=1.3$.
On the other hand,  the CMS inclusive jet data is included for
all values of $\alpha_{\rm max}$; note that for this dataset the
experimental $\chi^2$ shown in Tab.~\ref{tab:chi2tab_conservative} is
significantly worse than the $t_0$ value used for minimization and the
determination of $P(\alpha)$.

The maximally consistent dataset, found with $\alpha_{\rm max}=1.1$,
includes
the NMC $d/p$ data, the NuTeV
and HERA-I DIS data, the Drell-Yan data from E866 and E605, the D0 Z rapidity,
the ATLAS and CMS inclusive jets and the CMS $W$ electron asymmetry.

In Tab.~\ref{tab:chi2tab_conservative_v2}, we furthermore compare
the mean, mode and median of the
$P\lp \alpha\rp$ distributions for the experiments
excluded from the NNLO conservative fit with $\alpha_{\rm max}=1.1$
when the global fit is used as prior (i.e. the same numbers for the
corresponding entries in Tab.~\ref{tab:chi2tab_conservative_v3}), to
the same quantities computed using as a prior the conservative fit
itself. All the peak values of $P\lp \alpha\rp$ deteriorate when using the
conservative set as a prior, as they ought to. Clearly, this
deterioration will be maximal for datasets which are internally
consistent, but inconsistent with the rest, and more moderate for
experiments which are affected by internal inconsistencies, so that a
rescaling of uncertainties is needed in order to describe them,
regardless of what one takes as a prior. This is the case for instance
for the
NMC $\sigma^{\rm NC,p}$ which are affected by internal inconsistencies
as already mentioned.

\begin{figure}
\begin{center}
\epsfig{width=0.42\textwidth,figure=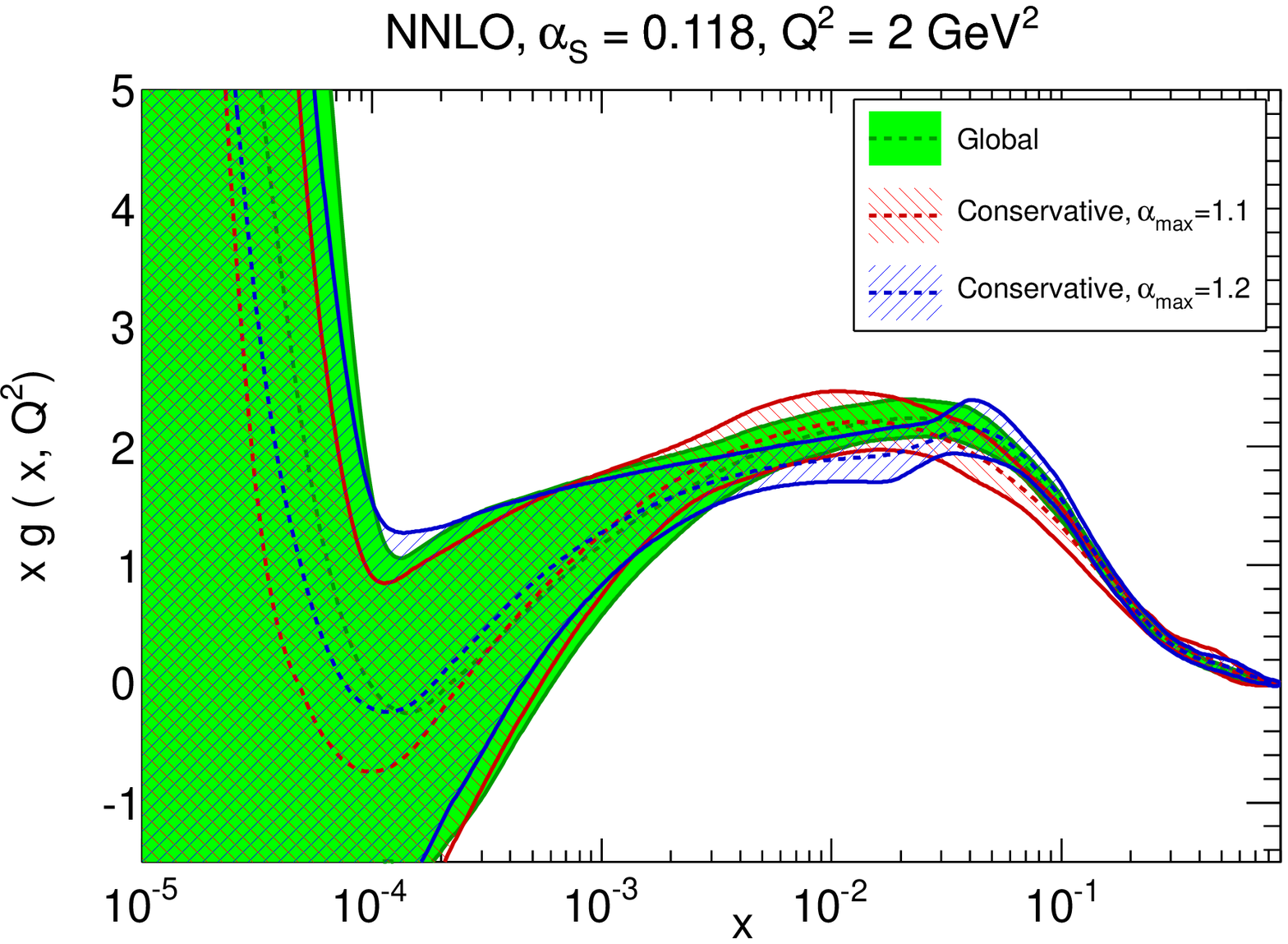}
\epsfig{width=0.42\textwidth,figure=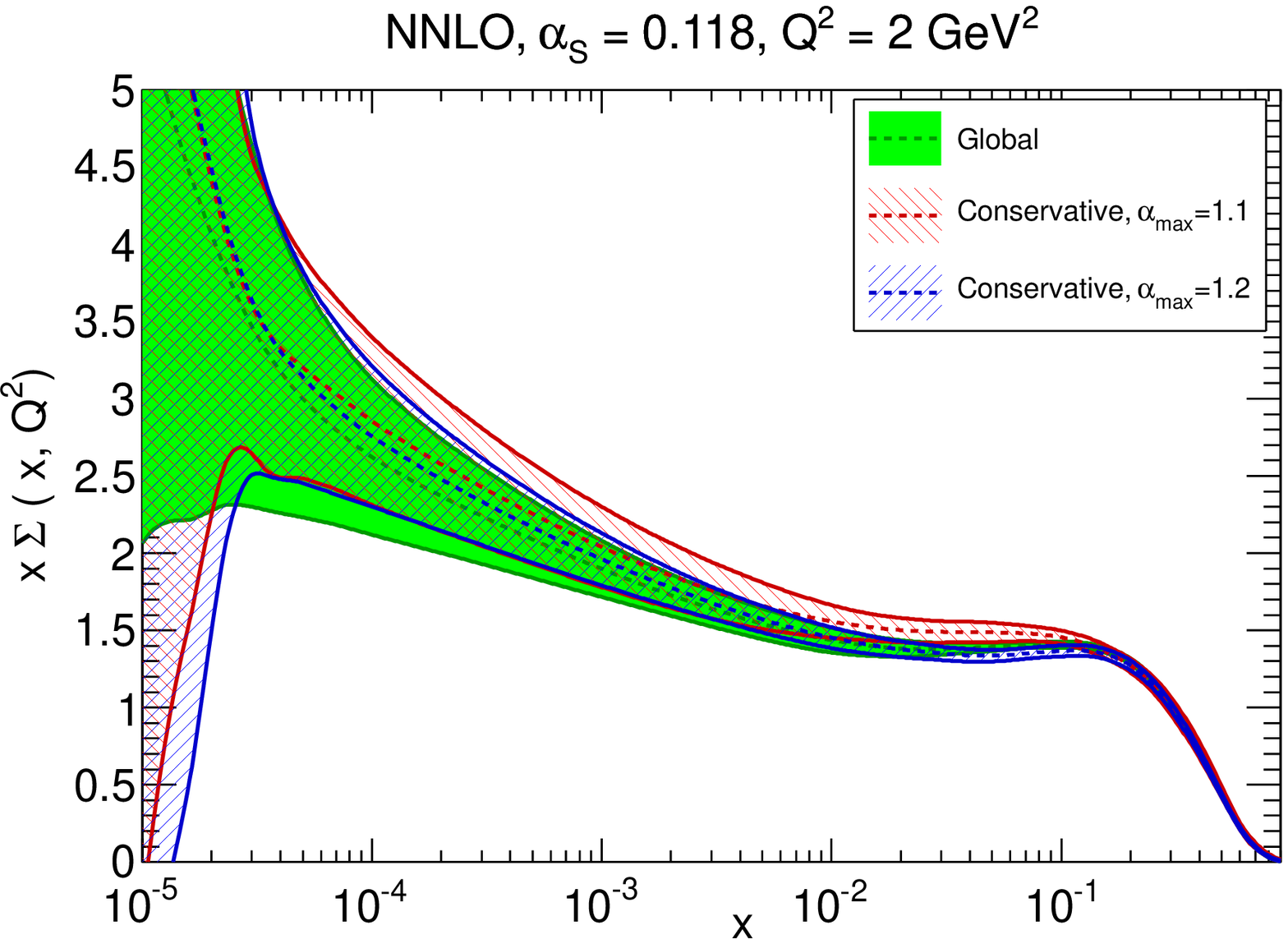}
\epsfig{width=0.42\textwidth,figure=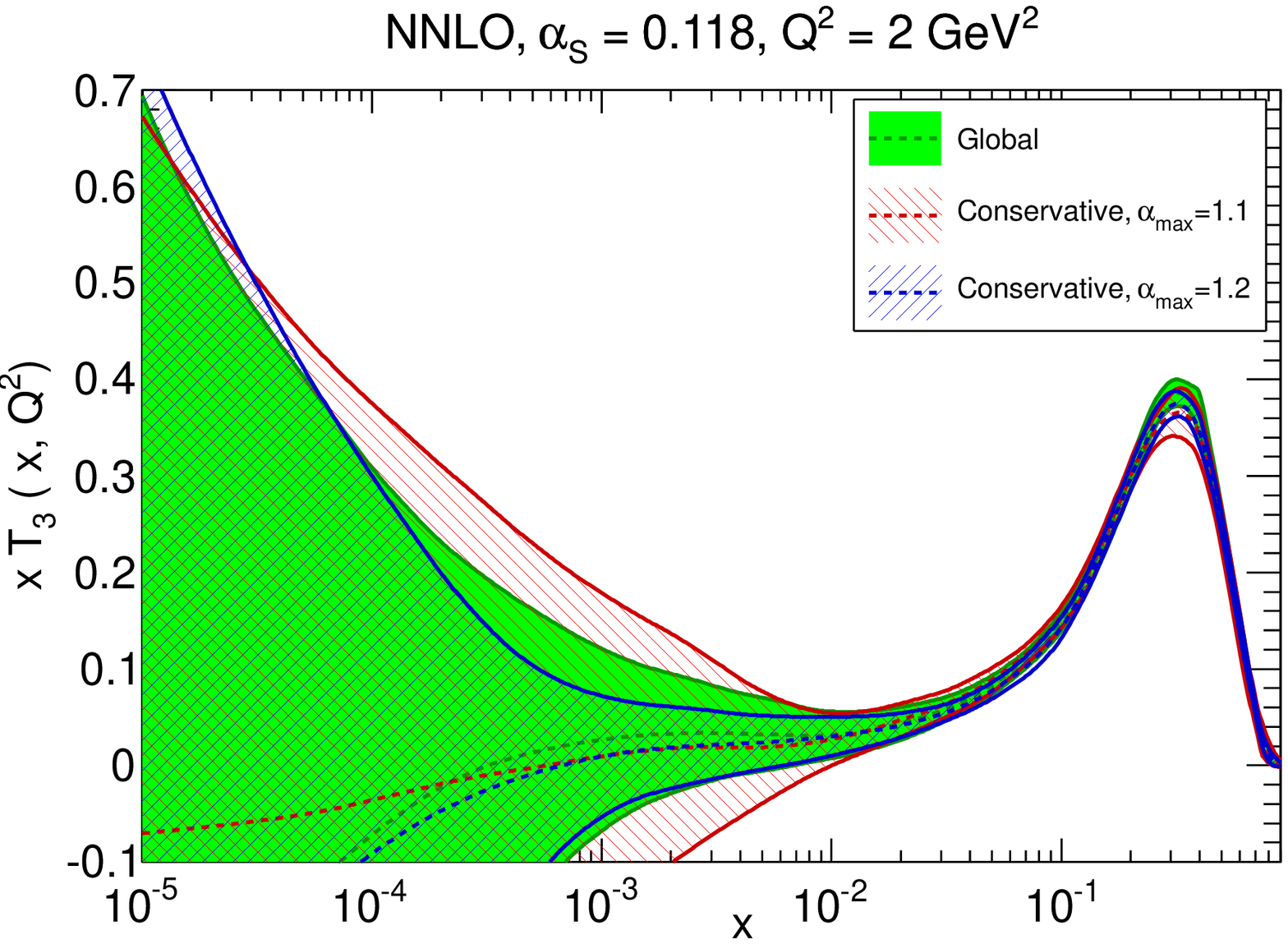}
\epsfig{width=0.42\textwidth,figure=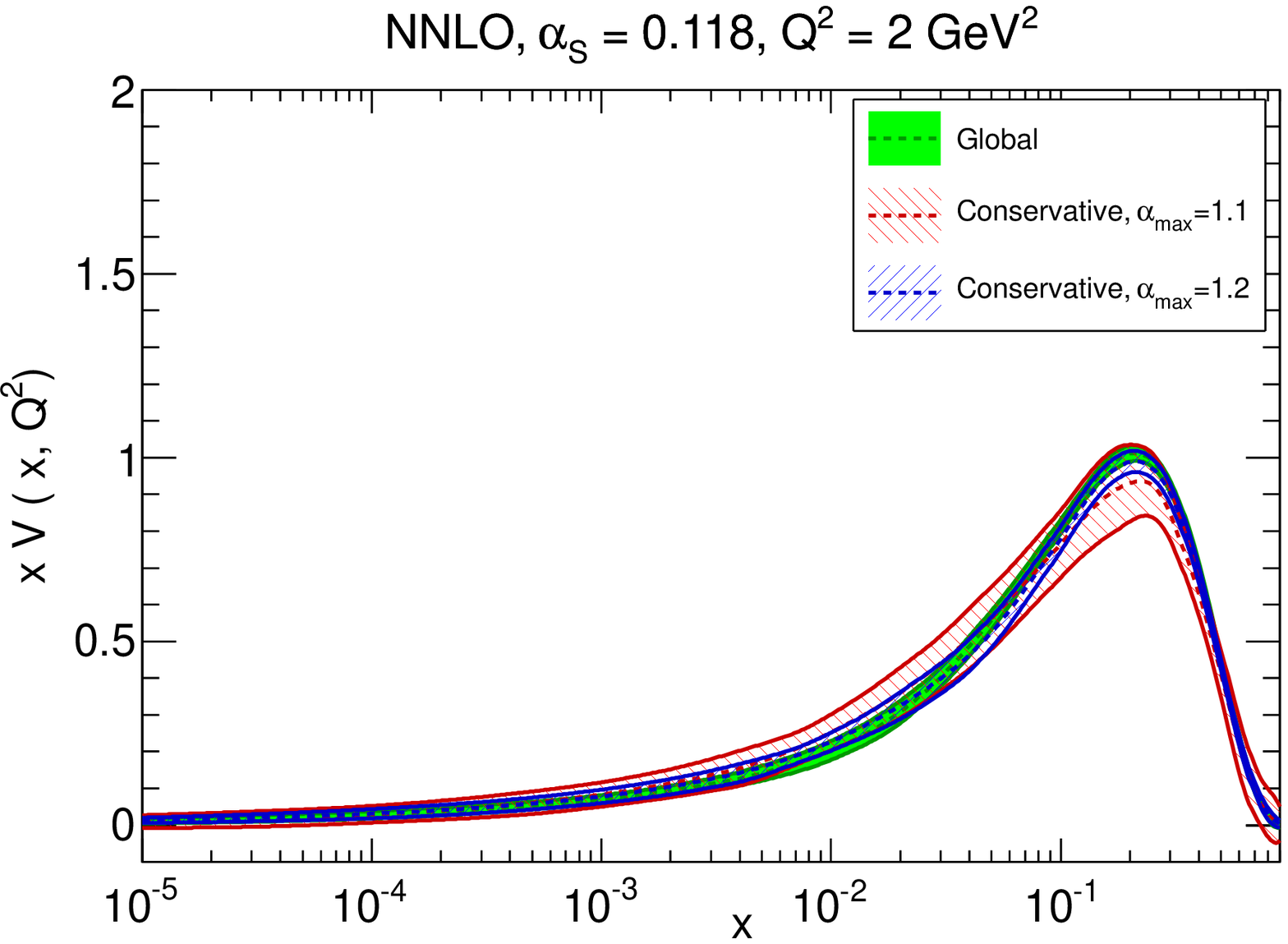}
\caption{\small
Same as Fig.~\ref{fig:th_unc_lo}, but now comparing the default global
NNLO fit to the two conservative fits with
 $\alpha_{\rm max}=1.1$ and $\alpha_{\rm max}=1.2$.
\label{fig:pdfs_conservative}}
\end{center}
\end{figure}

\begin{figure}
\begin{center}
\epsfig{width=0.42\textwidth,figure=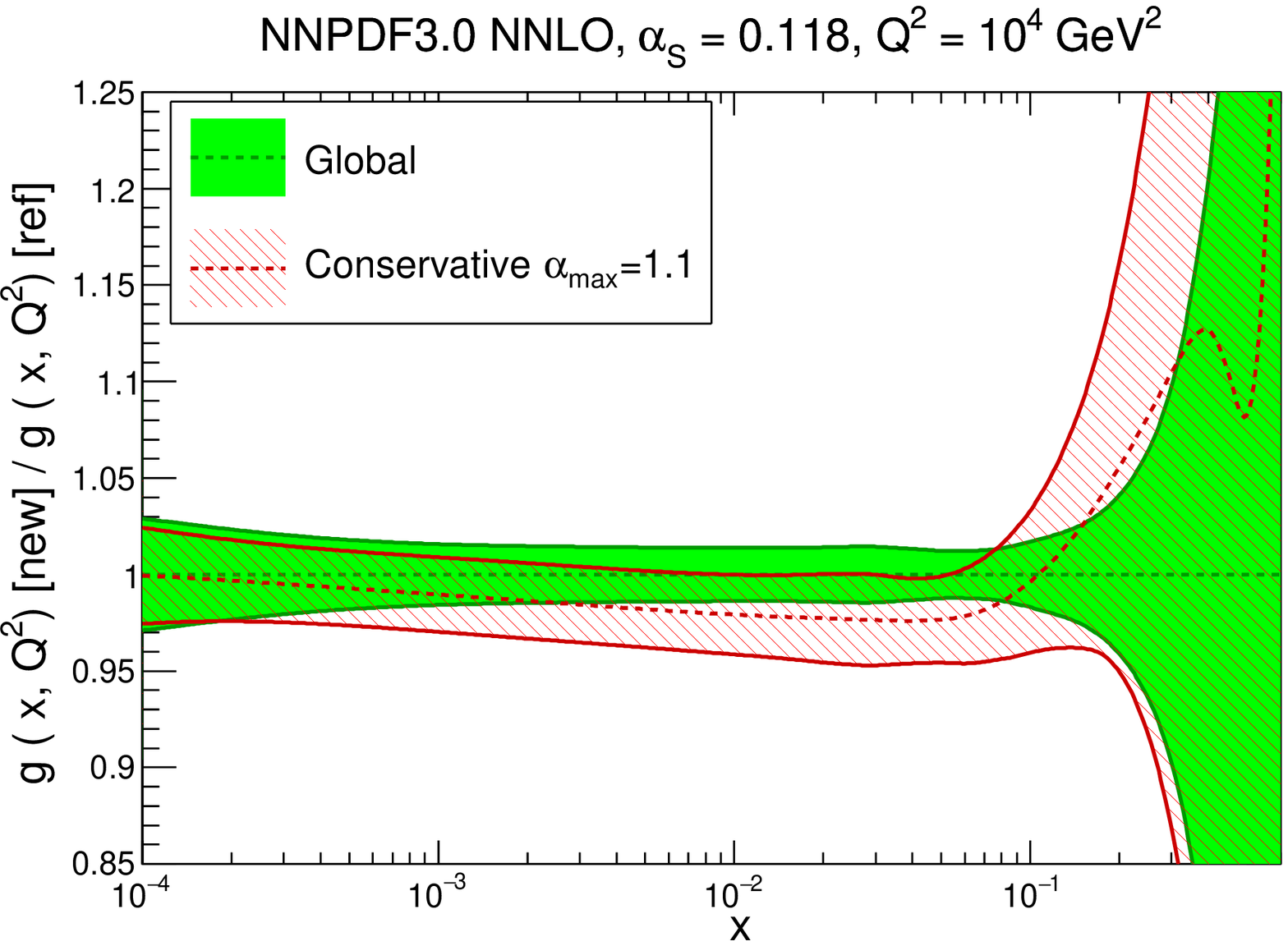}
\epsfig{width=0.42\textwidth,figure=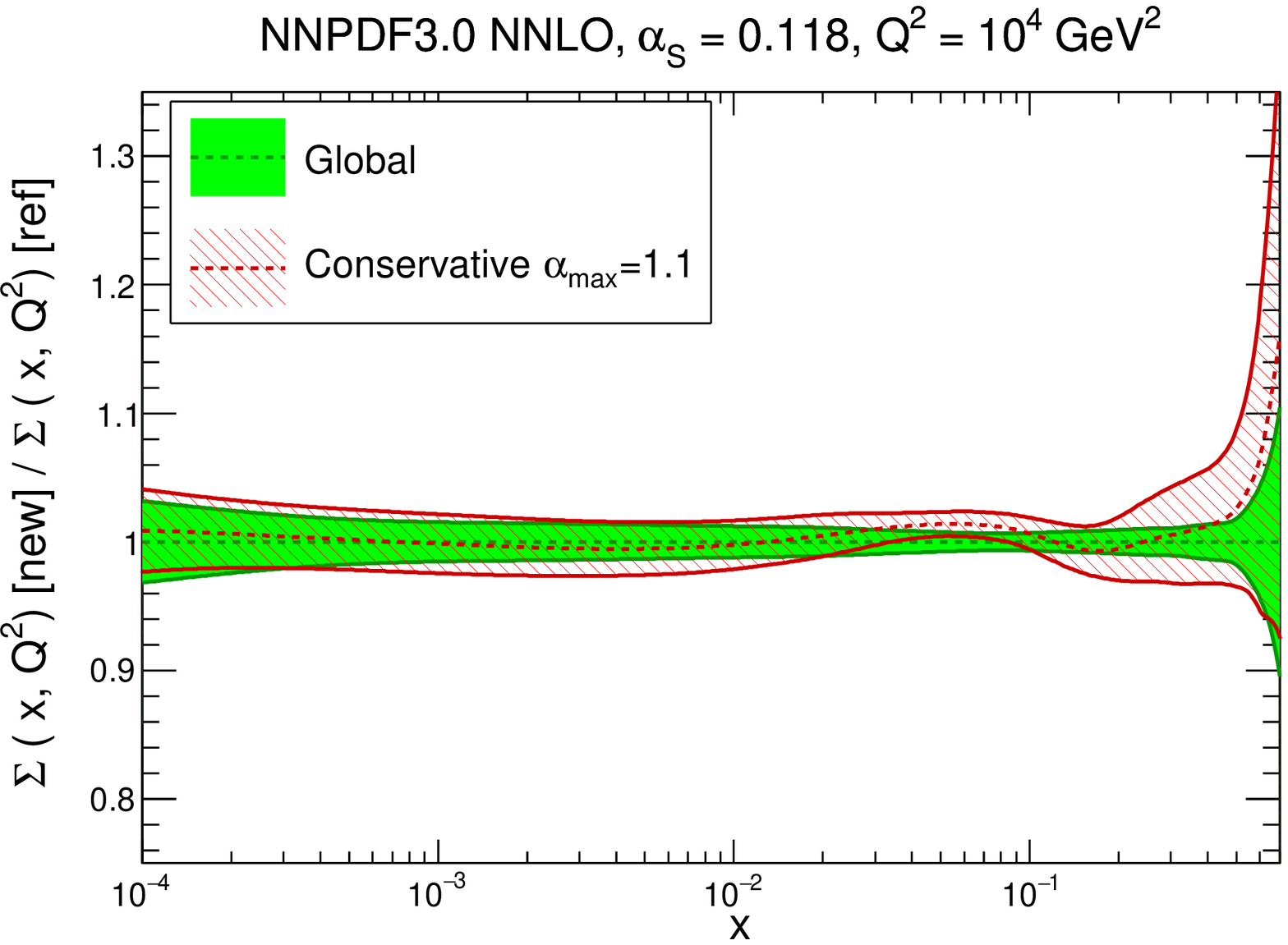}
\epsfig{width=0.42\textwidth,figure=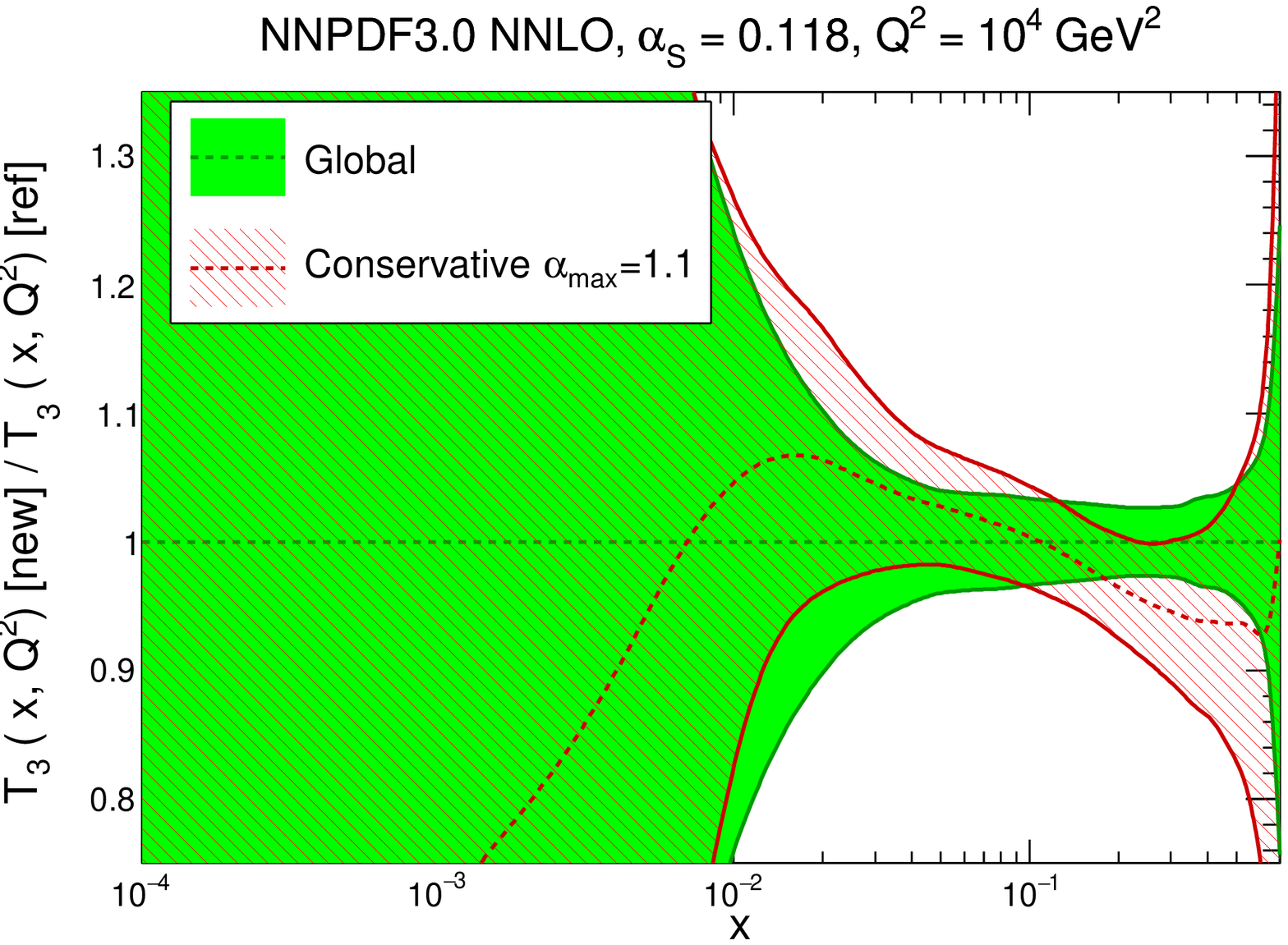}
\epsfig{width=0.42\textwidth,figure=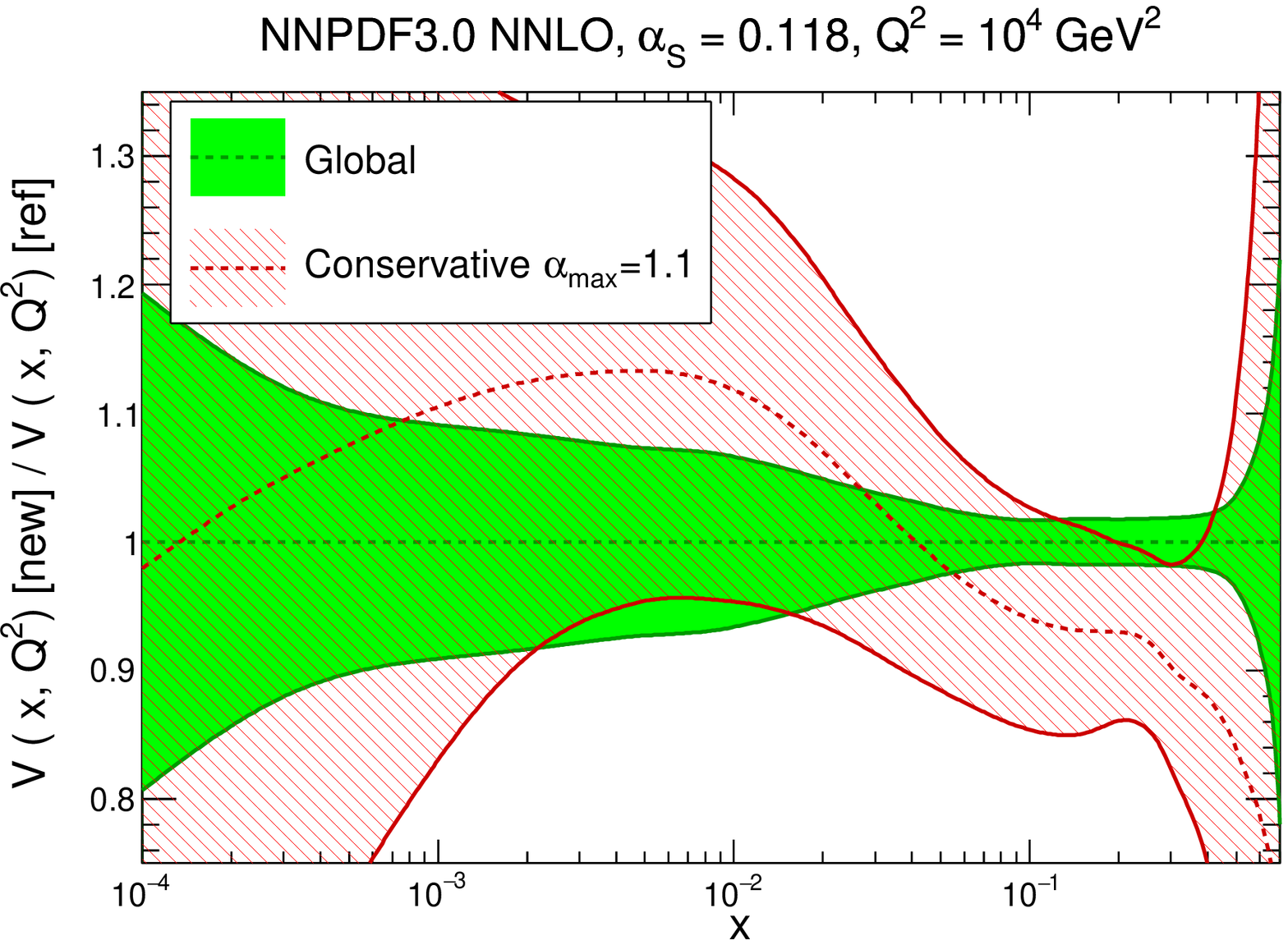}
\caption{\small
Same as Fig.~\ref{fig:pdfs_conservative}, but at $Q^2=10^4$ GeV$^2$,
and with results shown as ratios to the NNPDF3.0 default global fit.
\label{fig:pdfs_conservativelhcsc}}
\end{center}
\end{figure}

The  distance between the conservative sets and the baseline
NNPDF3.0 NNLO
global fit are show in
Fig.~\ref{fig:distances_global_vs_conservative},
while  PDFs are compared directly at $Q^2=2$~GeV$^2$ in Fig.~\ref{fig:pdfs_conservative},
where the NNLO  conservative fits with $\alpha_{\rm max}=1.1$ and
1.2 and the reference fit are shown, and at  $Q^2=10^4$~GeV$^2$ in Fig.~\ref{fig:pdfs_conservativelhcsc}, where
the   NNLO  conservative fit with $\alpha_{\rm max}=1.1$ is shown as a
ratio to the default global fit.
 All sets are consistent with the global fit,
with PDFs that differ at most at the one-sigma level, thereby
confirming the consistency of the procedure, though of course
PDF uncertainties are larger in the fits to reduced
datasets.
The small-$x$ gluon is similar in all cases because is driven by the
HERA-I data, while there is more dependence on the choice of
$\alpha_{\rm max}$ at medium and large $x$:
interestingly, in the region relevant for
Higgs production in gluon fusion the gluon is significantly affected by the
choice of $\alpha_{\rm max}$, though not beyond the one-sigma level.
The quarks are also in good agreement, with the main
differences seen at medium $x$.
The  set with $\alpha_{\rm max}=1.1$ has of course
the largest PDF uncertainties, though even with the correspondingly
restricted dataset of Tab.~\ref{tab:chi2tab_conservative} they are
not much worse than those of the global fit, with the valence and
triplet, and thus the quark-antiquark
flavor separation becoming rather more uncertain.

%

These conservative parton sets may be used for studies aimed at
assessing how individual datasets affect LHC observables, by studying
their effect on a maximally self-consistent dataset, such as performed
in Ref.~\cite{LHcons}. In the future, as more and more data will become
available,
this approach might also be used in
deciding for an optimal dataset on which a global fit should be based.

\subsubsection{Impact of the new HERA and LHC data}
\label{sec:impdata}

We now examine in detail the impact of the new HERA and LHC data in NNPDF3.0.
Results will be shown for NNLO fits, by investigating the
impact of the new data both on the global
fit, and also on a  HERA-only fit: while the former is more realistic,
the latter allows for an assessment of the specific impact of each
individual piece of data (though of course it over-estimates their
impact in a realistic setting). Jet data will be specifically
discussed in  Sect.~\ref{sec:jetless} below.
In all these fits, exactly the same theory and the same methodology of
the default set will be used, with only the dataset changing, so that
the impact of the dataset is specifically assessed.
This will eventually allow us to provide a quantitative assessment of
the dependence on the dataset of the uncertainty on our prediction.

In order to have a first overall assessment,
 we have produced a variant of the NNPDF3.0
fit using the same methodology, but using an NNPDF2.3-like
dataset.
We include in the dataset for this fit all, and only, the data
from Tables~\ref{tab:completedataset}-\ref{tab:completedataset2} which
were included in the NNPDF2.3 dataset. This is not quite identical to
the NNPDF2.3 dataset, because we include these data with the same cuts as in
NNPDF3.0 (which are sometimes slightly different than those
of
NNPDF2.3, as discussed in Sect.~\ref{sec:expdata}) and not the data of
Tables~\ref{tab:completedataset3}, despite the fact that these were
included in NNPDF2.3. It is however adequate to assess the (moderate) impact of
the new data as we now see.

\begin{figure}
\begin{center}
\vspace{-5cm}
\epsfig{width=0.99\textwidth,figure=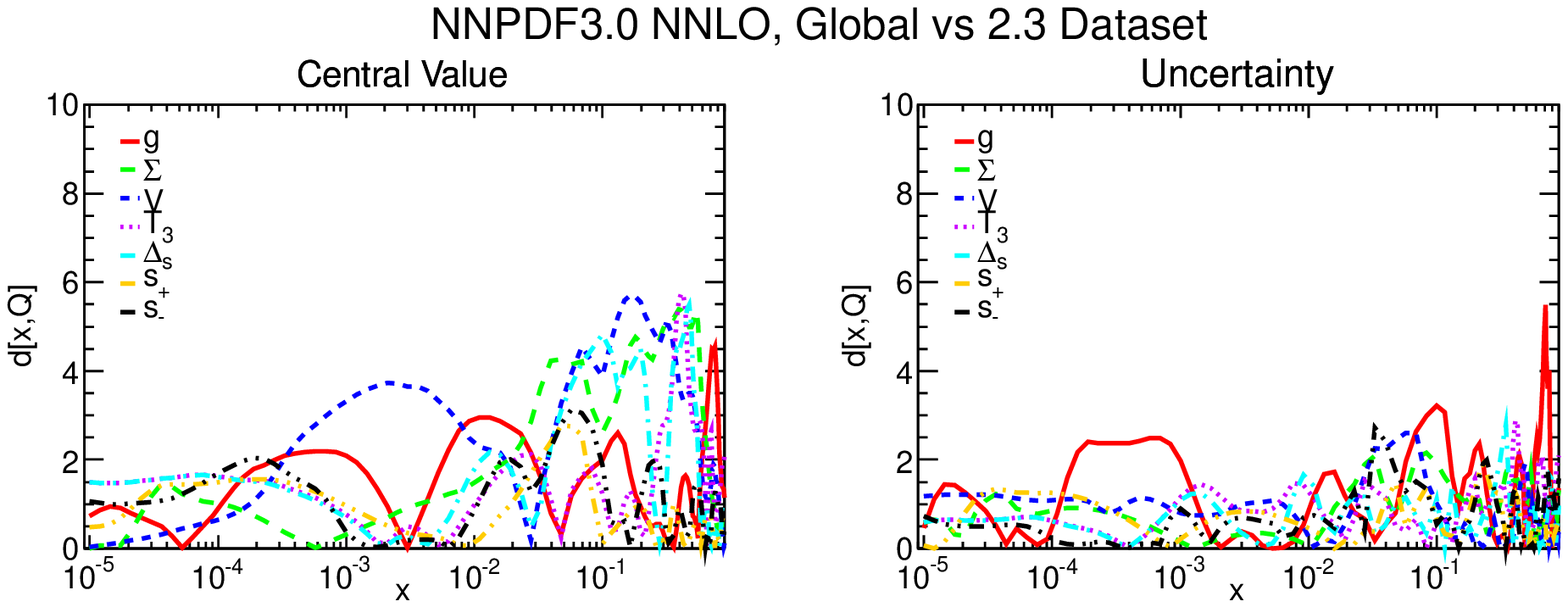}
\caption{\small  Same as Fig.~\ref{fig:distances_30_vs_23_nnlo}, but
  now comparing the default NNLO set to a set obtained using the same
  methodology but an  NNPDF2.3-like dataset.
 \label{fig:distances_global_vs_23dataset}}
\end{center}
\end{figure}

\begin{figure}
\begin{center}
\epsfig{width=0.42\textwidth,figure=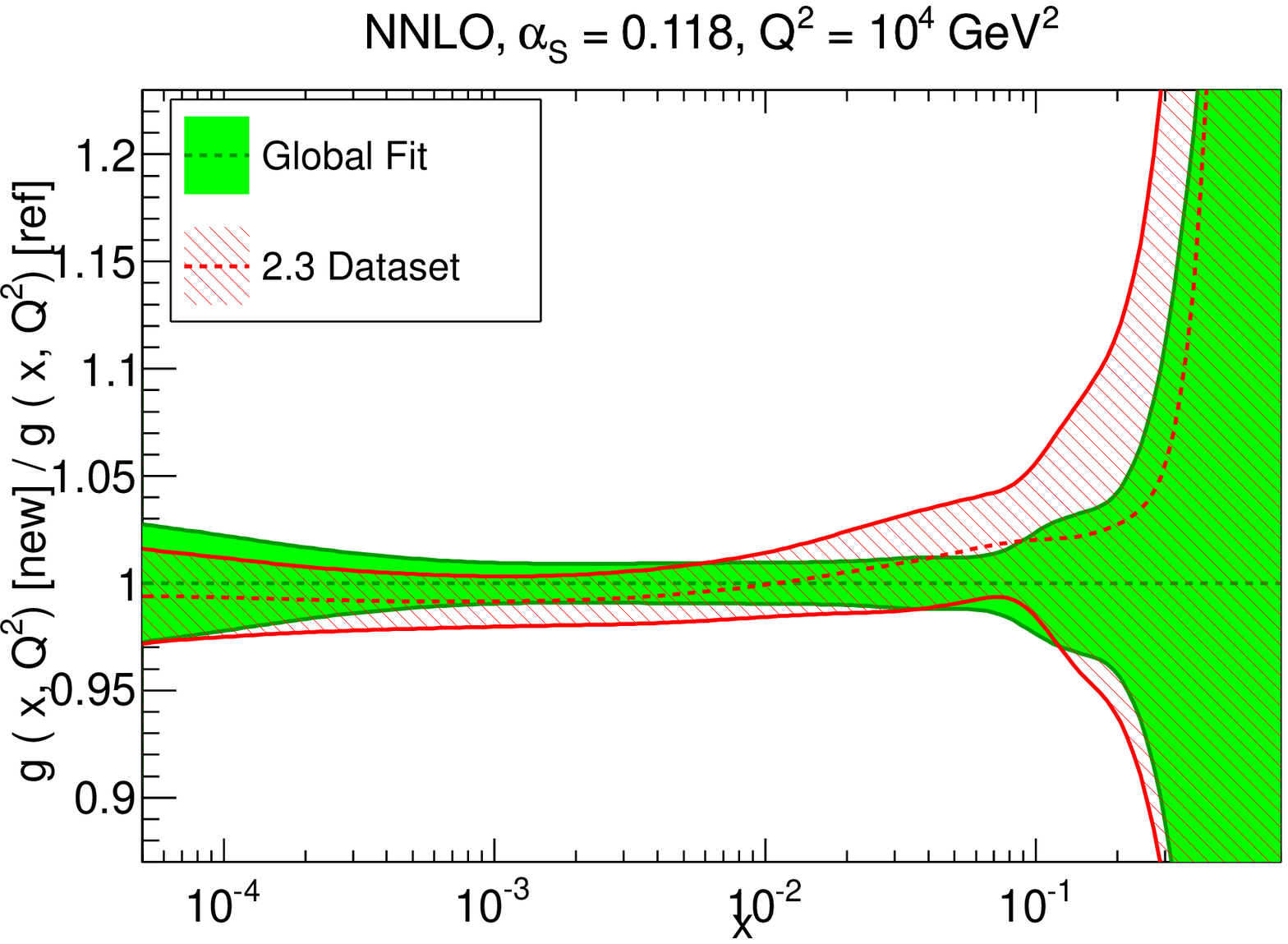}
\epsfig{width=0.42\textwidth,figure=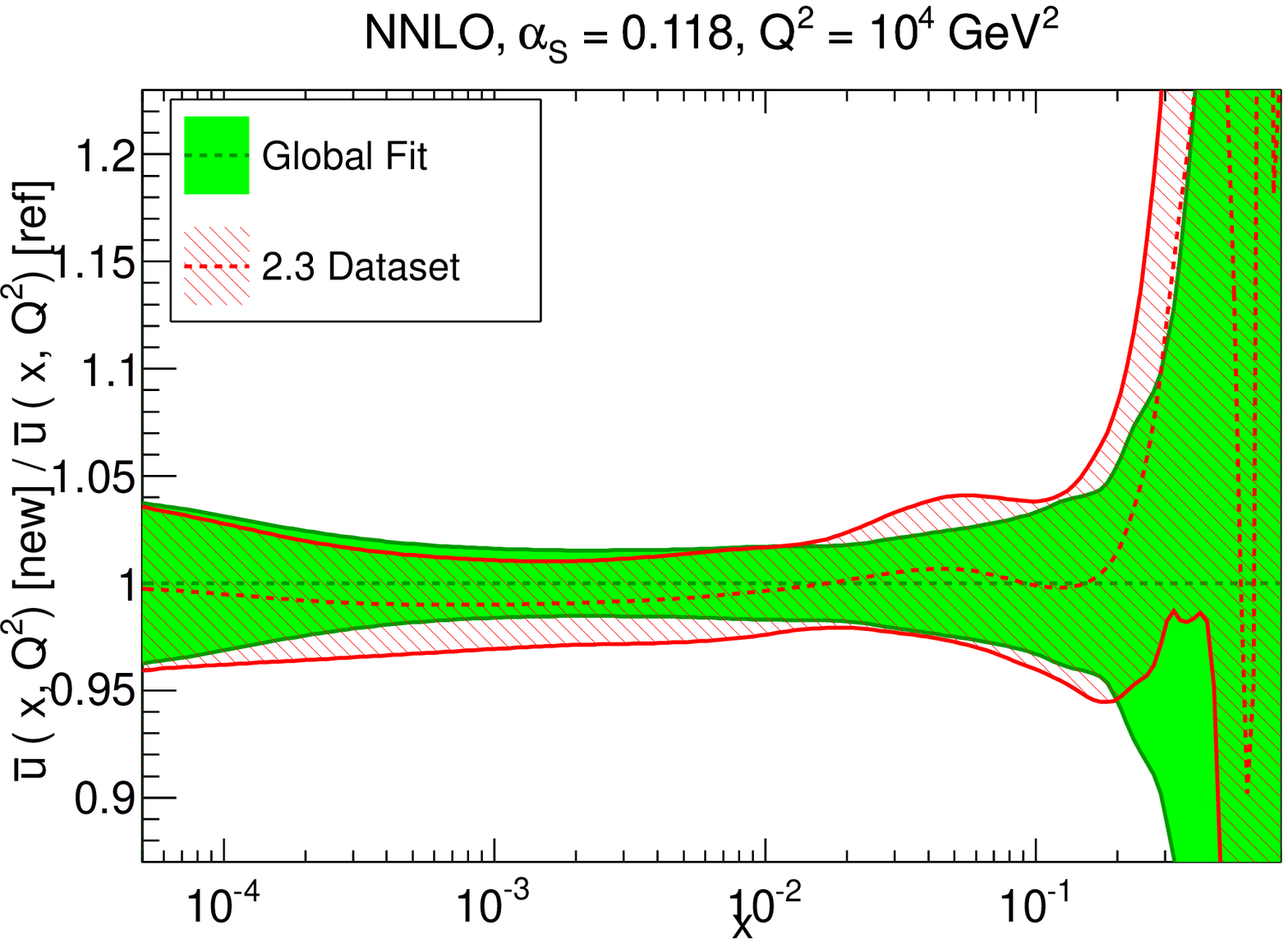}
\epsfig{width=0.42\textwidth,figure=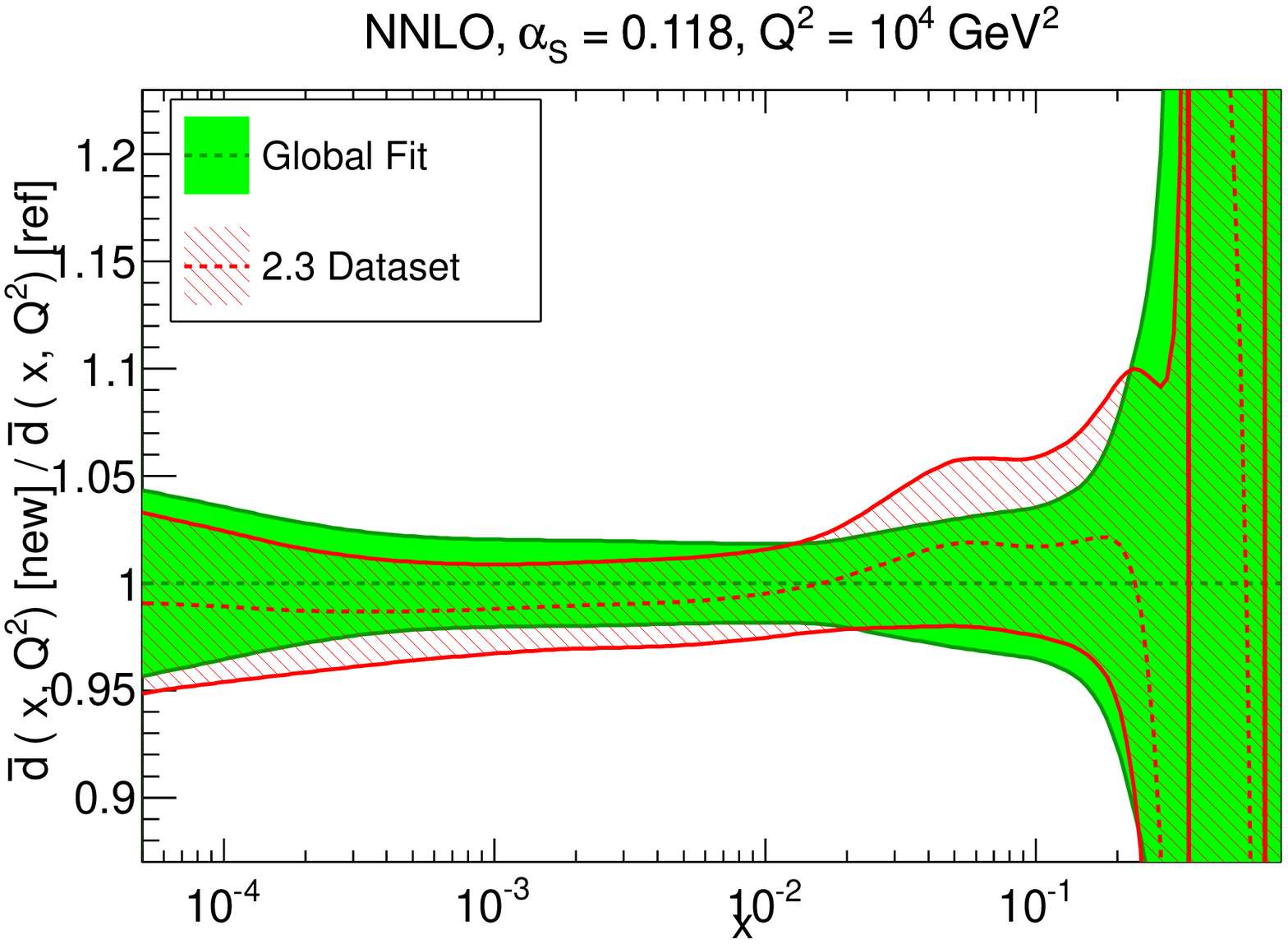}
\epsfig{width=0.42\textwidth,figure=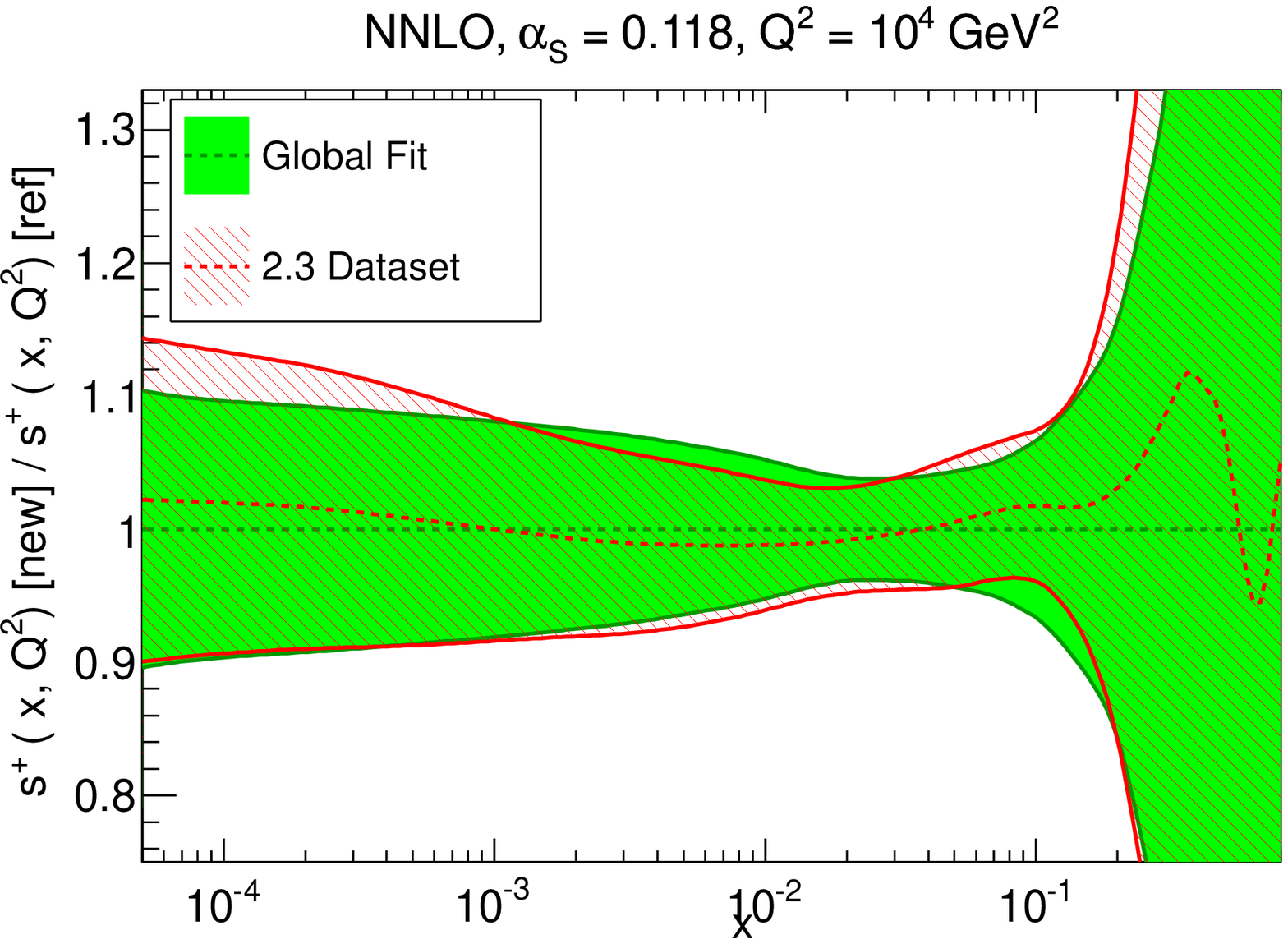}
\caption{\small
Comparison of  NNPDF3.0 NNLO PDFs at  $Q^2=10^4$ GeV$^2$ to PDFs
obtained using an NNPDF2.3-like dataset. Results are shown as ratio to
the default set.
From top to bottom and from left to right the
gluon, anti-up, anti-down quarks and  total
strangeness are shown. \label{fig:xpdf-30_vs_23dataset}}
\end{center}
\end{figure}

The distances between PDFs from this fit and their NNPDF3.0
counterparts  are shown in
Fig.~\ref{fig:distances_global_vs_23dataset},
while the PDFs at
$Q^2=10^4~\textrm{GeV}^2$ are compared in  Fig.~\ref{fig:xpdf-30_vs_23dataset}.
Is clear that the new  data affects moderately all PDFs:
central values vary within half a sigma of the PDF errors at most.
This was to be expected,
 since the NNPDF2.3 PDFs already described rather well all the
new experimental data that has been added in NNPDF3.0, so the main
impact of the new data is to reduce  uncertainties. Indeed, the PDF
comparison shows that the change in uncertainties, seen in the
distance plot again at a half-sigma level, always corresponds to a
reduction in uncertainty.
%


The largest effect on central values is seen for the large- and
medium-$x$ quarks,
followed by the gluon in the same region.
The small-$x$ gluons and quarks are quite stable since there is no
new data that affects them in this region.
Uncertainties mostly improve for the gluon PDF,
both at large $x$ thanks to the
 LHC  jet and top quark data,
and at medium and small $x$ from the new HERA-II data.
The new data favor a rather softer gluon
at large $x$ in comparison to  the NNPDF2.3-like dataset, though
differences are always within the PDF uncertainties.
Also for the antiquark sea there is a visible improvement, especially
at medium $x$, where the bulk of the LHC electroweak
vector boson production data is.
Finally, there are some improvements in strangeness; the role
of the LHC data in pinning down $s(x,Q)$ is discussed
in more detail in Sect.~\ref{sec:strangeness}.
\begin{figure}[h]
\begin{center}
\epsfig{width=0.92\textwidth,figure=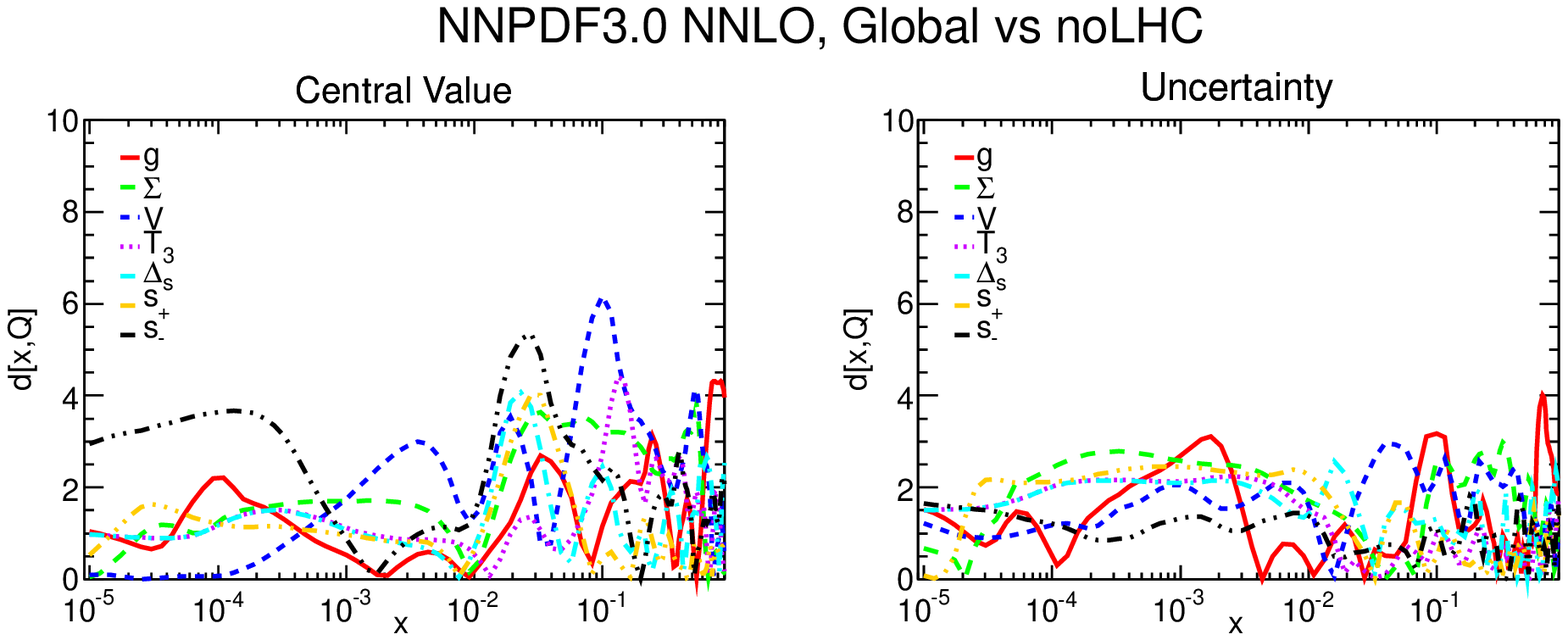}
\caption{\small
Same as Fig.~\ref{fig:distances_30_vs_23_nnlo}, but
  now comparing the default NNLO set to
PDFs
obtained using the same dataset, but with all
  LHC data excluded.
 \label{fig:distances_global_vs_nolhc}}
\end{center}
\end{figure}

\begin{figure}
\begin{center}
\epsfig{width=0.42\textwidth,figure=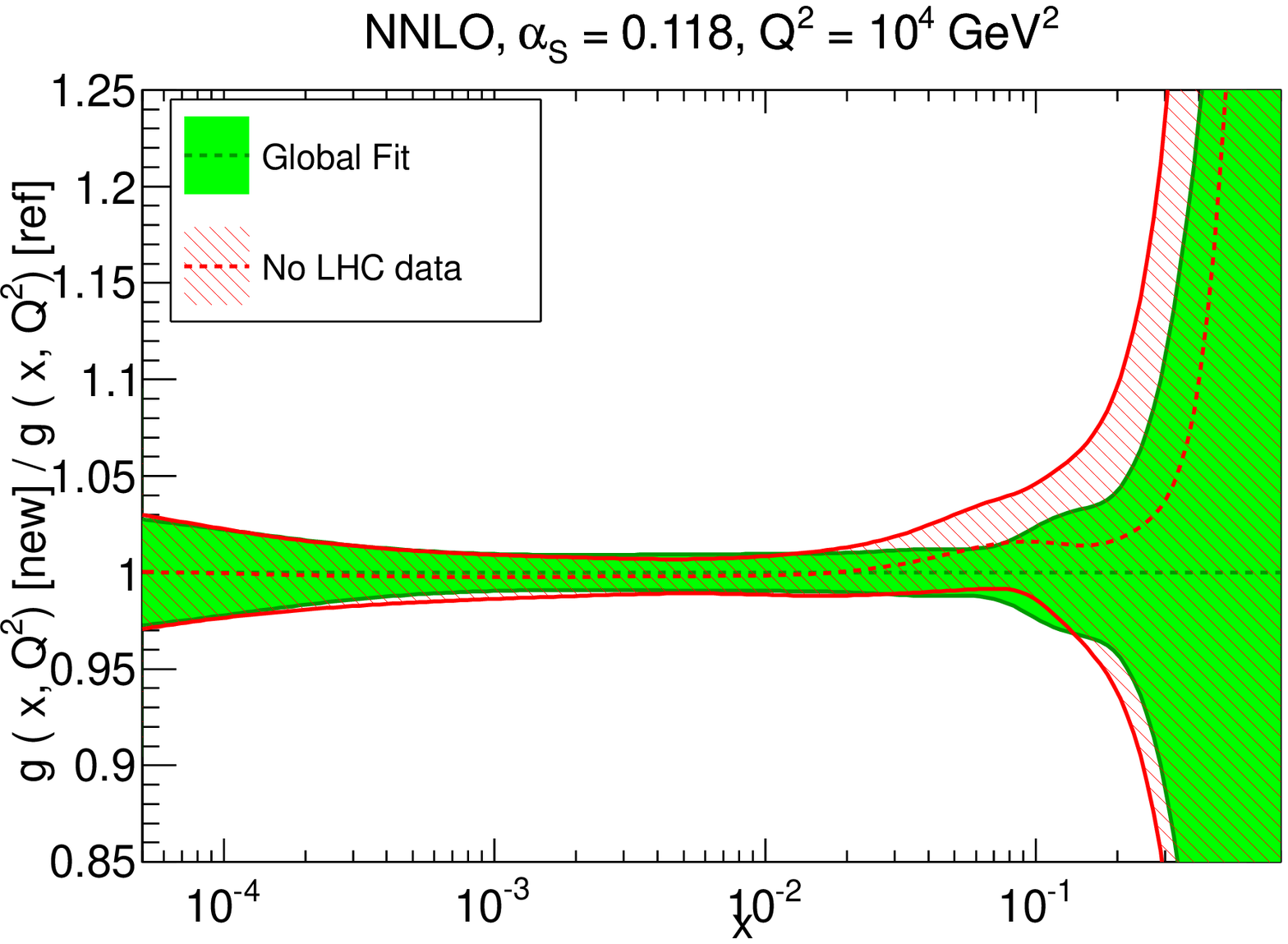}
\epsfig{width=0.42\textwidth,figure=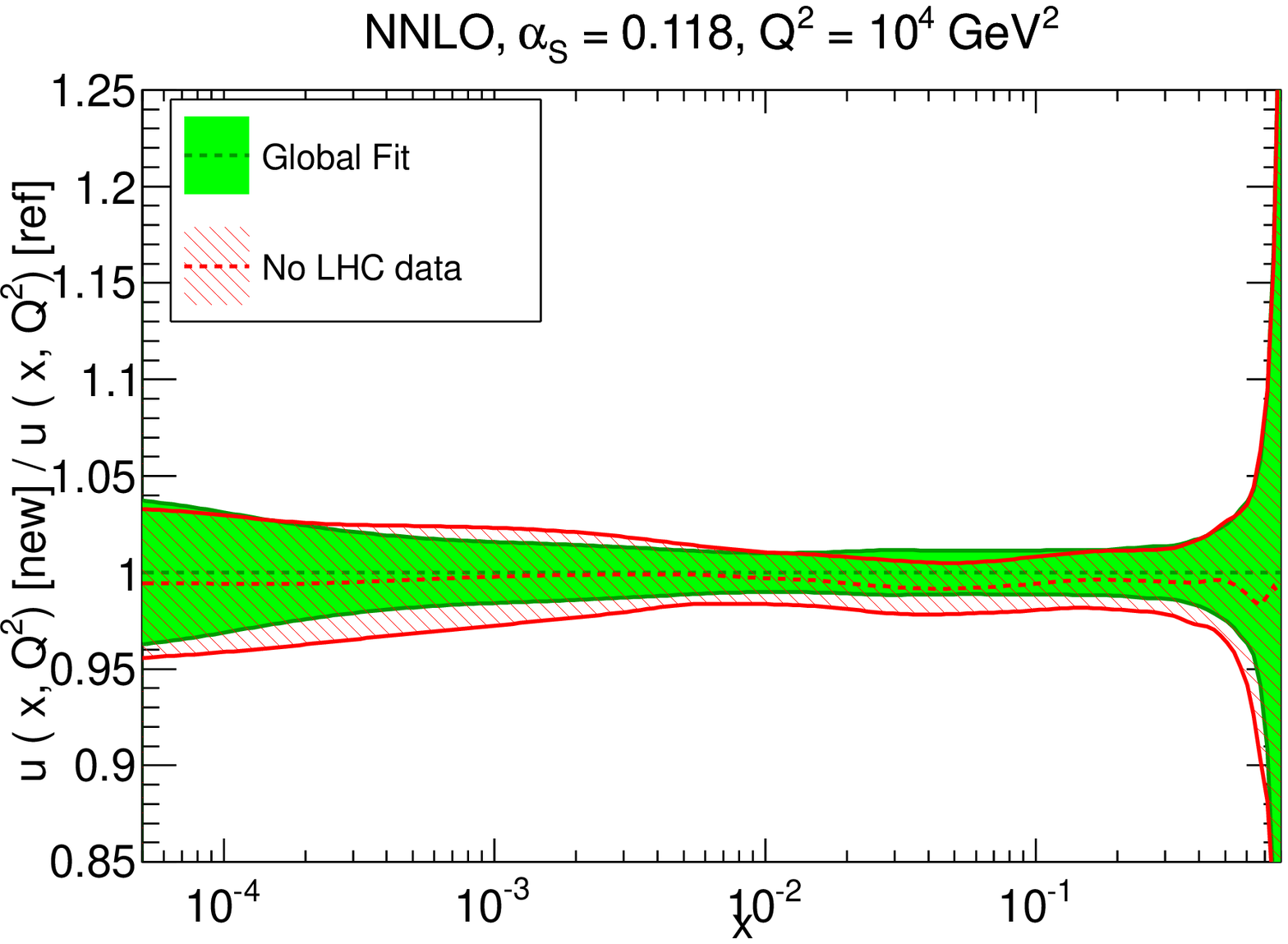}
\epsfig{width=0.42\textwidth,figure=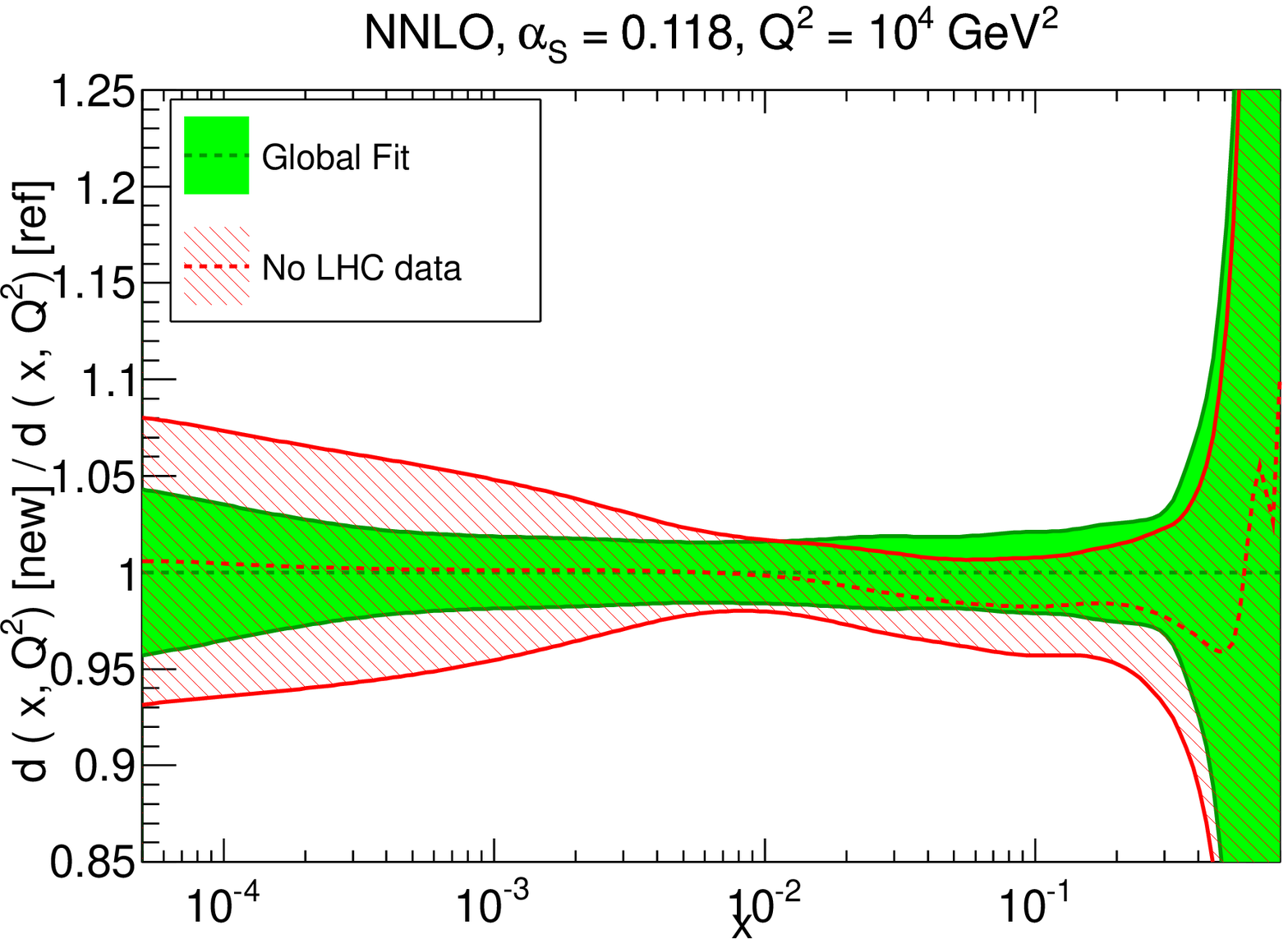}
\epsfig{width=0.42\textwidth,figure=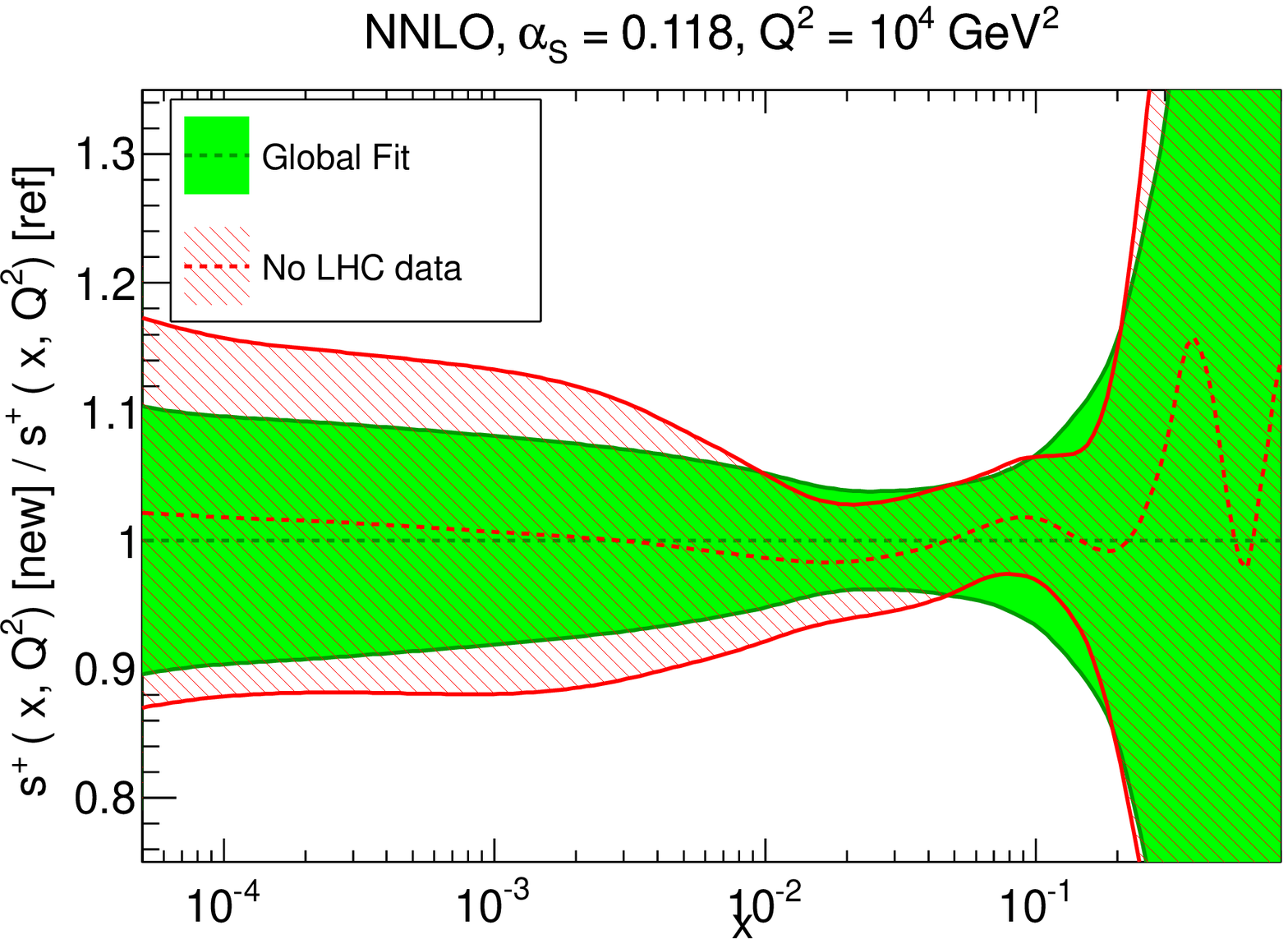}
\caption{\small
Same as Fig.~\ref{fig:xpdf-30_vs_23dataset} but
  now comparing the default NNLO set to
PDFs
obtained using the same dataset, but with all
  LHC data excluded.
 \label{fig:xpdf-30_vs_noLHC_highscale}}
\end{center}
\end{figure}

We now focus specifically on the impact of LHC data: to this purpose,
we produce a fit excluding all the LHC data from the dataset,
including those which were already in NNPDF2.3, and
keeping all the other data in the NNPDF3.0 fit (including the HERA-II data).
The corresponding distances are shown in
Fig.~\ref{fig:distances_global_vs_nolhc}, while PDFs are compared
in Fig.~\ref{fig:xpdf-30_vs_noLHC_highscale}.

The impact is seen to be moderate, at a
half-sigma level, both for central values and for
uncertainties, but it always leads to an improvement in uncertainties.
Central values are mostly affected  for
quarks at medium and large $x$, at to a lesser
extent for the gluon.

Reassuringly,
PDFs without LHC data are always within the one sigma uncertainty bands
of the global fit PDFs, confirming the
consistency of  LHC data with the previous dataset.
The gluon at medium and small $x$ is already well constrained by HERA and
Tevatron data, but the LHC improves uncertainties for  $x\ge 0.02$,
thanks
to  the ATLAS and CMS inclusive jet data and top quark production data.
The  down quark and  strange PDFs are also affected,
especially in the small-$x$ region, but also at medium $x$.

%

\begin{figure}
\begin{center}
\epsfig{width=0.42\textwidth,figure=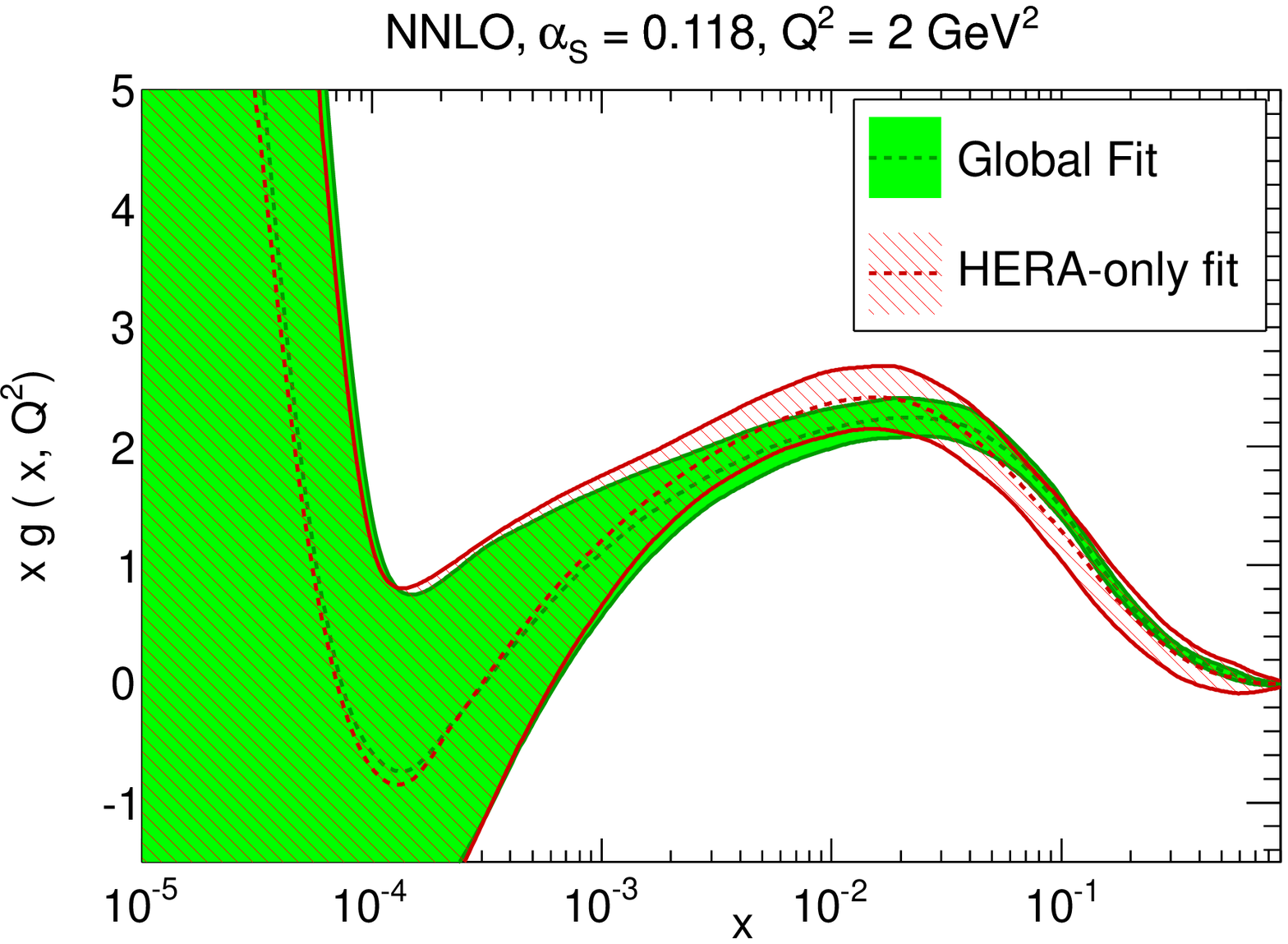}
\epsfig{width=0.42\textwidth,figure=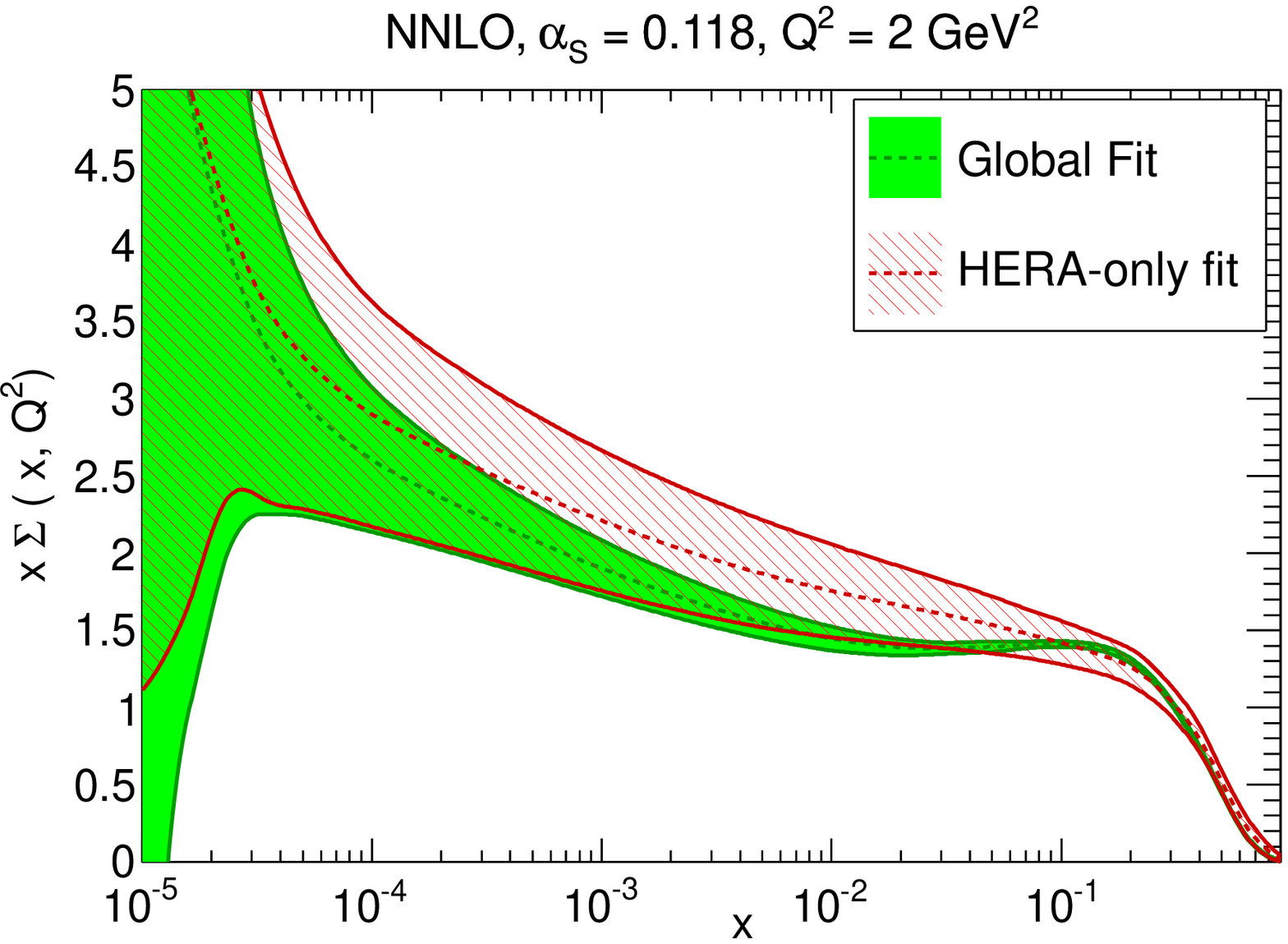}
\epsfig{width=0.42\textwidth,figure=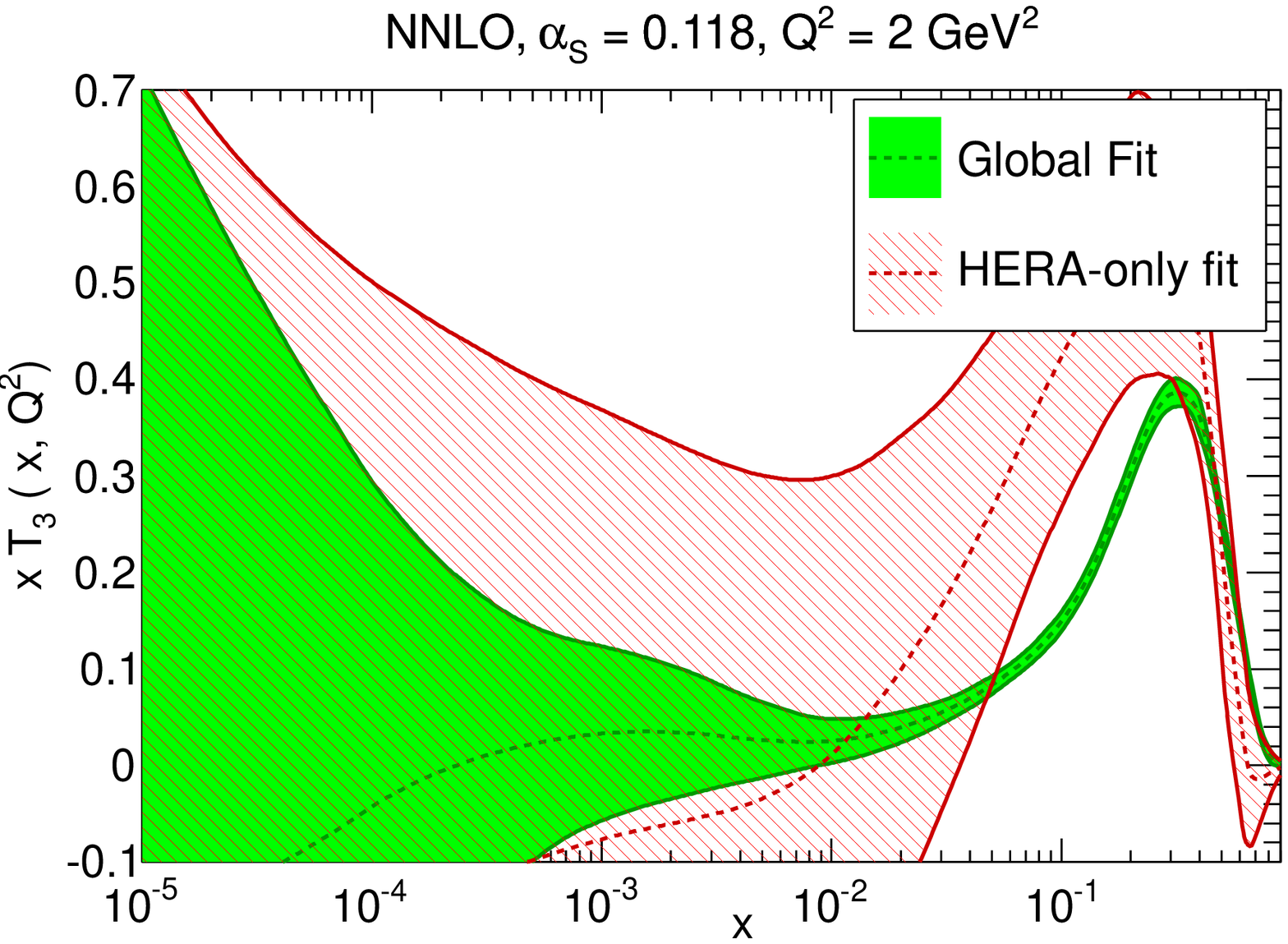}
\epsfig{width=0.42\textwidth,figure=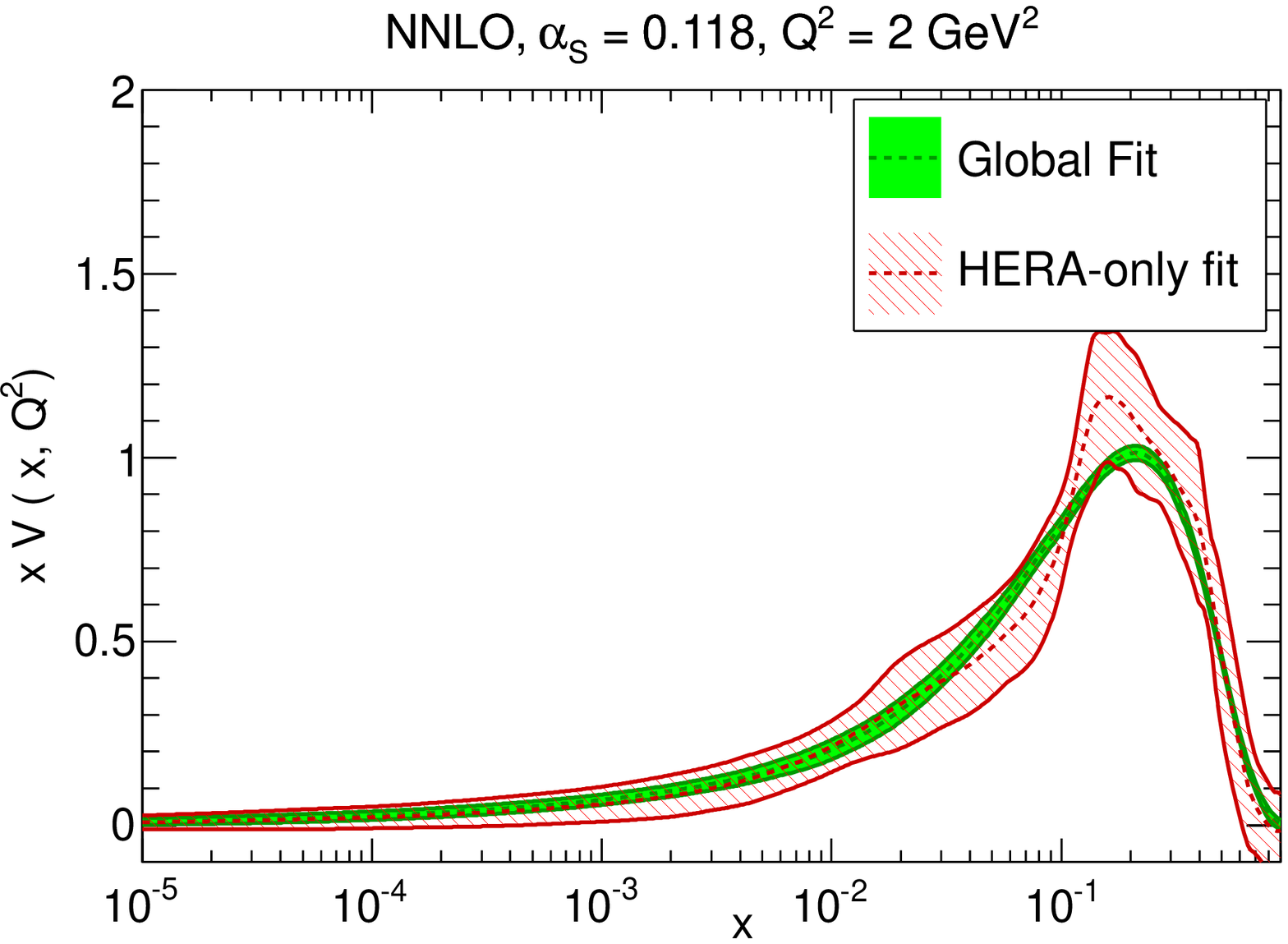}
\caption{\small \label{fig:pdfs_heraonly}
Same as Fig.~\ref{fig:pdfs_conservative}
but
  now comparing the default NNLO set to
PDFs
obtained using only HERA data.}
\end{center}
\end{figure}

Next, we construct a PDF set based on HERA data only. This is then
used to further assess the impact of the ATLAS and CMS data.
These HERA-only  PDFs are compared to the global fit at $Q^2=2$~GeV$^2$
  in Fig.~\ref{fig:pdfs_heraonly}.
Clearly, most PDFs, with the partial exception of  the small-$x$ gluon,
have much larger uncertainties than the global
fit: specifically, the quark flavor separation and the large-$x$ gluon
are very poorly constrained in a HERA-only fit, which is thus not
competitive with a global fit for phenomenology applications.

\begin{figure}
\begin{center}
\epsfig{width=0.42\textwidth,figure=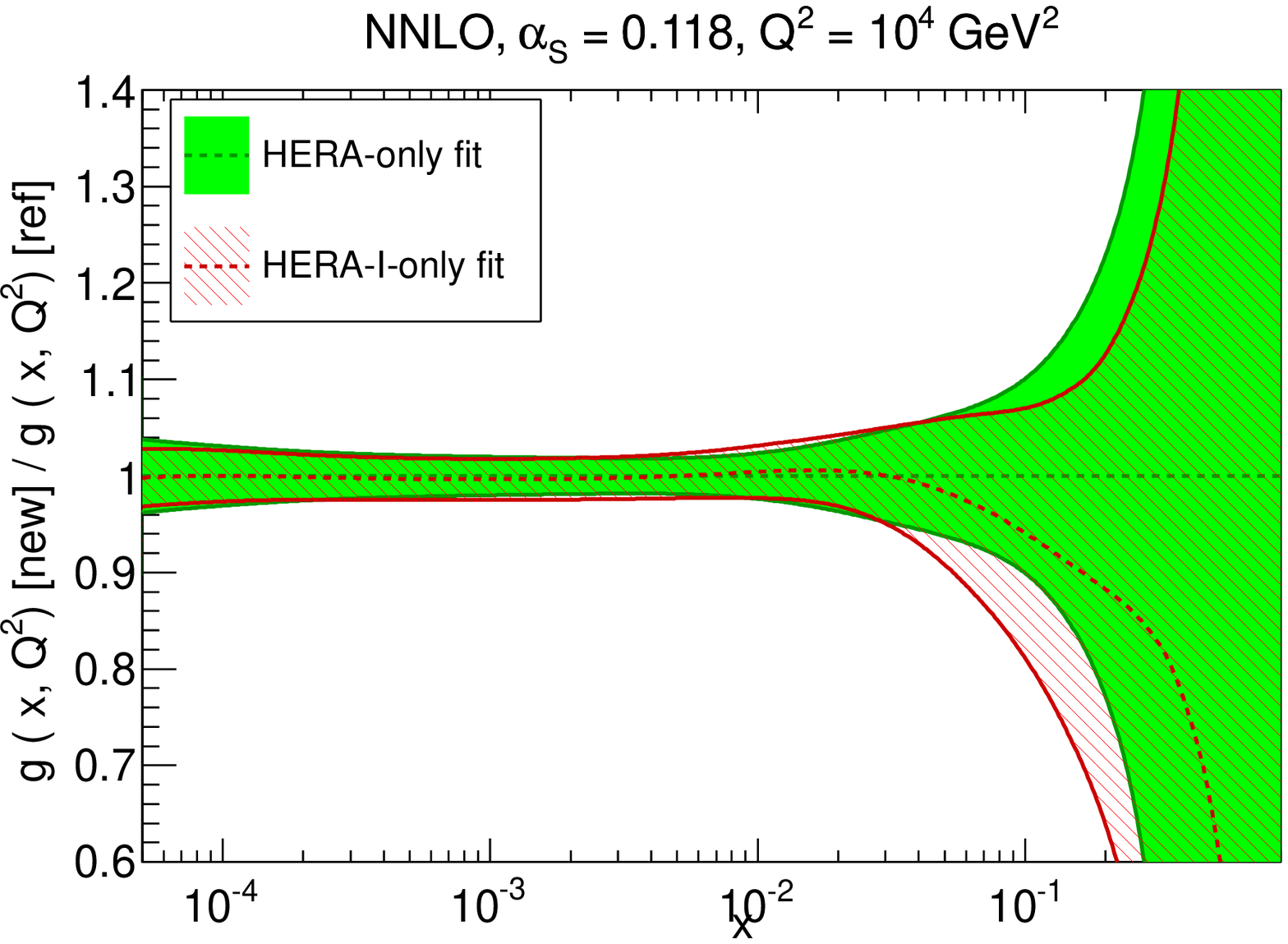}
\epsfig{width=0.42\textwidth,figure=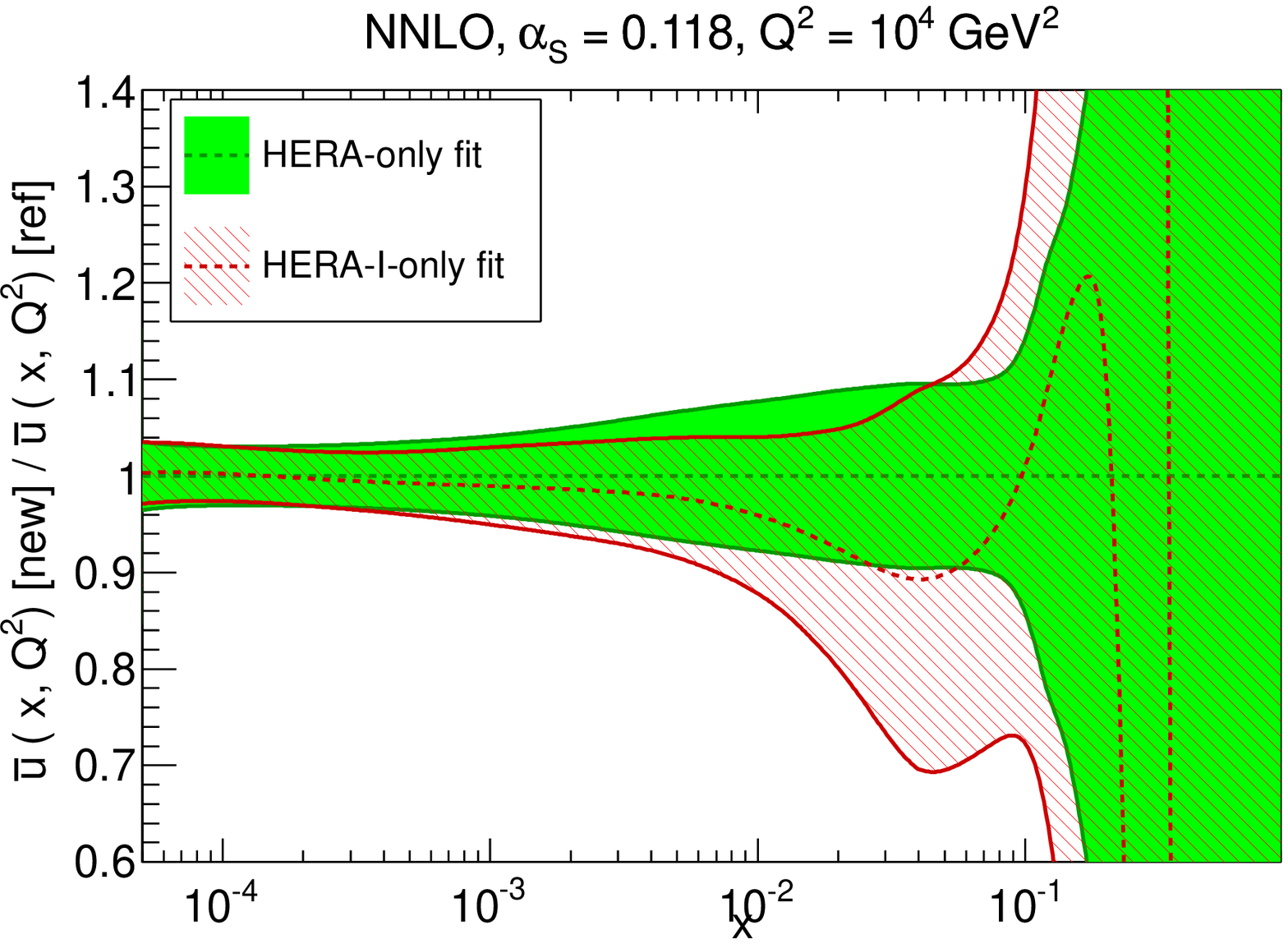}
\epsfig{width=0.42\textwidth,figure=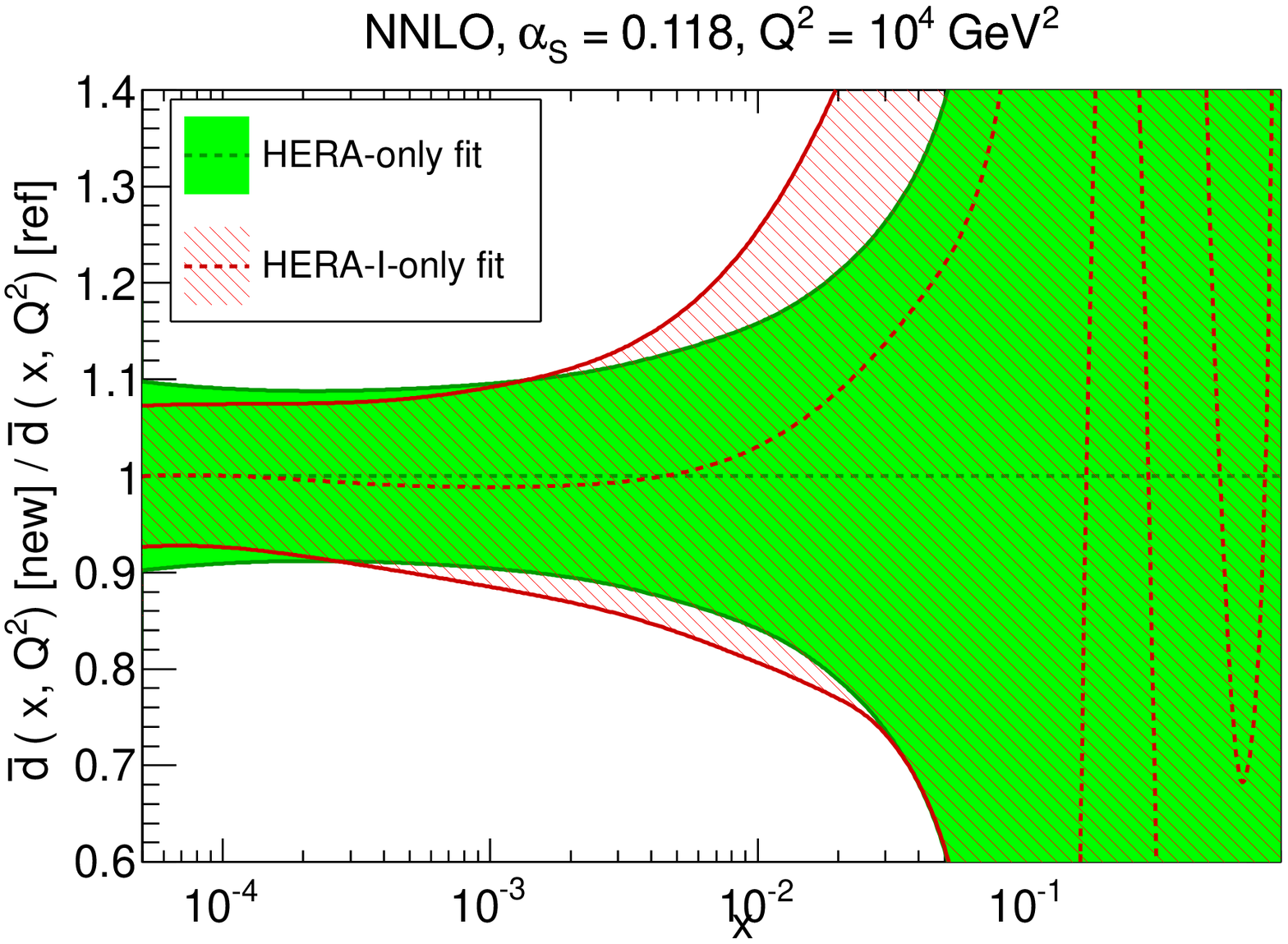}
\epsfig{width=0.42\textwidth,figure=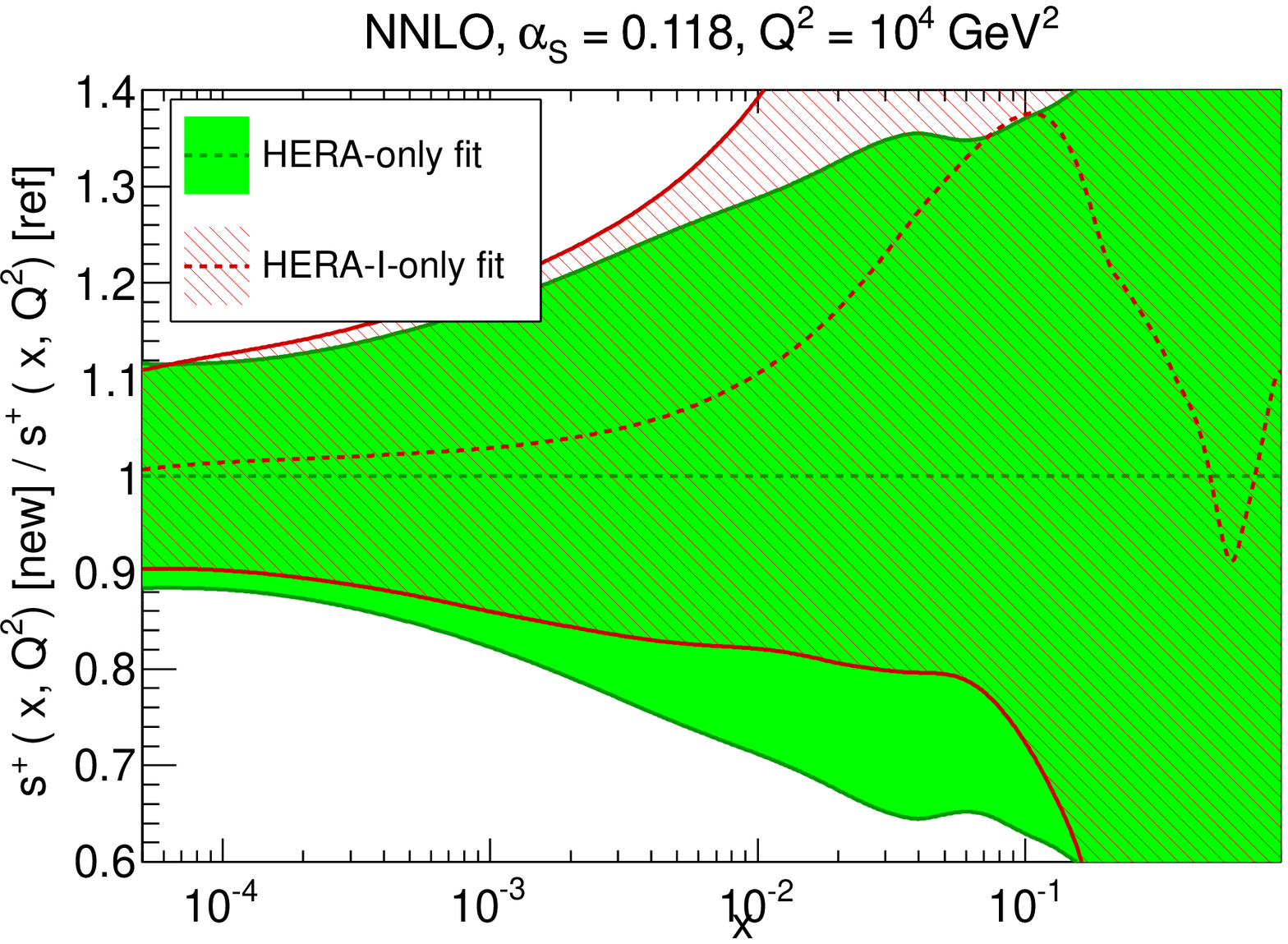}
\caption{\small
Same as Fig.~\ref{fig:xpdf-30_vs_23dataset} but now comparing
 HERA-only and HERA-I-only PDFs (see text).
 \label{fig:xpdf-heraIonly}}
\end{center}
\end{figure}

The HERA dataset has widened considerably with the
addition of the complete HERA-II
inclusive data
from H1 and ZEUS and  combined HERA charm production
data.
In order to study the impact of this data,
we have produced a version of the HERA-only fit
in which we have kept only the combined HERA-I data, i.e. a
HERA-I-only fit.
The NNLO PDFs of the HERA-only and HERA-I-only fits are compared
at $Q^2=10^4$ GeV$^2$ in Fig.~\ref{fig:xpdf-heraIonly}.
The additional information provided by HERA-II apparently has a
moderate impact:
the gluon is mostly unchanged, while the PDF uncertainties on the medium-
and large-$x$ up antiquarks and (to a lesser extent) on the
down antiquarks are moderately reduced.
We conclude that, while certainly beneficial, the
new HERA-II data does not change substantially the known fact
that HERA-only fits are affected by large PDF uncertainties.

We now study the response of  HERA-only fit of Fig.~\ref{fig:pdfs_conservative}
upon addition of various pieces of data.
In particular, we produce two fits, respectively
from a dataset obtained by adding to the
HERA data  all the ATLAS data or all the CMS data.
Specifically (see Tab.~\ref{tab:completedataset}) in the
HERA+CMS fit,
the HERA data is supplemented with  data on jet production,
$W$ asymmetries, Drell-Yan differential distributions, $W$+$c$ production
and top quark total cross-sections, while in the
HERA+ATLAS fit, the HERA measurements are
supplemented  with  $W,Z$ rapidity distributions
from the 2010 dataset,  inclusive jet data at 7 TeV and 2.76 TeV, and
high-mass Drell-Yan production.

\begin{figure}[t]
\begin{center}
\epsfig{width=0.89\textwidth,figure=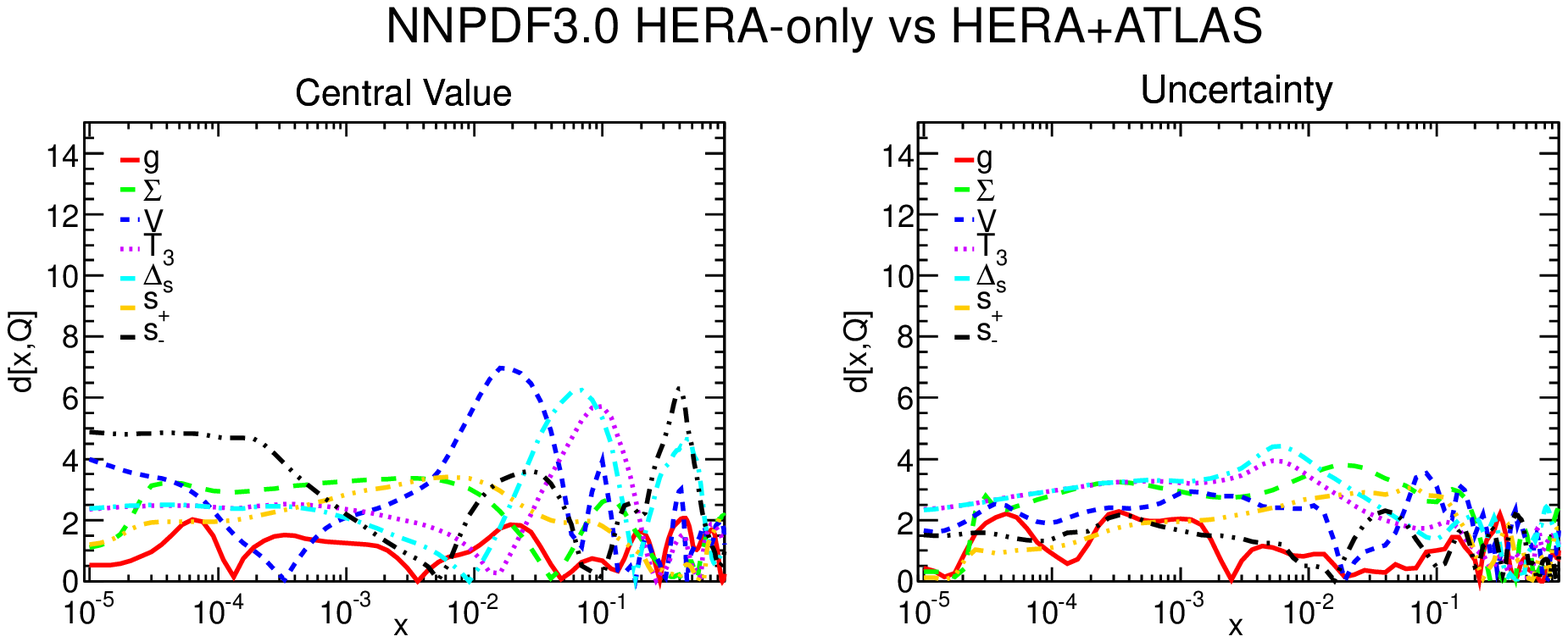}
\epsfig{width=0.89\textwidth,figure=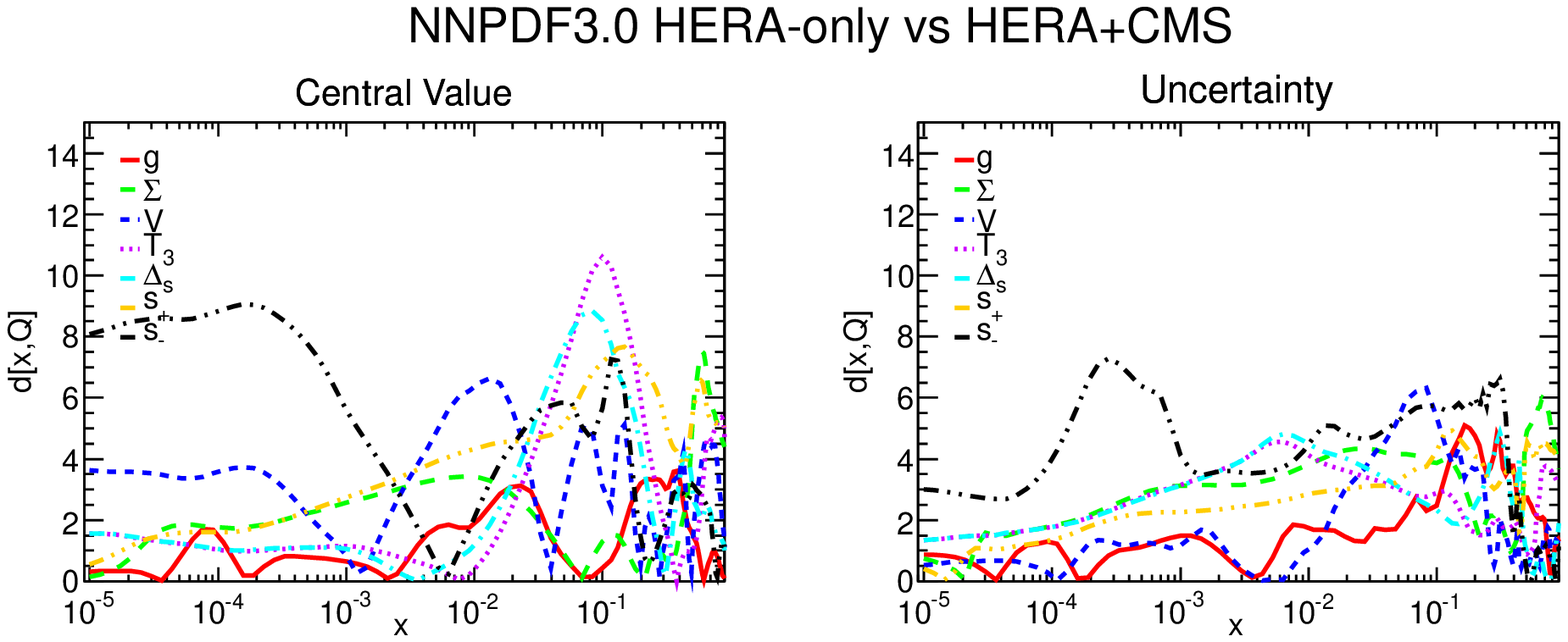}
\caption{\small
 Same as Fig.~\ref{fig:distances_30_vs_23_nnlo}, but
  now comparing the NNLO HERA-only and HERA+ATLAS PDFs (top) or
 the  HERA-only and HERA+CMS PDFs (bottom) (see text).
 \label{fig:distances_atlas_vs_cms}}
\end{center}
\end{figure}

\begin{figure}
\begin{center}
\epsfig{width=0.42\textwidth,figure=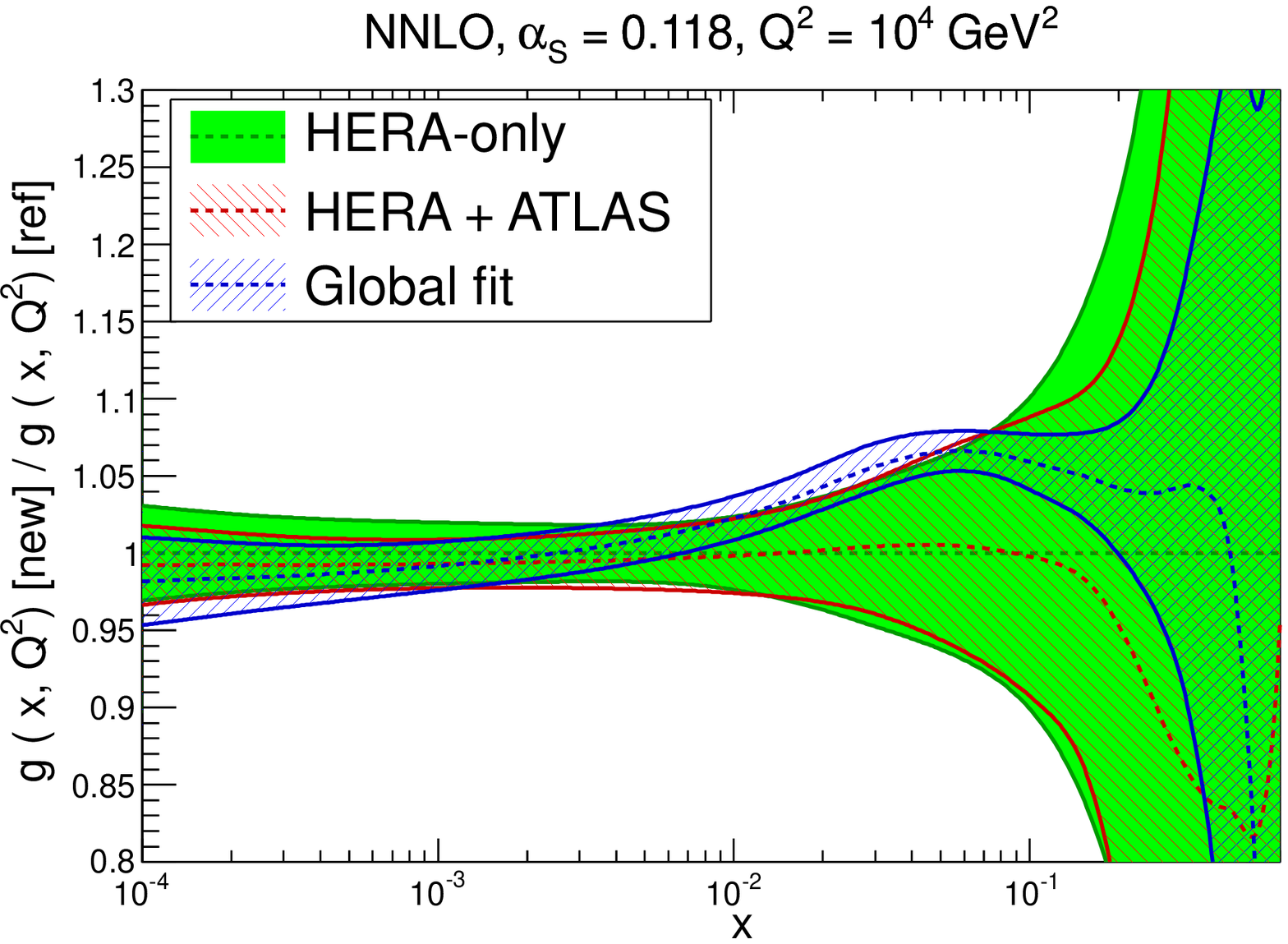}
\epsfig{width=0.42\textwidth,figure=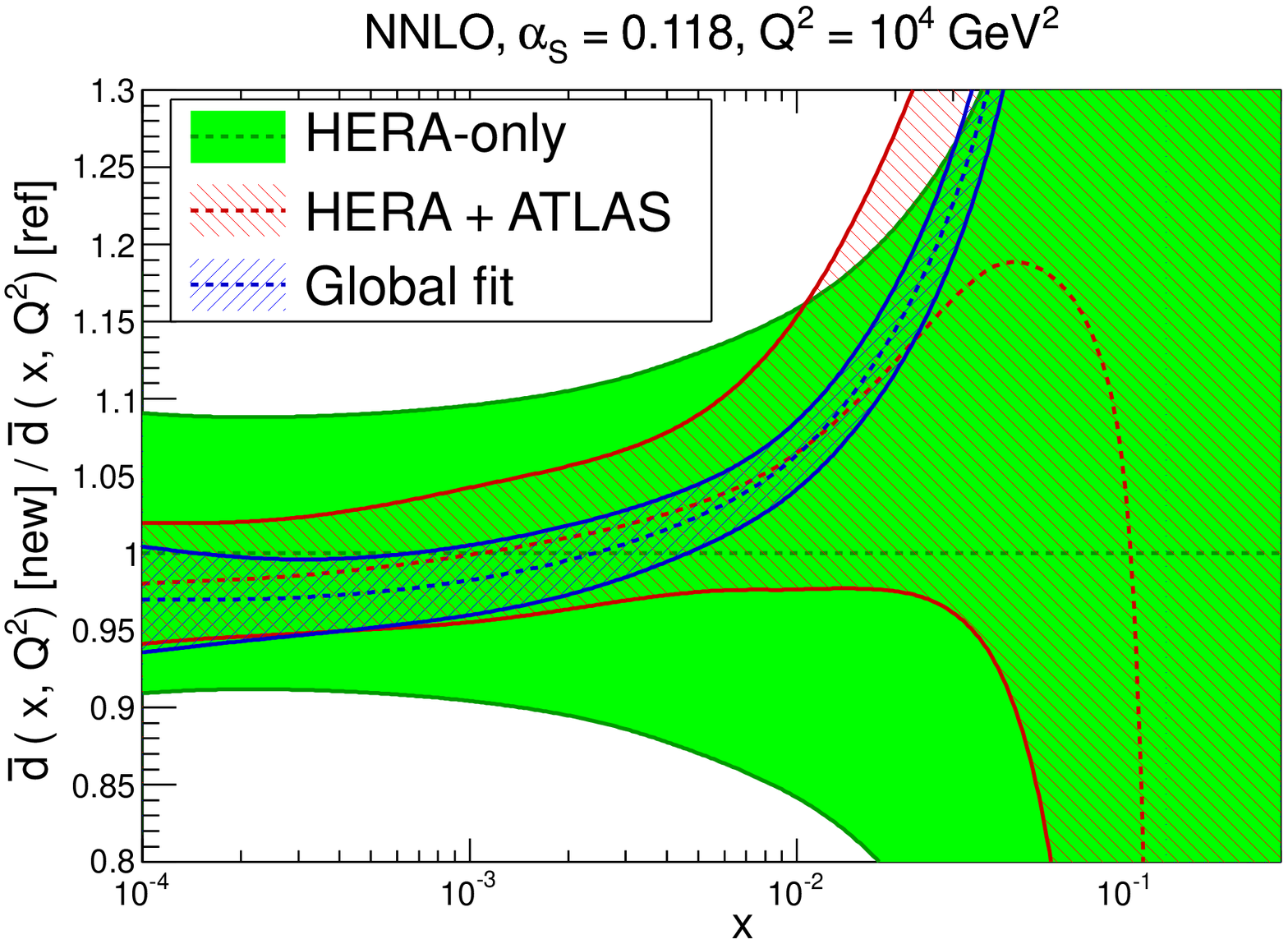}
\epsfig{width=0.42\textwidth,figure=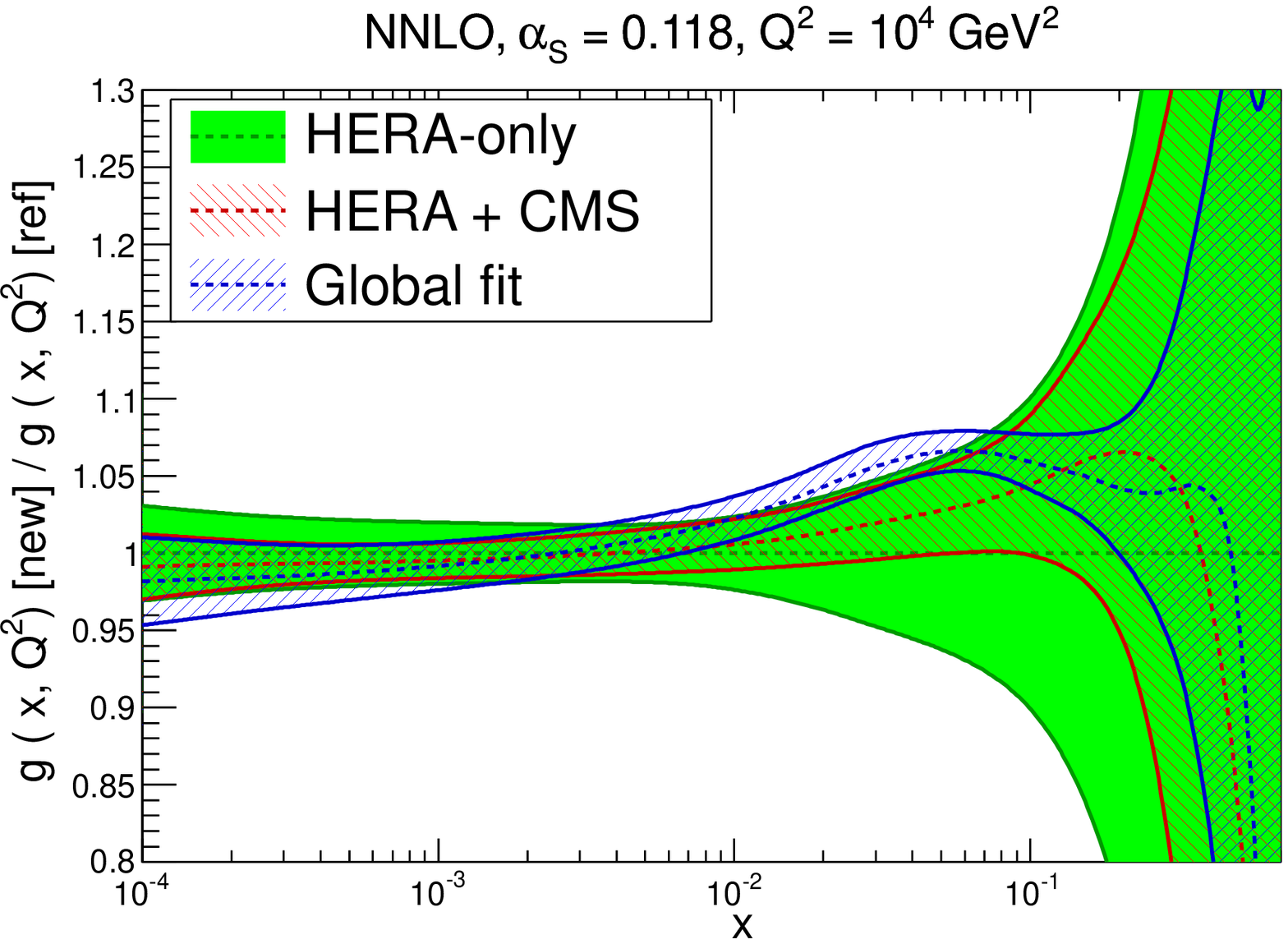}
\epsfig{width=0.42\textwidth,figure=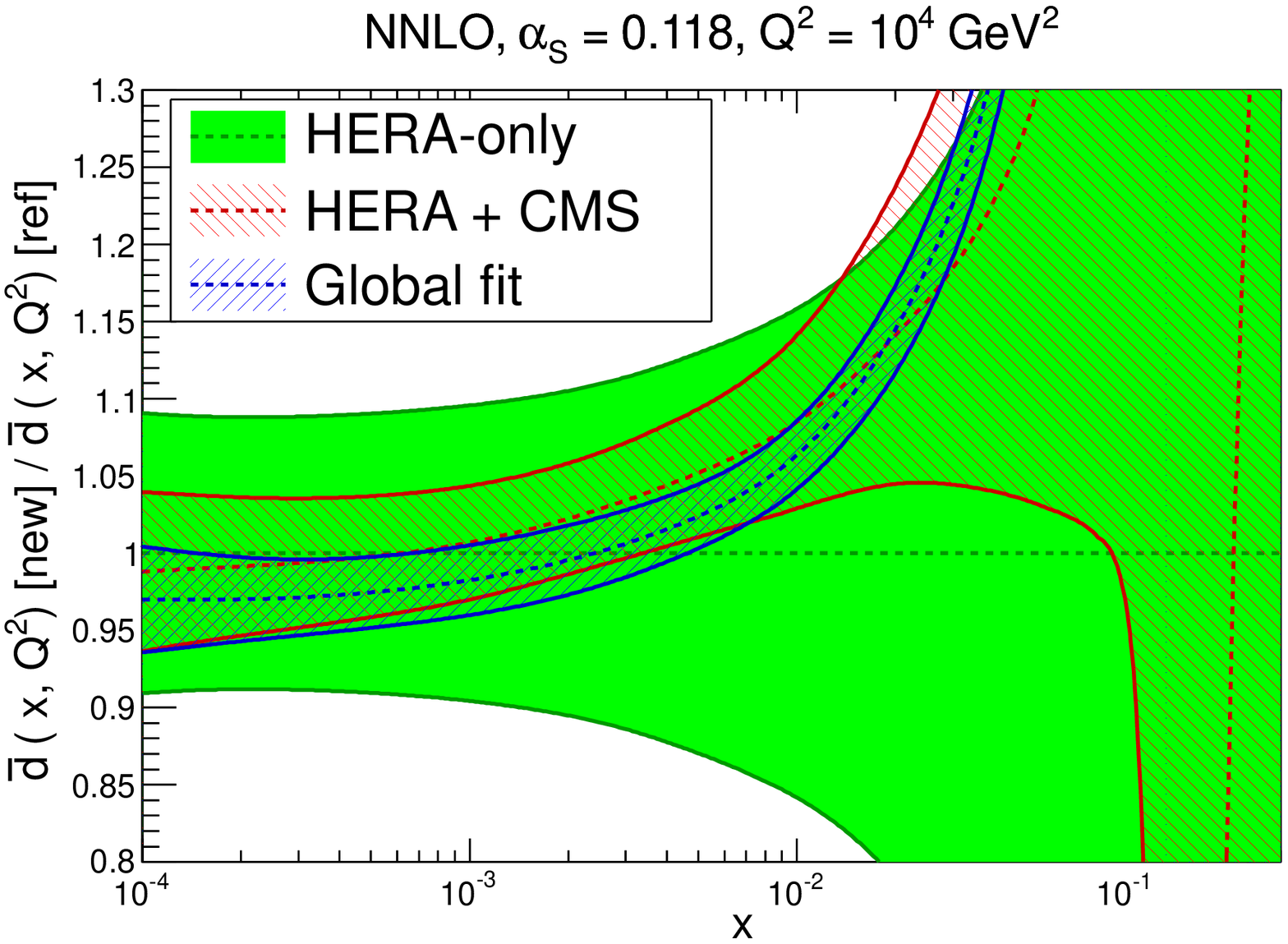}
\caption{\small
Comparison of the gluon and antidown  NNLO PDFs at  $Q^2=10^4$ GeV$^2$
of the HERA-only and HERA+ATLAS sets (top) or
the HERA-only and HERA+CMS sets (bottom), shown as rations to the
HERA-only PDFs. For reference, the PDFs from the default NNPDF3.0
global
set are also shown.
\label{fig:pdf-cmsonly}}
\end{center}
\end{figure}

The distances between the HERA-only and HERA+ATLAS, or HERA-only
vs. HERA+CMS fits are shown in Fig.~\ref{fig:distances_atlas_vs_cms},
while the gluon and $\bar{d}$ PDFs are shown
 in Fig.~\ref{fig:pdf-cmsonly}, with the  default global fit also shown
 for reference.
The impact of the LHC data is now apparent, in particular for
PDF combinations which are poorly
constrained by a fit to HERA data only, specifically the large-$x$
gluon and the  the quark flavor separation.
Note that
the CMS data provides more stringent constraints on the gluon at large $x$ since
it uses the 2011 inclusive jet data, which for ATLAS is still not available.
ATLAS and CMS have a similar constraining power for the medium and
large-$x$ quarks, with CMS slightly superior for the strangeness
PDFs thanks to the availability of the $W$+$c$ measurements, and also
for flavor separation (and thus for  $\bar{d}$) due to the fact that
the CMS electroweak dataset is somewhat more extensive.

%
On the other hand, comparison to the
global fit shows that neither of these HERA+ATLAS or HERA+CMS fits is
yet competitive. Also, comparison of these result to our previous
assessment of the overall impact of LHC data on the global fit
(Figs.~\ref{fig:distances_global_vs_nolhc} and
  \ref{fig:xpdf-30_vs_noLHC_highscale}) shows that gauging the impact
  of LHC data on PDFs by their effect on a HERA-only fit might be
  somewhat misleading: because of the good consistency of LHC data
  with the pre-LHC global dataset, their impact in the global fit is
  rather less pronounced.

We finally come to a global assessment of the dependence of the PDF
uncertainty on the dataset. This can be done by using the estimator
$\varphi_{\chi^2}$, Eq.~(\ref{eq:frdbdef}), which, as discussed  in Sect.~\ref{sec:closure},
is essentially the ratio  of the PDF uncertainty on the prediction to
the original experimental uncertainties, averaged over all data, and
keeping into account all correlations. Its
value thus measures by how much the data uncertainty, propagated to
the PDFs, and then propagated back to the data, is reduced due to the
fact that the PDF fits combines many data points into a single set of
PDFs which are constrained both by theory (such as perturbative evolution,
which connects different values of $Q^2$, and sum rules, which relates
different values of $x$) and also by smoothness (the value of any PDF
at neighboring points in $x$ cannot differ by an arbitrarily large amount).
For a single data point, or if data points were independent, its value would
be one by definition,  while its deviation from one estimates the uncertainty
reduction (or increase, which could happen in principle when combining
inconsistent data). Note that this indicator only probes the PDF
uncertainty at the data points: so the reduction in uncertainty when
widening the dataset is actually somewhat underestimated by it, in
that for a bigger dataset the given value of $\varphi$ actually refers
to a wider kinematic region.

In Tab.~\ref{tab:varchi2} we show $\varphi_{\chi^2}$ for the global NNPDF3.0
NLO and NNLO fits, as well as for the fits based on reduced datasets, starting
from the HERA-I only fit and including the fit without LHC data, in increasing order
of the size of the fitted dataset.
The result for the conservative set refers to the fit with threshold $\alpha_{\rm max}=1.1$.

For the global fits, we find $\varphi_{\chi^2}=0.291$ and 0.302 for the NLO and NNLO
sets respectively, to be compared with the corresponding value at LO, $\varphi_{\chi^2}=0.512$.
The improvement between LO and NLO, almost by a factor of two in terms
of the reduction of the PDF uncertainties to the fitted data points,
is clear evidence of the better consistency of the NLO fit in comparison
to the LO one. On the other hand, the NNLO fit is very similar to the NLO one in this respect
(perhaps marginally worse), consistent with the observation that the quality of the NNLO fit
is not better than that of the NLO fit, as also measured by the value of the $\chi^2$, see
Tab.~\ref{tab:chi2tab_exp_vs_t0}.

The monotonic decrease of the values of  $\varphi_{\chi^2}$ for the fits to reduced datasets,
from HERA-I to HERA-all, to HERA+ATLAS or HERA+CMS, to the global fit, shows the constant
uncertainty reduction as more data are combined.
The uncertainty on the conservative fit is larger than that of any fit except the HERA-I fit.
This provides evidence in favor of using our current default as reference, rather than the conservative
set, as the substantial decrease in $\varphi_{\chi^2}$ in the global fit in comparison to the conservative
one suggests that the overall consistency of the global fit is still quite good. Finally, the
uncertainty on the no-LHC fit is larger than that of a global fit, and in fact comparable to that of the
HERA+CMS fit: this shows that even though the impact of the LHC data is moderate, it is visible, and
it is comparable to the impact of all the other non-HERA data.

\begin{table}[h]
\centering
\begin{tabular}{l|c|c}
\hline
Dataset &  $\varphi_{\chi^2}$ NLO  & $\varphi_{\chi^2}$ NNLO \\
\hline
\hline
Global & 0.291 & 0.302 \\
\hline
HERA-I & 0.453 & 0.439 \\
HERA all & 0.375 &  0.343 \\
HERA+ATLAS & 0.391  & 0.318 \\
HERA+CMS & 0.315  & 0.345  \\
Conservative & 0.422 & 0.478 \\
no LHC & 0.312 & 0.316 \\
\hline
\end{tabular}
\caption{\small \label{tab:varchi2} The value of the fractional
  uncertainty
$\varphi_{\chi^2}$, Eq.~(\ref{eq:frdbdef}),
for the default NNPDF3.0 NLO and NNLO fits compared to that obtained
in various fits to reduced datasets.
For the LO global fit, we find $\varphi_{\chi^2}=0.512$.
The result for the conservative set refers to the fit
with  $\alpha_{\rm max}=1.1$.
 }
\end{table}


\clearpage

\subsubsection{Impact of jet data on the global fit}
\label{sec:jetless}

We now explore the impact of jet data in the NLO
and NNLO NNPDF3.0 fits, with the motivation of making sure that
theoretical limitations in the description of jet data, and in
particular the
current lack of full knowledge of NNLO corrections,
does not bias the fit results.

\begin{figure}
\begin{center}
\epsfig{width=0.42\textwidth,figure=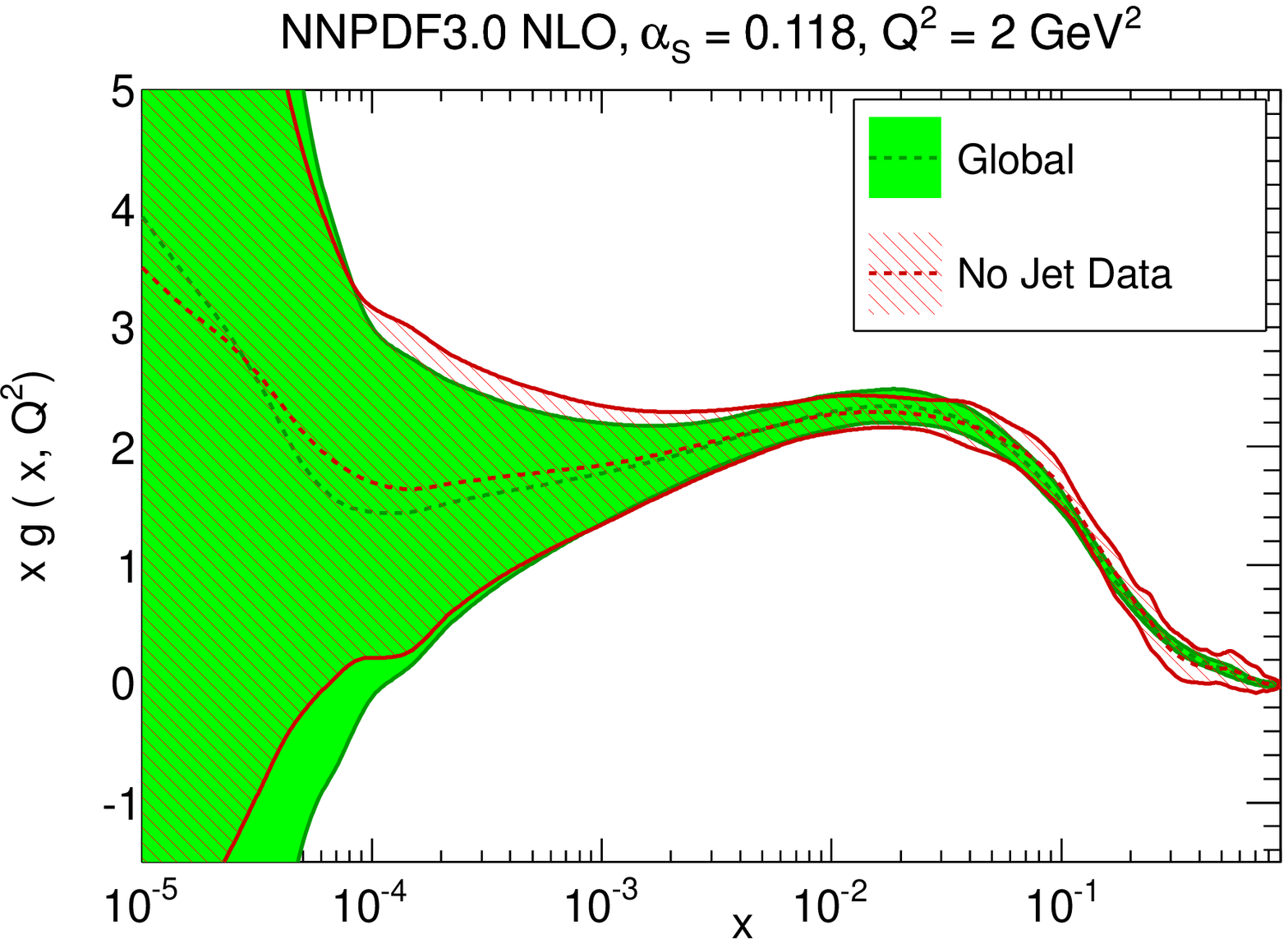}\epsfig{width=0.42\textwidth,figure=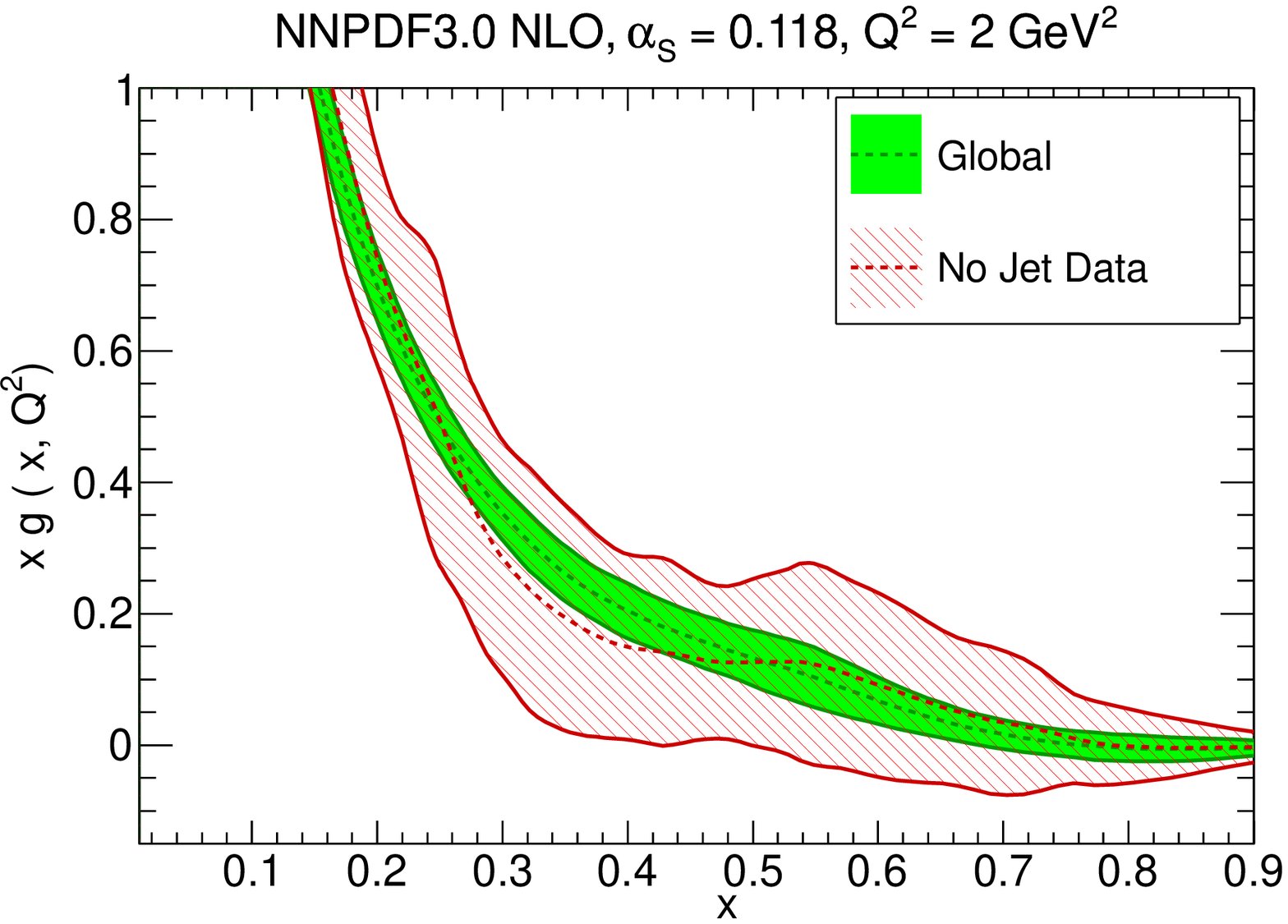}
 \epsfig{width=0.42\textwidth,figure=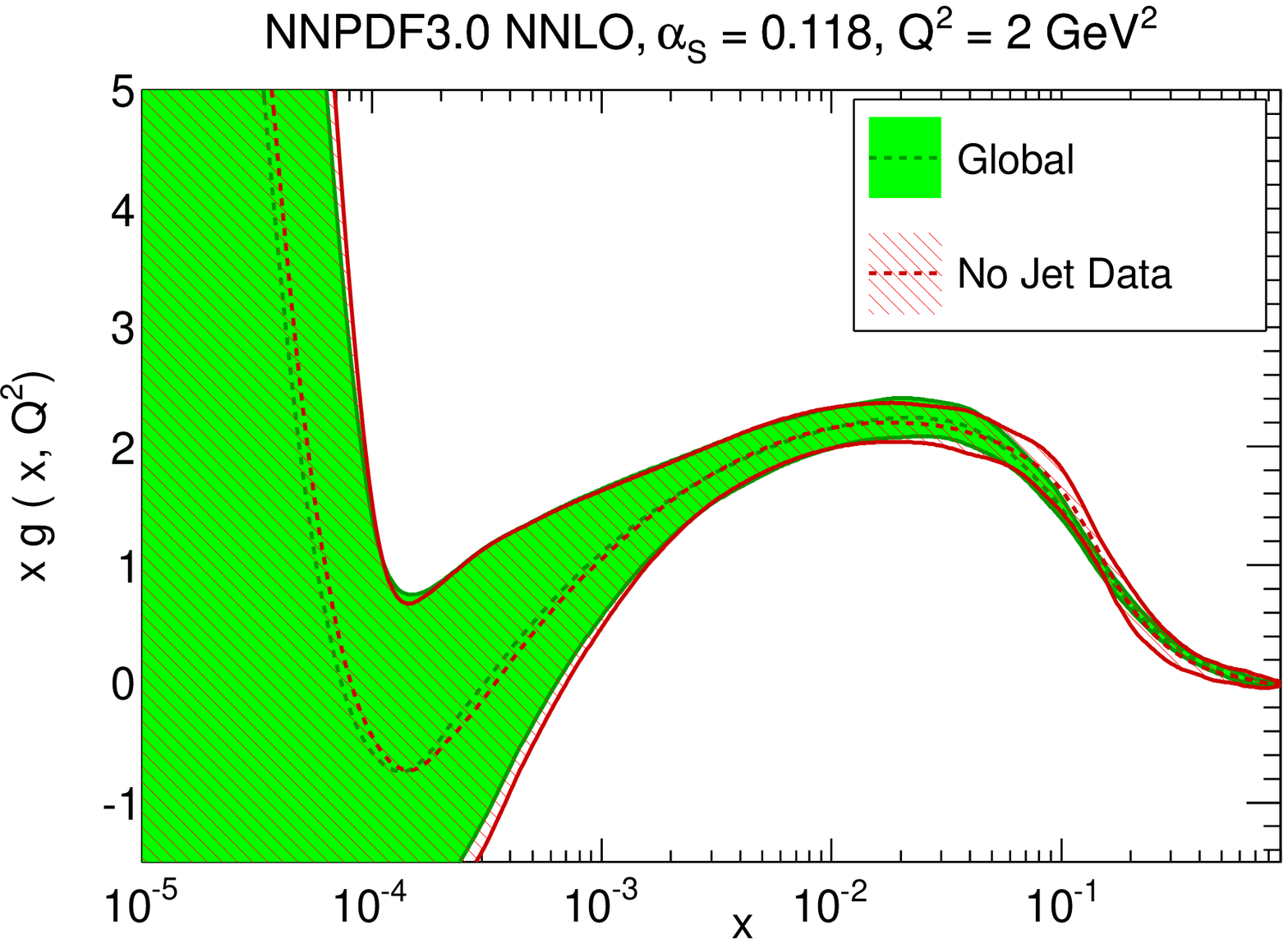}
\epsfig{width=0.42\textwidth,figure=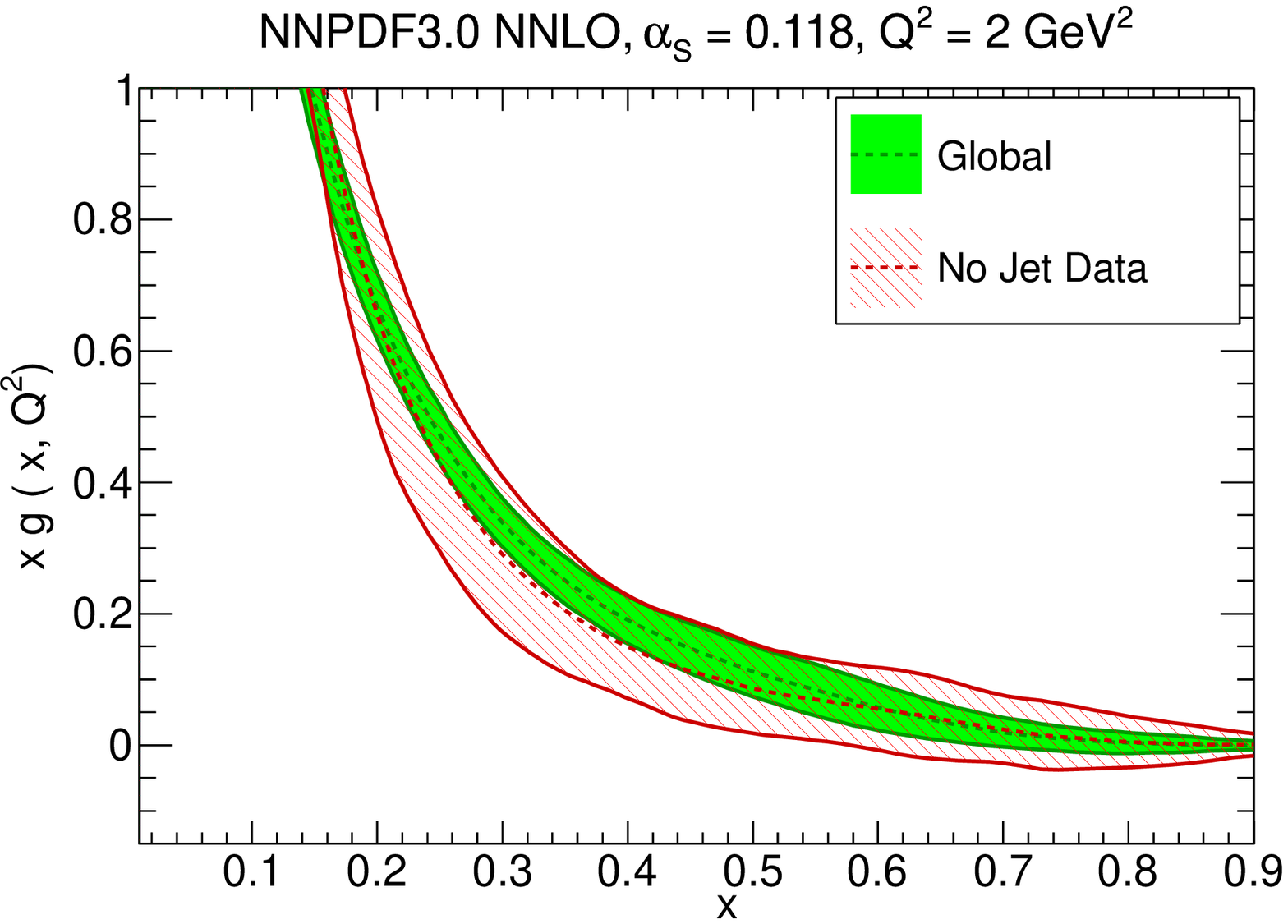}\caption{\small
Comparison of the gluon in a fit
to a dataset without jet data and in the global fit at NLO (top)
and NNLO (bottom), plotted at $Q^2=2$~GeV$^2$  vs. $x$ on a
logarithmic (left) and linear (right) scale.
\label{fig:abs-30_vs_30nojet_lowscale}}
\end{center}
\end{figure}

%
To  this purpose, we have produced versions of the NNPDF3.0 PDF fit
in which
all jet data are removed from the global dataset: the gluon from these
sets is compared to that from the default global fit at
$Q^2=2$~GeV$^2$ in Fig.~\ref{fig:abs-30_vs_30nojet_lowscale}:
Other PDFs are essentially unchanged upon removing jet data.
It is clear that
 removing jet data from the global fit leads to a substantial
increase of the PDF uncertainties on the gluon at medium-
and large-$x$. However, when jet data are included, the uncertainties
are very similar at NLO and NNLO, despite the fact that at NNLO the
jet dataset is significantly smaller due to the more restrictive cuts
which we have introduced in order to account for the incomplete
knowledge of NNLO corrections to jet production (see
Sect.~\ref{sec:thjetdata}): in fact, if anything, the uncertainties
are somewhat smaller at NNLO.
This is reassuring in that it is consistent with the expectation that
no instabilities are introduced by jet data in the NNLO fit despite
potentially large perturbative corrections, and in fact the fit
becomes tighter at NNLO.

In  Tab.~\ref{tab:jetless} we compare at NLO and NNLO the
$\chi^2$ to the collider jet data, both in the
reference NNPDF3.0 fit and in the fit without jet
data.
We provide the results using both the experimental and the $t_0$ $\chi^2$
definitions, whose values can differ significantly, especially at NNLO.
The description of jet data turns out to be reasonably good even when
they are not included in the fit, especially at NNLO. This  is evidence for consistency,
and it explains why they help in reducing the gluon uncertainty.
We also show the value of
the $\chi^2$ for top pair production, which is also sensitive to the gluon.
The fact that this value changes very little upon inclusion of jet
data is also evidence for general consistency.

We conclude that based on all evidence not including jet data (or not
including them at NNLO) would not lead to a significant change of the
central value of the extracted gluon distribution but it would lead to a deterioration
of its uncertainty. Given our conservative treatment of NNLO
perturbative corrections, described in Sect.~\ref{sec:thjetdata}, and
in the absence of indications of instability or inconsistency, we
believe that the determination of the gluon is most reliable when jet
data are kept in the dataset, as we do for our default fit.

\begin{table}
\centering
\begin{tabular}{c|cc|cc}
\hline
 & \multicolumn{4}{|c}{NLO} \\
\hline
  & \multicolumn{2}{c|}{Exp $\chi^2$} &  \multicolumn{2}{c}{$t_0$ $\chi^2$} \\
Dataset  & Global & No Jets
 &  Global & No Jets \\
\hline
\hline
CDF Run II &   0.95  &{\it 1.51 } &  1.05 & {\it 1.62   } \\
ATLAS 7 TeV + 2.76 TeV &  1.58  & {\it 1.88   } &  0.86 & {\it 0.96  } \\
CMS 7 TeV 2011 & 0.96  & {\it 1.32 } & 0.90   &{\it 1.17  } \\
\hline
Top quark pair-production & 1.43 & 1.26  & 1.67 & 1.49   \\
\hline
\end{tabular}
 \\
\vspace{0.5cm}
\begin{tabular}{c|cc|cc}
\hline
 & \multicolumn{4}{|c}{NNLO} \\
\hline
  & \multicolumn{2}{c|}{Exp $\chi^2$} &  \multicolumn{2}{c}{$t_0$ $\chi^2$} \\
Dataset  & Global & No Jets
 &  Global & No Jets \\
\hline
\hline
CDF Run II &  1.84 &{\it 1.85} & 1.20   & {\it 1.58  } \\
ATLAS 7 TeV + 2.76 TeV & 1.17 & {\it 1.00   } & 0.72  & {\it 0.65  } \\
CMS 7 TeV 2011 &  1.91 & {\it 2.23 } &  1.07  &{\it 1.37  } \\
\hline
Top quark pair-production &  0.73 &  0.43  & 0.61 & 0.42   \\
\hline
\end{tabular}
\caption{\small The  $\chi^2$ to
 jet data,  computed using
the default NNPDF3.0 PDFs, or PDFs based on a dataset without jet data;
values in italics correspond to data which have not been included
in the fit. Values are provided using both the experimental and $t_0$
definition of the $\chi^2$ (see
Sect.~\ref{sec:quality}). The value for top data (included in all
fits) is also shown.
\label{tab:jetless}
}
\end{table}

\clearpage

\subsubsection{Nucleon strangeness}
\label{sec:strangeness}

Recently the size of the strange distribution has been the object of
experimental and phenomenological debate.
In global fits, the strange PDF is mostly
 constrained by the neutrino-induced deep-inelastic scattering data,
 such as
CHORUS, NuTeV and NOMAD~\cite{MasonPhD,Mason:2007zz,
Goncharov:2001qe,Samoylov:2013xoa}.
While also inclusive data is sensitive to strangeness, the strongest constraint come
from the so-called dimuon process, charm production in charged-current DIS.
However, the theoretical treatment of this data is affected by various
sources of theoretical uncertainty, such as the need
to model charm fragmentation, the treatment of charm quark
mass effects at low scales, and effects related to the use of nuclear targets.
Recently, LHC data which constrain the strange PDF have become
available: on top of inclusive  $W$ and $Z$ production on- and
off-shell production,  $W$ production
in association with charm quarks which directly probes strangeness at
leading order.

In PDF global fits, with strangeness  mostly based on neutrino data,
the strange sea is typically smaller than
the up and down quark sea by a factor of order  $\sim \frac{1}{2}$.
In 2012, a QCD analysis of the ATLAS data on $W$ and $Z$ rapidity
distributions at 7 TeV~\cite{Aad:2012sb} suggested that the size of
the strange and  up and down sea is comparable, at least for $x\sim0.01$.
This analysis was revisited in the NNPDF2.3
framework~\cite{Ball:2012cx}, with the conclusion that while the ATLAS
data in isolation do favor a central value of $s(x,Q^2)$ similar in size to
$\bar{u}(x,Q^2)$ and $\bar{d}(x,Q^2)$, the uncertainties involved are
  so large that it is difficult to make a clear-cut statement, and in
  particular the central value of the strangeness fraction in the
global NNPDF2.3 fit is compatible with that of a HERA+ATLAS fit at the
one-sigma level. Also, it was found that when including the ATLAS data
in the global fit they would have little impact, and strangeness would
still be suppressed.

As discussed in Sect.~\ref{sec:dataover}, in NNPDF3.0 we have also
included (both at NLO and NNLO, see Sect.~\ref{sec:exclusion})
the CMS data~\cite{Chatrchyan:2013uja} for $W$+$c$ which have become available, and which
directly constrain the strange distribution.
The CMS $W$+$c$ data have been recently used in a QCD
analysis~\cite{Chatrchyan:2013mza}, together with HERA data, to show
that the strange
PDF $s(x,Q^2)$ from collider-only data can be determined with a precision
comparable to that of global fits which include neutrino data.
The CMS data favors a suppressed strangeness, consistent with the indications
from the neutrino data.
ATLAS $W$+$c$ data (not  included in NNPDF3.0 because
they are only available at the hadron level)
seem instead to favor a less suppressed  strangeness~\cite{Aad:2014xca}.
Fits including LHC $W,Z$ and $W$+$c$ data along  with fixed target
deep-inelastic scattering and Drell-Yan data have also been studied in
Ref.~\cite{Alekhin:2014sya}, with the conclusion that
a good fit to all these datasets can be obtained and finding
again a suppressed strangeness.

We now study this issue in light of the NNPDF3.0 global PDF
determination, by constructing PDF sets based on data which include or
exclude in turn various pieces of experimental information which are
sensitive to strangeness.
Specifically we have produced PDF sets based on the following
modifications of the NNPDF3.0 default dataset of Tables~\ref{tab:completedataset}-\ref{tab:completedataset2}:
\begin{itemize}
\item all neutrino data  (CHORUS and NuTeV) removed, CMS $W$+$c$ data included;
\item all neutrino data  (CHORUS and NuTeV) included, CMS $W$+$c$ data removed;
\item both neutrino data  (CHORUS and NuTeV) and CMS $W$+$c$ data removed.
\end{itemize}

\begin{figure}[t]
\begin{center}
\epsfig{width=0.48\textwidth,figure=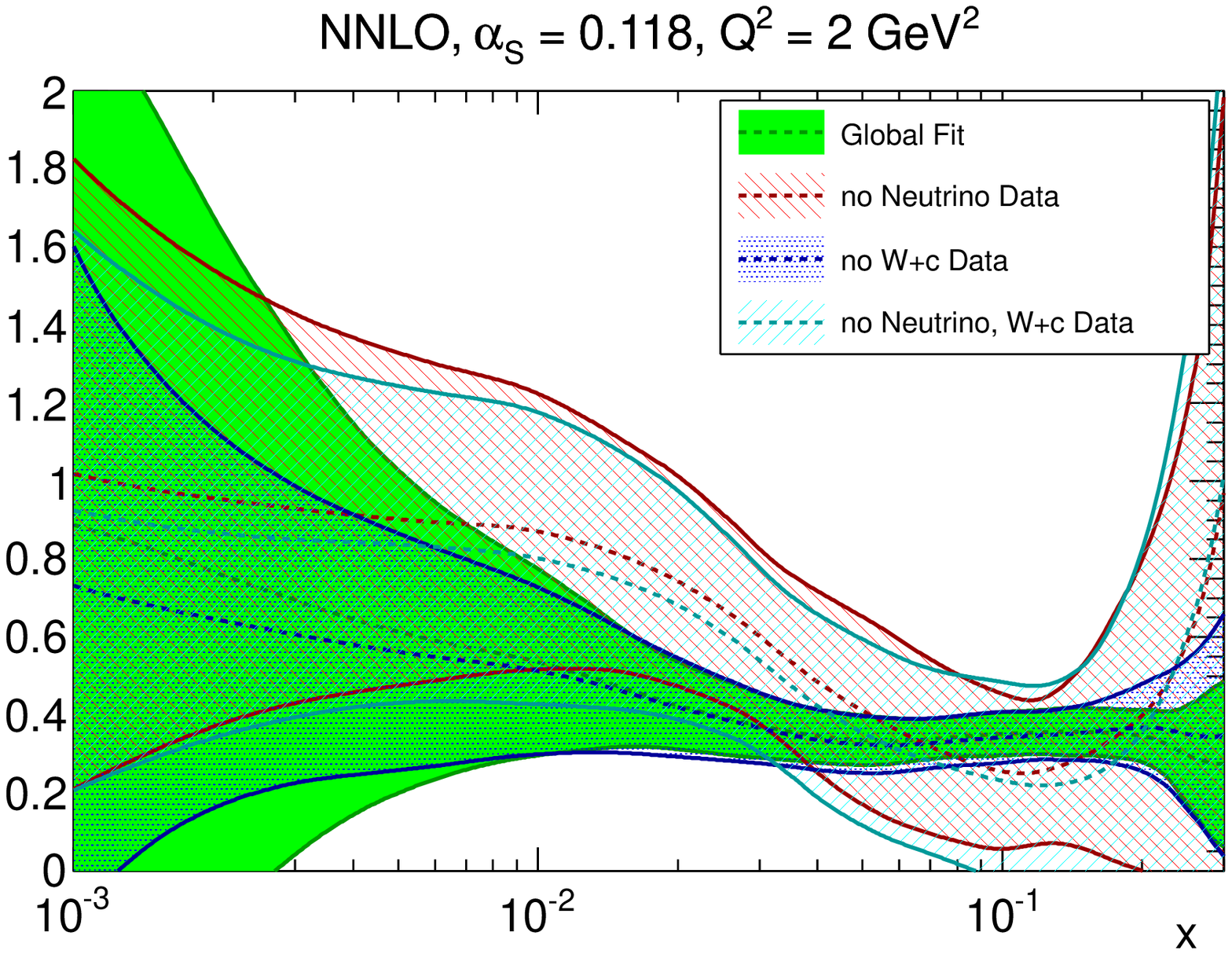}
\epsfig{width=0.48\textwidth,figure=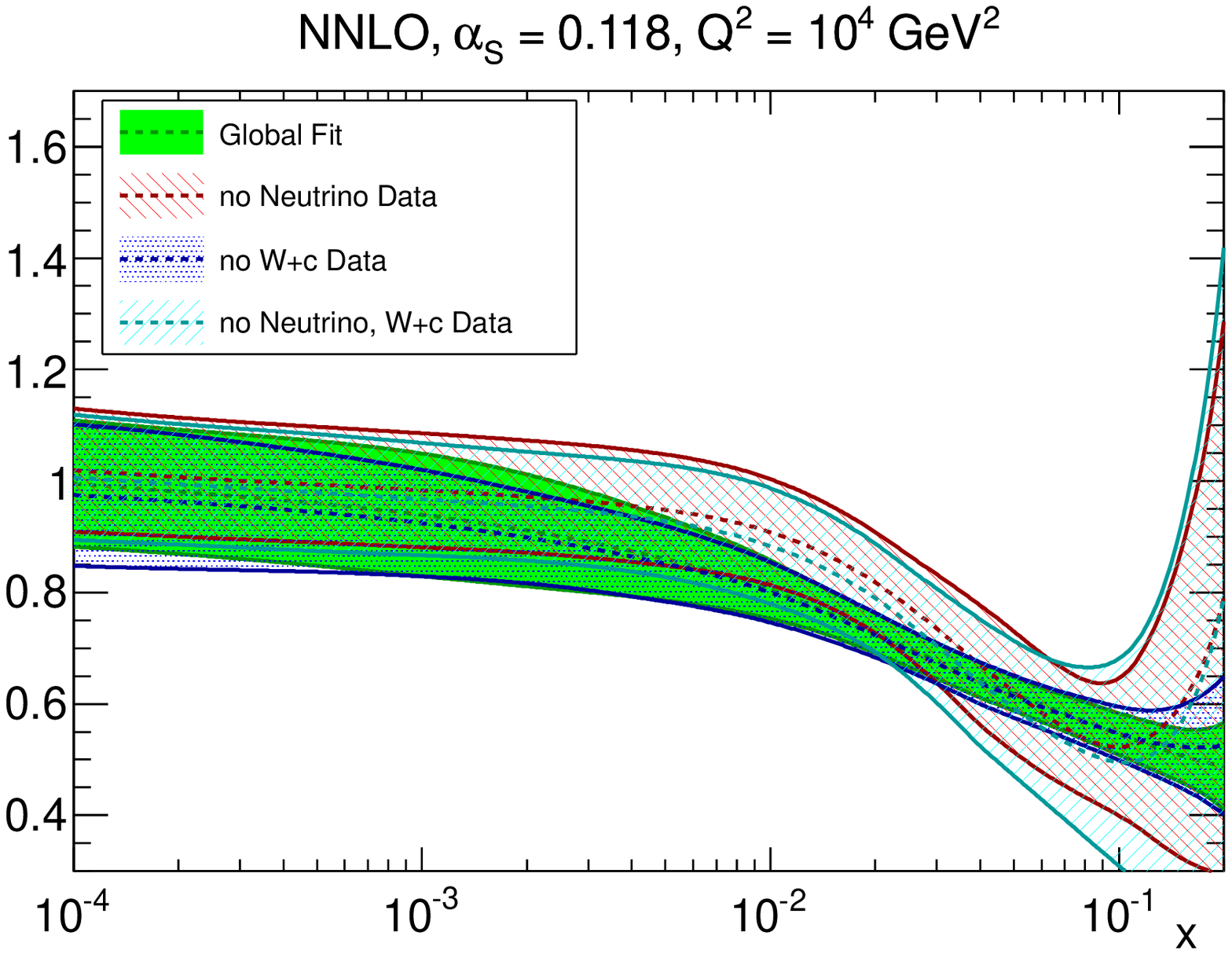}
\caption{\small The strangeness ratio $r_s$ Eq.~(\ref{eq:rs}),
at NNLO sets  with $\alpha_s(M_Z)=0.118$ plotted vs. $x$
at  $Q^2=2$ GeV$^2$ (left) and
$Q^2=10^4$ GeV$^2$ (right)
 for the default
NNPDF3.0 PDF set compared to set obtained excluding from the fitted
dataset
either neutrino data, or  $W$+$c$, or both neutrino and $W$+$c$ data. and a fit with
neither of the two datasets.
 \label{fig:strangeness}
}
\end{center}
\end{figure}

We now compare results obtained in each case, specifically for the
strangeness fraction $r_s$, defined as
\be
\label{eq:rs}
r_s (x,Q^2)=\frac{ s(x,Q^2)+\bar{s}(x,Q^2) }{  \bar{d}(x,Q^2) +
\bar{u}(x,Q^2) } \, .
\ee
In Fig.~\ref{fig:strangeness} $r_s$ is shown for the default NNPDF3.0
fit and the three above fits, plotted
as a function of $x$, for
 $Q^2=2$ GeV$^2$  and
$Q^2=10^4$ GeV$^2$.
%

First, we observe the remarkable compatibility of the
various fits (with, as usual, smaller uncertainty at a higher scale due
to asymptotic freedom): for all fits and all $x$ the one-sigma
PDF uncertainty bands overlap.
The global fit is always the most accurate, though at very low $x\lsim
10^{-3}$ and high scale the uncertainty is of similar size in all
fits  (in fact, it is even somewhat smaller in the fit without
$W$+$c$ than in the global fit, though this is a statistical
fluctuation due to the large uncertainty on the uncertainty in this
region).
While when removing neutrino data the uncertainty blows up,
removing the  $W$+$c$ has very little impact, leading to a moderate
uncertainty reduction for $x \ge 0.05$ when added to the fit without
neutrino data. The fits without neutrino data do lead to a slightly
higher central value of $r_s$ in the region of $x\sim0.01$, but the
 effect is not significant  on the scale of the uncertainty on
 these fits, though it would be significant on the scale of the
 uncertainty of the global fit. It must therefore be considered a
 statistical fluctuation due to the large uncertainty in the fit
 without neutrino data.

The $\chi^2$ for the relevant experiments in these various fits are
collected
in Tab.~\ref{tab:strangeness}, thereby allowing us to compare how
well each experiment is described when included or excluded from the
fit. We see that $W$+$c$ data are well described regardless of whether
they are included in the fit or not, while the neutrino data are not
well described unless they are included in the fit. This again
shows that the impact of the $W$+$c$ can only be moderate.

We conclude that the $W$+$c$ data alone are not yet competitive with the
neutrino data in determining strangeness, and that their inclusion
does not modify significantly the assessment of the size of the
nucleon strangeness in previous global fits, for which there is thus
clear evidence for suppression in comparison to light quarks
by a factor of between
two and three at low scales.

\begin{table}
\centering
\begin{tabular}{c|cccc}
\hline
  & \multicolumn{4}{c}{$\chi^2_{\rm exp}$}  \\
\hline
 & Global & No neutrino & No $W$+$c$  & No neutrino/$W$+$c$ \\
\hline
\hline
CHORUS & 1.13  & {\it 3.87 }  & 1.09  &  {\it 3.45 }   \\
NuTeV  & 0.62   & {\it 4.31 }  & 0.66  &   {\it 6.45}  \\
\hline
ATLAS $W,Z$ 2010  & 1.21   & 1.05   &  1.24   &  1.08   \\
CMS $W$+$c$ 2011    &  0.86  & 0.50   &  {\it 0.90 }  & {\it 0.61 }    \\
\hline
\end{tabular}
\caption{\small Values of the $\chi^2$ (experimental definition, see
Sect.~\ref{sec:quality})  to different data
with sensitivity to strangeness, using as input PDFs obtained from
fits in which these data are included or excluded in turn; values in italics
denotes cases in which the given data was not included in the fit.
See text for more details.
\label{tab:strangeness}
}
\end{table}
\clearpage

\subsection{Stability}
\label{sec:stability}

We will now check the dependence and
stability of our results upon  our
methodology and its variations.
Some of the issues which we address here were already
studied in Sect.~\ref{sec:robustness} in the context of closure tests,
but others are specific to the fit to actual data.

Specifically, we will study  the impact on the NNPDF3.0 results
of the new minimization and stopping
methodology discussed in Sect.~\ref{sec:minim}
in comparison to that previously used
in NNPDF2.3, the impact of the improved treatment of positivity
discussed in Sect.~\ref{sec:positivity}, the impact of a multiplicative vs. additive
treatment of systematic uncertainties (see Sect.~\ref{sec:chi2definition}),
our improved self-consistent determination of preprocessing ranges presented
in Sect.~\ref{sec:preproc}, and finally the independence of fitting basis (already
tested in closure tests in Sect.~\ref{sec:basis}).

\begin{figure}
\begin{center}
\epsfig{width=0.89\textwidth,figure=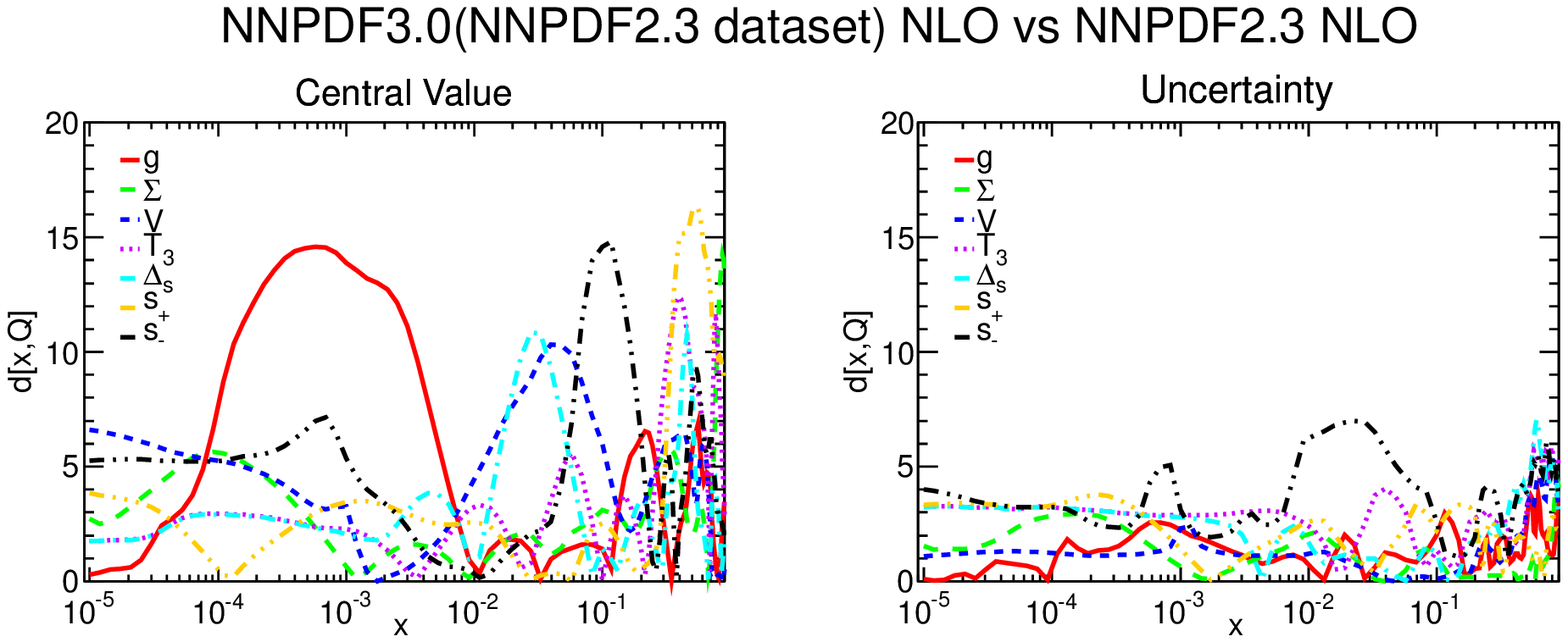}
\epsfig{width=0.89\textwidth,figure=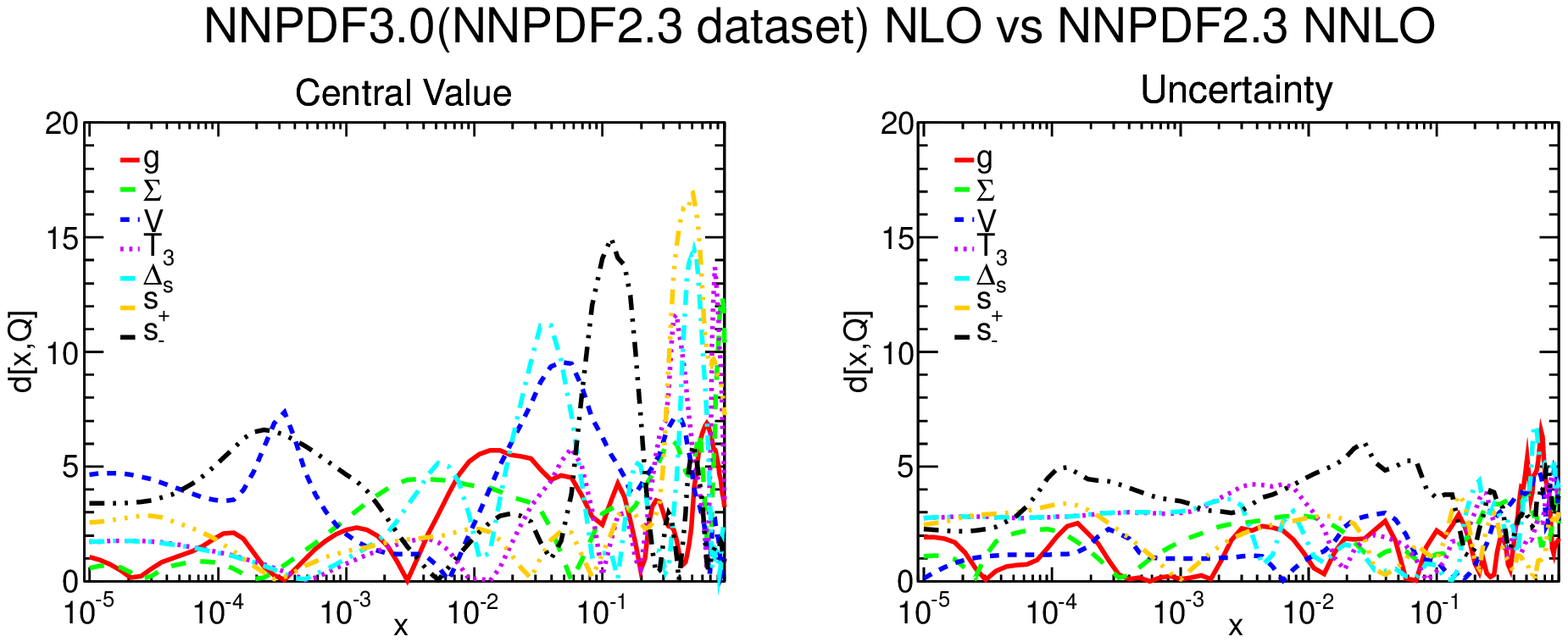}
\caption{\small
Same as Fig.~\ref{fig:distances_global_vs_23dataset}, but now
comparing the PDFs obtained from an NNPDF2.3-like dataset with
NNPDF3.0 methodology and theory to the published~\cite{Ball:2012cx} NNPDF2.3 sets at
NLO(top) and NNLO (bottom).
 \label{fig:distances_nnpdf30dat23}}
\end{center}
\end{figure}

\subsubsection{Impact of the NNPDF3.0 methodology}

First, we discuss the impact of the NNPDF3.0
methodology, by comparing the NNPDF2.3 global fit with the fit
discussed in Sect.~\ref{sec:impdata} and shown in
Fig.~\ref{fig:xpdf-30_vs_23dataset}, based on an NNPDF2.3-like
dataset, but with NNPDF3.0 methodology and theory settings
(specifically the use of FONLL-B), i.e. by
varying the methodology with fixed dataset.
This is the complementary comparison to that which we performed in
Sect.~\ref{sec:impdata}, where instead we varied the dataset with
fixed methodology.
With this comparison we can fully  disentangle the effects
of the new experimental data in NNPDF3.0 from that of
the improved fitting methodology and the new theoretical settings.

\begin{figure}
\begin{center}
\epsfig{width=0.42\textwidth,figure=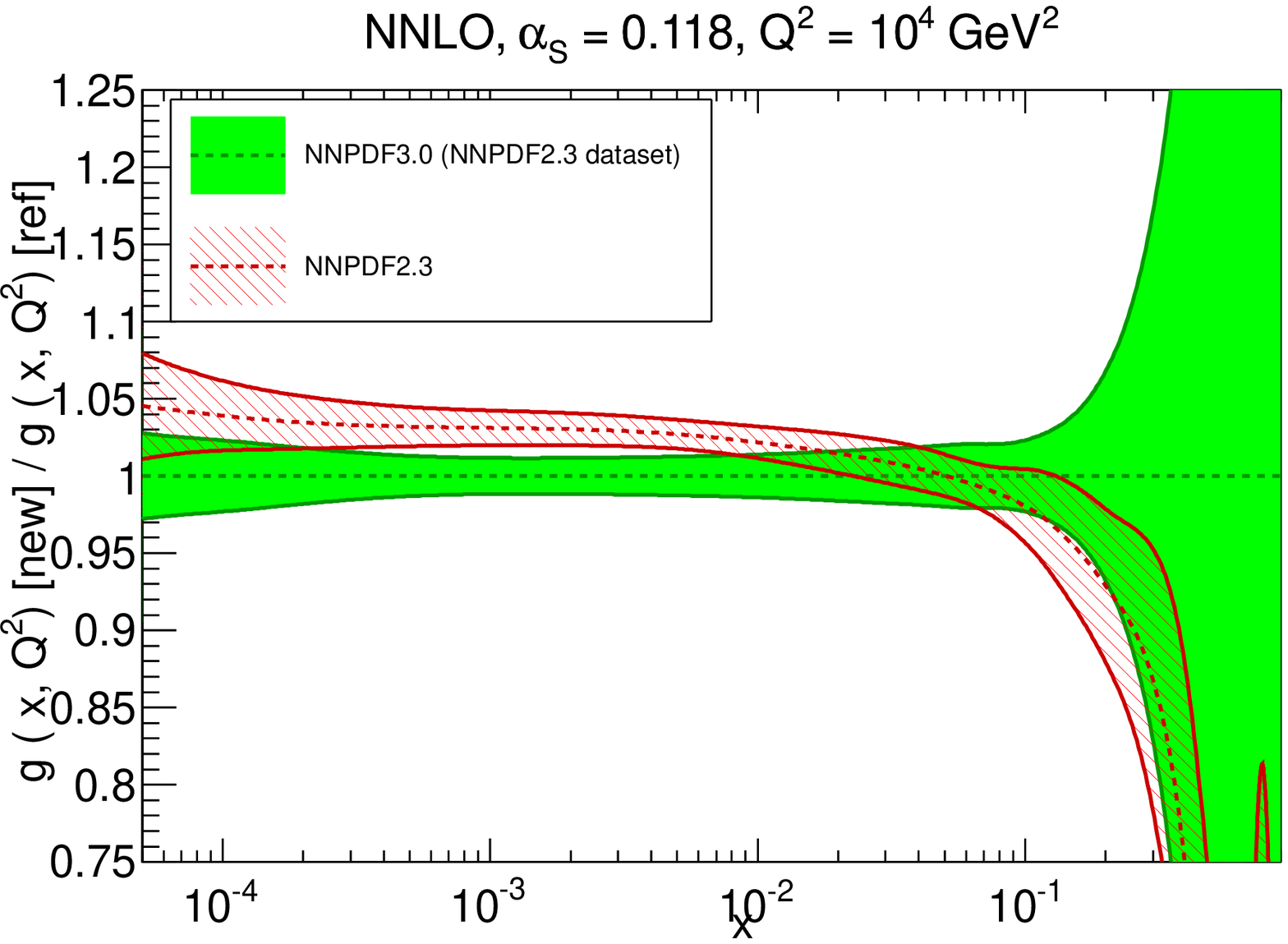}
\epsfig{width=0.42\textwidth,figure=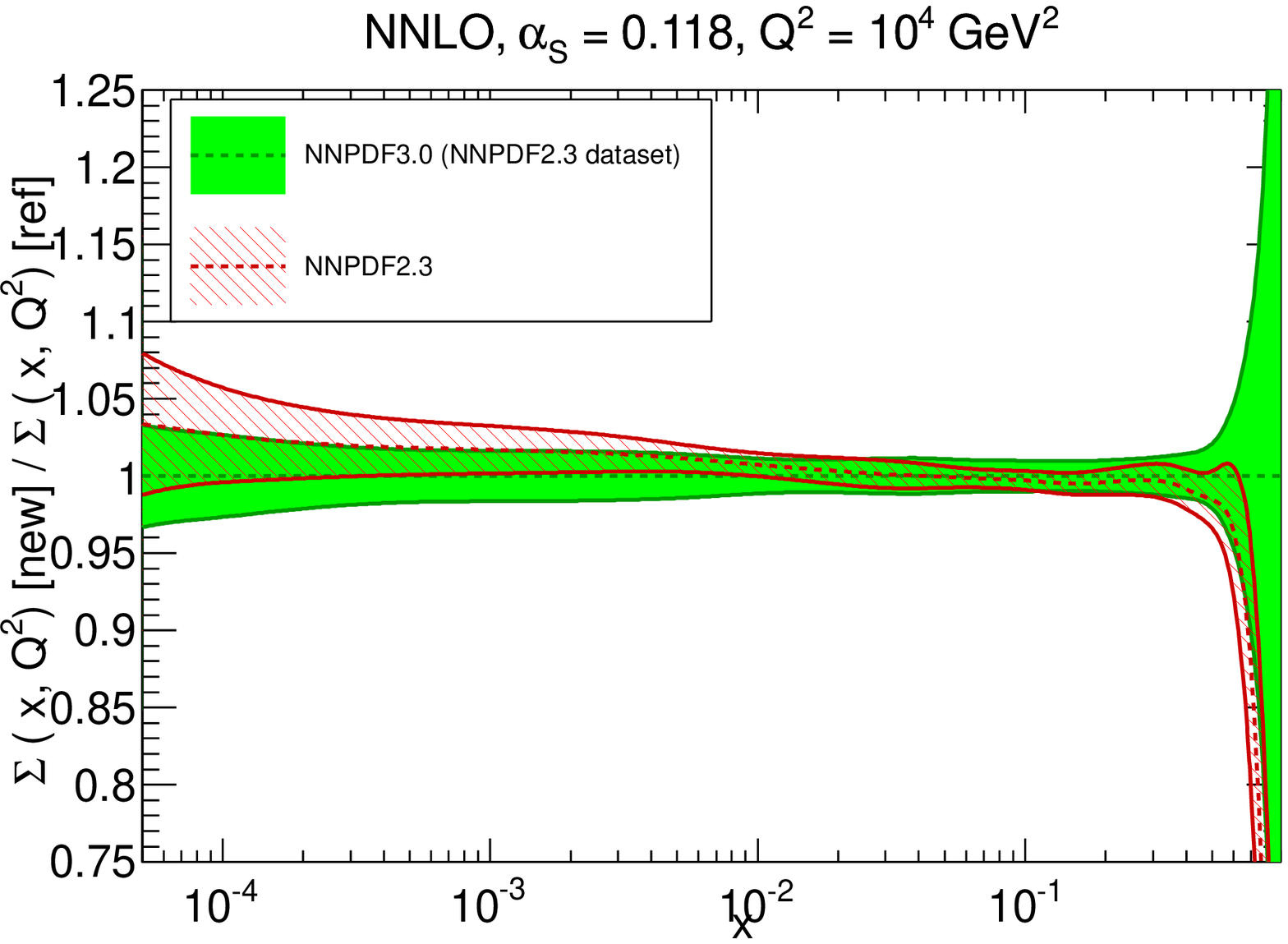}
\epsfig{width=0.42\textwidth,figure=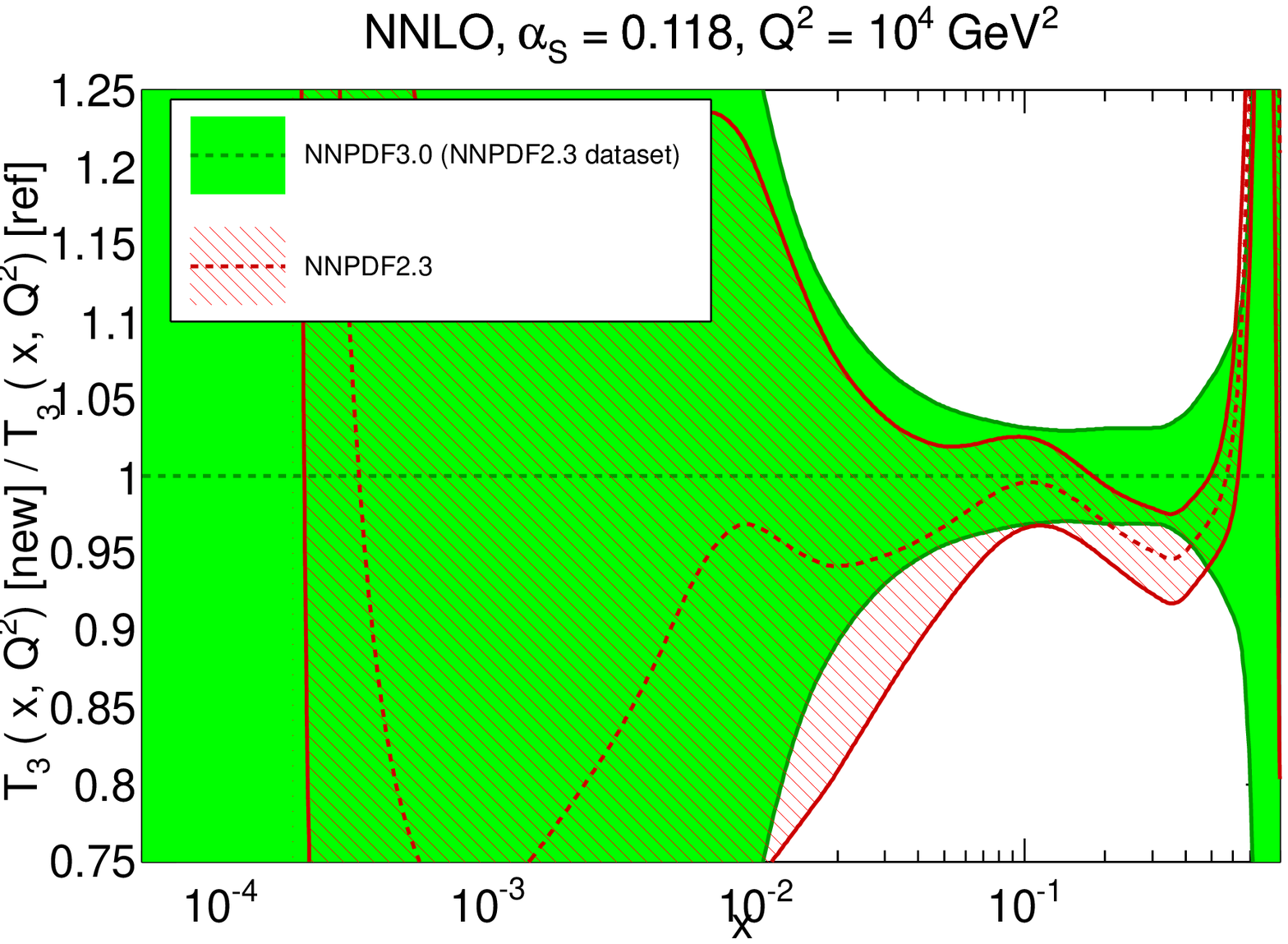}
\epsfig{width=0.42\textwidth,figure=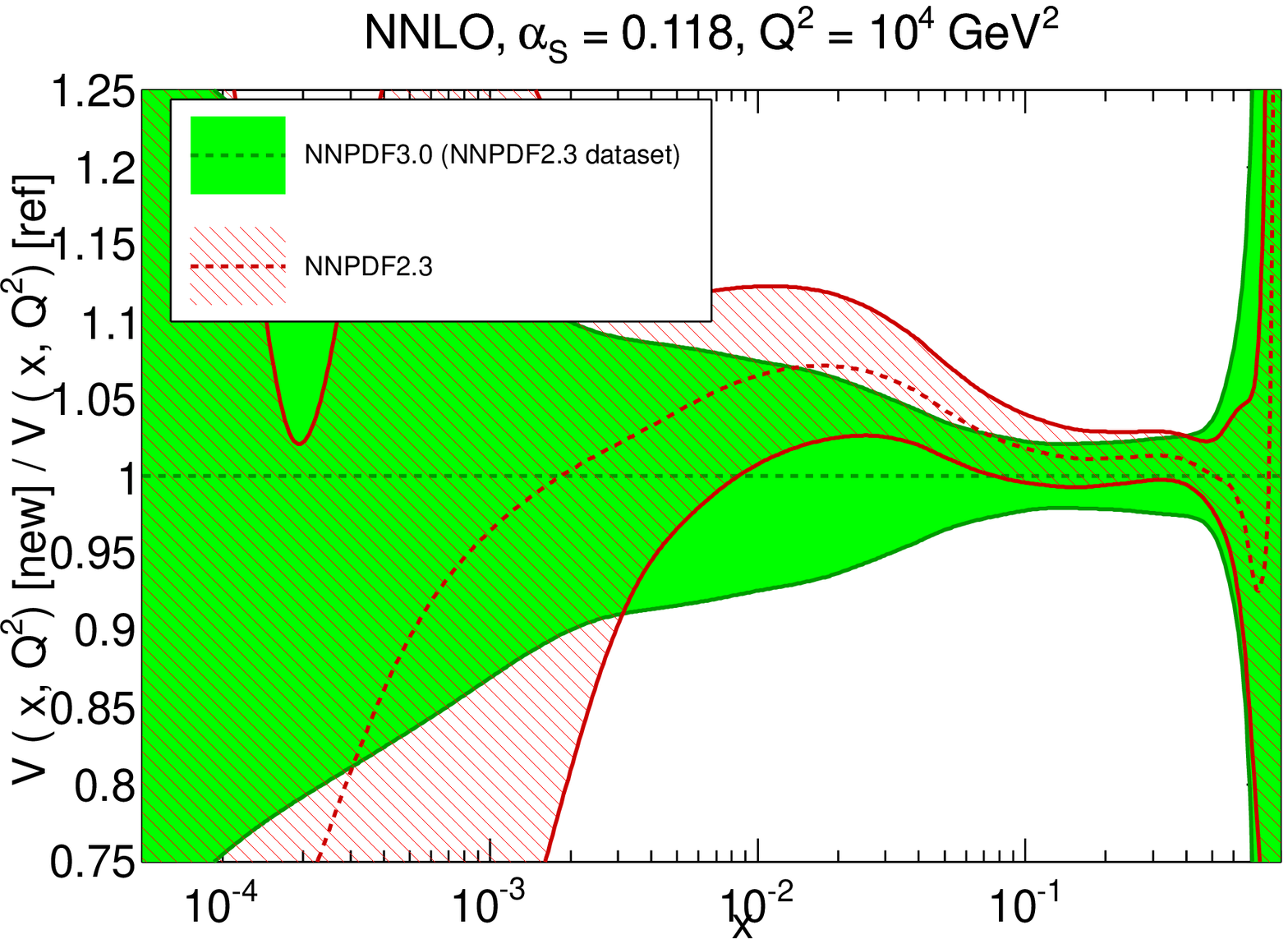}
\caption{\small \label{fig:pdfs_30dat23}
Same as Fig.~\ref{fig:xpdf-30_vs_23dataset}, but now
comparing the PDFs obtained from an NNPDF2.3-like dataset with
NNPDF3.0 methodology and theory to the published~\cite{Ball:2012cx} NNPDF2.3 sets at
NLO(top) and NNLO (bottom).}
\end{center}
\end{figure}

The distances between the original
NNPDF2.3 PDFs of Ref.~\cite{Ball:2012cx} and the NNPDF3.0 fit with
NNPDF2.3 data are shown in
Fig.~\ref{fig:distances_nnpdf30dat23} both at NLO and NNLO, while the
NNLO PDFs are compared
in Fig.~\ref{fig:pdfs_30dat23}.

In the NLO fit the new methodology and theory  settings have an impact
on the small-$x$ gluon and large-$x$ quarks at the one and a half sigma level.
The differences in the gluon can be understood  as a consequence of
having switched from the FONLL-A heavy quark scheme used in NNPDF2.3 to
the more accurate FONLL-B adopted in NNPDF3.0, while the differences seen for
quarks are necessarily a consequence of the more efficient methodology
and extended
positivity constraints (see Sect.~\ref{sec:positivityresults} below).
At NNLO the non-insignificant differences seen in all PDFs reflect the
improved methodology and positivity, as the NNLO theory used in 2.3 and 3.0 is the same.
At high scale the most noticeable difference is the softening of the
small-$x$ gluon seen in  Fig.~\ref{fig:pdfs_30dat23}.

The main conclusion of this comparison is that a significant part of
the differences between the final NNPDF2.3 and NNPDF3.0 sets, as seen
specifically at high scale in Fig.~\ref{fig:30_vs_23_highscale} and at
low scale   in Fig.~\ref{fig:30_vs_23_lowscale}, are due to the
improved methodology (minimization and generalized positivity). This
is consistent with the conclusion of Sect.~\ref{sec:impdata} (see in particular
Figs.~\ref{fig:distances_global_vs_23dataset}-\ref{fig:xpdf-30_vs_23dataset}) that the new data added in NNPDF3.0 have
generally a
moderate impact. In fact, we may cross-check the impact of the new
methodology by comparing the $\chi^2$ of the fit to the NNPDF2.3-like
dataset discussed in  Sect.~\ref{sec:impdata}, to the $\chi^2$ of the
original NNPDF2.3 fit to the same dataset: up to minor differences in
the dataset, the difference in $\chi^2$ are then due to the
methodology. With NNPDF2.3 PDFs we get  $\chi^2=1.1701$, while with
NNPDF3.0 methodology and an NNPDF2.3-like dataset we get
$\chi^2=1.1536$, which is a small but non-negligible improvement, of
more than 50 units in total $\chi^2$, which must be entirely
attributed to the new methodology.

\subsubsection{Constraints from positivity}
\label{sec:positivityresults}
As explained in Sect.~\ref{sec:positivity}, in NNPDF3.0 we adopt a
more extensive set of positivity constraints: specifically now the
number of constraints is equal to the number of PDFs (in fact, it
exceeds them because the gluon is constrained by two different
pseudo-observables), thereby ensuring not only positivity of the
observables used in PDF fitting, but also of potential new
observables such as cross-sections for new physics processes used in
searches (see Sect.~\ref{sec:highmassBSM} below).

\begin{figure}
\begin{center}
\epsfig{width=0.99\textwidth,figure=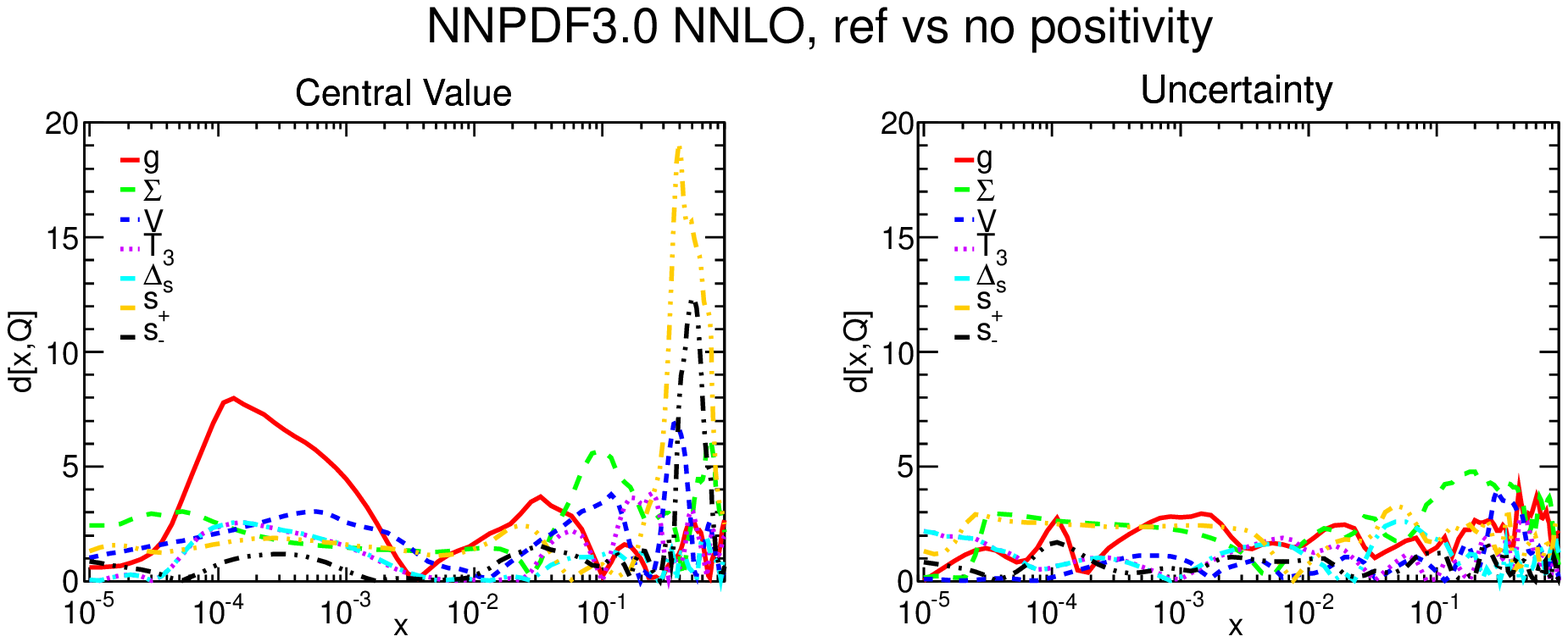}
\caption{\small Same as Fig.~\ref{fig:distances_global_vs_23dataset}, but now
comparing the default set to its counterpart obtained without imposing
positivity.
 \label{fig:distances_nopositivity}}
\end{center}
\end{figure}

\begin{figure}[t]
\begin{center}
\epsfig{width=0.42\textwidth,figure=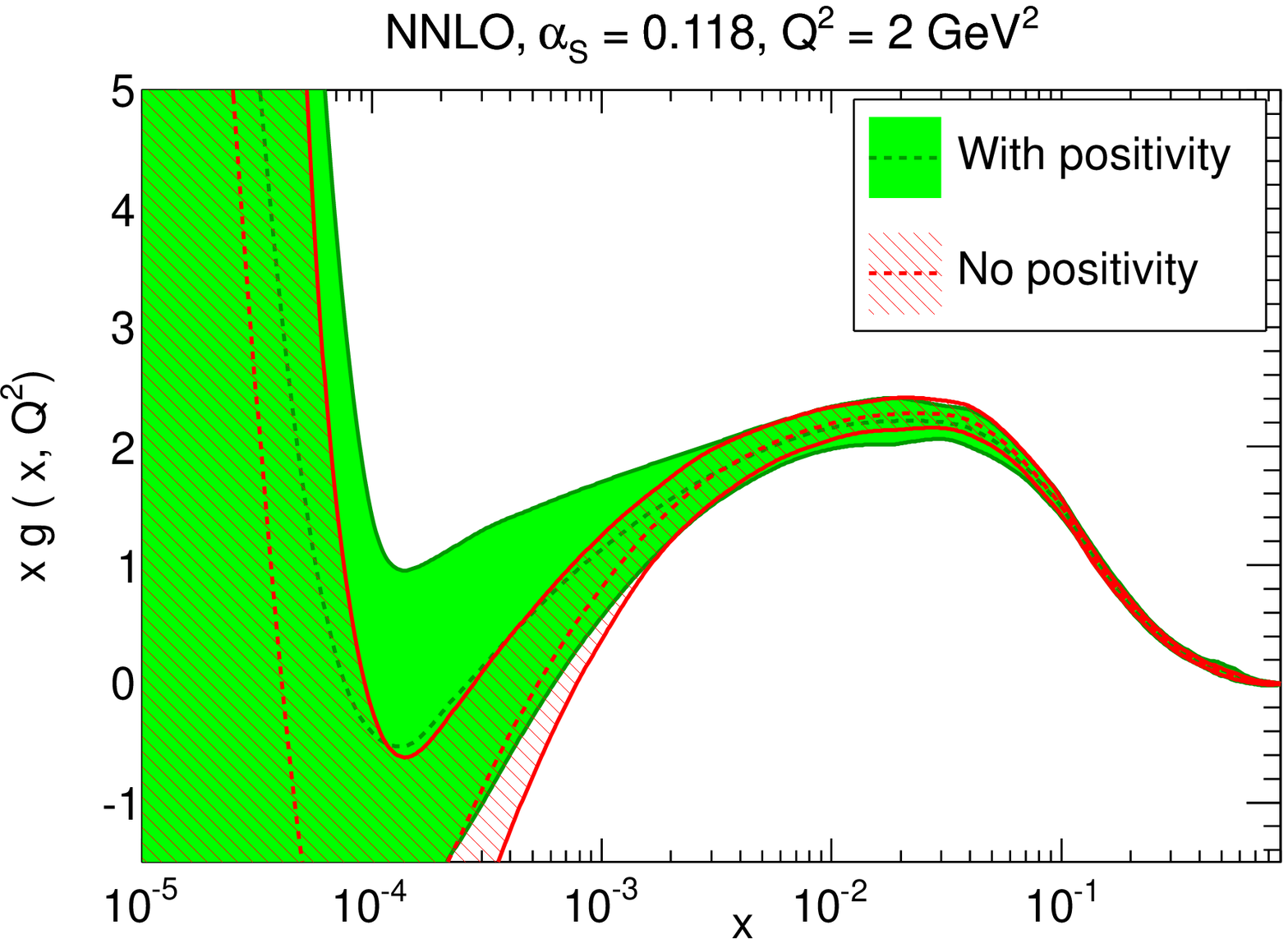}
\epsfig{width=0.42\textwidth,figure=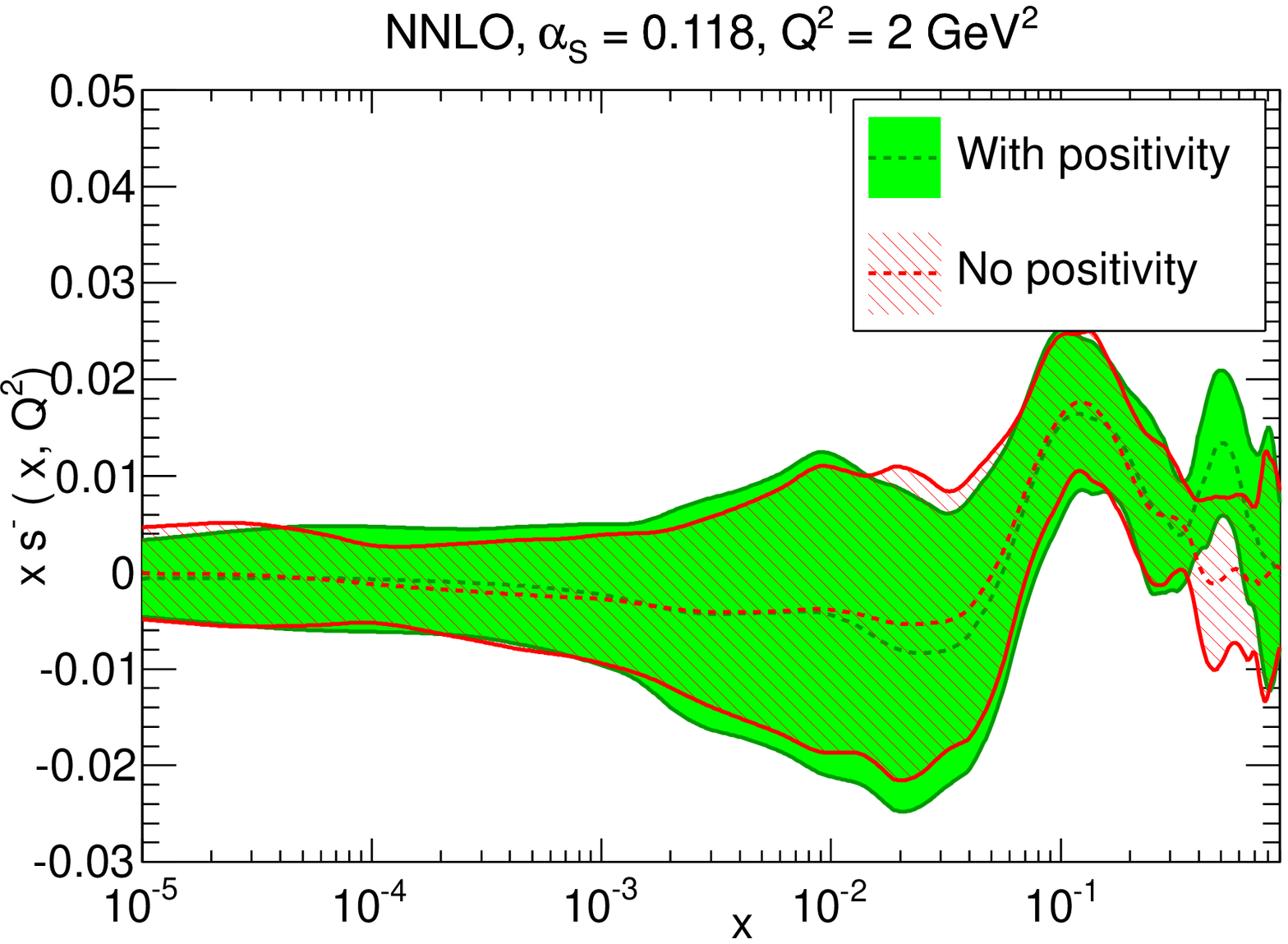}
\caption{\small
Comparison of the default NNPDF3.0 NNLO PDFs at $Q^2=$ 2 GeV$^2$  with
$\alpha_s(M_Z)=0.118$ to their counterpart obtained without imposing
positivity; the gluon (left) and strangeness asymmetry (right) are shown.
\label{fig:pdfs-nopositivity}}
\end{center}
\end{figure}

In order to quantify the impact of these positivity constraint, we have produced
an unphysical variant of the NNPDF3.0 NNLO in which positivity constraints are
removed (so in particular  physical cross-sections could become  negative).
The distances between the default fit and the fit without positivity are shown
in Fig.~\ref{fig:distances_nopositivity}, while some of the PDFs which
change most are compared in Fig.~\ref{fig:pdfs-nopositivity}.

The impact of positivity is relatively mild except
for the small-$x$ gluon and the large-$x$ strangeness, for which there
is little  direct experimental information. For all other PDFs and
$x$ ranges the impact of positivity is below the one sigma level.
Note that even so, strict positivity is necessary if one wishes to
obtain meaningful predictions, e.g. in new physics searches.
For the gluon,  the effects of the positivity
can be noticed already at $x \lsim 10^{-3}$, while at even smaller
$x$ the gluon would become  much more negative if positivity were not
imposed: so the impact of positivity is to make the gluon less
negative at small $x$. For the strangeness asymmetry, interestingly, the dip-bump
structure seen in the global fit is seen to be a consequence of positivity.

%

\begin{figure}[t]
\begin{center}
\epsfig{width=0.42\textwidth,figure=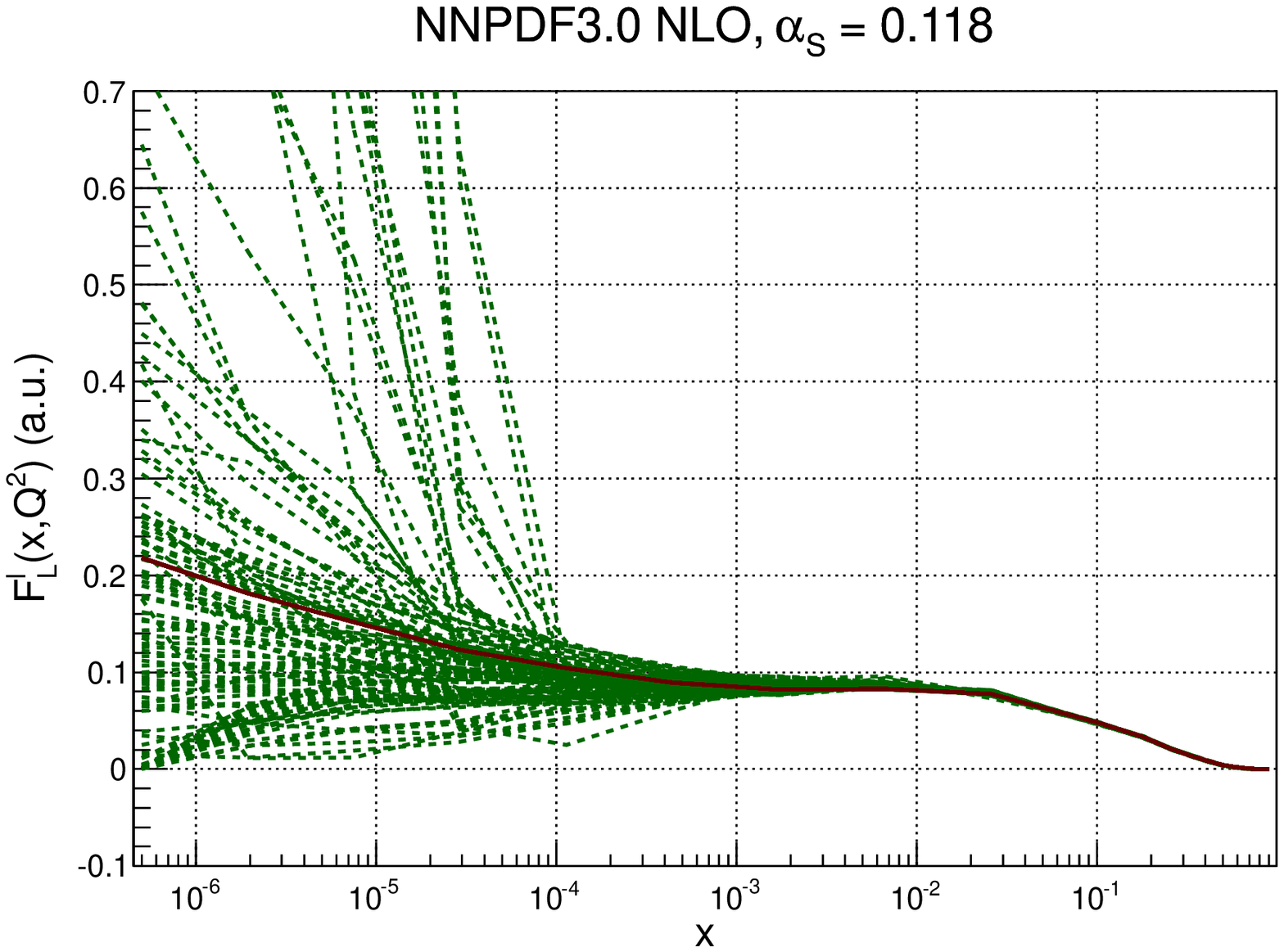}
\epsfig{width=0.42\textwidth,figure=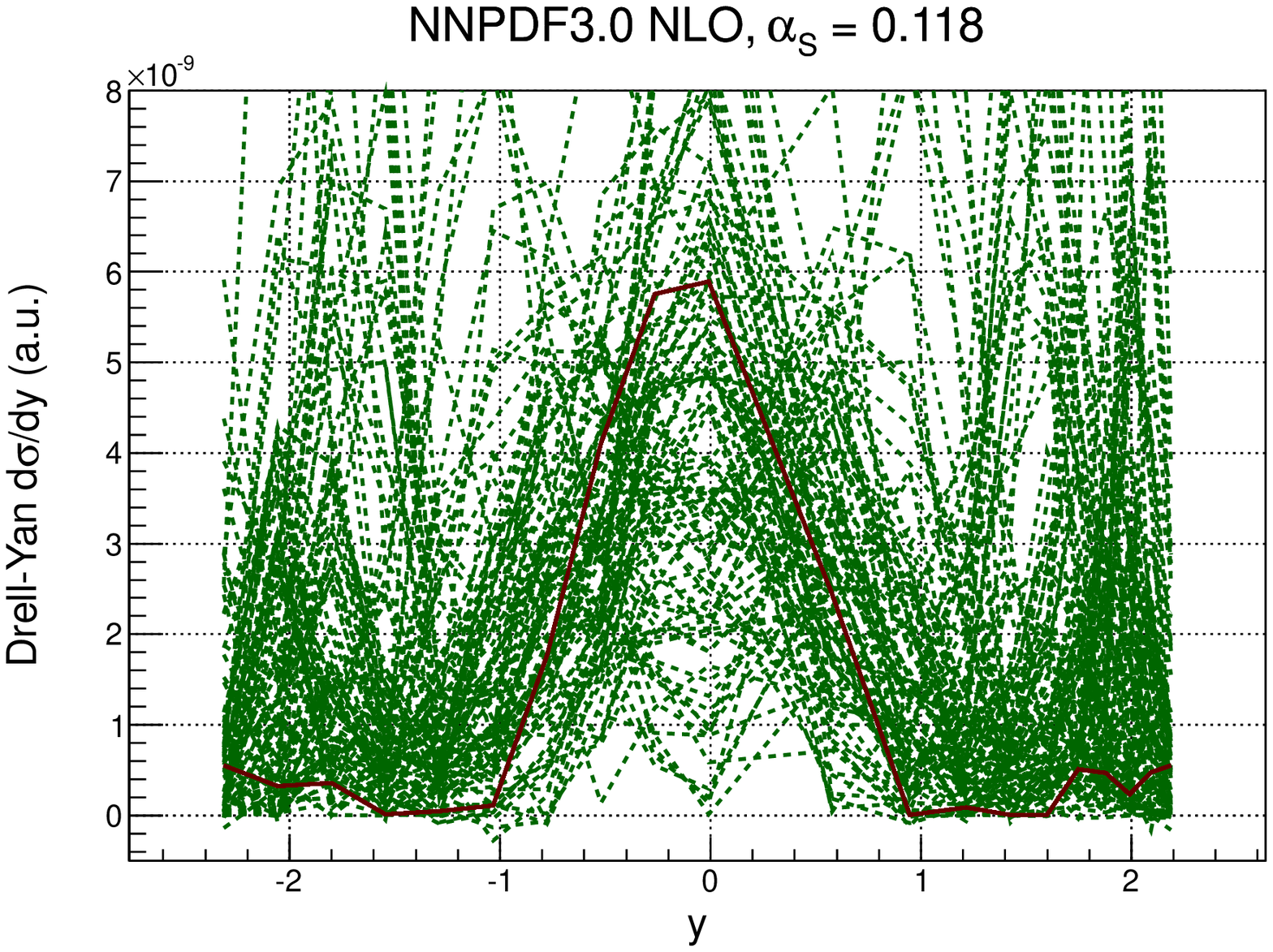}
\caption{\small
The light quark contribution to  $F_L$ (left), and the $s\bar{s}$ Drell-Yan
rapidity distribution (right) plotted in arbitrary units at
$Q^2=$ 5 GeV$^2$ for individual replicas in the NNPDF3.0 NLO set
(dashed green lines). The reference set used in the positivity
implementation  (see  Sect.~\ref{sec:positivity}) is also shown (red line).
\label{fig:obs-nopositivity}}
\end{center}
\end{figure}

As a test of the efficiency of the Lagrange multiplier method we use
to impose positivity,
we have explicitly checked a posteriori that physical cross sections
at NLO and NNLO are indeed not negative (note that because positivity
is always imposed at NLO, as discussed in Sect.~\ref{sec:positivity},
small violations of positivity at NNLO are
in principle possible).
This is illustrated  in Fig.~\ref{fig:obs-nopositivity}, where we plot
two pseudo-observables which are used to impose positivity, namely
the light component of $F_L$, and the $s\bar{s}$ Drell-Yan
rapidity distribution: individual replicas are shown (green dashed
curves) as well as the reference set in the positivity implementation
(see  Sect.~\ref{sec:positivity}), and the effectiveness of positivity
(especially for the Drell-Yan distribution) is clearly seen.

\subsubsection{Additive versus multiplicative systematics}
\label{sec:additive}

As discussed in Sect.~\ref{sec:chi2definition}, there is a certain
latitude in the treatment of correlated systematics, in particular
whether they are treated additively or multiplicatively (only
normalization uncertainties being certainly multiplicative).
In order to test the impact of the additive vs. multiplicative
treatment of systematics,
we have produced two modified version of the NNPDF3.0 fit, which only
differ in the treatment of the systematics: in the first one we treat all
systematics but the normalization as additive, and in the second, all
systematics but the normalization is randomized, and treated as either
additive or multiplicative at random for each replica. The
default treatment (multiplicative or additive) of systematics is given in
Tables~\ref{tab:completedataset} and~\ref{tab:completedataset2}
(fourth column).

\begin{figure}
\begin{center}
\epsfig{width=0.99\textwidth,figure=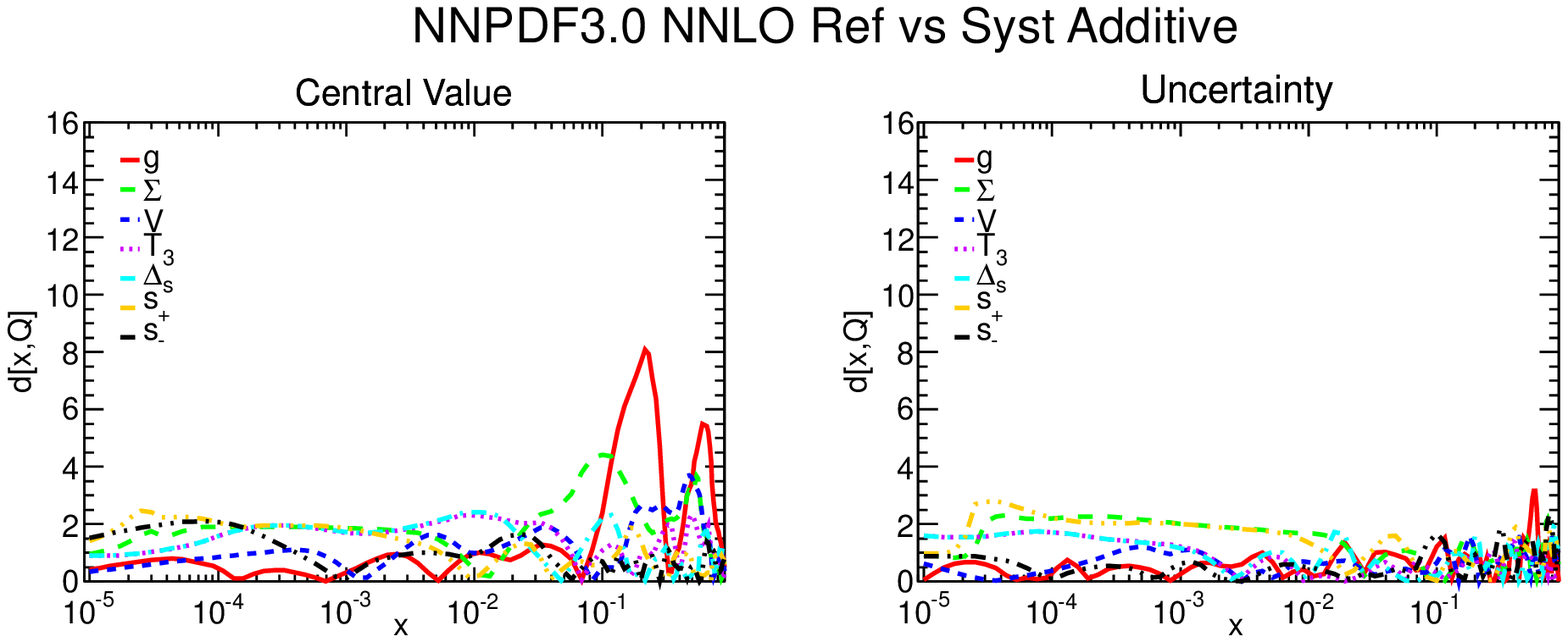}
\epsfig{width=0.99\textwidth,figure=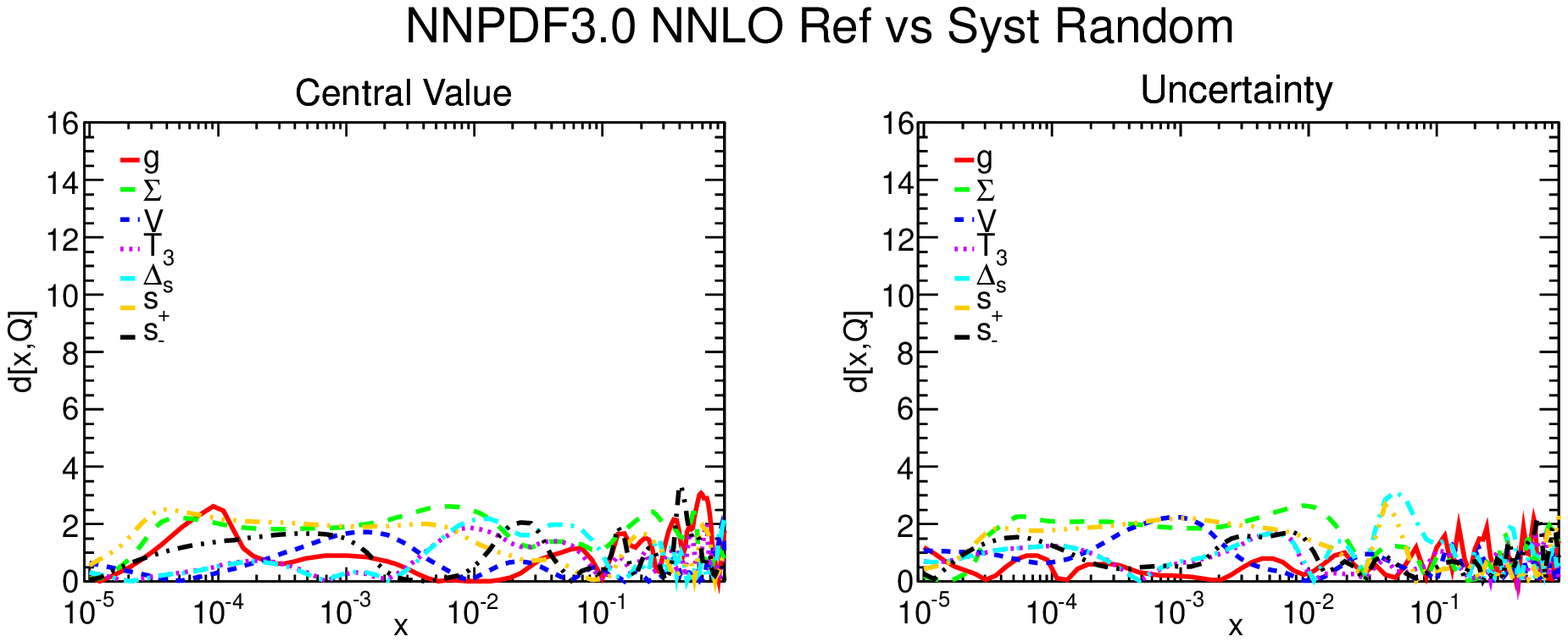}
\caption{\small
Same as Fig.~\ref{fig:distances_global_vs_23dataset}, but now
comparing the default set to its counterpart in which systematics
(except normalization) is treated as additive (top) or in which the
treatment of systematics (except normalization) is randomized (bottom).
 \label{fig:distances_additive}}
\end{center}
\end{figure}

\begin{figure}
\begin{center}
\epsfig{width=0.44\textwidth,figure=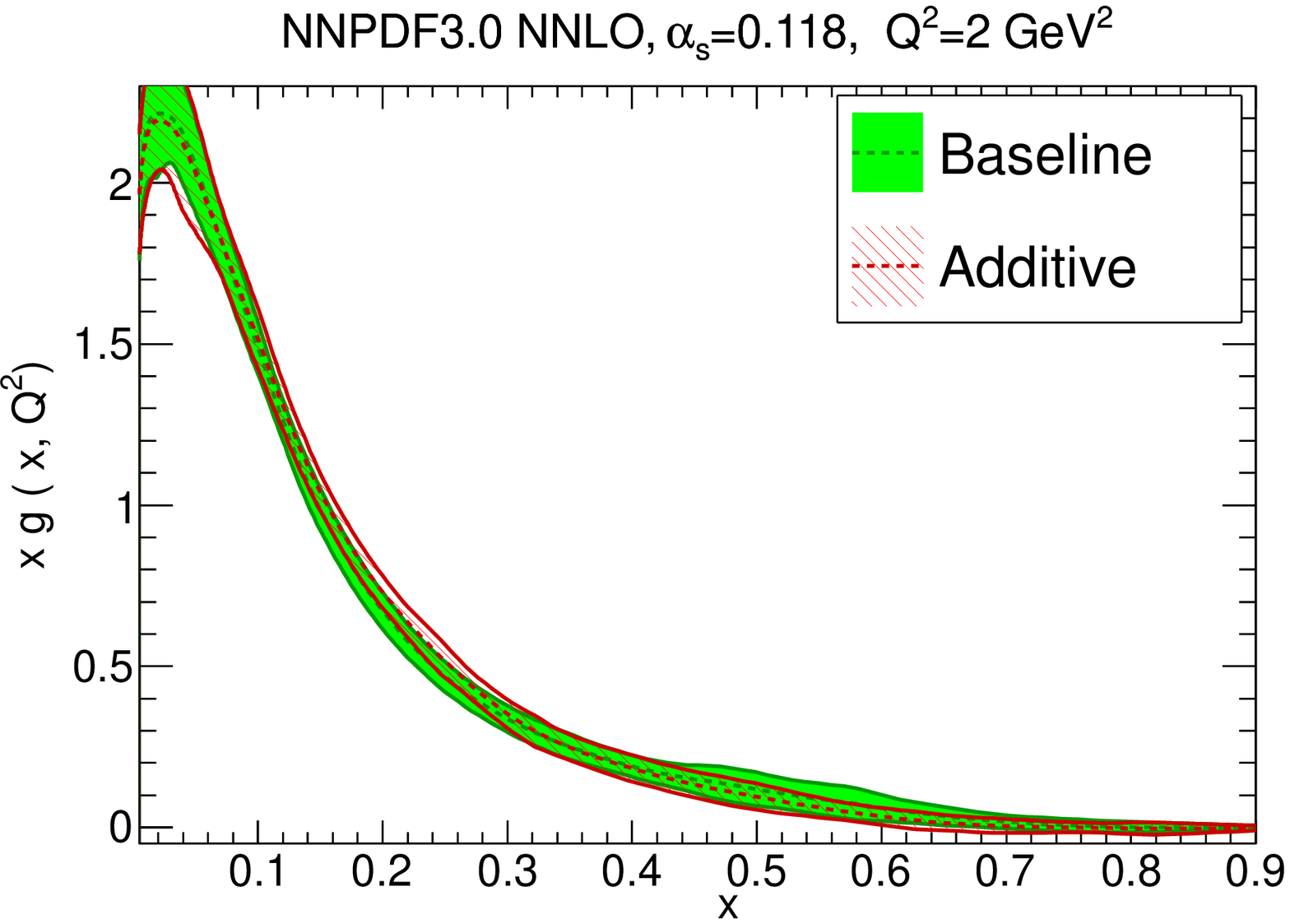}
\epsfig{width=0.44\textwidth,figure=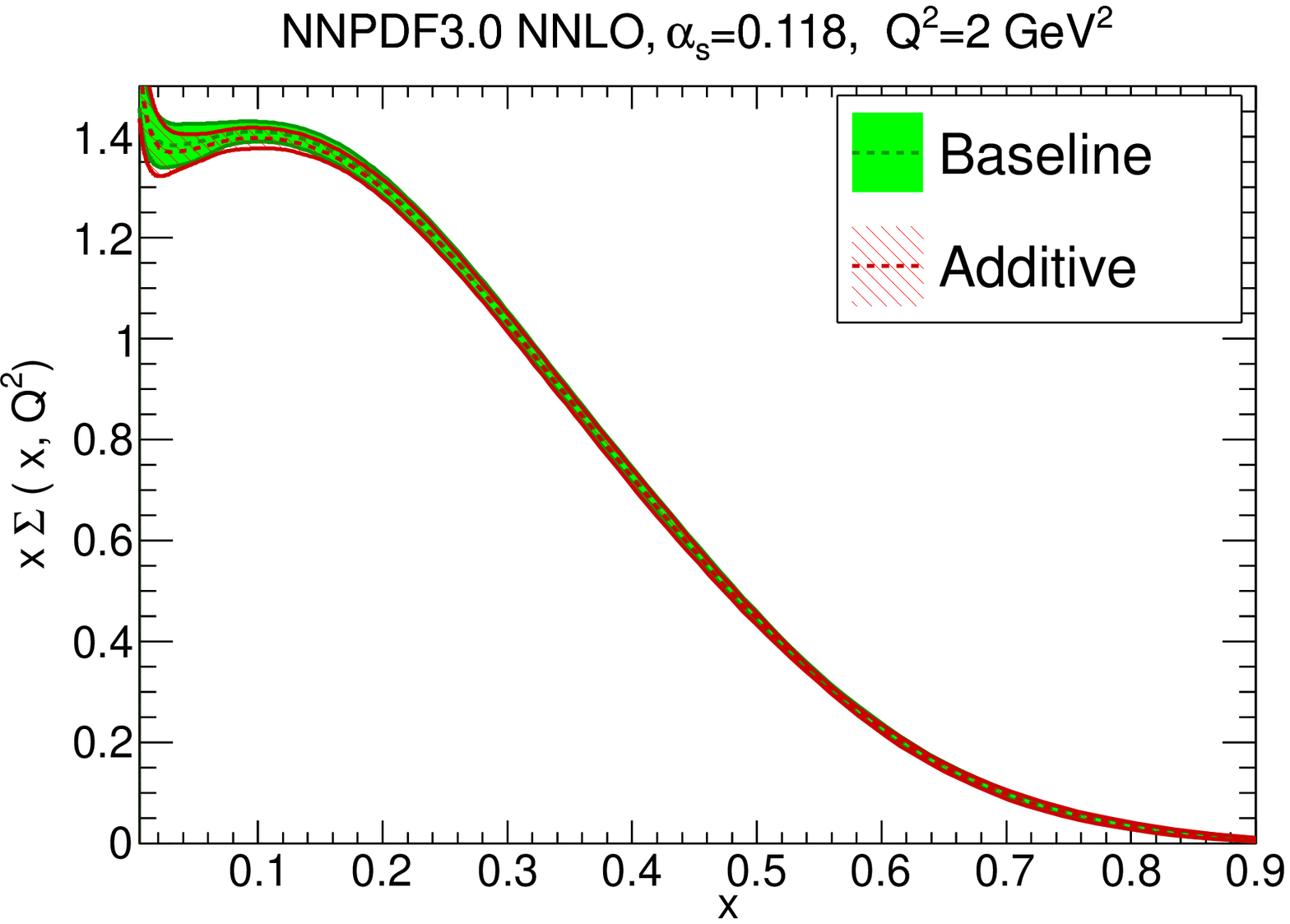}
\caption{\small
Comparison of the NNLO gluon (left) and singlet (right) in a
fit in which all systematic uncertainties
but normalization are treated as additive, to the baseline NNPDF3.0 fit,
where systematic uncertainties are treated as specified in
Tables~\ref{tab:completedataset}--\ref{tab:completedataset2}.
The PDFs are
plotted at $Q^2=2$~GeV$^2$ with a linear scale, since
the difference between baseline and additive is most
important at large-$x$.
\label{fig:pdfs-additive}}
\end{center}
\end{figure}

The distances between these two fits and the default
are shown in
Fig.~\ref{fig:distances_additive}, while two PDFs for which the effect
of the change is most noticeable are shown in
Fig.~\ref{fig:pdfs-additive} for the additive case, compared to the
NNPDF3.0 default.
The impact of the treatment of systematics turns out to be
essentially indistinguishable from statistical fluctuations for all
PDFs except the large-$x$ gluon, for which it can have an effect at
the half-sigma level if all systematics is treated as additive.
This can be understood as a consequence of the
fact that the gluon depends strongly on jet
data, which are  affected by large systematics. The impact on the gluon
is explicitly shown in
Fig.~\ref{fig:pdfs-additive}. The singlet is also shown: in this case,
the change in
uncertainty at small $x$ seen in the plot is actually compatible with
a statistical fluctuations, as the distance plot
Fig.~\ref{fig:distances_additive} demonstrates.
When systematics are randomized the effect is diluted and the changes
are always compatible with statistical fluctuations.

We conclude that the treatment of systematics, while an issue in
principle, in practice affects results at a level which is compatible
with fluctuations: in the absence of information on how to treat the
systematics this would be randomized, but then results are compatible
with the default. Even when all systematics are treated as additive,
which is an extreme case, only the gluon changes significantly: but
then the gluon is mostly affected by HERA and jet data, for both of
which the preferred treatment of systematics is
multiplicative~\cite{glazovpriv,goupriv}.
The default treatment of systematics in the NNPDF3.0 fit thus appears
to be both reliable and robust.

\subsubsection{Independence of preprocessing}
\label{sec:robustnesspreproc}

As discussed in Sect.~\ref{sec:preproc}, in NNPDF3.0 the preprocessing exponents
which characterize the function  that relates PDFs to their neural
network parametrization according to Eq.~(\ref{eq:preproc}) are
determined self-consistently. This method has been already used in
NNPDF polarized PDF fits~\cite{Ball:2013lla,Nocera:2014gqa}.
We now check that indeed the method works, and specifically that the
ranges of the effective exponents Eq.~(\ref{effalpha}-\ref{effbeta}) are
well within the preprocessing ranges, and thus not biased by it.

\begin{figure}[h]
\begin{center}
\epsfig{width=0.42\textwidth,figure=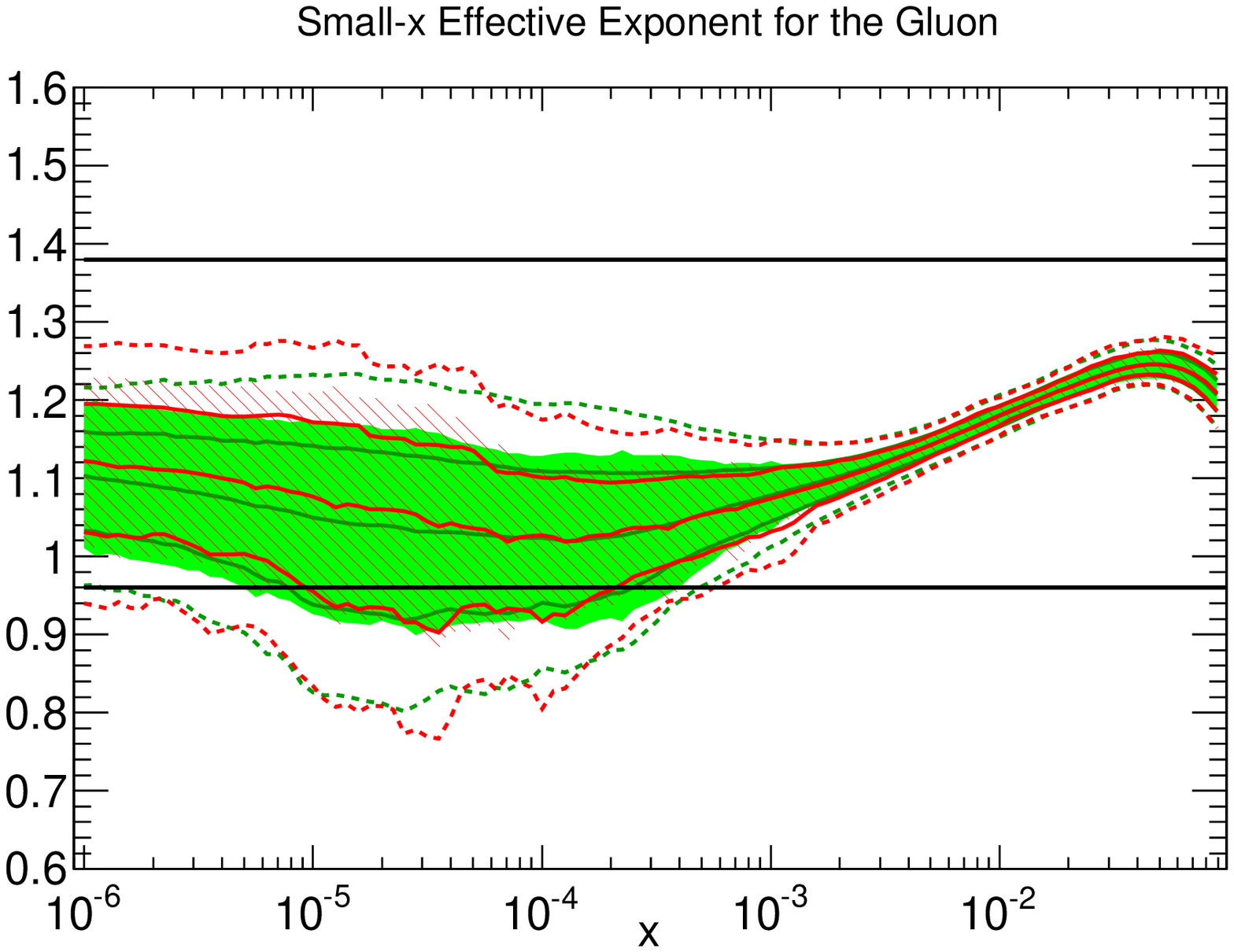}
\epsfig{width=0.42\textwidth,figure=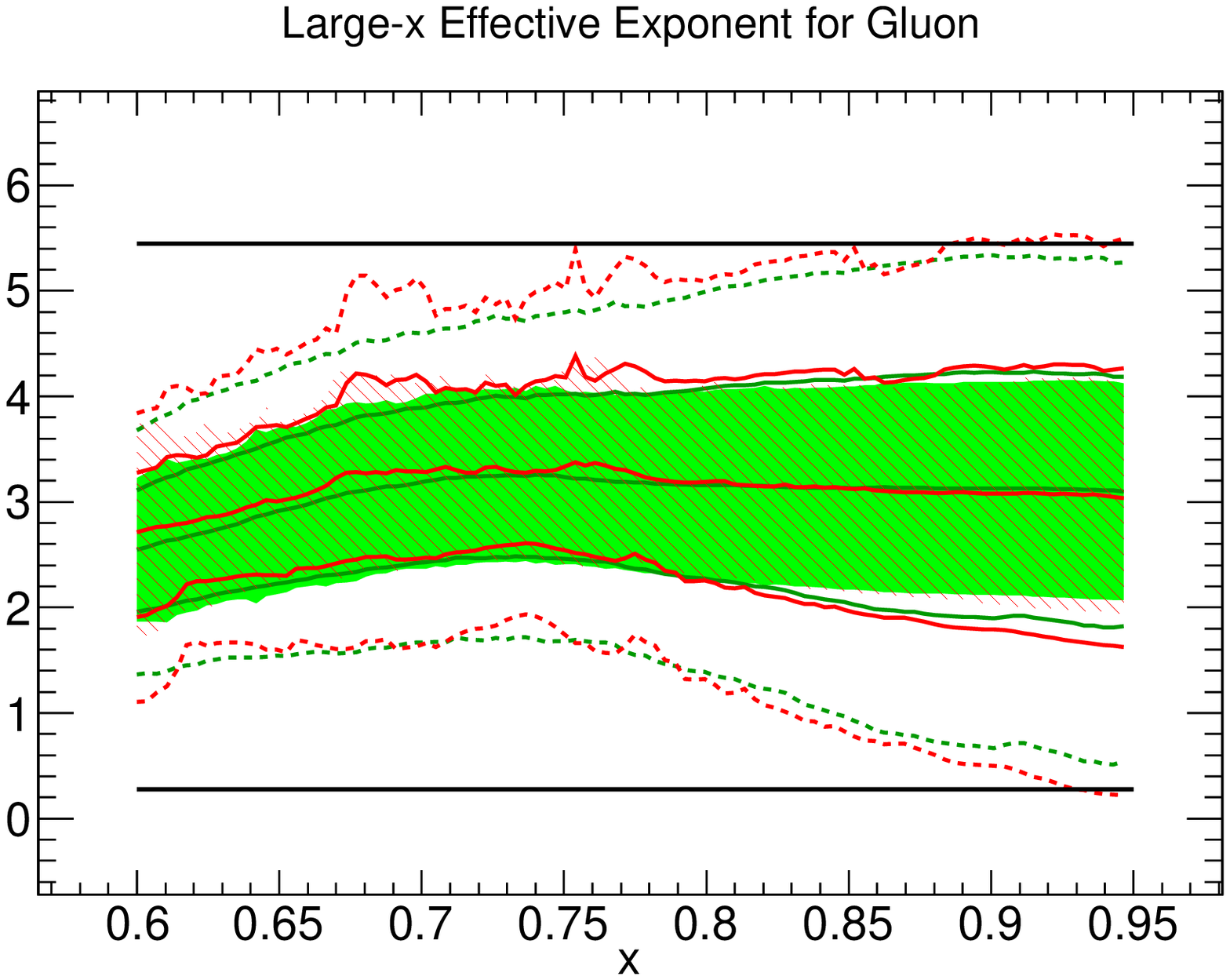}
\epsfig{width=0.42\textwidth,figure=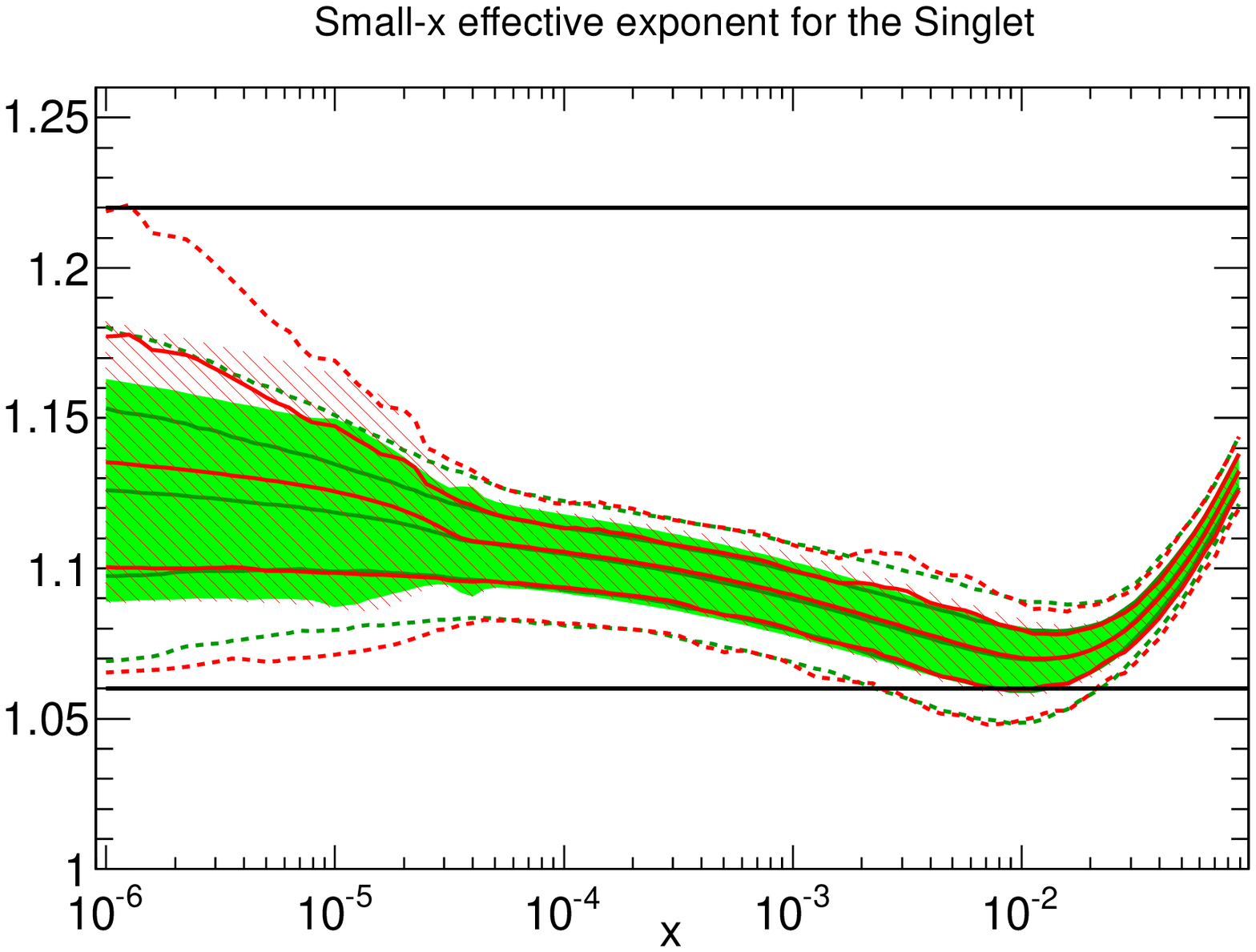}
\epsfig{width=0.42\textwidth,figure=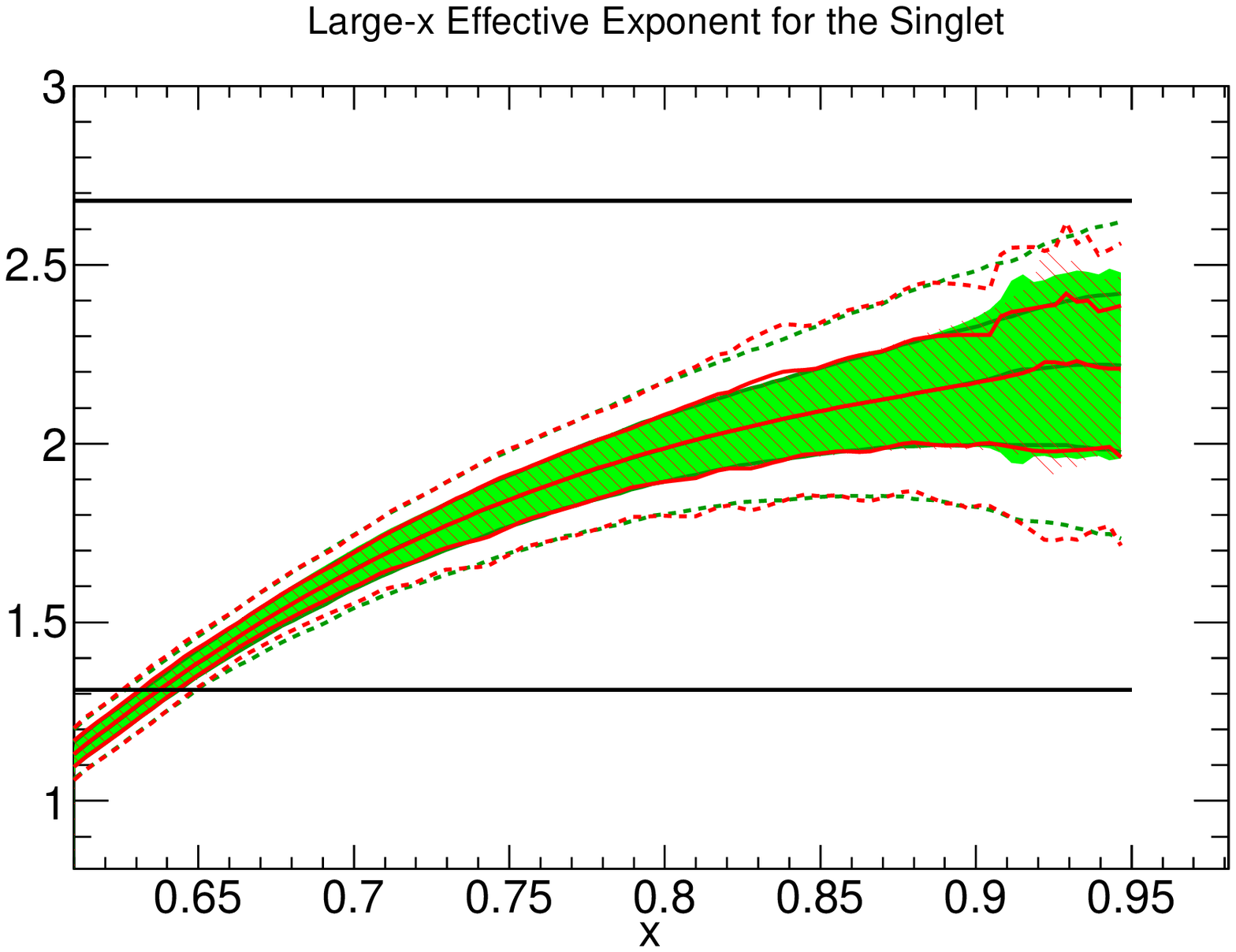}
\caption{\small The small-$x$ and large-$x$ effective
asymptotic exponents, $\alpha_{\rm eff}$, Eq.~(\ref{effalpha})
(left) and
$\beta_{\rm eff}$, Eq. (\ref{effbeta}) (right), for the gluon
 (top) and  singlet (bottom) in the NNPDF3.0 NLO set.
The green solid band is the 68\% confidence level
interval for each effective exponent as a function of $x$, for
the $N_{\rm rep}=1000$ replica fit, and the green dashed line
is twice this interval.
The red hatched band (and the red dashed lines)
are the corresponding results obtained from the
$N_{\rm rep}=100$ replica fit.
In each case, the black solid horizontal lines provide the
corresponding  range for the preprocessing exponents
in the input PDF parametrization Eq.~(\ref{eq:preproc}).
\label{fig:prepexps1}}
\end{center}
\end{figure}

The effective exponents and preprocessing ranges
 for the gluon and singlet in the global NNPDF3.0 NLO fit are shown
in Fig.~\ref{fig:prepexps1}:
the green solid band is the 68\% confidence level
interval for each effective exponent as a function of $x$, for
the $N_{\rm rep}=1000$ replica fit, and the green dashed line
is twice this interval.
The red hatched band (and the red dashed lines)
are the corresponding results obtained from the
$N_{\rm rep}=100$ replica fit.
In each case, the black solid horizontal lines are the extremes of the
preprocessing range, determined from the effective exponents of a
previous fit according to the procedure discussed in Sect.~\ref{sec:preproc}.
The preprocessing range is self-consistent in the sense that if the
range is determined again from the values of the effective exponents
shown in Fig.~\ref{fig:prepexps1} it is essentially unchanged, as it can be seen
from visual inspection of the plot.

It is furthermore clear that indeed the effective range is always well
within the allowed preprocessing range, thereby ensuring that the former is
not influenced or distorted by the latter. We have checked that this
is the case for all PDFs and all PDF sets discussed in this paper. Of
course, when changing substantially the dataset, such as, for example,
when constructing the HERA-only PDFs of Sect.~\ref{sec:resdataset},
the preprocessing ranges may change significantly (with fewer data the
acceptable ranges are generally wider) and thus the procedure has to
be iterated again for convergence.

\subsubsection{Independence of the PDF fitting basis}
\label{sec:bastability}

We finally wish to test for
independence of the fit results of the choice of PDF basis.
This was already tested explicitly in
 Sect.~\ref{sec:closure} in the framework of a closure test, and it is
 now verified for the fit to actual data, which is possibly more
 complicated, both because the final NNPDF3.0 PDFs have sometimes  more
 structure than the MSTW2008 PDFs used in that closure test, and also
 because of potential inconsistencies between data which are
 by construction absent in a closure test. It is worth pointing out
 that this independence is usually not achieved, or at least not
 completely,
by PDF determinations
 based on a standard functional forms, as shown in Ref.~\cite{JimenezDelgado:2012zx}.

\begin{figure}
\begin{center}
\epsfig{width=0.99\textwidth,figure=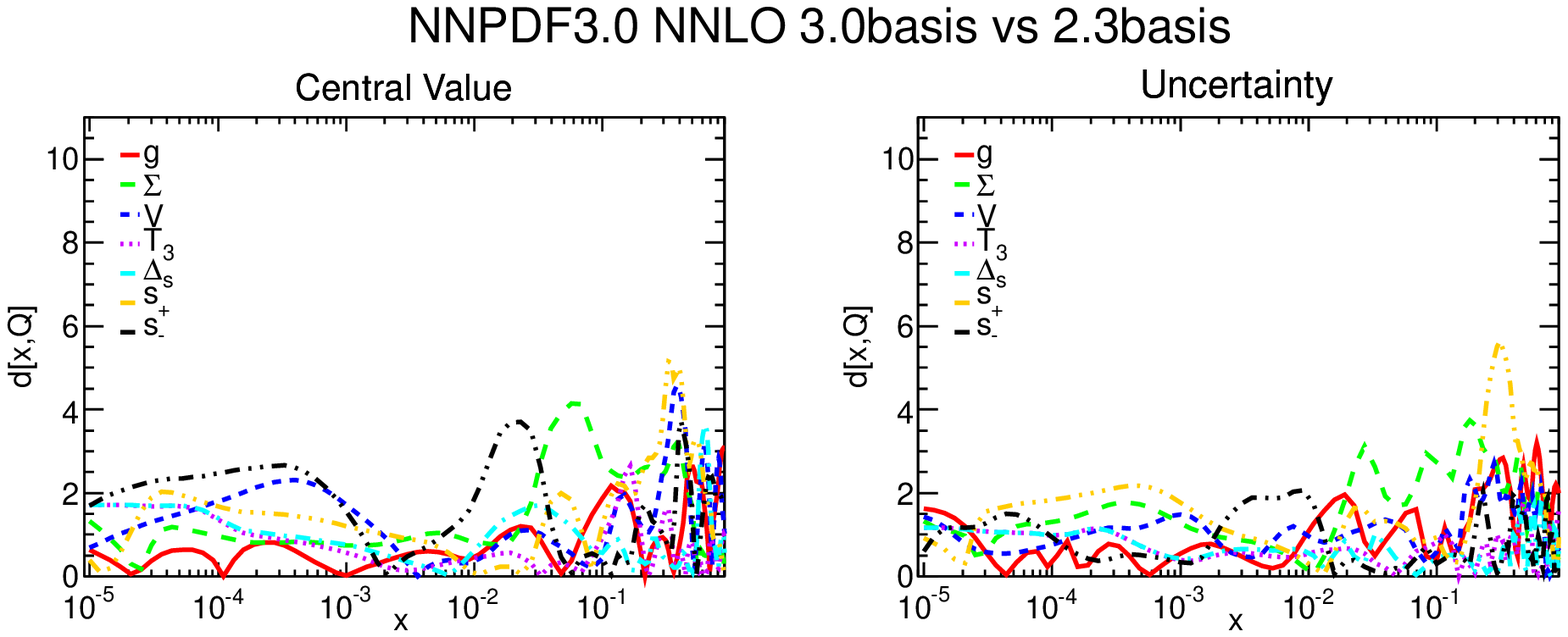}
\caption{\small Same as Fig.~\ref{fig:distances_global_vs_23dataset}, but now
comparing the default NNPDF3.0 fit to one in which the NNPDF2.3 basis
Eq.~(\ref{eq:nnpdf23basis}) instead of the basis
Eq.~(\ref{eq:nnpdf30basis})
has been used for PDF parametrization.
 \label{fig:distances_changebasis}}
\end{center}
\end{figure}

In Fig.~\ref{fig:distances_changebasis} we show the distances between
the default NNPDF3.0 NNLO PDFs and the same fit but using the NNPDF2.3
fitting PDF basis Eq.~(\ref{eq:nnpdf23basis}) instead of the default
NNPDF3.0 basis Eq.~(\ref{eq:nnpdf30basis}) for the parametrization of
input PDFs.
Results are consistent to what was found in the closure test:
distances are mostly compatible with statistical equivalence, with
only strangeness (whose parametrization is affected by the change of
basis) deviating at the half-sigma level at the valence peak,
which is above the threshold of statistical indistinguishability. Note
that the dip-bump structure in $s^-$ seen in
Fig.~\ref{fig:pdfs-nopositivity} (and related to positivity) arises
due to interference of different PDFs in the NNPDF2.3 basis, yet it is
perfectly reproduced, thereby confirming the remarkable stability of
the NNPDF3.0 results.


\clearpage

\subsection{Implications for LHC phenomenology}
\label{sec:phenomenology}

We now turn to a brief preliminary investigation of
the implications of NNPDF3.0 PDFs
for LHC phenomenology.
We start  by comparing  the NNLO PDF luminosities at 13 TeV, both to
NNPDF2.3,  with CT10 and MMHT.
Then we move to  predictions for a variety of LHC cross-sections
at 13 TeV, which we compute  at NLO using the automated
{\sc\small MadGraph5\_aMC@NLO} program~\cite{Alwall:2014hca}, and for
which we compare results obtained using  NNPDF2.3 and NNPDF3.0 PDFs:
specifically, vector boson, top, and
 Higgs production.
Then we turn to the implications of NNPDF3.0 PDFs
for the dominant Higgs production channel at the LHC,
gluon-fusion, and provide NNLO cross-sections
computed with the {\sc\small iHixs} code~\cite{Anastasiou:2011pi}, including a study the
dependence of  results on the
dataset used for PDF determination.
Finally, we study the production of high-mass states,
close to the LHC kinematic
threshold, as relevant for searches of massive New Physics
at the energy frontier.

\subsubsection{PDF luminosities}

\begin{figure}
\begin{center}
\epsfig{width=0.42\textwidth,figure=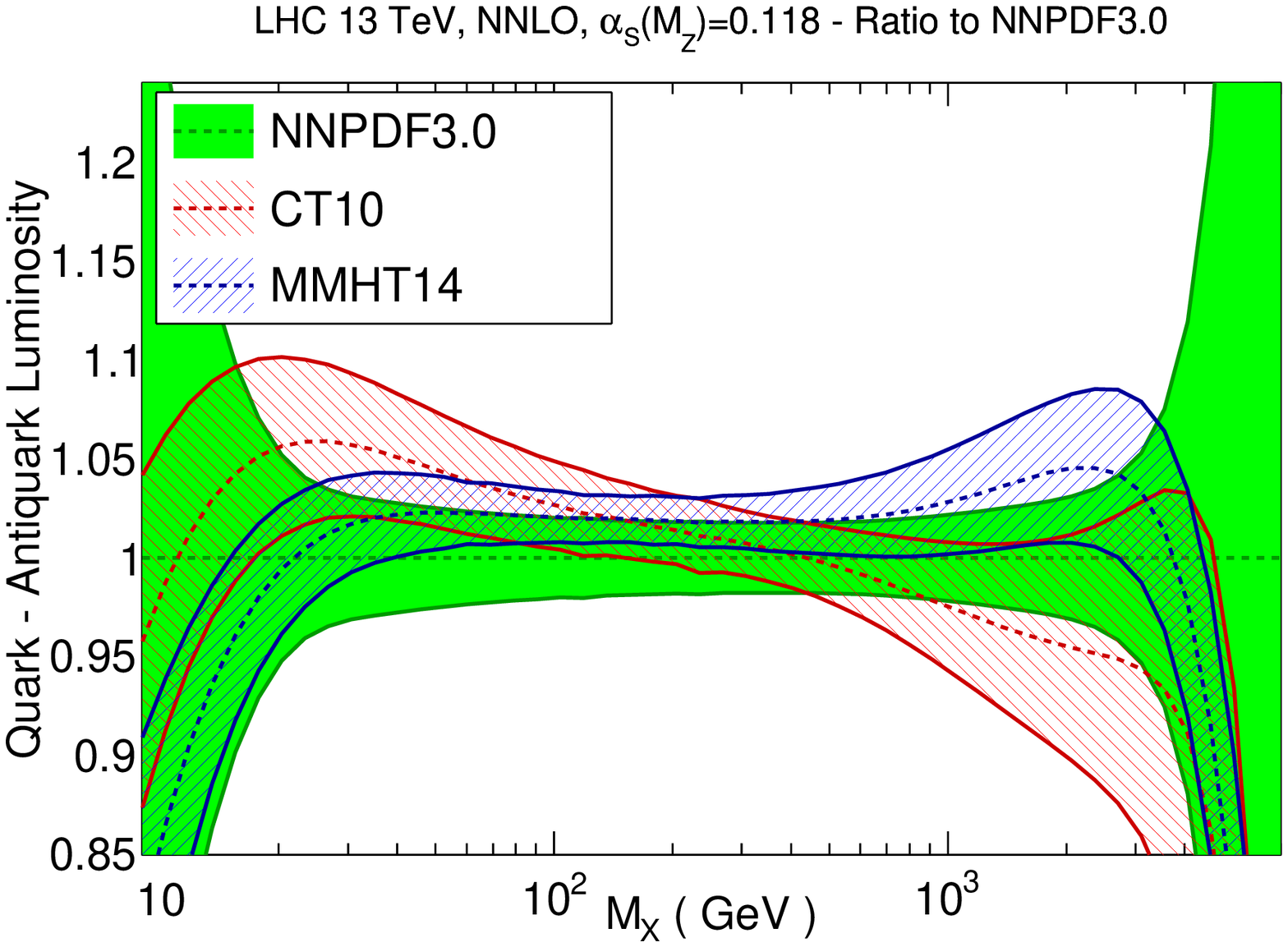}
\epsfig{width=0.42\textwidth,figure=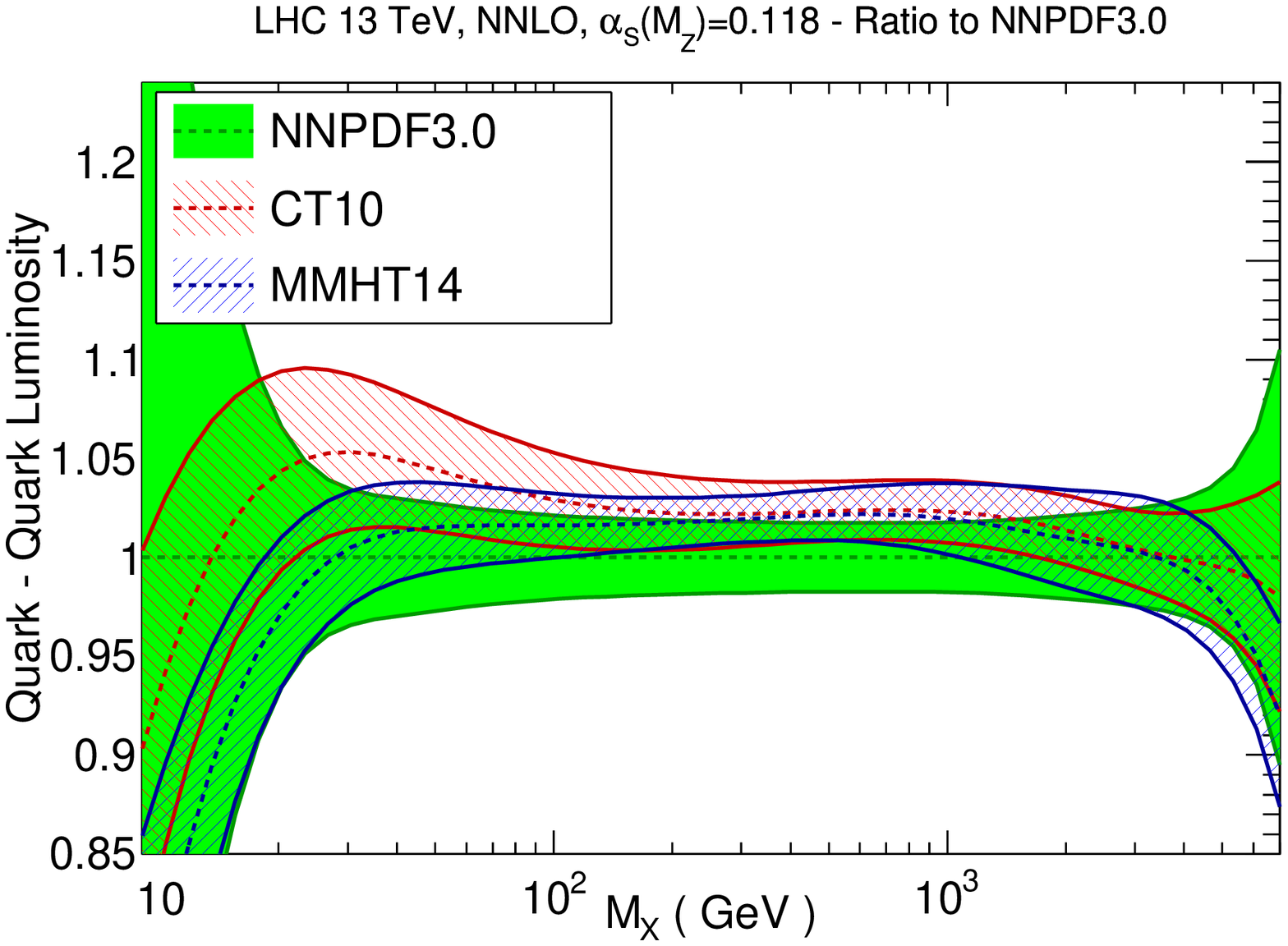}
\epsfig{width=0.42\textwidth,figure=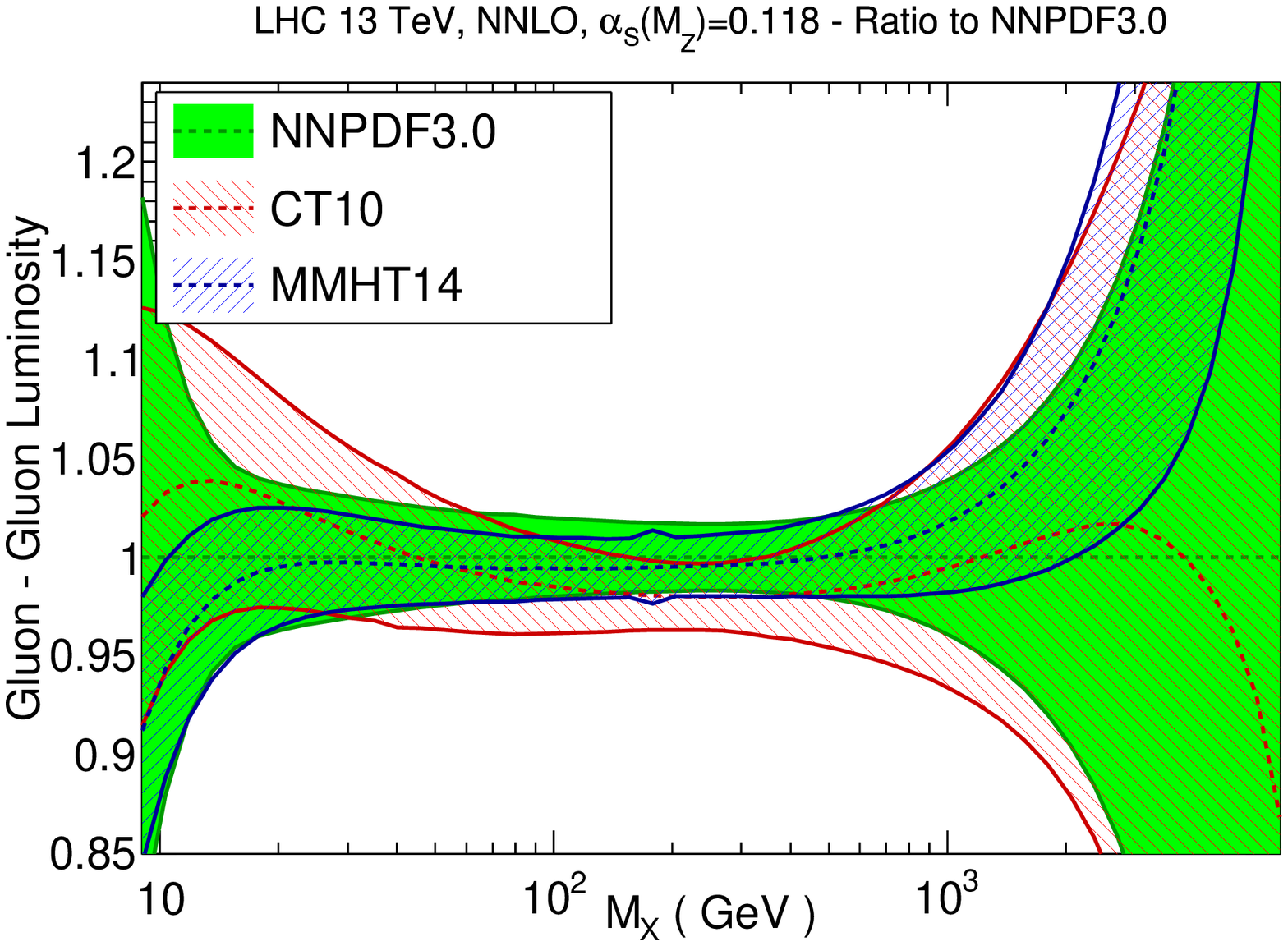}
\epsfig{width=0.42\textwidth,figure=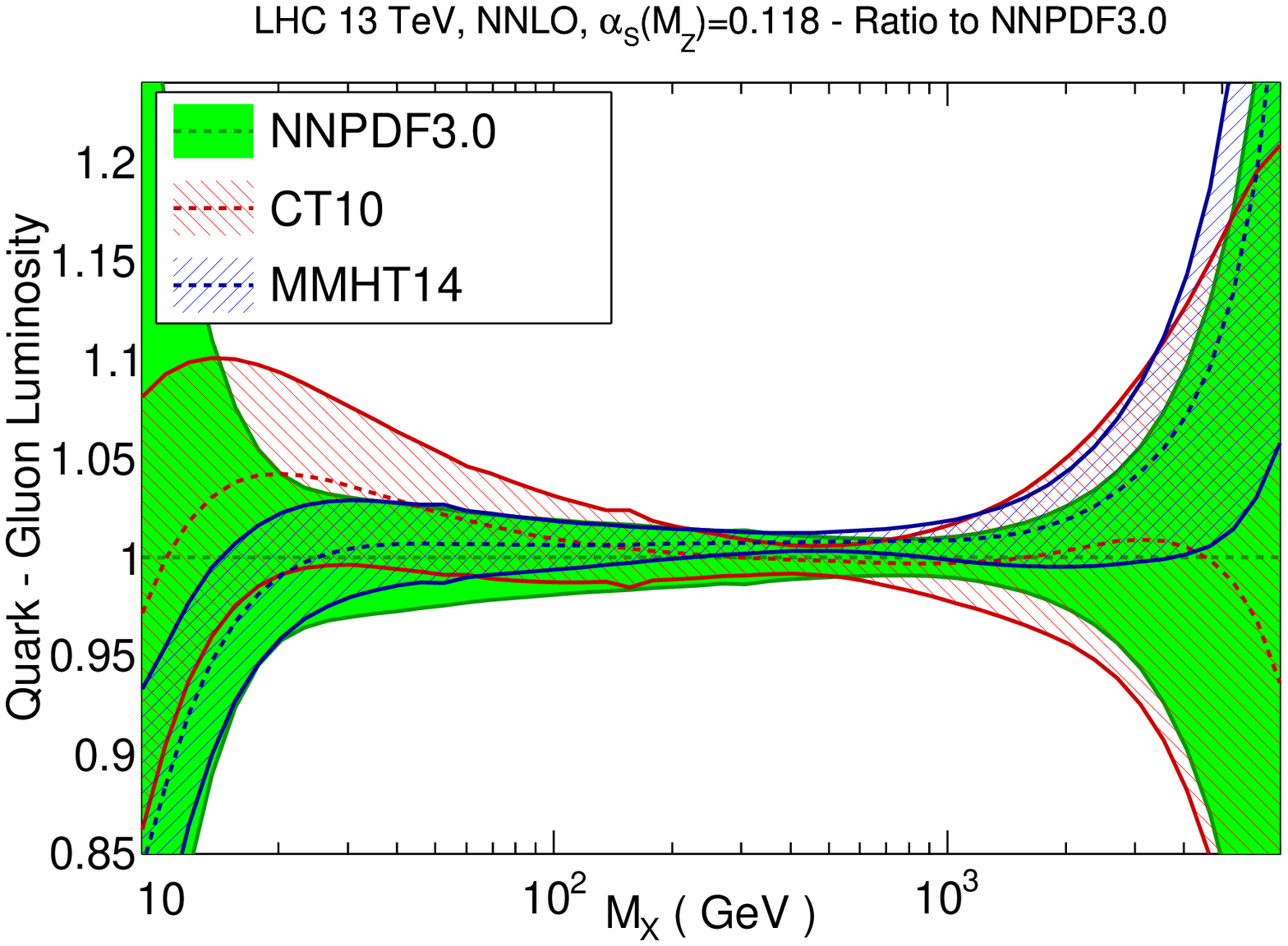}
\caption{\small
Same as Fig.~\ref{fig:lumi1nnlo} but now comparing NNPDF3.0,
MMHT and CT10 NNLO (all with $\alpha_s(M_Z)=0.118$).
Results are shown as ratios to NNPDF3.0.
 \label{fig:lumi2nnlo}}
\end{center}
\end{figure}

In Fig.~\ref{fig:lumi2nnlo}
we compare the PDF luminosities obtained using the  NNPDF3.0 set and
discussed in Sect.~\ref{sec:nnpdf30set}
to  CT10~\cite{Lai:2010vv,Gao:2013xoa} and
MMHT14~\cite{Harland-Lang:2014zoa}.
Note that these comparisons might become
obsolete once CT10 PDFs are updated, though they will be easily
updated using the recent {\tt APFEL} tool~\cite{Carrazza:2014gfa}.
The three sets are consistent within their uncertainties.
Quite in general NNPDF3.0 has
smaller uncertainties in the data region, but larger uncertainties in the
extrapolation regions.
For the $gg$ luminosity in the region relevant for Higgs production,
the agreement between the three sets has improved in comparison
to previous benchmarks using NNPDF2.3 and MSTW2008~\cite{Ball:2012wy}.
Large differences in central values are found for large values of $M_X$, relevant for the
production of very massive New Physics particles, though all sets are
compatible within their very large uncertainties.

\subsubsection{Implications for $\sqrt{s}$=13 TeV LHC processes}

An extensive phenomenological
comparison of NNPDF3.0 for all processes and a wide range of PDF sets
is beyond the scope of this work. We will thus only compare
NNPDF2.3 and NNPDF3.0, for a limited set of LHC observables.
Unless otherwise stated, we use the NLO sets with
$\alpha_s=0.118$ and work in the $N_f=5$ variable-flavor-number
scheme with massless bottom quarks.
Theoretical predictions at NLO  are computed using the
{\sc\small MadGraph5\_aMC@NLO} program~\cite{Alwall:2014hca}, version 2.1.2,
interfaced to {\sc\small LHAPDF6}.
The NLO results are sufficient to assess the  PDF dependence
of these observables, as typically the NNLO/NLO
$K$--factor have a weak PDF dependence.
In addition to NNPDF3.0 global fit results,
we also provide predictions using the conservative
parton set with $\alpha_{\rm max}=1.1$ as an illustration of results
found using a maximally
consistent dataset (see~\cite{Rojo:2014ana}).

Even though  we do not aim to a direct comparison to the raw data,
cross-sections have been computed at the fiducial level,
including resonance decays for some processes, and using
realistic generation cuts.
Jets are reconstructed with the anti-$k_T$ algorithm~\cite{Cacciari:2008gp} with
radius $R=0.5$, and the following cuts are applied to all jets in the final state:
\be
 |\eta_{\rm jet}| \le 4.5 \, , \quad p_{T,\rm jet} \ge 25~{\rm GeV} \, .
\ee
For final-state leptons, the following cuts are applied:
\be
|\eta_l| \le 2.5 \, , \quad p_{T,l} \ge 25~{\rm GeV}
\, , \quad m_{l^+l^-} \ge 30~{\rm GeV} \, .
\ee
Finally, for photons we impose
\be
|\eta_\gamma| \le 2.5 \, , \quad p_{T,\gamma} \ge 25~{\rm GeV}
\, ,
\ee
and use the Frixione isolation
criterion~\cite{Frixione:1998jh},
with
$\epsilon_{\gamma}=1.0$ and $n=1$ and
an isolation cone radius $R_0=0.4$.
No further analysis cuts are applied.
Renormalization and factorization scales are set dynamically
event by event to $\mu_f=\mu_r=H_T/2$, with $H_T$
the scalar sum of the transverse energies of all the final-state particles.
Within each run, PDF and scale uncertainties in
{\sc\small MadGraph5\_aMC@NLO} are obtained
at no extra cost using the reweighting technique introduced
in Ref.~\cite{Frederix:2011ss}: we thus provide in each case
both the central value and the PDF uncertainty.
%

\begin{table}
\centering
\small
\begin{tabular}{c|l|l|l||l}
\hline
Process  & NNPDF2.3  & NNPDF3.0 & RelDiff  & NNPDF3.0 $\alpha_{\rm max}=1.1$  \\
\hline
\hline
$p+p\to Z^0 \to e^+ e^-$    & 1.403 nb  ($\pm$ 1.5\%)  & 1.404 nb ($\pm$ 2.0\%)&  +0.1\%  &
1.45 nb ($\pm$ 2.0\%)  \\
$p+p\to W^+ \to e^+ \nu_e$  & 10.30 nb  ($\pm$ 1.3\%)  & 10.21 nb  ($\pm$ 1.9\%)&  -0.9\% & 10.29 nb ($\pm$ 2.3\%) \\
$p+p\to W^+ \to e^- \bar{\nu}_e$   &  7.67 nb ($\pm$ 1.3\%)  & 7.75 nb   ($\pm$ 1.9\%)& +1.1\% &
7.96 nb   ($\pm$ 1.9\%)
\\
$p+p\to W^+\bar{c}$   &  2.665 nb ($\pm$ 3.5\%) & 2.680 nb  ($\pm$ 4.2\%) &   +0.56\%  & 2.807 nb
 ($\pm$ 8.8\%)  \\
\hline
\hline
$p+p\to e^+\nu_e+{\rm jet}$    & 2.353 nb ($\pm$ 1.2\%) & 2.332 nb ($\pm$ 1.5\%) &  -0.9\% &
2.325 nb ($\pm$ 1.6\%)
  \\
$p+p\to \gamma+{\rm jet}$   & 62.24 nb   ($\pm$ 1.2\%) & 63.85 ($\pm$ 1.8\%)      &  +2.6\% &
61.51 ($\pm$ 1.9\%)
 \\
$p+p\to t\bar{t}$   & 678 pb ($\pm$ 1.7\%)   &  672 pb ($\pm$ 1.6\%)     & -0.9\%  &  655 pb
($\pm$ 3.3\%)  \\
\hline
\hline
$p+p\to H e^+ e^-$   & 26.48 fb ($\pm$ 1.4\%)  &  26.58 fb ($\pm$ 1.5\%)  & +0.4\% &
 27.07 fb ($\pm$ 2.3\%)
  \\
$p+p\to H e^+ \nu_e$   & 0.134 pb   ($\pm$ 1.6\%) & 0.131 pb   ($\pm$ 1.6\%)& -2.2\% &
0.137 pb  ($\pm$ 2.6\%)
\\
$p+p\to H t\bar{t}$   & 0.458 pb ($\pm$ 2.2\%)  &  0.4595 pb ($\pm$ 1.7\%)     & +0.6\% & 0.459 pb
 ($\pm$ 4.0\%)   \\
\hline
\end{tabular}
\caption{\small Cross-sections for LHC at 13~TeV,
computed at NLO using {\sc\small MadGraph5\_aMC@NLO} using
 NNPDF2.3 and NNPDF3.0 NLO PDFs, with $N_f=5$  and
$\alpha_s(M_Z)=0.118$.
In each case,  central values and the
one-sigma PDF uncertainty (in parenthesis) are given.
We also show the percentage difference between the
central values using the two PDF sets and
the prediction using conservative partons  with $\alpha_{\rm max}=1.1$.
\label{tab:amcxsec}
}
\end{table}

\begin{figure}
\begin{center}
\epsfig{width=0.99\textwidth,figure=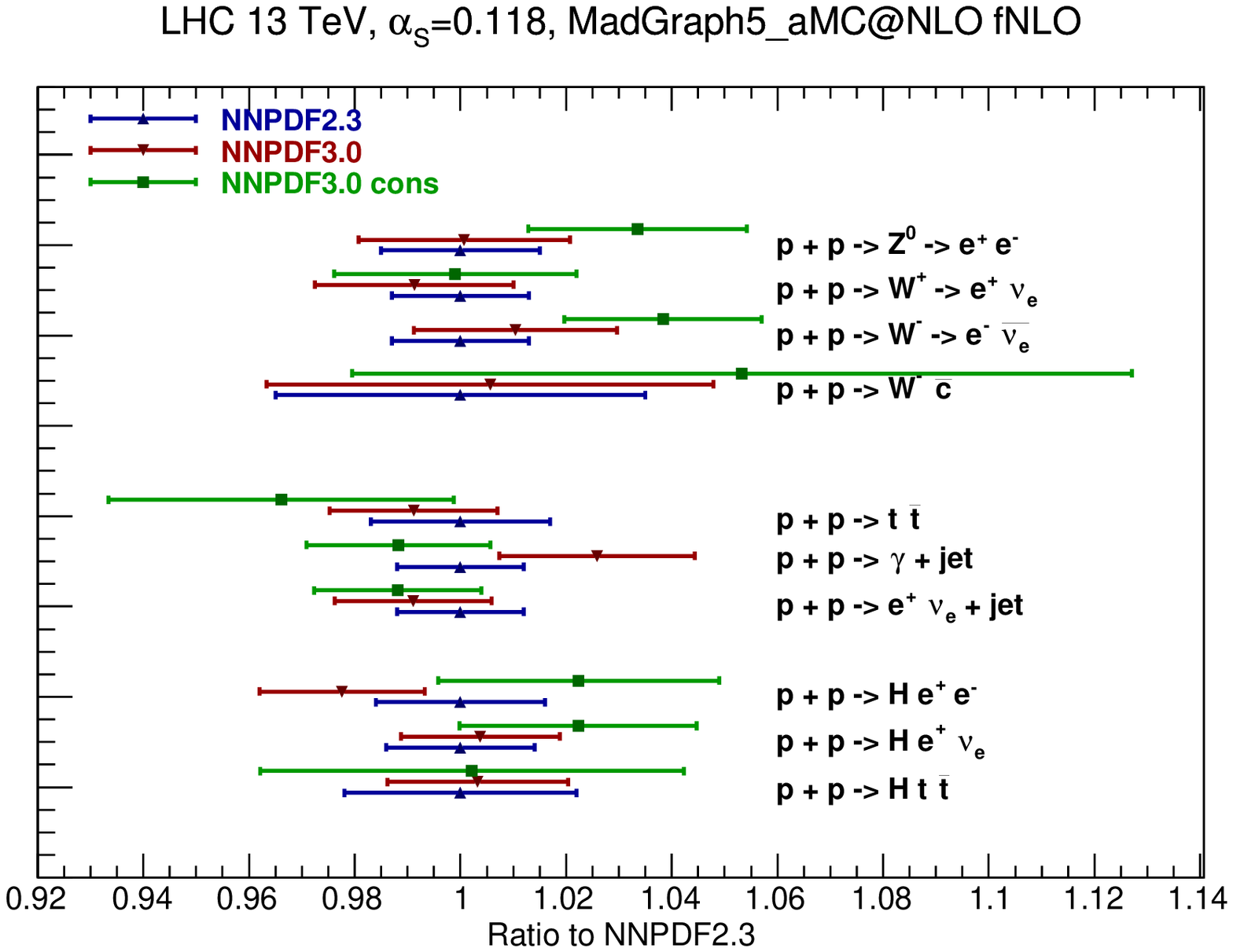}
\caption{\small
Graphical comparison of the results of Tab.~\ref{tab:amcxsec}.
Results are shown  normalized to the NNPDF2.3 central value.
\label{fig:amcxsec}
}
\end{center}
\end{figure}

Results are collected in Tab.~\ref{tab:amcxsec}, where
processes are grouped
into three subsets:
processes which are sensitive to  quark
and antiquarks, processes which are sensitive
to the gluon PDF, and Higgs production processes, except
gluon fusion which is discussed in the next section.
The results of Tab.~\ref{tab:amcxsec} are also represented
in Fig.~\ref{fig:amcxsec}, normalized
to NNPDF2.3.

For all these cross-section a remarkable stability between NNPDF2.3 and NNPDF3.0
is observed, with all results varying by no more that the size of the
corresponding PDF uncertainty.
For top-quark pair production, going from NNPDF2.3 to NNPDF3.0
the cross-section increases by about 1\%, about half the
PDF uncertainty. This can be understood recalling that thee
NNPDF3.0 $gg$ luminosity
is harder than is
NNPDF2.3 counterpart for $M_X \sim 400$ GeV; this also explains
the behaviour of the closely related $ttH$ production process.
Note that NNPDF2.3 already gave a very good description of all
available ATLAS and CMS 7 TeV and 8 TeV production
data~\cite{Czakon:2013tha}, even though they were not included in the fit.
For $W$ production in association with charm quarks, we use a $N_F=3$
scheme in order to retain the full charm mass dependence.
For this observable, results are very stable when moving
from NNPDF2.3 to NNPDF3.0.

Coming now to Higgs production,
for  $ttH$ the NNPDF3.0 result is about 2.3\% larger than the
NNPDF2.3 prediction, consistent with the expectation from
the $gg$ luminosity comparisons in Fig.~\ref{fig:lumi1nlo} for
$M_X \sim 500=700$ GeV (recall that the calculation uses a dynamical
setting of the factorization scale).
For the associate production channels, $hW$ and $hZ$, driven by the
$q\bar{q}$ luminosities, differences are well within one sigma,
 as  expected from the luminosities of
Fig.~\ref{fig:lumi1nlo}.

Coming finally to the  comparison with the  conservative PDFs, we note
that prediction obtained using the latter are generally
consistent at the one sigma level with the default ones,
occasionally with differences at the two sigma level, such as for example for $hl^+\nu$.

Of course, predictions from  conservative partons are generally affected by larger PDF
uncertainties because of the smaller dataset, though in some cases they are only slightly
less accurate than the global fit, such as for example for inclusive $W$ and $Z$ production.
On the other hand, for processes that depend on strangeness (like $W$+$c$) or that are gluon-driven
(like $t\bar{t}$ and $t\bar{t}h$) the PDF uncertainties are substantially larger
in the conservative partons than in the global fit.

\subsubsection{Higgs production in gluon fusion}
\label{sec:higgsgg}

We now focus specifically on  Higgs production in
gluon fusion,
the dominant channel at the LHC, for which
 theoretical uncertainties are a limiting factor for the determination
of  Higgs properties.
We provide predictions for the total cross-section at
NLO and NNLO
for the LHC 13 TeV, comparing the default NNPDF3.0 set to NNPDF2.3 and
to the various sets based on alternative datasets discussed in
Sect.~\ref{sec:resdataset}, with the main goal of studying the
dependence of this prediction on the underlying PDFs, along the lines
of the study presented in Ref.~\cite{LHcons}. The uncertainties shown are
pure PDF uncertainties, with $\alpha_s(M_Z)=0.118$, i.e. the
$\alpha_s$ uncertainty is not included.
The inclusive cross-section is computed using
{\sc\small iHixs}~1.3.3, with  $m_h=125$ GeV,
renormalization and factorization scales set to
$\mu_r=\mu_f=m_h$ and the infinite top mass (effective theory)
approach. Clearly, our predictions are not meant to be realistic,
however,
the effects which we do not include (such as for instance electroweak
corrections, or finite top, bottom and charm mass contributions) have
a negligible PDF dependence, while $\alpha_s$ uncertainties are
completely independent of the PDF uncertainty, given that the PDF and
$\alpha_s$ uncertainties combine in quadrature in the gaussian case
even when correlated~\cite{Lai:2010nw}. Hence our results do provide
an accurate assessment of the PDF dependence of the cross-section and its
uncertainty.

\begin{figure}
\begin{center}
\epsfig{width=0.70\textwidth,figure=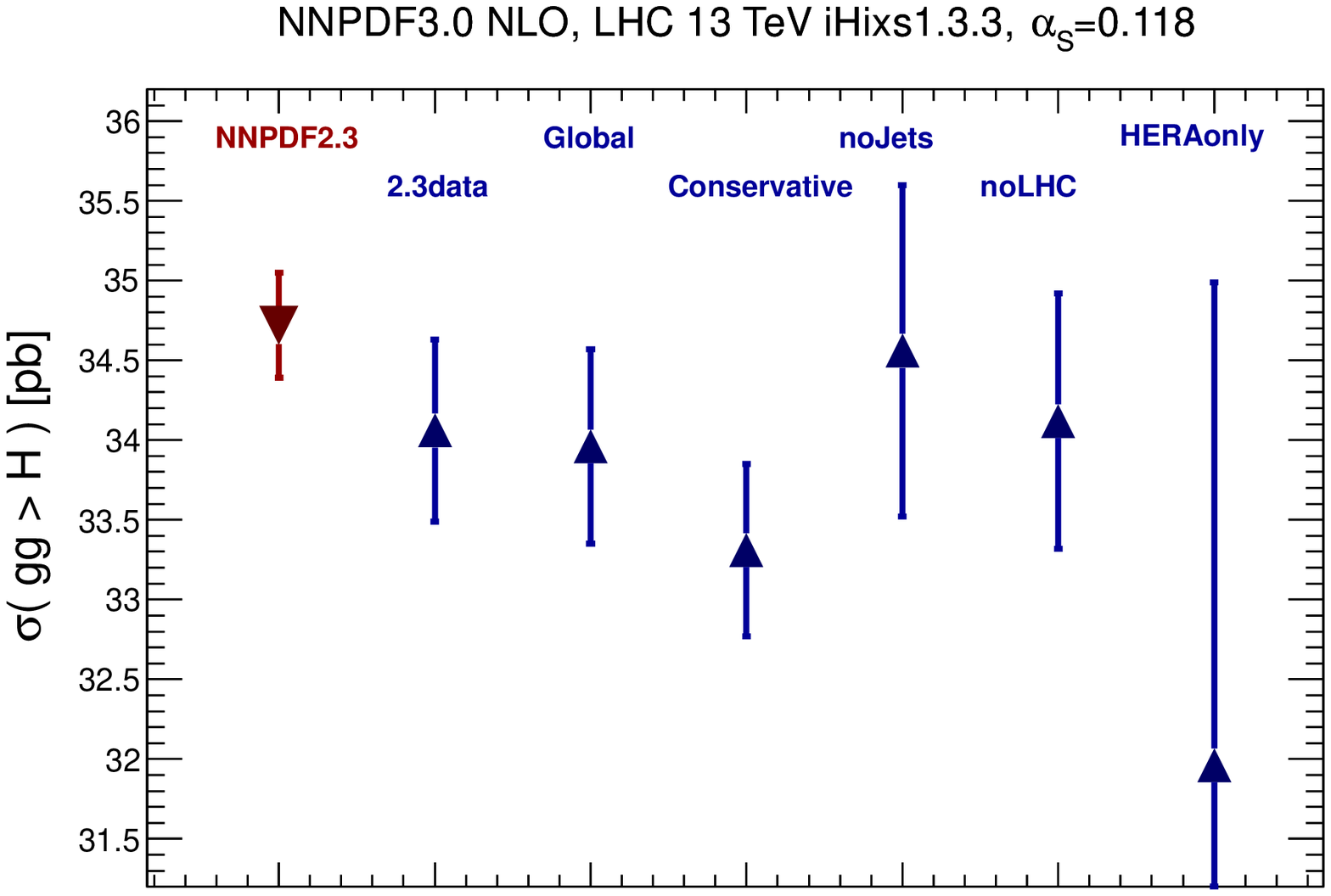}
\caption{\small
Graphical representation of the NLO results of
Tab.~\ref{tab:xsechiggs}.
\label{fig:higgsnnlo2}
}
\end{center}
\end{figure}

\begin{figure}
\begin{center}
\epsfig{width=0.70\textwidth,figure=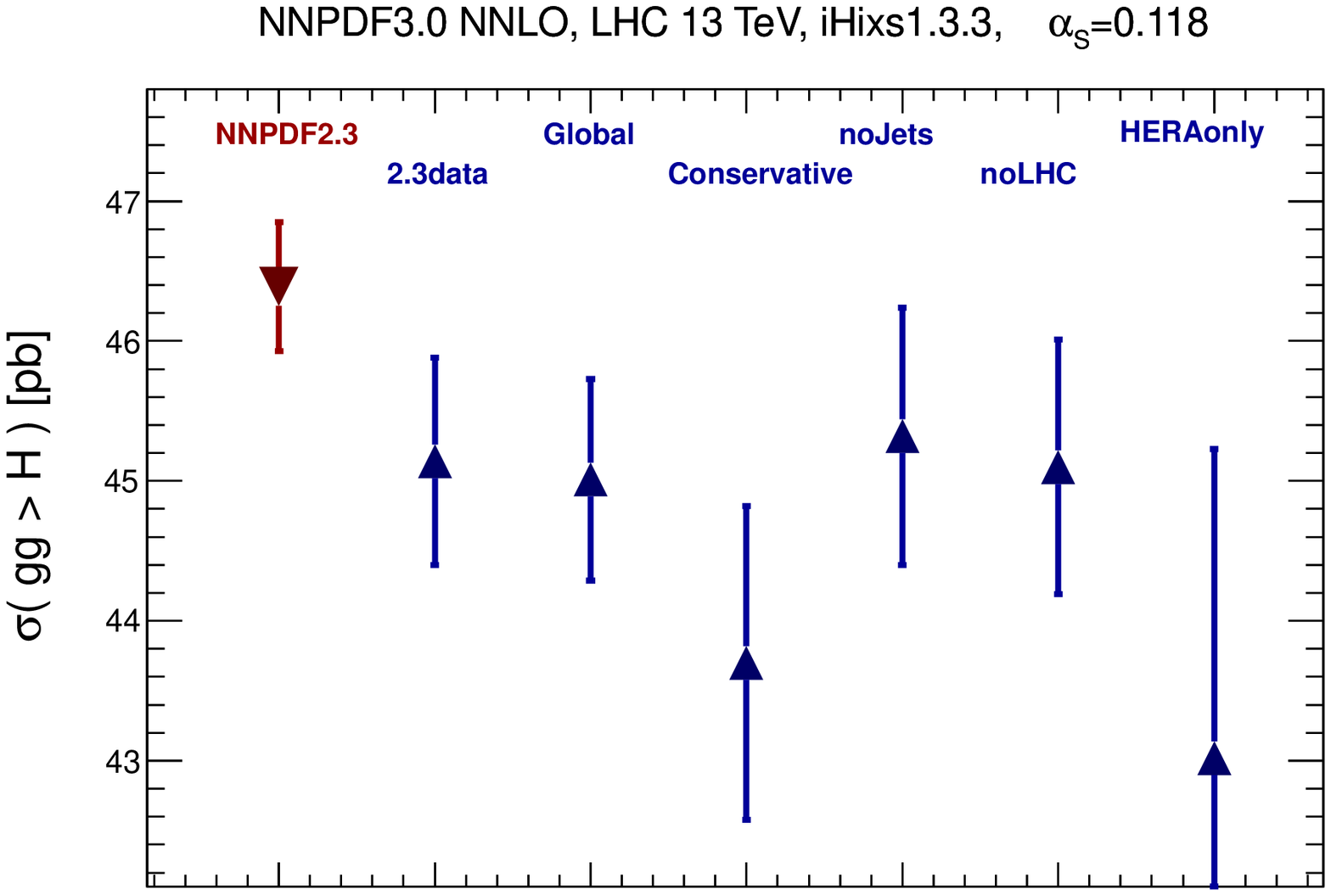}
\caption{\small
Graphical representation of the NNLO results of
Tab.~\ref{tab:xsechiggs}.
\label{fig:higgsnnlo}
}
\end{center}
\end{figure}

Results are collected in
Tab.~\ref{tab:xsechiggs}, and summarized graphically in Figs. \ref{fig:higgsnnlo2}-\ref{fig:higgsnnlo}.
We show results obtained using NNPDF2.3, and the following NNPDF3.0 sets: 2.3-like dataset,
default, conservative with $\alpha_{\rm max}=1.1$, no jet data, no LHC data and HERA-only,
all of which have been discussed in detail in Sect.~\ref{sec:resdataset}.
In Tab.~\ref{tab:xsechiggs} we also show the pull of each prediction compared to the NNPDF2.3 result,
defined as
\be
P \equiv \frac{\lp \sigma_{ggh}(2.3)-\sigma_{ggh}(3.0) \rp}{\sqrt{\Delta\sigma^2_{ggh}(3.0)+
\Delta\sigma^2_{ggh}(2.3)}} \, ,
\label{eq:pull}
\ee
where $\Delta\sigma_{ggh}$ is the one-sigma PDF uncertainty.
As  expected from the comparison of the gluon-gluon luminosities
in Fig.~\ref{fig:lumi1nnlo}, the NNLO cross-section
decreases by about 2-sigma when going from NNPDF2.3 to NNPDF3.0, while
the PDF uncertainty increases substantially.
At NLO the effect is less marked, with the NNPDF2.3 and NNPDF3.0
in agreement at the one-sigma level. Because a similar result is found
using NNPDF3.0 PDFs based on a 2.3-like dataset we must conclude that
the change is mostly due to methodological improvements, rather than
the underlying data. Because of the validation from closure testing,
the NNPDF3.0 results are more reliable, both in terms of central
values and uncertainties.

\begin{table}
\centering
\begin{tabular}{c|c|c|c|c}
\hline
&  $\sigma_{ggh}$ (pb) NLO  &  Pull &  $\sigma_{ggh}$ (pb) NNLO  &  Pull \\
\hline
\hline
NNPDF2.3  & 34.72 $\pm$ 0.33 & - & 46.39 $\pm$ 0.46  &   -  \\
\hline
NNPDF3.0 with 2.3 data  & 34.06 $\pm$ 0.57  & 1.0 & 45.14 $\pm$ 0.74  & 1.4   \\
NNPDF3.0 global  & 33.96 $\pm$ 0.61 & 1.1 & 45.01 $\pm$ 0.72  & 1.6 \\
\hline
NNPDF3.0 conservative $\alpha_{\rm max}=1.1$  & 33.31 $\pm$ 0.54  & 2.2 &
 43.70 $\pm$ 1.12  & 2.2  \\
NNPDF3.0 no Jets & 34.56 $\pm$ 1.04 & 0.2  &  45.32 $\pm$ 0.92 & 1.0  \\
NNPDF3.0 noLHC  &  34.12 $\pm$ 0.80  & 0.7 & 45.10 $\pm$ 0.91   & 1.3   \\
NNPDF3.0 HERA-only  &  31.96 $\pm$ 3.03 & 0.9 & 43.02 $\pm$ 2.21   & 1.5 \\
\hline
\end{tabular}
\caption{\label{tab:xsechiggs} \small
The total cross-section for Higgs production in gluon fusion at the
LHC 13 TeV at NLO (left) and NNLO (right) for $\alpha_s(M_z)=0.118$.
The  pull $P$ Eq.~(\ref{eq:pull}) is also given.
}
\end{table}

Results obtained using NNPDF3.0 sets based on alternative datasets
are always mutually
consistent at the one sigma level, with the conservative partons
leading to a lower result and the fit with no jets to a slightly
higher one but with significantly larger uncertainty. The lowest
central value is found using the  HERA-only set, which however
is affected by a PDF uncertainty which is by a factor three larger
than the default.

Despite these variations in results, all pulls are of similar size, between 0.7 and 1.1 at NLO
and between 1.3 and 1.6 at NNLO, with the only
exception of conservative partons, which have a larger pull of 2.2
both at NLO and NNLO. This means that the dependence of the results on
the dataset appears to be consistent with statistical fluctuations.

\begin{figure}[h]
\begin{center}
\epsfig{width=0.42\textwidth,figure=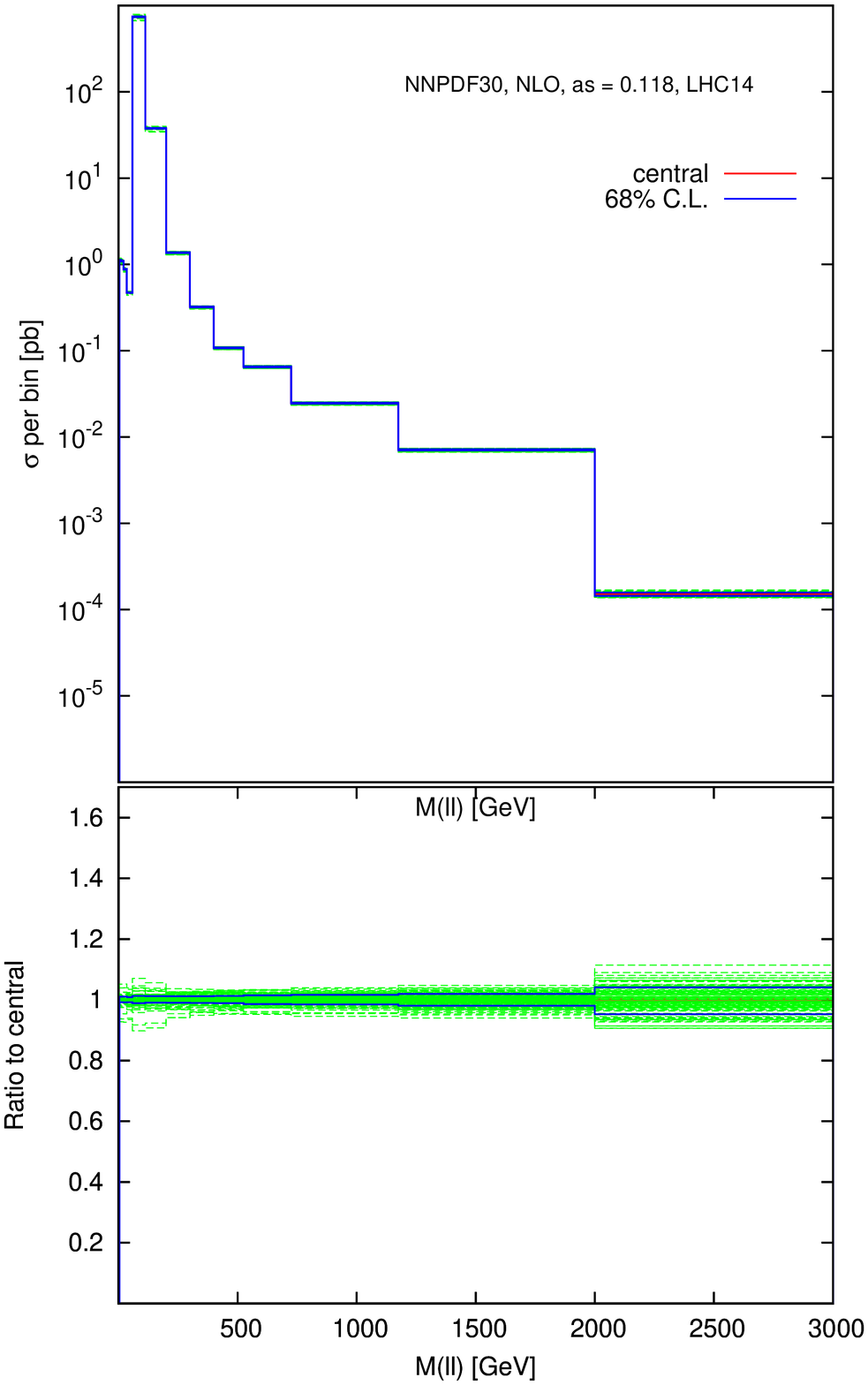}
\epsfig{width=0.42\textwidth,figure=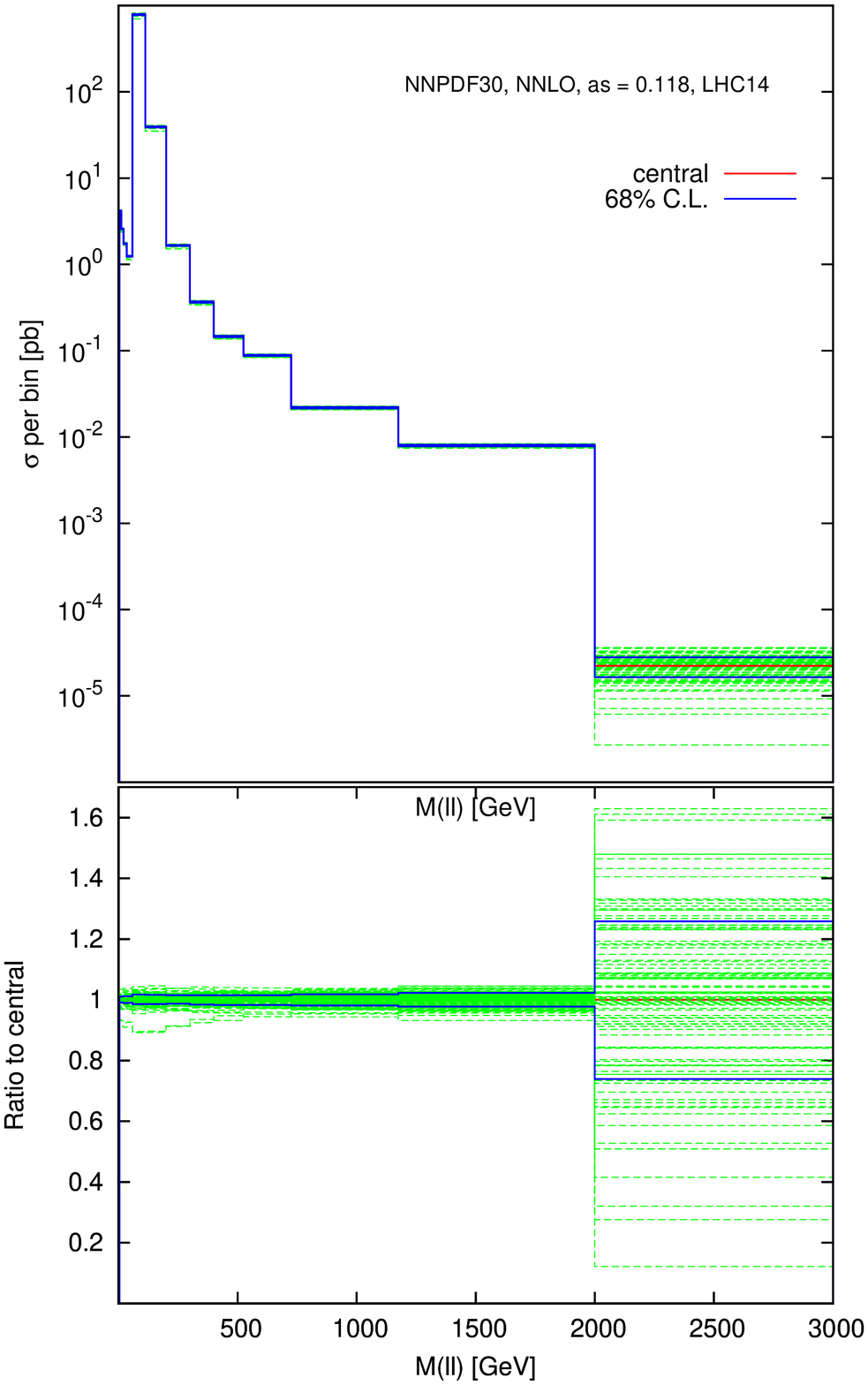}
\vspace{0.3cm}
\caption{\small
\label{fig:NLOposDY}
The dilepton invariant mass distribution in $pp \to \gamma^*/Z \to l^+l^-$
at the LHC 14 TeV with NNPDF3.0 at NLO (left) and NNLO (right) using {\sc\small FEWZ}.
Each (green) dashed line is one of the $N_{\rm rep}=100$ Monte
Carlo replicas, while the solid (red) line is the average and the outer
solid
(blue) lines give the edges of the  one-sigma interval. Both the
absolute result (top) and the ratio to the central value (bottom) are shown.
}
\end{center}
\end{figure}

\subsubsection{New Physics particle production at high masses}
\label{sec:highmassBSM}

A wide variety of scenarios of physics beyond the standard model include
heavy particles at the TeV scale that could be within the reach of the
LHC.
Production of very massive particles probes PDFs at
large $x$, where they are poorly known due to  the lack of direct experimental
information, and thus the corresponding predictions are  affected by substantial PDF uncertainties
(see e.g.  Refs.~\cite{Kramer:2012bx,Borschensky:2014cia}).
Consequently, PDFs can be a  limiting factor in  the determination of exclusion regions,
and improving their knowledge can lead to an increase in the search potential. In this context
an accurate assessment of their uncertainties is therefore crucial. The unbiased NNPDF approach is
advantageous in this respect in that it leads to uncertainty estimates
which are not biased by assumptions on the functional form of
PDFs. The only significant constraint on PDFs close to threshold comes
from positivity, which is now implemented in an optimal way as we
discussed in Sects.~\ref{sec:positivity}-\ref{sec:positivityresults} above.

As an example, we consider  high-mass Drell-Yan production and the
pair production of supersymmetric particles.
High-mass dilepton production is frequently used to search for new physics that couples
to the electroweak sector, and thus it is important to provide precise predictions
for the SM production mechanisms.

We have computed  the
dilepton invariant mass distribution in $pp \to \gamma^*/Z \to l^+l^-$ events
at the LHC 14 TeV with NNPDF3.0 at NLO and NNLO  using {\sc\small FEWZ}.
Recall from  Sect.~\ref{sec:positivity} that positivity is always
imposed at NLO, so an explicit check of positivity of the NNLO result
is nontrivial.
Results are shown in Fig.~\ref{fig:NLOposDY}, in different $M_{ll}$ bins:
 each of the $N_{\rm rep}=100$ NLO (left) or NNLO (right) Monte
Carlo replicas is shown as a green dashed line,  together with
the corresponding central values and one-sigma intervals.
All MC replicas are  positive up to the highest
invariant mass bins.

As a second example, we provide predictions for the pair production
of supersymmetric particles at the LHC 14 TeV.
The computation has been performed using {\sc\small Prospino}~\cite{Beenakker:1996ch,Prospino2} with NNPDF3.0 NLO,
using settings as close as possible to those of Refs.~\cite{Kramer:2012bx,Borschensky:2014cia},
though the only relevant physical input for our illustrative study are the sparticle masses.
Note that for this processes NNLO calculations are not available.
We have produced results for squark-squark, squark-antisquark and gluino-gluino production,
for three different values of the sparticle masses, namely 1 TeV, 2 TeV and 3 TeV.
We show the  predictions for the $N_{\rm rep}=100$ Monte
Carlo replicas of NNPDF3.0 for the  squark-antisquark
and gluino-gluino channels; the  squark-squark cross-section (not shown)
is  always positive in the
mass range that we are considering.
In each case, we also provide the average result and the 68\% confidence
level interval.
For comparison, predictions using the NNPDF2.3 NLO PDFs  are also shown.

\begin{figure}
\begin{center}
\epsfig{width=0.42\textwidth,figure=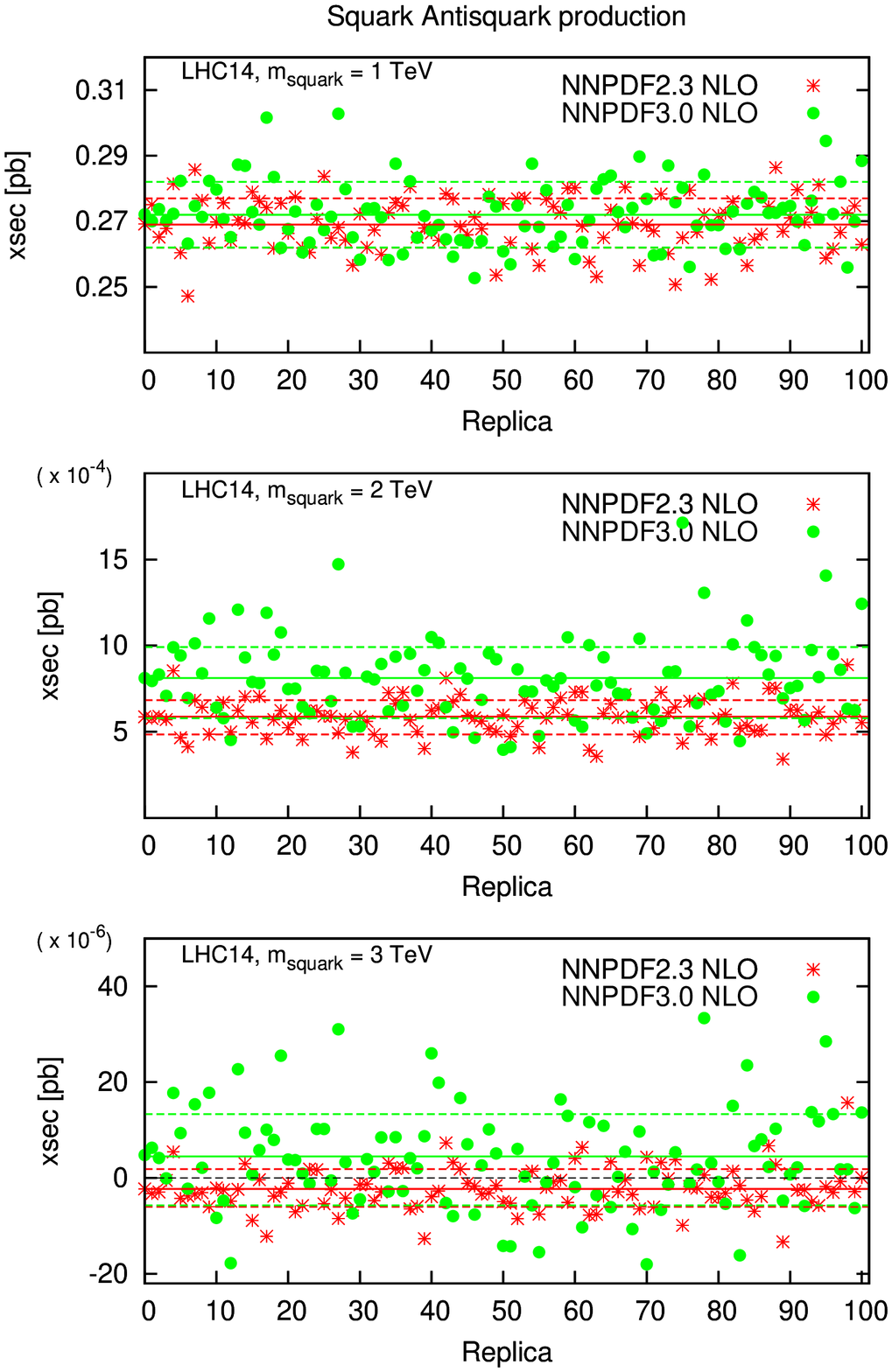}
\epsfig{width=0.42\textwidth,figure=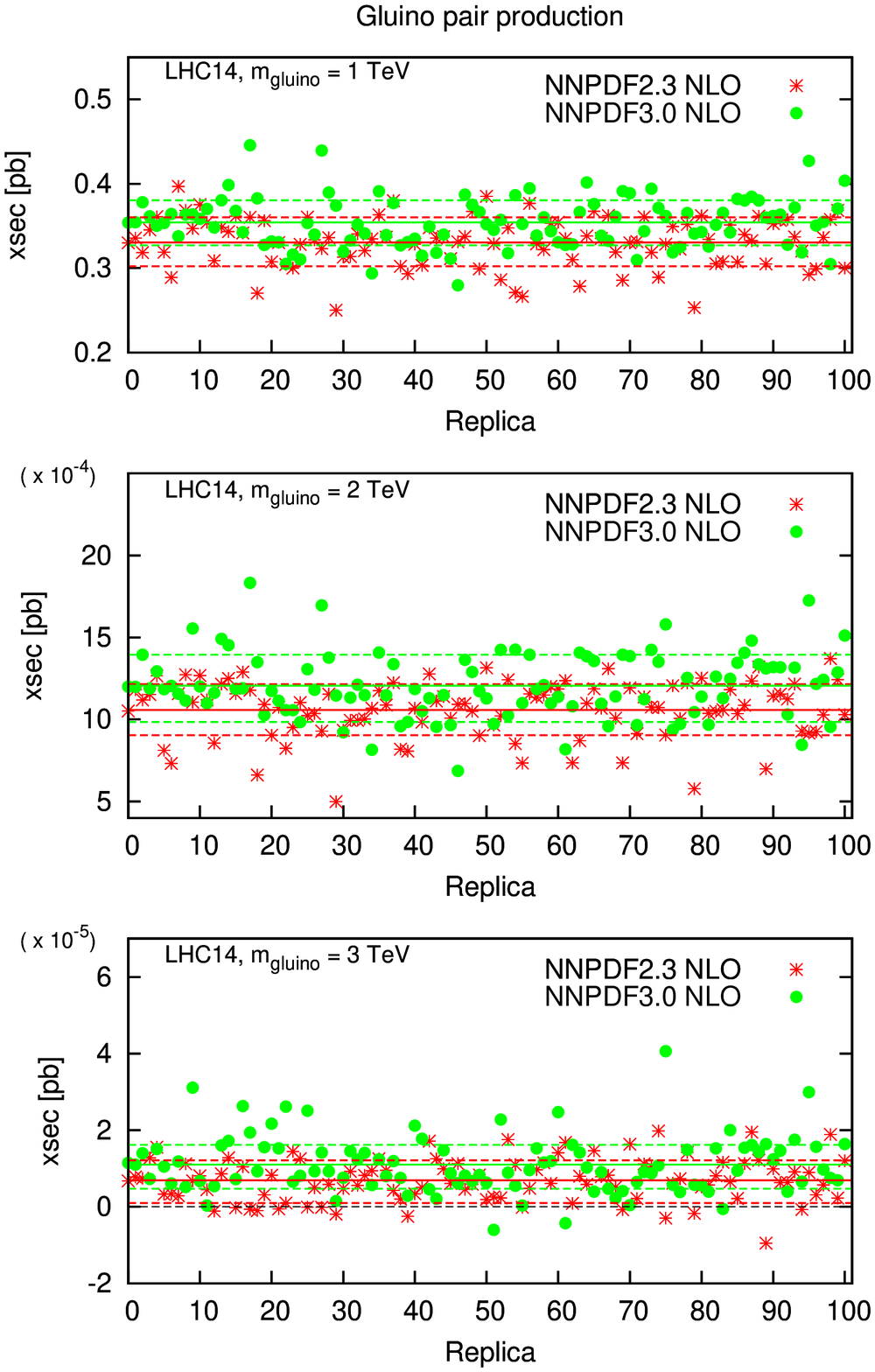}
\caption{\small
\label{fig:SUSYggNLO.eps}
Cross-sections for NLO squark-antisquark (left) and gluino-gluino (right)
pair production at  the LHC 14 TeV with NNPDF3.0 and NNPDF2.3, for
sparticle masses of 1 TeV, 2 TeV and 3 TeV.
In each case, we show the predictions for the $N_{\rm rep}=100$ Monte
Carlo replicas as well as by the average result and the 68\% confidence
level interval.
}
\end{center}
\end{figure}

Results are shown in Fig.~\ref{fig:SUSYggNLO.eps}.
In the case of gluino-gluino production (right), all replicas
are strictly positive up to $m_{\tilde{g}}=2$ TeV.
At 3 TeV, some replicas lead to slightly negative cross-sections:
15 in NNPDF2.3, and only 3 in NNPDF3.0.
In both cases, the 68\% confidence levels are always positive;
we conclude that the
occasional negative replica can be set to zero with no impact on the
central value or uncertainty.
For squark-antisquark production (left)
all cross sections are positive up to $m_{\tilde{q}}=2$ TeV,
while for $m_{\tilde{q}}=3$ TeV  some cross-sections
are slightly negative.
For NNPDF2.3, the central value was negative, while now for NNPDF3.0
the central value is positive and only a small part of the
68\% confidence level range is in the negative region. This means that
our improved positivity is still not fully efficient, and it allows
some replicas to lead to negative cross-section: partly because
positivity is imposed with a Lagrange multiplier which carries a large
but finite penalty, and also, because
positivity is only imposed for standard model processes, and not for
all possible processes.
Setting these negative cross-sections to zero would of course not modify the upper uncertainty range,
implying that for this particular subprocesses and sparticle masses arbitrarily small production
cross-sections are allowed within the large PDF uncertainties.
Note however the very large PDF uncertainties at the largest masses: for example, for
squark-antisquark production, the 68\% CL range for the cross-section for
$m_{\tilde{q}}= $ 3 TeV is $\sigma(\tilde{q}\bar{\tilde{q}}) \in \lc 0, 15\cdot10^{-5}\rc$ pb,
with a central value of $\sim 5\cdot10^{-5}$ pb.

We conclude that, thanks to the improved implementation of positivity
in NNPDF3.0, the cross-section for high-mass particle production is
positive, thereby improving over NNPDF2.3. Occasional replicas leading
to negative cross-sections very close to threshold can be set to zero without significantly
affecting central values and uncertainties.

\section{Summary and outlook}
\label{sec:delivery}
We conclude by summarizing the NNPDF3.0 fits that will be
made public, their use  in specific
phenomenological application, and future developments.

All  PDF sets are made available via the {\sc\small LHAPDF6}
library~\cite{Bourilkov:2006cj}, from version 6.1.4 onward,
\begin{center}
{\bf \url{http://lhapdf.hepforge.org/}~}
\end{center}
For the global fits, we also provide the corresponding grids in
 {\sc\small LHAPDF5} format via the the NNPDF {\sc\small HepForge}
webpage,
\begin{center}
{\bf \url{http://nnpdf.hepforge.org/}~}
\end{center}
though  whenever possible, the modern v6 interface
should be used.

The interpolation used in the delivery of NNPDF sets has been
substantially improved in  the transition
from {\sc\small LHAPDF5} to {\sc\small LHAPDF6}.
\begin{itemize}
\item The default interpolation algorithms,
of {\sc\small LHAPDF6} are now used, instead of the in-house interpolation
used by NNPDF sets in {\sc\small LHAPDF5}.
These algorithms are faster and more accurate, but we have checked
that in most of the relevant range of $x$ and $Q^2$ the
{\sc\small LHAPDF6} and {\sc\small LHAPDF5} algorithms
agree at the permille level.
\item A wider range for the $x$ and $Q$ grids is now adopted:
the NNPDF3.0 sets can be used
for
\be
\label{eq:range}
10^{-9} \le x \le 1 \, , \quad \, 1\,{\rm GeV} \le Q \le 100\,{\rm TeV} \, ,
\ee
which is adequate for future phenomenological applications,
including calculations of cross-section of ultra-high energy neutrinos,
which require PDFs down to $x\sim 10^{-8}$, and studies of a Future
Circular Collider with a center-of-mass energy of 100 TeV.
Outside the range in Eq.~(\ref{eq:range}), where no experimental
information is available,
the values of the PDFs are frozen.
\item Thanks to  the novel functionalities
in {\sc\small LHAPDF6}, for all the NNPDF3.0  sets
the $Q^2$ interpolation grid has been
divided into different subgrids separated by heavy quark thresholds.
This  ensures that the correct threshold behavior of the
heavy quark PDFs is satisfied also by the interpolation, and in
particular that the heavy quark PDFs are exactly zero below the
threshold.

\end{itemize}

The NNPDF3.0 PDF sets to be made available via {\tt LHAPDF6} are the following:

\begin{itemize}

\item Global NNPDF3.0 sets - baseline fits

The baseline LO, NLO and NNLO NNPDF3.0 sets are based on the global
dataset, with $\alpha_s(M_Z)=0.118$ and a variable-flavor number with
up to five active flavors, the convention for the corresponding
{\sc\small LHAPDF6} grid files is
\begin{flushleft}
\tt NNPDF30\_lo\_as\_0118 \\
\tt NNPDF30\_nlo\_as\_0118 \\
\tt NNPDF30\_nnlo\_as\_0118
\end{flushleft}
The leading order NNPDF3.0 allows, among other
applications, to compare
theoretical predictions for cross-sections
computed at different perturbative orders.
In {\sc\small LHAPDF6},
the output of the NNPDF3.0 LO PDFs is forced to be
positive using the suitable meta-data instructions.
For uses in leading-order Monte Carlo event generators,
we still recommend the use NNPDF2.3LO, since, as part of the updated
Monash 2013 Tune~\cite{Skands:2014pea} of {\sc\small Pythia8}~\cite{Sjostrand:2007gs}, it
has been shown to provide
an excellent description of a variety of soft and semi-hard data.

\item Global NNPDF3.0 sets - fits with $\alpha_s$ variations

Using the same global dataset as above, for the NLO and NNLO
fits we have produced fits
with different values of $\alpha_s(M_Z)$, that can be used to
evaluate the combined PDF+$\alpha_s$ uncertainties in cross-sections:
\begin{flushleft}
\tt NNPDF30\_nlo\_as\_0115,   NNPDF30\_nlo\_as\_0117,  NNPDF30\_nlo\_as\_0119,  NNPDF30\_nlo\_as\_0121\\
\tt NNPDF30\_nnlo\_as\_0115,   NNPDF30\_nnlo\_as\_0117,  NNPDF30\_nnlo\_as\_0119,  NNPDF30\_nnlo\_as\_0121
\end{flushleft}
The procedure to combine PDF and $\alpha_s$ uncertainties in a Monte Carlo set
like NNPDF3.0 is explained in~\cite{Demartin:2010er} (see also
Ref.~\cite{Alekhin:2011sk}).
The choice of the uncertainty $\delta\,\alpha_s$ to be assigned
to the strong coupling is left to the PDF users. For LO, we also provide
a fit with   $\alpha_s(M_Z)=0.130$, as this is the value typically
required in LO calculations:
\begin{flushleft}
\tt NNPDF30\_lo\_as\_0130
\end{flushleft}
Note that from the NLO and NNLO range of $\alpha_s$ values, the predictions
for any other value of $\alpha_s$ not included in the list can be
easily obtained from either interpolation or extrapolation.

\item Global NNPDF3.0 - fits with different maximum number of active flavors

For various important phenomenological applications,
PDF sets in which either the charm quark or both the charm and bottom quarks
are treated as massive are required.
In the following we denote these two schemes as
the $N_f=3$ and $N_f=4$ schemes.
In order to obtain $N_f=3$ and $N_f=4$ versions of the NNPDF3.0 sets,
we follow the procedure outlined in~\cite{Ball:2011mu}.
We start from  the baseline NNPDF3.0 fits, where the charm and bottom quarks are
massless parton ($N_f=5$ scheme), at the input scale, $Q_0^2=1$ GeV$^2$.
This boundary condition is then evolved upwards in $Q^2$, but using
the $N_f=3$ and $N_f=4$ schemes,  to produce
the desired sets with different maximum number of active partons.
This has been done for a reduced range of $\alpha_s(M_Z)$ values, to
be able to compute the combined PDF+$\alpha_s$ uncertainties.
These fits are:
\begin{flushleft}
\tt NNPDF30\_nlo\_as\_0117\_nf3, NNPDF30\_nlo\_as\_0118\_nf3, NNPDF30\_nlo\_as\_0119\_nf3\\
\tt NNPDF30\_nnlo\_as\_0117\_nf3, NNPDF30\_nnlo\_as\_0118\_nf3, NNPDF30\_nnlo\_as\_0119\_nf3\\
\tt NNPDF30\_nlo\_as\_0117\_nf4, NNPDF30\_nlo\_as\_0118\_nf4, NNPDF30\_nlo\_as\_0119\_nf4\\
\tt NNPDF30\_nnlo\_as\_0117\_nf4, NNPDF30\_nnlo\_as\_0118\_nf4, NNPDF30\_nnlo\_as\_0119\_nf4\\
\end{flushleft}
In addition, using a similar procedure we have provided sets in the
$N_f=6$ scheme, that is, treating the top quark as a massless parton.
While treating the top quark as massless is not justified at the LHC, it
might become appropriate at higher energy colliders, see for example Ref.~\cite{Dawson:2014pea}.
These sets that include the top PDF are denoted as
\begin{flushleft}
\tt NNPDF30\_nlo\_as\_0117\_nf6, NNPDF30\_nlo\_as\_0118\_nf6, NNPDF30\_nlo\_as\_0119\_nf6\\
\tt NNPDF30\_nnlo\_as\_0117\_nf6, NNPDF30\_nnlo\_as\_0118\_nf6, NNPDF30\_nnlo\_as\_0119\_nf6\\
\end{flushleft}
Likewise, we also provide similar fits with different maximum
number of active flavors for the LO fits, for the two
values of $\alpha_s$ used:
\begin{flushleft}
\tt NNPDF30\_lo\_as\_0118\_nf3,
NNPDF30\_lo\_as\_0118\_nf4,
NNPDF30\_lo\_as\_0118\_nf6, \\
NNPDF30\_lo\_as\_0130\_nf3,
NNPDF30\_lo\_as\_0130\_nf4,
NNPDF30\_lo\_as\_0130\_nf6 \\
\end{flushleft}

\item Fits based on reduced datasets

We have also released fits
based on reduced datasets,
as discussed in detail in Sect.~\ref{sec:resdataset}.
First,  conservative partons, for which the
fit with the choice of threshold value of $\alpha_{\rm max}=1.1$ is used.
We provide results in the restricted range of $\alpha_s$ values:
\begin{flushleft}
\small
\tt NNPDF30\_nlo\_as\_0117\_cons,
NNPDF30\_nlo\_as\_0118\_cons,
NNPDF30\_nlo\_as\_0119\_cons \\
\tt NNPDF30\_nnlo\_as\_0117\_cons,
NNPDF30\_nnlo\_as\_0118\_cons,
NNPDF30\_nnlo\_as\_0119\_cons \\
\end{flushleft}

Then, HERA-only PDFs, which may be useful to gauge
the impact of new LHC measurements:
\begin{flushleft}
\small
\tt NNPDF30\_nlo\_as\_0117\_hera,
NNPDF30\_nlo\_as\_0118\_hera,
NNPDF30\_nlo\_as\_0119\_hera \\
\tt NNPDF30\_nnlo\_as\_0117\_hera,
NNPDF30\_nnlo\_as\_0118\_hera,
NNPDF30\_nnlo\_as\_0119\_hera \\
\end{flushleft}
In view of possible reweighting applications, for the central value of
$\alpha_s(M_Z)=0.118$ also sets $N_{\rm rep}=1000$ are provided,
\begin{flushleft}
\tt NNPDF30\_nlo\_as\_0118\_hera\_1000\\
\tt NNPDF30\_nnlo\_as\_0118\_hera\_1000
\end{flushleft}

Next,  HERA+ATLAS and HERA+CMS fits.
These can be useful to compare with analogous fits presented by the LHC
collaborations, and are labeled as
\begin{flushleft}
\small
\tt NNPDF30\_nlo\_as\_0117\_atlas,
NNPDF30\_nlo\_as\_0118\_atlas,
NNPDF30\_nlo\_as\_0119\_atlas \\
\tt NNPDF30\_nnlo\_as\_0117\_atlas,
NNPDF30\_nnlo\_as\_0118\_atlas,
NNPDF30\_nnlo\_as\_0119\_atlas \\
\tt NNPDF30\_nlo\_as\_0117\_cms,
NNPDF30\_nlo\_as\_0118\_cms,
NNPDF30\_nlo\_as\_0119\_cms \\
\tt NNPDF30\_nnlo\_as\_0117\_cms,
NNPDF30\_nnlo\_as\_0118\_cms,
NNPDF30\_nnlo\_as\_0119\_cms
\end{flushleft}

Then, sets without LHC data,
sometimes used for  theory
comparisons with LHC measurements.
Also here, in view of possible reweighting exercises with LHC data,
we provide for the central value of $\alpha_s(M_Z)=0.118$ a set
of $N_{\rm rep}=1000$ replicas:
\begin{flushleft}
\small
\tt NNPDF30\_nlo\_as\_0117\_nolhc,
NNPDF30\_nlo\_as\_0118\_nolhc,
NNPDF30\_nlo\_as\_0119\_nolhc\\
NNPDF30\_nlo\_as\_0118\_nolhc\_1000 \\
\tt NNPDF30\_nnlo\_as\_0117\_nolhc,
NNPDF30\_nnlo\_as\_0118\_nolhc,
NNPDF30\_nnlo\_as\_0119\_nolhc\\
NNPDF30\_nnlo\_as\_0118\_nolhc\_1000 \\
\end{flushleft}

Finally, a fit excluding all jet data, which avoids the use of
approximate NNLO results.
Again, these are provided in a range of $\alpha_s$ values, and denoted by
\begin{flushleft}
\small
\tt NNPDF30\_nlo\_as\_0117\_nojet,
NNPDF30\_nlo\_as\_0118\_nojet,
NNPDF30\_nlo\_as\_0119\_nojet\\
\tt NNPDF30\_nnlo\_as\_0117\_nojet,
NNPDF30\_nnlo\_as\_0118\_nojet,
NNPDF30\_nnlo\_as\_0119\_nojet\\
\end{flushleft}

\end{itemize}
Note that for all  fits for which $N_{\rm rep}=1000$ replicas
are available, the corresponding $N_{\rm rep}=100$ replica sets have been
obtained using independent random seeds, so that the $N_{\rm
  rep}=1000$ and $N_{\rm rep}=100$ replica sets are fully independent.

Now that PDFs validated by a closure test, and fitted to a wide
variety of LHC data are available, three main directions of progress
are foreseen.
From a methodological point of view, some optimization is still
possible: specifically by producing  sets with a reduced number
of PDF replicas which retain as much as possible of the information
from the full dataset.
From the phenomenological point of view, new data will be added to the
dataset as
soon as they become available. These include
the final
combined HERA dataset, jet data from ATLAS and CMS from the 2011 and
2012 runs, and electroweak gauge boson production at 8 TeV, and, in
the longer term, LHC data from Run II will also be included.
Finally, from the theory point of view we plan to improve
the treatment of heavy quarks by introducing an intrinsic charm PDFs,
 to provide PDF sets based on resummed QCD theory, and to produce
 QCD+QED sets including a full fit of the photon PDF.

With data, methodology and theory under control the main outstanding
issue is a full characterization of theoretical
uncertainties on PDFs, which remains an open problem. It will have to
be addressed in order for parton distributions to become a controlled
tool for precision physics at present and future colliders.

\bigskip
\bigskip
\begin{center}
\rule{5cm}{.1pt}
\end{center}
\bigskip
\bigskip

{\bf\noindent  Acknowledgments \\}
We thank L.~Lyons for stressing the importance of closure testing of
parton distributions, and A.~de~Roeck, A.~David, J.~Huston,
M.~Mangano, A.~Mitov, P.~Nadolsky,
G.~Passarino, G.~Salam, R.~Tanaka,
R.~Thorne and G.~Watt for many illuminating discussions.
We thank F.~Petriello for his help in running {\sc\small FEWZ}
and comparing the results and G.~Ferrera
for his help in setting up {\sc\small DYNNLO}.
We are grateful to S.~Glazov for support with the HERA-II H1 data;
A.~Cooper-Sarkar and I.~Abt for support with the ZEUS
HERA-II data; S.~Glazov, C.~Gwenlan and A.~Tricoli
for information about the
ATLAS measurements; J.~Alcaraz,
J.~Berryhill, M.~Cepeda, M.~Gouzevitch, I.~Josa, K.~Lipka and K.~Rabbertz for
assistance with the CMS data; and D.~Ward,
J.~Anderson, K.~Mueller, R.~McNulty
and T.~Shears for help with the LHCb electroweak measurements.

\noindent
V.~B. is supported by the ERC grant 291377,
“LHCtheory: Theoretical predictions and analyses of LHC physics:
advancing the precision frontier”.
S.~F. and S.~C. are  supported in part by an Italian PRIN2010 grant,
    by a
    European Investment Bank  EIBURS grant, and by the European Commission
    through
    the HiggsTools Initial Training
    Network PITN-GA-2012-316704.
J.~R. is supported by an STFC Rutherford Fellowship ST/K005227/1.
J.~R. and N.~H. are
supported by an European Research Council Starting Grant "PDF4BSM".
R.~D.~B. and L.~D.~D. are funded by an STFC Consolidated Grant
ST/J000329/1.



\appendix

\section{QCD and weak corrections to vector boson production data}
\label{app-ew}

In this appendix we provide a detailed overview of the theoretical
calculations that we have used to include
LHC vector boson production data in the NNPDF3.0 analysis.
For each datapoint, we provide the size of the NNLO QCD
and of the  NLO weak corrections,
comparing with the total experimental uncertainty.
Results for the QCD and weak corrections are given for all the
LHC neutral current Drell-Yan  measurements included in NNPDF3.0,
namely for the CMS double differential distributions from the 2011
run~\cite{CMSDY},
the ATLAS high mass Drell-Yan distributions from the 2011 run
~\cite{Aad:2013iua}, the ATLAS
$Z$ boson rapidity distribution from the 2010 run~\cite{Aad:2011dm},
and for LHCb $Z$ rapidity distributions
in the forward region from the 2011 run~\cite{Aaij:2012mda}.
Using these calculations, here we also provide, for the neutral-current
Drell-Yan data, the NNLO and weak $C$-factors,
defined as in Sect.~\ref{sec:theorytools}.

As we will show now,
the effect of the pure weak corrections is negligible for
neutral-current Drell-Yan data
around $Z$ peak region, and it becomes numerically important
for small and large values of the dilepton mass $M_{ll}$.
For charged-current Drell-Yan production, weak corrections are small
outside the $W$ peak region (where all the available
data lies) and thus can be safely neglected
in the present analysis.
For charged-current DY, NNLO QCD corrections are moderate;
they have been included in the NNPDF3.0 analysis but
are not discussed explicitly in this Appendix.

In this work,
theoretical predictions
are obtained with {\sc\small FEWZ3.1}~\cite{Li:2012wna}
and have been cross-checked against {\sc\small DYNNLO1.3}~\cite{DYNNLOurl}.
The NNPDF2.3 NLO set with $\alpha_s=0.118$ is used as  input
for the NLO theoretical predictions,
and  NNPDF2.3 NNLO is used for the corresponding NNLO calculations,
as well as to determine the $C$--factors.
The $G_{\mu}$ scheme is adopted for all electroweak computations, where the input parameters
are $M_Z$, $M_W$ and $G_F$, while $\alpha_e$
and $s_W$ may be obtained from these using:
\begin{equation}
\label{eq:ew}
s_W=1-\frac{M_W^2}{M_Z^2} \qquad \alpha_e = R_t^2\,G_F M_W^2\,\frac{s_W}{\pi},
\end{equation}
where $R_t^2=1.4142135624$. In this analysis we use $G_F=1.16637 \cdot 10^{-5}$ GeV$^{-2}$ , $M_W=80.398$ GeV,
$M_Z=91.1876$ GeV and do not adopt the narrow-width approximation.

\subsection{CMS double differential distributions}

The CMS experiment has measured~\cite{CMSDY}
the lepton pair rapidity distributions $y_{ll}$ for the Drell-Yan process
in six bins of the invariant mass of the final state lepton pair, $M_{ll}$.
 For each bin in $M_{ll}$, the  $y_{ll}$
distribution is divided in 24 equally spaced bins
in rapidity, up to  $y_{ll}<2.4$,
except for the last bin in $M_{ll}$ that is divided in
12 bins.
In the theoretical calculations,
we set $\mu_F$ and $\mu_R$ to the average
value  $\la M_{ll}\ra$ of each dilepton invariant mass bin.
The cuts on the final state leptons are the following:
\begin{eqnarray*}
\centering
&&p_T^{\rm min}>9 \,{\rm GeV} \,\, \wedge \,\,p_T^{\rm max}> 14 \, {\rm GeV} \, ,\\
&&20\,{\rm GeV}\, < M_{ll} < \, 1500 \,{\rm GeV} \, , \\
&& |\eta^{1,2}_{l}| < 2.4 \, ,
\end{eqnarray*}
with $p_T^{\rm min}$ ($p_T^{\rm max}$) being the transverse momentum of the
softer (harder) lepton, and $|\eta_{l}|$ are the rapidities
of the two leptons.
In Tabs.~\ref{tab:bin1}-\ref{tab:bin6} we
provide the results of our calculations, and
 compare
the total experimental uncertainty with the corresponding NNLO
QCD and the NLO pure weak corrections, for the six
bins in invariant mass of this measurement.

The NNLO corrections to this measurement are found
to be substantial, specially in the bins with $M_{ll}$ below
the $Z$ peak region.
For example, the average NNLO correction is around 10\% in the first bin,
$\sim$6\% in the second bin, and and $\sim$4\% in the $Z$ peak
region.
These corrections are also found to be relatively constant
as a function of the dilepton rapidity $|y_{ll}|$, except
for the last bin of the distribution, where the correction
is substantially larger.
The comparison between data and theory predictions for the first
bin of this measurement was shown in Fig~\ref{fig:cfactEWandNNLO}.
In the region with dilepton mass $M_{ll}\le 45$ GeV, the NNLO
QCD corrections are larger than the experimental uncertainties,
and therefore their inclusion is essential to achieve
a good fit quality.

Concerning the pure weak corrections, 6th column of  Tabs.~\ref{tab:bin1}-\ref{tab:bin6},
we note that it is found to be at most at the few percent level,
for example, the average over all rapidity bins is $\sim 3$\% for the
second invariant mass bin and $\sim 4$\% for the third.
Above the $Z$ peak mass we find these weak corrections to be below the
1\% range.
Note that in the highest invariant mass bin, with $200\le M_{ll} \le$ 1500 GeV,
the cross-sections are dominated by the region around $M_{ll}\sim 200$ GeV,
while the contribution from higher masses, where weak effects are known
to be more substantial, has less weight.

\begin{table}[htb]
\small
\begin{center}
\begin{tabular}{|c|c|c|c|c|c|}
\hline
\multicolumn{6}{|c|}{20 GeV $<M_{ll}<30$ GeV}\\
\hline
$|y_{ll}|$ & $d\sigma^{\rm exp}/dy_{ll}/dM_{ll}$ (pb)  & $d\sigma^{\rm NLO}/dy_{ll}/dM_{ll}$ (pb) & $\Delta_{\rm exp}$ (\%)  &  $\Delta_{\rm NNLO}$ (\%) & $\Delta_{\rm pure EW}$ (\%) \\
\hline
 $[0.0,0.1]$ &	17.84  &  14.83 &  6.5	&  8.1	&  0.4	 \\
 $[0.1,0.2]$ &  17.68 &  14.78 &  6.2	&  8.9	&  0.6	 \\
 $[0.2,0.3]$ &	17.21  &  14.80 &  6.4	&  9.0	&  0.3	 \\
 $[0.3,0.4]$ &	17.63  &  14.78 &  6.0	&  8.9	&  0.4	 \\
 $[0.4,0.5]$ &	17.84  &  14.70 &  5.6	&  8.8	&  0.4	 \\
 $[0.5,0.6]$ &	18.10 &  14.71 &  5.2  &  8.8	&  0.2	 \\
 $[0.6,0.7]$ &	18.41 &  14.59 &  4.8  &  8.9  &  0.4	 \\
 $[0.7,0.8]$ &	18.16 &  14.54 &  4.3	&  9.2	&  0.2	 \\
 $[0.8,0.9]$ &	18.05  &  14.46 &  3.8	&  9.4	&  0.6	 \\
 $[0.9,1.0]$ &	17.84  &  14.36 &  3.5  &  9.6	&  0.5	 \\
 $[1.0,1.1]$ &	17.52 &  14.31 &  3.3  &  9.7	&  0.3	 \\
 $[1.1,1.2]$ &	17.47 &  14.16 &  3.3  &  9.8	&  0.4	 \\
 $[1.2,1.3]$ &	16.90 &  14.07 &  3.4  &  9.9	&  0.4	 \\
 $[1.3,1.4]$ &	16.95 &  13.93 &  4.0  &  10.1 &  0.1	 \\
 $[1.4,1.5]$ &	16.32 &  13.75 &  4.5  &  10.4 &  0.5	 \\
 $[1.5,1.6]$ &	16.48 &  13.60 &  4.5  &  10.7 &  0.1	 \\
 $[1.6,1.7]$ &	15.48 &  13.25 &  4.7  &  10.7 &  0.2	 \\
 $[1.7,1.8]$ &	15.22 &  12.84 &  5.2  &  10.3 &  0.4	 \\
 $[1.8,1.9]$ &	14.06 &  12.25 &  5.2  &  9.2	&  0.1	 \\
 $[1.9,2.0]$ &	12.59 &  11.24 &  5.4  &  7.6	&  0.8	 \\
 $[2.0,2.1]$ &	11.02 &  9.58	&  5.2  &  6.1	&  0.8	 \\
 $[2.1,2.2]$ & 8.39  &   7.28  &  5.6  &  6.0  &  0.6   \\
 $[2.2,2.3]$ & 5.46  &   4.58  &  7.7  &  9.5  &  0.3  \\
 $[2.3,2.4]$ & 2.05  &   1.48  & 10.3  &  13.4 &  0.4 \\
\hline
 Average	  &            &	& 5.2  & 9.3 &  0.6 \\
\hline
\end{tabular}
\end{center}
\caption{\small \label{tab:bin1}
Experimental measurements and theoretical
predictions for the CMS Drell-Yan double differential
distributions in the dilepton invariant mass bin with
 20 GeV $<M_{ll}<30$ GeV, as a function of the
dilepton rapidity $|y_{ll}|$.
For each rapidity bin we provide the experimental central value
(2nd column),
the NLO theoretical predictions obtained with
NNPDF2.3NLO with $\alpha_s=0.118$ (3rd column),
the total percentage experimental uncertainty (4th column),
the NNLO QCD correction (5th column) and the NLO pure weak
correction (6th column).
 In the last row the average over
all the rapidity bins is provided.
See text for more details.
}
\end{table}


\begin{table}[htb]
\small
\begin{center}
\begin{tabular}{|c|c|c|c|c|c|}
\hline
\multicolumn{6}{|c|}{30 GeV $<M_{ll}<45$ GeV}\\
\hline
$|y_{ll}|$ & $d\sigma^{\rm exp}/dy_{ll}/dM_{ll}$ (pb)  & $d\sigma^{\rm NLO}/dy_{ll}/dM_{ll}$ (pb) & $\Delta_{\rm exp}$ (\%)  &  $\Delta_{\rm NNLO}$ (\%) & $\Delta_{\rm pure EW}$ (\%) \\
\hline
  $[0.0,0.1]$ &  26.87 & 25.36 &  3.5 &  6.7 &  3.1\\
  $[0.1,0.2]$ &  25.66 & 25.42 &  3.3 &  6.4 &  3.0\\
  $[0.2,0.3]$ &  25.87 & 25.37 &  3.0 &  6.2 &  2.7\\
  $[0.3,0.4]$ &  25.66 & 25.45 &  3.1 &  6.1 &  2.6\\
  $[0.4,0.5]$ &  26.13 & 25.28 &  2.8 &  6.1 &  3.0\\
  $[0.5,0.6]$ &  27.08 & 25.22 &  2.7 &  6.0 &  2.9\\
  $[0.6,0.7]$ &  26.18 & 25.29 &  2.8 &  5.8 &  2.7\\
  $[0.7,0.8]$ &  25.66 & 25.19 &  3.1 &  5.6 &  2.8\\
  $[0.8,0.9]$ &  26.29 & 25.06 &  3.0 &  5.3 &  2.8\\
  $[0.9,1.0]$ &  25.61 & 24.95 &  2.9 &  5.2 &  3.0\\
  $[1.0,1.1]$ &  25.71 & 24.93 &  2.9 &  5.2 &  2.6\\
  $[1.1,1.2]$ &  25.55 & 24.72 &  2.7 &  5.3 &  2.7\\
  $[1.2,1.3]$ &  24.50 & 24.33 &  2.8 &  5.5 &  3.6\\
  $[1.3,1.4]$ &  25.76 & 24.21 &  2.6 &  5.8 &  2.4\\
  $[1.4,1.5]$ &  24.08 & 23.30 &  2.6 &  5.8 &  3.0\\
  $[1.5,1.6]$ &  23.35 & 22.19 &  2.7 &  5.6 &  2.9\\
  $[1.6,1.7]$ &  21.67 & 20.58 &  2.7 &  4.9 &  3.8\\
  $[1.7,1.8]$ &  19.26 & 18.63 &  2.7 &  3.6 &  2.8\\
  $[1.8,1.9]$ &  16.58 & 16.13 &  3.2 &  1.6 &  3.1\\
  $[1.9,2.0]$ &  13.80 & 13.21 &  3.4 &  0.9 &  3.7\\
  $[2.0,2.1]$ &  11.02 & 10.30 &  4.3 &  3.7 &  3.4\\
  $[2.1,2.2]$ &   7.87 & 7.41 &  5.3 &  5.4 &  3.7\\
  $[2.2,2.3]$ &   4.67 & 4.43 &  6.7 &  3.2 &  4.3\\
  $[2.3,2.4]$ &   1.63 & 1.49 &  9.6 & 17.5 &  6.2\\
\hline
 Average  &     & 	& 3.5  &  5.6 & 3.2 \\
\hline
\end{tabular}
\end{center}
\caption{\small \label{tab:bin2}
Same as Tab.~\ref{tab:bin1}  for the dilepton invariant mass bin with
 30 GeV $<M_{ll}<45$ GeV.}
\end{table}


\begin{table}[htb]
\small
\begin{center}
\begin{tabular}{|c|c|c|c|c|c|}
\hline
\multicolumn{6}{|c|}{45 GeV $<M_{ll}<60$ GeV}\\
\hline
$|y_{ll}|$ & $d\sigma^{\rm exp}/dy_{ll}/dM_{ll}$ (pb)  & $d\sigma^{\rm NLO}/dy_{ll}/dM_{ll}$ (pb) & $\Delta_{\rm exp}$ (\%)  &  $\Delta_{\rm NNLO}$ (\%) & $\Delta_{\rm pure EW}$ (\%) \\
\hline
  $[0.0,0.1]$ &  11.23 & 10.51 &  2.8 &  5.7 &  2.9\\
  $[0.1,0.2]$ &  11.18 & 10.51 &  2.8 &  4.6 &  3.3\\
  $[0.2,0.3]$ &  11.07 & 10.51 &  3.3 &  3.9 &  2.7\\
  $[0.3,0.4]$ &  11.33 & 10.50 &  3.2 &  3.4 &  3.3\\
  $[0.4,0.5]$ &  10.65 & 10.46 &  3.9 &  3.1 &  2.8\\
  $[0.5,0.6]$ &  11.70 & 10.50 &  2.7 &  3.0 &  2.4\\
  $[0.6,0.7]$ &  11.49 & 10.41 &  2.7 &  3.0 &  3.4\\
  $[0.7,0.8]$ &  11.28 & 10.39 &  3.3 &  3.0 &  3.0\\
  $[0.8,0.9]$ &  11.28 & 10.30 &  2.8 &  3.0 &  3.4\\
  $[0.9,1.0]$ &  11.18 & 10.21 &  2.8 &  2.9 &  2.9\\
  $[1.0,1.1]$ &  10.76 & 10.02 &  2.9 &  2.8 &  2.7\\
  $[1.1,1.2]$ &  10.28 & 9.73 &  3.1 &  2.6 &  3.0\\
  $[1.2,1.3]$ &   9.76 & 9.25 &  3.2 &  2.2 &  3.3\\
  $[1.3,1.4]$ &   9.81 & 8.69 &  2.7 &  1.8 &  3.5\\
  $[1.4,1.5]$ &   8.92 & 8.15 &  2.9 &  1.2 &  3.0\\
  $[1.5,1.6]$ &   8.40 & 7.48 &  3.8 &  0.5 &  3.3\\
  $[1.6,1.7]$ &   7.35 & 6.79 &  2.9 &  0.5 &  2.6\\
  $[1.7,1.8]$ &   6.40 & 5.99 &  3.3 &  1.6 &  3.0\\
  $[1.8,1.9]$ &   5.93 & 5.14 &  4.4 &  3.1 &  3.2\\
  $[1.9,2.0]$ &   4.93 & 4.28 &  5.3 &  4.8 &  3.0\\
  $[2.0,2.1]$ &   4.04 & 3.34 &  6.5 &  6.9 &  3.6\\
  $[2.1,2.2]$ &   2.57 & 2.42 &  6.1 &  9.4 &  2.5\\
  $[2.2,2.3]$ &   1.78 & 1.44 & 11.8 & 12.7 &  3.7\\
  $[2.3,2.4]$ &   0.63 & 0.50 & 16.7 & 18.1 &  0.4\\
\hline
 Average	  &     & 	& 4.4  &  4.3 & 3.9 \\
\hline
\end{tabular}
\end{center}
\caption{\small \label{tab:bin3}
Same as Tab.~\ref{tab:bin1}  for the dilepton invariant mass bin with
 45 GeV $<M_{ll}<60$ GeV.}
\end{table}


\begin{table}[htb]
\small
\begin{center}
\begin{tabular}{|c|c|c|c|c|c|}
\hline
\multicolumn{6}{|c|}{60 GeV $<M_{ll}<120$ GeV}\\
\hline
$|y_{ll}|$ & $d\sigma^{\rm exp}/dy_{ll}/dM_{ll}$ (pb)  & $d\sigma^{\rm NLO}/dy_{ll}/dM_{ll}$ (pb) & $\Delta_{\rm exp}$ (\%)  &  $\Delta_{\rm NNLO}$ (\%) & $\Delta_{\rm pure EW}$ (\%) \\
\hline
  $[0.0,0.1]$ & 317.44 & 294.30 &  1.3 &  2.2 &  1.1\\
  $[0.1,0.2]$ & 315.35 & 291.35 &  1.3 &  2.1 &  2.1\\
  $[0.2,0.3]$ & 315.87 & 291.02 &  1.3 &  2.0 &  1.6\\
  $[0.3,0.4]$ & 312.72 & 289.59 &  1.3 &  1.9 &  1.3\\
  $[0.4,0.5]$ & 312.72 & 286.10 &  1.3 &  1.8 &  1.6\\
  $[0.5,0.6]$ & 309.57 & 282.63 &  1.4 &  1.7 &  1.8\\
  $[0.6,0.7]$ & 306.95 & 276.41 &  1.4 &  1.5 &  1.6\\
  $[0.7,0.8]$ & 299.60 & 270.50 &  1.4 &  1.4 &  2.1\\
  $[0.8,0.9]$ & 292.26 & 264.98 &  1.4 &  1.2 &  0.9\\
  $[0.9,1.0]$ & 285.96 & 255.89 &  1.5 &  1.0 &  1.8\\
  $[1.0,1.1]$ & 275.99 & 245.52 &  1.5 &  0.8 &  2.3\\
  $[1.1,1.2]$ & 264.45 & 237.54 &  1.6 &  0.6 &  0.9\\
  $[1.2,1.3]$ & 249.76 & 223.67 &  1.7 &  0.3 &  1.4\\
  $[1.3,1.4]$ & 234.02 & 208.77 &  1.6 &  0.1 &  2.2\\
  $[1.4,1.5]$ & 219.33 & 193.84 &  1.7 &  0.2 &  1.5\\
  $[1.5,1.6]$ & 196.76 & 176.08 &  1.6 &  0.5 &  2.1\\
  $[1.6,1.7]$ & 177.87 & 156.71 &  1.8 &  0.9 &  2.3\\
  $[1.7,1.8]$ & 153.74 & 139.25 &  2.0 &  1.3 &  1.2\\
  $[1.8,1.9]$ & 131.70 & 117.22 &  2.0 &  1.8 &  1.8\\
  $[1.9,2.0]$ & 106.51 & 96.52 &  2.5 &  2.4 &  1.4\\
  $[2.0,2.1]$ &  82.38 & 74.07 &  3.2 &  3.3 &  2.5\\
  $[2.1,2.2]$ &  57.19 & 52.84 &  3.7 &  4.8 &  0.3\\
  $[2.2,2.3]$ &  33.58 & 31.05 &  6.3 &  8.5 &  1.9\\
  $[2.3,2.4]$ &  10.49 & 10.40 &  5.0 & 44.5 &  0.1\\
\hline
Average	  &    & & 2.1  &  3.6 & 1.6 \\
\hline
\end{tabular}
\end{center}
\caption{\small \label{tab:bin4}
Same as Tab.~\ref{tab:bin1}  for the dilepton invariant mass bin with
 60 GeV $<M_{ll}<120$ GeV.}
\end{table}


\begin{table}[htb]
\small
\begin{center}
\begin{tabular}{|c|c|c|c|c|c|}
\hline
\multicolumn{6}{|c|}{120 GeV $<M_{ll}<200$ GeV}\\
\hline
$|y_{ll}|$ & $d\sigma^{\rm exp}/dy_{ll}/dM_{ll}$ (pb)  & $d\sigma^{\rm NLO}/dy_{ll}/dM_{ll}$ (pb) & $\Delta_{\rm exp}$ (\%)  &  $\Delta_{\rm NNLO}$ (\%) & $\Delta_{\rm pure EW}$ (\%) \\
\hline
  $[0.0,0.1]$ &   3.47 & 3.24 &  4.7 &  1.0 &   0.8\\
  $[0.1,0.2]$ &   3.61 & 3.22 &  4.5 &  0.5 &   0.3\\
  $[0.2,0.3]$ &   3.54 & 3.21 &  4.8 &  0.0 &   0.8\\
  $[0.3,0.4]$ &   3.52 & 3.17 &  4.5 &  0.3 &   0.0\\
  $[0.4,0.5]$ &   3.38 & 3.15 &  4.2 &  0.5 &   0.7\\
  $[0.5,0.6]$ &   3.52 & 3.10 &  4.2 &  0.5 &   0.2\\
  $[0.6,0.7]$ &   3.31 & 3.05 &  4.6 &  0.4 &   0.4\\
  $[0.7,0.8]$ &   3.32 & 2.99 &  4.1 &  0.2 &   0.4\\
  $[0.8,0.9]$ &   3.22 & 2.92 &  4.6 &  0.1 &   0.1\\
  $[0.9,1.0]$ &   3.05 & 2.83 &  4.1 &  0.0 &   0.4\\
  $[1.0,1.1]$ &   3.03 & 2.74 &  4.0 &  0.1 &   0.7\\
  $[1.1,1.2]$ &   2.79 & 2.61 &  4.1 &  0.2 &   0.2\\
  $[1.2,1.3]$ &   2.71 & 2.49 &  4.2 &  0.5 &   0.4\\
  $[1.3,1.4]$ &   2.52 & 2.34 &  4.4 &  0.8 &   0.4\\
  $[1.4,1.5]$ &   2.44 & 2.16 &  4.5 &  1.3 &   0.0\\
  $[1.5,1.6]$ &   2.19 & 1.98 &  5.0 &  1.7 &   0.5\\
  $[1.6,1.7]$ &   1.95 & 1.76 &  5.4 &  2.2 &   0.5\\
  $[1.7,1.8]$ &   1.58 & 1.54 &  6.3 &  2.4 &   0.4\\
  $[1.8,1.9]$ &   1.25 & 1.29 &  7.5 &  2.4 &   0.6\\
  $[1.9,2.0]$ &   1.16 & 1.05 &  8.6 &  1.8 &   0.4\\
  $[2.0,2.1]$ &   0.73 & 0.80 & 10.8 &  0.4 &   0.2\\
  $[2.1,2.2]$ &   0.53 & 0.56 & 13.9 &  2.3 &   0.5\\
  $[2.2,2.3]$ &   0.28 & 0.32 & 18.7 &  6.8 &   2.1\\
  $[2.3,2.4]$ &   0.11 & 0.10 & 24.8 & 13.4 &   0.6\\
\hline
 Average  &           & 	& 6.9  &  1.7 & 0.5 \\
\hline
\end{tabular}
\end{center}
\caption{\small \label{tab:bin5}
Same as Tab.~\ref{tab:bin1}  for the dilepton invariant mass bin with
 120 GeV $<M_{ll}<200$ GeV.}
\end{table}


\begin{table}[htb]
\small
\begin{center}
\begin{tabular}{|c|c|c|c|c|c|}
\hline
\multicolumn{6}{|c|}{200 GeV $<M_{ll}<1500$ GeV}\\
\hline
$|y_{ll}|$ & $d\sigma^{\rm exp}/dy_{ll}/dM_{ll}$ (pb)  & $d\sigma^{\rm NLO}/dy_{ll}/dM_{ll}$ (pb) & $\Delta_{\rm exp}$ (\%)  &  $\Delta_{\rm NNLO}$ (\%) & $\Delta_{\rm pure EW}$ (\%) \\
\hline
  $[0.0,0.2]$ &  0.530 & 0.533 & 13.8 &  1.3 &  0.0\\
  $[0.2,0.4]$ &  0.609 & 0.527 & 10.3 &  1.6 &  0.0\\
  $[0.4,0.6]$ &  0.651 & 0.518 &  8.0 &  1.5 &  0.1\\
  $[0.6,0.8]$ &  0.522 & 0.501 &  9.2 &  2.0 &  0.0\\
  $[0.8,1.0]$ &  0.534 & 0.476 &  7.7 &  2.5 &  0.1\\
  $[1.0,1.2]$ &  0.527 & 0.439 &  7.4 &  0.8 &  0.2\\
  $[1.2,1.4]$ &  0.476 & 0.388 &  6.9 &  3.4 &  0.1\\
  $[1.4,1.6]$ &  0.323 & 0.319 &  9.0 &  2.6 &  0.1\\
  $[1.6,1.8]$ &  0.247 & 0.238 & 10.5 &  0.5 &  0.3\\
  $[1.8,2.0]$ &  0.152 & 0.154 & 14.5 &  2.2 &  0.0\\
  $[2.0,2.2]$ &  0.097 & 0.078 & 20.6 &  4.4 &  0.5\\
  $[2.2,2.4]$ &  0.022 & 0.021 & 36.4 & 11.4 &  0.6\\
\hline
 Average	  &    & 	& 12.9  &  2.9 & 0.2 \\
\hline
\end{tabular}
\end{center}
\caption{\small \label{tab:bin6}
Same as Tab.~\ref{tab:bin1}  for the dilepton invariant mass bin with
 200 GeV $<M_{ll}<1200$ GeV.}
\end{table}

\clearpage

\subsection{ATLAS high-mass Drell-Yan differential cross section}

Now we turn to present the results for
the ATLAS high-mass Drell-Yan production data
from the 2011 run~\cite{Aad:2013iua}, based on an integrated luminosity of 4.9 fb$^{-1}$.
This measurement is presented
in terms of the invariant mass of the electron pairs produced
in the range $116 \,{\rm GeV} < M_{ll} <1.5$~TeV.
Theoretical predictions are computed using the same codes
and settings described above, we have used as renormalization
and factorization scales
 $\mu_F=\mu_R=M_Z$ and the following cuts to the final
states leptons have been applied:
\begin{eqnarray*}
\centering
&&p_T^{\rm l}>25 \,{\rm GeV} \, ,\\
&& |\eta^{1,2}_{l}| < 2.5 \, .
\end{eqnarray*}
In Tab.~\ref{tab:atlasZmass}, with the same structure as
Tab.~\ref{tab:bin1}, we provide the experimental central value,
the NLO theoretical predictions obtained with
NNPDF2.3NLO with $\alpha_s=0.118$,
the total percentage experimental uncertainty,
the NNLO QCD correction, and the NLO pure weak
correction.
 In the last row, the average over
all the data points in the measurement is given.

As expected, the size of the pure weak corrections increases
monotonically with the value of $M_{ll}$, reaching up to $\sim 7$\%
in the highest mass bin.
While the uncertainties in experimental data are still rather higher
due to the limited statistics, is clear that for 8 TeV data and
even more for Run II measurements, including electroweak corrections
in PDf fits will be mandatory.
On the other hand, QCD NNLO corrections are found to be at the few
percent level, typically smaller than the total
experimental uncertainty.

\begin{table}[ht]
\small
 \begin{tabular}{|c|c|c|c|c|c|}
 \hline
 \multicolumn{6}{|c|}{ATLAS 2011 DY invariant mass distribution}\\
\hline
$M_{ll}$ & $d\sigma^{\rm exp}/dM_{ll}$ (pb)  & $d\sigma^{\rm NLO}/dM_{ll}$ (pb) & $\Delta_{\rm exp}$ (\%)  &  $\Delta_{\rm NNLO}$ (\%) & $\Delta_{\rm pure EW}$ (\%) \\
\hline
 $[116, 130]$    &     224.00   &    212.78  &       4.3   &       0.9  &        0.7\\
 $[130, 150]$    &     102.00   &     91.60  &       4.5   &       1.3  &        0.4\\
 $[150, 170]$    &      51.20   &     45.15  &       5.0   &       1.5  &        0.1\\
 $[170, 190]$    &      28.40   &     25.55  &       5.4   &       2.8  &        0.1\\
 $[190, 210]$    &      18.70   &     15.76  &       6.1   &       3.4  &        0.2\\
 $[210, 230]$    &      10.70   &     10.28  &       7.5   &       1.6  &        0.1\\
 $[230, 250]$    &       8.23   &      7.22  &       7.9   &       0.4  &        0.2\\
 $[250, 300]$    &       4.66   &      3.93  &       7.2   &       0.0  &        0.0\\
 $[300, 400]$    &       1.70   &      1.40  &       7.8   &       3.4  &        0.3\\
 $[400, 500]$    &       0.47   &      0.45  &      11.3   &       4.7  &        1.0\\
 $[500,700]$    &       0.15   &      0.12  &      12.4   &       5.0  &        1.5\\
 $[700,1000]$    &     0.0221   &    0.0196  &      25.1   &       9.0  &        3.4\\
$[1000, 1500]$   &     0.0029   &    0.0022  &      51.0   &       1.0  &        6.8\\
\hline
 Average       &              & 	   & 12.0        &  2.7 & 1.2 \\
\hline
 \end{tabular}
  \caption{\label{tab:atlasZmass}
Same as Tab.~\ref{tab:bin1} for the ATLAS measurement of the high-mass
Drell-Yan invariant dilepton mass distribution from the 2011 dataset.
}
 \end{table}

\subsection{ATLAS Z rapidity distribution}

Now we consider the
ATLAS measurements of the $Z$ rapidity
distribution from an integrated
luminosity of $36~{\rm pb}^{-1}$~\cite{Aad:2011dm}.
This dataset was already included in our previous
NNPDF2.3 analysis~\cite{Ball:2012cx}.
In the theoretical calculations, we have used
the following cuts for the final-state lepton kinematics:
\begin{align*}
\centering
& p_T^{l} \ge 20~ {\rm GeV},\\
&66~{\rm GeV} \le M_{ll} \le 116~{\rm GeV}, \\
&|\eta^{1,2}_{l}| \le 4.9.
\end{align*}
In Tab.~\ref{tab:atlasZrap}, with again the same structure as
Tab.~\ref{tab:bin1}, we provide
for this measurement
the experimental central value,
the NLO theoretical predictions obtained with
NNPDF2.3NLO with $\alpha_s=0.118$,
the total percentage experimental uncertainty,
the NNLO QCD correction, and the NLO pure weak
correction.

From  Tab.~\ref{tab:atlasZrap} we see that
in the $Z$ peak region, both the NNLO QCD and
NLO weak corrections are small, of the order
1\% at most, and slowly varying with the lepton rapidity.
On the other hand, the precision of the data is high and therefore
it is necessary to include these corrections in the theory
calculation, as has been done in this work.
This is specially true in the central region, with
$|y_{ll}|\lsim 1.6$.

\begin{table}[ht]
\begin{center}
\small
 \begin{tabular}{|c|c|c|c|c|c|}
 \hline
 \multicolumn{6}{|c|}{ATLAS $Z$ rapidity distribution}\\
\hline
$|y_{ll}|$ & $d\sigma^{\rm exp}/dy_{ll}$ (pb)  & $d\sigma^{\rm NLO}/dy_{ll}$ (pb) & $\Delta_{\rm exp}$ (\%)  &  $\Delta_{\rm NNLO}$ (\%) & $\Delta_{\rm pure EW}$ (\%) \\
\hline
 $[0.0, 0.4]$     &    129.27    &   123.44  & 1.9 &       0.7  &        0.9\\
 $[0.4, 0.8]$     &    129.44    &   122.22  & 1.9 &       0.9  &        0.7\\
 $[0.8, 1.2]$     &    125.81    &   119.98  & 1.8 &       0.8  &        0.7\\
 $[1.2, 1.6]$     &    118.23    &   116.58  & 1.9 &       0.7  &        0.8\\
 $[1.6, 2.0]$     &    113.37    &   112.07  & 2.3 &       0.6  &        0.6\\
 $[2.0, 2.4]$     &    105.26    &   105.32  & 3.7 &       0.6  &        0.9\\
 $[2.4, 2.8]$     &     92.18    &    92.63  & 6.3 &      0.9  &        0.7\\
 $[2.8, 3.6]$     &     53.38    &    54.93  & 10.1 &     2.2  &        0.7\\
\hline
 Average	& 	       &           & 3.7 &        0.9 &        0.8 \\
\hline
 \end{tabular}
 \caption{\label{tab:atlasZrap}
Same as Tab.~\ref{tab:bin1} for the ATLAS measurement of
the rapidity distributions of $Z$ bosons from the 2010 dataset.
}
\end{center}
 \end{table}

\subsection{LHCb Z forward rapidity distribution}

Finally, we present here the  theoretical predictions for the
LHCb $Z\to ee$ rapidity distributions in the forward region from the 2011
dataset~\cite{Aaij:2012mda}.
In the calculation,  the following cuts on the lepton kinematics have been used,
\begin{align*}
\centering
& p_T^{l} \ge 20~ {\rm GeV},\\
& 60~{\rm GeV} \le M_{ll} \le 120~{\rm GeV},\\
& 2 \le \eta^{1,2}_{l} \le 4.5.
\end{align*}
Results are shown in Tab.~\ref{tab:lhcbZrap}, with the usual
structure.
As in the previous case of the
ATLAS measurement of the $Z$ rapidity
distribution, and as we expect for calculations
in the the $Z$ peak region, the NLO pure
weak corrections  are found to be quite small, of order
1\% at most, and slowly varying with the lepton rapidity.
On the the hand the
NNLO QCD corrections are more important, and interestingly
they increase monotonically with the rapidity of the
dilepton system, reaching up to $\sim 4$\% in the most forward bin.
The experimental uncertainties for this measurement in any case are
larger than the NNLO QCD corrections.

\begin{table}[ht]
\begin{center}
\small
 \begin{tabular}{|c|c|c|c|c|c|}
 \hline
 \multicolumn{6}{|c|}{LHCb $Z\to e^+e^-$ forward rapidity distribution}\\
\hline
$y_{ll}$ & $d\sigma^{\rm exp}/dy_{ll}$ (pb)  & $d\sigma^{\rm NLO}/dy_{ll}$ (pb) & $\Delta_{\rm exp}$ (\%)  &  $\Delta_{\rm NNLO}$ (\%) & $\Delta_{\rm pure EW}$ (\%) \\
\hline
$[2.00, 2.25]$ &         13.6   &      13.2   & 6.6  &    1.2   &       1.2\\
$[2.25, 2.50]$ &         39.4   &      36.7   & 3.6  &    0.8   &       0.4\\
$[2.50, 2.75]$ &         56.7   &      51.8   & 3.4  &    0.2   &       0.6\\
$[2.75, 3.00]$ &         63.2   &      61.0   & 3.3  &    0.5   &       0.7\\
$[3.00, 3.25]$ &         59.9   &      56.1   & 3.8  &    1.2   &       0.6\\
$[3.25, 3.50]$ &         43.8   &      38.0   & 4.3  &    1.9   &       0.3\\
$[3.50, 3.75]$ &         20.5   &      17.3   & 6.8  &    2.6   &       0.6\\
$[3.75, 4.00]$ &          5.9   &       4.6   & 16.9  &    3.0   &       0.1\\
$[4.00, 4.25]$ &         0.66   &       0.5   & 80.3  &    4.0   &       0.1\\
\hline
 Average	& 	       &           & 14.3 &        1.7 &        0.5 \\
\hline
 \end{tabular}
 \caption{\label{tab:lhcbZrap}
\small Same as Tab.~\ref{tab:bin1} for the LHCb measurement of
the rapidity distributions of $Z$ bosons in
the forward region from the 2011 dataset.}
\end{center}
 \end{table}

\section{Distance estimators}
\label{app:distances}

In previous publications we have extensively used, for comparisons
between PDFs, the distance between two
NNPDF fits, each represented by a sample of Monte Carlo replicas.
This distance estimator was first introduced in Ref.~\cite{Ball:2010de}, and
can be used both to assess the compatibility between two
PDFs sets, and to test whether two PDF sets are statistically equivalent.
In the course of the present work we have updated the definition of this distance estimator,
as we discuss the details of this new definition below.
We also explain how the distance estimator can be suitably modified to be used
in the validation of the closure test fits.

Given a Monte Carlo sample of $N_{\text{rep}}$ replicas representing the probability distribution
of a given PDF set, $\left\{f^{(k)}\right\}$,
the expectation value of the distribution as a function of $x$ and $Q^2$ is given by
\begin{equation}
  \bar{f}(x,Q^2)\equiv \left< f (x,Q^2) \right>_{\rm rep}
= \frac{1}{N_{\text{rep}}}\sum_k^{N_{\text{rep}}} f^{(k)}(x,Q^2)\, ,
\end{equation}
where the index $(k)$ runs over all the replicas in the sample.
The variance of the sample is estimated as
\begin{equation}
\label{eq:varsample}
  \sigma^2\left[ f (x,Q^2) \right] = \frac{1}{N_{\text{rep}}-1}\sum_k^{N_{\text{rep}}} \left( f^{(k)}(x,Q^2) - \left<f(x,Q^2)\right>_{\rm rep}\right)^2\,.
\end{equation}
The variance of the mean is, in turn, defined in
terms of the variance of the sample by
\begin{equation}
\label{variancemean}
  \sigma^2\left[\left<f(x,Q^2)\right>_{\rm rep}\right] = \frac{1}{N_{\text{rep}}} \sigma^2\left[ f (x,Q^2) \right]\,.
\end{equation}
The variance of the variance itself can be computed using
\begin{equation}
\label{eq:variancevariance}
   \sigma^2\left[\sigma^2\left[f(x,Q^2)\right]\right] =
     \frac{1}{N_{\text{rep}}} \left[ m_4\left[f(x,Q^2)\right]
   - \frac{N_{\text{rep}} - 3}{N_{\text{rep}} - 1}
     \left( \sigma^2\left[f(x,Q^2)\right] \right)^2 \right]\,,
\end{equation}
where $m_4\left[f(x,Q^2)\right]$ denotes the fourth moment of the probability
distribution for $f(x,Q^2)$, namely
\begin{equation}
  m_4\left[f(x,Q^2)\right]=\frac{1}{N_{\mathrm{rep}}}
  \left[\sum_{k=1}^{N_\mathrm{rep}}\left(f^{(k)}(x,Q^2)- 
  \left< f (x,Q^2) \right>_{\rm rep} \right)^4\right]\,.
\end{equation}

Given the definitions above, the distance between two
sets of PDFs, each characterized by a given distribution of the
Monte Carlo replicas, denoted by
$\left\{f^{(k)}\right\}$ and $\left\{g^{(k)}\right\}$,
 can be defined as the square
root of the square difference of the PDF central values in
units of the uncertainty of the mean, that is
\begin{equation}
  d_{\bar{f},\bar{g}}(x,Q^2) =
  \sqrt{\frac{\left(\bar{f}-\bar{g} \right)^2}
  {\sigma^2 \left[\bar{f}\right] + \sigma^2 \left[\bar{g}\right]}}\,.
  \label{eq:CVdistance}
\end{equation}
In Eq.~(\ref{eq:CVdistance}), the denominator uses
the variance of the mean of the distribution, defined
as in Eq.~(\ref{variancemean}).
An analogous distance can be defined for the variances of
the two samples:
\begin{equation}
   d_{\sigma\left[ f\right],\sigma\left[ g\right]}(x,Q^2) =
  \sqrt{\frac{\left(\sigma^2\left[ f\right] -
\sigma^2\left[ g\right]\right)^2}
  {\sigma^2 \left[\sigma^2 \left[f\right]\right] + \sigma^2 \left[\sigma^2 \left[g\right]\right]}}\,.
  \label{eq:Vardistance}
\end{equation}
where now in the denominator we have the variance of the variance,
Eq.~(\ref{eq:variancevariance}).

The distances for the central values
and for the variances defined in Eqs.~(\ref{eq:CVdistance}) and~(\ref{eq:Vardistance})
test whether the underlying distributions
from which the two Monte Carlo samples
$\left\{f^{(k)}\right\}$ and $\left\{g^{(k)}\right\}$ are drawn have respectively
the same mean and the same standard deviation.
In particular, it is possible to show that one expects these distances to fluctuate around
$d \sim 1$ if the two samples do indeed come from the same
distribution.
On the other hand, values of the distances around $d\sim \sqrt{N_\mathrm{rep}}$
indicates that the central values (the variances) of the two PDF sets differ by one standard deviation
in units of the variance of the distribution Eq.~(\ref{eq:varsample})
(in units of the variance of the variance Eq.~(\ref{eq:variancevariance})).

In this paper, when producing the distances between two NNPDF sets,
Eqs.~(\ref{eq:CVdistance}) and~(\ref{eq:Vardistance}),
we always compare sets of $N_{\rm rep}=100$ replicas, and therefore
a value of the distance around $d\sim 10$ indicates that two sets
differ by one standard deviation (this applies both to central values
and to variances).
The PDFs are sampled at the scale $Q_0^2=2$ GeV$^2$ for $50$ points in $x$ equally
spaced in logarithmic scale in the interval [$10^{-5}$, $0.1$] and then 50 more
points linearly spaced over the interval [$0.1$, $0.95$].

The main difference between the new definition of the distance estimators
and the previous one introduced in Ref.~\cite{Ball:2010de}, and used in previous
NNPDF publications, is that we have now removed the
additional bootstrap sampling of the distances distributions.
For this reason, with the new definition,  distances can become arbitrarily small,
for instance when the central values of the two distributions coincide.
Other than this, the interpretation of the distance plots is very similar
both with the old and new definitions.

The new definition for the distance introduced above is used in this paper
whenever we are comparing two NNPDF fits.
In the context of closure tests described in Sect.~\ref{sec:closure},  it is useful
to use a slightly different definition.
The motivation is that we want to compare the closure test fitted
PDFs with the input PDF, and therefore
we need to remove the scaling factor $1/{\sqrt{N_{\rep}}}$, which is only
required when comparing two Monte Carlo sets.
In addition, in the denominator we should include only the variance of the fitted
PDFs, since for the input PDF only the central value is used in the definition
of the pseudo data.
Therefore, when comparing closure test fitted PDFs with the corresponding
input PDF in Sect.~\ref{sec:closure},
we define the distance of the fitted PDF set, $f_\mathrm{fit}$, with respect to the initial PDF set
used for generating the pseudodata, $f_\mathrm{in}$, as follows
\begin{equation}
\label{eq:distct}
  d_{\rm ct}\left[f_{i,\mathrm{fit}},f_{i,\mathrm{in}}\right](x,Q) \equiv
  \sqrt{\frac{\left(\bar{f}_{i,\mathrm{fit}}(x,Q) - f_{i,\mathrm{in}}(x,Q) \right)^2}
  {\sigma^2 \left[ f_{i,\mathrm{fit}} \right](x,Q)}}\, .
\end{equation}
With this definition, the distance Eq.~(\ref{eq:distct}) between the closure test PDF and the underlying theory is measured in units of the standard deviation of the fitted PDFs.
In this work, the distances
between closure test fits
will be always calculated at the initial parametrisation scale of $Q^2=1$ GeV$^2$.


\providecommand{\href}[2]{#2}\begingroup\raggedright\endgroup

\end{document}